\documentclass[11pt]{article}
\usepackage{makeidx}
\usepackage{amsfonts}
\usepackage{amssymb}
\usepackage{amsmath}
\usepackage{geometry}
\usepackage{appendix}

\DeclareMathSizes{10.95}{10}{7}{6}
\begin{document}

\title{Financial Interactions and Collective States: \\
Part II. Banks, Investors and Firms}
\author{Pierre Gosselin\thanks{%
Pierre Gosselin : Institut Fourier, UMR 5582 CNRS-UGA, Universit\'{e}
Grenoble Alpes, BP 74, 38402 St Martin d'H\`{e}res, France.\ E-Mail:
Pierre.Gosselin@univ-grenoble-alpes.fr} \and A\"{\i}leen Lotz\thanks{%
A\"{\i}leen Lotz: Cerca Trova, BP 114, 38001 Grenoble Cedex 1, France.\
E-mail: a.lotz@cercatrova.eu}}
\date{October 2025}
\maketitle

\begin{abstract}
In a previous paper, we applied a field formalism to analyze capital
allocation and accumulation within a microeconomic framework of investors
and firms. The financial connections were modeled by a field of stakes,
representing the links between agents. We showed that the resulting
collective states were composed of interconnected groups of agents defined
by their connections, their returns and disposable capital. However, within
this framework, the collective states exhibited structural instability, as
capital shortages in specific sectors could trigger cascades of defaults.

The present model refines this framework by introducing a third type of
agent, banks, a type of investor that can create money through loans. We
show that money creation neither eliminates systemic instability nor
prevents the emergence of defaults. In fact, the effect of banks on system
stability and defaults is ambiguous: When banks favor firms over investors,
money creation stabilizes the system by providing the necessary capital to
prevent initial defaults, whereas when banks favor investors over firms,
investors' influence is strengthened, potentially amplifying instability and
defaults. Moreover, regardless of whether they favor investors or firms,
banks may facilitate the propagation of defaults once they have started.
Ultimately, because banks are themselves investors, the emergence of highly
capitalized, high-return banks can directly generate instability in the
system.

Beyond these mechanisms, the analysis reveals the structural limits of
macroprudential regulation. Highly capitalized, high-return investors and
banks may appear more diversified and resilient, yet they constitute the
primary source of endogenous instability. The model thus highlights that
systemic fragility is inherent to the very structure of financial
interdependence and capital flows.


\end{abstract}

\section{Introduction}

In a series of papers\footnote{%
See Gosselin and Lotz (2020--2024).}, we developed a field formalism to
describe how multiple collective states can emerge from a landscape of
microeconomic interactions. These collective states constitute structural
equilibria that shape individual dynamics and may appear or disappear
through transitions between states.

In our earlier work (Gosselin and Lotz 2025), we analyzed the diffusion of
capital across sectors between two types of agents, firms and investors,
with endogenized connections---namely, their investment stakes. In this
setting, agents tend to organize into several weakly interacting groups of
investors. The size and shape of these groups depend on cross-sectoral
investment uncertainty, while their stability depends on the degree of
homogeneity in the capital distribution within each group.

The present paper extends this analysis by introducing a third type of
agent: banks. Banks represent a distinctive class of investors that may hold
stakes in other banks, investors, and firms, while also being able to create
money through loans. To investigate the impact of endogenous participation
and lending, we follow the same methodology as in Gosselin and Lotz (2025).
We develop a field model describing the connections among the three types of
agents---investors, banks, and firms. Collective states continue to emerge
from the network of agents' connections, now determined by the optimization
of each agent's objective function. In terms of fields, this introduces two
additional fields---one for investors and one for banks---whose variables
are the stakes, and whose respective action functionals are derived from the
underlying microeconomic setup\footnote{%
See Gosselin and Lotz (2024) for the translation method.}. Minimizing these
two action functionals, together with the investors' return equations from
Gosselin and Lotz (2024), yields the collective states of the system.

In the initial framework of Gosselin and Lotz (2024), banks impacted
stability in two opposing ways: they stabilized the system while
simultaneously increasing investors' leverage. The balance between these
effects depended on the relative size of the banking sector within each
group.

Reducing the types of agents to solely investors and firms, Gosselin and
Lotz (2025) showed that investors split into two categories, high- and
low-return investors, with high-return investors holding larger capital
endowments and ultimately demanding higher returns, and some firms led to
default for lack of capital, thereby triggering potential contagion.
However, the introduction of banks may modify this picture.

The results of the present paper refine and extend the analysis of Gosselin
and Lotz (2024) and (2025). We recover the infinite possibilities of
collective states, each collective state organized into groups that may
themselves exhibit multiple sub-collective states, but also the ambiguous
role played by banks in the system.

Indeed, banks reduce disparities between investors but may nonetheless
generate instability when they channel their resources primarily toward
investors rather than firms. Moreover, banks themselves divide into high-
and low-return groups, thereby shifting disparities to the banking level. In
this setting, instability remains possible, together with associated
defaults, but for different reasons linked to the perception of risk by
banks, or any factor directing capital primarily toward investors rather
than directly to firms.

This paper is organized into six main parts. Following a brief literature
review (Section 2), Part I introduces the basic notations, including general
notations in Section 3 and the description of investments and stakes in
Section 4. Part II, the general setup, presents the field return equations,
the fields of stakes, and the modeling of uncertainty (Sections 5--7). Part
III formalizes the equations of the field model, combining the return
equations (Section 8), the field equations (Section 9), and the uncertainty
equations (Section 10). Part IV, the resolution of the model, details the
methodology (Section 11), analyzes the no-default scenario (Section 12), and
presents the default states (Section 13). Part V focuses on the stability of
sub-collective states, examining their internal dynamics (Section 14),
internal stability (Section 15), and possible transitions between them
(Section 16). Finally, Part VI synthsizes the results (Section 17), and
discusses the results (Section 18).\ Section 19 concludes.

\section{Literature review}

Four major directions are related to our approach.

The first direction addresses heterogeneity among agents through
distributions of agents modeled by probability densities. In mean field
games (MFG) and mean field control, individual agents are negligible
compared with the population but interact through aggregate variables (see
Bensoussan, Frehse, and Yam (2018) and Lasry and Lions (2007, 2010a,
2010b)).\ This approach has been applied in the dynamic programming
framework developed by Gomes, Vilanova, and others (see Gomes et al., 2015;
Achdou et al., 2014). Heterogeneous Agent New Keynesian (HANK) models
incorporate similar heterogeneity into macroeconomic structures.\ An
equilibrium probability distribution is derived from a set of optimizing
heterogeneous agents\ in a new Keynesian context (see Kaplan and Violante
(2018), as well as Kaplan, Moll, and Violante (2018) for quantitative
implementations). Information-theoretic approaches build on Sims' (2003,
2006) rational inattention theory to model agents optimizing under
informational constraints (see Yang (2018) and Matejka and McKay (2015)).
This information theoretic approach considers probabilistic states around
the equilibrium and replaces the Walrasian equilibrium with a statistical
equilibrium derived from an entropy maximisation program.\ In these three
types of models,\ probability distributions can be seen as particular types
of collective states postulated \textit{a priori} as equilibria of the
microeconomics setup.

Field economics, on the contrary, builds on the interactions between agents
at the microeconomic level. We do not postulate an equilibrium probability
distribution for each type of agent.\ Rather we consider any possible
probability density for the entire system of $N$ agents and their
interactions, and translate these probability densities in terms of fields.
Since the fields encompass all the possible densities of agents as their
realizations\footnote{%
In our formalism, the notion of fields refers to some abstract complex
functions defined on the state space and is similar to the
second-quantized-wave functions of quantum theory.}, the state-space in
field economics is thus much larger than those considered in the above
approaches. This allows us to study the agents' economic structural
relations and the emergence of the collective states induced by these
specific micro-relations, which in turn impact each agent's stochastic
dynamics at the microeconomic level. These emerging collective states are in
general multiple with their own characteristics, average quantities, and
distributions of agents, etc.

Interacting agents with heterogeneous behavioral rules have been dealt with
by multi-agent systems, particularly agent-based models (ABMs), with an
emphasis on non-equilibrium dynamics and bounded rationality (Gaffard and
Napoletano (2012), Delli Gatti et al. (2005)). Mandel, Landini, and
Gallegati (2010, 2012) further develop ABMs to capture innovation, sectoral
dynamics, and macroeconomic fluctuations. The field of economic networks,
notably Jackson (2010, 2014), focuses on the structural properties of agent
interactions within economic systems. Both approaches are highly numerical
and model-dependent.\ They also rely on microeconomic relations, such as 
\textit{ad hoc }reaction functions, that may be too simplistic. Field
economics, on the contrary, accounts for transitions between scales.
Macroeconomic patterns do not emerge solely from the dynamics of a large
population of agents: they are grounded in behaviours and interaction
structures. Describing these structures in terms of field theory allows for
the emergence of phases at the macro scale, and the study of their impact at
the individual level.

Econophysics applies statistical physics methods to socio-economic systems
(see Abergel et al. (2011a, 2011b) for reviews of the field's developments).
Lux (2008, 2016) explores stylized facts and agent-based dynamics, while
Kleinert (2009) uses path integral formulations to analyze financial
markets. Other relevant works include Bardoscia et al. (2017), Bouchaud and M%
\'{e}zard (2000), Chakraborti et al. (2011), Chakraborti et al. (2013).
However, Econophysics does not apply the full potential of field theory to
economic systems.\ Instead, it seeks to reveal empirical laws in economic
systems. Field economics, in contrast, keeps track of usual microeconomic
concepts, such as utility functions, expectations, and forward-looking
behaviors.\ It integrates these behaviours into the analytical treatment of
multi-agent systems by translating the main characteristics of optimizing
agents in terms of statistical systems. Closer to our approach, Bardoscia et
al. (2017) study a general equilibrium model for a large economy in the
context of statistical mechanics, and show that phase transitions may occur
in the system. Our issue is similar, but our use of field theory addresses a
larger class of dynamic models.

The second direction addresses interactions between Finance and the Real
Economy. The interactions between financial frictions and capital
accumulation have typically been modeled within DSGE frameworks, enriched
with credit constraints, incomplete markets, or firm-level heterogeneity.
Cochrane (2006) provides a review of asset pricing in macro models.
Bernanke, Gertler, and Gilchrist (1999) model the financial accelerator,
linking firm balance sheets to investment dynamics. Holmstrom and Tirole
(1997) study liquidity provision under moral hazard. More recent work
includes Campello et al. (2010) and Quadrini (2012). Other contributions
extend this framework to heterogeneous agents and endogenous risk (Grassetti
et al. (2022), Grosshans and Zeisberger (2018), B\"{o}hm, Kuehn, and
Schmedders (2008), Khan and Thomas (2013), Monacelli et al. (2011) and Moll
(2014)).

Field economics differs from DSGE models in that their models are micro
models that stand for the entire set of agents.\ This does not allow the
study of the diffusion and circulation of capital among agents. Field
economics, on the contrary, studies the entire system of agents.\ When
dealing with capital, it describes the different states that result from
this diffusion and circulation of capital. Besides, the relative
(un)stability of capital allocation can be assessed globally: it is the
relative distribution of returns or external conditions that determine
investors' allocation of capital to firms. We can study how a local interest
rate or a change in sectoral returns would impact the equilibrium of one
sector, and the whole system.

The third direction covers the literature on default and systemic risk, and
their contagion within financial networks. Early theoretical models were
developed by Allen and Gale (2000), Cifuentes, Ferrucci, and Shin (2005),
and Gai and Kapadia (2010). The network-based approach has been formalized
by Acemoglu, Ozdaglar, and Tahbaz-Salehi (2015) and extended by Bardoscia et
al. (2019) and Glasserman and Young (2015) using graph-theoretic
stress-testing techniques and feedback loop dynamics (see also Battiston et
al. (2012, 2020) for recursive losses in interconnected systems and Haldane
and May (2011) for an ecology and epidemiology approach to financial
instability). Empirically, contributions include Reinhart and Rogoff (2009),
Gennaioli, Martin, and Rossi (2012, 2018), Adrian and Brunnermeier (2016),
Langfield, Liu, and Ota (2020).

This paper adds to these studies by considering firms, investors and banks
in the analysis. Moreover, collective states are described both in terms of
global averages and sectoral quantities,\ allowing disparities in firms'
returns and borrowing conditions across agents to be explicitly accounted
for. Our analysis therefore moves back and forth between the micro and macro
levels to identify the conditions under which micro-level defaults propagate
to the macroeconomic level.

A fourth strand of the literature emphasizes the intermediary role of banks
and the regulatory frameworks---especially macroprudential
policies---designed to stabilize the financial system. It includes theories
of banking (Diamond and Dybvig (1983); Diamond and Rajan (2001); Freixas,
Parigi and Rochet (2000); Freixas and Rochet (2008)) and modern
macro-financial models (Gertler and Kiyotaki (2010); Brunnermeier and
Sannikov (2014); Adrian and Brunnermeier (2016)). Macroprudential tools are
explicitly aimed at reducing the likelihood that idiosyncratic shocks
trigger widespread defaults. Empirical studies and policy evaluations (Borio
and Drehmann (2009), Brunnermeier, Crockett, Goodhart, Persaud, and Shin
(2009); Acharya and Richardson (2012); Galati and Moessner (2013); Drehmann
and Juselius (2014); IMF (2014); Cerutti, Claessens and Laeven (2017))
indicate that well-designed macroprudential measures can lower the
probability of default, although their effectiveness depends on timing,
calibration, and cross-border coordination.

Our work differs from these approaches in that, in our framework,
instability is inherent to collective states, and the apparently more
capitalized, high-yield investors or banks are themselves sources of
systemic fragility. Since risk is an ad-hoc concept rather than an absolute
measure, restricting access to credit to low-return investors and firms in
the name of prudence may in fact strengthen high-return investors and banks,
thereby increasing structural instability.

\part{Basic notations}

\section{General notations}

We briefly recall the notations and the basic assumptions of our model. The
economic space comprises an infinite number of \emph{sectors}. Each sector
is labeled by a position, denoted by $X$, $X^{\prime }$, $X^{\prime \prime }$
and so on. Each sector includes banks, investors and firms.

Since we are working within a field model, agents are, among other things,
characterized by the sector they occupy. Agents are not indexed
individually; rather, they are distinguished by their economic variables and
by their sector. We consider three types of agents: banks, investors and
firms. We denote the banks and investors located in sector $X$ as \emph{bank}
$X$ and \emph{investor }$X$ respectively, and the firms located in sector $X$
as firm $X$. Variables pertaining to banks and investors will be denoted
with a bar and a hat respectively. For any type of agent, the \emph{agent's
disposable capital} is the capital used to produce or invest, denoted $K$.
It is the sum of the agent's private capital along with the capital
allocated to them, either through shares or loans.

The \emph{agents' average disposable capital in sector }$X$, for any type of
agent, is the average of the agents' disposable capital in sector $X$.\ It
is denoted $\bar{K}_{X}$ for banks, $\hat{K}_{X}$ for investors, and $K_{X}$
for firms.\ 

The \emph{aggregate disposable capital in sector} $X$ will be denoted by $%
\bar{K}\left[ X\right] $, $\hat{K}\left[ X\right] $ and $K\left[ X\right] $
for banks, investors and firms, respectively.

We will denote the \emph{background fields}, the fields minimizing the
action functionals of banks, investors and firms $\bar{\Psi}\left( X\right) $%
, $\hat{\Psi}\left( X\right) $ and $\Psi \left( X\right) $,\ respectively.\
The \emph{densities }of banks, investors and of firms $X$ will be written $%
\left\Vert \bar{\Psi}\left( X\right) \right\Vert ^{2}$,$\ \left\Vert \hat{%
\Psi}\left( X\right) \right\Vert ^{2}$and $\left\Vert \Psi \left( X\right)
\right\Vert ^{2}$, respectively. The total number, across all sectors, of
investors and of firms will be denoted $\left\Vert \bar{\Psi}\right\Vert
^{2} $, $\left\Vert \hat{\Psi}\right\Vert ^{2}$ and $\left\Vert \Psi
\right\Vert ^{2}$, respectively. Note that, by construction, the disposable
capital per sector for banks investors and firms\footnote{%
Expressions of $\bar{K}\left[ \hat{X}^{\prime }\right] $, $\hat{K}\left[ 
\hat{X}^{\prime }\right] $\ and $K^{\prime }\left[ X^{\prime }\right] $\ as
functions of returns and shares are given in Appendix 1.} satisfy the
identities:%
\begin{equation*}
\bar{K}\left[ X\right] =\bar{K}_{X}\left\Vert \bar{\Psi}\left( X\right)
\right\Vert ^{2}
\end{equation*}%
\begin{equation*}
\hat{K}\left[ X\right] =\hat{K}_{\hat{X}}\left\Vert \hat{\Psi}\left(
X\right) \right\Vert ^{2}
\end{equation*}%
\begin{equation*}
K\left[ X\right] =K_{X}\left\Vert \Psi \left( X\right) \right\Vert ^{2}
\end{equation*}

\section{Investments and stakes}

In this model, only investors and banks engage in investment. Their
investments or \emph{stakes} can take two forms: equities (i.e., the
purchase of shares) or loans. A bank or an investor is thus characterized by
the stakes it owns. We denote the stakes by $S$, and use the index $\eta $,
which takes the value $E$ for equities and $L$ for loans. We will present
the notations for investors and banks successively.

\subsection{Investors}

Investors may invest either in other investors or in firms, whether within
their own sector or in other sectors.

Investments made by one investor in another investor are denoted $\hat{S}$,
and are the sum of $\hat{S}_{E}$ and $\hat{S}_{L}$, and investments made in
a firm, denoted $S$, which is similarly the sum of $S_{E}$ and $S_{L}$.

In our field model of investment, two sectors must be distinguished: the
sector in which the investor is located (the origin sector of the
investment), and the sector in which the investment is made (the destination
sector).

To differentiate origin and destination sectors, we will use notations such
as $X$ and $X^{\prime }$ , or $X^{\prime }$ and $X^{\prime \prime }$,
respectively.\ When talking about a stake, \$\$ we may write in parenthesis,
on the right, the sector of origin, on the left, the sector of destination,
and in between, when needed, the capital of the beneficiary. For instance,
the stake $\hat{S}_{\eta }\left( X^{\prime },\hat{K}^{\prime },X\right) $ is
read: the stake of type $\eta $ made by investor $X$ (located in sector $X$)
towards investor $X^{\prime }$, who has capital $K^{\prime }$. When capital
is not specified, as in $\hat{S}_{\eta }\left( X^{\prime },X\right) $, we
refer to the average stake from investor $X$ to investor $X^{\prime }$, the
average being taken over the capital of the beneficiaries of the sector $%
X^{\prime }$

An \emph{average stake} is defined as the inter- and intra-sectoral average
over all sectors $X^{\prime }$ and $X$ of the stakes taken by investors in
those sectors. It is denoted by $\left\langle \hat{S}_{\eta }\left(
X^{\prime },X\right) \right\rangle $ and $\left\langle S_{\eta }\left(
X^{\prime },X\right) \right\rangle $.

We will use the term \emph{aggregate} specifically to denote a sum over all
sectors connected to one specific sector.

The \emph{inward aggregate stake} is the total amount of stakes invested by
all investors in the economic space into a given investor or firm in sector $%
X^{\prime }$, relative to the capital available to the investors or firms of
that sector, respectively. It is denoted $\hat{S}_{\eta }\left( X^{\prime
}\right) $ for investors and $S_{\eta }\left( X^{\prime }\right) $ for firms.

The\textbf{\ }\emph{outward aggregate stakes}\textbf{\ }$\left\langle \hat{S}%
_{\eta }\left( X^{\prime },X\right) \right\rangle _{X^{\prime }}$\textbf{\ }%
compute the average stakes taken by an investor $X$\ in investors of all
sectors.

The \emph{average aggregate stake} is the average, across all sectors, of
the aggregate stakes taken in investors and firms in each sector, denoted $%
\left\langle \hat{S}_{\eta }\left( X^{\prime }\right) \right\rangle $, and $%
\left\langle S_{\eta }\left( X^{\prime }\right) \right\rangle $
respectively. Summing both equity and loan investments into investors is
denoted $\left\langle \hat{S}\left( X^{\prime }\right) \right\rangle $ and $%
\left\langle S\left( X^{\prime }\right) \right\rangle $ in firms.

Average stakes are not necessarily equal to average aggregate stakes,
because the latter are measured relative to the capital of the destination
sectors.

The connections between agents are proportions of investment.\ They are
denoted $k$. These factors will be denoted according to the same notation
rules as the stakes, at the exception of the proportion of aggregated stake $%
\hat{S}\left( X^{\prime }\right) \ $that will be denoted $\underline{\hat{k}}%
\left( X^{\prime }\right) $.

The \emph{inverse uncertainty }of investor $X$\ about its stakes in
investors $X^{\prime }$\ and firms $X^{\prime }$\ will be measured by $\hat{w%
}_{\eta }\left( X^{\prime },X\right) $\ and $w_{\eta }\left( X^{\prime
},X\right) $\ respectively. When averaged across sectors $X^{\prime }$, they
measure the \emph{average uncertainty }among investors $X$ about their
stakes. They will be denoted $\hat{w}_{\eta }\left( X\right) $\ and $w_{\eta
}\left( X\right) $.

\subsection{Banks}

Banks may invest either in other banks, in investors or in firms, whether
within their own sector or in other sectors. Investments made by one bank in
another bank are denoted $\bar{S}$, and is the sum of $\bar{S}_{E}$ and $%
\bar{S}_{L}$. Total stakes in investors are written $\hat{S}^{B}$ which is
the sum of $\hat{S}_{E}^{B}$ and $\hat{S}_{L}^{B}$. Investments made in a
firm are denoted $S^{B}$, which is similarly the sum of $S_{E}^{B}$ and $%
S_{L}^{B}$.

The conventions are the same as for investors. When talking about a stake,
two sectors must be distinguished: the sector in which the investor is
located (the origin sector of the investment), and the sector in which the
investment is made (the destination sector). We may write in parenthesis, on
the right, the sector of origin, on the left, the sector of destination, and
in between, when needed, the capital of the beneficiary. For instance, the
stake $\bar{S}_{\eta }\left( X^{\prime },K^{\prime },X\right) $ is read: the
stake of type $\eta $ made by bank $X$ (located in sector $X$) towards bank $%
X^{\prime }$, who has capital $K^{\prime }$. When capital is not specified,
as in $\bar{S}_{\eta }\left( X^{\prime },X\right) $, we refer to the average
stake from bank $X$ to bank $X^{\prime }$, the average being taken over the
capital of the beneficiaries of the sector $X^{\prime }$

An \emph{average stake} is defined as the inter- and intra-sectoral average
over all sectors $X^{\prime }$ and $X$, of the stakes taken by banks in
those sectors. It is denoted by $\left\langle \bar{S}_{\eta }\left(
X^{\prime },X\right) \right\rangle $, $\left\langle \hat{S}_{\eta
}^{B}\left( X^{\prime },X\right) \right\rangle $ and $\left\langle S_{\eta
}^{B}\left( X^{\prime },X\right) \right\rangle $.

We will use the term \emph{aggregate} specifically to denote a sum over all
sectors connected to one specific sector.

The \emph{inward aggregate stake} is the total amount of stakes invested by
all banks in the economic space into a given bank, investor or firm in
sector $X^{\prime }$, relative to the capital available to the banks,
investors or firms of that sector, respectively. It is denoted $\bar{S}%
_{\eta }\left( X^{\prime }\right) $ for banks $\hat{S}_{\eta }^{B}\left(
X^{\prime }\right) $ for investors and $S_{\eta }\left( X^{\prime }\right) $
for firms.

The\textbf{\ }\emph{outward aggregate stakes}\textbf{\ }$\left\langle \bar{S}%
_{\eta }\left( X^{\prime },X\right) \right\rangle _{X^{\prime }}$ $%
\left\langle \hat{S}_{\eta }^{B}\left( X^{\prime },X\right) \right\rangle
_{X^{\prime }}$ compute the average stakes taken by a bank $X$\ in banks and
investors respectively of all sectors.

The \emph{average aggregate stake} is the average, across all sectors, of
the aggregate stakes taken in banks, investors and firms in each sector,
denoted $\left\langle \bar{S}_{\eta }\left( X^{\prime }\right) \right\rangle 
$, $\left\langle \hat{S}_{\eta }^{B}\left( X^{\prime }\right) \right\rangle $%
, and $\left\langle S_{\eta }^{B}\left( X^{\prime }\right) \right\rangle $
respectively. Summing both equity and loan investments is written $%
\left\langle \bar{S}\left( X^{\prime }\right) \right\rangle $, $\left\langle 
\hat{S}^{B}\left( X^{\prime }\right) \right\rangle $ and $\left\langle
S^{B}\left( X^{\prime }\right) \right\rangle $.

The connections between agents are proportions of investment.\ They are
denoted $k$. These factors will be denoted according to the same notation
rules as the stakes, at the exception of the proportion of aggregated stakes 
$\bar{S}\left( X^{\prime }\right) $ and $\hat{S}^{B}\left( X^{\prime
}\right) \ $that will be denoted $\underline{\bar{k}}\left( X^{\prime
}\right) $ and $\underline{\hat{k}}^{B}\left( X^{\prime }\right) $.

The \emph{inverse uncertainty }of bank $X$\ about its stakes in banks $%
X^{\prime }$, investors $X^{\prime }$\ and firms $X^{\prime }$\ will be
measured by $\bar{w}_{\eta }\left( X^{\prime },X\right) $, $\hat{w}_{\eta
}\left( X^{\prime },X\right) $\ and $w_{\eta }\left( X^{\prime },X\right) $\
respectively. When averaged across sectors $X^{\prime }$, they measure the 
\emph{average uncertainty }among investors $X$ about their stakes. They will
be denoted $\bar{w}_{\eta }\left( X\right) $, $\hat{w}_{\eta }\left(
X\right) $\ and $w_{\eta }\left( X\right) $.

\part*{The general set-up}

In Gosselin and Lotz (2024), we designed a field model for banks, investors,
and firms spread within a sector space, the sectors interacting through
exogenous connections. Three fields, one for each type of agent, were used
to derive the banks and investors' field return equation. We extend this
model by endogenizing the connections between agents. This is done by
introducing two fields in the model, the banks' field of stakes and the
investors' field of stakes.

We follow the structure of Gosselin and Lotz (2025) and include banks in the
system, as a particular type of investor that can create monetary capital by
loans to firms and other investors. We first reconsider the return equations
for investors and banks. Then, we rewrite these equations in terms of shares
of investment, or stakes, in order to reinterpret these equations in the
context of fields of stakes.

\section{The field return equations}

We start with the field version of return equations for investors and banks.
The presence of this second type of investor modifies the return equations
for the investors described in Gosselin and Lotz (2024, 2025).

\subsection{Basic formulation}

We denote $\hat{f}\left( X\right) $ the average return of an investor $X$, $%
\bar{f}\left( X\right) $ the average return of a bank $X$, and define the
excess return of these investors $X$ and banks $X$\ with respect to the
interest rate as $\hat{f}\left( X\right) -\bar{r}$ and $\bar{f}\left(
X\right) -\bar{r}$, respectively. We also denote $f\left( X\right) $\ the
return of a firm $X$\ and $f\left( X\right) -\bar{r}$\ its excess.

Since these returns are computed over the disposable capital, each type of
agent - bank, investor and firm, loans included, the banks' excess return
must be scaled by the factor $1+\underline{\bar{k}}_{L}\left( \hat{X}\right) 
$ to account for the impact of loan repayments to other banks and the real
average excess return of a bank $X$ is: 
\begin{equation*}
\frac{\bar{f}\left( X\right) -\bar{r}}{1+\underline{\overline{\bar{k}}}%
_{L}\left( X\right) }
\end{equation*}%
Similarly, we must divide an investor's excess return by the factor:%
\begin{equation*}
1+\underline{\hat{k}}_{L}\left( \hat{X}\right) +\underline{\hat{k}}%
_{L}^{B}\left( \hat{X}\right) +\kappa \left[ \frac{\underline{\hat{k}}%
_{L}^{B}}{1+\bar{k}}\right] \left( X\right)
\end{equation*}%
This factor accounts for the impact of loan repayments to investors and
banks on the overall return of the investor\footnote{%
See Gosselin and Lotz (2024) for details.}. The coefficient $\kappa \left[ 
\frac{\underline{\hat{k}}_{L}^{B}}{1+\bar{k}}\right] \left( X\right) $\
describes the amount of loans from banks.\ It is proportional to the credit
multiplier that measures the level of bank loans. It involves the share of
bank private capital $\frac{\underline{\hat{k}}_{L}^{B}}{1+\bar{k}}$\
multiplied by the money multiplier $\kappa $. Its precise definition will be
given below. The real average excess return of an investor $X$ is therefore:%
\begin{equation*}
\frac{\hat{f}\left( X\right) -\bar{r}}{1+\underline{\hat{k}}_{L}\left(
X\right) +\kappa \left[ \frac{\underline{\hat{k}}_{L}^{B}}{1+\bar{k}}\right]
\left( X\right) }
\end{equation*}%
Similarly, the real excess return of a firm is given by:%
\begin{equation*}
\frac{f\left( X\right) -\bar{r}}{1+\underline{k}_{L}\left( X\right) +\kappa %
\left[ \frac{\underline{k}_{L}^{B}}{1+\bar{k}}\right] \left( X\right) }
\end{equation*}%
It is the return of the firm, computed relative to its disposable capital.
The denominator represents the total volume of loans to the firm.

\subsubsection{Investors' return equation}

The investors' return equation in each sector $X$ relates all the excess
returns of investors and firms across all sectors. Under a no-default
scenario, they can be written as\footnote{%
See Gosselin and Lotz (2024) for details.}:%
\begin{eqnarray}
&&\left( \delta \left( X-X^{\prime }\right) -\frac{\hat{K}\left[ X^{\prime }%
\right] \hat{k}_{E}\left( X^{\prime },X\right) }{1+\underline{\hat{k}}\left(
X^{\prime }\right) +\underline{\hat{k}}_{E}^{B}\left( X^{\prime }\right)
+\kappa \left[ \frac{\underline{\hat{k}}_{L}^{B}}{1+\bar{k}}\right] \left(
X^{\prime }\right) }\right) \left( \frac{\hat{f}\left( X^{\prime }\right) -%
\bar{r}}{1+\underline{\hat{k}}_{L}\left( X^{\prime }\right) +\kappa \left[ 
\frac{\underline{\hat{k}}_{L}^{B}}{1+\bar{k}}\right] \left( X^{\prime
}\right) }\right)  \label{GN} \\
&=&\frac{k_{E}\left( X^{\prime },X\right) K\left[ X\right] }{1+\underline{k}%
\left( X^{\prime }\right) +\underline{k}_{E}^{\left( B\right) }\left(
X^{\prime }\right) +\kappa \left[ \frac{\underline{k}_{L}^{B}}{1+\bar{k}}%
\right] \left( X^{\prime }\right) }\frac{f_{1}\left( X\right) -\bar{r}}{1+%
\underline{k}_{L}\left( X^{\prime }\right) +\kappa \left[ \frac{\underline{k}%
_{L}^{B}}{1+\bar{k}}\right] \left( X^{\prime }\right) }  \notag
\end{eqnarray}%
with:%
\begin{equation*}
\kappa \left[ \frac{\underline{\hat{k}}_{L}^{B}}{1+\bar{k}}\right] \left(
X^{\prime }\right) =\kappa \int \frac{\underline{\hat{k}}_{L}^{B}\left(
X^{\prime },X\right) }{1+\bar{k}\left( X\right) }\bar{K}\left\vert \bar{\Psi}%
\left( \bar{K},X\right) \right\vert ^{2}d\bar{K}dX
\end{equation*}%
\begin{equation*}
\kappa \left[ \frac{\underline{k}_{L}^{B}}{1+\bar{k}}\right] \left(
X^{\prime }\right) =\kappa \int \frac{\underline{k}_{L}^{B}\left( X^{\prime
},X\right) }{1+\bar{k}\left( X\right) }\bar{K}\left\vert \bar{\Psi}\left( 
\bar{K},X\right) \right\vert ^{2}d\bar{K}dX
\end{equation*}%
The excess return: 
\begin{equation*}
\frac{\hat{f}\left( X\right) -\bar{r}}{1+\underline{\hat{k}}_{L}\left(
X\right) +\kappa \left[ \frac{\underline{\hat{k}}_{L}^{B}}{1+\bar{k}}\right]
\left( X\right) }
\end{equation*}%
can be decomposed into two components.\ First, the amount of stakes taken by
an investor $X$ in other investors\footnote{%
The factor $\frac{\hat{K}\left[ X^{\prime }\right] }{1+\underline{\hat{k}}%
\left( X^{\prime }\right) +\underline{\hat{k}}_{1}^{B}\left( X^{\prime
}\right) +\kappa \left[ \frac{\underline{\hat{k}}_{2}^{B}}{1+\bar{k}}\right]
\left( X^{\prime }\right) }$ measures the firm private capital, that is the
part of the disposable capital belonging to the investors. Participations
are proportional to this amount.},\footnote{%
The coefficient $\underline{k}\left( X\right) $ is defined by $\underline{%
\hat{k}}\left( \hat{X}^{\prime }\right) =\hat{k}\left( \hat{X}^{\prime
}\right) \frac{\hat{K}\left[ \hat{X}^{\prime }\right] }{\left\langle \hat{K}%
\right\rangle \left\vert \hat{\Psi}_{0}\left( \hat{X}\right) \right\vert ^{2}%
}$. It represents the ratio of invested capital by investors in a sector
with respect to the level of private capital of firms in this sector. (see
Gosselin and Lotz (2024) for details).
\par
{}}:%
\begin{equation*}
\frac{\hat{k}_{E}\left( X^{\prime },X\right) \hat{K}\left[ X^{\prime }\right]
}{1+\underline{\hat{k}}\left( X^{\prime }\right) }
\end{equation*}%
multiplied by their corresponding returns $\frac{\hat{f}\left( X^{\prime
}\right) -\bar{r}}{1+\underline{\hat{k}}_{L}\left( X^{\prime }\right)
+\kappa \left[ \frac{\underline{\hat{k}}_{L}^{B}}{1+\bar{k}}\right] \left(
X^{\prime }\right) }$; second, the contribution from returns of firms $%
X^{\prime }$ :%
\begin{equation*}
\frac{f\left( X\right) -\bar{r}}{1+\underline{k}_{L}\left( X^{\prime
}\right) +\kappa \left[ \frac{\underline{k}_{L}^{B}}{1+\bar{k}}\right]
\left( X^{\prime }\right) }
\end{equation*}%
multiplied by:%
\begin{equation*}
\frac{k_{E}\left( X^{\prime },X\right) K\left[ X\right] }{1+\underline{k}%
\left( X^{\prime }\right) +\underline{k}_{E}^{\left( B\right) }\left(
X^{\prime }\right) +\kappa \left[ \frac{\underline{k}_{L}^{B}}{1+\bar{k}}%
\right] \left( X^{\prime }\right) }
\end{equation*}%
that computes the share invested by investor $X$. This share is proportional
to $k_{1}\left( X^{\prime },X\right) $ the proportion of connections%
\footnote{%
The coefficient $\underline{k}\left( X\right) $ is defined by:%
\begin{equation*}
\underline{k}\left( X\right) =k\left( X\right) \frac{\hat{K}_{X}\left\vert 
\hat{\Psi}\left( \hat{X}\right) \right\vert ^{2}}{\left\langle
K\right\rangle \left\vert \Psi _{0}\left( X\right) \right\vert ^{2}}
\end{equation*}%
$\ $It represents the ratio of invested capital by investors in a sector
with respect to the level of private capital of firms in this sector.}
between an investor $X$ and a firm $X^{\prime }$, multiplied by the firm's
private capital:%
\begin{equation*}
\frac{k_{E}\left( X^{\prime },X\right) K\left[ X\right] }{1+\underline{k}%
\left( X^{\prime }\right) +\underline{k}_{E}^{\left( B\right) }\left(
X^{\prime }\right) +\kappa \left[ \frac{\underline{k}_{L}^{B}}{1+\bar{k}}%
\right] \left( X^{\prime }\right) }
\end{equation*}%
In this basic formulation, the intersectoral links $k_{E}$, $\hat{k}_{E}$, $%
k_{L}$ and $\hat{k}_{L}$ are modeled as exogenous and non-normalized.
However, investors' return equation can also be expressed in terms of stakes
owned by each agent.

These coefficients represent the total loans provided by the banks in the
space of sectors $X$ to investors $X^{\prime }$ and firms $X^{\prime }$.

\subsubsection{Banks return equation}

The field expression for the investors' return equation were derived under a
no-default scenario in Gosselin and Lotz (2024). They are expressed as
follows:%
\begin{eqnarray}
&&\left( \delta \left( X^{\prime }-X\right) -\frac{\bar{K}\left[ X^{\prime }%
\right] \bar{k}_{E}\left( X^{\prime },X\right) }{1+\underline{\overline{\bar{%
k}}}\left( X^{\prime }\right) }\right) \frac{\bar{f}\left( X^{\prime
}\right) -\left( 1+\kappa \right) \bar{r}}{1+\underline{\overline{\bar{k}}}%
_{L}\left( X^{\prime }\right) }  \label{RGB} \\
&&-\int \frac{\hat{K}\left[ X^{\prime }\right] \underline{\hat{k}}%
_{E}^{B}\left( X^{\prime },X\right) }{1+\underline{\hat{k}}\left( X^{\prime
}\right) +\underline{\hat{k}}_{E}^{B}\left( X^{\prime }\right) +\kappa \left[
\frac{\underline{\hat{k}}_{L}^{B}}{1+\bar{k}}\right] \left( X^{\prime
}\right) }\frac{\hat{f}\left( X^{\prime }\right) -\bar{r}}{1+\underline{\hat{%
k}}_{L}\left( X^{\prime }\right) +\kappa \left[ \frac{\underline{\hat{k}}%
_{L}^{B}\left( X^{\prime }\right) }{1+\bar{k}\left( X\right) }\right] }%
dX^{\prime }  \notag \\
&=&\frac{K\left[ X\right] \underline{k}_{E}^{\left( B\right) }\left(
X^{\prime },X\right) }{1+\underline{k}\left( X^{\prime }\right) +\underline{k%
}_{E}^{\left( B\right) }\left( X^{\prime }\right) +\kappa \left[ \frac{%
\underline{k}_{L}^{\left( B\right) }\left( X^{\prime }\right) }{1+\underline{%
\bar{k}}}\right] }\frac{f\left( X\right) -\bar{r}}{1+\underline{k}_{L}\left(
X^{\prime }\right) +\kappa \left[ \frac{\underline{k}_{L}^{B}}{1+\bar{k}}%
\right] \left( X^{\prime }\right) }  \notag
\end{eqnarray}%
Equation (\ref{RGB}) decomposes the excess returns of banks:%
\begin{equation*}
\frac{\bar{f}\left( X\right) -\left( 1+\kappa \right) \bar{r}}{1+\underline{%
\bar{k}}_{L}\left( X\right) }
\end{equation*}%
into three components: first, the stakes in other banks\footnote{%
The factor $\frac{\bar{K}\left[ \bar{X}^{\prime }\right] }{1+\underline{%
\overline{\bar{k}}}\left( \bar{X}^{\prime }\right) }$ measures the private
capital of a bank, that is the part of the disposable capital belonging to
the banks. Participations are proportional to this quantity. The coefficient 
$\underline{k}\left( X\right) $ is defined by $\underline{\overline{\bar{k}}}%
\left( \bar{X}^{\prime }\right) =\bar{k}\left( \bar{X}^{\prime }\right) 
\frac{\bar{K}\left[ \bar{X}^{\prime }\right] }{\left\langle \bar{K}%
\right\rangle \left\vert \bar{\Psi}_{0}\left( \bar{X}\right) \right\vert ^{2}%
}$.It represents the ratio of invested capital by banks in a sector with
respect to the level of private capital of banks in this sector. (see
Gosselin and Lotz (2024) for details).
\par
{}}:%
\begin{equation*}
\frac{\bar{K}\left[ X^{\prime }\right] \bar{k}_{E}\left( X^{\prime
},X\right) }{1+\underline{\bar{k}}\left( X^{\prime }\right) }
\end{equation*}%
scaled by their respective returns $\frac{\bar{f}\left( X^{\prime }\right) -%
\bar{r}}{1+\underline{\overline{\bar{k}}}_{L}\left( X^{\prime }\right) }$;
second, the banks stakes in investors\footnote{%
The factor $\frac{\hat{K}\left[ \hat{X}^{\prime }\right] }{1+\underline{\hat{%
k}}\left( \hat{X}^{\prime }\right) }$ measures the private capital of an
investor, that is the part of the disposable capital belonging to the
investors. Participations are proportional to this quantity. The coefficient 
$\underline{k}\left( X\right) $ is defined by $\underline{\hat{k}}\left( 
\hat{X}^{\prime }\right) =\hat{k}\left( \hat{X}^{\prime }\right) \frac{\hat{K%
}\left[ \hat{X}^{\prime }\right] }{\left\langle \hat{K}\right\rangle
\left\vert \hat{\Psi}_{0}\left( \hat{X}\right) \right\vert ^{2}}$.It
represents the ratio of invested capital by investors in a sector with
respect to the level of private capital of investors in this sector. (see GL
for details).
\par
Similarly $\underline{\hat{k}}_{\eta }^{B}\left( \hat{X}^{\prime }\right) =%
\hat{k}_{\eta }^{B}\left( \hat{X}^{\prime }\right) \frac{\hat{K}\left[ \hat{X%
}^{\prime }\right] }{\left\langle \hat{K}\right\rangle \left\vert \hat{\Psi}%
_{0}\left( \hat{X}\right) \right\vert ^{2}}$ for $\eta =1,2$, represents the
ratio of invested capital (shares or loans) by banks in a sector with
respect to the level of private capital of investors in this sector.
\par
{}}:%
\begin{equation*}
\frac{\hat{K}\left[ X^{\prime }\right] \underline{\hat{k}}_{E}^{B}\left(
X^{\prime },X\right) }{1+\underline{\hat{k}}\left( X^{\prime }\right) +%
\underline{\hat{k}}_{E}^{B}\left( X^{\prime }\right) +\kappa \left[ \frac{%
\underline{\hat{k}}_{L}^{B}}{1+\bar{k}}\right] \left( X^{\prime }\right) }
\end{equation*}%
multiplied by their return:%
\begin{equation*}
\frac{\hat{f}\left( X^{\prime }\right) -\bar{r}}{1+\underline{\hat{k}}%
_{L}\left( X^{\prime }\right) +\kappa \left[ \frac{\underline{\hat{k}}%
_{L}^{B}\left( X^{\prime }\right) }{1+\bar{k}\left( X\right) }\right] }
\end{equation*}%
and third, the returns from firms:%
\begin{equation*}
\frac{\hat{f}\left( X^{\prime }\right) -\bar{r}}{1+\underline{\hat{k}}%
_{L}\left( X^{\prime }\right) +\kappa \left[ \frac{\underline{\hat{k}}%
_{L}^{B}\left( X^{\prime }\right) }{1+\bar{k}\left( X\right) }\right] }
\end{equation*}%
weighted by the shares held by banks in these firms\footnote{%
The factor $\frac{K\left[ \bar{X}\right] }{1+\underline{k}\left( \hat{X}%
^{\prime }\right) +\underline{k}_{1}^{\left( B\right) }\left( \bar{X}%
^{\prime }\right) +\kappa \left[ \frac{\underline{k}_{2}^{\left( B\right)
}\left( \bar{X}^{\prime }\right) }{1+\underline{\bar{k}}}\right] }$ measures
the private capital of a firm, that is the part of the disposable capital
belonging to the firm. The interpretation of the coefficients arising in the
sum $\underline{k}\left( \hat{X}^{\prime }\right) +\underline{k}_{1}^{\left(
B\right) }\left( \bar{X}^{\prime }\right) +\kappa \left[ \frac{\underline{k}%
_{2}^{\left( B\right) }\left( \bar{X}^{\prime }\right) }{1+\underline{\bar{k}%
}}\right] $ is the same as for investors.}:%
\begin{equation*}
\frac{K\left[ X\right] \underline{k}_{E}^{\left( B\right) }\left( X^{\prime
},X\right) }{1+\underline{k}\left( X^{\prime }\right) +\underline{k}%
_{E}^{\left( B\right) }\left( X^{\prime }\right) +\kappa \left[ \frac{%
\underline{k}_{L}^{\left( B\right) }\left( X^{\prime }\right) }{1+\underline{%
\bar{k}}}\right] }
\end{equation*}%
This share is proportional to a factor $\underline{k}_{E}^{\left( B\right)
}\left( X^{\prime },X\right) $, measuring the connection between investor $X$
and a firm $X^{\prime }$, multiplied by the private capital of the firm%
\footnote{%
The coefficient $\underline{k}\left( X\right) $ is defined by:%
\begin{equation*}
\underline{k}\left( X\right) =k\left( X\right) \frac{\hat{K}_{X}\left\vert 
\hat{\Psi}\left( \hat{X}\right) \right\vert ^{2}}{\left\langle
K\right\rangle \left\vert \Psi _{0}\left( X\right) \right\vert ^{2}}
\end{equation*}%
It represents the ratio of invested capital by investors in a sector with
respect to the level of private capital of firms in this sector.}:%
\begin{equation*}
\frac{K\left[ X\right] }{1+\underline{k}\left( X^{\prime }\right) +%
\underline{k}_{E}^{\left( B\right) }\left( X^{\prime }\right) +\kappa \left[ 
\frac{\underline{k}_{L}^{\left( B\right) }\left( X^{\prime }\right) }{1+%
\underline{\bar{k}}}\right] }
\end{equation*}%
Both return equations for investors and banks involve the average disposable
capitals $\hat{K}_{X^{\prime }}$, $\bar{K}_{X^{\prime }}$ \ $K_{X^{\prime }}$%
, along with the disposable capitals per sector:%
\begin{equation*}
K^{\prime }\left[ X^{\prime }\right] =K_{X^{\prime }}\left\Vert \Psi \left(
X^{\prime }\right) \right\Vert ^{2}\text{, }\hat{K}\left[ X^{\prime }\right]
=\hat{K}_{X^{\prime }}\left\Vert \hat{\Psi}\left( X^{\prime }\right)
\right\Vert ^{2}\text{, }\bar{K}\left[ X^{\prime }\right] =\bar{K}%
_{X^{\prime }}\left\Vert \bar{\Psi}\left( X^{\prime }\right) \right\Vert ^{2}
\end{equation*}%
Following Gosselin and Lotz (2025), we will rewrite the investors' and
banks' equations in terms of stakes.

\subsection{Formulation in terms of stakes}

To study the dynamics of interacting groups of investors, we require an
alternative formulation based on endogenous stakes. In this formulation,
several levels of stakes must be considered: (i) the stakes of an investor
(or bank) $X$ invested in sector $X^{\prime }$; (ii) the outward aggregate
stakes, obtained by summing these stakes over all sectors $X^{\prime }$,
representing the total stakes originating from $X$; and (iii) the inward
aggregate stakes, obtained by summing over all sectors $X$ and weighting by
the disposable capital of each sector $X$, which yields the total stakes
directed toward sector $X^{\prime }$. We will define these three types for
both investors and banks.

\subsubsection{Investors' stakes}

\paragraph{Stakes between two sectors}

We define the stakes invested by an investor $X$ in investor $X^{\prime }$
with capital $\hat{K}^{\prime }$ as:%
\begin{equation}
\hat{S}_{\eta }\left( X^{\prime },\hat{K}^{\prime },X\right) =\frac{\hat{K}%
^{\prime }\hat{k}_{\eta }\left( X^{\prime },X\right) \left\vert \hat{\Psi}%
\left( \hat{K}^{\prime },X^{\prime }\right) \right\vert ^{2}}{1+\underline{%
\hat{k}}\left( X^{\prime }\right) +\underline{\hat{k}}_{E}^{B}\left(
X^{\prime }\right) +\kappa \left[ \frac{\underline{\hat{k}}_{L}^{B}}{1+\bar{k%
}}\right] \left( X^{\prime }\right) }  \label{CFF1}
\end{equation}%
and the stakes invested by an investor $X$ in a firm $X^{\prime }$ with
capital $K^{\prime }$ as:%
\begin{equation}
S_{\eta }\left( X^{\prime },K^{\prime },X\right) =\frac{k_{\eta }\left(
X^{\prime },X\right) K^{\prime }\left\vert \Psi \left( K^{\prime },X^{\prime
}\right) \right\vert ^{2}}{1+\underline{k}\left( X^{\prime }\right) +%
\underline{k}_{E}^{\left( B\right) }\left( X^{\prime }\right) +\kappa \left[ 
\frac{\underline{k}_{L}^{B}}{1+\bar{k}}\right] \left( X^{\prime }\right) }
\label{CFF2}
\end{equation}%
The total stakes invested by an investor $X$ in both investors\ $X^{\prime }$
and firms $X^{\prime }$, written $\hat{S}_{\eta }\left( X^{\prime },X\right) 
$ and $S_{\eta }\left( X^{\prime },X\right) $, respectively, will thus be
defined as the integrals $\hat{S}_{\eta }\left( X^{\prime },\hat{K}^{\prime
},X\right) $ and $S_{\eta }\left( X^{\prime },K^{\prime },X\right) $ over $%
\hat{K}^{\prime }$, yielding the expressions:%
\begin{equation*}
\hat{S}_{\eta }\left( X^{\prime },X\right) \equiv \int \frac{\hat{k}_{\eta
}\left( X^{\prime },X\right) \hat{K}^{\prime }\left\vert \hat{\Psi}\left(
X^{\prime }\right) \right\vert ^{2}}{1+\underline{\hat{k}}\left( X^{\prime
}\right) +\underline{\hat{k}}_{E}^{B}\left( X^{\prime }\right) +\kappa \left[
\frac{\underline{\hat{k}}_{L}^{B}}{1+\bar{k}}\right] \left( X^{\prime
}\right) }=\frac{\hat{k}_{\eta }\left( X^{\prime },X\right) \hat{K}_{\hat{X}%
^{\prime }}\left\vert \hat{\Psi}\left( X^{\prime }\right) \right\vert ^{2}}{%
1+\underline{\hat{k}}\left( X^{\prime }\right) +\underline{\hat{k}}%
_{E}^{B}\left( X^{\prime }\right) +\kappa \left[ \frac{\underline{\hat{k}}%
_{L}^{B}}{1+\bar{k}}\right] \left( X^{\prime }\right) }
\end{equation*}%
and:%
\begin{equation*}
S_{\eta }\left( X^{\prime },X\right) \equiv \int \frac{k_{\eta }\left(
X^{\prime },X\right) K^{\prime }\left\vert \Psi \left( K^{\prime },X^{\prime
}\right) \right\vert ^{2}}{1+\underline{k}\left( X^{\prime }\right) +%
\underline{k}_{E}^{\left( B\right) }\left( X^{\prime }\right) +\kappa \left[ 
\frac{\underline{k}_{L}^{B}}{1+\bar{k}}\right] \left( X^{\prime }\right) }%
dK^{\prime }=\frac{k_{\eta }\left( X^{\prime },X\right) K_{X^{\prime
}}\left\vert \Psi \left( X^{\prime }\right) \right\vert ^{2}}{1+\underline{k}%
\left( X^{\prime }\right) +\underline{k}_{E}^{\left( B\right) }\left(
X^{\prime }\right) +\kappa \left[ \frac{\underline{k}_{L}^{B}}{1+\bar{k}}%
\right] \left( X^{\prime }\right) }
\end{equation*}%
The total stakes allocated by sector $X$ in investors $X^{\prime }$, denoted 
$\hat{S}\left( X^{\prime },X\right) $, are given by the sums of shares and
loans invested: 
\begin{equation*}
\hat{S}\left( X^{\prime },X\right) =\hat{S}_{E}\left( X^{\prime },X\right) +%
\hat{S}_{L}\left( X^{\prime },X\right)
\end{equation*}%
and similarly, the total stakes allocated by investors $X$ in firms $%
X^{\prime }$, denoted $S\left( X^{\prime },X\right) $, are given by: 
\begin{equation*}
S\left( X^{\prime },X\right) =S_{E}\left( X^{\prime },X\right) +S_{L}\left(
X^{\prime },X\right)
\end{equation*}%
Since all disposable capital is invested, these quantities satisfy the
constraint:%
\begin{equation}
\int \left( \hat{S}_{R}\left( X^{\prime },X\right) +\hat{S}_{L}\left(
X^{\prime },X\right) \right) dX^{\prime }+\int \left( S_{E}\left( X^{\prime
},X\right) +S_{L}\left( X^{\prime },X\right) \right) dX^{\prime }=1
\label{CSt}
\end{equation}

\paragraph{ Stakes according to the investor's sector of origin: outward
aggregate stakes}

The outward aggregate stakes represent the average stakes of an investor $X$
in investors $X^{\prime }$. They are defined by:%
\begin{equation}
\left\langle \hat{S}_{\eta }\left( X^{\prime },X\right) \right\rangle
_{X^{\prime }}  \label{Twn}
\end{equation}%
and:%
\begin{equation}
\left\langle \hat{S}\left( X^{\prime },X\right) \right\rangle _{X^{\prime }}
\label{Tws}
\end{equation}%
where the bracket $\left\langle {}\right\rangle _{X^{\prime }}$ denotes the
average over investors $X^{\prime }$.

\paragraph{Stakes according to their sector of destination: inward aggregate
stakes}

We compute the inward aggregate stakes $\hat{S}_{\eta }\left( X^{\prime
}\right) $, that is the aggregate stakes allocated in investors $X^{\prime }$
with respect to the disposable capital in sector $X^{\prime }$. Similarly,
we compute the inward aggregate stakes allocated in firms of sector $%
X^{\prime }$ with respect to the disposable capital in sector $X^{\prime }$, 
$S_{\eta }\left( X^{\prime }\right) $. They are defined as:%
\begin{eqnarray}
\hat{S}_{\eta }\left( X^{\prime }\right) &=&\int \hat{S}_{\eta }\left(
X^{\prime },X\right) \frac{\hat{K}_{X}\left\vert \hat{\Psi}\left( X\right)
\right\vert ^{2}}{\hat{K}_{X^{\prime }}\left\vert \hat{\Psi}\left( X^{\prime
}\right) \right\vert ^{2}}dX  \label{Gsn} \\
S_{\eta }\left( X^{\prime }\right) &=&\int S_{\eta }\left( X^{\prime
},X\right) \frac{\hat{K}_{X}\left\vert \hat{\Psi}\left( X\right) \right\vert
^{2}}{K_{X^{\prime }}\left\vert \Psi \left( X^{\prime }\right) \right\vert
^{2}}dX  \label{Gst}
\end{eqnarray}%
so that $\hat{S}_{E}\left( X^{\prime }\right) $ and $\hat{S}_{L}\left(
X^{\prime }\right) $ are the proportions of aggregate cross-sectoral equity
or debt invested into investors $\hat{X}^{\prime }$, within the sector
disposable capital $\hat{K}_{\hat{X}^{\prime }}\left\vert \hat{\Psi}\left(
X^{\prime }\right) \right\vert ^{2}$, and $S_{E}\left( X^{\prime }\right) $
and $S_{L}\left( X^{\prime }\right) $ are the same proportions invested into
firms $X^{\prime }$ within their own disposable capital $K_{X^{\prime
}}\left\vert \Psi \left( X^{\prime }\right) \right\vert ^{2}$.

Ultimately, we define the global proportion of aggregate cross-sectoral
equity- or debt- investment in each sector: 
\begin{equation}
\hat{S}\left( X^{\prime }\right) =\hat{S}_{E}\left( X^{\prime }\right) +\hat{%
S}_{L}\left( X^{\prime }\right)  \label{Gsv}
\end{equation}%
and:%
\begin{equation}
S\left( X^{\prime }\right) =S_{E}\left( X^{\prime }\right) +S_{L}\left(
X^{\prime }\right)  \label{Gsw}
\end{equation}

\subsubsection{Banks' stakes}

\paragraph{Stakes between two sectors}

The total stakes invested by banks $X$ in banks $X^{\prime }$, investors $%
X^{\prime }$, and firms $X^{\prime }$, denoted $\bar{S}_{\eta }\left(
X^{\prime },X\right) $, $\hat{S}_{\eta }\left( X^{\prime },X\right) $ and $%
S_{\eta }\left( X^{\prime },X\right) $, respectively, are:%
\begin{equation*}
\bar{S}_{\eta }\left( X^{\prime },X\right) =\int \frac{\bar{K}^{\prime }\bar{%
k}_{\eta }\left( X^{\prime },X\right) \left\vert \bar{\Psi}\left( \bar{K}%
^{\prime },X^{\prime }\right) \right\vert ^{2}}{1+\underline{\bar{k}}\left(
X^{\prime }\right) }d\bar{K}^{\prime }=\frac{\bar{K}_{X^{\prime }}\bar{k}%
_{\eta }\left( X^{\prime },X\right) \left\vert \bar{\Psi}\left( X^{\prime
}\right) \right\vert ^{2}}{1+\underline{\bar{k}}\left( X^{\prime }\right) }
\end{equation*}%
\begin{eqnarray*}
\hat{S}_{E}^{B}\left( X^{\prime },X\right) &=&\frac{\underline{\hat{k}}%
_{E}^{B}\left( X^{\prime },X\right) \hat{K}_{X^{\prime }}\left\vert \hat{\Psi%
}\left( X^{\prime }\right) \right\vert ^{2}}{1+\underline{\hat{k}}\left(
X^{\prime }\right) +\underline{\hat{k}}_{E}^{B}\left( X^{\prime }\right)
+\kappa \left[ \frac{\underline{\hat{k}}_{L}^{B}}{1+\bar{k}}\right] \left(
X^{\prime }\right) } \\
\hat{S}_{L}^{B}\left( X^{\prime },X\right) &=&\frac{\kappa \underline{\hat{k}%
}_{L}^{B}\left( X^{\prime },X\right) }{1+\underline{\bar{k}}\left( X\right) }%
\frac{\hat{K}_{X^{\prime }}\left\vert \hat{\Psi}\left( X^{\prime }\right)
\right\vert ^{2}}{1+\underline{\hat{k}}\left( X^{\prime }\right) +\underline{%
\hat{k}}_{E}^{B}\left( X^{\prime }\right) +\kappa \frac{\underline{\hat{k}}%
_{L}^{B}\left( X^{\prime }\right) }{1+\bar{k}}}
\end{eqnarray*}%
with:%
\begin{equation*}
\left[ \frac{\underline{k}_{L}^{B}}{1+\bar{k}}\right] \left( X^{\prime
}\right) =\int \frac{\underline{k}_{L}^{B}\left( X^{\prime },X\right) }{1+%
\bar{k}\left( X\right) }\bar{K}\left\vert \bar{\Psi}\left( \bar{K},X\right)
\right\vert ^{2}d\bar{K}dX\text{,}
\end{equation*}%
\begin{eqnarray}
S_{E}^{B}\left( X^{\prime },X\right) &=&\frac{K_{X^{\prime }}\underline{k}%
_{E}^{\left( B\right) }\left( X^{\prime },X\right) \left\vert \Psi \left(
X^{\prime }\right) \right\vert ^{2}}{1+\underline{k}\left( X^{\prime
}\right) +\underline{k}_{E}^{\left( B\right) }\left( X^{\prime }\right)
+\kappa \left[ \frac{\underline{k}_{L}^{\left( B\right) }}{1+\underline{\bar{%
k}}}\right] \left( X^{\prime }\right) }  \label{DFr} \\
S_{L}^{B}\left( X^{\prime },X\right) &=&\frac{\kappa \underline{k}%
_{L}^{\left( B\right) }\left( X^{\prime },X\right) }{1+\underline{\bar{k}}%
\left( X\right) }\frac{K_{X^{\prime }}\left\vert \Psi \left( X^{\prime
}\right) \right\vert ^{2}}{1+\underline{k}\left( X^{\prime }\right) +%
\underline{k}_{E}^{\left( B\right) }\left( X^{\prime }\right) +\kappa \left[ 
\frac{\underline{k}_{L}^{\left( B\right) }}{1+\underline{\bar{k}}}\right]
\left( X^{\prime }\right) }  \notag
\end{eqnarray}%
and the global proportion of aggregate cross-sectoral equity- or debt-
investment in banks:%
\begin{equation*}
\bar{S}\left( X^{\prime },X\right) =\bar{S}_{E}\left( X^{\prime },X\right) +%
\bar{S}_{L}\left( X^{\prime },X\right)
\end{equation*}%
These coefficients satisfy the constraints:%
\begin{equation}
\int \left( \bar{S}_{E}\left( X^{\prime },X\right) +\bar{S}_{L}\left(
X^{\prime },X\right) \right) dX^{\prime }+\int \hat{S}_{E}^{B}\left(
X^{\prime },X\right) dX^{\prime }+\int S_{E}^{B}\left( X^{\prime },X\right)
dX^{\prime }=1  \label{CSr}
\end{equation}%
\begin{eqnarray*}
&&\int \hat{S}_{L}^{B}\left( X^{\prime },X\right) dX^{\prime }+\int
S_{L}^{B}\left( X^{\prime },X\right) dX^{\prime } \\
&=&\frac{\kappa }{1+\bar{k}\left( X\right) } \\
&=&\kappa \left( 1-\int \bar{S}\left( X,\bar{Y}\right) \frac{\bar{K}_{\bar{Y}%
}\left\vert \bar{\Psi}\left( \bar{Y}\right) \right\vert ^{2}}{\bar{K}%
_{X}\left\vert \bar{\Psi}\left( X\right) \right\vert ^{2}}d\bar{Y}\right)
\rightarrow \kappa \left( 1-\bar{S}\left( X\right) \right)
\end{eqnarray*}

\paragraph{Stakes according to their sector of origin: outward aggregate
stakes}

The outward aggregate stakes are defined as the average stakes of a bank $X$
in banks $X^{\prime }$ and in investors $X^{\prime }$. They are defined by:%
\begin{equation}
\left\langle \bar{S}_{\eta }\left( X^{\prime },X\right) \right\rangle
_{X^{\prime }}\text{, }\left\langle \hat{S}_{\eta }^{B}\left( X^{\prime
},X\right) \right\rangle _{X^{\prime }}  \label{TWr}
\end{equation}%
and:%
\begin{equation}
\left\langle \bar{S}\left( X^{\prime },X\right) \right\rangle _{X^{\prime }}%
\text{, }\left\langle \hat{S}^{B}\left( X^{\prime },X\right) \right\rangle
_{X^{\prime }}  \label{Twh}
\end{equation}%
where the bracket $\left\langle {}\right\rangle _{X^{\prime }}$ denotes the
average over investors $X^{\prime }$.

\paragraph{Stakes according to their sector of destination: inward aggregate
stakes}

The inward aggregate stakes $\bar{S}_{\eta }\left( X^{\prime }\right) $
allocated in banks of sector $X^{\prime }$ with respect to their disposable
capital, The aggregate stakes $\bar{S}_{\eta }\left( X^{\prime }\right) $
allocated by banks in investors $X^{\prime }$ with respect to their sector\
disposable capital \ and the aggregate stakes $S_{\eta }\left( X^{\prime
}\right) $ allocated by banks in firms $X^{\prime }$ with respect to their
disposable capital in sector $X^{\prime }$ are, respectively:%
\begin{eqnarray*}
\bar{S}_{\eta }\left( X^{\prime }\right) &=&\int \bar{S}_{\eta }\left(
X^{\prime },X\right) \frac{\bar{K}_{X}\left\vert \bar{\Psi}\left( X\right)
\right\vert ^{2}}{\bar{K}_{X^{\prime }}\left\vert \bar{\Psi}\left( X^{\prime
}\right) \right\vert ^{2}}dX \\
\bar{S}\left( X^{\prime }\right) &=&\bar{S}_{E}\left( X^{\prime }\right) +%
\bar{S}_{L}\left( X^{\prime }\right)
\end{eqnarray*}%
\begin{eqnarray*}
\hat{S}_{\eta }^{B}\left( X^{\prime }\right) &=&\int \frac{\hat{S}_{\eta
}^{B}\left( X^{\prime },X\right) \bar{K}_{X}\left\vert \bar{\Psi}\left(
X\right) \right\vert ^{2}}{K_{X^{\prime }}\left\vert \Psi \left( X^{\prime
}\right) \right\vert ^{2}}dX \\
\hat{S}^{B}\left( X^{\prime }\right) &=&\hat{S}_{E}^{B}\left( X^{\prime
}\right) +\hat{S}_{L}^{B}\left( X^{\prime }\right)
\end{eqnarray*}%
\begin{eqnarray*}
S_{\eta }^{B}\left( X^{\prime }\right) &=&\int \frac{S_{\eta }^{B}\left(
X^{\prime },X\right) \bar{K}_{X}\left\vert \bar{\Psi}\left( X\right)
\right\vert ^{2}}{K_{X^{\prime }}\left\vert \Psi \left( X^{\prime }\right)
\right\vert ^{2}}dX \\
S^{B}\left( X^{\prime }\right) &=&S_{E}^{B}\left( X^{\prime }\right)
+S_{L}^{B}\left( X^{\prime }\right)
\end{eqnarray*}

\subsubsection{Return equation in terms of stakes}

Assuming\footnote{%
As in Gosselin and Lotz (2024).} that investors invest in neighbouring
firms, we can write:%
\begin{eqnarray*}
S_{E}\left( X^{\prime },X\right) &=&S_{E}\left( X,X\right) \delta \left(
X^{\prime }-X\right) \\
S_{L}\left( X^{\prime },X\right) &=&S_{L}\left( X,X\right) \delta \left(
X^{\prime }-X\right)
\end{eqnarray*}%
where $\delta \left( X^{\prime }-X\right) $\ is the Dirac function. It is
equal to $0$\ for $X^{\prime }\neq X$. Under these assumptions, the
constraint (\ref{CSt}) simplifies as:%
\begin{equation*}
\int \left( \hat{S}_{E}\left( X^{\prime },X\right) +\hat{S}_{L}\left(
X^{\prime },X\right) \right) dX^{\prime }+S_{E}\left( X,X\right)
+S_{L}\left( X,X\right) =1
\end{equation*}%
Similarly, we assume that for banks:%
\begin{eqnarray*}
S_{E}^{B}\left( X^{\prime },X\right) &=&S_{E}^{B}\left( X,X\right) \delta
\left( X^{\prime }-X\right) \\
S_{L}^{B}\left( X^{\prime },X\right) &=&S_{L}^{B}\left( X,X\right) \delta
\left( X^{\prime }-X\right)
\end{eqnarray*}%
and the constraint (\ref{CSr}) simplifies as:%
\begin{equation*}
\int \left( \hat{S}_{E}\left( X^{\prime },X\right) +\hat{S}_{L}\left(
X^{\prime },X\right) \right) dX^{\prime }+S_{E}\left( X,X\right)
+S_{L}\left( X,X\right) =1
\end{equation*}%
\begin{equation*}
\int \left( \bar{S}_{E}\left( X^{\prime },X\right) +\bar{S}_{L}\left(
X^{\prime },X\right) \right) dX^{\prime }+\int \hat{S}_{E}^{B}\left(
X^{\prime },X\right) dX^{\prime }+S_{E}^{B}\left( X,X\right) =1
\end{equation*}

\paragraph{Investors' return equation}

The return equation (\ref{GN}) can thus be written as\footnote{%
See Appendix 2 for details.}:%
\begin{equation}
0=\int \left( \delta \left( X^{\prime }-X\right) -\hat{S}_{E}\left(
X^{\prime },X\right) \right) \widehat{DF}\left( X^{\prime }\right) \hat{R}%
_{exc}\left( X^{\prime }\right) dX^{\prime }-S_{E}\left( X,X\right)
R_{exc}\left( X\right)  \label{QDL}
\end{equation}%
where $\widehat{DF}\left( X\right) $ is the investor's discount factor for
debt repayment to investors and banks and is defined as:%
\begin{equation*}
\widehat{DF}\left( X\right) =\frac{1-\left( \hat{S}\left( X\right) +\hat{S}%
_{E}^{B}\left( X\right) +\hat{S}_{L}^{B}\left( X\right) \right) }{1-\left( 
\hat{S}_{E}\left( X\right) +\hat{S}_{E}^{B}\left( X\right) \right) }
\end{equation*}%
and where investors' and firms' excess returns are respectively defined by:%
\begin{eqnarray*}
\hat{R}_{exc}\left( X\right) &=&\hat{f}\left( X\right) -\hat{r}\left(
X\right) \\
R_{exc}\left( X\right) &=&f\left( X\right) -\bar{r}\left( X\right)
\end{eqnarray*}

\paragraph{Banks' return equation}

Similarly, the banks' return equation (\ref{RGB}) without default can be
written as:%
\begin{eqnarray}
0 &=&\int \left( \delta \left( X^{\prime }-X\right) -\bar{S}_{E}\left(
X^{\prime },X\right) \right) \overline{DF}\left( X^{\prime }\right) \bar{R}%
_{exc}\left( X^{\prime }\right)  \label{SCn} \\
&&-\int \hat{S}_{E}^{B}\left( X^{\prime },X\right) \widehat{DF}\left(
X^{\prime }\right) \hat{R}_{exc}\left( X^{\prime }\right) -S_{E}^{B}\left(
X^{\prime },X\right) R_{exc}\left( X\right)  \notag
\end{eqnarray}%
where $\overline{DF}\left( X\right) $ is the bank's discount factor for debt
repayment and is defined as:%
\begin{equation*}
\overline{DF}\left( X\right) =\frac{1-\bar{S}\left( X\right) }{1-\bar{S}%
_{E}\left( X\right) }
\end{equation*}%
The banks' excess returns, $\bar{R}_{exc}\left( X^{\prime }\right) $, are
given by:%
\begin{equation*}
\bar{R}_{exc}\left( X^{\prime }\right) =\bar{f}\left( X^{\prime }\right)
-\left( 1+\kappa \right) \bar{r}
\end{equation*}%
The full equations with default\footnote{%
Derived in Appendix 2.} will be presented later in the text. Note that, in
the following, we will also encounter the firm's discount factor for debt
repayment: 
\begin{equation*}
DF\left( X\right) =\frac{1-\left( S\left( X\right) +S^{B}\left( X\right)
\right) }{1-S_{E}\left( X\right) -S_{E}^{B}\left( X\right) }
\end{equation*}

\section{The field of stakes}

Now that the return equations are written in terms of stakes, we can
endogenize these variables by considering a field theory for stakes. \ To do
so, we follow the same approach as in Gosselin and Lotz (2025\textbf{)}.\ We
first formulate a micro-level model in which investment decisions, i.e. the
investors' stakes, result from the optimization of a benefit/uncertainty
trade-off. We will then translate this microeconomic framework into a
field-theoretic formalism. This formalism will be characterized by two
fields $\Gamma $ and $\bar{\Gamma}$\ whose arguments will be the investors'
and banks' stakes respectively. To these fields, we associate their field
action functionals, that encompass the micro-level behaviors.

\subsection{Micro setup for endogenous stakes}

Let us consider a set of investors, banks and firms, indexed by Latin
letters to distinguish these types of agents.\ The positions of these agents
within the sector space will be denoted $X_{i}$, $X_{j}$, $X_{k}$...
depending on the number of agents involved in an expression.\ We will
further denote $\hat{S}_{\eta ij}$\ the stakes taken by an investor $i$\ in
another investor $j$,\ and $S_{\eta ik}$\ the stakes taken by an investor $i$%
\ in a firm $k$. The stakes taken by bank $i$ in an other bank $j$ will be
written $\bar{S}_{\eta ij}$, its stakes in investor $j$ and firm $k$ are
denoted $\hat{S}_{\eta ij}^{B}$ and $S_{\eta ik}^{B}$, respectively. This
difference notwithstanding, in what follows, the notation will be similar to
that previously adopted.

\subsubsection{Investors}

In a classical framework, each investor optimizes their investments based on
their respective expected returns, and the uncertainty, or risk, associated
with each investment. Assuming agents adjust instantaneously to any change
in their environment, an investor indexed by $j$ maximizes the objective
function:%
\begin{equation*}
\sum_{j}\hat{S}_{Eij}\hat{f}_{j}+\sum_{j}\hat{S}_{Lij}\hat{r}_{j}-\frac{1}{2}%
\sum \frac{\left( \hat{S}_{\eta ij}\right) ^{2}}{\hat{w}_{\eta _{i}}\left( 
\hat{X}_{j}\right) }+\sum_{k}S_{Eik}f_{k}+\sum_{k}S_{Lik}\bar{r}_{k}-\frac{1%
}{2}\sum_{k}\frac{\left( S_{\eta ik}\right) ^{2}}{w_{\eta ik}\left(
X_{k}\right) }
\end{equation*}%
where the coefficients $\hat{w}_{\eta }$ and $w_{\eta }$ represent the
inverse uncertainties associated with returns.

This objective function is maximized under the constraint:%
\begin{equation*}
\sum_{j}\left( \hat{S}_{Eij}+\hat{S}_{Lij}\right) +\sum_{k}\left(
S_{Eik}+S_{Lik}\right) =1
\end{equation*}%
which is implemented by a Lagrange multiplier $\lambda _{i}$. The solutions
of the instantaneous optimization are given by:%
\begin{eqnarray*}
\hat{S}_{Eij} &=&\hat{w}_{Eij}\left( \hat{f}_{j}+\lambda _{i}\right) \\
\hat{S}_{Lij} &=&\hat{w}_{Lij}\left( \hat{r}_{j}+\lambda _{i}\right) \\
S_{Eik} &=&w_{Eik}\left( f_{k}+\lambda _{i}\right) \\
S_{Lik} &=&w_{Lik}\left( \bar{r}_{k}+\lambda _{i}\right)
\end{eqnarray*}%
or, taking into account some inertia in the allocation:%
\begin{eqnarray*}
\alpha \frac{d^{2}}{dt^{2}}\hat{S}_{Eij} &=&-\frac{\hat{S}_{Eij}}{\hat{w}%
_{Eij}}+\left( \hat{f}_{j}+\lambda _{i}\right) \\
\alpha \frac{d^{2}}{dt^{2}}\hat{S}_{Lij} &=&-\frac{\hat{S}_{Lij}}{\hat{w}%
_{Lij}}+\left( \hat{r}_{j}+\lambda _{i}\right) \\
\alpha \frac{d^{2}}{dt^{2}}S_{Eik} &=&-\frac{S_{Eik}}{w_{Eik}}+\left( \hat{r}%
_{j}+\lambda _{i}\right) \\
\alpha \frac{d^{2}}{dt^{2}}S_{Lik} &=&-\frac{S_{Lik}}{w_{Lik}}+\left( \hat{r}%
_{j}+\lambda _{i}\right)
\end{eqnarray*}%
with $\alpha <1$ in general\footnote{%
Note that in the optimization equations, the uncertainty coefficients $\hat{h%
}_{\eta }$\ and $h_{\eta }$\ are perceived as exogenous by any individual
agent. We will see below that, in the context of the field description,
these coefficients are in fact endogenous to the system as a whole.}.

\subsubsection{Banks}

Similarly, banks will optimize their investments based on their respective
expected returns, and the uncertainty, or risk, associated with each
investment. However, unlike investors, banks can create money, which leads
us to consider their activities as investor and lender independently. In
terms of loans, banks are not constrained by an overall disposable capital.
Instead, they lend independently of their participations, with the amount of
loans determined solely as a multiple of their private capital.

As investors, we consider that banks take shares in banks, investors and
firms, and grant loans to other banks. We thus consider the objective
function for bank $i$:%
\begin{eqnarray*}
&&\sum_{j}\bar{S}_{Eij}^{B}\bar{f}_{j}+\sum_{j}\bar{S}_{Lij}^{B}\bar{r}_{j}-%
\frac{1}{2}\sum \frac{\left( \sum_{j}\bar{S}_{\eta ij}\right) ^{2}}{\bar{w}%
_{\eta _{i}}\left( X_{j}\right) } \\
&&+\sum_{j}\hat{S}_{Eij}^{B}\hat{f}_{j}-\frac{1}{2}\sum \frac{\left( \sum_{j}%
\hat{S}_{Eij}^{B}\right) ^{2}}{\hat{w}_{Ei}^{B}\left( X_{j}\right) }%
+\sum_{k}S_{Eik}^{B}f_{k}-\frac{1}{2}\sum \frac{\left(
\sum_{k}S_{Eik}^{B}\right) ^{2}}{w_{\eta ik}^{B}\left( X_{k}\right) }
\end{eqnarray*}%
As a lender, the banks are described by the following objective function for
loans:%
\begin{equation*}
\sum_{j}\hat{S}_{Lij}^{B}\hat{r}_{j}-\frac{1}{2}\sum \frac{\left( \sum_{j}%
\hat{S}_{Lij}^{B}\right) ^{2}}{\hat{w}_{\eta _{Li}}\left( X_{j}\right) }%
+\sum_{k}S_{Lik}^{B}\bar{r}_{k}-\frac{1}{2}\sum \frac{\left(
\sum_{k}S_{Lik}^{B}\right) ^{2}}{w_{\eta ik}\left( X_{k}\right) }
\end{equation*}%
where the coefficients $\bar{w}_{\eta _{i}}$ $\hat{w}_{\eta }^{B}$ and $%
w_{\eta }^{B}$ measure the system-dependent return uncertainties.

The solutions of the instantaneous optimization equations are given by:%
\begin{equation*}
\bar{S}_{Eij}=\bar{w}_{Eij}\bar{f}_{j}\text{, }\bar{S}_{Lij}=\bar{w}_{Lij}%
\bar{r}_{j}
\end{equation*}%
\begin{equation*}
\hat{S}_{Eij}^{B}=\hat{w}_{Eij}^{B}\hat{f}_{j}\text{, }\hat{S}_{Lij}^{B}=%
\hat{w}_{Lij}^{B}\hat{r}_{j}
\end{equation*}%
\begin{equation*}
S_{Eik}^{B}=w_{Eik}^{B}f_{k}\text{, }S_{Lik}^{B}=w_{Lik}r_{k}
\end{equation*}%
or, taking into account some inertia in the allocation:%
\begin{eqnarray*}
\alpha \frac{d}{dt}\bar{S}_{Eij} &=&-\bar{S}_{Eij}+\bar{w}_{Eij}\bar{f}_{j}
\\
\alpha \frac{d}{dt}\bar{S}_{Eij} &=&-\bar{S}_{Eij}+\bar{w}_{Eij}\hat{f}_{j}
\\
\alpha \frac{d}{dt}\hat{S}_{Eij}^{B} &=&-\hat{S}_{Eij}^{B}+\hat{w}_{Eij}^{B}%
\hat{f}_{j} \\
\alpha \frac{d}{dt}\hat{S}_{Lij}^{B} &=&-\hat{S}_{Lij}^{B}+\hat{w}_{Lij}^{B}%
\hat{r}_{j} \\
\alpha \frac{d}{dt}S_{Eik} &=&-S_{Eik}^{B}+w_{Eik}^{B}f_{k} \\
\alpha \frac{d}{dt}S_{Lik}^{B} &=&-S_{Lik}^{B}+w_{Lik}^{B}\bar{r}_{k}
\end{eqnarray*}%
with $\alpha <1$ in general.

\subsection{Field translation of the set-up}

To translate the above micro setup into a field representation, we consider
two fields - one for investors, one for banks - whose variables are the
stakes of investors and banks, respectively, together with the positions of
both the agent and their stakes within the sector space.

\subsubsection{Investors}

The investors' field of stakes is: 
\begin{equation*}
\Gamma \left( \hat{S}^{\left( T\right) },X^{\prime },X\right) \equiv \Gamma
\left( S_{E},\hat{S}_{E},S_{L},\hat{S}_{L},X^{\prime },X\right)
\end{equation*}%
where $S_{E}$ and $S_{L}$\ denote the stakes of investor $X$ in firms of the
same sector $X$ through shares and loans, respectively, while $\hat{S}_{E}$
and $\hat{S}_{L}$ denote the stakes in another investor $X^{\prime }$.\ We
gather the four possible types of stakes in a vector $\hat{S}^{\left(
T\right) }$:%
\begin{equation*}
\hat{S}^{\left( T\right) }=\left( S_{E},\hat{S}_{E},S_{L},\hat{S}_{L}\right)
\end{equation*}%
The translation of the micro set-up in terms of fields yields the action
functional $S\left( \Gamma \right) $\footnote{%
See Appendix 2 of Gosselin and Lotz (2025) for the details of the
computation.},\footnote{%
An equivalent formulation of (\ref{GCt}) is given in Appendix 2.2 in
Gosselin and Lotz (2025).}:%
\begin{eqnarray}
S\left( \Gamma \right) &=&-\int \sigma _{\hat{K}}^{2}\Gamma ^{\dag }\left( 
\hat{S}^{\left( T\right) },X^{\prime },X\right) \nabla _{\hat{S}_{\eta
}^{\left( T\right) }}^{2}\Gamma \left( \hat{S}^{\left( T\right) },X^{\prime
},X\right) d\left( \hat{S}^{\left( T\right) },X^{\prime },X\right)
\label{GCt} \\
&&-\int \beta \left\vert \Gamma \left( \hat{S}^{\left( T\right) },X^{\prime
},X\right) \right\vert ^{2}d\left( \hat{S}^{\left( T\right) },X^{\prime
},X\right)  \notag \\
&&+\sum_{\eta }\int \left( \frac{\left( \hat{S}_{\eta }^{\left( T\right)
}\right) ^{2}}{2\hat{w}_{\eta }^{T}\left( X^{\prime },X\right) }-\hat{V}%
_{\eta }\hat{S}_{\eta }^{\left( T\right) }-\beta \right) \left\vert \Gamma
\left( \hat{S}^{\left( T\right) },X^{\prime },X\right) \right\vert
^{2}d\left( \hat{S}^{\left( T\right) },X^{\prime },X\right)  \notag \\
&&+\int \lambda \left( X\right) \left( \sum_{\eta }\int \hat{S}_{\eta
}^{\left( T\right) }\left\vert \Gamma \left( \hat{S}^{\left( T\right)
},X^{\prime },X\right) \right\vert ^{2}dX^{\prime }d\hat{S}^{\left( T\right)
}-1\right) \left\vert \Gamma \left( \hat{S}^{\left( T\right) },X^{\prime
},X\right) \right\vert ^{2}d\left( \hat{S}^{\left( T\right) },X^{\prime
},X\right)  \notag
\end{eqnarray}%
where the components of the vector $\left( \hat{w}_{\eta }^{T}\left(
X^{\prime },X\right) \right) $ are the field-translated (inverse)
uncertainty coefficients $\hat{w}_{\eta ij}$ and $w_{\eta ik}$:%
\begin{equation*}
\left( \hat{w}_{\eta }^{T}\left( X^{\prime },X\right) \right) =\left( \hat{w}%
_{E}\left( X^{\prime },X\right) ,\hat{w}_{L}\left( X^{\prime },X\right)
,w_{E}\left( X,X\right) ,w_{L}\left( X,X\right) \right)
\end{equation*}%
which reflect the uncertainty perceived by investors $X$ regarding their
stakes (equity or loans) in investors $X^{\prime }$\ and firms $X$. The
functions $\hat{V}_{\eta }\left( \hat{S}_{\eta }\right) $ involved in (\ref%
{GCt}) are defined in Appendix 3.\ 

The Lagrange multipliers $\lambda \left( X\right) $ implement the constraint:%
\begin{equation*}
\sum_{\eta }\int \hat{S}_{\eta }^{\left( T\right) }\left\vert \Gamma \left( 
\hat{S}^{\left( T\right) },X^{\prime },X\right) \right\vert ^{2}dX^{\prime }d%
\hat{S}^{\left( T\right) }=1
\end{equation*}%
where the integral sums the stakes of an investor $X$ across various sectors 
$X^{\prime }$, and the factor $\left\vert \Gamma \left( \hat{S}^{\left(
T\right) },X^{\prime },X\right) \right\vert ^{2}$ weights the stakes $\hat{S}%
_{\eta }^{\left( T\right) }$ according to the number of investors $X^{\prime
}$ benefiting from them.

We define the partial averages:%
\begin{eqnarray*}
S_{\eta }\left( X,X\right) &=&\int S_{\eta }\left\vert \Gamma \left( S_{E},%
\hat{S}_{E},S_{L},\hat{S}_{L},X^{\prime },X\right) \right\vert ^{2}d\left(
S_{E},\hat{S}_{E},S_{L},\hat{S}_{L},X^{\prime }\right) \\
\hat{S}_{\eta }\left( X^{\prime },X\right) &=&\int \hat{S}_{\eta }\left\vert
\Gamma \left( S_{E},\hat{S}_{E},S_{L},\hat{S}_{L},X^{\prime },X\right)
\right\vert ^{2}d\left( S_{E},\hat{S}_{E},S_{L},\hat{S}_{L}\right)
\end{eqnarray*}%
so that the allocation constraint can be written as\footnote{%
Appendix 2 in Gosselin and Lotz (2025) derives the minimization equation of (%
\ref{GCt}) in terms of the sectoral averages:%
\begin{equation*}
\hat{S}_{\eta }^{\left( T\right) }\left( \hat{X}^{\prime },\hat{X}\right) =%
\frac{\int \hat{S}_{\eta }^{\left( T\right) }\left\vert \Gamma \left( \hat{S}%
^{\left( T\right) },\hat{X}^{\prime },\hat{X}\right) \right\vert ^{2}d\hat{S}%
^{\left( T\right) }}{\int \left\vert \Gamma \left( \hat{S}^{\left( T\right)
},\hat{X}^{\prime },\hat{X}\right) \right\vert ^{2}d\hat{S}^{\left( T\right)
}}
\end{equation*}%
}:%
\begin{equation*}
\int \left( \hat{S}_{E}\left( X^{\prime },X\right) +\hat{S}_{L}\left(
X^{\prime },X\right) \right) dX^{\prime }+\int \left( S_{E}\left( X^{\prime
},X\right) +S_{L}\left( X^{\prime },X\right) \right) dX^{\prime }=1
\end{equation*}

\subsubsection{Banks}

The field of banks is defined as:%
\begin{equation*}
\bar{\Gamma}\left( \bar{S}^{\left( T\right) },\bar{X}^{\prime },\hat{X}%
^{\prime },\bar{X}\right) =\bar{\Gamma}\left( \bar{S}_{E},S_{E}^{B},\hat{S}%
_{E}^{B},\bar{S}_{L},S_{L}^{B},\hat{S}_{L}^{B},X^{\prime },X^{\prime
},X\right)
\end{equation*}%
with:%
\begin{equation*}
\bar{S}^{\left( T\right) }=\left( \bar{S}_{E},S_{E}^{B},\hat{S}_{E}^{B},\bar{%
S}_{L},S_{L}^{B},\hat{S}_{L}^{B}\right)
\end{equation*}%
The arguments $S_{E}^{B}$ and $S_{L}^{B}$\ are the stakes taken by a bank $X$
in firms of the same sector, through shares and loans respectively, while $%
\hat{S}_{E}^{B}$ and $S_{L}^{B}$ are those taken in an investor $X^{\prime }$%
, and $\bar{S}_{E}$ and $\bar{S}_{L}$ are the stakes taken in a bank $%
X^{\prime }$.

The action functional for banks is given by\footnote{%
See Appendix 3.}:

\begin{eqnarray}
S\left( \bar{\Gamma}\right) &=&-\sigma _{\hat{K}}^{2}\sum_{\eta }\int \bar{%
\Gamma}^{\dag }\left( \bar{S}^{\left( T\right) },X^{\prime },X\right) \nabla
_{\bar{S}_{\eta }^{\left( T\right) }}^{2}\bar{\Gamma}\left( \bar{S}^{\left(
T\right) },X^{\prime },X\right) d\left( \bar{S}^{\left( T\right) },X^{\prime
},X\right)  \label{GT} \\
&&+\int \left( \sum_{\eta }\left( \frac{\left( \bar{S}_{\eta }^{\left(
T\right) }\right) ^{2}}{2\bar{w}_{\eta }\left( X^{\prime },X\right) }-\bar{V}%
_{\eta }\bar{S}_{\eta }^{\left( T\right) }\right) -\beta \right) \left\vert 
\bar{\Gamma}\left( \bar{S}^{\left( T\right) },X^{\prime },X\right)
\right\vert ^{2}d\left( \bar{S}^{\left( T\right) },X^{\prime },X\right) 
\notag \\
&&+\int \lambda \left( \bar{X}\right) \left( \sum_{\eta =E,L}\int \bar{S}%
_{\eta }^{B}\left( X^{\prime },X\right) d\bar{X}^{\prime }+\int \hat{S}%
_{E}^{B}\left( X^{\prime },X\right) d\hat{X}^{\prime }+S_{E}^{B}\left(
X,X\right) -1\right) \left\vert \bar{\Gamma}\left( \bar{S}^{\left( T\right)
},X^{\prime },X\right) \right\vert ^{2}  \notag \\
&&+\int \lambda ^{\prime }\left( \bar{X}\right) \left( \int \hat{S}%
_{L}^{B}\left( X^{\prime },X\right) d\hat{X}^{\prime }+S_{L}^{B}\left(
X,X\right) -\kappa \left( 1-\bar{S}_{\eta }^{B}\left( \bar{X}\right) \right)
\right) \left\vert \bar{\Gamma}\left( \bar{S}^{\left( T\right) },X^{\prime
},X\right) \right\vert ^{2}  \notag
\end{eqnarray}%
where we defined the various averages in the states defined by the fields:%
\begin{eqnarray*}
\bar{S}_{\eta }\left( X^{\prime },X\right) &=&\int \bar{S}_{\eta }\left\vert 
\bar{\Gamma}\right\vert ^{2}d\left( \bar{S}^{\left( T\right) },X^{\prime
}\right) \\
S_{\eta }^{B}\left( X,X\right) &=&\int S_{\eta }^{B}\left\vert \bar{\Gamma}%
\right\vert ^{2}d\left( \bar{S}^{\left( T\right) },X^{\prime }\right) \\
\hat{S}_{\eta }^{B}\left( X^{\prime },X\right) &=&\int \hat{S}_{\eta
}^{B}\left\vert \bar{\Gamma}\right\vert ^{2}d\left( \bar{S}^{\left( T\right)
},X^{\prime }\right)
\end{eqnarray*}%
and the coefficients:%
\begin{equation*}
\bar{w}_{E}^{T}\left( X^{\prime },X\right) =\bar{w}_{E}\left( X^{\prime
},X\right) \text{, }\bar{w}_{L}^{T}\left( X^{\prime },X\right) =\bar{w}%
_{L}\left( X^{\prime },X\right)
\end{equation*}%
\begin{equation*}
\hat{w}_{3}^{T}\left( X^{\prime },X\right) =\hat{w}_{E}\left( X^{\prime
},X\right) \text{, }\bar{w}_{4}^{T}\left( X^{\prime },X\right) =\hat{w}%
_{L}\left( X^{\prime },X\right)
\end{equation*}%
\begin{equation*}
\hat{w}_{5}^{T}\left( X^{\prime },X\right) =w_{E}\left( X,X\right) \text{, }%
\bar{w}_{6}^{T}\left( X^{\prime },X\right) =w_{L}\left( X,X\right)
\end{equation*}%
are the field translation of the inter- and intra-sectoral uncertainty
coefficients, which were defined at the micro level.

The Lagrange multipliers $\lambda \left( X\right) $ and $\lambda ^{\prime
}\left( X\right) $ enforce the allocation constraints for the stakes of
banks:%
\begin{equation*}
\sum_{\eta }\int \bar{S}_{\eta }^{B}\left( X^{\prime },X\right) dX^{\prime
}+\int \hat{S}_{E}^{B}\left( X^{\prime },X\right) dX^{\prime
}+S_{E}^{B}\left( X,X\right) =1
\end{equation*}%
\begin{equation*}
\int \hat{S}_{L}^{B}\left( X^{\prime },X\right) dX^{\prime }+S_{L}^{B}\left(
X,X\right) =\kappa \left( 1-\bar{S}_{\eta }^{B}\left( X\right) \right)
\end{equation*}%
\begin{equation*}
\bar{S}^{\left( T\right) }\left( X^{\prime },X\right) \left\vert \Gamma
\left( X^{\prime },X\right) \right\vert ^{2}=\int \bar{S}^{\left( T\right)
}\left\vert \Gamma \left( \bar{S}^{\left( T\right) },X^{\prime },X\right)
\right\vert ^{2}
\end{equation*}%
The functions $\bar{V}_{\eta }\left( \bar{S}\right) $, involved in the
expression of $S\left( \bar{\Gamma}\right) $ are defined in Appendix 3.

\section{Modeling the uncertainty}

In Gosselin and Lotz (2024), investments risk was supposed to be exogenous
and the nature of risk was unspecified. To fully endogenize the model, we
must now fill this gap by deriving the investment risk coefficients $\hat{w}%
\left( X^{\prime },X\right) $, $w\left( X\right) $, $\bar{w}\left( \hat{X}%
^{\prime },\hat{X}\right) $, $\hat{w}^{B}\left( X^{\prime },X\right) $ and $%
w^{B}\left( X\right) $ from some additional assumptions about risk
propagation\footnote{%
To be fully precise, the coefficients $\hat{w}\left( \hat{X}^{\prime },\hat{X%
}\right) $ and $w\left( \hat{X}\right) $ are a measure of the inverse of
risk.}. Because investors invest in in one another, investment risk is
inherently non-local, and emerges endogenously from the intermediation
structure: risk is composed through layers of exposure and depends on the
investment decisions of others.

\subsection{Nature of uncertainty}

In the present model, firms' returns are observed by all investors, albeit
with a delay. This lag generates anticipation errors, thereby introducing
uncertainty in the model.

All investors are assumed to have a minimal and uniform level of uncertainty
regarding firms within their own sector. That is, investors will always be
more confident about firms in their own sector than firms in other sectors.
Furthermore, no investor is assumed more confident in their own judgment
than any other investor within the same sector. However, this assumption can
be easily relaxed.

Uncertainty is measured by the variance an investor assigns to his
investment. It depends on the model's parameters and on the investor's
allocation choices. Uncertainty may relate, on the one hand, to other
investors' potential returns: these are \emph{errors of appraisal} about
other investors' strategy, acumen, etc. On the other hand, they can relate
to firms' returns: these are investors' \emph{errors of anticipation }about
firms, first within their own sector --- which is set minimal and uniform by
assumption --- and second about firms in other sectors. Finally, uncertainty
may extend to the overall stability of the whole set of returns.

We assume that investors' uncertainty regarding firms within their own
sector is minimal and uniformly distributed across those firms, i.e. that
the average anticipation error regarding intra-sector firms is on average
zero. In contrast, both the evaluation errors investors from one sector make
about the returns of investors in other sectors, and their anticipation
errors regarding these returns, depend on the model's parameters and the
system's current state.

\subsection{Structure of investment risk}

We analyze the risk structure for investors and banks in turn. Starting from
the return equation of each agent, we derive the investment risk associated
with its stakes. This yields the expressions for the inverse uncertainty
that weight each investment.

\subsubsection{Investors structure of investment risk}

To analyze the interconnected structure of investment risk, we consider the
return equation (\ref{QDL}) in the absence of default:%
\begin{equation*}
\left( \delta \left( X-X^{\prime }\right) -\hat{S}_{E}\left( X^{\prime
},X\right) \right) \widehat{DF}\left( X^{\prime }\right) \hat{R}_{exc}\left(
X^{\prime }\right) =S_{E}\left( X,X\right) R_{exc}\left( X\right)
\end{equation*}%
and its series expansion:%
\begin{eqnarray*}
\hat{R}_{exc}\left( X\right) &=&\left( \widehat{DF}\left( X\right) \right)
^{-1}\left[ S_{E}\left( X,X\right) R_{exc}\left( X\right) \right] \\
&&+\left( \widehat{DF}\left( X\right) \right) ^{-1}\sum_{m}\hat{S}%
_{E}^{m}\left( X^{\prime },X\right) \left[ S_{E}\left( X,X\right)
R_{exc}\left( X^{\prime }\right) \right]
\end{eqnarray*}%
This last equation reveals a diffusion effect: investing in other investors
initiates a chain of increasingly remote investments, in which the return $%
\hat{f}\left( X^{\prime }\right) $ of investor $X^{\prime }$ comes with an
ever increasing risk to investor $X$. Assuming investment risk increases
multiplicatively along each investment path, the risk associated with each
path, written $\rho _{p}$ can be represented as the product of local risks
along the path of length $m$:%
\begin{eqnarray}
&&\rho _{p}\left( \left( \widehat{DF}\left( X\right) \right) ^{-1}\hat{S}%
_{E}^{m}\left( \left( X^{\prime }\right) ^{\prime },X^{\prime }\right) \left[
\hat{S}_{E}\left( \left( X^{\prime }\right) ^{\prime }\right) R_{exc}\left(
\left( X^{\prime }\right) ^{\prime }\right) \right] \right)  \label{NC} \\
&\rightarrow &\zeta ^{2}\left( \hat{w}_{E}^{\left( 0\right) }\left( \left(
X^{\prime }\right) ^{\prime },X_{m-1}\right) ...\hat{w}_{E}^{\left( 0\right)
}\left( X_{1},X^{\prime }\right) \right) ^{-1}\hat{S}_{E}^{2m}\left(
X^{\prime },X\right)  \notag
\end{eqnarray}%
where the term $\zeta ^{2}$ denotes the variance of the term:%
\begin{equation*}
\hat{S}_{E}\left( \left( X^{\prime }\right) ^{\prime }\right) R_{exc}\left(
\left( X^{\prime }\right) ^{\prime }\right)
\end{equation*}%
and represents the uncertainty attached to a direct investment into a firm $%
X^{\prime }$.\ This parameter $\zeta ^{2}$ will be considered constant in
first-order approximation to focus on the propagation of risk rather than on
the intrinsic variance of firm-level returns.

The coefficients $\hat{w}_{E}^{\left( 0\right) }\left( X_{1},X^{\prime
}\right) $ are the local investment risk between neighboring agents. They
depend on the distance between sectors, but in first approximation, we will
consider them constant.

We further assume that investment risk is additive for disconnected paths,
up to a normalization factor, so that first-order estimate of the total
investment risk $\rho \left( \hat{X},\hat{f}\left( X^{\prime }\right)
\right) $ can be obtained by summing the contributions of all distinct paths.%
\begin{eqnarray}
\rho \left( \hat{X},\hat{f}\left( X^{\prime }\right) \right)
&=&\sum_{paths}\rho _{p}  \label{Ch} \\
&\rightarrow &\sum \zeta ^{2}\left( \hat{w}_{E}^{\left( 0\right) }\left(
X^{\prime },X_{m-1}\right) ...\hat{w}_{E}^{\left( 0\right) }\left( X_{m-1},%
\hat{X}\right) \right) ^{-1}\hat{S}_{E}^{2m}\left( X^{\prime },X\right) 
\notag
\end{eqnarray}%
Using these assumptions and under a minimal investment condition, agents tend%
\footnote{%
See Appendices 3.2 and 3.3 in Gosselin and Lotz (2025).} to self-organize
into relatively closed investment groups.\ 

In this setting, and assuming that the uncertainty beared by shares and
loans are identical\footnote{%
See Gosselin and Lotz (2025) and section 5.2 for details.}, an investor $X$\
investing directly in firms $X$\ faces an uncertainty $\zeta ^{2}$\ that
depends on the intrinsic risk of the firms and on the shares invested by the
investor. When investor $X$ chooses to invest in an investor $X^{\prime }$,
the risk it faces\ is given by:%
\begin{equation}
\widehat{IR}\left( X^{\prime }\right) \zeta ^{2}  \label{Invrisk0}
\end{equation}%
where the expression $\widehat{IR}\left( X^{\prime }\right) $\textbf{\ }%
accounts for the summation of the various chain of investments (\ref{Ch})
across investors $X_{1}$, that writes:%
\begin{equation}
\widehat{IR}\left( X^{\prime }\right) =\frac{1}{\hat{w}_{E}^{\left( 0\right)
}\left( X^{\prime },X\right) }\frac{\left( \gamma \left\langle \hat{S}%
_{E}\left( X_{1},X^{\prime }\right) \right\rangle _{X_{1}}\right) ^{2}}{%
1-\left( \gamma \left\langle \hat{S}_{E}\left( X\right) \right\rangle
\right) ^{2}}  \label{hf}
\end{equation}%
and $\zeta ^{2}$,\ that accounts for the risk pertaining to the firms that
will be the ultimate recipients of these chains of investments.

The coefficient $\gamma $ represents the average uncertainty associated with
distance-dependent investment paths, and is given by: 
\begin{equation*}
\gamma ^{2}\simeq \left( \hat{w}_{E}^{\left( 0\right) }\left( \left(
X^{\prime }\right) ^{\prime },X_{m-1}\right) ...\hat{w}_{E}^{\left( 0\right)
}\left( X_{1},X^{\prime }\right) \right) ^{-\frac{1}{m}}
\end{equation*}%
In formula (\ref{hf}), the term:%
\begin{equation*}
\left( \gamma \left\langle \hat{S}_{E}\left( X\right) \right\rangle \right)
^{2}
\end{equation*}%
describes the risk associated to an average share of $\left\langle \hat{S}%
_{E}\left( X\right) \right\rangle $ between investors. Higher shares
correspond to higher risk and, consequently, to a lower value of the
coefficient $\hat{w}\left( X^{\prime },X\right) $. The coefficient $\hat{w}%
_{E}^{\left( 0\right) }\left( X^{\prime },X\right) $\ is a factor of local
inverse uncertainty: it is a factor of confidence of investors $X$\ in
investors $X^{\prime }$. The higher this factor, the higher the investments
in sector $X^{\prime }$. As such, this coefficient captures the
characteristics of local uncertainty that induce deviations from average
behavior.

Comparing the risks of investing in firms and investors, Bayes' rule yields
the inverse risk coefficient for investors $X$ investing in investors $%
X^{\prime }$:%
\begin{equation}
\hat{w}\left( X^{\prime },X\right) =\frac{1}{1+\widehat{IR}\left( X^{\prime
}\right) }  \label{hp}
\end{equation}%
while the risk coefficients for investments in firms are given by:%
\begin{equation}
w\left( X,X\right) =1-\left\langle w\left( X^{\prime },X\right)
\right\rangle _{X^{\prime }}  \label{hc}
\end{equation}%
where the average $\gamma $ is taken over the set of sectors in which sector 
$X$ allocates capital.

To the first approximation, the expanded form for the risk coefficient for
investment in investors is given by:%
\begin{equation}
\hat{w}\left( X^{\prime },X\right) \simeq \frac{2\left( 1-\left( \gamma
\left\langle \hat{S}_{E}\left( X\right) \right\rangle \right) ^{2}\right) 
\hat{w}_{E}^{\left( 0\right) }\left( X^{\prime },X\right) }{1+\hat{w}%
_{E}^{\left( 0\right) }\left( X^{\prime },X\right) \left( 1-\left( \gamma
\left\langle \hat{S}_{E}\left( X\right) \right\rangle \right) ^{2}\right)
+\Delta \left( \gamma \left\langle \hat{S}_{E}\left( X_{1},X^{\prime
}\right) \right\rangle _{X_{1}}\right) ^{2}}  \label{hb}
\end{equation}%
The contribution:\textbf{\ }%
\begin{equation*}
\Delta \left( \gamma \left\langle \hat{S}_{E}\left( X_{1},X^{\prime }\right)
\right\rangle _{\hat{X}_{1}}\right) ^{2}=\left( \gamma \left\langle \hat{S}%
_{E}\left( X_{1},X^{\prime }\right) \right\rangle _{\hat{X}_{1}}\right)
^{2}-\left( \gamma \left\langle \hat{S}_{E}\left( X\right) \right\rangle
\right) ^{2}
\end{equation*}%
in equation (\ref{hb}) is the gap between the risk perception of investor $%
X^{\prime }$\ and that of the market. It increases with the risk perception
of investor $X^{\prime }$, so that the higher this gap, the lower the
coefficient $\hat{w}\left( X^{\prime },X\right) $, and the lower the
investment of investor $X^{\prime }$ in investor $X$.

The functions $\hat{w}$ and $w$\ defined in equations (\ref{hb}) and (\ref%
{hc}) depend on exogeneous parameters but also on the stakes $\hat{S}%
_{E}\left( X^{\prime },X\right) $, $\hat{S}\left( X^{\prime },X\right) $.

\subsubsection{Banks structure of investment risk}

Similarly, for banks, we begin with the return equations of banks:%
\begin{eqnarray}
0 &=&\int \left( \delta \left( X^{\prime }-X\right) -\bar{S}_{E}\left(
X^{\prime },X\right) \right) \overline{DF}\left( X^{\prime }\right) \bar{R}%
_{exc}\left( X^{\prime }\right) \\
&&-\int \hat{S}_{E}^{B}\left( X^{\prime },X\right) \widehat{DF}\left(
X^{\prime }\right) \hat{R}_{exc}\left( X^{\prime }\right) -S_{E}^{B}\left(
X^{\prime },X\right) R_{exc}\left( X\right)  \notag
\end{eqnarray}%
and their series expansion:%
\begin{eqnarray*}
\hat{R}_{exc}\left( X\right) &=&\left( \overline{DF}\left( X\right) \right)
^{-1}\left( \left[ S_{E}\left( X,X\right) R_{exc}\left( X\right) \right]
+\int \hat{S}_{E}^{B}\left( X^{\prime },X\right) \widehat{DF}\left(
X^{\prime }\right) \hat{R}_{exc}\left( X^{\prime }\right) \right) \\
&&+\left( \overline{DF}\left( X\right) \right) ^{-1}\sum_{m}\hat{S}%
_{E}^{m}\left( X^{\prime },X\right) \left( \left[ S_{E}\left( X^{\prime
},X^{\prime }\right) R_{exc}\left( X^{\prime }\right) \right] +\int \hat{S}%
_{E}^{B}\left( X^{\prime \prime },X^{\prime }\right) \widehat{DF}\left(
X^{\prime }\right) \hat{R}_{exc}\left( X^{\prime \prime }\right) \right)
\end{eqnarray*}%
The treatment is similar to that for investors, and we present only the
results here\footnote{%
See Appendix 5 for the derivation.}.

While investing in firms $X^{\prime }$, investors $X^{\prime }$, and banks $%
X^{\prime }$, a bank $X$ faces three types of uncertainties measured by the
coefficients $\bar{w}\left( X^{\prime },X\right) $, $\hat{w}^{B}\left(
X^{\prime },X\right) $ and $w^{B}\left( X,X\right) $. The derivation of
these coefficients is similar to the derivation for investors.

The uncertainty $\xi ^{2}$ measures the risk faced by a bank $X$ when
investing in a firm $X^{\prime }$. It depends both on the intrinsic risk of
the firm, and on the level of bank shares held in that firm.

The uncertainty faced by bank $X$ in investment in investors $X^{\prime }$
is given by:%
\begin{equation}
\zeta ^{2}\widehat{IR^{B}}\left( X^{\prime }\right)  \label{Invrisk}
\end{equation}%
with:%
\begin{equation*}
\widehat{IR^{B}}\left( X^{\prime },X\right) =\frac{1}{\hat{w}_{E}^{\left(
0\right) B}\left( X^{\prime },X\right) }\left( 1+\frac{\left( \gamma
\left\langle \hat{S}_{E}\left( X_{1},X^{\prime }\right) \right\rangle
_{X_{1}}\right) ^{2}}{1-\left( \gamma \left\langle \hat{S}_{E}\left(
X^{\prime },\left( X^{\prime }\right) ^{\prime }\right) \right\rangle
\right) ^{2}}\right)
\end{equation*}%
and $\zeta ^{2}$ represents the average level of uncertainty in firms.

This is the total uncertainty faced by investors in the previous paragraph,
except for the factor:%
\begin{equation*}
\frac{1}{\hat{w}_{E}^{\left( 0\right) B}\left( X^{\prime },X\right) }
\end{equation*}%
measuring the local perception of bank $X$ with respect to risk of investors 
$X^{\prime }$. The total risk factor is weighted by $\zeta ^{2}$ to account
for the global level of risk in firms where investors will invest.

The third type of uncertainty is faced by bank $X$ in investing in banks $%
X^{\prime }$. It is given by:

\begin{equation}
\overline{IRG}\left( X^{\prime },X\right) =\overline{IRM}\left( X^{\prime
}\right) \overline{IR}\left( X^{\prime },X\right)  \label{ncr}
\end{equation}%
with:%
\begin{equation*}
\overline{IR}\left( X^{\prime },X\right) =\frac{1}{\bar{w}_{E}^{\left(
0\right) }\left( X^{\prime },X\right) }\left( 1+\frac{\left( \bar{\gamma}%
\left\langle \bar{S}_{E}\left( X_{1},X^{\prime }\right) \right\rangle
_{X_{1}}\right) ^{2}}{1-\left( \bar{\gamma}\left\langle \bar{S}_{E}\left(
X_{1},X^{\prime }\right) \right\rangle _{X_{1}}\right) ^{2}}\right)
\end{equation*}%
\begin{equation*}
\overline{IRM}\left( X^{\prime }\right) =\bar{\zeta}^{2}\zeta
^{2}\left\langle \widehat{IR^{B}}\left( \left( X^{\prime }\right) ^{\prime
},X^{\prime }\right) \right\rangle _{\left( X^{\prime }\right) ^{\prime
}}+\xi ^{2}
\end{equation*}%
The term $\overline{IR}\left( X^{\prime },X\right) $ represents the level of
uncertainty for a bank investing in banks $X^{\prime }$, including the local
factor $\frac{1}{\bar{w}_{E}^{\left( 0\right) }\left( X^{\prime },X\right) }$%
, which reflects banks $X$ perception of risk in sector $X^{\prime }$. In
the uncertainty formula (\ref{ncr}), this term is multiplied by the factor $%
\overline{IRM}\left( X^{\prime }\right) $ that measures the uncertainty a
bank $X^{\prime }$ will face: a risk $\xi ^{2}$ for its shares in firms and
a risk $\zeta ^{2}\left\langle \widehat{IR^{B}}\left( \left( X^{\prime
}\right) ^{\prime },X^{\prime }\right) \right\rangle _{\left( X^{\prime
}\right) ^{\prime }}$ for bank $X^{\prime }$ investment in investors. This
risk is weighted by a factor $\bar{\zeta}^{2}$ representing the shares
invested in investors by the bank $X^{\prime }$.

The weights of inverse uncertainty between banks for both loans and stakes%
\footnote{%
See section 5.2 for details.} are given by:%
\begin{equation}
\left( \bar{w}\left( X^{\prime },X\right) \right) ^{-1}=1+\frac{1}{2}\left( 
\frac{\overline{IRG}\left( X^{\prime },X\right) }{\widehat{IR^{B}}\left(
X^{\prime },X\right) }+\frac{\overline{IRG}\left( X^{\prime },X\right) }{\xi
^{2}}\right)  \label{cfn}
\end{equation}%
and those for banks investing investors are given by:%
\begin{equation}
\left( \hat{w}_{E}^{B}\left( X^{\prime },X\right) \right) ^{-1}=1+2\frac{%
\widehat{IR^{B}}\left( X^{\prime },X\right) }{\overline{IRG}\left( X^{\prime
},X\right) }+\frac{\widehat{IR^{B}}\left( X^{\prime },X\right) }{\xi ^{2}}
\label{cft}
\end{equation}%
where:%
\begin{equation}
w_{E}^{B}\left( X^{\prime },X\right) =1-\bar{w}\left( X^{\prime },X\right) -%
\hat{w}_{E}^{B}\left( X^{\prime },X\right)  \label{cfv}
\end{equation}%
are the risk of investing in investors $X^{\prime }$ and banks $X^{\prime }$%
, respectively. These risks depend on the level of investments these agents
will themselves realize. These level are weighted by the level of
uncertainty $\gamma $ and $\bar{\gamma}$.

The factor $\overline{IRG}\left( X^{\prime }\right) $ : 
\begin{equation*}
\overline{IRG}\left( X^{\prime }\right) =\left( \frac{\bar{\zeta}^{2}%
\overline{IR}\left( X^{\prime }\right) }{\left\langle \hat{w}_{E}^{\left(
0\right) B}\left( \left( X^{\prime }\right) ^{\prime },X^{\prime }\right)
\right\rangle _{\left( X^{\prime }\right) ^{\prime }}}+\frac{\xi ^{2}}{\zeta
^{2}}\frac{\overline{IR}\left( X^{\prime }\right) }{\widehat{IR^{B}}\left(
X^{\prime }\right) }\right)
\end{equation*}%
stands for the global risk of investing. It combines the relative risk to
invest in a bank $X^{\prime }$ rather than in an investor $X^{\prime }$, $%
\frac{\overline{IR}\left( X^{\prime }\right) }{\widehat{IR^{B}}\left(
X^{\prime }\right) }$, and the general propensity for banks to invest in $%
X^{\prime }$ compared to any other investment $\frac{\bar{\zeta}^{2}\zeta
^{2}\overline{IR}\left( X^{\prime }\right) }{\left\langle \hat{w}%
_{E}^{\left( 0\right) B}\left( \left( X^{\prime }\right) ^{\prime
},X^{\prime }\right) \right\rangle _{\left( X^{\prime }\right) ^{\prime }}}$.

In the above, the coefficient $\bar{\gamma}$ is the average uncertainty
associated with distance-dependent investment paths for banks: 
\begin{equation*}
\bar{\gamma}^{2}\simeq \left( \frac{1}{\bar{w}_{1}^{\left( 0\right) }\left(
\left( X^{\prime }\right) ^{\prime },X_{m-1}^{\prime }\right) ...\bar{w}%
_{1}^{\left( 0\right) }\left( X_{1}^{\prime },X^{\prime }\right) }\right) ^{%
\frac{1}{m}}
\end{equation*}%
the coefficients $\bar{w}_{1}^{\left( 0\right) }\left( X_{1}^{\prime
},X^{\prime }\right) $\ capture the local component of uncertainty perceived
in neighboring banks, whereas $\zeta ^{2}$\ and $\xi ^{2}$\ capture the
global uncertainty in investors and firms returns respectively.Since they
are considered constant across agents, they characterize the two types of
agents' uncertainty.

\part*{The equations of the field model}

Three types of equations characterize the full model with endogenous stakes.
First, the field return equations (\ref{QDL}) and (\ref{SCn}), expressed in
terms of excess return and discount factors. Second, the stakes' field
equation, which we will first derive by minimizing the field of stakes'
action functional, and then express in terms of stakes. And third, the
endogenous equations governing uncertainties, as measured by the
coefficients $\hat{w}_{\eta }\left( X^{\prime },X\right) $, $\hat{w}\left(
X^{\prime },X\right) $, $\bar{w}_{\eta }\left( X^{\prime },X\right) $ and $%
\bar{w}\left( X^{\prime },X\right) $.

\section{The field return equations for investors and banks}

\subsection{Investors' field return equations}

The minimization equations for the investor field are the return equations (%
\ref{QDL}).\ They link the distribution of stakes and the disposable capital
across the various sectors.\ Under a no-default scenario, they are\footnote{%
The default scenario will be examined later, in section 6.}:%
\begin{equation}
0=\int \left( \delta \left( X^{\prime }-X\right) -\hat{S}_{E}\left(
X^{\prime },X\right) \right) \widehat{DF}\left( X^{\prime }\right) \hat{R}%
_{exc}\left( X^{\prime }\right) dX^{\prime }-S_{E}\left( X,X\right)
R_{exc}\left( X\right)  \label{QDM}
\end{equation}%
\ Note that the firm's excess returns $R_{exc}\left( X\right) $ include the
dividends and the share price variations, which themselves depend on
productivity and some exogenous factors\footnote{%
See Gosselin and Lotz (2024).}.

\subsection{Banks return equation}

Similarly, we consider equation (\ref{SCn}) for banks' returns:%
\begin{eqnarray}
0 &=&\int \left( \delta \left( X^{\prime }-X\right) -\bar{S}_{E}\left(
X^{\prime },X\right) \right) \overline{DF}\left( X^{\prime }\right) \bar{R}%
_{exc}\left( X^{\prime }\right)  \label{QDB} \\
&&-\int \hat{S}_{E}^{B}\left( X^{\prime },X\right) \widehat{DF}\left(
X^{\prime }\right) \hat{R}_{exc}\left( X^{\prime }\right) -S_{E}^{B}\left(
X^{\prime },X\right) R_{exc}\left( X\right)  \notag
\end{eqnarray}%
which constitutes, with (\ref{QDM}), the first block of equations for the
model.

\section{The equations for the fields of stakes}

We will first derive the minimization equation for the field of stakes, then
re-express this equation in terms of sectoral stakes in banks, investors and
firms.

\subsection{Stakes' field minimization equation for investors and banks}

\subsubsection{Investors}

The collective states satisfy the minimization equations associated with the
action functional $S\left( \Gamma \right) $. To a first approximation, these
variations in stakes can be neglected\footnote{%
This corresponds to neglecting the gradient terms in $X$ in the minimization
equation.}.\ Moreover, loans and participations in any given investment are
assumed to carry the same level of uncertainty\footnote{%
See Appendix 3.1 in Gosselin and Lotz (2025) for details.}, so that the
uncertainty-dependent coefficients for participations and loans are equal,
so that:%
\begin{equation}
\hat{w}_{E}\left( X^{\prime },X\right) =\hat{w}_{L}\left( X^{\prime
},X\right) =\frac{1}{2}\hat{w}\left( X^{\prime },X\right)  \label{Sn}
\end{equation}%
\begin{equation}
w_{E}\left( X,X\right) =w_{L}\left( X,X\right) =\frac{w\left( X,X\right) }{2}
\label{Sd}
\end{equation}%
The various stakes taken by investors will be proportional to these
coefficients. Assuming capital is fully invested, we can normalize these
coefficients and impose:%
\begin{equation}
\hat{w}_{E}\left( X\right) +\hat{w}_{L}\left( X\right) +w_{E}\left( X\right)
+w_{L}\left( X\right) =1  \label{CTn}
\end{equation}%
where:%
\begin{equation}
\hat{w}_{\eta }\left( X\right) =\int \hat{w}_{\eta }\left( X^{\prime
},X\right) dX^{\prime }  \label{Krf}
\end{equation}%
and:%
\begin{equation}
w_{\eta }\left( X\right) =\int w_{\eta }\left( X^{\prime },X\right)
dX^{\prime }  \label{Krh}
\end{equation}%
Nevertheless, the coefficients depend on the uncertainty in returns, which
themselves are functions of the shares $\underline{\hat{S}}_{\eta }$and $%
S_{\eta }$. The coefficients $\hat{w}_{\alpha }\left( X^{\prime },X\right) $
and $w_{\alpha }\left( X,X\right) $ are therefore endogenous, and the
resolution of the system will thus require a detailed characterization of
these uncertainties.

Using the above simplifying assumptions, the minimization equations for the
field of stakes $\Gamma \left( \hat{S}^{\left( T\right) },X^{\prime
},X\right) $ satisfy\footnote{%
See Gosselin and Lotz (2025).}:%
\begin{eqnarray}
0 &=&\left( \sum_{\eta }\left( -\sigma _{\hat{K}}^{2}\nabla _{\hat{S}_{\eta
}^{\left( T\right) }}^{2}+\frac{\left( \hat{S}_{\eta }^{\left( T\right)
}\right) ^{2}}{2\hat{w}_{\eta }}-\hat{V}_{\eta }\hat{S}_{\eta }^{\left(
T\right) }+\lambda \left( X\right) \left\Vert \Gamma \left( \hat{S}^{\left(
T\right) },X^{\prime },X\right) \right\Vert _{\hat{X}}^{2}\hat{S}_{\eta
}^{\left( T\right) }\right) -\beta \right)  \label{abc} \\
&&\times \Gamma \left( \hat{S}^{\left( T\right) },X^{\prime },X\right) 
\notag
\end{eqnarray}%
with $\lambda \left( X\right) $ a Lagrange multiplier implementing the
constraint (\ref{CTn}) on stakes.

In first approximation, the solutions take the form:%
\begin{equation*}
\Gamma _{0,X^{\prime },X}\left( \hat{S}^{\left( T\right) }\right) =N\exp
\left( -\sum_{\eta }\frac{\left( \hat{S}_{\eta }^{\left( T\right) }-\hat{S}%
_{\eta }^{\left( T\right) }\left( X^{\prime },X\right) \right) ^{2}}{2\sigma
_{\hat{K}}^{2}}\right)
\end{equation*}%
where $N$\ is a normalization factor. Solving the minimization equations
thus reduces to finding the stakes invested by sector $X$\ into sector $%
X^{\prime }$, i.e. the sectoral averages $\hat{S}_{\eta }^{\left( T\right)
}\left( X^{\prime },X\right) $.\ They satisfy\footnote{%
See Appendix 2 in Gosselin and Lotz (2025).}:%
\begin{equation*}
\hat{S}_{\eta }^{\left( T\right) }\left( X^{\prime },X\right) =\hat{w}_{\eta
}\left( X^{\prime },X\right) \left( \hat{V}_{\eta }\left( X^{\prime
},X\right) +\lambda \left( X\right) \right)
\end{equation*}%
and depend on the Lagrange multipliers\footnote{%
See Appendix 2 in Gosselin and Lotz (2025).}.

\subsubsection{Banks}

As for investors, as a first approximation, we assume that loans and
participations in any given investment carry the same level of uncertainty%
\footnote{%
See Appendix 3.1 for details.}, so that the uncertainty-dependent
coefficients for participations and loans are equal, and:%
\begin{equation*}
\bar{w}_{E}\left( X^{\prime },X\right) =\bar{w}_{L}\left( X^{\prime
},X\right) =\frac{1}{2}\bar{w}\left( X^{\prime },X\right)
\end{equation*}%
\begin{equation*}
\hat{w}_{E}^{B}\left( X^{\prime },X\right) =\hat{w}_{L}^{B}\left( X^{\prime
},X\right) =\frac{1}{2}\hat{w}^{B}\left( X^{\prime },X\right) \text{, }%
w_{E}^{B}\left( X,X\right) =w_{L}^{B}\left( X,X\right) =\frac{1}{2}%
w^{B}\left( X,X\right)
\end{equation*}%
The various stakes taken by investors and banks will be proportional to
these coefficients. Assuming capital is fully invested, we can normalize
these coefficients and impose: 
\begin{equation}
\bar{w}_{E}\left( X\right) +\bar{w}_{L}\left( X^{\prime },X\right) +\hat{w}%
_{E}^{B}\left( X^{\prime },X\right) +w_{E}^{B}\left( X,X\right) =1
\label{CTd}
\end{equation}%
where:%
\begin{equation*}
\bar{w}_{E}\left( X\right) =\int \bar{w}_{E}\left( X^{\prime },X\right)
dX^{\prime }
\end{equation*}%
and:%
\begin{equation*}
\bar{w}_{L}\left( X\right) =\int \bar{w}_{L}\left( X^{\prime },X\right)
dX^{\prime }
\end{equation*}%
These coefficients depend on the uncertainty in returns, which themselves
are functions of the shares $\hat{S}_{\eta }$, $S_{\eta }$, $\bar{S}_{\eta }$%
, $\hat{S}_{\eta }^{B}$, $S_{\eta }^{B}$. The coefficients $\hat{w}_{\alpha
}\left( X^{\prime },X\right) $, $w_{\alpha }\left( X,X\right) $, $\bar{w}%
_{\eta }\left( X^{\prime },X\right) $, $\hat{w}_{\eta }^{B}\left( X^{\prime
},X\right) $, $w_{\eta }^{B}\left( X,X\right) $ are therefore endogeneous,
and the resolution of the system will thus require a detailed
characterization of these uncertainties.

The minimization equation for banks is similar to the investors' equation,
but involves the field for banks' stakes $\bar{\Gamma}\left( \bar{S}^{\left(
T\right) },\bar{X}^{\prime },\hat{X}^{\prime },\bar{X}\right) $. We find%
\footnote{%
See Appendix 3 and 4 for the derivation of the following formulas.}:%
\begin{eqnarray}
&&0=-\sigma _{\hat{K}}^{2}\sum_{\eta }\nabla _{\bar{S}_{\eta }^{\left(
T\right) }}^{2}\bar{\Gamma}\left( \bar{S}^{\left( T\right) },\bar{X}^{\prime
},\hat{X}^{\prime },\bar{X}\right)  \label{bcd} \\
&&+\left( \sum_{\eta }\left( \frac{\left( \bar{S}_{\eta }^{\left( T\right)
}\right) ^{2}}{2\bar{w}_{\eta }}-\bar{V}_{\eta }^{\left( T\right) }\bar{S}%
_{\eta }^{T}+\lambda _{\eta }\left( X\right) \bar{S}_{\eta }^{T}\right)
-\beta \right) \bar{\Gamma}\left( \bar{S}^{\left( T\right) },\bar{X}^{\prime
},\hat{X}^{\prime },\bar{X}\right)  \notag
\end{eqnarray}%
with:%
\begin{eqnarray*}
\lambda _{\eta }\left( X\right) &=&\lambda \left( X\right) \text{ for }\eta
=1,2,3,5 \\
\lambda _{\eta }\left( X\right) &=&\lambda ^{\prime }\left( X\right) \text{
for }\eta =5,6
\end{eqnarray*}%
are Lagrange multiplier implementing the constraints (\ref{CTn}) and (\ref%
{CTd}). The solutions to the minimization equations have the form\footnote{%
See Appendix 4.}:%
\begin{equation*}
\Gamma _{0,\bar{X}^{\prime },\hat{X}^{\prime },\bar{X}}\left( \bar{S}%
^{\left( T\right) }\right) \Gamma \left( \bar{X}^{\prime },\hat{X}^{\prime },%
\bar{X}\right)
\end{equation*}%
where:%
\begin{equation*}
\Gamma _{0,\bar{X}^{\prime },\hat{X}^{\prime },\bar{X}}\left( \bar{S}%
^{\left( T\right) }\right) =\bar{N}\exp \left( -\sum_{\eta }\frac{\left( 
\bar{S}_{\eta }^{\left( T\right) }-\bar{S}_{\eta }^{\left( T\right) }\left( 
\bar{X}^{\prime },\hat{X}^{\prime },\bar{X}\right) \right) ^{2}}{2\sigma _{%
\hat{K}}^{2}}\right)
\end{equation*}%
where $\bar{N}$\ is a normalization factor. Solving the minimization
equations thus reduces to finding the stakes invested by sector $X$\ into
sector $X^{\prime }$, i.e.the sectoral averages $\bar{S}_{\eta }^{\left(
T\right) }\left( X^{\prime },X\right) $. We show in Appendix 16 that $\bar{S}%
_{\eta }^{\left( T\right) }\left( \bar{X}^{\prime },\hat{X}^{\prime },\bar{X}%
\right) $ satisfies:%
\begin{equation}
\bar{S}_{\eta }^{\left( T\right) }\left( \bar{X}^{\prime },\hat{X}^{\prime },%
\bar{X}\right) =\bar{w}_{\eta }\left( \bar{X}^{\prime },\hat{X}^{\prime },%
\bar{X}\right) \left( \bar{V}_{\eta }\left( \bar{X}^{\prime },\hat{X}%
^{\prime },\bar{X}\right) +\lambda _{\eta }\left( \hat{X}\right) \right)
\end{equation}%
and depends on the Lagrange multipliers $\lambda _{\eta }\left( \hat{X}%
\right) $.

\subsection{Equations in terms of sectoral stakes}

Solving for the Lagrange multiplier, equations (\ref{abc}) and (\ref{bcd})\
can be written in expanded form, as functions of the sectoral stakes. To do
so, we define several average quantities for investors and banks.

For investors, we define the coefficients $\hat{w}\left( X\right) $\ and $%
w\left( X\right) $\ as the average uncertainties faced by an investor $X$\
over its stakes in other investors and in firms $X$,\ respectively:%
\begin{equation}
\hat{w}\left( X\right) =\int \hat{w}\left( X^{\prime },X\right) dX^{\prime }
\label{Grf}
\end{equation}%
and:%
\begin{equation}
w\left( X\right) =1-\hat{w}\left( X\right)  \label{Grh}
\end{equation}%
We further define $R\left( X\right) $, $\hat{R}\left( X^{\prime }\right) $,
and $\hat{R}$ as the returns from all stakes (shares and loans) in firms $X$%
, in investors $X^{\prime }$, and across all investors, respectively:%
\begin{equation}
R\left( X\right) =\frac{1}{2}\left( f\left( X\right) +r\left( X\right)
\right)  \label{Frl}
\end{equation}%
\begin{equation*}
\hat{R}\left( X^{\prime }\right) =\frac{1}{2}\left( \hat{f}\left( X^{\prime
}\right) +\hat{r}\left( X^{\prime }\right) \right)
\end{equation*}%
and:%
\begin{equation}
\hat{R}_{X}=\frac{1}{2}\left( \left\langle \hat{f}\left( X^{\prime }\right)
\right\rangle _{\hat{w}\left( X\right) }+\left\langle \hat{r}\left(
X^{\prime }\right) \right\rangle _{\hat{w}\left( X\right) }\right)
\label{Frh}
\end{equation}%
The computation of $\hat{R}_{X}$ involves a risk-weighted average of
returns, whose definition, for any function $F\left( X^{\prime }\right) $,
is:%
\begin{equation}
\left\langle F\left( X^{\prime }\right) \right\rangle _{\hat{w}\left(
X\right) }=\frac{\int F\left( X^{\prime }\right) \hat{w}\left( X^{\prime
},X\right) }{\hat{w}\left( X\right) }  \label{Wv}
\end{equation}%
Given these definitions, the expression:\textbf{\ }%
\begin{equation*}
R^{w,\hat{w}}\left( X\right) =\hat{w}\left( X\right) \hat{R}_{X}+w\left(
X\right) R\left( X\right) \mathbf{\ }
\end{equation*}%
computes the return that investor $X$\ can expect from a diversified
investment between firms $X$\ and the market.

Considering banks, we define similarly the coefficients $\bar{w}\left(
X\right) $\ and $\hat{w}_{E}^{B}\left( X\right) $ and $w_{E}^{B}\left(
X\right) $\ as the average uncertainty of bank $X$\ over its stakes in other
banks, investors and firms $X$\ respectively:%
\begin{eqnarray}
\bar{w}\left( X\right) &=&\int \bar{w}\left( X^{\prime },X\right) dX^{\prime
} \\
\hat{w}_{E}^{B}\left( X\right) &=&\int \hat{w}_{E}^{B}\left( X^{\prime
},X\right) dX^{\prime }
\end{eqnarray}%
and:%
\begin{equation}
w_{E}^{B}\left( X\right) =1-\bar{w}\left( X\right) -\hat{w}_{E}^{B}\left(
X\right)
\end{equation}%
We also define the returns from all stakes (shares and loans) taken in banks 
$X^{\prime }$ by a bank $X$ and across all banks, respectively as:%
\begin{equation*}
\bar{R}\left( X^{\prime }\right) =\frac{1}{2}\left( \bar{f}\left( X^{\prime
}\right) +\bar{r}\left( X^{\prime }\right) \right)
\end{equation*}%
and:%
\begin{equation}
\bar{R}_{X}=\frac{1}{2}\left( \left\langle \bar{f}\left( X^{\prime }\right)
\right\rangle _{\bar{w}_{E}}+\left\langle \bar{r}\left( X^{\prime }\right)
\right\rangle _{\bar{w}_{L}}\right)
\end{equation}%
where the risk-weighted average of returns, for any function $F\left(
X^{\prime }\right) $, is:%
\begin{equation}
\left\langle F\left( X^{\prime }\right) \right\rangle _{\bar{w}_{\eta
}\left( X\right) }=\frac{\int F\left( X^{\prime }\right) \bar{w}_{\eta
}\left( X^{\prime },X\right) }{\bar{w}_{\eta }\left( X\right) }
\end{equation}%
Ultimately, the expressions:\textbf{\ }%
\begin{equation}
\bar{R}^{\bar{w},\hat{w}_{E}^{B},w_{E}^{B}}\left( X\right) =\bar{w}\left(
X\right) \bar{R}_{X}+\hat{w}_{E}^{B}\left( X\right) \left\langle \hat{f}%
\left( X^{\prime }\right) \right\rangle _{\hat{w}_{E}}+w_{E}^{B}\left(
X\right) f\left( X\right) \mathbf{\ }  \label{rwe}
\end{equation}%
\begin{equation}
\bar{R}^{\hat{w}_{L}^{B},w_{L}^{B}}\left( X\right) =\hat{w}_{L}^{B}\left(
X\right) \left\langle \hat{r}\left( X\right) \right\rangle +w_{L}^{B}\left(
X\right) r\left( X\right) \mathbf{\ }  \label{rwl}
\end{equation}%
and:%
\begin{equation}
\bar{R}^{\bar{w},\hat{w}^{B},w^{B}}\left( X\right) =\frac{1}{2}\left( \bar{R}%
^{\bar{w},\hat{w}_{E}^{B},w_{E}^{B}}\left( X\right) +\bar{R}^{\hat{w}%
_{L}^{B},w_{L}^{B}}\left( X\right) \right)  \label{rw}
\end{equation}%
compute respectively the return that investor $X$\ can expect from a
diversified investment between firms $X$\ and the market (banks and
investors) and the return due to loans to firms $X$ and investors.

\subsubsection{Stakes taken by investors in investors}

Using the above notations, the stakes taken by investors in other investors
can be expressed as:%
\begin{equation}
\hat{S}_{E}\left( X^{\prime },X\right) =\frac{1}{2}\underline{\hat{S}}\left(
X^{\prime },X\right) +\frac{1}{2}\hat{w}\left( X^{\prime },X\right) \hat{%
\Delta}^{E}\left( X^{\prime },X\right)  \label{Shn}
\end{equation}%
\begin{equation}
\hat{S}_{L}\left( X^{\prime },X\right) =\frac{1}{2}\underline{\hat{S}}\left(
X^{\prime },X\right) +\frac{1}{2}\hat{w}\left( X^{\prime },X\right) \hat{%
\Delta}^{L}\left( X^{\prime },X\right)  \label{Sht}
\end{equation}%
and:%
\begin{equation}
\hat{S}\left( X^{\prime },X\right) =\underline{\hat{S}}\left( X^{\prime
},X\right) +\hat{w}\left( X^{\prime },X\right) \hat{\Delta}\left( X^{\prime
},X\right)  \label{Shv}
\end{equation}%
where:%
\begin{eqnarray*}
\hat{\Delta}^{E}\left( X^{\prime },X\right) &=&\hat{f}\left( X^{\prime
}\right) -R^{w,\hat{w}}\left( X\right) \\
\hat{\Delta}^{L}\left( X^{\prime },X\right) &=&\hat{r}\left( X^{\prime
}\right) -R^{w,\hat{w}}\left( X\right) \\
\hat{\Delta}\left( X^{\prime },X\right) &=&\hat{R}\left( X^{\prime }\right)
-R^{w,\hat{w}}\left( X\right)
\end{eqnarray*}%
and:%
\begin{equation*}
\underline{\hat{S}}_{E}\left( X^{\prime },X\right) =\underline{\hat{S}}%
_{L}\left( X^{\prime },X\right) =\frac{1}{2}\underline{\hat{S}}\left(
X^{\prime },X\right) =\frac{1}{2}\hat{w}\left( X^{\prime },X\right)
\end{equation*}%
As expected, equations (\ref{Shn}) and (\ref{Sht})\ for $\hat{S}_{E}\left(
X^{\prime },X\right) $ and $\hat{S}_{L}\left( X^{\prime },X\right) $
indicate that investments are inversely proportional to the uncertainty
coefficients $\hat{w}\left( X^{\prime },X\right) $, and vary linearly with
the difference between the expected returns on shares, $\hat{f}\left(
X^{\prime }\right) $, or loans, $\hat{r}\left( X^{\prime }\right) $, and the
corresponding risk-weighted returns $R^{w,\hat{w}}\left( X\right) $.

\subsubsection{Stakes taken by investors in firms}

Similarly, the stakes equations taken by investors in firms are derived
using the uncertainty coefficient is $w\left( X\right) $: 
\begin{equation}
S_{E}\left( X,X\right) =\frac{1}{2}\underline{S}\left( X,X\right) +\frac{1}{2%
}w\left( X\right) \Delta ^{E}\left( X\right)  \label{SNp}
\end{equation}%
\begin{equation*}
S_{L}\left( X,X\right) =\frac{1}{2}\underline{S}\left( X,X\right) +\frac{1}{2%
}w\left( X\right) \Delta ^{L}\left( X\right)
\end{equation*}%
and:%
\begin{equation}
S\left( X,X\right) =\underline{S}\left( X,X\right) +w\left( X\right) \left( 
\hat{w}\left( X\right) \Delta \left( X\right) \right)  \label{STp}
\end{equation}%
with:%
\begin{equation*}
\underline{S}_{E}\left( X,X\right) =\underline{S}_{L}\left( X,X\right) =%
\frac{1}{2}\underline{S}\left( X,X\right) =\frac{1}{2}w\left( X,X\right)
\end{equation*}%
and:%
\begin{equation}
\Delta ^{E}\left( X\right) =f\left( X\right) -R^{w,\hat{w}}\left( X\right)
\label{DTp}
\end{equation}%
\begin{equation}
\Delta ^{L}\left( X\right) =\bar{r}\left( X\right) -R^{w,\hat{w}}\left(
X\right)  \label{DTr}
\end{equation}%
\begin{equation}
\Delta \left( X\right) =R\left( X\right) -\hat{R}_{X}  \label{DTs}
\end{equation}%
Here too, the stakes are inversely proportional to the uncertainty of the
investments and depend on the difference between the expected returns of
these shares or loans and the risk-weighted returns expected from the market
and from firms.

The equations (\ref{Shn}), (\ref{Sht}), (\ref{Shv}), (\ref{SNp}) and (\ref%
{STp}) relate the stakes, the returns, and each sector average disposable
capital, along with the coefficients $\hat{w}\left( X^{\prime },X\right) $, $%
w\left( X\right) $ and their averages. The cofficients $\hat{w}\left(
X^{\prime },X\right) $ and $w\left( X\right) $ are partially endogenous,
reflecting subjective uncertainties both between sectors, and between
individual agents.\ As investors invest in one another, these uncertainties
propagate throughout the system. Furthermore, the disposable capital of each
sector depends on the structure of connections, as derived in Gosselin and
Lotz (2024).\ Their specific form is recalled below.

\subsubsection{Stakes taken by banks in banks}

\begin{equation}
\bar{S}_{E}\left( X^{\prime },X\right) =\frac{1}{2}\underline{\bar{S}}\left(
X^{\prime },X\right) +\frac{1}{2}\bar{w}\left( X^{\prime },X\right) \bar{%
\Delta}^{E}\left( X^{\prime },X\right)  \label{SBN}
\end{equation}%
\begin{equation}
\bar{S}_{L}\left( X^{\prime },X\right) =\frac{1}{2}\underline{\bar{S}}\left(
X^{\prime },X\right) +\frac{1}{2}\bar{w}\left( X^{\prime },X\right) \bar{%
\Delta}^{L}\left( X^{\prime },X\right)  \label{SBT}
\end{equation}%
\begin{equation}
\bar{S}\left( X^{\prime },X\right) =\underline{\bar{S}}\left( X^{\prime
},X\right) +\bar{w}\left( X^{\prime },X\right) \bar{\Delta}\left( X^{\prime
},X\right)  \label{SBF}
\end{equation}%
where:%
\begin{eqnarray}
\bar{\Delta}^{E}\left( X^{\prime },X\right) &=&\bar{f}\left( X^{\prime
}\right) -\bar{R}^{\bar{w},\hat{w}_{E}^{B},w_{E}^{B}}\left( X\right)
\label{DBn} \\
\bar{\Delta}^{L}\left( X^{\prime },X\right) &=&\bar{r}\left( X^{\prime
}\right) -\bar{R}^{\hat{w}_{L}^{B},w_{L}^{B}}\left( X\right)  \label{DBt} \\
\bar{\Delta}\left( X^{\prime },X\right) &=&\bar{R}_{X}-\bar{R}^{\bar{w},\hat{%
w}^{B},w^{B}}\left( X\right)  \label{DBr}
\end{eqnarray}%
and:%
\begin{equation*}
\underline{\bar{S}}_{E}\left( X^{\prime },X\right) =\underline{\bar{S}}%
_{L}\left( X^{\prime },X\right) =\frac{1}{2}\underline{\bar{S}}\left(
X^{\prime },X\right) =\frac{1}{2}\bar{w}\left( X^{\prime },X\right)
\end{equation*}

\subsubsection{Stakes taken by banks in investors}

The level of shares and loans of a bank $X$ in investor $X^{\prime }$ is
given by: 
\begin{equation}
\hat{S}_{E}^{B}\left( X^{\prime },X\right) =\underline{\hat{S}}%
_{E}^{B}\left( X^{\prime },X\right) +\hat{w}_{E}^{B}\left( X^{\prime
},X\right) \left( \hat{\Delta}^{E}\right) ^{B}\left( X^{\prime },X\right)
\label{scb}
\end{equation}%
and:%
\begin{equation}
\frac{\hat{S}_{L}^{B}\left( X^{\prime },X\right) }{\kappa \left( 1-\bar{S}%
\left( X\right) \right) }=\underline{\hat{S}}_{L}^{B}\left( X^{\prime
},X\right) +\hat{w}_{L}^{B}\left( X^{\prime },X\right) \left( \hat{\Delta}%
^{L}\right) ^{B}\left( X^{\prime },X\right)  \label{sdb}
\end{equation}%
with:%
\begin{eqnarray*}
\left( \hat{\Delta}^{E}\right) ^{B}\left( X^{\prime },X\right) &=&\hat{f}%
\left( X^{\prime }\right) -\bar{R}^{\bar{w},\hat{w}_{E}^{B},w_{E}^{B}}\left(
X\right) \\
\left( \hat{\Delta}^{L}\right) ^{B}\left( X^{\prime },X\right) &=&\bar{r}%
\left( X^{\prime }\right) -\bar{R}^{\hat{w}_{L}^{B},w_{L}^{B}}\left( X\right)
\end{eqnarray*}%
and:%
\begin{equation*}
\underline{\hat{S}}_{E}^{B}\left( X^{\prime },X\right) =\hat{w}%
_{E}^{B}\left( X^{\prime },X\right)
\end{equation*}%
\begin{equation*}
\underline{\hat{S}}_{L}^{B}\left( X^{\prime },X\right) =\hat{w}%
_{L}^{B}\left( X^{\prime },X\right)
\end{equation*}

\subsubsection{Stakes taken by banks in firms}

Ultimately, the level of stakes of bank $X$ in firm $X$ are given by:%
\begin{equation}
S_{E}^{B}\left( X,X\right) =w_{E}^{B}\left( X\right) +w_{E}^{B}\left(
X\right) \left( \Delta ^{E}\right) ^{B}\left( X\right)  \label{SFN}
\end{equation}%
\begin{equation}
\frac{S_{L}^{B}\left( X,X\right) }{\kappa \left( 1-\bar{S}\left( X\right)
\right) }=w_{L}^{B}\left( X\right) +w_{L}^{B}\left( X\right) \Delta
^{L}\left( X\right)  \label{sdt}
\end{equation}%
and:%
\begin{equation*}
\left( \Delta ^{E}\right) ^{B}\left( X\right) =f\left( X\right) -\bar{R}^{%
\bar{w},\hat{w}_{E}^{B},w_{E}^{B}}\left( X\right)
\end{equation*}%
\begin{equation*}
\Delta ^{L}\left( X\right) =\bar{r}\left( X\right) -R^{w,\hat{w}}\left(
X\right)
\end{equation*}

\subsection{Inward aggregate stakes}

Equations (\ref{QDM}) and (\ref{QDB}) describe the stakes of investors and
banks. These equations also involve inward aggregate stakes, which partially
aggregate the stakes directed toward each sector. The formulas for these
aggregate quantities are presented below\footnote{%
\ See Appendix 9.}.

\subsubsection{Inward aggregate stakes of investors in investors}

The return equations (\ref{QDM}) involve the inward aggregate stakes (\ref%
{Gsn}), (\ref{Gst}), (\ref{Gsv}), and (\ref{Gsw}). Defining the coefficients 
$\hat{w}$\ and $w$\ as the averages over the sector space of the coefficients%
\footnote{%
\ See equations (\ref{Grf}) and (\ref{Grh}).} $\hat{w}\left( X^{\prime
}\right) $\ and $w\left( X\right) $, these inward aggregate stakes can be
expressed as follows:%
\begin{equation}
\hat{S}_{E}\left( X^{\prime }\right) \simeq \left( \frac{1}{2}\underline{%
\hat{S}}\left( X^{\prime }\right) +\frac{1}{2}\hat{w}\left( X^{\prime
}\right) \hat{\Delta}^{E}\left( X^{\prime }\right) \right) \frac{%
\left\langle \hat{K}\right\rangle \left\Vert \hat{\Psi}\right\Vert ^{2}}{%
\hat{K}_{X^{\prime }}\left\vert \hat{\Psi}\left( X^{\prime }\right)
\right\vert ^{2}}  \label{Grstn}
\end{equation}%
\begin{equation}
\hat{S}_{L}\left( X^{\prime }\right) \simeq \left( \frac{1}{2}\underline{%
\hat{S}}\left( X^{\prime }\right) +\frac{1}{2}\hat{w}\left( X^{\prime
}\right) \hat{\Delta}^{L}\left( X^{\prime }\right) \right) \frac{%
\left\langle \hat{K}\right\rangle \left\Vert \hat{\Psi}\right\Vert ^{2}}{%
\hat{K}_{X^{\prime }}\left\vert \hat{\Psi}\left( X^{\prime }\right)
\right\vert ^{2}}  \label{Grstw}
\end{equation}%
and:%
\begin{equation}
\hat{S}\left( X^{\prime }\right) \simeq \left( \underline{\hat{S}}\left(
X^{\prime }\right) +\hat{w}\left( X^{\prime }\right) \hat{\Delta}\left(
X^{\prime }\right) \right) \frac{\left\langle \hat{K}\right\rangle
\left\Vert \hat{\Psi}\right\Vert ^{2}}{\hat{K}_{X^{\prime }}\left\vert \hat{%
\Psi}\left( X^{\prime }\right) \right\vert ^{2}}  \label{Grstd}
\end{equation}%
where $\hat{\Delta}^{E}\left( X^{\prime }\right) $, $\hat{\Delta}^{L}\left(
X^{\prime }\right) $, and $\hat{\Delta}\left( X^{\prime }\right) $ are the
aggregates of $\hat{\Delta}^{E}\left( X^{\prime },X\right) $, $\hat{\Delta}%
^{L}\left( X^{\prime },X\right) $, and $\hat{\Delta}\left( X^{\prime
},X\right) $ over the variable $X$. They can be written as follows:%
\begin{equation}
\hat{\Delta}^{E}\left( X^{\prime }\right) =\hat{f}\left( X^{\prime }\right)
-\left\langle R^{w,\hat{w}}\left( X\right) \right\rangle  \label{dtf}
\end{equation}%
\begin{equation}
\hat{\Delta}^{L}\left( X^{\prime }\right) =\hat{r}\left( X^{\prime }\right)
-\left\langle R^{w,\hat{w}}\left( X\right) \right\rangle  \label{dtr}
\end{equation}%
and:%
\begin{equation}
\hat{\Delta}\left( X^{\prime }\right) =\hat{R}\left( X^{\prime }\right)
-\left\langle R^{w,\hat{w}}\left( X\right) \right\rangle  \label{dtR}
\end{equation}%
with:%
\begin{eqnarray*}
\hat{R} &=&\left\langle \hat{R}_{X}\right\rangle \\
R &=&\left\langle R\left( X\right) \right\rangle
\end{eqnarray*}%
and:%
\begin{equation*}
\left\langle R^{w,\hat{w}}\left( X\right) \right\rangle =\left\langle \hat{w}%
\left( X^{\prime }\right) \right\rangle \hat{R}+\left\langle w\left(
X\right) \right\rangle R
\end{equation*}%
which gives the average of the weighted return $R^{w,\hat{w}}\left( X\right) 
$.

\subsubsection{Inward aggregate stakes of investors in firms}

Similarly, the inward aggregate stakes\ (\ref{Gst}) invested in firms $X$
are given by: 
\begin{equation}
S_{\eta }\left( X\right) =S_{\eta }\left( X,X\right) \frac{\hat{K}%
_{X}\left\vert \hat{\Psi}\left( X\right) \right\vert ^{2}}{K_{X}\left\vert
\Psi \left( X\right) \right\vert ^{2}}  \label{Grstf}
\end{equation}%
thus, using the expressions for the stakes invested in firms (\ref{SNp}) and
(\ref{STp}), we obtain:%
\begin{equation}
S_{E}\left( X\right) =\left( \frac{1}{2}\underline{S}\left( X,X\right) +%
\frac{1}{2}w\left( X\right) \Delta ^{E}\left( X\right) \right) \frac{\hat{K}%
_{X}\left\vert \hat{\Psi}\left( X\right) \right\vert ^{2}}{K_{X}\left\vert
\Psi \left( X\right) \right\vert ^{2}}
\end{equation}%
\begin{equation*}
S_{L}\left( X\right) =\left( \frac{1}{2}\underline{S}\left( X,X\right) +%
\frac{1}{2}w\left( X\right) \Delta ^{E}\left( X\right) \right) \frac{\hat{K}%
_{X}\left\vert \hat{\Psi}\left( X\right) \right\vert ^{2}}{K_{X}\left\vert
\Psi \left( X\right) \right\vert ^{2}}
\end{equation*}%
and:%
\begin{equation}
S\left( X,X\right) =\left( \underline{S}\left( X,X\right) +w\left( X\right)
\left( \hat{w}\left( X\right) \Delta \left( X\right) \right) \right) \frac{%
\hat{K}_{X}\left\vert \hat{\Psi}\left( X\right) \right\vert ^{2}}{%
K_{X}\left\vert \Psi \left( X\right) \right\vert ^{2}}
\end{equation}

\subsubsection{Inward aggregate stakes of banks in banks}

The return equations (\ref{QDB}) involve the inward aggregate stakes (\ref%
{Gsn}), (\ref{Gst}), (\ref{Gsv}), and (\ref{Gsw}). Defining the coefficients 
$\hat{w}$\ \ and $w$\ as the averages over the sector space of the
coefficients\footnote{%
\ See equations (\ref{Grf}) and (\ref{Grh}).} $\hat{w}\left( X^{\prime
}\right) $\ and $w\left( X\right) $, these inward cross-stakes between banks
are given by: 
\begin{equation}
\left\langle \bar{S}_{E}\left( X^{\prime },X\right) \right\rangle _{X}=\frac{%
1}{2}\left\langle \underline{\bar{S}}\left( X^{\prime },X\right)
\right\rangle _{X}+\frac{1}{2}\left\langle \bar{w}\left( X^{\prime
},X\right) \right\rangle _{X}\bar{\Delta}^{E}\left( X^{\prime }\right)
\label{BCSn}
\end{equation}%
\begin{equation}
\left\langle \bar{S}_{L}\left( X^{\prime },X\right) \right\rangle _{X}=\frac{%
1}{2}\left\langle \underline{\bar{S}}\left( X^{\prime },X\right)
\right\rangle _{X}+\frac{1}{2}\left\langle \bar{w}\left( X^{\prime
},X\right) \right\rangle _{X}\bar{\Delta}^{L}\left( X^{\prime },X\right)
\label{BCSpt}
\end{equation}%
\begin{eqnarray}
\left\langle \bar{S}\left( X^{\prime },X\right) \right\rangle _{X}
&=&\left\langle \bar{S}_{E}\left( X^{\prime },X\right) \right\rangle
_{X}+\left\langle \bar{S}_{L}\left( X^{\prime },X\right) \right\rangle _{X}
\label{BCSt} \\
&=&\left\langle \underline{\bar{S}}\left( X^{\prime },X\right) \right\rangle
_{X}+\left\langle \bar{w}\left( X^{\prime },X\right) \right\rangle \bar{%
\Delta}\left( X\right)  \notag
\end{eqnarray}%
where $\hat{\Delta}^{E}\left( X^{\prime }\right) $, $\hat{\Delta}^{L}\left(
X^{\prime }\right) $, and $\hat{\Delta}\left( X^{\prime }\right) $ are the
aggregates of $\hat{\Delta}^{E}\left( X^{\prime },X\right) $, $\hat{\Delta}%
^{L}\left( X^{\prime },X\right) $, and $\hat{\Delta}\left( X^{\prime
},X\right) $ over the variable $X$. They can be written as:%
\begin{eqnarray*}
\bar{\Delta}^{E}\left( X^{\prime }\right) &=&\bar{f}\left( X^{\prime
}\right) -\left\langle \bar{R}^{\bar{w},\hat{w}^{B},w^{B}}\left( X\right)
\right\rangle \\
&& \\
\bar{\Delta}^{L}\left( X^{\prime }\right) &=&\bar{r}\left( X^{\prime
}\right) -\left\langle \bar{R}^{\bar{w},\hat{w}^{B},w^{B}}\left( X\right)
\right\rangle \\
&& \\
\bar{\Delta}\left( X\right) &=&\bar{R}_{X}-\left\langle \bar{R}^{\bar{w},%
\hat{w}^{B},w^{B}}\left( X\right) \right\rangle
\end{eqnarray*}%
and:%
\begin{equation*}
\left\langle \bar{R}^{\bar{w},\hat{w}^{B},w^{B}}\left( X\right)
\right\rangle =\left\langle \bar{w}\left( X\right) \right\rangle \bar{R}%
_{X}+\left\langle \hat{w}_{E}^{B}\left( X\right) \right\rangle \left\langle 
\hat{f}\left( X^{\prime }\right) \right\rangle _{\hat{w}_{E}}+\left\langle
w_{E}^{B}\left( X\right) \right\rangle \left\langle f\left( X\right)
\right\rangle
\end{equation*}%
Consequently, the inward aggregate stakes are obtained by weighting (\ref%
{BCSn}), (\ref{BCSpt}), and (\ref{BCSt}) with the corresponding scapital
ratios:

\begin{equation}
\bar{S}_{\eta }\left( X^{\prime }\right) =\left\langle \bar{S}_{\eta }\left(
X^{\prime },X\right) \right\rangle _{X}\frac{\left\langle \bar{K}%
\right\rangle \left\Vert \bar{\Psi}\right\Vert ^{2}}{\bar{K}_{X}\left\vert 
\bar{\Psi}\left( X\right) \right\vert ^{2}}  \label{BCStn}
\end{equation}%
and:%
\begin{equation}
\bar{S}\left( X^{\prime }\right) =\left\langle \bar{S}\left( X^{\prime
},X\right) \right\rangle _{X}\frac{\left\langle \bar{K}\right\rangle
\left\Vert \bar{\Psi}\right\Vert ^{2}}{\bar{K}_{X}\left\vert \bar{\Psi}%
\left( X\right) \right\vert ^{2}}  \label{BCStd}
\end{equation}

\subsubsection{Inward aggregate stakes of banks in investors}

The inward stakes of banks in investors $X^{\prime }$ are given by:%
\begin{equation*}
\left\langle \hat{S}_{E}^{B}\left( X^{\prime },X\right) \right\rangle
_{X}=\left\langle \underline{\hat{S}}_{E}^{B}\left( X^{\prime },X\right)
\right\rangle _{X}+\left\langle \hat{w}_{E}^{B}\left( X^{\prime },X\right)
\right\rangle _{X}\left( \hat{\Delta}^{E}\right) ^{B}\left( X^{\prime
}\right)
\end{equation*}%
and:%
\begin{equation}
\frac{\left\langle \hat{S}_{L}^{B}\left( X^{\prime },X\right) \right\rangle
_{X}}{\kappa \left( 1-\bar{S}\left( X\right) \right) }=\left\langle 
\underline{\hat{S}}_{L}^{B}\left( X^{\prime },X\right) \right\rangle
_{X}+\left\langle \hat{w}_{L}^{B}\left( X^{\prime },X\right) \right\rangle
_{X}\left( \hat{\Delta}^{L}\right) ^{B}\left( X^{\prime }\right)
\label{BCSHd}
\end{equation}%
where $\hat{\Delta}^{E}\left( X^{\prime }\right) $, $\hat{\Delta}^{L}\left(
X^{\prime }\right) $ are the aggregates of $\hat{\Delta}^{E}\left( X^{\prime
},X\right) $, $\hat{\Delta}^{L}\left( X^{\prime },X\right) $, and $\hat{%
\Delta}\left( X^{\prime },X\right) $ over the variable $X$. They can be
written as:%
\begin{eqnarray*}
\left( \hat{\Delta}^{E}\right) ^{B}\left( X^{\prime }\right) &=&\hat{f}%
\left( X^{\prime }\right) -\left\langle \bar{R}^{\bar{w},\hat{w}%
_{E}^{B},w_{E}^{B}}\left( X\right) \right\rangle \\
\left( \hat{\Delta}^{L}\right) ^{B}\left( X^{\prime }\right) &=&\bar{r}%
\left( X^{\prime }\right) -\left\langle \bar{R}^{\hat{w}_{L}^{B},w_{L}^{B}}%
\left( X\right) \right\rangle
\end{eqnarray*}%
and:%
\begin{equation*}
\left\langle \bar{R}^{\hat{w}_{L}^{B},w_{L}^{B}}\left( X\right)
\right\rangle =\left\langle \hat{w}_{L}^{B}\left( X\right) \right\rangle
\left\langle \hat{r}\left( X\right) \right\rangle +\left\langle
w_{L}^{B}\left( X\right) \right\rangle \left\langle r\left( X\right)
\right\rangle
\end{equation*}%
which computes the average of the weighted return $\bar{R}^{\hat{w}%
_{L}^{B},w_{L}^{B}}\left( X\right) $.

The corresponding agregate stakes are given by:%
\begin{eqnarray*}
\hat{S}_{E}^{B}\left( X^{\prime }\right) &=&\left\langle \hat{S}%
_{E}^{B}\left( X^{\prime },X\right) \right\rangle _{X}\frac{\left\langle 
\bar{K}\right\rangle \left\Vert \bar{\Psi}\right\Vert ^{2}}{\hat{K}%
_{X^{\prime }}\left\vert \hat{\Psi}\left( X^{\prime }\right) \right\vert ^{2}%
} \\
\hat{S}_{L}^{B}\left( X^{\prime }\right) &=&\left\langle \hat{S}%
_{L}^{B}\left( X^{\prime },X\right) \right\rangle _{X}\frac{\left\langle 
\bar{K}\right\rangle \left\Vert \bar{\Psi}\right\Vert ^{2}}{\hat{K}%
_{X^{\prime }}\left\vert \hat{\Psi}\left( X^{\prime }\right) \right\vert ^{2}%
}
\end{eqnarray*}

\subsubsection{Inward aggregate stakes of banks in firms}

Ultimately, the inward banks' stakes in firms are given by (\ref{SFN}) and (%
\ref{sdt}), with the associated inward aggregate stakes:%
\begin{equation}
S_{E}^{B}\left( X,X\right) =S_{E}^{B}\left( X,X\right) \frac{\left\langle 
\bar{K}_{X}\right\rangle \left\vert \bar{\Psi}\left( X\right) \right\vert
^{2}}{K_{X}\left\vert \Psi \left( X\right) \right\vert ^{2}}  \label{BCStbn}
\end{equation}%
and:%
\begin{equation}
\frac{S_{L}^{B}\left( X,X\right) }{\kappa \left( 1-\bar{S}\left( X\right)
\right) }=\frac{S_{L}^{B}\left( X,X\right) }{\kappa \left( 1-\bar{S}\left(
X\right) \right) }\frac{\left\langle \bar{K}_{X}\right\rangle \left\vert 
\bar{\Psi}\left( X\right) \right\vert ^{2}}{K_{X}\left\vert \Psi \left(
X\right) \right\vert ^{2}}  \label{sFB}
\end{equation}

\section{The equations for uncertainty}

\subsection{Investors}

The equation defining the weight of inverse uncertainty for cross-investment
between investors, derived in section 4, is:%
\begin{equation}
\hat{w}\left( X^{\prime },X\right) =\frac{2\left( 1-\left( \gamma
\left\langle \hat{S}_{E}\left( X\right) \right\rangle \right) ^{2}\right) 
\hat{w}_{E}^{\left( 0\right) }\left( X^{\prime },X\right) }{1+\hat{w}%
_{E}^{\left( 0\right) }\left( X^{\prime },X\right) \left( 1-\left( \gamma
\left\langle \hat{S}_{E}\left( X\right) \right\rangle \right) ^{2}\right)
+\Delta \left( \gamma \left\langle \hat{S}_{E}\left( X_{1},X^{\prime
}\right) \right\rangle _{\hat{X}_{1}}\right) ^{2}}  \label{cfh}
\end{equation}%
Recall that $\gamma $ is the average uncertainty of the distance-dependent
investment paths and is given by: 
\begin{equation*}
\gamma ^{2}\simeq \left( \hat{w}_{E}^{\left( 0\right) }\left( \left(
X^{\prime }\right) ^{\prime },X_{m-1}\right) ...\hat{w}_{E}^{\left( 0\right)
}\left( X_{1},X^{\prime }\right) \right) ^{-\frac{1}{m}}
\end{equation*}%
where the coefficient $\hat{w}_{E}^{\left( 0\right) }\left( X^{\prime
},X\right) $\ was defined in section 4.

The weight of inverse uncertainty for investments in firms is given by:%
\begin{equation}
w\left( X,X\right) =1-\left\langle w\left( X^{\prime },X\right)
\right\rangle _{X^{\prime }}  \label{NCH}
\end{equation}

\subsection{Banks}

Similarly, the weights for inverse uncertainty of banks are:%
\begin{equation}
\left( \bar{w}\left( X^{\prime },X\right) \right) ^{-1}=1+\frac{1}{2}\left( 
\frac{\overline{IRG}\left( X^{\prime },X\right) }{\widehat{IR^{B}}\left(
X^{\prime },X\right) }+\frac{\overline{IRG}\left( X^{\prime },X\right) }{\xi
^{2}}\right)  \label{Cfn}
\end{equation}%
and those for banks investing in investors are given by:%
\begin{equation}
\left( \hat{w}_{E}^{B}\left( X^{\prime },X\right) \right) ^{-1}=1+2\frac{%
\widehat{IR^{B}}\left( X^{\prime },X\right) }{\overline{IRG}\left( X^{\prime
},X\right) }+\frac{\widehat{IR^{B}}\left( X^{\prime },X\right) }{\xi ^{2}}
\label{Cft}
\end{equation}%
\begin{equation}
w_{E}^{B}\left( X^{\prime },X\right) =1-\bar{w}\left( X^{\prime },X\right) -%
\hat{w}_{E}^{B}\left( X^{\prime },X\right)  \label{chv}
\end{equation}%
Since the functions $\hat{w}$ and $w$\ both depend on the stakes, their
expressions complete the minimization equations and yield $\hat{S}_{E}\left(
X^{\prime },X\right) $, $\hat{S}\left( X^{\prime },X\right) $, $S_{E}\left(
X,X\right) $, $S\left( X,X\right) $.

Note, incidentally, that the above equations hold for a fixed value of the
uncertainty $\gamma $; however, since $\gamma $ can vary across the sector
space, the groups of investors may be relatively interconnected. In
first-order approximation, we will assume them to be independent, and solve
the model for each group separately\footnote{%
We will examine the interactions between groups in greater detail in the
third paper of this series.}.

\part*{Resolution of the model}

The resolution of the model consists of solving the return equations (\ref%
{QDM}) and (\ref{QDB}) using the relationships between stakes and returns
established in equations in (\ref{Shn}), (\ref{Sht}), (\ref{Shv}), (\ref{SNp}%
), and (\ref{STp}) for investors, and (\ref{SBN}), (\ref{SBT}), (\ref{SBF}),
(\ref{scb}), (\ref{sdb}), (\ref{SFN}) for banks, along with the uncertainty
coefficients (\ref{cfh}), (\ref{Cfn}) and (\ref{Cft}).\textbf{\ }The
solutions of these minimization equations define the collective states of
the system. We will detail the resolution process, implement it under a
no-default scenario, and study the possibility of defaults\footnote{%
Default states will be discussed subsequently.}.

\section{Methodology}

\subsection{Groups and sub-collective states}

To solve the full system, we must determine the levels of capital and
returns for each type of agent in each sector of the sector space $S$, along
with the distribution of stakes across all these sectors. Taken together,
these quantities define a collective state of the system.

Since capital and returns can be derived from the levels of stakes between
investors, a collective state can be described by a set of stake values for
investors: 
\begin{equation*}
\left\{ \hat{S}_{E}\left( X^{\prime },X\right) ,\hat{S}\left( X^{\prime
},X\right) ,S_{E}\left( X,X\right) ,S\left( X,X\right) \right\} _{\left(
X^{\prime },X\right) \in S}
\end{equation*}%
and for banks:%
\begin{equation*}
\left\{ \hat{S}_{E}^{B}\left( X^{\prime },X\right) ,\hat{S}^{B}\left(
X^{\prime },X\right) ,S_{E}^{B}\left( X,X\right) ,S^{B}\left( X,X\right) ,%
\bar{S}_{E}\left( X^{\prime },X\right) ,\bar{S}^{B}\left( X^{\prime
},X\right) \right\} _{\left( X^{\prime },X\right) \in S}
\end{equation*}%
These quantities are governed by equations (\ref{QDM}) and (\ref{QDB}).\
These are non-local equations that involve the network of connections within
each group.

As we have seen, under uncertainty, agents connect to only a finite number
of neighbours, so that they are organized into several loosely-connected
groups. Each of these groups, $G_{\alpha }$, can display one among several
possible phases, each defined by a distribution of stakes among the agents
of $G_{\alpha }$:%
\begin{eqnarray*}
&&\left\{ \hat{S}_{E}\left( X^{\prime },X\right) ,\hat{S}\left( X^{\prime
},X\right) ,S_{E}\left( X,X\right) ,S\left( X,X\right) ,\right. \\
&&\left. \hat{S}_{E}^{B}\left( X^{\prime },X\right) ,\hat{S}^{B}\left(
X^{\prime },X\right) ,S_{E}^{B}\left( X,X\right) ,S^{B}\left( X,X\right) ,%
\bar{S}_{E}\left( X^{\prime },X\right) ,\bar{S}^{B}\left( X^{\prime
},X\right) \right\} _{\left( X^{\prime },X\right) \in G_{\alpha }}
\end{eqnarray*}%
We define a \emph{sub-collective state} as the combination of one group, $%
G_{\alpha }$, and one of its possible phases. Thus, collective states are
organized into several distinct, weakly interacting sub-collective states,
so that, in a first-order approximation, a collective state can be seen as a
set of sub-collective states, expressed as:%
\begin{eqnarray*}
\cup _{\alpha } &&\left\{ \hat{S}_{E}\left( X^{\prime },X\right) ,\hat{S}%
\left( X^{\prime },X\right) ,S_{E}\left( X,X\right) ,S\left( X,X\right)
,\right. \\
&&\left. \hat{S}_{E}^{B}\left( X^{\prime },X\right) ,\hat{S}^{B}\left(
X^{\prime },X\right) ,S_{E}^{B}\left( X,X\right) ,S^{B}\left( X,X\right) ,%
\bar{S}_{E}\left( X^{\prime },X\right) ,\bar{S}^{B}\left( X^{\prime
},X\right) \right\} _{\left( X^{\prime },X\right) \in G_{\alpha }}
\end{eqnarray*}%
where the groups $G_{\alpha }$ describe a particular organization of the
sector space $S$.

The organization of the sector space into groups $\left\{ G_{\alpha
}\right\} $ is not unique; there are an infinite number of such
organizations. Moreover, for each organization, multiple distinct phases are
possible for each group: the combination of uncertainty and
interconnectivity of sectors leads to an infinite number of possible
solutions, some of which correspond to default states\footnote{%
See Gosselin and Lotz (2024).}. Therefore, there is an infinite number of
possible collective states, each representing a potential configuration of
the full system, among an infinite number of such configurations.\ Each of
these configurations is itself composed of multiple groups, each in a
possible phase.

Any group together with its phase is called a \emph{sub-collective state}.\
For any group, several sub-collective states may exist.

\subsection{Principle of resolution}

The return equations in sector $X$, equations (\ref{QDM}) and (\ref{QDB}),
connect all investors in the sector space and, as such, comprise three types
of variables: the inward aggregate stakes on investors and banks in sector $%
X $; the returns of investors $X$, banks $X$, and firms $X$, considered
independently; and the average returns of all investors across the sector
space.

The resolution follows a four-step process:

1. We average the return equations over all sectors.\ This yields the
averages of all the variables in the equations: average stakes, levels of
capital and returns.

2. We then substitute these results back into the return equations, and
further replace the aggregate stakes (\ref{Grstn}), (\ref{Grstd}), and (\ref%
{Grstf}) in terms of returns, so that the return equations depend solely on
returns.

3. Solving this equations yields the returns per sector and, in turn, the
aggregate stakes and disposable capital per sector.

4. Using equations (\ref{Shn}), (\ref{Sht}), and (\ref{Shv}), we reconstruct
the remaining quantities of the model.

\section{Solving the model: no-default scenario}

\subsection{Step 1: solving for average stakes, capital, and returns}

\subsubsection{General resolution}

The return equations (\ref{QDM}) and (\ref{QDB}), averaged, can be written
as:%
\begin{equation}
\left( 1-\left\langle \hat{S}\left( X^{\prime },X\right) \right\rangle
\right) \left\langle \hat{R}_{exc}\left( X^{\prime }\right) \right\rangle
\simeq \left\langle S_{E}\left( X,X\right) \right\rangle \left\langle
R_{exc}\left( X\right) \right\rangle  \label{QDN}
\end{equation}%
and:%
\begin{eqnarray}
0 &=&\left( 1-\left\langle \bar{S}\left( X^{\prime },X\right) \right\rangle
\right) \left\langle \bar{R}_{exc}\left( X^{\prime }\right) \right\rangle
\label{QDP} \\
&&-\left\langle \widehat{DF}\left( X\right) \right\rangle \left\langle \hat{S%
}_{E}^{B}\right\rangle \frac{\left\langle \bar{K}\right\rangle \left\Vert 
\bar{\Psi}\right\Vert ^{2}}{\left\langle \hat{K}\right\rangle \left\Vert 
\hat{\Psi}\right\Vert ^{2}}\left\langle \hat{R}_{exc}\left( X^{\prime
}\right) \right\rangle -\left\langle S_{E}^{B}\left( X^{\prime },X\right)
\right\rangle \left\langle R_{exc}\left( X\right) \right\rangle  \notag
\end{eqnarray}%
The solutions of (\ref{QDN}) and (\ref{QDP}) ultimately depend on the
assumptions regarding the firms' returns per unit of capital, $\left\langle
f\left( X\right) \right\rangle $.\ As a first approximation, and to simplify
the computations, we assume constant returns to scale, so that:%
\begin{equation*}
\left\langle f\left( X\right) \right\rangle =\left\langle f_{1}\left(
X\right) \right\rangle
\end{equation*}%
\textbf{\ }where $\left\langle f_{1}\left( X\right) \right\rangle $\ is the
net average productivity. Decreasing returns to scale will then be
introduced as corrections, particularly when computing disposable capital,
in which case the firms' returns will be\footnote{%
See Appendix 8 in Gosselin and Lotz (2025).}:%
\begin{equation}
\left\langle f\left( X\right) \right\rangle =\frac{\left\langle f_{1}\left(
X\right) \right\rangle }{\left\langle K\right\rangle ^{r}}-\frac{C}{%
\left\langle K\right\rangle }-C_{0}  \label{DCr}
\end{equation}%
where $C$ represents a fixed cost and $C_{0\text{ }}$ a cost per unit of
capital\footnote{%
More precisely, considering the marginal productivity $\frac{1}{1-r}%
\left\langle f_{1}\left( X\right) \right\rangle \left\langle K\right\rangle
^{1-r}$, the average return per unit of capital, including costs, is:%
\begin{equation*}
\frac{1}{1-r}\left\langle f_{1}\left( X\right) \right\rangle \left\langle
K\right\rangle ^{1-r}-C-C_{0}\left\langle K\right\rangle
\end{equation*}%
Including the factor $\frac{1}{1-r}$\ in the definition of $\left\langle
f_{1}\left( X\right) \right\rangle $\ yields the formula presented in the
text.}.

\paragraph{Resolution for investors}

The global average stakes in investors, $\left\langle \hat{S}\left(
X^{\prime }\right) \right\rangle $, and in firms, $\left\langle S_{E}\left(
X,X\right) \right\rangle $ and $\left\langle S\left( X,X\right)
\right\rangle $, as well as the ratio of the average disposable capital of
investors to that of firms, $\frac{\left\langle \hat{K}\right\rangle
\left\Vert \hat{\Psi}\right\Vert ^{2}}{\left\langle K\right\rangle
\left\Vert \Psi \right\Vert ^{2}}$, are expressed as functions of the
average shares in investors, $\left\langle \hat{S}_{E}\left( X^{\prime
}\right) \right\rangle $, and the total stakes in firms, $\left\langle
S\left( X\right) \right\rangle $. To solve the average return equation (\ref%
{QDN}), we use equations (\ref{Shn}), (\ref{Sht}), (\ref{Shv}), (\ref{SNp}),
and (\ref{STp}) to reexpress all variables - namely, the global average
stakes in firms and the average returns in firms, $\left\langle S\left(
X\right) \right\rangle $ and $\left\langle f\left( X\right) \right\rangle $,
respectively - as functions of the average shares in investors, $%
\left\langle \hat{S}_{E}\left( X^{\prime }\right) \right\rangle $.\ In this
way, the full system is expressed as a function of a single variable: the
average shares in investors.\ 

\paragraph{Resolution for banks}

The method is similar to that used for the investors' equations, using (\ref%
{SBN}), (\ref{SBT}), (\ref{SBF}), (\ref{scb}), (\ref{sdb}), (\ref{SFN}), (%
\ref{sdt}), (\ref{cfh}), (\ref{NCH}), (\ref{Cfn}), and (\ref{Cft}) to
express stakes, and the capital ratios, $\frac{\left\langle \bar{K}%
\right\rangle \left\Vert \bar{\Psi}\right\Vert ^{2}}{\left\langle
K\right\rangle \left\Vert \Psi \right\Vert ^{2}}$, $\frac{\left\langle \bar{K%
}\right\rangle \left\Vert \bar{\Psi}\right\Vert ^{2}}{\left\langle \hat{K}%
\right\rangle \left\Vert \hat{\Psi}\right\Vert ^{2}}$, $\frac{\left\langle 
\hat{K}\right\rangle \left\Vert \hat{\Psi}\right\Vert ^{2}}{\left\langle
K\right\rangle \left\Vert \Psi \right\Vert ^{2}}$, as functions of the
average shares $\left\langle \hat{S}_{E}\left( X^{\prime },X\right)
\right\rangle $ and $\left\langle \bar{S}_{E}\left( X^{\prime },X\right)
\right\rangle $. These expressions, together with the solution of (\ref{QDN}%
) for the investors' average excess return, $\left\langle \hat{f}\left(
X^{\prime }\right) \right\rangle -\bar{r}$, are then substituted into (\ref%
{QDP}), yielding an equation for $\left\langle \bar{S}_{E}\left( X^{\prime
},X\right) \right\rangle $.

For later purposes, note that, using (\ref{QDN}),\ equation (\ref{QDP}) can
be rewritten as an equation for bank returns only:%
\begin{equation}
0=\left( 1-\left\langle \bar{S}\right\rangle \right) \left\langle \bar{R}%
_{exc}\left( X^{\prime }\right) \right\rangle -\left\langle S_{E}^{\left(
e\right) }\right\rangle _{0}\left\langle R_{exc}\left( X\right) \right\rangle
\label{Rf}
\end{equation}%
where the effective stakes, $\left\langle S_{E}^{\left( e\right)
}\right\rangle _{0}$, encompass indirect investment in firms through the
shares of banks in intermediate investors:%
\begin{equation*}
\left\langle S_{E}^{\left( e\right) }\left( X,X\right) \right\rangle
_{0}=\left( \left\langle \widehat{DF}\left( X^{\prime }\right) \right\rangle 
\frac{\left\langle \hat{S}_{E}^{B}\right\rangle \left\langle \bar{K}%
\right\rangle \left\Vert \bar{\Psi}\right\Vert ^{2}}{\left\langle \hat{K}%
\right\rangle \left\Vert \hat{\Psi}\right\Vert ^{2}}\frac{\left\langle
S_{E}\left( X,X\right) \right\rangle }{\left( 1-\left\langle \hat{S}\left(
X^{\prime },X\right) \right\rangle \right) }+\left\langle S_{E}^{B}\left(
X,X\right) \right\rangle \right)
\end{equation*}

\subsubsection{Average variables}

Numerical analyses of the solutions reveal that investors' and banks'
cross-investments, $\left\langle \hat{S}_{E}\left( X^{\prime }\right)
\right\rangle $ and $\left\langle \bar{S}_{E}\left( X^{\prime },X\right)
\right\rangle $, increase with the excess average firm return $\left\langle
f\left( X\right) \right\rangle $ and decrease with uncertainty $\gamma $.

When average firm returns exceed the interest rate, $\left\langle f\left(
X\right) \right\rangle >r$, higher firm returns foster sectoral firm capital
accumulation.\ The average capital held by firms, $\left\langle
K\right\rangle \left\Vert \Psi \right\Vert ^{2}$, rises with returns.
Investors allocate capital within their own sector, and in firms
specifically.\ Consequently, disposable capital for investors diminishes,
and the ratios of average disposable capital of investors and banks over
that of firms, $\frac{\left\langle \hat{K}\right\rangle \left\Vert \hat{\Psi}%
\right\Vert ^{2}}{\left\langle K\right\rangle \left\Vert \Psi \right\Vert
^{2}}$ and $\frac{\left\langle \bar{K}\right\rangle \left\Vert \bar{\Psi}%
\right\Vert ^{2}}{\left\langle K\right\rangle \left\Vert \Psi \right\Vert
^{2}}$, respectively,\ decrease as returns increase.

Conversely, when average firm returns do not exceed the interest rate,\ $%
\left\langle f\left( X\right) \right\rangle <r$, they fail to offset the
higher investment risk, leading to a decline in capital invested in firms.
As global uncertainty $\gamma $\ increases, investments become more
diversified, resulting in higher capital ratios $\frac{\left\langle \hat{K}%
\right\rangle \left\Vert \hat{\Psi}\right\Vert ^{2}}{\left\langle
K\right\rangle \left\Vert \Psi \right\Vert ^{2}}$ and $\frac{\left\langle 
\bar{K}\right\rangle \left\Vert \bar{\Psi}\right\Vert ^{2}}{\left\langle
K\right\rangle \left\Vert \Psi \right\Vert ^{2}}$.

When firms' returns are high, uncertainty tends to favor direct investment
in firms; conversely, when uncertainty is low, it promotes investment
diversification among investors and banks.

Overall, incorporating banks into the model increases the disposable capital
of both investors and firms.\ Under decreasing returns to scale, it also
smoothes excess average returns, which in turn affects the stability of the
system, as will be shown below.

\subsubsection{Average variables under constant return to scale}

Solutions can be approximated when firms' average returns, $\left\langle
f\left( X\right) \right\rangle $, are close to the average interest rate, $%
\left\langle \bar{r}\left( X\right) \right\rangle $.

\paragraph{Benchmark case}

In the benchmark case, average returns are equal the average interest rate,
that is, $\left\langle R_{exc}\left( X\right) \right\rangle =0$.\ In this
situation, the value of $\left\langle \hat{S}_{E}\left( X^{\prime },X\right)
\right\rangle $, denoted $z_{0}$, is a function of the level of uncertainty $%
\gamma $.\ 

When there is no uncertainty, $\gamma $ $=$ $0$, we have:%
\begin{equation*}
\left\langle \hat{S}_{E}\left( X^{\prime },X\right) \right\rangle
=\left\langle \hat{S}_{L}\left( X^{\prime },X\right) \right\rangle
=\left\langle S_{E}\left( X,X\right) \right\rangle =\left\langle S_{L}\left(
X,X\right) \right\rangle =z_{0}=\frac{1}{4}
\end{equation*}%
and:%
\begin{equation*}
\left\langle \bar{S}_{E}\left( X^{\prime },X\right) \right\rangle
=\left\langle \bar{S}_{E}\left( X^{\prime },X\right) \right\rangle
=\left\langle \hat{S}_{E}^{B}\left( X^{\prime },X\right) \right\rangle
=\left\langle S_{E}^{B}\left( X^{\prime },X\right) \right\rangle =x=\frac{1}{%
4}
\end{equation*}%
In the absence of both excess returns and uncertainty, portfolio
diversification is maximal.\ 

However, when uncertainty is present ($\gamma >0$), stakes remain equivalent
in terms of expected returns, but not in terms of risk.\ In this case, $%
z_{0} $ measures the effect of uncertainty on investment allocation. When
uncertainty is high, the level of average shares in investors $\left\langle 
\hat{S}_{E}\left( X^{\prime },X\right) \right\rangle $ is correspondingly
low: for equivalent expected returns, investors prefer direct over
intermediated investment in firms. Conversely, when uncertainty is low, the
system reverts to the benchmark configuration, where both investors and
banks diversify their investments.

\paragraph{Impact of differential returns on investors' cross-sectoral stakes%
}

When average returns differ from the interest rate, average stakes depend on
both the differential return, $\left\langle f_{1}\left( X\right)
\right\rangle -\left\langle \bar{r}\left( X\right) \right\rangle $, and $%
z_{0}$, the measure of uncertainty.

The average shares held in, and the loans granted by, investors to other
investors are respectively given by:%
\begin{eqnarray}
\left\langle \hat{S}_{E}\left( X^{\prime },X\right) \right\rangle
&=&\left\langle \hat{S}_{E}\left( X^{\prime }\right) \right\rangle
=z_{0}\left( 1+\frac{1}{2}\frac{z_{0}^{2}}{D}\left\langle R_{exc}\left(
X\right) \right\rangle \right) \\
&&  \notag \\
\left\langle \hat{S}_{L}\left( X^{\prime },X\right) \right\rangle
&=&z_{0}\left( 1-\frac{1}{2}\left( 1-\frac{z_{0}^{2}}{D}\right) \left\langle
R_{exc}\left( X\right) \right\rangle \right)  \notag
\end{eqnarray}%
where:%
\begin{equation*}
D=1-5z_{0}+8z_{0}^{2}
\end{equation*}%
and the total average stake is:%
\begin{eqnarray}
\left\langle \hat{S}\left( X^{\prime },X\right) \right\rangle
&=&\left\langle \hat{S}_{E}\left( X^{\prime },X\right) \right\rangle
+\left\langle \hat{S}_{L}\left( X^{\prime },X\right) \right\rangle \\
&=&2z_{0}\left( 1-\frac{1}{4}\frac{\left( 1-3z_{0}\right) \left(
1-2z_{0}\right) }{D}\left\langle R_{exc}\left( X\right) \right\rangle \right)
\notag
\end{eqnarray}%
For a given level of uncertainty, any increase in firms' excess returns
induces investors to allocate more capital toward firms.\ They do so not
only within their own sector, but also across sectors through the
intermediation of other investors. As a result, average cross-sectoral
equity investments, $\left\langle \hat{S}_{E}\left( X^{\prime },X\right)
\right\rangle $, increase. By contrast, the attractiveness and relative
profitability of loans decreases, leading to a reduction in $\left\langle 
\hat{S}_{L}\left( X^{\prime },X\right) \right\rangle $. The overall
cross-investors volume of stakes $\left\langle \hat{S}\left( X^{\prime
},X\right) \right\rangle $ will nonetheless decline, since direct investment
in firms remains comparatively more profitable.

A relative increase in uncertainty dampens cross-sectoral intermediation.
Portfolio management under uncertainty implies diversification through
loans, which limits investors' ability to fully capture higher returns.
Consequently, investors' aggregate returns increase more slowly than firms'
returns:

\begin{equation*}
\left\langle \hat{R}_{exc}\left( X\right) \right\rangle =\frac{\left\langle
R_{exc}\left( X\right) \right\rangle }{2}
\end{equation*}%
Hence, an increase in firms' returns, $\left\langle f_{1}\left( X\right)
\right\rangle $, results only in a partial rise in average investors'
returns, $\left\langle \hat{f}\left( X\right) \right\rangle $, by an amount
of $\left\langle f_{1}\left( X\right) \right\rangle /2$.

\paragraph{Impact of differential returns on investors' stakes in firms}

The stakes in firms are given by: 
\begin{eqnarray}
\left\langle S_{E}\left( X,X\right) \right\rangle &=&\frac{1-2z_{0}}{2}%
\left( 1+\left( z_{0}\varepsilon \left( z_{0}\right) +\left( \frac{3}{4}%
-z_{0}\right) \right) \left\langle R_{exc}\left( X\right) \right\rangle
\right) \\
&&  \notag \\
\left\langle S_{L}\left( X,X\right) \right\rangle &=&\frac{1-2z_{0}}{2}%
\left( 1+\left( z_{0}-\varepsilon \left( z_{0}\right) \left( \frac{1}{2}%
-z_{0}\right) -\frac{3}{8}\right) \left\langle R_{exc}\left( X\right)
\right\rangle \right)  \notag \\
&&  \notag \\
\left\langle S\left( X,X\right) \right\rangle &=&\left\langle S_{E}\left(
X,X\right) \right\rangle +\left\langle S_{L}\left( X,X\right) \right\rangle 
\notag \\
&=&\left( 1-2z_{0}\right) \left( 1+\left( \frac{1}{2}+\varepsilon \left(
z_{0}\right) \right) z_{0}\left\langle R_{exc}\left( X\right) \right\rangle
\right)  \notag
\end{eqnarray}%
where:%
\begin{equation*}
\varepsilon \left( z_{0}\right) =\frac{z_{0}^{2}\left( 1-4z_{0}\right) }{%
\left( 1-5z_{0}+8z_{0}^{2}\right) }
\end{equation*}%
When firms experience excess returns, they tend to raise capital through
equity issuance rather than debt. Consequently, shares in firms, $%
\left\langle S_{E}\left( X,X\right) \right\rangle $, increase, while loans, $%
\left\langle S_{L}\left( X,X\right) \right\rangle $, decline. Nonetheless,
total investment in firms, $\left\langle S\left( X,X\right) \right\rangle $,
rises. In addition, a relative increase in uncertainty further increase
direct investment in firms.

\paragraph{Impact of differential returns on banks' shares}

We define the reference values $\left\langle S_{E}^{\left( e\right)
}\right\rangle _{0}$, $\left\langle \bar{S}_{E}^{B}\right\rangle _{0}$,\ and 
$\left\langle \hat{S}_{E}^{B}\right\rangle _{0}$\ as the values of the
effective share $\left\langle S_{E}^{\left( e\right) }\right\rangle $, the
shares of banks in firms $\left\langle \bar{S}_{E}^{B}\right\rangle $\ and
investors $\left\langle \hat{S}_{E}^{B}\right\rangle $\ for $\left\langle
f_{1}\left( X\right) \right\rangle -\left\langle \bar{r}\left( X\right)
\right\rangle =0$.

Defining $x=\left\langle \bar{S}_{E}\right\rangle _{0}$, we show\footnote{%
See Appendix 6.} that, under high uncertainty, we have, to a first
approximation\footnote{%
Quadratic corrections in $x$ to these formula are given in Appendix 6.5.1.4.}%
:%
\begin{equation*}
\left\langle \bar{S}_{E}^{B}\right\rangle _{0}\simeq 1-6\left\langle \bar{S}%
_{E}\right\rangle _{0}=1-6x
\end{equation*}%
and:%
\begin{equation*}
\left\langle \hat{S}_{E}^{B}\right\rangle _{0}\simeq 4\left\langle \bar{S}%
_{E}\right\rangle _{0}=4x
\end{equation*}%
That is, investments in banks and in investors are proportional, as both are
subject to the same uncertainty of financial circulation. Investment in
investors, however, is higher than in banks, since under our assumptions,
investing in other banks lengthens the path from the banks, seen as
investors, and the real returns provided by firms.

For banks, the cross-shares in banks are given by:%
\begin{equation}
\left\langle \bar{S}_{E}\left( X^{\prime },X\right) \right\rangle =\left(
1+\left( \frac{1-\left\langle \bar{S}_{E}\right\rangle _{0}}{\left(
1-2\left\langle \bar{S}_{E}\right\rangle \right) }\left\langle S_{E}^{\left(
e\right) }\right\rangle _{0}-\left\langle \bar{S}_{E}^{B}\right\rangle _{0}-%
\frac{\left\langle \hat{S}_{E}^{B}\right\rangle _{0}}{2}\right) \left\langle
R_{exc}\left( X\right) \right\rangle \right) \left\langle \bar{S}%
_{E}\right\rangle _{0}  \label{FRP}
\end{equation}%
and cross-loans by:%
\begin{equation}
\left\langle \bar{S}_{L}\left( X^{\prime },X\right) \right\rangle =\left(
1+\left( \frac{\left\langle \bar{S}_{E}\right\rangle _{0}}{\left(
1-2\left\langle \bar{S}_{E}\right\rangle \right) }\left\langle S_{E}^{\left(
e\right) }\right\rangle _{0}-\left\langle \bar{S}_{E}^{B}\right\rangle _{0}-%
\frac{\left\langle \hat{S}_{E}^{B}\right\rangle _{0}}{2}\right) \left\langle
R_{exc}\left( X\right) \right\rangle \right) \left\langle \bar{S}%
_{E}\right\rangle _{0}  \label{FRDm}
\end{equation}%
where the reference values $\left\langle S_{E}^{\left( e\right)
}\right\rangle _{0}$, $\left\langle \bar{S}_{E}^{B}\right\rangle _{0}$,\ and 
$\left\langle \hat{S}_{E}^{B}\right\rangle _{0}$\ are the values of the
effective share $\left\langle S_{E}^{\left( e\right) }\right\rangle $, the
shares of banks in firms $\left\langle \bar{S}_{E}^{B}\right\rangle $\ and
investors $\left\langle \hat{S}_{E}^{B}\right\rangle $\ for $\left\langle
f_{1}\left( X\right) \right\rangle -\left\langle \bar{r}\left( X\right)
\right\rangle =0$. In first approximation\footnote{%
See Appendix 6.}, under relatively high uncertainty, we have:%
\begin{equation*}
\left\langle \bar{S}_{E}^{B}\right\rangle _{0}\simeq 1-6\left\langle \bar{S}%
_{E}\right\rangle _{0}
\end{equation*}%
and:%
\begin{equation*}
\left\langle \hat{S}_{E}^{B}\right\rangle _{0}\simeq 6\left\langle \bar{S}%
_{E}\right\rangle _{0}
\end{equation*}%
Again, investments in banks and investors remain proportional, as both are
subject to uncertainty in capital circulation. Investment in investors is
higher than in banks, since, given our assumptions, investing in other banks
lengthens the path from the banks, seen as investors, and the real returns
provided by firms.

Formulas (\ref{FRP}) and (\ref{FRDm}) show that $\left\langle \bar{S}%
_{E}\right\rangle $ and $\left\langle \bar{S}_{L}\right\rangle $ depend
negatively\footnote{%
See Appendix 8.} on $\left\langle f_{1}\left( X\right) \right\rangle $. The
same result holds for the total cross-stakes between banks, given by:%
\begin{eqnarray}
\left\langle \bar{S}\left( X^{\prime },X\right) \right\rangle
&=&2\left\langle \bar{S}_{E}\left( X^{\prime },X\right) \right\rangle -\frac{%
\left\langle \bar{S}_{E}\left( X^{\prime },X\right) \right\rangle }{\left(
1-2\left\langle \bar{S}_{E}\left( X^{\prime },X\right) \right\rangle \right) 
}\left\langle S_{E}^{\left( e\right) }\left( X,X\right) \right\rangle
_{0}\left\langle R_{exc}\left( X\right) \right\rangle  \label{FRS} \\
&\simeq &2x-2x\left( \frac{\left\langle \hat{S}_{E}^{B}\right\rangle _{0}}{2}%
+\frac{\left\langle \bar{S}_{E}^{B}\right\rangle _{0}}{2\left(
1-\left\langle \bar{S}_{E}\right\rangle _{0}\right) }\right) \left\langle
R_{exc}\left( X\right) \right\rangle  \notag
\end{eqnarray}%
These results differ from the investors' case. When firms' returns increase,
banks\ gain to directly invest in firms or in investors.\ This is explicit
in the expressions for banks' shares in firms and investors:%
\begin{eqnarray}
\left\langle S_{E}^{B}\right\rangle &\simeq &1-6x+\frac{7}{4}x^{2}\left(
1+5x\right) \left\langle R_{exc}\left( X\right) \right\rangle \\
&&-\frac{1}{2}\left( 1-x\right) ^{2}x\left\langle \hat{R}_{exc}\left(
X\right) \right\rangle -\frac{1}{4}\left( 1-x\right) x^{2}\left\langle \bar{R%
}_{exc}\left( X\right) \right\rangle  \notag
\end{eqnarray}%
and similarly:%
\begin{eqnarray*}
\left\langle \hat{S}_{E}^{B}\left( X^{\prime },X\right) \right\rangle
&=&1-\left\langle S_{E}^{B}\right\rangle -\left\langle \bar{S}\left(
X^{\prime },X\right) \right\rangle \\
&\simeq &4x+2x\left( \frac{4-11x}{2}\right) \left\langle R_{exc}\left(
X\right) \right\rangle \\
&&+\frac{1}{2}\left( 1-x\right) ^{2}x\left\langle \hat{R}_{exc}\left(
X\right) \right\rangle +\frac{1}{4}\left( 1-x\right) x^{2}\left\langle \bar{R%
}_{exc}\left( X\right) \right\rangle
\end{eqnarray*}%
which depends positively on $\left\langle f_{1}\left( X\right) \right\rangle 
$.

The stake $\left\langle S_{E}^{B}\right\rangle $ depends negatively on both $%
\left\langle \hat{R}_{exc}\left( X\right) \right\rangle $ and $\left\langle 
\bar{R}_{exc}\left( X\right) \right\rangle $, as expected: when the returns
of banks or investors increase, banks prefer to invest in these two types of
agents.

The stake $\left\langle \hat{S}_{E}^{B}\left( X^{\prime },X\right)
\right\rangle $ depends positively on $\left\langle \hat{R}_{exc}\left(
X\right) \right\rangle $, but also indirectly on $\left\langle \bar{R}%
_{exc}\left( X\right) \right\rangle $: when banks' returns rise, so do their
investments in investors.

Finally, as uncertainty increases, banks' stakes in firms, $\left\langle
S_{E}^{B}\right\rangle $, rise, while their stakes in other banks and
investors decrease.

\paragraph{Impact of differential returns on banks loans}

The banks' average perception of investors cross-stakes, $z^{B}$,$\ $%
measures the banks' valuation of risks associated with cross-stakes between
investors. It is defined as:%
\begin{equation*}
z^{B}=E^{B}\left\langle \hat{S}_{E}\right\rangle
\end{equation*}%
As in the investors' case, we can also define $z_{0}^{B}$, the value of $%
E^{B}\left\langle \hat{S}_{E}\right\rangle $ for $\left\langle f\left(
X\right) \right\rangle -\left\langle \bar{r}\left( X\right) \right\rangle =0$%
. For simplicity, we assume that $z^{B}=z$ and $z_{0}^{B}=$ $z_{0}$,
implying that the banks' perception of investor's risk assessment is unbiased%
\footnote{%
This assumption can be relaxed if needed.}.

As expected, banks' loans to investors and firms depend positively on firms'
returns, $\left\langle f\left( X\right) \right\rangle $, and on investors'
shares,\ $z_{0}$. They can be written as: 
\begin{eqnarray}
&&\left\langle \hat{S}_{L}^{B}\left( X^{\prime }\right) \right\rangle
\label{LNP} \\
&=&\kappa \left( 1-2z_{0}\right) 2z_{0}\left( 1+z_{0}\left( 1-\frac{\left(
1-4z_{0}\right) \left( 1-2z_{0}\right) }{2\left( 1-5z_{0}+8z_{0}^{2}\right) }%
\right) \left\langle R_{exc}\left( X\right) \right\rangle \right.  \notag \\
&&\left. +\left( 1-2z_{0}\right) \left( \left\langle \hat{r}\left( X^{\prime
}\right) \right\rangle -\left\langle r\left( X\right) \right\rangle \right)
\right)  \notag
\end{eqnarray}%
\begin{eqnarray}
&&\left\langle S_{L}^{B}\left( X,X\right) \right\rangle  \label{LNDM} \\
&=&\kappa \left( 1-2z_{0}\right) ^{2}  \notag \\
&&\times \left( 1+z_{0}\left( 1+\frac{\left( 1-4z_{0}\right) z_{0}}{\left(
1-5z_{0}+8z_{0}^{2}\right) }\right) \left\langle R_{exc}\left( X\right)
\right\rangle +2z_{0}\left( \left\langle r\left( X\right) \right\rangle
-\left\langle \hat{r}\left( X^{\prime }\right) \right\rangle _{\hat{w}%
_{2}}\right) \right)  \notag
\end{eqnarray}%
An increase in firms' returns calls for additional capital, both directly
and indirectly, so that loans expand in proportion to the multiplier $\kappa 
$. However, this expansion in loans remains bounded by the assumption of
decreasing returns to scale.

\subsubsection{Adjusting for decreasing return to scale}

Accounting for decreasing returns to scale for firms amounts to replacing $%
\left\langle f_{1}\left( X\right) \right\rangle $ in the formulas for stakes
by:%
\begin{equation*}
\frac{\left\langle f_{1}\left( X\right) \right\rangle }{\left\langle
K\right\rangle ^{r}}-\frac{C}{\left\langle K\right\rangle }-C_{0}
\end{equation*}%
The previous interpretations remain valid. However, the impact of
productivity, $\left\langle f_{1}\left( X\right) \right\rangle $, is now
dampened by the level of disposable capital available to both firms and
investors.

\paragraph{Disposable capital for investors and associated agregate stakes}

The disposable capital available to investors is given by:%
\begin{equation}
\left\langle \hat{K}\right\rangle \left\Vert \hat{\Psi}\right\Vert
^{2}\simeq \frac{\hat{\mu}V\sigma _{\hat{K}}^{2}}{2}\left( \frac{\frac{%
\left\Vert \hat{\Psi}_{0}\right\Vert ^{2}}{\hat{\mu}}\left( 1-\hat{S}\right)
+\tau \left\langle \hat{f}\right\rangle \hat{S}}{\left\langle \hat{f}%
\right\rangle \left( 1-\hat{S}\right) }\right) ^{2}  \label{above}
\end{equation}%
where $\left\langle \hat{f}\right\rangle $ denotes the average return for
investors:%
\begin{equation}
\left\langle \hat{f}\right\rangle \simeq \left\langle \hat{r}\left(
X^{\prime }\right) \right\rangle +\frac{f_{a}-f_{b}\left( \frac{C_{0}+\bar{r}%
}{f_{1}\left( X\right) }\right) ^{\frac{2}{r}}}{2}  \label{Rpl}
\end{equation}%
and where $f_{a}$, $f_{b}$, and $\tau $ are some coefficients\footnote{%
These coefficients are defined in Apppendix 5.4 in Gosselin and Lotz (2025).
\par
.}, while the parameter $\left\Vert \hat{\Psi}_{0}\right\Vert ^{2}$
represents the average number of investors per sector.

In expression (\ref{above}), the parameter $\hat{\mu}$\ represents the
fluctuations in the number of agents within each sector. When $\hat{\mu}$\
is large, the number of agents can vary significantly.\ In such a case, the
average cease to be a reliable indicator for the level of disposable capital.

The associated average aggregate stakes are given by:%
\begin{eqnarray*}
\left\langle \hat{S}_{E}\left( X^{\prime }\right) \right\rangle
&=&\left\langle \hat{S}_{E}\left( X^{\prime },X\right) \right\rangle \\
\left\langle \hat{S}_{L}\left( X^{\prime }\right) \right\rangle
&=&\left\langle \hat{S}_{L}\left( X^{\prime },X\right) \right\rangle
\end{eqnarray*}%
\begin{eqnarray*}
\left\langle S_{E}\left( X\right) \right\rangle &=&\left\langle \hat{S}%
_{E}\left( X,X\right) \right\rangle \frac{\left\langle \hat{K}\right\rangle
\left\Vert \hat{\Psi}\right\Vert ^{2}}{\left\langle K\right\rangle
\left\Vert \Psi \right\Vert ^{2}} \\
\left\langle S_{L}\left( X\right) \right\rangle &=&\left\langle S_{L}\left(
X,X\right) \right\rangle \frac{\left\langle \hat{K}\right\rangle \left\Vert 
\hat{\Psi}\right\Vert ^{2}}{\left\langle K\right\rangle \left\Vert \Psi
\right\Vert ^{2}}
\end{eqnarray*}

\paragraph{Disposable capital for banks and associated aggregate stakes}

The disposable capital available to banks, $\left\langle \bar{K}%
\right\rangle \left\Vert \bar{\Psi}\right\Vert ^{2}$, is given by:%
\begin{equation*}
\left\langle \bar{K}\right\rangle \left\Vert \bar{\Psi}\right\Vert ^{2}=9%
\frac{\sigma _{\hat{K}}^{2}V\left\Vert \bar{\Psi}_{0}\right\Vert ^{4}\left(
1-\bar{S}\right) ^{2}}{2\hat{\mu}\left\langle \bar{f}\right\rangle ^{2}}
\end{equation*}%
where banks' returns are, to a first approximation, determined by their
returns on loans:%
\begin{equation*}
\left\langle \bar{f}\right\rangle \simeq \left( 1+\kappa \right) \bar{r}
\end{equation*}%
The associated average aggregate stakes are given by:%
\begin{eqnarray*}
\left\langle \bar{S}_{E}\left( X^{\prime }\right) \right\rangle
&=&\left\langle \bar{S}_{E}\left( X^{\prime },X\right) \right\rangle \\
\left\langle \bar{S}_{L}\left( X^{\prime }\right) \right\rangle
&=&\left\langle \bar{S}_{L}\left( X^{\prime },X\right) \right\rangle \frac{%
\left\langle \bar{K}\right\rangle \left\Vert \bar{\Psi}\right\Vert ^{2}}{%
\left\langle K\right\rangle \left\Vert \Psi \right\Vert ^{2}}
\end{eqnarray*}%
\begin{equation*}
\left\langle \hat{S}_{E}^{B}\left( X^{\prime },X\right) \right\rangle
=\left\langle \hat{S}_{E}^{B}\left( X^{\prime },X\right) \right\rangle \frac{%
\left\langle \bar{K}\right\rangle \left\Vert \bar{\Psi}\right\Vert ^{2}}{%
\left\langle \hat{K}\right\rangle \left\Vert \hat{\Psi}\right\Vert ^{2}}
\end{equation*}%
\begin{equation*}
\left\langle \hat{S}_{L}^{B}\left( X^{\prime },X\right) \right\rangle
=\left\langle \hat{S}_{L}^{B}\left( X^{\prime },X\right) \right\rangle \frac{%
\left\langle \bar{K}\right\rangle \left\Vert \bar{\Psi}\right\Vert ^{2}}{%
\left\langle \hat{K}\right\rangle \left\Vert \hat{\Psi}\right\Vert ^{2}}
\end{equation*}%
\begin{eqnarray*}
\left\langle S_{E}^{B}\left( X\right) \right\rangle &=&\left\langle
S_{E}^{B}\left( X,X\right) \right\rangle \frac{\left\langle \bar{K}%
\right\rangle \left\Vert \bar{\Psi}\right\Vert ^{2}}{\left\langle
K\right\rangle \left\Vert \Psi \right\Vert ^{2}} \\
\left\langle S_{L}^{B}\left( X\right) \right\rangle &=&\left\langle
S_{L}^{B}\left( X,X\right) \right\rangle \frac{\left\langle \bar{K}%
\right\rangle \left\Vert \bar{\Psi}\right\Vert ^{2}}{\left\langle
K\right\rangle \left\Vert \Psi \right\Vert ^{2}}
\end{eqnarray*}

\paragraph{Disposable capital for firms}

The disposable capital available to firms, $\left\langle K\right\rangle
\left\Vert \Psi \right\Vert ^{2}$, is given by: 
\begin{equation}
\left\langle K\right\rangle \left\Vert \Psi \right\Vert ^{2}\simeq \left( 1-%
\frac{\left\langle S\left( X,X\right) \right\rangle \left\langle \hat{K}%
\right\rangle \left\Vert \hat{\Psi}\right\Vert ^{2}+\left\langle S^{B}\left(
X,X\right) \right\rangle \left\langle \bar{K}\right\rangle \left\Vert \bar{%
\Psi}\right\Vert ^{2}}{\frac{2\epsilon }{3\sigma _{\hat{K}}^{2}}\left( \frac{%
\left\langle f_{1}\right\rangle }{C_{0}+\bar{r}}\right) ^{\frac{2}{r}}}%
\right) \left( \left\langle K\right\rangle \left\Vert \Psi \right\Vert
^{2}\right) _{0}
\end{equation}%
where the factor:%
\begin{equation*}
\left( \left\langle K\right\rangle \left\Vert \Psi \right\Vert ^{2}\right)
_{0}=\left( \left( \frac{2\epsilon }{3\sigma _{\hat{K}}^{2}}\right) ^{\frac{r%
}{2}}\frac{\left\langle f_{1}\left( X\right) \right\rangle }{C_{0}+\bar{r}}%
\right) ^{\frac{2}{r}}
\end{equation*}%
represents the firms' average disposable capital in the absence of
investors' stakes, i.e. when $\left\langle S\left( X,X\right) \right\rangle
=\left\langle S^{B}\left( X,X\right) \right\rangle =0$.

The presence of investors and banks reduces the share of firms' disposable
capital, producing an eviction effect. Due to decreasing returns, capital is
limited and caped at a certain level, a part of which is held by investors
and banks. Even if firms could reach this capital level without investors,
doing so would require considerable cost and a longer time span.

In general, firms' disposable capital, $\left\langle K\right\rangle
\left\Vert \Psi \right\Vert ^{2}$,\ increases with $\left\langle
f_{1}\right\rangle $: a global rise in firms' productivity, $\left\langle
f_{1}\left( X\right) \right\rangle $,\ increases their capital levels and
favours direct capital accumulation in firms, while reducing intermediation.
Overall, investors' total disposable capital, $\left\langle \hat{K}%
\right\rangle \left\Vert \hat{\Psi}\right\Vert ^{2}$,\ decreases with $%
\left\langle f_{1}\right\rangle $, although their private capital\footnote{%
Defined as $\left( 1-\left\langle \hat{S}\left( X^{\prime },X\right)
\right\rangle \right) \left\langle \hat{K}\right\rangle \left\Vert \hat{\Psi}%
\right\Vert ^{2}$.} increases.

Ultimately, introducing decreasing returns to scale amounts to replacing, in
all formulas for stakes, the excess return under constant returns to scale, $%
\left( \left\langle f_{1}\left( X\right) \right\rangle -\left\langle \hat{r}%
\left( X\right) \right\rangle \right) $,\ by its expression under decreasing
returns, $\left\langle f\right\rangle -\left\langle \hat{r}\left( X^{\prime
}\right) \right\rangle $, as\ defined in equation (\ref{Rpl}). The
interpretations remain valid, but the results are dampened by decreasing
returns.

Note that in all the above formulas, we assumed the same perception of
uncertainty for banks and investors.\ This implies that the parameters $x$
and $z$, measuring the impact of uncertainty on banks' and investors'
investments, are approximatively equal. This assumption can be relaxed by
treating $x$ and $z$ as independent parameters. The formulas derived above
remain valid in that case.

\subsection{Step 2: Writing the return equations in terms of returns}

As a first approximation, the integrals in equations (\ref{QDM}) and (\ref%
{QDB}) can be replaced by their averages.\ This allows the return equations
to be expressed as:%
\begin{eqnarray}
0 &=&\widehat{DF}\left( X\right) \hat{R}_{exc}\left( X\right) -\left\langle 
\hat{S}_{E}\left( X^{\prime },X\right) \right\rangle _{X^{\prime
}}\left\langle \widehat{DF}\left( X^{\prime }\right) \right\rangle
\left\langle \hat{R}_{exc}\left( X^{\prime }\right) \right\rangle
\label{PRm} \\
&&-S_{E}\left( X,X\right) R_{exc}\left( X\right)  \notag
\end{eqnarray}%
\begin{eqnarray}
0 &=&\frac{1-\bar{S}\left( X\right) }{1-\bar{S}_{E}\left( X\right) }\bar{R}%
_{exc}\left( X\right) -\left\langle \bar{S}_{E}\left( X^{\prime },X\right)
\right\rangle _{X^{\prime }}\left\langle \widehat{DF}\left( X^{\prime
}\right) \right\rangle \left\langle \bar{R}_{exc}\left( X^{\prime }\right)
\right\rangle  \label{SCm} \\
&&-\left\langle \hat{S}_{E}^{B}\left( X^{\prime },X\right) \right\rangle
_{X^{\prime }}\left\langle \widehat{DF}\left( X^{\prime }\right)
\right\rangle \left\langle \hat{R}_{exc}\left( X^{\prime }\right)
\right\rangle -S_{E}^{B}\left( X,X\right) R_{exc}\left( X\right)  \notag
\end{eqnarray}%
The various parameters are defined in Appendix 10.

Equation (\ref{PRm}) involves several averages computed in step 1, namely $%
\left\langle \hat{S}\left( X^{\prime }\right) \right\rangle $, $\left\langle 
\hat{S}\left( X^{\prime }\right) \right\rangle $, and $\left\langle \hat{f}%
\left( X^{\prime }\right) \right\rangle $, as well as the returns $\hat{f}%
\left( X\right) $ and $f\left( X\right) $, the aggregate stakes $\hat{S}%
\left( X\right) $, $\hat{S}_{E}\left( X\right) $, and $\left\langle \hat{S}%
_{E}\left( X^{\prime },X\right) \right\rangle _{X^{\prime }}$,\ and the
shares in firms, $S_{E}\left( X,X\right) $, $\bar{S}_{E}\left( X\right) $, $%
\bar{S}\left( X\right) $, and $\left\langle \bar{S}_{E}\left( X^{\prime
},X\right) \right\rangle _{X^{\prime }}$, as given by equations (\ref{BCSn}%
), (\ref{BCSt}), and (\ref{TWr}), respectively.

\paragraph{Investors' return equations}

Using equations (\ref{Shn}), (\ref{Sht}), (\ref{Shv}), along with the risk
coefficients (\ref{hb}) and (\ref{hc}),\ we can replace the average stakes
by functions of returns and average variables, and recast explicitly
equation (\ref{PRm}) as a return equation for $\hat{f}\left( X\right) $%
\footnote{%
See Appendix 7 in Gosselin and Lotz (2025).}:%
\begin{equation}
0=\frac{\left\langle \hat{f}\right\rangle ^{2}\left( \frac{\Xi }{1-\Xi }%
\right) ^{2}-\Xi \left( \hat{f}^{\prime }\left( X\right) \right) ^{2}}{%
\left\langle \hat{f}\right\rangle ^{2}\left( \frac{\Xi }{1-\Xi }\right) ^{2}-%
\frac{1}{2}A\left( 1+\hat{\Delta}^{E}\left( X\right) \right) \left( \hat{f}%
^{\prime }\left( X\right) \right) ^{2}}\hat{R}_{exc}\left( X\right) -H\left(
X\right)  \label{DTR}
\end{equation}%
where:%
\begin{equation*}
\hat{f}^{\prime }\left( X\right) =\hat{f}\left( X\right) \left(
1-\left\langle \hat{S}\right\rangle \right) +\left\langle \hat{S}\left(
X^{\prime },X\right) \right\rangle _{X^{\prime }}\left\langle \hat{f}%
\right\rangle
\end{equation*}%
\begin{equation*}
\Xi =A\left( 1+\hat{\Delta}\left( X\right) \right)
\end{equation*}%
and:%
\begin{eqnarray*}
H\left( X\right) &=&\hat{w}\left( X\right) \left( 1-w\left( X\right) \hat{%
\Delta}\left( X\right) +\frac{1}{2}\left\langle \hat{R}_{exc}\left(
X^{\prime }\right) \right\rangle \right) \\
&&\times \left\langle \widehat{DF}\left( X^{\prime }\right) \right\rangle
\left\langle \hat{R}_{exc}\left( X^{\prime }\right) \right\rangle
+S_{E}\left( X,X\right) R_{exc}\left( X\right)
\end{eqnarray*}%
Here, $\hat{\Delta}^{E}\left( X\right) $ and $\hat{\Delta}\left( X\right) $
are defined by\ (\ref{dtf}) and (\ref{dtR}), respectively,\ and the
parameter $A$\ is defined in Appendix 7 in Gosselin and Lotz (2025).

\paragraph{Banks' return equations}

The same method applies to banks. Using equations (\ref{SBN}), (\ref{SBT}), (%
\ref{SBF}), (\ref{scb}), (\ref{sdb}), and (\ref{SFN}) for banks, along with
the risk coefficients (\ref{cfh}) and (\ref{Cfn}), the return equation for $%
\bar{f}\left( X\right) $ becomes: 
\begin{equation*}
0=\frac{1-\bar{A}\left( 1+\bar{\Delta}\left( X\right) \right) \frac{%
\left\langle \bar{K}\right\rangle \left\Vert \bar{\Psi}\right\Vert ^{2}}{%
\bar{K}_{X}\left\vert \bar{\Psi}\left( X\right) \right\vert ^{2}}}{1-\frac{1%
}{2}\bar{A}\left( 1+\bar{\Delta}^{E}\left( X\right) \right) \frac{%
\left\langle \bar{K}\right\rangle \left\Vert \bar{\Psi}\right\Vert ^{2}}{%
\bar{K}_{X}\left\vert \bar{\Psi}\left( X\right) \right\vert ^{2}}}\bar{R}%
_{exc}\left( X\right) -\bar{G}
\end{equation*}%
This equation can also be expressed in expanded form: 
\begin{equation}
0=\frac{\left\langle \bar{f}\right\rangle ^{2}D-\bar{A}\left( 1+\bar{\Delta}%
\left( X\right) \right) \left( \bar{f}\left( X\right) \left( 1-\left\langle 
\bar{S}\right\rangle \right) +\left\langle \bar{S}\left( X^{\prime
},X\right) \right\rangle _{X^{\prime }}\left\langle \bar{f}\right\rangle
\right) ^{2}}{\left\langle \bar{f}\right\rangle ^{2}D-\frac{1}{2}\bar{A}%
\left( 1+\bar{\Delta}^{E}\left( X^{\prime }\right) \right) \left( \bar{f}%
\left( X\right) \left( 1-\left\langle \bar{S}\right\rangle \right)
+\left\langle \bar{S}\left( X^{\prime },X\right) \right\rangle _{X^{\prime
}}\left\langle \bar{f}\right\rangle \right) ^{2}}\bar{R}_{exc}\left(
X\right) -\bar{G}  \label{DTS}
\end{equation}%
with: 
\begin{equation*}
D=\frac{\bar{A}\left( 1+\bar{\Delta}\left( X\right) \right) }{1-\bar{A}%
\left( 1+\bar{\Delta}\left( X\right) \right) }
\end{equation*}%
and:%
\begin{eqnarray}
\bar{G} &=&\left\langle \bar{S}_{E}\left( X^{\prime },X\right) \right\rangle
_{X^{\prime }}\frac{1-\left\langle \bar{S}\left( X^{\prime }\right)
\right\rangle }{1-\left\langle \bar{S}_{E}\left( X^{\prime }\right)
\right\rangle }\left\langle \bar{R}_{exc}\left( X^{\prime }\right)
\right\rangle \\
&&+\left\langle \hat{S}_{E}^{B}\left( X^{\prime },X\right) \right\rangle
_{X^{\prime }}\left\langle \widehat{DF}\left( X^{\prime }\right)
\right\rangle \left\langle \hat{R}_{exc}\left( X^{\prime }\right)
\right\rangle +S_{E}^{B}\left( X,X\right) R_{exc}\left( X\right)  \notag
\end{eqnarray}%
The formula for $\bar{A}$ is given in Appendix 9.

\subsection{Step 3: solving for sectoral returns and aggregate stakes}

Having established the average solutions, we now refine the analysis to
identify sector-specific equilibria. In contrast with the previous study,
the incorporation of banks modifies the distribution of investors' returns.
The principle of a dual solution---where high- and low- return states
coexist within the same sectors---remains valid for both banks and investors.

For investors, however, such a configuration arises only under specific
conditions. When these conditions are not satisfied, the system converges
toward a unique solution, characterized by returns clustered around the
overall average across all investors. In the following, we examine
separately the return structures of investors and of banks.

\subsubsection{Investors' returns}

Investors' returns per sector depend on three coefficients. First, the
investors' estimation of risks in investors stakes, denoted $z$ and defined
by\footnote{%
The link between $z$ and $z_{0}$, the value of $\left\langle \hat{S}%
_{E}\left( X^{\prime },X\right) \right\rangle $ under the constraint $%
\left\langle f\left( X\right) \right\rangle -\left\langle \bar{r}\left(
X\right) \right\rangle =0$, is given in Appendix 5.4.1 of Gosselin and Lotz
(2025). In first approximation, we can consider $z=z_{0}$.}:%
\begin{equation*}
z=\left\langle \hat{S}_{E}\left( X^{\prime },X\right) \right\rangle
\end{equation*}%
Second, the relative strength of banks' investments in investors, denoted%
\footnote{%
To the first approximation we identify $x=\left\langle \bar{S}_{E}\left(
X^{\prime },X\right) \right\rangle =\left\langle \bar{S}_{E}\right\rangle
_{0}$ with $\left\langle \bar{S}_{E}\right\rangle _{0}$ defined in section
12.1.3.. First order corrections to this approximation are defined in
Appendix 6.5.2.} $x$:%
\begin{equation*}
x=\left\langle \bar{S}_{E}\left( X^{\prime },X\right) \right\rangle
\end{equation*}%
Third, the relative average number of banks compared to investors, denoted $%
P $, defined by:%
\begin{equation*}
P=\frac{\left\Vert \bar{\Psi}_{0}\right\Vert ^{2}}{\left\Vert \hat{\Psi}%
_{0}\right\Vert ^{2}}
\end{equation*}%
The higher the ratio $P$, the higher the importance of the banking system
and its impact on intermediation. The higher the uncertainty of banks
regarding investors, the lower are $x$ and the impact of banks on investors.
Under high uncertainty, banks reduce both participations and loans.
Conversely, the higher the coefficient $z$, the higher is the intermediation
among investors, and thereby the higher the impact of banks on investors.

Two types of solutions emerge depending on the three parameters $P$, $x$,
and $z$.

\paragraph{First case: a unique solution}

We show that when $P$ exceeds a certain threshold\footnote{%
See Appendix 9.}:%
\begin{equation*}
P>Th
\end{equation*}%
with:%
\begin{equation*}
Th=\frac{2\left( 2z-1\right) ^{2}\left( A-1\right) }{\left( 1-2x\right) }+%
\frac{\sqrt{\left( 2\left( 2z-1\right) ^{2}\left( A-1\right) \right)
^{2}-\left( 1-2x\right) \left( \left( A-1\right) \left( 4-4z-\frac{1}{z}%
\right) \right) }}{\left( 1-2x\right) }
\end{equation*}%
and:%
\begin{equation*}
A=\frac{2-24z+112z^{2}-224z^{3}+160z^{4}}{1-9z+26z^{2}-12z^{3}+8z^{4}}
\end{equation*}%
In this case, the investors' return equations per sector (\ref{DTR}) have
only a unique solution, which is close to the average solution:%
\begin{equation}
\hat{R}_{exc}\left( X\right) \simeq z\left( \left\langle \hat{R}_{exc}\left(
X^{\prime }\right) \right\rangle \right) +\left( 1-z\right) R_{exc}\left(
X\right)  \label{UniqInv}
\end{equation}

\paragraph{Second case: two solutions}

Conversely, when the relative average number of banks is below the threshold:%
\begin{equation*}
P<Th
\end{equation*}%
the investors' return equations per sector (\ref{DTR}) have two solutions,
as in Gosselin and Lotz (2025). These two solutions\ for investors' excess
returns depend on firms' excess returns per sector as well as the average
returns in the system. They are given by\footnote{%
See Appendix 7 in Gosselin and Lotz (2025) for the derivation.}:%
\begin{equation}
\hat{R}_{exc}^{H}\left( X\right) =\frac{\hat{R}_{exc}^{H,0}\left( X\right) }{%
2}+\sqrt{\left( \frac{\hat{R}_{exc}^{H,0}\left( X\right) }{2}\right) ^{2}-%
\frac{1-2z}{1-z}\frac{\left\langle \bar{r}\left( X\right) \right\rangle }{%
\left\vert a\left( z,P\right) \right\vert }\hat{R}_{exc}^{L,0}\left(
X\right) }  \label{SLn}
\end{equation}%
and:%
\begin{equation}
\hat{R}_{exc}^{L}\left( X\right) =\frac{\hat{R}_{exc}^{H,0}\left( X\right) }{%
2}-\sqrt{\left( \frac{\hat{R}_{exc}^{H,0}\left( X\right) }{2}\right) ^{2}-%
\frac{1-2z}{1-z}\frac{\left\langle \bar{r}\left( X\right) \right\rangle }{%
\left\vert a\left( z,P\right) \right\vert }\hat{R}_{exc}^{L,0}\left(
X\right) }  \label{Slp}
\end{equation}%
with:%
\begin{equation}
\hat{R}_{exc}^{H,0}\left( X\right) =\frac{\frac{1-2z}{1-z}\left\langle \bar{r%
}\left( X\right) \right\rangle +z\left( b\left( z,L\right) \left\langle \hat{%
R}_{exc}\left( X^{\prime }\right) \right\rangle +\frac{1}{2}\frac{1-2z}{%
\left( z-1\right) ^{2}}R_{exc}\left( X\right) \right) }{2\left\vert a\left(
z,P\right) \right\vert }  \label{dernier1}
\end{equation}%
and:%
\begin{equation}
\hat{R}_{exc}^{L,0}\left( X\right) =z\left\langle \hat{R}_{exc}\left(
X^{\prime }\right) \right\rangle +\frac{\left( 1-z\right) }{2}R_{exc}\left(
X\right)  \label{dernier2}
\end{equation}%
Here, the functions $a\left( z,P\right) $\ and $b\left( z,L\right) $\ are
defined in Appendix 10.5. Note that the firms' excess return, $R_{exc}\left(
X\right) $,\ is given by $\left( f_{1}\left( X\right) -\left\langle \hat{r}%
\left( X^{\prime }\right) \right\rangle \right) $\ under constant returns to
scale, and by $\frac{1}{2}f_{a}-\frac{1}{2}f_{b}\left( \frac{C_{0}+\bar{r}}{%
f_{1}\left( X\right) }\right) ^{\frac{1}{r}}$\ under decreasing returns to
scale\footnote{%
See Appendix 8.2.}. The interpretations are similar in both cases, but with
dampened amplitudes under decreasing returns.

Obviously, the two solutions (\ref{SLn}) and (\ref{Slp}) differ whether the
second term is added or substracted:%
\begin{equation}
\sqrt{\left( \frac{\hat{R}_{exc}^{H,0}\left( X\right) }{2}\right) ^{2}-\frac{%
1-2z}{1-z}\frac{\left\langle \bar{r}\left( X\right) \right\rangle }{%
\left\vert a\left( z,P\right) \right\vert }\hat{R}_{exc}^{L,0}\left(
X\right) }  \label{dernier3}
\end{equation}%
When it is added, the high-return solution (\ref{SLn}) arises, in which
returns are relatively higher. When it is substracted, we obtain the
low-return solution (\ref{Slp}).

When excess returns are relatively small compared to the interest rate,
these two solutions can be expressed as:%
\begin{equation}
\hat{R}_{exc}^{H}\left( X\right) =\hat{R}_{exc}^{H,0}\left( X\right) -\hat{R}%
_{exc}^{L,0}\left( X\right)  \label{HRet}
\end{equation}%
and:%
\begin{equation}
\hat{R}_{exc}^{L}\left( X\right) =\hat{R}_{exc}^{L,0}\left( X\right) \left(
1-\frac{2\left\vert a\left( z,P\right) \right\vert \left( 1-z\right) }{%
\left( 1-2z\right) }\left( \frac{\left\langle \hat{R}_{exc}\left( X\right)
\right\rangle }{\left\langle \bar{r}\right\rangle }+\frac{R_{exc}\left(
X\right) }{\left\langle \bar{r}\right\rangle }\right) \right)  \label{LRet}
\end{equation}%
Note that the high-return solution occurs only under certain conditions. For
instance, when uncertainty is high, $\hat{R}_{exc}^{H}\left( X\right) $\
would need to be very high to offset the uncertainty, which rules out this
solution practically. Therefore, the high-return solution arises only when
investors' risk perception is low.

\subsubsection{Banks' returns}

For banks, there are two solutions that have a form similar to those of
investors:%
\begin{equation*}
\bar{R}_{exc}\left( X\right) =\frac{\bar{R}_{exc}^{H,0}\left( X\right) }{2}%
\pm \sqrt{\left( \left( \frac{\bar{R}_{exc}^{H,0}\left( X\right) }{2}\right)
^{2}-\left( \kappa +2\right) \left( \kappa +1\right) \frac{1-x}{x}\bar{R}%
_{exc}^{L,0}\left( X\right) \right) ^{2}}
\end{equation*}%
with:%
\begin{equation*}
\bar{R}_{exc}^{H,0}\left( X\right) =\left( \kappa +2\right) \left( \kappa
+1\right) \left( \frac{1-x}{2x}\left\langle \bar{r}\left( X^{\prime }\right)
\right\rangle +\frac{\left( \left\langle f\left( X\right) \right\rangle
-\left\langle \bar{r}\left( X^{\prime }\right) \right\rangle \right) }{4}%
+\left( \left\langle \hat{f}\left( X^{\prime }\right) \right\rangle
-\left\langle \bar{r}\left( X^{\prime }\right) \right\rangle \right) x\right)
\end{equation*}%
and:%
\begin{eqnarray*}
&&\bar{R}_{exc}^{L,0}\left( X\right) =\left\langle \bar{S}_{E}\left(
X^{\prime },X\right) \right\rangle _{X^{\prime }}\left\langle \overline{DF}%
\left( X^{\prime }\right) \right\rangle \left( \left\langle \bar{f}\left(
X^{\prime }\right) \right\rangle -\bar{r}\right) \\
&&+\left\langle \hat{S}_{E}^{B}\left( X^{\prime },X\right) \right\rangle
_{X^{\prime }}\left\langle \widehat{DF}\left( X^{\prime }\right)
\right\rangle \left( \left\langle \hat{f}\left( X^{\prime }\right)
\right\rangle -\bar{r}\right) +S_{E}^{B}\left( X,X\right) \left( f_{1}\left(
X\right) _{dr}-\bar{r}\right)
\end{eqnarray*}%
Since $\frac{2x}{\left( \kappa +2\right) \left( \kappa +1\right) }<<1$, the
low-return solution is close to the average:%
\begin{eqnarray*}
&&\bar{R}_{exc}^{L}\left( X\right) \\
&\simeq &\hat{R}_{exc}^{L,0}\left( X\right) \left( 1-\frac{2x}{\left(
1-x\right) \left( \kappa +2\right) \left( \kappa +1\right) }\bar{R}%
_{exc}^{H,0}\left( X\right) \right) \\
&\simeq &\bar{R}_{exc}^{L,0}\left( X\right)
\end{eqnarray*}%
This can be rewritten in terms of banks' uncertainty $x$ as:%
\begin{equation*}
\bar{R}_{exc}\left( X\right) \simeq x\left( \left\langle \bar{f}\left(
X^{\prime }\right) \right\rangle -\left( \kappa +1\right) \bar{r}\right)
+4x\left( \left\langle \hat{f}\left( X^{\prime }\right) \right\rangle -\bar{r%
}\right) +\left( 1-6x\right) \left( f\left( X\right) -\bar{r}\right)
\end{equation*}%
Acting as an investor, that is loans excluded, banks' returns combine their
investments in other banks, in investors, and in firms. The corresponding
coefficients depend on uncertainty. Under high uncertainty, banks primarily
invest in firms, whereas under low uncertainty, they diversify. In this
case, the shares in investors are larger than in other banks, since
intermediation through investors provides a more direct path than the
bank-investor-firm path.

The high-return solution is expressed as:%
\begin{equation*}
\bar{R}_{exc}^{H}\left( X\right) \simeq \bar{R}_{exc}^{H,0}\left( X\right)
\end{equation*}

\subsubsection{Interpretation of investors' sectoral returns}

As mentioned above, the introduction of banks modifies the structure of
investors' returns. By allocating capital between firms and investors, banks
facilitate access to capital and reduce the pressure that investors exert on
returns. When banks are able to perform this reallocative role effectively,
investors' returns tend to cluster around an average, with variations
reflecting sector-specific differences. Conversely, when banks do not
fulfill this function, or when they are too few in number to provide firms
with the necessary capital, the dual outcome of high and low returns
described in Gosselin and Lotz (2025) persists, albeit in a mitigated form.
In the following, we examine these two types of investor equilibria in turn.

\paragraph{First case: unique solution}

When the ratio of banks to investors is sufficiently high, bank loans and
participations are sufficient to smooth capital allocation. Recall that the
unique solution is given by equation (\ref{UniqInv}):%
\begin{equation*}
\hat{R}_{exc}\left( X\right) \simeq z\left( \left\langle \hat{R}_{exc}\left(
X^{\prime }\right) \right\rangle \right) +\left( 1-z\right) R_{exc}\left(
X\right)
\end{equation*}%
The upward pressure on returns is dampened: investors' returns converge
toward the average solution, aside from sector-specific discrepancies among
investors. This induces investors to seek higher returns through
diversification between direct investments $\left( 1-z\right) R_{exc}\left(
X\right) $ and cross-holdings with other investors $z\left( \left\langle 
\hat{R}_{exc}\left( X^{\prime }\right) \right\rangle \right) $.

The threshold governing this outcome depends on the relative number of banks
to investors and on the parameters $z$ and $x$, which represent the inverse
perception of risk by investors and banks, respectively. This threshold
increases with $x$: the greater the confidence of banks in investors, the
larger the banking system must be to smooth returns. In other words, for a
given relative number of banks, an increase in $x$ may shift the system
toward a dual-solution regime for the investors' return equations.

\paragraph{Second case: two solutions}

When banks are scarce, or fail to reallocate capital between firms and
investors, the dual outcome of high and low returns identified in Gosselin
and Lotz (2025) persists, albeit in an attenuated form.

\subparagraph{Low-return solution}

Under this solution, investors' excess returns are given, in last analysis,
by equation (\ref{LRet}). Using the definition (\ref{dernier2}) of $\hat{R}%
_{exc}^{L,0}\left( X\right) $, this solution can be analyzed as a sum of
returns over investors:%
\begin{equation}
z\left\langle \hat{R}_{exc}\left( X^{\prime }\right) \right\rangle
\end{equation}%
and over firms:%
\begin{equation}
\frac{\left( 1-z\right) }{2}R_{exc}\left( X\right)  \label{CT2}
\end{equation}%
both terms are dampened by the factor:%
\begin{equation}
1-\frac{2\left\vert a\left( z,P\right) \right\vert \left( 1-z\right) }{%
\left( 1-2z\right) }\left( \frac{\left\langle \hat{R}_{exc}\left( X\right)
\right\rangle }{\left\langle \bar{r}\right\rangle }+\frac{R_{exc}\left(
X\right) }{\left\langle \bar{r}\right\rangle }\right)  \label{dfc}
\end{equation}%
The contributions (\ref{CTn}) and (\ref{CT2}) can be interpreted as follows:
when uncertainty is high, the parameter $z$ is low, and investors primarily
allocate capital to firms within their own sector, as expected. When
uncertainty is low, investors diversify their portfolios - and stakes -
between direct investments in firms and cross-holdings with other investors.
Under the low-return solution, however, investors' returns are largely
determined by their sectoral firm investments, and only marginally by
intermediation through other investors:\ their exposure to risk is thus
reduced. Sectoral returns differ from the average only in that average firm
returns are replaced by sector-specific ones.

As noted above, these contributions are dampened by the factor (\ref{dfc}),
which implies that investors' average excess returns under the low-return
solution are systematically lower than the overall average. Specifically,
low-return investors experience a loss equal to a fraction of their average
excess returns, $\frac{\left\langle \hat{R}_{exc}\left( X\right)
\right\rangle }{\left\langle \bar{r}\right\rangle }+\frac{R_{exc}\left(
X\right) }{\left\langle \bar{r}\right\rangle }$. This fraction is directly
linked to the existence of the high-return solution, in which high-return
investors achieve above-average returns at the expense of low-return
investors and firms. Since the high-return solution is facilitated by low
uncertainty, this fraction decreases as uncertainty rises.

\subparagraph{High-return solution}

Under this solution, cross-sectoral investments - loans and participations
in investors - drive the returns through the term:%
\begin{equation*}
\hat{R}_{exc}^{H,0}\left( X\right)
\end{equation*}%
However, to properly assess these higher returns, one must substract the
contribution from the low-return solution:%
\begin{equation*}
\hat{R}_{exc}^{L,0}\left( X\right)
\end{equation*}%
Overall, the solution $\hat{R}_{exc}^{H}\left( X\right) $ decomposes into
three terms:%
\begin{equation*}
\hat{R}_{exc}^{H}\left( X\right) =\left( 1-z\right) \left\langle \hat{R}%
_{exc}\left( X^{\prime }\right) \right\rangle +\left( \frac{z\left(
1-2z\right) }{\left( 1-z\right) ^{2}\left\vert a\left( z,P\right)
\right\vert }-\frac{1-z}{2}\right) \left\langle R_{exc}\left( X^{\prime
}\right) \right\rangle +\frac{\left( 1-z\right) }{\left\vert a\left(
z,P\right) \right\vert }\left\langle \bar{r}\left( X\right) \right\rangle
\end{equation*}%
When the perception of risk is low, cross-investments (the first term)
exceed investments in firms (the second term). \ Under this solution, due to
firms' low productivity and returns, investors allocate more capital to
other investors than to firms in their own sector. The third term captures
the additional profit that investors with high disposable capital can
negociate from loans to other investors.

Structurally, the high-return solution is systematically greater than the
low-return solution, since investors with high disposable capital can demand
higher interest rates and returns. As noted above, these additional returns
are realized at the expense of low-return investors; therefore, for any
given average, it is the discrepancy between high- and low-return investors
that determines the overall average solution.

The coefficient $\left\vert a\left( z,P\right) \right\vert $\ is increasing
in $P$, so that any increase in the relative number of banks reduces the
high-return solution $\hat{R}_{exc}^{H}\left( X\right) $\ and increases the
low-return solution $\hat{R}_{exc}^{L}\left( X\right) $. In this way, the
relative size of the banking system mitigates the discrepancy between the
two solutions.

\subsubsection{Interpretation of banks' sectoral returns}

The solution for banks reflects the fact that, when acting as investors,
they diversify their allocations across other banks, investors, and firms.
The level of investment in investors is higher than in other banks due to
uncertainty, as the intermediation of investors is more direct than that of
banks. However, when uncertainty is very high, banks---like other
investors---tend to invest directly in firms. A second solution, analogous
to the high-return strategy, emerges under conditions of very high excess
returns\footnote{%
See Appendix 9.}.

\subsubsection{Implications for the collective state}

The collective state thus consists of two types of solutions. The low-return
solution represents the standard outcome, similar to the results obtained
for average returns, except for the dampening factor induced by the presence
of high-return solutions. In contrast, the high-return solution corresponds
to a deviation from these standard returns. It describes investors who
maintain very high level of capital by attracting other investors and
extracting higher returns from both firms and other investors.\ Other
investors in the system, by comparison, receive lower-than-average returns.\
Regarding loans, firms and investors face varying interest rates depending
on whether they borrow from high-return or low-return investors.

This suggests that, in general, returns are primarily driven by firms'
returns, whereas superior returns are largely determined by cross-sectoral
investments made by investors with high levels of disposable capital.

As we will show later, these two solutions - high and low - can affect the
stability of the collective states. Even small changes in returns may
trigger large capital flows, potentially destabilizing the collective state
and shifting the system toward alternative states, including possibly
default states.

\subsubsection{Outward aggregate stakes}

Having established the possible solutions for sectoral returns, we can now
compute the associated levels of stakes and capital.

\paragraph{Outward aggregate stakes in investors}

The outward aggregate stakes (\ref{Twn}) and (\ref{Twh}) compute the stakes
of investors $X$\ outside their own sector.

The proportion of capital invested by investor $X$\ in other investors, $%
\left\langle \hat{S}_{E}\left( X^{\prime },X\right) \right\rangle
_{X^{\prime }}$, is:\textbf{\ }%
\begin{equation}
\left\langle \hat{S}_{E}\left( X^{\prime },X\right) \right\rangle
_{X^{\prime }}\simeq \frac{1}{2}\left\langle \hat{w}\left( X^{\prime
},X\right) \right\rangle _{X^{\prime }}\left( 1+\hat{\Delta}^{E}\left(
X\right) \right)  \label{Shg}
\end{equation}%
Similarly, the proportion of loans granted by investor $X$\ to other
investors, $X^{\prime }$ $\left\langle \hat{S}_{E}\left( X^{\prime
},X\right) \right\rangle _{X^{\prime }}$, is:%
\begin{equation}
\left\langle \hat{S}_{L}\left( X^{\prime },X\right) \right\rangle
_{X^{\prime }}\simeq \frac{1}{2}\left\langle \hat{w}\left( X^{\prime
},X\right) \right\rangle _{X^{\prime }}\left( 1+\hat{\Delta}^{L}\left(
X\right) \right)  \notag
\end{equation}%
where the coefficients of proportionality of investors $X$ investments in
other investors and in firms are given by the averages of formula (\ref{hb}%
), and can be written, respectively, as:%
\begin{equation}
\left\langle \hat{w}\left( X^{\prime },X\right) \right\rangle =\frac{\left(
1-\left( \gamma \left\langle \hat{S}_{E}\left( X\right) \right\rangle
\right) ^{2}\right) }{2-\left( \gamma \left\langle \hat{S}_{E}\left(
X\right) \right\rangle \right) ^{2}}  \label{CFt}
\end{equation}%
and:%
\begin{equation}
w\left( X\right) =1-\left\langle \hat{w}\left( X^{\prime },X\right)
\right\rangle  \label{CFtbis}
\end{equation}%
From formula (\ref{CFt}), we infer that the higher the average perceived
risk, $\left( \gamma \left\langle \hat{S}_{E}\left( X\right) \right\rangle
\right) ^{2}$, the lower the coefficient $\left\langle \hat{w}\left(
X^{\prime },X\right) \right\rangle $\ and, consequently, the lower the
participations between investors.

Recall that the quantity $\Delta \left( X\right) $\ measures the relative
return of firms $X$\ across equity and loans.\ Hence, the higher the returns
of firms $X$, the less investors $X$ allocate capital externally, and this
capital is further diversified among multiple investors $X^{\prime }$. This
corresponds to the low-return solution, where firms' productivity is
sufficient to attract investments and sustain global levels of returns.
Under the high-return solution, due to the high level of returns extracted
from investors $\left\langle \hat{f}\left( X^{\prime }\right) \right\rangle $%
, investors will rather invest externally.

\paragraph{Outward aggregate banks cross-stakes}

The outward aggregate stakes of banks were defined in (\ref{TWr}) and (\ref%
{Twh}). For aggregate banks' cross-stakes, we find:

\begin{equation}
\left\langle \bar{S}_{E}\left( X^{\prime },X\right) \right\rangle
_{X^{\prime }}\simeq \frac{\left\langle \bar{w}\left( X^{\prime },X\right)
\right\rangle }{2}\left( 1+\left\langle \bar{\Delta}^{E}\left( X^{\prime
},X\right) \right\rangle _{X^{\prime }}\right)
\end{equation}%
\begin{equation*}
\left\langle \bar{S}_{L}\left( X^{\prime },X\right) \right\rangle
_{X^{\prime }}\simeq \frac{\left\langle \bar{w}\left( X^{\prime },X\right)
\right\rangle }{2}\left( 1+\left\langle \bar{\Delta}^{L}\left( X^{\prime
},X\right) \right\rangle _{X^{\prime }}\right)
\end{equation*}%
Consequently, we have:%
\begin{equation*}
\left\langle \bar{S}\left( X^{\prime },X\right) \right\rangle _{X^{\prime }}=%
\frac{\left\langle \bar{w}\left( X^{\prime },X\right) \right\rangle }{2}%
\left( 1+\left\langle \bar{\Delta}\left( X^{\prime },X\right) \right\rangle
_{X^{\prime }}\right)
\end{equation*}%
where:%
\begin{equation*}
\left\langle \bar{\Delta}^{E}\left( X^{\prime },X\right) \right\rangle
_{X^{\prime }}=\left\langle \bar{f}\left( X^{\prime }\right) \right\rangle -%
\bar{w}_{L}\left( X\right) \left\langle \bar{r}\left( X^{\prime }\right)
\right\rangle _{\bar{w}_{L}}-\hat{w}_{E}^{B}\left( X\right) \left\langle 
\hat{f}\left( X^{\prime }\right) \right\rangle _{\hat{w}_{E}}-w_{E}^{B}%
\left( X\right) f\left( X\right)
\end{equation*}%
\begin{equation*}
\left\langle \bar{\Delta}^{L}\left( X^{\prime },X\right) \right\rangle
_{X^{\prime }}=\left\langle \bar{r}\left( X^{\prime }\right) \right\rangle -%
\bar{w}_{E}\left( X\right) \left\langle \bar{f}\left( X^{\prime }\right)
\right\rangle _{\bar{w}_{E}}-\hat{w}_{E}^{B}\left( X\right) \left\langle 
\hat{f}\left( X^{\prime }\right) \right\rangle _{\hat{w}_{E}}-w_{E}^{B}%
\left( X\right) f\left( X\right)
\end{equation*}%
and:%
\begin{equation*}
\left\langle \bar{\Delta}\left( X^{\prime },X\right) \right\rangle
_{X^{\prime }}=\frac{\left\langle \bar{\Delta}^{E}\left( X^{\prime
},X\right) \right\rangle _{X^{\prime }}+\left\langle \bar{\Delta}^{L}\left(
X^{\prime },X\right) \right\rangle _{X^{\prime }}}{2}
\end{equation*}

\paragraph{Outward aggregate banks' stakes in investors}

Similarly, we obtain:%
\begin{equation*}
\left\langle \hat{S}_{E}^{B}\left( X^{\prime },X\right) \right\rangle
_{X^{\prime }}\simeq \left\langle \hat{w}_{E}^{B}\left( X^{\prime },X\right)
\right\rangle _{X^{\prime }}\left( 1+\left\langle \left( \hat{\Delta}%
^{E}\right) ^{B}\left( X^{\prime },X\right) \right\rangle _{X^{\prime
}}\right)
\end{equation*}%
\begin{equation*}
\frac{\left\langle \hat{S}_{L}^{B}\left( X^{\prime },X\right) \right\rangle
_{X^{\prime }}}{\kappa \left( 1-\bar{S}\left( X\right) \right) }\simeq
\left\langle \hat{w}_{L}^{B}\left( X^{\prime },X\right) \right\rangle
_{X^{\prime }}\left( 1+\left\langle \left( \hat{\Delta}L\right) ^{B}\left(
X^{\prime },X\right) \right\rangle _{X^{\prime }}\right)
\end{equation*}%
with:%
\begin{equation*}
\left\langle \hat{\Delta}^{E}\left( X^{\prime },X\right) \right\rangle
_{X^{\prime }}=\left\langle \hat{f}\left( X^{\prime }\right) \right\rangle -%
\frac{1}{2}\left( \left\langle \bar{f}\left( X^{\prime }\right)
\right\rangle _{\bar{w}_{E}}+\left\langle \bar{r}\left( X^{\prime }\right)
\right\rangle _{\bar{w}_{L}}\right) -\left\langle w_{E}^{B}\left( X\right)
\right\rangle f\left( X\right)
\end{equation*}%
and:%
\begin{equation*}
\left\langle \hat{\Delta}^{L}\left( X^{\prime },X\right) \right\rangle
_{X^{\prime }}=\left\langle \hat{r}\left( X^{\prime }\right) \right\rangle
-\left\langle \hat{f}\left( X^{\prime }\right) \right\rangle _{\hat{w}%
_{E}}-\left\langle w_{E}\left( X\right) \right\rangle f\left( X\right)
\end{equation*}

\subsubsection{Capital level per sector}

Under decreasing returns to scale, firms' returns are:%
\begin{equation*}
f\left( X\right) =\frac{f_{1}\left( X\right) }{\left( K_{X}\left\vert \hat{%
\Psi}\left( X\right) \right\vert ^{2}\right) ^{r}}-\frac{C}{K_{X}}-C_{0}
\end{equation*}%
For each solution (\ref{SLn}), the disposable capital for banks is given by:%
\begin{equation}
\bar{K}_{X}\left\vert \bar{\Psi}\left( X\right) \right\vert ^{2}\simeq
\left\langle \bar{K}\right\rangle \left\Vert \bar{\Psi}\right\Vert
^{2}\left( \frac{\left\langle \bar{f}\right\rangle }{\left( \left(
1-\left\langle \bar{S}\right\rangle \right) \bar{f}\left( X\right)
+\left\langle \bar{S}\left( X^{\prime },X\right) \right\rangle _{X^{\prime
}}\left\langle \bar{f}\right\rangle \right) }\bar{I}_{X/\left\langle
X^{\prime }\right\rangle }\right) ^{2}  \label{Dcf}
\end{equation}%
where:%
\begin{equation*}
\bar{I}_{X/\left\langle X^{\prime }\right\rangle }=\frac{\frac{\left\langle 
\bar{S}\left( X^{\prime },X\right) \right\rangle _{X}}{1-\left\langle \bar{S}%
\left( X^{\prime },X\right) \right\rangle _{X^{\prime }}}}{\frac{%
\left\langle \bar{S}\left( X^{\prime },X\right) \right\rangle }{%
1-\left\langle \bar{S}\left( X^{\prime },X\right) \right\rangle }}
\end{equation*}%
This quantity measures the level of investment in investor $X$\ by other
sectors, $\frac{\left\langle \bar{S}\left( X^{\prime },X\right)
\right\rangle _{X}}{1-\left\langle \bar{S}\left( X^{\prime },X\right)
\right\rangle _{X^{\prime }}}$, relative to the level of investment in the
rest of the market, $\frac{\left\langle \bar{S}\left( X^{\prime },X\right)
\right\rangle }{1-\left\langle \bar{S}\left( X^{\prime },X\right)
\right\rangle }$. It is also weighted by the ratio:%
\begin{equation*}
\frac{\left\langle \bar{f}\right\rangle }{\left( \left( 1-\left\langle \bar{S%
}\right\rangle \right) \bar{f}\left( X\right) +\left\langle \bar{S}\left(
X^{\prime },X\right) \right\rangle _{X^{\prime }}\left\langle \bar{f}%
\right\rangle \right) }
\end{equation*}%
which measures a saturation effect: a high level of returns limits the
amount of capital that generates this level of return. The return $\bar{f}%
\left( X\right) $ alone does not determine the capital level; rather, it is
the weighted combination with the average returns $\left\langle \bar{S}%
\left( X^{\prime },X\right) \right\rangle _{X^{\prime }}\left\langle \bar{f}%
\right\rangle $,\ reflecting the share of capital reinvested in the market
by banks $X$.

The capital levels per sector for investors are:%
\begin{equation}
\hat{K}_{X}\left\vert \hat{\Psi}\left( X\right) \right\vert ^{2}\simeq
\left\langle \hat{K}\right\rangle \left\Vert \hat{\Psi}\right\Vert
^{2}\left( \frac{\left\langle \hat{g}\right\rangle }{\hat{g}\left( X\right) }%
I_{X/\left\langle X^{\prime }\right\rangle }\right) ^{2}  \label{Dcn}
\end{equation}%
where $\left\langle \hat{g}\right\rangle $ and $\hat{g}\left( X\right) $\
are given by:%
\begin{equation*}
\left\langle \hat{g}\right\rangle =\left\langle \hat{f}\right\rangle +\frac{%
\left( \hat{S}_{E}^{B}+\hat{S}_{L}^{B}\right) }{1-\bar{S}}\frac{\left\langle 
\bar{K}\right\rangle \left\Vert \bar{\Psi}\right\Vert ^{2}}{\left\langle 
\hat{K}\right\rangle \left\Vert \hat{\Psi}\right\Vert ^{2}}\left\langle \bar{%
f}\right\rangle
\end{equation*}%
\begin{equation*}
\hat{g}\left( X\right) =\left( \hat{f}\left( X\right) +\frac{\left(
\left\langle \hat{S}_{E}^{B}\left( X,X^{\prime }\right) \right\rangle
_{X^{\prime }}+\left\langle \hat{S}_{L}^{B}\left( X,X^{\prime }\right)
\right\rangle _{X^{\prime }}\right) \frac{\left\langle \bar{K}\right\rangle
\left\Vert \bar{\Psi}\right\Vert ^{2}}{\left\langle \hat{K}\right\rangle
\left\Vert \hat{\Psi}\right\Vert ^{2}}\left\langle \bar{f}\right\rangle }{%
1-\left\langle \bar{S}\right\rangle }\right) +\left\langle \hat{S}\left(
X^{\prime },X\right) \right\rangle _{X^{\prime }}\left\langle \hat{g}%
\right\rangle
\end{equation*}%
The average investors' return is defined by\ (\ref{Rpl}), and investors $X$\
returns $\hat{f}\left( X\right) $ are given by:%
\begin{equation*}
\hat{f}\left( X\right) \simeq \hat{r}\left( X\right) +\frac{1}{2}f_{a}-\frac{%
1}{2}f_{b}\left( \frac{C_{0}+\hat{r}\left( X\right) }{f_{1}\left( X\right) }%
\right) ^{\frac{1}{r}}
\end{equation*}%
Disposable capital (\ref{Dcn}) is proportional to the ratio:%
\begin{equation}
\hat{I}_{X/\left\langle X^{\prime }\right\rangle }=\frac{\frac{\left\langle 
\hat{S}\left( X,X^{\prime }\right) \right\rangle _{X^{\prime }}}{%
1-\left\langle \hat{S}\left( X,X^{\prime }\right) \right\rangle _{X^{\prime
}}}}{\frac{\left\langle \hat{S}\left( X,X^{\prime }\right) \right\rangle }{%
1-\left\langle \hat{S}\left( X,X^{\prime }\right) \right\rangle }}
\label{Idef}
\end{equation}%
which measures the level of investment in investor $X$\ by other sectors, $%
\frac{\left\langle \hat{S}\left( X,X^{\prime }\right) \right\rangle
_{X^{\prime }}}{1-\left\langle \hat{S}\left( X,X^{\prime }\right)
\right\rangle _{X^{\prime }}}$, relative to the level of investment in the
rest of the market, $\frac{\left\langle \hat{S}\left( X,X^{\prime }\right)
\right\rangle }{1-\left\langle \hat{S}\left( X,X^{\prime }\right)
\right\rangle }$.

It is further weighted by the ratio $\frac{\left\langle \hat{g}\right\rangle 
}{\hat{g}\left( X\right) }$, which measures the saturation effect: high
returns limit the capital that can generate these returns. Thus, the
effective capital depends not only on the sector-specific return, $\hat{f}%
\left( X\right) $, but also on its weighted combination with $\left\langle 
\hat{S}\left( X^{\prime },X\right) \right\rangle _{X^{\prime }}\left\langle 
\hat{f}\right\rangle $,\ which translates the share of capital reinvested in
the market by investors $X$.

The disposable capital for firms is given by:%
\begin{equation}
K_{X}\left\vert \Psi \left( X\right) \right\vert ^{2}\simeq \left( 1-\left(
S\left( X,X\right) \frac{\hat{K}_{X}\left\vert \hat{\Psi}\left( X\right)
\right\vert ^{2}}{\left( K_{X}\left\Vert \Psi \left( X\right) \right\Vert
^{2}\right) _{0}}+S^{B}\left( X,X\right) \frac{\bar{K}_{X}\left\vert \bar{%
\Psi}\left( X\right) \right\vert ^{2}}{\left( K_{X}\left\Vert \Psi \left(
X\right) \right\Vert ^{2}\right) _{0}}\right) \right) \left( K_{X}\left\Vert
\Psi \left( X\right) \right\Vert ^{2}\right) _{0}  \label{Dct}
\end{equation}%
where:%
\begin{equation*}
\left( K_{X}\left\Vert \Psi \left( X\right) \right\Vert ^{2}\right)
_{0}=\left( \left( \frac{2\epsilon }{3\sigma _{\hat{K}}^{2}}\right) ^{\frac{r%
}{2}}\frac{f_{1}\left( X\right) }{C_{0}+\bar{r}}\right) ^{\frac{2}{r}}
\end{equation*}%
represents the disposable capital of firms $X$ in the absence of investor
participation, i.e. when\textbf{\ }$S\left( X,X\right) =0$.

Under both solutions, firms' capital increases with firms' productivity $%
f_{1}\left( X\right) $, and investors $X$ disposable capital increases with
firms' productivity $f_{1}\left( X\right) $. This refines our results for
the average capital obtained in step 1, since\ this increase in firms'
productivity in any given sector $X$ increases investments in investors $X$
from other sectors.\textbf{\ }However, investors' disposable capital is
higher under the high-return solution than under the low-return solution,
since investors attract more capital; under this solution, the ratio $%
I_{X/\left\langle X^{\prime }\right\rangle }$\textbf{\ }is particularly high.

\subsubsection{Capital ratios per sector}

The ratio between the banks' average disposable capital and the banks'
disposable capital per sector is:

\begin{equation}
\frac{\left\langle \bar{K}\right\rangle \left\Vert \bar{\Psi}\right\Vert ^{2}%
}{\bar{K}_{X}\left\vert \bar{\Psi}\left( X\right) \right\vert ^{2}}\simeq
\left( \frac{\left( \left( 1-\left\langle \bar{S}\right\rangle \right) \bar{f%
}\left( X\right) +\left\langle \bar{S}\left( X^{\prime },X\right)
\right\rangle _{X^{\prime }}\left\langle \bar{f}\right\rangle \right) }{%
\left\langle \bar{f}\right\rangle }\bar{I}_{X/\left\langle X^{\prime
}\right\rangle }\right) ^{2}  \label{CR1}
\end{equation}%
Using (\ref{FRV}), the ratio of bank's average disposable capital to
investors' sector disposable capital is: 
\begin{equation}
\frac{\left\langle \bar{K}\right\rangle \left\Vert \bar{\Psi}\right\Vert ^{2}%
}{\hat{K}_{X}\left\vert \hat{\Psi}\left( X\right) \right\vert ^{2}}\simeq
\left( \frac{\hat{g}\left( X\right) }{\left\langle \bar{g}\right\rangle }%
\right) ^{2}\frac{\left\Vert \bar{\Psi}_{0}\right\Vert ^{4}}{\left\Vert \hat{%
\Psi}_{0}\right\Vert ^{4}\left( 1+\frac{\left\Vert \bar{\Psi}_{0}\right\Vert
^{2}}{\left\Vert \hat{\Psi}_{0}\right\Vert ^{2}}\left\langle \hat{S}%
_{L}^{B}\right\rangle \right) }  \label{CR2}
\end{equation}%
The ratio of banks disposable capital to firms' disposable capital is:%
\begin{equation}
\frac{\bar{K}_{X}\left\vert \bar{\Psi}\left( X\right) \right\vert ^{2}}{%
K_{X}\left\vert \Psi \left( X\right) \right\vert ^{2}}\simeq \frac{18\sigma
_{\hat{K}}^{2}V\left\Vert \bar{\Psi}_{0}\left( X\right) \right\Vert ^{4}}{%
\left( \bar{f}\left( X\right) +\frac{\left\langle \bar{S}\left( X^{\prime
},X\right) \right\rangle _{X^{\prime }}}{\left( 1-\left\langle \bar{S}%
\right\rangle \right) }\left\langle \bar{f}\right\rangle \right) ^{2}\hat{\mu%
}\left( \left( \frac{2\epsilon }{3\sigma _{\hat{K}}^{2}}\right) ^{\frac{r}{2}%
}\frac{f_{1}\left( X\right) }{C_{0}+\frac{S_{L}\left( X\right) }{%
1-S_{E}\left( X\right) }\bar{r}}\right) ^{\frac{2}{r}}}  \label{CR3}
\end{equation}%
and the ratio of investors' disposable capital to firms' disposable capital
is:%
\begin{equation*}
\frac{\hat{K}_{X}\left\vert \hat{\Psi}\left( X\right) \right\vert ^{2}}{%
K_{X}\left\vert \Psi \left( X\right) \right\vert ^{2}}\simeq \frac{18\sigma
_{\hat{K}}^{2}V\left\Vert \hat{\Psi}_{0}\left( X\right) \right\Vert ^{4}}{%
\hat{\mu}F_{1}^{2}\left( \left( \frac{2\epsilon }{3\sigma _{\hat{K}}^{2}}%
\right) ^{\frac{r}{2}}\frac{f_{1}\left( X\right) }{C_{0}+\frac{S_{L}\left(
X\right) }{1-S_{E}\left( X\right) }\bar{r}}\right) ^{\frac{2}{r}}}
\end{equation*}%
where $F_{1}$ is defined in Appendix 9.

\subsubsection{Investors' inward aggregate stakes}

Having computed the outward aggregate stakes in investors and the level of
disposable capital per sector, we can now derive the inward aggregate stakes
- the stakes held by all investors and banks in a given sector - to better
understand the distribution of investments.\ Since firms $X$ can only be
invested in by investors $X$, the stakes in firms - which represent the
proportion invested in firms $X$ - and the aggregate stakes in firms - which
represent the share of total capital invested in firms - differ only by a
single factor: the ratio of investors' disposable capital, both private and
intermediated, to the firms' disposable capital.

\paragraph{Inward aggregate stakes in investors}

Recall that the general formulas for the inward aggregate stakes in
investors are given by (\ref{Grstn}), (\ref{Grstw}), and (\ref{Grstd}):%
\begin{eqnarray*}
\hat{S}_{E}\left( X^{\prime }\right) &=&\frac{1}{2}\hat{w}\left( X^{\prime
}\right) \left( 1+\hat{\Delta}^{E}\left( X^{\prime }\right) \right) Q_{\hat{K%
}}\left( X^{\prime }\right) \\
\hat{S}_{L}\left( X^{\prime }\right) &=&\frac{1}{2}\hat{w}\left( X^{\prime
}\right) \left( 1+\hat{\Delta}^{L}\left( X^{\prime }\right) \right) Q_{\hat{K%
}}\left( X^{\prime }\right) \\
\hat{S}\left( X^{\prime }\right) &=&\hat{w}\left( X^{\prime }\right) \left(
1+\Delta \left( X^{\prime }\right) \right) Q_{\hat{K}}\left( X^{\prime
}\right)
\end{eqnarray*}%
These formulas involve the ratio of the aggregate disposable capital of all
investors $X^{\prime }$ to that of investors $X$, which is computed by (\ref%
{Dcn}):%
\begin{equation*}
Q_{\hat{K}}\left( X^{\prime }\right) =\frac{\left\langle \hat{K}%
\right\rangle \left\Vert \hat{\Psi}\right\Vert ^{2}}{\hat{K}_{X^{\prime
}}\left\vert \hat{\Psi}\left( X^{\prime }\right) \right\vert ^{2}}=\left( 
\frac{\hat{f}\left( X\right) \left( 1-\left\langle \hat{S}\right\rangle
\right) +\left\langle \hat{S}\left( X^{\prime },X\right) \right\rangle
_{X^{\prime }}\left\langle \hat{f}\right\rangle }{\left\langle \hat{f}%
\right\rangle I_{X/\left\langle X^{\prime }\right\rangle }}\right) ^{2}
\end{equation*}%
and the proportionality coefficient:%
\begin{equation*}
\hat{w}\left( X^{\prime }\right) =\frac{\left( 1-\left( \gamma \left\langle 
\hat{S}_{E}\left( X\right) \right\rangle \right) ^{2}\right) }{2-\left(
\gamma \left\langle \hat{S}_{E}\left( X\right) \right\rangle \right)
^{2}-\gamma ^{2}\left\langle \hat{S}_{E}\left( X\right) \right\rangle
\left\langle \hat{w}\left( X^{\prime },X\right) \right\rangle \left\langle
w\left( X\right) \right\rangle \Delta \left( X^{\prime }\right) }
\end{equation*}%
which depends directly on uncertainty and returns.

The capital invested in the investors of a given sector decreases with
uncertainty and increases with the relative returns of both investors and
firms in that sector. This reflects the fact that investors $X^{\prime }$\
also invest in $X$\ to gain access to the returns of the firms in that
sector.

\paragraph{Inward stakes in firms}

As expected, the proportion of shares invested in firms $X$, $S_{E}\left(
X,X\right) $, increases with their return $f_{1}\left( X\right) $:%
\begin{equation}
S_{E}\left( X,X\right) =\frac{1}{2}w\left( X\right) \left( 1+\left( \hat{w}%
\left( X\right) \left( f\left( X\right) -\hat{R}\right) +\frac{1}{2}w\left(
X\right) R_{exc}\left( X\right) \right) \right)  \label{Shr}
\end{equation}%
\begin{equation}
S_{L}\left( X,X\right) =\frac{1}{2}w\left( X\right) \left( 1+\left( \hat{w}%
\left( X\right) \left( r\left( X\right) -\hat{R}\right) \right) \right)
\label{LN}
\end{equation}%
The decision to invest in firms $X$\ depends negatively on the average
returns of cross-investments $\left\langle \hat{f}\left( X^{\prime }\right)
\right\rangle _{\hat{w}_{E}}$\ and the interest rates on loans between
investors $\left\langle \hat{r}\left( X^{\prime }\right) \right\rangle _{%
\hat{w}_{L}}$. Consequently, stakes in firms are higher under the low-return
solution than under the high-return one.

The inward total stake in firms, which is the sum of equations (\ref{Shr})
and (\ref{LN}), is given by:%
\begin{equation}
S\left( X,X\right) =w\left( X\right) \left( 1+\left( \hat{w}\left( X\right)
\left( R\left( X\right) -\hat{R}\right) \right) \right)  \label{Shl}
\end{equation}

\paragraph{Inward aggregate stakes in firms}

The aggregate stakes of investors in firms $X$ are obtained by multiplying (%
\ref{Shr}), (\ref{LN}), and (\ref{LN}) by the capital ratio, which yields:%
\begin{equation*}
S_{E}\left( X\right) \equiv S_{E}\left( X,X\right) \frac{\hat{K}%
_{X}\left\vert \hat{\Psi}\left( X\right) \right\vert ^{2}}{K_{X}\left\vert
\Psi \left( X\right) \right\vert ^{2}}
\end{equation*}%
\begin{equation*}
S_{L}\left( X\right) \equiv S_{L}\left( X,X\right) \frac{\hat{K}%
_{X}\left\vert \hat{\Psi}\left( X\right) \right\vert ^{2}}{K_{X}\left\vert
\Psi \left( X\right) \right\vert ^{2}}
\end{equation*}%
and:%
\begin{equation*}
S\left( X\right) =S\left( X,X\right) \frac{\hat{K}_{\hat{X}}\left\vert \hat{%
\Psi}\left( X\right) \right\vert ^{2}}{K_{\hat{X}}\left\vert \Psi \left(
X\right) \right\vert ^{2}}
\end{equation*}%
These formula reflect two opposing effects: increased investment due to
higher returns $f_{1}\left( X\right) $, and the reduction in the capital
density ratio $\frac{\hat{K}_{X}\left\vert \hat{\Psi}\left( X\right)
\right\vert ^{2}}{K_{X}\left\vert \Psi \left( X\right) \right\vert ^{2}}$
resulting from decreasing marginal returns. Thus, the overall proportion of
investors' capital in firms depends on the structural parameters of the
model.

\subsubsection{Banks inward aggregate stakes per sector}

\paragraph{Inward aggregate cross-stakes between banks}

Similarly, once the returns and capital ratios per sector are determined,
the inward aggregate stakes $\bar{S}_{E}\left( X\right) $, $\bar{S}%
_{L}\left( X\right) $, and $\bar{S}\left( X\right) $ are computed using (\ref%
{BCSn}), (\ref{BCSpt}), (\ref{BCSt}), (\ref{BCStn}), and (\ref{BCStd}), with
the capital ratio given by (\ref{CR1}).

The corresponding coefficients are given by averages of (\ref{Cfn}) and (\ref%
{Cft}) over $X$:%
\begin{equation*}
\left( \left\langle \bar{w}\left( X^{\prime },X\right) \right\rangle
_{X}\right) ^{-1}=1+\frac{1}{2}\left( \frac{\left\langle \overline{IRG}%
\left( X^{\prime },X\right) \right\rangle _{X}}{\left\langle \widehat{IR^{B}}%
\left( X^{\prime },X\right) \right\rangle _{X}}+\frac{\left\langle \overline{%
IRG}\left( X^{\prime },X\right) \right\rangle _{X}}{\xi ^{2}}\right)
\end{equation*}%
and:%
\begin{equation*}
\left( \left\langle \bar{w}\left( X^{\prime },X\right) \right\rangle
_{X}\right) ^{-1}\simeq 1+2\frac{\left\langle \widehat{IR^{B}}\left(
X^{\prime },X\right) \right\rangle _{X}}{\left\langle \overline{IRG}\left(
X^{\prime },X\right) \right\rangle _{X}}+\frac{\left\langle \widehat{IR^{B}}%
\left( X^{\prime },X\right) \right\rangle _{X}}{\xi ^{2}}
\end{equation*}

\paragraph{Inward aggregate banks' stakes in investors}

The inward aggregate stakes $\hat{S}_{E}^{B}\left( X\right) $ and $\hat{S}%
_{L}^{B}\left( X\right) $ are computed using (\ref{BCStd}) and (\ref{BCSHd}%
), with the capital ratio given by (\ref{CR2}), where $w_{E}^{B}\left(
X\right) $ is defined by (\ref{Cft}) and:%
\begin{eqnarray*}
\left\langle \hat{w}_{L}^{B}\left( X^{\prime },X\right) \right\rangle _{X}
&\simeq &\left\langle \hat{w}_{L}\left( X^{\prime },X\right) \right\rangle
_{X} \\
&\simeq &\left\langle \hat{w}_{E}\left( X^{\prime },X\right) \right\rangle
_{X} \\
&\rightarrow &\frac{\left( 1-\left( \gamma \left\langle \hat{S}_{E}\left(
X\right) \right\rangle \right) ^{2}\right) }{2-\left( \gamma \left\langle 
\hat{S}_{E}\left( X\right) \right\rangle \right) ^{2}+\left( \gamma
\left\langle \hat{S}_{E}\left( X_{1},X^{\prime }\right) \right\rangle
_{X_{1}}\right) ^{2}-\left( \gamma \left\langle \hat{S}_{E}\left( X\right)
\right\rangle \right) ^{2}}
\end{eqnarray*}

\paragraph{Inward aggregate banks' stakes in firms}

The inward aggregate stakes $S_{E}^{B}\left( X\right) $ and $S_{L}^{B}\left(
X\right) $ are obtained using (\ref{BCStbn}) and (\ref{sFB}), with the
capital ratio given by (\ref{CR3}), and the coefficients are given by:%
\begin{eqnarray*}
w_{E}^{B}\left( X\right) &=&\left\langle w_{E}^{B}\left( X\right)
\right\rangle \\
&\rightarrow &\frac{1+\left( 1-\left( \bar{\gamma}\left\langle \bar{S}%
_{E}\right\rangle \right) ^{2}\right) \left( 4+3\left( 1-\left( \bar{\gamma}%
\left\langle \bar{S}_{E}\right\rangle \right) ^{2}\right) -\frac{%
\left\langle \hat{S}_{E}^{B}\right\rangle ^{2}}{\left\langle
S_{E}\right\rangle ^{2}}\left( 1-\left( \gamma \left\langle \hat{S}%
_{E}\right\rangle \right) ^{2}\right) \right) }{\left( 1+4\left( 1-\left( 
\bar{\gamma}\left\langle \bar{S}_{E}\right\rangle \right) ^{2}\right)
\right) \left( \frac{\left\langle \hat{S}_{E}^{B}\right\rangle ^{2}}{%
\left\langle S_{E}\right\rangle ^{2}}\left( 1-\left( \gamma \left\langle 
\hat{S}_{E}\right\rangle \right) ^{2}\right) +\left( 1-\left( \bar{\gamma}%
\left\langle \bar{S}_{E}\right\rangle \right) ^{2}\right) +1\right) }
\end{eqnarray*}%
\begin{eqnarray*}
w_{L}^{B}\left( X\right) &\simeq &\left\langle w_{L}^{B}\left( X\right)
\right\rangle =\frac{1}{2-\left( \gamma \left\langle \hat{S}_{E}\left(
X^{\prime },X\right) \right\rangle \right) ^{2}} \\
\hat{w}_{L}^{B}\left( X\right) &\simeq &1-w_{L}^{B}\left( X\right)
=1-\left\langle w_{L}^{B}\left( X\right) \right\rangle \frac{1-\left( \gamma
\left\langle \hat{S}_{E}\left( X^{\prime },X\right) \right\rangle \right)
^{2}}{2-\left( \gamma \left\langle \hat{S}_{E}\left( X^{\prime },X\right)
\right\rangle \right) ^{2}}
\end{eqnarray*}

\subsection{Step 4: solving for stakes}

Once the sectoral returns $\bar{f}\left( X\right) $ and $\hat{f}\left(
X\right) $ are established, we close the resolution by computing the stakes
invested sector by sector.

\paragraph{Investors}

We first compute cross-investment stakes between investors.

The uncertainty $\frac{\hat{w}\left( X^{\prime },X\right) }{2}$, defined in
equation (\ref{hb}), impacts investments in shares such that:%
\begin{eqnarray}
\hat{S}_{E}\left( X^{\prime },X\right) &=&\frac{1}{2}\hat{w}\left( X^{\prime
},X\right) \left( 1+\Delta ^{E}\left( X^{\prime },X\right) \right)
\label{SNXPX} \\
\hat{S}_{L}\left( X^{\prime },X\right) &=&\frac{1}{2}\hat{w}\left( X^{\prime
},X\right) \left( 1+\Delta ^{L}\left( X^{\prime },X\right) \right)  \notag \\
&&  \notag \\
\hat{S}\left( X^{\prime },X\right) &\simeq &\hat{w}\left( X^{\prime
},X\right) \left( 1+\Delta \left( X^{\prime },X\right) \right)  \label{SNP}
\end{eqnarray}%
where $\Delta ^{E}\left( X^{\prime },X\right) $ and $\Delta \left( X^{\prime
},X\right) $ are defined by (\ref{DTp}) and (\ref{DTs}), and the expressions
for the stakes invested in firms $S_{E}\left( X,X\right) $ and $S\left(
X,X\right) $, are provided in (\ref{Shr}) and (\ref{Shl}), respectively.

Any given collective state is characterized by two stakes, (\ref{SNXPX}) and
(\ref{SNP}). These solutions do not fundamentaly differ from the aggregate
stakes described in (\ref{Shg}). The behaviors of agents sector by sector
follow a two-solution pattern. Differences between sectoral stakes (\ref%
{SNXPX}) and (\ref{SNP}) and aggregate stakes result from local variations
in the perception of risks $\hat{w}\left( X^{\prime },X\right) $ and $%
w\left( X\right) $. The sector dependence of these two parameters impact the
solutions and induce local variations around the aggregate stakes, returns,
and capital levels. We show\footnote{%
See Appendix 23.} that:%
\begin{equation*}
\hat{w}\left( X^{\prime },X\right) =\frac{2\left( 1-\left( \gamma
\left\langle \hat{S}_{E}\left( X\right) \right\rangle \right) ^{2}\right) 
\hat{w}_{1}^{\left( 0\right) }\left( X^{\prime },X\right) }{1+\hat{w}%
_{1}^{\left( 0\right) }\left( X^{\prime },X\right) \left( 1-\left( \gamma
\left\langle \hat{S}_{E}\left( X\right) \right\rangle \right) ^{2}\right)
+\left( \gamma \left\langle \hat{S}_{E}\left( X_{1},X^{\prime }\right)
\right\rangle _{X_{1}}\right) ^{2}-\left( \gamma \left\langle \hat{S}%
_{E}\left( X\right) \right\rangle \right) ^{2}}
\end{equation*}%
and:%
\begin{equation*}
w\left( X^{\prime },X\right) =1-\left\langle \hat{w}\left( X^{\prime
},X\right) \right\rangle _{X^{\prime }}
\end{equation*}%
Here, the coefficients $\hat{w}_{1}^{\left( 0\right) }\left( X^{\prime
},X\right) $ computes a local factor of inverse uncertainty. When this
factor is high, $\hat{w}\left( X^{\prime },X\right) $ is high and sector $%
X^{\prime }$ attracts investment from sector $X$. This coefficient thus
captures the local characteristics of uncertainty that induce the deviations
from the average behavior.

\subparagraph{Banks}

Ultimately, we close the resolution by computing the cross-investment stakes
among banks and stakes in investors (\ref{SBN}), (\ref{SBT}), and (\ref{SBF}%
).

The cross-stakes between banks are:%
\begin{eqnarray*}
\bar{S}_{E}\left( X^{\prime },X\right) &=&\frac{1}{2}\bar{w}\left( X^{\prime
},X\right) \left( 1+\bar{\Delta}^{E}\left( X^{\prime },X\right) \right) \\
\bar{S}_{L}\left( X^{\prime },X\right) &=&\frac{1}{2}\bar{w}\left( X^{\prime
},X\right) \left( 1+\bar{\Delta}^{L}\left( X^{\prime },X\right) \right) \\
\bar{S}\left( X^{\prime },X\right) &=&\bar{w}\left( X^{\prime },X\right)
\left( 1+\bar{\Delta}\left( X^{\prime },X\right) \right)
\end{eqnarray*}%
where $\bar{\Delta}^{E}\left( X^{\prime }\right) $, $\bar{\Delta}^{L}\left(
X^{\prime }\right) $, and $\bar{\Delta}\left( X^{\prime }\right) $ are the
averages of (\ref{DBn}), (\ref{DBt}), and (\ref{DBr}) over $X$, with $\bar{w}%
\left( X^{\prime },X\right) $ defined by (\ref{Cfn}).

Banks' stakes in investors are given by (\ref{scb}) and (\ref{sdb}):%
\begin{equation}
\hat{S}_{E}^{B}\left( X^{\prime },X\right) =\underline{\hat{S}}%
_{E}^{B}\left( X^{\prime },X\right) +\hat{w}_{E}^{B}\left( X^{\prime
},X\right) \left( \hat{\Delta}^{E}\right) ^{B}\left( X^{\prime },X\right)
\end{equation}%
\begin{equation}
\frac{\hat{S}_{L}^{B}\left( X^{\prime },X\right) }{\kappa \left( 1-\bar{S}%
\left( X\right) \right) }=\underline{\hat{S}}_{L}^{B}\left( X^{\prime
},X\right) +\hat{w}_{L}^{B}\left( X^{\prime },X\right) \left( \hat{\Delta}%
^{L}\right) ^{B}\left( X^{\prime },X\right)
\end{equation}%
and Banks' stakes in \ firms are given by (\ref{SFN}) and (\ref{sdt}):%
\begin{equation}
S_{E}^{B}\left( X,X\right) =w_{E}^{B}\left( X\right) +w_{E}^{B}\left(
X\right) \left( \Delta ^{E}\right) ^{B}\left( X\right)
\end{equation}%
\begin{equation}
\frac{S_{L}^{B}\left( X,X\right) }{\kappa \left( 1-\bar{S}\left( X\right)
\right) }=w_{L}^{B}\left( X\right) +w_{L}^{B}\left( X\right) \Delta
^{L}\left( X\right)
\end{equation}%
with $\bar{w}\left( X\right) $, $\hat{w}_{E}^{B}\left( X\right) $, and $%
w_{E}^{B}\left( X\right) $\ defined by the averages over $X^{\prime }$ of (%
\ref{Cfn}), (\ref{Cft}), and (\ref{chv}), respectively. The coefficients for
banks investing in banks are:

\begin{equation}
\left( \bar{w}\left( X\right) \right) ^{-1}=1+\frac{1}{2}\left( \frac{%
\left\langle \overline{IRG}\left( X^{\prime },X\right) \right\rangle
_{X^{\prime }}}{\left\langle \widehat{IR^{B}}\left( X^{\prime },X\right)
\right\rangle _{X^{\prime }}}+\frac{\left\langle \overline{IRG}\left(
X^{\prime },X\right) \right\rangle _{X^{\prime }}}{\xi ^{2}}\right)
\end{equation}%
Those for banks investing investors are given by:%
\begin{equation}
\left( \hat{w}_{E}^{B}\left( X\right) \right) ^{-1}=1+2\frac{\left\langle 
\widehat{IR^{B}}\left( X^{\prime },X\right) \right\rangle _{X^{\prime }}}{%
\left\langle \overline{IRG}\left( X^{\prime },X\right) \right\rangle
_{X^{\prime }}}+\frac{\left\langle \widehat{IR^{B}}\left( X^{\prime
},X\right) \right\rangle _{X^{\prime }}}{\xi ^{2}}
\end{equation}%
and those for banks investing in firms are:%
\begin{equation*}
\bar{w}_{E}^{B}\left( X,X\right) =1-\left\langle \bar{w}\left( X^{\prime
},X\right) \right\rangle _{X^{\prime }}-\left\langle \hat{w}_{E}^{B}\left(
X^{\prime },X\right) \right\rangle _{X^{\prime }}
\end{equation*}%
The local coefficients $\hat{w}_{L}^{B}\left( X^{\prime },X\right) $ and $%
w_{L}^{B}\left( X\right) $ are defined by\footnote{%
See Appendix 11.}:%
\begin{equation}
\hat{w}_{L}^{B}\left( X^{\prime },X\right) =\frac{\left( 1-\left( \gamma
\left\langle \hat{S}_{E}\left( X_{1},X^{\prime }\right) \right\rangle
\right) ^{2}\right) \hat{w}_{L}^{\left( 0\right) }\left( X^{\prime
},X\right) \left( 1+\Delta \hat{r}\left( X^{\prime }\right) \right) }{1+\hat{%
w}_{L}^{\left( 0\right) }\left( X^{\prime },X\right) \left( 1-\left( \gamma
\left\langle \hat{S}_{E}\left( X_{1},X^{\prime }\right) \right\rangle
\right) ^{2}\right) +\Delta \left( \gamma \left\langle \hat{S}_{E}\left(
X_{1},X^{\prime }\right) \right\rangle _{X_{1}}\right) ^{2}}
\end{equation}%
and:%
\begin{equation}
w_{L}^{B}\left( X\right) =1-\hat{w}_{L}^{B}\left( X^{\prime },X\right)
\end{equation}%
with:%
\begin{equation*}
\Delta \left( \gamma \left\langle \hat{S}_{E}\left( X_{1},X^{\prime }\right)
\right\rangle _{X_{1}}\right) ^{2}=\left( \gamma \left\langle \hat{S}%
_{E}\left( X_{1},X^{\prime }\right) \right\rangle _{X_{1}}\right)
^{2}-\left( \gamma \left\langle \hat{S}_{E}\left( X_{1},X^{\prime }\right)
\right\rangle \right) ^{2}
\end{equation*}%
The coefficients $\bar{w}_{E}^{\left( 0\right) }\left( X^{\prime },X\right) $%
, $w_{E}^{\left( 0\right) B}\left( X^{\prime },X\right) $, $\left\langle 
\hat{w}_{E}^{\left( 0\right) B}\left( \left( X^{\prime }\right) ^{\prime
},X^{\prime }\right) \right\rangle _{\left( X^{\prime }\right) ^{\prime }}$%
,\ and $\hat{w}_{L}^{\left( 0\right) }\left( X^{\prime },X\right) $\
represent local inverse uncertainties in shares in banks, investors and
firms.\ Higher values of $\bar{w}_{E}^{\left( 0\right) }\left( X^{\prime
},X\right) $\ and $w_{E}^{\left( 0\right) B}\left( X^{\prime },X\right) $
indicate higher investments of sector $X$\ in sector $X^{\prime }$.

The factor $\hat{w}_{L}^{B}\left( X^{\prime },X\right) $\textbf{\ }captures
the relative inverse uncertainty of investors relatively to sector $%
X^{\prime }$: the higher this coefficient, the more relatively risky are
investors $X^{\prime }$\ and the lower the shares in investors $X^{\prime }$.

The factor $\hat{w}_{2}^{\left( 0\right) }\left( X^{\prime },X\right) $\ is
a local factor of inverse uncertainty in loans to investors. The higher $%
\hat{w}_{2}^{\left( 0\right) }\left( X^{\prime },X\right) $, the higher the
loans of bank $X$\ in investors $X^{\prime }$.

Thus, the local characteristics of uncertainty conditions the investments.
These factors measure any condition that locally modify investments and
perception of risk.

\section{Default states}

Defaults states can be defined as modifications of non-default states.
Assuming one or several defaulting firms initially in one sector, the
default state is built recursively from this initial impact.\ Once the
initial default has propagated, the returns in each sector will be
deviations from what would have been the returns in a non-default scenario,
equation (\ref{QDM}).

We will detail below the propagation mechanism and the condition for the
default to materialize and spread. The loss incured by investors and the
fraction of defaulting investors will then be computed.

\subsection{Uncertainty and default}

Note that the role of uncertainty changes depending on the agent who bares
the risk of default. When uncertainty increases, the threshold diminishes
and the risk of default increases for the agent invested in, since
uncertainty deters other investors.

In our model, uncertainty is defined very broadly, and corresponds to a
global perception of risk. The parameter $\gamma $, when not otherwise
mentionned, refers to an average perception of risk of the whole system on
all the sectors of the system. It would nonetheless be easy to render this
parameter sector-dependent, and as a matter of fact, we will do so in our
interpretations.

The fact that the uncertainty is high does not necessarily imply that the
risk has risen, and inversely.\ Uncertainty and risk are not correlated.
Risk can exist without necessarily being perceived. In this case, Investors
may act under a false presomption of full certainty, and risk is all the
more great.

In the study of defaults, the notion of risk perception is fundamental, and
its role depends on the perception agents have of it and where and from
where it is seen. The default of a firm will be less likely to propagate to
other investors in a climate of confidence, when $\gamma $ is small.
Alternately, the system as a whole is in a greater zone of default when it
benefits from a false climate of confidence.\ The intermediation being
stronger, if a default occurs, intermediation will favor its propagation.
The propagation of a default is more likely when, locally, one sector is
being perceived as risky, and thus locally $\gamma $ increases, investors in
the sector default and this default propagates because the rest of the
system is impaired by an intermediation corresponding to a $\gamma $ small.
Default is more likely when the differential between the $\gamma $ of the
system and the $\gamma $ of a specific sector is wider.

\subsection{Principle of propagation}

The default states for investors have been described in Gosselin and Lotz
(2025). The principle of propagation is similar if we include banks. The
default states are found recursively. We first assume some realized initial
default states $S_{0}$, $\hat{S}_{0}$ and $\bar{S}_{0}$, when a set of
firms, investors and banks\ experience a complete loss of their private
capital. In terms of returns this corresponds to:%
\begin{eqnarray}
\hat{f}\left( X^{\prime }\right) &=&-1  \label{Cdf} \\
\bar{f}\left( X^{\prime }\right) &=&-1  \notag \\
f_{1}^{\prime }\left( \hat{X}\right) &=&-1  \notag
\end{eqnarray}%
for investors, banks and firms in $\hat{S}_{0}$, $\bar{S}_{0}$ and $S_{0}$
respectively. The number $-1$ stands for a return of $-100$ percent, so that
the private capital for defaulting firms, banks and investors are $0$.

The possible default sets are then obtained from this initial set by the
limit of a sequence of equations. We assume that after $n$ iterations, the
sets of defaulting agents are $S_{n}$, $\hat{S}_{n}$ and $\bar{S}_{n}$. We
define the remaining sets of agents as $S/\hat{S}_{n}$, $S/\bar{S}_{n}$ and $%
S/S_{n}$.

Then these sets are included in the return equations for investors and banks:

\paragraph{Investors}

\begin{eqnarray}
0 &=&\int \left( \delta \left( X-X^{\prime }\right) -\hat{S}_{E}\left(
X^{\prime },X\right) \right) \widehat{DF}\left( X^{\prime }\right) \left( 
\hat{f}_{n+1}\left( X^{\prime }\right) -\bar{r}\right) dX^{\prime }
\label{dfn} \\
&&+\int_{\hat{S}_{n}}\frac{1-\left( \hat{S}\left( X^{\prime }\right) +\hat{S}%
_{E}^{B}\left( X^{\prime }\right) +\hat{S}_{L}^{B}\left( X^{\prime }\right)
\right) }{\hat{S}_{L}\left( X^{\prime }\right) }\hat{S}_{L}\left( X^{\prime
},X\right) dX^{\prime }  \notag \\
&&+\int_{S_{n}}\frac{1-\left( S\left( X^{\prime }\right) +S_{E}^{B}\left(
X^{\prime }\right) +S_{L}^{B}\left( X^{\prime }\right) \right) }{S_{L}\left(
X^{\prime }\right) }S_{L}\left( X^{\prime },X\right) dX^{\prime }-\int
S_{E}\left( X^{\prime },X\right) \left( \left( f\left( X^{\prime }\right) -%
\bar{r}\right) \right) dX^{\prime }  \notag
\end{eqnarray}

\paragraph{Banks}

\begin{eqnarray}
0 &=&\left( \delta \left( X-X^{\prime }\right) -\bar{S}_{E}\left( X^{\prime
},X\right) \right) \left( \bar{f}_{n+1}\left( X^{\prime }\right) -\bar{r}%
\right) \overline{DF}\left( X^{\prime }\right) -\widehat{DF}\left( X^{\prime
}\right) \hat{S}_{E}^{B}\left( X^{\prime },X\right) \left( \hat{f}\left(
X^{\prime }\right) -\bar{r}\right) \\
&&+\int_{\bar{S}_{n}}\bar{S}_{L}\left( X^{\prime },X\right) \frac{\left( 1-%
\bar{S}\left( X^{\prime }\right) \right) }{\bar{S}_{L}\left( X^{\prime
}\right) }+\int_{\hat{S}_{n}}\hat{S}_{L}^{B}\left( X^{\prime },X\right) 
\frac{1-\left( S\left( X^{\prime }\right) +S_{E}^{B}\left( X^{\prime
}\right) +S_{L}^{B}\left( X^{\prime }\right) \right) }{S_{L}\left( X^{\prime
}\right) +S_{L}^{B}\left( X^{\prime }\right) }  \notag \\
&&+\int_{S_{n}}S_{L}^{B}\left( X^{\prime },X\right) \frac{1-\left( \hat{S}%
\left( X^{\prime },X\right) +\hat{S}_{E}^{B}\left( X^{\prime }\right) +\hat{S%
}_{L}^{B}\left( X^{\prime }\right) \right) }{\hat{S}_{L}\left( X^{\prime
}\right) +\hat{S}_{L}^{B}\left( X^{\prime }\right) }-S_{E}^{B}\left(
X^{\prime },X\right) \left( f\left( X^{\prime }\right) -\bar{r}\right) 
\notag
\end{eqnarray}%
These equations define new sets of returns $\hat{f}_{n+1}\left( X^{\prime
}\right) $ and $\bar{f}_{n+1}\left( X^{\prime }\right) $ which in turn may
define new default sets $S_{n+1}$, $\hat{S}_{n+1}$ and $\bar{S}_{n+1}$. Once
the iteration stabilizes, we find a remaining sets of - non-defaulted -
agents $\left( \hat{S}_{\infty },\bar{S}_{\infty },S_{\infty }\right) $,
defined by the limit $\left( S/\hat{S}_{n},\bar{S}/\bar{S}%
_{n},S/S_{n}\right) \underset{n\rightarrow \infty }{\rightarrow }\left( \hat{%
S}_{\infty },\bar{S}_{\infty },S_{\infty }\right) $ with returns $\left\{ 
\hat{f}_{n}\left( X\right) \rightarrow \hat{f}\left( X\right) ,\bar{f}%
_{n}\left( X\right) \rightarrow \bar{f}\left( X\right) \right\} $ for which
the resulting disposable capitals are given by (\ref{Dcn}), (\ref{Dcf}) and (%
\ref{Dct}).

\subsection{Conditions for propagation}

Firms default when they meet the conditions for initial defaults. But, for
this default to propagate to the whole set of investors and banks, some
additional conditions must be met. First, the defaut must propagate to the
firm's immediate investors or banks. Then, it must propagate from these
investors to other investors and banks. The case of investors default was
considered in Gosselin and Lotz 2025 and we focus on the banks default only%
\footnote{%
See Appendix 11.3.}.

\subsubsection{Propagation from firms to banks}

Some firms may lack the disposable capital to face their costs, pushing them
into default. Whether these initial defaults may push into default their
intra-sectoral banks depends on the magnitude of loss of the defaulting
firms.

For the default to spread to the firm's banks, the level of loss a firm has
to experience, by lack of capital or increase in costs, must be below a
negative threshold\footnote{%
See Appendix 8 in Gosselin and Lotz (2025).}:%
\begin{equation*}
f\left( X\right) -r<D_{Th}
\end{equation*}%
where the threshold $D_{Th}$ is:%
\begin{eqnarray*}
D_{Th} &=&-\frac{1}{S_{E}^{B}\left( X,X\right) }\left( 1+\bar{r}%
+\left\langle \bar{S}_{E}\left( X^{\prime },X\right) \right\rangle
_{X^{\prime }}\overline{DF}\left( X^{\prime }\right) \left( \left\langle 
\bar{f}\left( X^{\prime }\right) \right\rangle -\bar{r}\right) \right. \\
&&+\left. \left\langle \hat{S}_{E}^{B}\left( X^{\prime },X\right)
\right\rangle _{X^{\prime }}\widehat{DF}\left( X^{\prime }\right) \left(
\left\langle \hat{f}\left( X^{\prime }\right) \right\rangle -\bar{r}\right)
\right)
\end{eqnarray*}%
This threshold is negative, and the default zone includes all returns that
are below it.

When average excess returns $\left( \left\langle \bar{f}\left( X^{\prime
}\right) \right\rangle -\bar{r}\right) $ among the whole set of banks or the
set of investors $\left\langle \hat{f}\left( X^{\prime }\right)
\right\rangle -\left\langle \bar{r}\right\rangle $ increase the default zone
decreases, and so does the risk of default in the sector. The sector is
safer financially. On the contrary, when average excess returns among the
whole set of banks or investors decrease increases, the sector as a whole
becomes more risky, and investors diversify away from the sector and in
other investors, which further increases the default risk. For a given level
of returns among banks or investors, any increase in interest rates or in
the level of loans between investors increases the default zone and the risk
of default for investors.

Assuming that uncertainty $\gamma $\ increases, the average banks' shares in
other banks $\left\langle \bar{S}_{E}\left( X^{\prime },X\right)
\right\rangle _{X^{\prime }}$\ or in investors $\left\langle \hat{S}%
_{E}^{B}\left( X^{\prime },X\right) \right\rangle _{X^{\prime }}$\ diminish,
and the default zone increases. This also applies if the bank mere
perception of risk increases, while the investors' perception remains
unchanged.

The introduction of banks induces a particular effect. Actually, when
banks's loans are essentially invested in investors, the coefficient $%
\left\langle \widehat{DF}\left( X^{\prime }\right) \right\rangle $
diminishes and may even tend to $0$, which increases $D_{Th}$\ and
consequently the default zone.

\subsubsection{Propagation from investors and banks to other sectors}

Given an initial default in investors and banks in certain sectors, the
condition for default to propagate to investors is the same as in part 1 and
the condition under which the default may spread to other banks is:%
\begin{eqnarray}
&&\mathbf{H}\left( \left\langle \left( \hat{S}_{L}\left( X^{\prime
},X\right) \right) \right\rangle _{X^{\prime }}+S_{L}\left( X,X\right)
\right)  \label{DFLTw} \\
&&+\mathbf{G}\left( \left\langle \bar{S}_{L}\left( X^{\prime },X\right)
\right\rangle _{X^{\prime }}+\left\langle \hat{S}_{L}^{B}\left( X^{\prime
},X\right) \right\rangle _{X^{\prime }}+S_{L}^{B}\left( X,X\right) \right)
>2\left\langle \bar{f}\right\rangle  \notag
\end{eqnarray}%
where $\mathbf{H}$ and $\mathbf{G}$ are functions of stakes, uncertainty and
capital ratios under the non-default scenario\footnote{%
Their formula are in Appendix 11.2.}.

\paragraph{Interpretation of the condition}

Equation (\ref{DFLTw}) expresses the potential for systemic default as a
global property of a group of connected agents. While a default may begin in
a single sector, its full propagation depends on the structural
configuration of the system.

Actually, equation (\ref{DFLTw}) involves the \ banks average loans, $%
\left\langle \bar{S}_{L}\left( X^{\prime },X\right) \right\rangle
_{X^{\prime }}+\left\langle \hat{S}_{L}^{B}\left( X^{\prime },X\right)
\right\rangle _{X^{\prime }}+S_{L}^{B}\left( X,X\right) $, and the average
investors loans, $\left\langle \hat{S}_{L}\left( X^{\prime },X\right)
\right\rangle +\left\langle S_{L}\left( X,X\right) \right\rangle $.

First, (\ref{DFLTw}) indicates that a high volume of investors loans, $%
\left\langle \hat{S}_{L}\left( X^{\prime },X\right) \right\rangle
+\left\langle S_{L}\left( X,X\right) \right\rangle $, or of banks loans $%
\left( \left\langle \bar{S}_{L}\left( X^{\prime },X\right) \right\rangle
_{X^{\prime }}+\left\langle \hat{S}_{L}^{B}\left( X^{\prime },X\right)
\right\rangle _{X^{\prime }}+S_{L}^{B}\left( X,X\right) \right) $ combined
with a high level of global inter-bank stakes---i.e., $\bar{S}\left(
X^{\prime }\right) $ close to $1$ ---places the system in the default zone.
Any further increase in average loan exposure across the system facilitates
the transition into default. Changes in the banks' average expected return $%
\left\langle \bar{f}\right\rangle $, has a similar effect.

\paragraph{Role of the coefficients}

Coefficients $\mathbf{G}$ and $\mathbf{H}$ modify the threshold equation (%
\ref{DFLTw}). The higher these coefficients and the lower the threshold for
the loans to enter the default.

We show\footnote{%
See Appendix 11.} that coefficient $\mathbf{G}$\ increase as function of $%
\bar{S}\left( X^{\prime }\right) $, $\left\langle \hat{S}\left( X\right)
\right\rangle $, with capital ratios $\frac{\hat{K}_{X}\left\vert \hat{\Psi}%
\left( X\right) \right\vert ^{2}}{K_{X}\left\vert \Psi \left( X\right)
\right\vert ^{2}}$ $\frac{\bar{K}_{X}\left\vert \bar{\Psi}\left( X\right)
\right\vert ^{2}}{K_{X}\left\vert \Psi \left( X\right) \right\vert ^{2}}$ \
and decreases as a function of $S_{E}^{B}\left( X,X\right) $ and uncertaint $%
\gamma $. Coefficient $\mathbf{H}$\ increases as function of $\bar{S}\left(
X^{\prime }\right) $, $\left\langle \hat{S}\left( X\right) \right\rangle $, $%
\hat{S}_{E}\left( X^{\prime }\right) $, $\hat{S}_{E}^{B}\left( X^{\prime
}\right) $, with capital ratios $\frac{\hat{K}_{X}\left\vert \hat{\Psi}%
\left( X\right) \right\vert ^{2}}{K_{X}\left\vert \Psi \left( X\right)
\right\vert ^{2}}$, $\frac{\bar{K}_{X}\left\vert \bar{\Psi}\left( X\right)
\right\vert ^{2}}{K_{X}\left\vert \Psi \left( X\right) \right\vert ^{2}}$
while it decreases with loans $\hat{S}_{L}\left( X^{\prime }\right) +$ $\hat{%
S}_{L}^{B}\left( X^{\prime }\right) $\ and decrease as a function of $%
S_{E}^{B}\left( X,X\right) $, $S_{E}\left( X,X\right) $ and uncertaint $%
\gamma $.

As a consequence, when banks stakes in banks and investors increases, the
threshold decreases which amplifies the likelihood of systemic default.
Similarly Thie threshold for default decreases with higher investor
disposable capital and lower perceived uncertainty. In other words, the
higher the lower the global risk perception for investors or banks, the more
prone the system is to cascading failure.

This tendency is further magnified by the ratio $\frac{\left\langle \hat{K}%
_{X}\left\vert \hat{\Psi}\left( X\right) \right\vert ^{2}\right\rangle }{%
\left\langle K_{X}\left\vert \Psi \left( X\right) \right\vert
^{2}\right\rangle }$ and $\frac{\bar{K}_{X}\left\vert \bar{\Psi}\left(
X\right) \right\vert ^{2}}{K_{X}\left\vert \Psi \left( X\right) \right\vert
^{2}}$: When they are high, defaults are more probable, structurally.

Moreover, the expression for $H$\ given in Appendix 11.2 shows that this
coefficient depends on $\widehat{DF}\left( X^{\prime }\right) $, which is an
inverse measure of the rate of investments of banks in investors. When banks
take large stakes in investors, $\widehat{DF}\left( X^{\prime }\right) $
diminishes, and so does $H$: when banks'perception of investors uncertainty
leads them to invest in investors rather than in firms, the risk of
propagation increases.

\subparagraph{Correlation between factors of fragility}

These various causes of structural fragility tend to reinforce one another.
A high level of capital invested in the financial sector typically coincides
with a low perception of risk and higher inter-investor stakes $\left\langle 
\hat{S}\left( X\right) \right\rangle $ or banks stakes in investors $%
\left\langle \hat{S}^{B}\left( X\right) \right\rangle $. Taken together,
these factors lower the threshold of average loan exposure ---$\left\langle 
\bar{S}_{L}\left( X^{\prime },X\right) \right\rangle _{X^{\prime
}}+\left\langle \hat{S}_{L}^{B}\left( X^{\prime },X\right) \right\rangle
_{X^{\prime }}+S_{L}^{B}\left( X,X\right) $--- required to trigger a
structural default.

Note that the banks default condition is similar to the condition for
investors, excpt that it involves the various loans: loans between banks $%
\left\langle \bar{S}_{L}\left( X^{\prime },X\right) \right\rangle
_{X^{\prime }}$ \ and loans to investors and firms $\left\langle \hat{S}%
_{L}^{B}\left( X^{\prime },X\right) \right\rangle _{X^{\prime
}}+S_{L}^{B}\left( X,X\right) $, and also also involves the loans of
investors $\left\langle \left( \hat{S}_{L}\left( X^{\prime },X\right)
\right) \right\rangle _{X^{\prime }}+S_{L}\left( X,X\right) $, consequence
of the interconnections between banks and investors. The default propagate
if a combination of loans of banks and loans of investors is above some
threshold.

\subsection{Description of the default state}

Once the default has materialized, the system is in a default state. This
state can be analyzed as a deviation from a no-default state. To do so, we
will compute, for the default state, both the average loss and the fraction
of investors affected.

When defaults occur, banks' returns are shifted by:

\begin{equation}
\bar{f}\left( X\right) \rightarrow \bar{f}\left( X\right) -d\bar{f}
\end{equation}%
where $d\bar{f}$ is the loss incured by the remaining investors for each
investor defaulting. We define, $\mu $ as the fraction of investors impacted
by the default. Appendix 11 first computes $\frac{\left\langle d\bar{f}%
\right\rangle }{\mu }$, representing the loss incurred by the remaining
banks per defaulting sector.

We find that:%
\begin{eqnarray}
&&\frac{\left\langle d\bar{f}\right\rangle }{\mu }=-\left( \frac{%
S_{E}^{B}\left( X,X\right) }{1-\bar{S}\left( X^{\prime }\right) }%
+S_{E}\left( X,X\right) \frac{S_{E}^{B}\left( X,X\right) }{1-\bar{S}\left(
X^{\prime }\right) }\frac{\hat{S}_{E}^{B}\left( X^{\prime },X\right) }{%
1-\left\langle \hat{S}\left( X^{\prime }\right) \right\rangle }\widehat{DF}%
\left( X^{\prime }\right) \right) \frac{A}{B}dC  \label{Lss} \\
&&+\left( \frac{1}{1-\bar{S}\left( X^{\prime }\right) }+\frac{%
S_{E}^{B}\left( X,X\right) }{1-\bar{S}\left( X^{\prime }\right) }\frac{\hat{S%
}_{E}^{B}\left( X^{\prime },X\right) }{1-\left\langle \hat{S}\left(
X^{\prime }\right) \right\rangle }\widehat{DF}\left( X^{\prime }\right)
\right) \left( \left\langle \left( \hat{S}_{L}\left( X^{\prime },X\right)
\right) \right\rangle _{X^{\prime }}+S_{L}\left( X,X\right) \right)   \notag
\\
&&+\frac{1}{1-\bar{S}\left( X^{\prime }\right) }\left( \left\langle \bar{S}%
_{L}\left( X^{\prime },X\right) \right\rangle _{X^{\prime }}+\left\langle 
\hat{S}_{L}^{B}\left( X^{\prime },X\right) \right\rangle _{X^{\prime
}}+S_{L}^{B}\left( X,X\right) \right)   \notag
\end{eqnarray}%
with:%
\begin{equation*}
A=A_{1}\left( \left\langle \left( \hat{S}_{L}\left( X^{\prime },X\right)
\right) \right\rangle _{X^{\prime }}+S_{L}\left( X,X\right) \right)
+A_{2}\left( \left\langle \bar{S}_{L}\left( X^{\prime },X\right)
\right\rangle _{X^{\prime }}+\left\langle \hat{S}_{L}^{B}\left( X^{\prime
},X\right) \right\rangle _{X^{\prime }}+S_{L}^{B}\left( X,X\right) \right) 
\end{equation*}%
\begin{equation*}
A_{1}=\frac{\frac{w\left( X\right) \hat{w}\left( X\right) }{2}\frac{\hat{K}%
_{X}\left\vert \hat{\Psi}\left( X\right) \right\vert ^{2}}{K_{X}\left\vert
\Psi \left( X\right) \right\vert ^{2}}+w_{E}^{B}\left( X\right) \left\langle 
\hat{w}_{E}^{B}\left( X\right) \right\rangle \frac{\bar{K}_{X}\left\vert 
\bar{\Psi}\left( X\right) \right\vert ^{2}}{K_{X}\left\vert \Psi \left(
X\right) \right\vert ^{2}}}{1-\left\langle \hat{S}\left( X\right)
\right\rangle }\text{, }A_{2}=w_{E}^{B}\left( X\right) \left\langle \bar{w}%
\left( X\right) \right\rangle _{\bar{S}}\frac{\bar{K}_{X}\left\vert \bar{\Psi%
}\left( X\right) \right\vert ^{2}}{K_{X}\left\vert \Psi \left( X\right)
\right\vert ^{2}}
\end{equation*}%
\begin{eqnarray*}
B &=&\left( 1-2\frac{S\left( X\right) +S^{B}\left( X\right) }{1-\left(
S\left( X\right) +S^{B}\left( X\right) \right) }\right)  \\
&&\times \left( 1-dC\frac{\left( \frac{w\left( X\right) \hat{w}\left(
X\right) }{2}\left( 1-\frac{\left\langle S_{E}\left( X,X\right)
\right\rangle }{1-\left\langle \hat{S}\left( X\right) \right\rangle }\right) 
\frac{\hat{K}_{X}\left\vert \hat{\Psi}\left( X\right) \right\vert ^{2}}{%
K_{X}\left\vert \Psi \left( X\right) \right\vert ^{2}}+w_{E}^{B}\left(
X\right) \left\langle \bar{w}\left( X\right) \right\rangle _{f\left(
X\right) }\frac{\bar{K}_{X}\left\vert \bar{\Psi}\left( X\right) \right\vert
^{2}}{K_{X}\left\vert \Psi \left( X\right) \right\vert ^{2}}\right) }{1-2%
\frac{S\left( X\right) +S^{B}\left( X\right) }{1-\left( S\left( X\right)
+S^{B}\left( X\right) \right) }}\right) 
\end{eqnarray*}%
where:%
\begin{equation}
dC=C-\frac{r\left( f_{1}\left( X\right) +\tau \Delta F_{\tau }\left( \bar{R}%
\left( K,X\right) \right) \right) }{\left( K_{X}\right) ^{r-1}}  \label{Dcr}
\end{equation}%
This parameter reflects the impact of defaults on the firms' cost
structure.\ It measures the loss in returns per unit of stake divested from
firm $X$. Thus, higher values of $dC$ are associated with larger capital
loss and an increased likelihood of default. 

The above expressions reveal that, the higher the investment stake $S\left(
X\right) $ in sector $X$, the smaller the value of $dC$. Financially
vulnerable states---characterized by a high $dC$---are more prone to
initiate a cascade of defaults. 

The fraction of defaulting investors is given by:%
\begin{equation}
\mu =\frac{1}{\frac{\left\langle d\bar{f}\right\rangle }{\mu }-2\left\langle 
\bar{f}\right\rangle }  \label{frc}
\end{equation}%
where all parameter values are those of a non-default scenario. Ultimately,
expressions (\ref{Lss}) and (\ref{frc}) allow to find the loss incurred by
the remaining banks:%
\begin{equation*}
\left\langle d\bar{f}\right\rangle =\frac{1}{\frac{\left\langle d\bar{f}%
\right\rangle }{\mu }-2\left\langle \bar{f}\right\rangle }\frac{\left\langle
d\bar{f}\right\rangle }{\mu }
\end{equation*}

The average loss due to default for the entire group of firms can also be
derived:%
\begin{equation}
\left\langle df\left( X\right) \right\rangle =-\frac{A}{B}\mu dC
\end{equation}

\part*{Stability of sub-collective states}

We define a sub-collective state as a specific group of sectors in the
system (and hence their agents) that are in a particular phase---that is,
the set of stakes, available capital, returns per sector, and all other
relevant variables.

Until now, we have treated sub-collective states as static. However, for any
group of investors, multiple sub-collective states can exist, including
several default states. Moreover, our analysis of default states has shown
that the set of defaulting agents ultimately results from a cascade of
defaults among firms and investors. This suggests that shocks could trigger
new phases in the collective states, and that there exist underlying
dynamics between these phases. Instabilities within a phase's dynamics may
thus reveal the possibility of transitions between phases.

To study these dynamics and the potential for phase transitions, we must
transform the static return equation (\ref{QDM}) into a dynamic one by
introducing a time parameter. We can then use this dynamic return equation
to analyze the deviations of any sub-group from a given static phase, and,
in turn, assess its stability or instability.

\section{Internal dynamics}

To reveal the internal dynamics of the system, we transform the static
return equations (\ref{QDM}) and (\ref{QDB}) for each group into a dynamic
system. To do so, we perform a first-order perturbation of these return
equations.

\subsection{Dynamic return equation}

To study the distribution of stakes within groups dynamically, we must
specify the temporal sequence of investments by replacing the static return
equations (\ref{QDM}) and (\ref{QDB}) with a dynamic formulation. For
investors, this can be expressed as:%
\begin{eqnarray}
0 &=&\widehat{DF}\left( X^{\prime },\theta -1\right) \left( \hat{f}\left(
X^{\prime },\theta \right) -\bar{r}\right)  \label{CPR} \\
&&-\left\langle \hat{S}_{E}\left( X^{\prime },X,\theta \right) \right\rangle
_{X^{\prime }}\left\langle \widehat{DF}\left( X,\theta -1\right)
\right\rangle \left( \left\langle \hat{f}\left( X^{\prime },\theta \right)
\right\rangle -\left\langle \hat{r}\left( X^{\prime }\right) \right\rangle
\right)  \notag \\
&&-S_{E}\left( X,X,\theta -1\right) \left( f\left( X,\theta -1\right)
-\left\langle \bar{r}\left( X\right) \right\rangle \right)  \notag
\end{eqnarray}%
where $\widehat{DF}\left( X,\theta -1\right) $ denotes the dynamic discount
factor of investor $X$ for debt repayment:

\begin{equation*}
\widehat{DF}\left( X,\theta -1\right) =\frac{1-\left\langle \hat{S}\left(
X^{\prime },\theta -1\right) \right\rangle +\left\langle \hat{S}%
_{E}^{B}\left( X^{\prime },\theta -1\right) \right\rangle +\left\langle \hat{%
S}_{L}^{B}\left( X^{\prime },\theta -1\right) \right\rangle }{1-\left\langle 
\hat{S}_{E}\left( X^{\prime },\theta -1\right) \right\rangle +\left\langle 
\hat{S}_{E}^{B}\left( X^{\prime },\theta -1\right) \right\rangle }
\end{equation*}%
and the firms' returns are assumed to exhibit decreasing returns to scale:%
\begin{equation*}
f\left( X,\theta -1\right) -\left\langle \bar{r}\left( X\right)
\right\rangle =\frac{f_{1}\left( X,\theta -1\right) +\tau \left(
\left\langle f_{1}\left( X\right) \right\rangle -\left\langle f_{1}\left(
X^{\prime }\right) \right\rangle \right) }{\left( K\left( X,,\theta
-1\right) \left\Vert \Psi \left( \left( X,,\theta -1\right) \right)
\right\Vert ^{2}\right) ^{r}}-\frac{C}{K\left( X,,\theta -1\right) }%
-C_{0}-\left\langle \bar{r}\left( X\right) \right\rangle
\end{equation*}%
The dynamic equation for the bank's return is given by:%
\begin{eqnarray}
0 &=&\overline{DF}\left( X,\theta -1\right) \left( \bar{f}\left( X,\theta
\right) -\left( 1+\kappa \right) \bar{r}\right)  \label{CPT} \\
&&-\left\langle \bar{S}_{E}\left( X^{\prime },X,\theta -1\right)
\right\rangle _{X^{\prime }}\left\langle \overline{DF}\left( X^{\prime
},\theta -1\right) \right\rangle \left( \left\langle \bar{f}\left( X^{\prime
},\theta \right) \right\rangle -\left( 1+\kappa \right) \bar{r}\right) 
\notag \\
&&-\int \left\langle \hat{S}_{E}^{B}\left( X^{\prime },X,\theta -1\right)
\right\rangle _{X^{\prime }}\left\langle \widehat{DF}\left( X^{\prime
},\theta -1\right) \right\rangle \left( \left\langle \hat{f}\left( X^{\prime
},\theta \right) \right\rangle -\left\langle \hat{r}\left( X^{\prime
}\right) \right\rangle \right) dX^{\prime }  \notag \\
&&-S_{E}^{B}\left( X,X,\theta -1\right) \left( f\left( X,\theta -1\right)
-\left\langle \bar{r}\left( X\right) \right\rangle \right)  \notag
\end{eqnarray}%
where $\overline{DF}\left( X,\theta -1\right) $ denotes the dynamic discount
factor of investor $X$ for debt repayment:%
\begin{equation*}
\overline{DF}\left( X,\theta -1\right) =\frac{1-\bar{S}\left( X,\theta
-1\right) }{1-\bar{S}_{E}\left( X,\theta -1\right) }
\end{equation*}%
and the returns they generate:%
\begin{equation*}
\hat{f}\left( X^{\prime },\theta \right) -\bar{r}
\end{equation*}%
The returns generated by firms are themselves determined by lagged
productivity:%
\begin{equation*}
f\left( X,\theta -1\right) -\bar{r}
\end{equation*}%
so that capital reallocation delays the effect on firms' returns.\footnote{%
In this study, we focus on the case of decreasing returns to scale.}%
\footnote{%
The formula for \ $f\left( \hat{X},\theta -1\right) -\bar{r}$ is provided in
Appendix 9 in Gosselin and Lotz (2025).}.

\subsection{Dynamic fluctuations of stakes}

Let us consider a first-order fluctuation in the returns of investors and
banks, denoted by $\delta \hat{f}\left( X\right) $ and $\delta \bar{f}\left(
X\right) $ respectively. Expanding the dynamic return equation (\ref{CPR})
around a static solution yields\footnote{%
Derived in Appendix 9 in Gosselin and Lotz (2025).} a dynamic equation for $%
\delta \hat{f}\left( X\right) $:\ 
\begin{eqnarray}
0 &=&\int dX^{\prime }\left( \delta \left( X-X^{\prime }\right) -\hat{S}%
_{E}\left( X^{\prime },X,\theta -1\right) \right) \delta \widehat{DF}\left(
X^{\prime },\theta -1\right) \left( \hat{f}\left( X\right) +\delta \hat{f}%
\left( X\right) -\bar{r}\right) \\
&&-\int \delta \hat{S}_{E}\left( X^{\prime },X,\theta -1\right) \widehat{DF}%
\left( X^{\prime },\theta -1\right) \left( \hat{f}\left( X^{\prime }\right) -%
\bar{r}\right) dX^{\prime }  \notag \\
&&-\delta S_{E}\left( X,X,\theta -1\right) \left( f\left( X,\theta -1\right)
-\bar{r}\right) -S_{E}\left( X,X,\theta -1\right) \frac{\partial f\left(
X,\theta -1\right) }{\partial S\left( X,\theta -1\right) }  \notag
\end{eqnarray}%
where:%
\begin{equation*}
\delta \widehat{DF}\left( X^{\prime },\theta -1\right) =\frac{1-\left(
\left\langle \hat{S}^{\prime }\left( X^{\prime },\theta -1\right)
\right\rangle +\delta \left\langle \hat{S}^{\prime }\left( X^{\prime
},\theta -1\right) \right\rangle \right) }{1-\left( \left\langle \hat{S}%
_{E}^{\prime }\left( X^{\prime },\theta -1\right) \right\rangle +\delta
\left\langle \hat{S}_{E}^{\prime }\left( X^{\prime },\theta -1\right)
\right\rangle \right) }
\end{equation*}%
with:%
\begin{eqnarray*}
\left\langle \hat{S}^{\prime }\left( X^{\prime },\theta -1\right)
\right\rangle &=&\left\langle \hat{S}\left( X^{\prime },\theta -1\right)
\right\rangle +\left\langle \hat{S}_{E}^{B}\left( X^{\prime },\theta
-1\right) \right\rangle +\left\langle \hat{S}_{L}^{B}\left( X^{\prime
},\theta -1\right) \right\rangle \\
\left\langle \hat{S}_{E}^{\prime }\left( X^{\prime },\theta -1\right)
\right\rangle &=&\left\langle \hat{S}_{E}\left( X^{\prime },\theta -1\right)
\right\rangle +\left\langle \hat{S}_{E}^{B}\left( X^{\prime },\theta
-1\right) \right\rangle
\end{eqnarray*}%
and where the aggregate stake allocation, $\delta S\left( X\right) $ varies
both with intra-sectoral investments, $\delta S\left( X,X\right) $, and with
the total disposable capital available for investment: 
\begin{equation}
\delta S\left( X\right) =\delta S\left( X,X\right) \frac{\hat{K}_{\hat{X}%
}\left\vert \hat{\Psi}\left( X\right) \right\vert ^{2}}{K_{\hat{X}%
}\left\vert \Psi \left( X\right) \right\vert ^{2}}+S\left( X,X\right) \delta 
\frac{\hat{K}_{X}\left\vert \hat{\Psi}\left( X\right) \right\vert ^{2}}{%
K_{X}\left\vert \Psi \left( X\right) \right\vert ^{2}}
\end{equation}%
Similarly, expanding the dynamic return equation (\ref{CPT}) around a static
solution yields a dynamic equation for $\delta \bar{f}\left( X\right) $: 
\begin{eqnarray}
0 &=&\int dX^{\prime }\left( \delta \left( X-X^{\prime }\right)
-\left\langle \bar{S}_{E}\left( X^{\prime },X,\theta -1\right) \right\rangle
_{X^{\prime }}\right) \\
&&\times \delta \overline{DF}\left( X^{\prime },\theta -1\right) \left( \bar{%
f}\left( X^{\prime },\theta \right) +\delta \bar{f}\left( X^{\prime },\theta
\right) -\left( 1+\kappa \right) \bar{r}\right) dX^{\prime }  \notag \\
&&-\int \delta \bar{S}_{E}\left( X^{\prime },X,\theta -1\right) \overline{DF}%
\left( X^{\prime },\theta -1\right) \left( \bar{f}\left( X^{\prime },\theta
\right) -\left( 1+\kappa \right) \bar{r}\right) dX^{\prime }  \notag \\
&&-\int \hat{S}_{E}^{B}\left( X^{\prime },X,\theta -1\right) \delta \widehat{%
DF}\left( X^{\prime },\theta -1\right) \left( \hat{f}\left( X\right) +\delta 
\hat{f}\left( X\right) -\bar{r}\right) dX^{\prime }  \notag \\
&&-\int \delta \hat{S}_{E}^{B}\left( X^{\prime },X,\theta -1\right) \widehat{%
DF}\left( X^{\prime },\theta -1\right) \left( \hat{f}\left( X^{\prime
}\right) -\bar{r}\right) dX^{\prime }  \notag \\
&&-\delta S_{E}^{B}\left( X,X,\theta -1\right) \left( f\left( X,\theta
-1\right) -\left\langle \bar{r}\left( X\right) \right\rangle \right)
-S_{E}^{B}\left( X,X,\theta -1\right) \frac{\partial f\left( X,\theta
-1\right) }{\partial S^{B}\left( X,\theta -1\right) }  \notag
\end{eqnarray}%
where:%
\begin{equation*}
\delta \overline{DF}\left( X^{\prime },\theta -1\right) =\frac{1-\left( \bar{%
S}\left( X,\theta -1\right) +\delta \bar{S}\left( X,\theta -1\right) \right) 
}{1-\left( \bar{S}_{E}\left( X,\theta -1\right) +\delta \bar{S}_{E}\left(
X,\theta -1\right) \right) }
\end{equation*}

The variations in stakes in each sector are described by a coupled system of
equations involving investor returns and aggregate sectoral stake
allocations $\left( \delta \bar{f}\left( X,\theta \right) ,\delta \hat{f}%
\left( X,\theta \right) ,\delta S^{T}\left( X,,\theta -1\right) \right) $
where $S^{T}\left( X,,\theta -1\right) $ denotes the total share of invested
capital in firms:%
\begin{equation*}
S^{T}\left( X,,\theta -1\right) =S\left( X,\theta -1\right) +S_{E}^{B}\left(
X,,\theta -1\right) +S_{L}^{B}\left( X,\theta -1\right)
\end{equation*}%
which captures the lag between the investments $\hat{S}_{E}\left( X^{\prime
},X,\theta -1\right) $, $\bar{S}_{E}\left( X^{\prime },X,\theta -1\right) $, 
$\hat{S}_{E}^{B}\left( X^{\prime },\theta -1\right) $ and $S_{E}\left(
X,X,\theta -1\right) $. We obtain\footnote{%
See Appendix 12.}:%
\begin{equation}
\left( 
\begin{array}{c}
\delta \bar{f}\left( X,\theta \right) \\ 
\delta \hat{f}\left( X,\theta \right) \\ 
\delta S^{T}\left( X,,\theta -1\right)%
\end{array}%
\right) =\left[ 
\begin{array}{ccc}
a & 0 & c \\ 
d & -1 & f \\ 
g & 0 & i%
\end{array}%
\right] \left( 
\begin{array}{c}
\delta \bar{f}\left( X,\theta -1\right) \\ 
\delta \hat{f}\left( X,\theta -1\right) \\ 
\delta S^{T}\left( X,,\theta -2\right)%
\end{array}%
\right)  \label{SV}
\end{equation}%
This system of equations can alternatively be reformulated entirely in terms
of stakes:%
\begin{equation*}
\left( \delta \bar{S}\left( X^{\prime },X,\theta \right) ,\delta \hat{S}%
\left( X^{\prime },X,\theta \right) ,\delta S^{T}\left( X,,\theta -1\right)
\right)
\end{equation*}%
in firms and other investors\footnote{%
See \ Appendix 12.}:%
\begin{equation*}
\left( 
\begin{array}{c}
\delta \bar{S}\left( X^{\prime },X,\theta \right) \\ 
\delta \hat{S}\left( X^{\prime },X,\theta \right) \\ 
\delta S^{T}\left( X,,\theta -1\right)%
\end{array}%
\right) =\left[ 
\begin{array}{ccc}
a & 0 & \frac{\bar{w}\left( X^{\prime },X\right) }{2}c \\ 
\frac{\hat{w}\left( X^{\prime },X\right) }{\bar{w}\left( X^{\prime
},X\right) }d & -1 & \frac{\hat{w}\left( X^{\prime },X\right) }{2}f \\ 
\frac{g}{\frac{\bar{w}\left( X^{\prime },X\right) }{2}} & 0 & i%
\end{array}%
\right] \left( 
\begin{array}{c}
\delta \bar{S}\left( X^{\prime },X,\theta -1\right) \\ 
\delta \hat{S}\left( X^{\prime },X,\theta -1\right) \\ 
\delta S^{T}\left( X,,\theta -2\right)%
\end{array}%
\right)
\end{equation*}%
The coefficients $\alpha $, $\beta $, $c$, and $h$ are defined explicitly in
Appendix 12.

\section{Internal stability}

To study the stability of any group under fluctuations, we first consider
that, for each sector $X^{\prime }$, the stakes of other investors in $%
X^{\prime }$can be aggregated. This amounts to replacing the stakes from
each alternate sector $X$ $\ $to $X^{\prime }$, $\hat{S}\left( X^{\prime
},X,\theta -1\right) $, by their average $\left\langle \hat{S}\left(
X^{\prime },X,\theta -1\right) \right\rangle _{X}$ over the entire sector
space. By doing so, the resulting return equations for $X^{\prime }$ depend
solely on $X^{\prime }$ and the inward aggregated averages. This approach
allows for the analysis of deviations in the return equation (\ref{SV})
sector by sector, independently. We will then provide a more detailed
account of the interactions between sectors to study the propagation of
perturbations throughout the sector space.

\subsection{Independent fluctuation of stakes}

\subsubsection{Eigenvalues and stability}

The eigenvalues of equation (\ref{SV}) determine the stability of each
sector\ under some perturbations. We have studied the stability resulting
specifically from investors'fluctuations in Gosselin and Lotz (2025) and in
this work we focus on banks.

When stakes are approximated by inward aggregate stakes, these eigenvalues
are expressed as:

\begin{equation}
\left( -1,\frac{1}{2}\left( a+i\right) -\frac{1}{2}\sqrt{\left( a-i\right)
^{2}+4cg},\frac{1}{2}\left( a+i\right) +\frac{1}{2}\sqrt{\left( a-i\right)
^{2}+4cg}\right)  \label{gnv}
\end{equation}%
They are typically negative\footnote{%
This stability analysis is conducted in Appendix 10 in Gosselin and Lotz
(2025).}, so that the system is generally stable. However, since the
dominant eigenvalue generally decreases with $\left\langle \bar{f}\left(
X\right) \right\rangle -\bar{r}$, instability arises when returns are high.
This corresponds to the high-return solution presented above and is
consistent with the higher level of capital circulation in this case.

A sector becomes unstable only when banks behave like investors, with $\bar{f%
}\left( X\right) -\left( 1+\kappa \right) \bar{r}>>1$ and $f\left( X\right)
=O\left( 1\right) $. In this situation, banks achieve returns higher than
those of firms. Such a case does not correspond to an equilibrium and is
ruled out in the long term.

Nonetheless, banks can also indirectly generate instability. As shown in
Gosselin and Lotz (2025), investors may become unstable under several
conditions. One case is of particular interest in the present work: when the
returns of firms $X$ do not differ significantly from the interest rates,
and when investors $X$ are heavily invested in these firms, that is: 
\begin{equation*}
f\left( X\right) -\bar{r}<<1\text{ and\textbf{\ }}S\left( X,\theta -1\right)
\rightarrow 1
\end{equation*}%
investors $X$ become unstable only if $\left\langle \hat{f}\left( X^{\prime
}\right) \right\rangle $\ is sufficiently low: even small shifts in expected
returns can induce significant reallocations of capital. This situation
arises, for instance, when uncertainty is very high and investors hold a
large amount of disposable capital, so that the potential returns across the
entire sector space, $\left\langle \hat{f}\left( X^{\prime }\right)
\right\rangle $ are too low to attract investors $X$ who primarily invest in
firms $X$. This leads to a high concentration of capital in firms\textbf{\ }$%
X$\textbf{\ , }which in turn results in low returns per unit of capital.
However, any fluctuation in returns\textbf{\ }$\left\langle \hat{f}\left(
X^{\prime }\right) \right\rangle $\textbf{\ }may shift the system toward
another equilibrium.

Without banks, these conditions are rarely met. However, when a large number
of banks invest predominantly in investors rather than directly in
firms---whether through shares or loans---these investments magnifies the
value of investors' stakes in firms' disposable capital, so that\textbf{\ }$%
S\left( X,\theta -1\right) \rightarrow 1$\textbf{,} thereby inducing
sectoral instability.

\subsubsection{Dynamics of perturbations}

For a small initial perturbation in investors' returns and stakes, the
dominant term in the dynamics is\footnote{%
See Appendix 12.}:%
\begin{equation}
\left( 
\begin{array}{c}
\delta \bar{f}\left( X,\theta \right) \\ 
\delta \hat{f}\left( X,\theta \right) \\ 
\delta S^{T}\left( X,,\theta -1\right)%
\end{array}%
\right) \simeq c\exp \left( \lambda _{+}\theta \right) V_{3}  \label{st}
\end{equation}%
where $\lambda _{+}$ denotes the largest eigenvalue of the system\footnote{%
The coefficients are detailed in the Appendix 12.}, $V_{3}$ the associated
eigenvector and where:%
\begin{equation*}
c=\left( 0,0,1\right) \left( \left( V_{1},V_{2},V_{3}\right) ^{-1}\left( 
\begin{array}{c}
\delta \bar{f} \\ 
\delta \hat{f} \\ 
\delta S%
\end{array}%
\right) \right)
\end{equation*}%
with $\left( V_{1},V_{2},V_{3}\right) $ the matrix of eigenvectors of the
system.

Depending on the initial disturbances $\delta \bar{f}$, $\delta \hat{f}$ and 
$\delta S^{T}$, two cases may arise: either banks $X$ and investors $X$
returns and total stakes in firms $X$ decay to zero, leading the system back
to its equilibrium; or they are amplified, inducing a transition of the
system towards another equilibrium.

In this second case, a positive perturbation either in $\delta \bar{f}$, $%
\delta \hat{f}$ or $\delta S^{T}$ pushes the sector towards a new
equilibrium of higher returns and larger capital accumulation in investors $%
X $: higher banks or investors' returns attract capital and increase the
firms' capitalization in the sector. Conversely, a negative perturbation
shifts the sector towards lower returns and lower firm capitalization.

\subsection{Interacting fluctuations between sectors}

In general, fluctuations in cross-sectoral investments are correlated, which
implies that the dynamics of several---or even all---sectors must be
considered in interaction. Consequently, system stability depends on the
specific interactions between sub-groups of agents, which should be analyzed
through sector-dependent stakes (\ref{SNp}) rather than through their
averages:%
\begin{eqnarray*}
\bar{S}\left( X^{\prime },X\right) &=&\bar{w}\left( X^{\prime },X\right)
\left( 1+\bar{w}\left( X\right) \left( \Delta \left( \bar{f}\left( X^{\prime
}\right) +\bar{r}\left( X^{\prime }\right) \right) \right) \right) \\
\hat{S}\left( X^{\prime },X\right) &=&\hat{w}\left( X^{\prime },X\right)
\left( 1+\frac{\hat{f}\left( X^{\prime }\right) +\hat{r}\left( X^{\prime
}\right) }{2}-\hat{w}\left( X\right) \hat{R}-w\left( X\right) R\left(
X\right) \right) \\
S\left( X,X\right) &=&w\left( X\right) \left( 1+\hat{w}\left( X\right)
\left( \frac{f\left( X\right) +\bar{r}\left( X\right) }{2}-\frac{%
\left\langle \hat{f}\left( X^{\prime }\right) \right\rangle _{\hat{w}%
_{E}}+\left\langle \hat{r}\left( X^{\prime }\right) \right\rangle _{\hat{w}%
_{L}}}{2}\right) \right)
\end{eqnarray*}%
with:%
\begin{equation*}
\hat{R}=\frac{\left\langle \hat{f}\left( X^{\prime }\right) \right\rangle _{%
\hat{w}_{E}}+\left\langle \hat{r}\left( X^{\prime }\right) \right\rangle _{%
\hat{w}_{L}}}{2}
\end{equation*}%
\begin{equation*}
R\left( X\right) =\frac{f\left( X\right) +r\left( X\right) }{2}
\end{equation*}%
\begin{equation*}
\Delta \bar{f}\left( X^{\prime }\right) =\bar{f}\left( X^{\prime }\right)
-\left( \bar{w}\left( X\right) \frac{\left\langle \bar{f}\left( X^{\prime
}\right) \right\rangle _{\bar{w}_{E}}+\left\langle \bar{r}\left( X^{\prime
}\right) \right\rangle _{\bar{w}_{L}}}{2}+\hat{w}_{E}^{B}\left( X\right)
\left\langle \hat{f}\left( X^{\prime }\right) \right\rangle _{\hat{w}%
_{E}}+w_{E}^{B}\left( X\right) f\left( X\right) \right)
\end{equation*}%
and:%
\begin{equation*}
\Delta \bar{r}\left( X^{\prime }\right) =\bar{r}\left( X^{\prime }\right)
-\left( \bar{w}\left( X\right) \frac{\left\langle \bar{f}\left( X^{\prime
}\right) \right\rangle _{\bar{w}_{E}}+\left\langle \bar{r}\left( X^{\prime
}\right) \right\rangle _{\bar{w}_{L}}}{2}+\hat{w}_{E}^{B}\left( X\right)
\left\langle \hat{f}\left( X^{\prime }\right) \right\rangle _{\hat{w}%
_{E}}+w_{E}^{B}\left( X\right) f\left( X\right) \right)
\end{equation*}%
The sector-dependent variations of the total stakes are given by:%
\begin{eqnarray*}
\delta \bar{S}\left( X^{\prime },X\right) &=&\bar{w}\left( X^{\prime
},X\right) \bar{w}\left( X\right) \left( \delta \bar{f}\left( X^{\prime
}\right) -w_{E}^{B}\left( X\right) \delta f\left( X\right) \right) \\
\delta \hat{S}\left( X^{\prime },X,\theta -1\right) &\rightarrow &\frac{\hat{%
w}\left( X^{\prime },X\right) }{2}\left( \frac{\delta \hat{f}\left(
X^{\prime }\right) }{2}-\left\langle w\left( X\right) \right\rangle \delta
R\left( X\right) \right) \\
\delta S\left( X,X,\theta -1\right) &\rightarrow &w\left( X\right) \hat{w}%
\left( X\right) \frac{\delta f\left( X\right) }{2}
\end{eqnarray*}%
These formulas show how a change in the return of sector $X^{\prime }$\
modifies the stakes $\hat{S}\left( X^{\prime },X,\theta -1\right) $\ held by
an investor $X$, thereby indicating that sectors $X$\ and $X^{\prime }$\ are
interrelated.

\begin{eqnarray*}
\hat{S}_{E}\left( X^{\prime },X\right) &\rightarrow &\frac{\hat{w}\left(
X^{\prime },X\right) }{2}\left( 1+\left( \hat{f}\left( X^{\prime }\right)
-\left\langle \hat{w}\left( X\right) \right\rangle \hat{R}-\left\langle
w\left( X\right) \right\rangle R\left( X\right) \right) \right) \\
\bar{S}_{E}\left( X^{\prime },X\right) &=&\frac{\bar{w}\left( X^{\prime
},X\right) }{2}\left( 1+\bar{w}\left( X\right) \Delta \bar{f}\left(
X^{\prime }\right) \right)
\end{eqnarray*}%
with:%
\begin{equation*}
\hat{R}=\frac{\left\langle \hat{f}\left( X^{\prime }\right) \right\rangle _{%
\hat{w}_{E}}+\left\langle \hat{r}\left( X^{\prime }\right) \right\rangle _{%
\hat{w}_{L}}}{2}
\end{equation*}%
\begin{equation*}
R\left( X\right) =\frac{f\left( X\right) +r\left( X\right) }{2}
\end{equation*}%
and:%
\begin{equation*}
\Delta \bar{f}\left( X^{\prime }\right) =\bar{f}\left( X^{\prime }\right)
-\left( \bar{w}\left( X\right) \frac{\left\langle \bar{f}\left( X^{\prime
}\right) \right\rangle _{\bar{w}_{E}}+\left\langle \bar{r}\left( X^{\prime
}\right) \right\rangle _{\bar{w}_{L}}}{2}+\hat{w}_{E}^{B}\left( X\right)
\left\langle \hat{f}\left( X^{\prime }\right) \right\rangle _{\hat{w}%
_{E}}+w_{E}^{B}\left( X\right) f\left( X\right) \right)
\end{equation*}%
and their sector-dependent variations:%
\begin{eqnarray*}
\delta \hat{S}_{E}\left( X^{\prime },X,\theta -1\right) &=&\frac{\hat{w}%
\left( X^{\prime },X\right) }{2}\left( 1+\left( \delta \hat{f}\left(
X^{\prime }\right) -\left\langle w\left( X\right) \right\rangle \delta
R\left( X\right) \right) \right) \\
\delta \bar{S}_{E}\left( X^{\prime },X\right) &=&\frac{\bar{w}\left(
X^{\prime },X\right) }{2}\left( 1+\bar{w}\left( X\right) \delta \bar{f}%
\left( X^{\prime }\right) -w_{E}^{B}\left( X\right) \delta f\left( X\right)
\right)
\end{eqnarray*}%
These corrections modify\footnote{%
See Appendix 11 in Gosselin and Lotz (2025).}, for each sector, the return
equations:%
\begin{eqnarray*}
\frac{1-\left( \hat{S}\left( X\right) \right) }{1-\left( \hat{S}_{E}\left(
X\right) \right) }\delta \hat{f}\left( X,\theta \right) &=&d\delta \bar{f}%
\left( X,\theta -1\right) +e\delta \hat{f}\left( X,\theta -1\right) +f\delta
S^{T}\left( X,\theta -2\right) \\
&&+\int T\left( X,X^{\prime }\right) \delta \bar{f}\left( X^{\prime },\theta
-1\right)
\end{eqnarray*}%
\begin{eqnarray*}
\frac{1-\bar{S}\left( X\right) }{1-\bar{S}_{E}\left( X\right) }\delta \bar{f}%
\left( X,\theta -1\right) &=&a\delta \bar{f}\left( X,\theta -1\right)
+b\delta \hat{f}\left( X,\theta -1\right) +c\delta S^{T}\left( X,\theta
-2\right) \\
&&+\int V\left( X,X^{\prime }\right) \delta \bar{f}\left( X^{\prime },\theta
-1\right) +\int W\left( X,X^{\prime }\right) \delta \hat{f}\left( X^{\prime
},\theta -1\right)
\end{eqnarray*}%
where the coefficients:%
\begin{eqnarray*}
V\left( X,X^{\prime }\right) &=&\bar{S}_{E}\left( X^{\prime },X\right) \frac{%
1-\left\langle \bar{S}\left( X^{\prime }\right) \right\rangle }{%
1-\left\langle \bar{S}_{E}\left( X^{\prime }\right) \right\rangle } \\
W\left( X,X^{\prime }\right) &=&\hat{S}_{E}^{B}\left( X^{\prime },X\right) 
\frac{1-\left\langle \hat{S}\left( X^{\prime }\right) \right\rangle
+\left\langle \hat{S}_{E}^{B}\left( X^{\prime }\right) \right\rangle
+\left\langle \hat{S}_{L}^{B}\left( X^{\prime }\right) \right\rangle }{%
1-\left\langle \hat{S}_{E}\left( X^{\prime }\right) \right\rangle
+\left\langle \hat{S}_{E}^{B}\left( X^{\prime }\right) \right\rangle }
\end{eqnarray*}%
and:%
\begin{equation*}
T\left( X,X^{\prime }\right) =\hat{S}_{E}\left( X^{\prime },X\right) \frac{%
1-\left\langle \hat{S}\left( X^{\prime }\right) \right\rangle }{%
1-\left\langle \hat{S}_{E}\left( X^{\prime }\right) \right\rangle }
\end{equation*}%
transcribe the interactions between sectors $X$ and $X^{\prime }$. These
corrections modify\footnote{%
See Appendix 12.}, for each sector, the system matrix by introducing an
additional term:%
\begin{equation*}
\left[ 
\begin{array}{ccc}
a & 0 & c \\ 
d & -1 & f \\ 
g & 0 & i%
\end{array}%
\right] \rightarrow \left[ 
\begin{array}{ccc}
a & 0 & c \\ 
d & -1 & f \\ 
g & 0 & i%
\end{array}%
\right] +\left[ 
\begin{array}{ccc}
0 & V\left( X,X^{\prime }\right) & W\left( X,X^{\prime }\right) \\ 
0 & 0 & T\left( X,X^{\prime }\right) \\ 
0 & 0 & 0%
\end{array}%
\right]
\end{equation*}%
and alters the dynamic coefficients at each point in the group, such that
equation (\ref{SV}) becomes:%
\begin{eqnarray}
\left( 
\begin{array}{c}
\delta \bar{f}\left( X,\theta \right) \\ 
\delta \hat{f}\left( X,\theta \right) \\ 
\delta S^{T}\left( X,\theta -1\right)%
\end{array}%
\right) &=&\left[ 
\begin{array}{ccc}
a & 0 & c \\ 
d & -1 & f \\ 
g & 0 & i%
\end{array}%
\right] \left( 
\begin{array}{c}
\delta \bar{f}\left( X,\theta -1\right) \\ 
\delta \hat{f}\left( X,\theta -1\right) \\ 
\delta S^{T}\left( X,\theta -2\right)%
\end{array}%
\right) \\
&&+\int dX^{\prime }\left[ 
\begin{array}{ccc}
0 & V\left( X,X^{\prime }\right) & W\left( X,X^{\prime }\right) \\ 
0 & 0 & T\left( X,X^{\prime }\right) \\ 
0 & 0 & 0%
\end{array}%
\right] \left( 
\begin{array}{c}
\delta \bar{f}\left( X,\theta -1\right) \\ 
\delta \hat{f}\left( X,\theta -1\right) \\ 
\delta S^{T}\left( X,\theta -2\right)%
\end{array}%
\right)  \notag
\end{eqnarray}%
The above modifications turn the eigenvalues (\ref{gnv}) into local
quantities, so that the stability of each sector now depends on the value of 
$c\left( X\right) $.

\subsubsection{Unilateral interactions: Modification of eigenvalues and
instability propagation}

To analyze how sub-groups of interacting agents diverge from non-interacting
ones, we assume that stakes are both sector-dependent and dynamically
interrelated. Under this assumption, it becomes possible to assess the
impact of sector-specific variations $W$ on neighboring sectors as well as
on the system as a whole.

A typical case of unilateral interactions between two sectors can be
expressed as a dynamical system of two distinct sectors $X_{1}$ and $X_{2}$,
in the following way:

\begin{equation}
\left( 
\begin{array}{c}
\delta \bar{f}\left( X_{1},\theta \right) \\ 
\delta \hat{f}\left( X_{1},\theta \right) \\ 
\delta S^{T}\left( X_{1},\theta -1\right)%
\end{array}%
\right) =\left[ 
\begin{array}{ccc}
a_{1} & 0 & c_{1} \\ 
d_{1} & -1 & f_{1} \\ 
g_{1} & 0 & i_{1}%
\end{array}%
\right] \left( 
\begin{array}{c}
\delta \bar{f}\left( X_{1},\theta -1\right) \\ 
\delta \hat{f}\left( X_{1},\theta -1\right) \\ 
\delta S^{T}\left( X_{1},\theta -2\right)%
\end{array}%
\right)  \label{Eqtn}
\end{equation}%
and:%
\begin{eqnarray}
\left( 
\begin{array}{c}
\delta \bar{f}\left( X_{2},\theta \right) \\ 
\delta \hat{f}\left( X_{2},\theta \right) \\ 
\delta S^{T}\left( X_{2},\theta -1\right)%
\end{array}%
\right) &=&\left[ 
\begin{array}{ccc}
a_{2} & 0 & c_{2} \\ 
d_{2} & -1 & f_{2} \\ 
g_{2} & 0 & i_{2}%
\end{array}%
\right] \left( 
\begin{array}{c}
\delta \bar{f}\left( X_{2},\theta -1\right) \\ 
\delta \hat{f}\left( X_{2},\theta -1\right) \\ 
\delta S^{T}\left( X_{2},\theta -2\right)%
\end{array}%
\right) \\
&&+\left[ 
\begin{array}{ccc}
0 & V\left( X_{2},X_{1}\right) & W\left( X_{2},X_{1}\right) \\ 
0 & 0 & T\left( X_{2},X_{1}\right) \\ 
0 & 0 & 0%
\end{array}%
\right] \left( 
\begin{array}{c}
\delta \bar{f}\left( X_{1},\theta -1\right) \\ 
\delta \hat{f}\left( X_{1},\theta -1\right) \\ 
\delta S^{T}\left( X_{1},\theta -2\right)%
\end{array}%
\right)  \notag
\end{eqnarray}%
This system exhibits non-reciprocal interactions: variations in one sector
unilaterally affect the other.\ Investment behavior in sector $1$ remains
unchanged, but causes investment in sector $2$ to deviate from its average.
Such interactions, by themselves, do not modify the eigenvalues or the
overall stability of the system. As long as sector $1$ remains stable, the
dynamics of sector $2$ is impacted but still stable: fluctuations within
both blocks remain damped and contained. Instability arises only when sector 
$1$ becomes unstable, triggering a reallocation of investment stakes and
disposable capital across sectors. This instability may then propagate to
adjacent sectors and, ultimately, to the system as a whole.

\subsubsection{Multilateral interactions: Modification of eigenvalues and
higher instability}

To study the reciprocal interactions between two sectors, we symmetrically
modify both dynamical matrices, replacing (\ref{Eqtn}) with:

\begin{eqnarray}
\left( 
\begin{array}{c}
\delta \bar{f}\left( X_{1},\theta \right) \\ 
\delta \hat{f}\left( X_{1},\theta \right) \\ 
\delta S^{T}\left( X_{1},\theta -1\right)%
\end{array}%
\right) &=&\left[ 
\begin{array}{ccc}
a_{1} & 0 & c_{1} \\ 
d_{1} & -1 & f_{1} \\ 
g_{1} & 0 & i_{1}%
\end{array}%
\right] \left( 
\begin{array}{c}
\delta \bar{f}\left( X_{1},\theta -1\right) \\ 
\delta \hat{f}\left( X_{1},\theta -1\right) \\ 
\delta S^{T}\left( X_{1},\theta -2\right)%
\end{array}%
\right)  \label{CD} \\
&&+\left[ 
\begin{array}{ccc}
0 & V\left( X_{1},X_{2}\right) & W\left( X_{1},X_{2}\right) \\ 
0 & 0 & T\left( X_{1},X_{2}\right) \\ 
0 & 0 & 0%
\end{array}%
\right] \left( 
\begin{array}{c}
\delta \bar{f}\left( X_{2},\theta -1\right) \\ 
\delta \hat{f}\left( X_{2},\theta -1\right) \\ 
\delta S^{T}\left( X_{2},\theta -2\right)%
\end{array}%
\right)  \notag
\end{eqnarray}%
and we refer to this configuration as a multilateral deviation from the
average. We then consider the modifications of each sector's eigenvalues and
the corresponding stability, under the simplifying hypothesis\footnote{%
See Appendix 11 in Gosselin and Lotz (2025).}: 
\begin{equation*}
\left( a_{1},c_{1},d_{1},f_{1},g_{1},i_{1}\right) \simeq \left(
a_{2},c_{2},d_{2},f_{2},g_{2},i_{2}\right)
\end{equation*}%
\begin{equation*}
V\left( X_{2},X_{1}\right) =T\left( X_{2},X_{1}\right) =V\left(
X_{1},X_{2}\right) =T\left( X_{1},X_{2}\right) =0
\end{equation*}%
to isolate the effect of the reciprocal extra-shares $W=W\left(
X_{1},X_{2}\right) $ and $W^{\prime }=W\left( X_{2},X_{1}\right) $. We then
determine the resulting modification of the dominant eigenvalue:%
\begin{equation*}
\Delta \lambda _{+}\left( X\right) =\frac{\hat{S}_{E}\left( X^{\prime
},X,\theta -1\right) \frac{1-\hat{S}\left( X^{\prime }\right) }{1-\hat{S}%
_{E}\left( X^{\prime }\right) }\hat{S}_{E}\left( X,X^{\prime },\theta
-1\right) \frac{1-\hat{S}\left( X\right) }{1-\hat{S}_{E}\left( X\right) }}{%
\lambda _{X}-\lambda _{X^{\prime }}}\frac{h^{2}}{\left( \alpha -\beta
\right) ^{2}+4ch}
\end{equation*}%
Under these conditions, the largest eigenvalue may indirectly drive the
system into an unstable zone, where $\lambda _{X}>>1$.

More generally, circular investment deviations---i.e., loops of altered
investment patterns involving multiple sectors---modify the eigenvalues
through sums of contributions of the type:%
\begin{equation*}
\Delta \lambda _{i\alpha }=\prod \frac{W_{j_{k+1}\alpha _{k+1},j_{k}\alpha
_{k}}}{\lambda _{i\alpha }-\lambda _{j_{k+1}\alpha _{k+1}}}
\end{equation*}%
where $\lambda _{i\alpha }${} denote the eigenvalues associated with sector $%
i$, and $W_{j_{k+1}\alpha _{k+1},j_{k}\alpha _{k}}${}represents the
interaction between $j_{k}\alpha _{k}$ and $j_{k+1}\alpha _{k+1}$.

The global effect is the sum over all possible paths and computed by: 
\begin{equation*}
\Delta \lambda _{i\alpha }=\sum_{\left( j_{k}\alpha _{k}\right) }\prod \frac{%
W_{j_{k+1}\alpha _{k+1},j_{k}\alpha _{k}}}{\lambda _{i\alpha }-\lambda
_{j_{k+1}\alpha _{k+1}}}
\end{equation*}%
These loops directly affect the eigenvalues and the stability of the
collective state. When the largest eigenvalue becomes positive, the system
enter an unstable zone and may transition to another equilibrium. In this
regime, any decline in returns toward zero or negative values can drive the
system into default. Overall, such looped interactions increase systemic
fragility and contribute to the propagation of instabilities throughout the
investment network.

\section{Transitions}

Within groups of stably interconnected agents, external fluctuations may
change connectivity patterns which may in turn generate circulation loops
among specific groups of agents and reinforce internal investments within
these groups.

Such feedback loops may trigger instability when fluctuations in investors'
returns divert capital away from firms, and push the system toward a default
state. This new equilibrium typically involves a smaller, more concentrated
group of investors.

\subsection{Group modification}

Under greater certainty, an anticipated increase in returns may induce new
investment toward new groups of agents. Consequently, several groups may
merge to form a larger collective structure.

Let us consider a shock affecting both returns and stakes within a single
group $G_{a}$:%
\begin{equation*}
\left\{ \left( 
\begin{array}{c}
\delta \bar{f}_{a}\left( X,\theta \right) \\ 
\delta \hat{f}_{a}\left( X,\theta \right) \\ 
\delta S_{a}^{T}\left( X,\theta -1\right)%
\end{array}%
\right) \right\} _{a}
\end{equation*}%
where $a$ is the group index. This may induce two groups $G_{a}${} and $%
G_{b} $ to merge, creating a joint vector of returns $\left( \hat{f}%
_{a}+\delta \hat{f}_{a}\right) \left( X,\theta \right) $ and stakes $\left(
S_{a}+\delta S_{a}\right) \left( X,\theta -1\right) $:%
\begin{equation*}
\left( 
\begin{array}{c}
\left( \mathbf{\bar{f}+\delta \bar{f}}\right) \left( X,\theta \right) \\ 
\left( \mathbf{\hat{f}+\delta \hat{f}}\right) \left( X,\theta \right) \\ 
\left( \mathbf{S}^{T}\mathbf{+\delta S}^{T}\right) \left( X,\theta -1\right)%
\end{array}%
\right)
\end{equation*}%
which represents the evolving state of the newly formed group.

Dynamically, the transition unfolds through the propagation mechanism:%
\begin{eqnarray*}
\delta \bar{f}_{a}\left( X_{a},\theta \right) &\rightarrow &\delta
S_{ab}^{T}\left( X_{a},\theta -1\right) \\
\delta \hat{f}_{a}\left( X_{a},\theta \right) &\rightarrow &\delta
S_{ab}^{T}\left( X_{a},\theta -1\right) \\
\delta S_{ab}^{T}\left( X_{a},\theta -1\right) &\rightarrow &\delta
S_{b}^{T}\left( X_{b},\theta -1\right) \rightarrow \delta \lambda \left(
X_{b},\theta -1\right)
\end{eqnarray*}%
Even if all elements of the original groups were incorporated into the new
structure, $\left\{ X_{a}\right\} \cup \left\{ X_{b}\right\} $, the
resulting group may still contain unstable components, potentially driving
the system toward new equilibria, including default configurations.

\subsection{Interpretation}

The evolution towards instability can be analyzed within the framework
described in (\ref{CD}). When new connections are introduced, the system
evolves according to $\left( \bar{f}\left( X\right) ,\hat{f}\left( X\right)
\right) \rightarrow \left( \bar{f}\left( X\right) ,\hat{f}\left( X\right)
\right) +\left( \delta \bar{f}\left( X\right) ,\delta \hat{f}\left( X\right)
\right) $, potentially triggering transitions toward instability in specific
sectors. While the impacted sector begins to exhibit fluctuations, the rest
of the system may remain stable in the short term. Gradually, however, the
real economic activity of the affected group contracts, and its financial
stability deteriorates under the cumulative effect of these fluctuations.
Over time, this instability can lead to default in some sectors, thereby
reshaping the structure and connectivity of the broader system\footnote{%
The parameters characterizing the emergence of such unstable regimes, as
well as the range of fluctuations capable of triggering these transitions,
are detailed in Appendix 10 in Gosselin and Lotz (2025).}.

\part*{Results and Discussion}

\section{Results}

Incorporating money creation into our framework refines and modifies our
previous results in several respects, particularly regarding the
distribution of capital between investors and firms, the circulation of
capital, and overall stability\footnote{%
The specific characteristics of investors have already been discussed in
Gosselin and Lotz (2025).}. The general structure of collective states as
collections of sub-collective states, already established in our earlier
work, is presented here for completeness.

\subsection{Sub-collective states}

Introducing a third type of agent does not alter the overall structure of
collective states. In a fully endogenized model, agents tend to
self-organize into relatively independent groups, which in turn leads to a
resolution at the group level.

Each group is characterized by the sectors it contains, the number of agents
per sector, and the phase the group is in. Each phase is defined by the
agents' interconnections, namely here their stakes, disposable capital
levels, and returns. Multiple phases exist, and among them, one or several
may actually correspond to default states, that are phases in which some
sectors within the group disappear due to a lack capital. We define a \emph{%
sub-collective state} as a group within a phase.\ Each group can thus convey
to multiple sub-collective states. Collective states thus exhibit a
two-dimensional multiplicity: first, in the decomposition of agents into
groups, and second, in the infinite combinations of sub-collective states
associated with a given group partition.

The collective states of the system are a collection of sub-collective
states, each characterized by three essential parameters, the agents'
returns, their capital levels, and their mutual stakes.\ Finding the
collective states of the system thus amounts to finding these three
variables for each group of the system that determine the phase of each
group.\ From these variables, we will also derive all the relevant
quantities of the system.

\subsection{Equilibria}

The model yields two possible types of equilibria: a no-default equilibrium
and a default equilibrium. These represent two distinct static
configurations, each associated with a different resolution mechanism. The
no-default equilibria are solved under the assumption that all firms have
access to all the capital they require. In the resolution involving default,
all the effects of default are taken into account---namely, both the loss of
capital by the firms and, consequently, by their investors. This highlights
the mechanism through which default propagates across the system.

\subsubsection{Equilibrium without default}

\paragraph{Average equilibrium}

When banks are introduced in the model, they provide an alternate, and
actually the prime source of loans in the system. However, this impact of
banks crucially depends on their relative weight on the system. An
underdevelopped banking system will not substantially impact the system. The
general results of Gosselin and Lotz (2025) will apply.\textbf{\ }In the
following, we will suppose the number of banks is large enough to have a
systemic impact. The following results must weighed by the size of the
banking system.

The introduction of banks in the system impose to refine the notion of
uncertainty by distinguishing their and investors' views of uncertainty.
This distinction may concern investments or information, but may also
reflect the specific regulations governing banks in their loans and
participation. These factors can alter the level of uncertainty faced by
banks and, in turn, exogenously modify their stakes.

A prominant banking system may alter the distribution of disposable capital.
When banks\ mainly finance firms, they can crowd out investors by providing
most of firms' disposable capital through loans. This capital will be less
volatile and subject to fluctuations in investors' decisions. However, to be
sustainable in the long run, the interest rates charged on firms must be
lower than firms' productivity. When banks rather choose finance investors,
they will further magnify investors' influence on the system, since
investors will benefit from both loans and banks' participation.

The allocation of capital depends on how banks compare firms' and investors'
uncertainty.

When banks and investors have a similar perception of uncertainty, we
recover the same results as Gosselin and Lotz (2025): under constant
productivity and interest rates, uncertainty reduces investment flows
between investors. By increasing intra-sectoral investment, uncertainty
decreases the capital available to investors and, indirectly, to firms. This
reduction in intermediation stabilizes the collective state: capital
fluctuations are dampened for firms, investors, and banks, albeit with lower
returns for both investors and banks.

When divergences in risk perception between banks and investors appear, they
distort capital allocation: an increase in banks' uncertainty regarding
firms' returns reduces lending to firms and shifts credit and capital flows
toward investors. This shift raises investors' disposable capital, thereby
increasing capital flows among them, while also channeling equity and loans
into firms, effectively transferring ownership of firms' disposable capital
to investors.

\paragraph{Equilibria per sector}

\subparagraph{Investors}

Investors' sectoral equilibria depend on whether the number of banks exceeds
a critical threshold. The number of banks can be above or below this
threshold.

When the number of banks is below this threshold, banks do not modify the
form of the solutions for investors. A double solution emerges, of high- and
low-returns, although mitigated by the impact of banks: both investors and
firms have an increased access to capital, which limits unequal
concentration of disposable capital and discrepancies between high- and
low-returns. Qualitatively, sector-level solutions do not differ
significantly from Gosselin and Lotz (2025): investors may adopt either
high-return or low-return profile relative to their sector, regardless of
firms' returns. Nevertheless, the difference between high- and low-return
investors is dampened by the loans and shares held by the banks in investors.

When the number of banks is above the threshold, the form of the solutions
is modified. There is only one solution for investors: banks do have an
impact and provide enough capital to equalize returns and capital levels
between investors. Investors have sufficient disposable capital to invest in
all available opportunities, and they do so evenly. As a result, all
investors benefit from high levels of disposable capital, and no individual
investor is able to extract above average returns. Moreover, because
investors' disposable capital is significantly increased and must be
invested, both firms and investors benefit from this capital expansion.
Since investors diversify across both firms and other investors, and all
investors have sufficient disposable capital, investors will adopt a
low-return profile: firms gain access to disposable capital, but this
capital remains under investors' control.

The effect of the number of banks is amplified by the way uncertainty is
perceived by banks and investors. First, the threshold is positively
correlated with banks' confidence in investors: as banks lose confidence in
investors, the threshold decreases, effectively restricting investors'
access to bank credit, thereby favoring firms and modifying the equilibria
towards the unique solution. Second the threshold is negatively correlated
to the confidence of the market, that is the level of intermediation. When
this confidence increases, investors tend to exchange with each other, which
reduces the dispcrepancy between high- and low-return solutions, enhancing
the impact of banks on the market.

Thus, considering the impact of banks number potentialized by uncertainty's
perception, the action of banks could be understood as stabilizing the
allocation of capital between investors and firms.

However, the way banks perceive uncertainty has further implications for the
system. When the number of banks is below the threshold and they favour
investors over firms, the dual solution may still be more unstable than in
the absence of banks: by funding investors, banks enable them to invest more
and exert control over firms' disposable capital, which renders the firms'
capital more dependent on the market and its fluctuations.

When banks disproportionately favor investors over firms but remain numerous
enough to remain above the threshold, they equalize investors' returns, but
most of the firms' disposable capital ends up being held by investors. The
level of investors' stakes being highly dependent on the returns of their
investments, a shock in the distribution of these returns can trigger larger
shifts in capital.

\subparagraph{Banks}

Banks may to some extent reduce the heterogeneity between high- and
low-return investors by providing capital evenly across the system. However,
nothing prevents banks themselves to being split into high- and low returns
profiles, regardless of the sector in which they operate; while also
benefiting from the capacity to extend substantial loans. This ability to
grant large loans reinforces the discrepancies between high and low returns:
a few banks achieve very high returns and hold significant capital, while
many others display comparatively low returns and limited capital.

The emergence of high- and low-return profiles among banks --- and
eventually among investors --- is a characteristic feature of the model. The
average resolution, although important, is not representative, and
optimization behaviors globally induce a strong dispersion among agents of
the same type. Furthermore, the impact of banks on investors shows that
reducing the difference between high- and low-return banks would require a
third type of agent, performing a form of regulation or following rules that
differ from standard optimization behaviors.

\subsubsection{Equilibrium with defaults}

A default state is a static configuration that includes the initially
defaulting firms, as well as all firms, banks, and investors affected by the
propagation of the default. Default states are an integral part of the
model's set of possible solutions.\ However, to be found, these states have
to be deemed possible in the resolution process. We must assume that one or
more firms are in default, to find the corresponding static solution under
this hypothesis using a recursive resolution. Thus the solutions under
default depend both on the set of initial defaults and on the conditions
governing the propagation dynamics.

The conditions for a default state depend on the structural parameters of
the system: its global level of loans, returns, intermediation, and
uncertainty. We can distinguish two set of conditions: one, for initial
defaults to propagate from the defaulting firm to its direct investors and
banks, and the second, for the propagation of default these investors and
banks to other investors and banks.

The global level of loans measures the total amount of credit granted to
firms, investors, and banks by investors and banks alike. It serves as a
measure of the degree of interconnection among agents. Accordingly, the
higher this level, the more easily an initial default can propagate
throughout the system.

The role of uncertainty is more ambiguous. When confidence in the investors
and banks directly connected to initially defaulting firms is high, it
reduces the likelihood of contagion from those firms to their direct
investors. However, a high average level of confidence among investors
favors the propagation of defaults to the entire system. Ultimately, low
average returns for investors weaken agents and increase the likelihood of
contagion.

In this context, the role of banks is ambivalent. By providing capital to
firms, banks may directly prevent some initial defaults in firms. However,
their loans may also indirectly propagate defaults to neighboring investors
and, ultimately, throughout the system. This second, indirect effect depends
on the overall level of loans within the system. A large number of banks
leads to a relatively higher proportion of loans compared to equity, thereby
increasing the likelihood of defaults emerging and spreading.

Moreover, banks are not immune to uncertainty. When banks are overly
confident about the financial health of investors, they tend to prioritize
investors over firms in their investment decisions. This, in turn, increases
the amount of loans extended to investors, thereby facilitating the
propagation of defaults from investors back to banks.

\subsection{Stability and transitions}

\subsubsection{Stability}

\paragraph{Average stability}

When considered on average, a sub-collective state has the apparearance of
stability. Average instability may arise in particular cases, for instance
when firms' returns are much lower than investors' and banks' returns%
\footnote{%
See section 15.1.}.\ But, on average, this situation does not occur for
sub-collective states, which, by assumption, do not interact with each
others: on average, investors and firms returns are similar, which excludes
any global instability phenomenon. Yet, some instability is present at the
sectoral level, confirming that averages do not properly describe collective
states.

\paragraph{Stability of sector equilibrium}

The distribution of firms' and investors' returns across a group of sectors
may give rise to sectoral instability that can propagate throughout the
group. This confirms that the relevant quantities to describe the collective
state are not the variables' averages but rather their distribution across
sectors, measured by the density functions, i.e., the fields.

\textbf{\ }The introduction of banks does not directly induce instability in
the system. At the sectoral level, banks do not increase fluctuations in
stakes and returns\footnote{%
See section 14.1.}, and at the global level, they do not significantly
modify the phenomenon of propagation and amplification of instability across
sectors\footnote{%
See Gosselin and Lotz (2025).}.\ 

High-return banks may drag the system toward instability.\ The mechanism is
similar to that of high-return investors detailed in Gosselin and Lotz
(2025): fluctuations in returns and stakes within one sector may propagate
to interconnected sectors, dragging the system towards instability.\ When a
sector experiences significant fluctuations in returns or capital
allocations, these disturbances may be transmited to other sectors,
potentially affecting the entire system.

When banks behave as lenders and provide capital to low- rather than
high-return investors, they may indirectly stabilize the system by modifying
the ratio of high-to-low strategies among investors. Indeed, high-return
strategies are more sensitive to fluctuations in global returns and, as
such, are more unstable. By contrast, low-return strategies, which
prioritize one's sector while diversifying only moderately, tends to be more
stable and are indeed the norm.

When banks unduly favor investors over firms in their investment decisions,
investors' disposable capital will increase, which will yield a proportional
rise in investors' stakes in firms. This shift in ownership of firms'
disposable capital may lead to the sector instability \footnote{%
See Gosselin and Lotz (2025).}: any deviation in firms' returns may trigger
large flows of capital in and out of the sector. Firms within this sector,
or in interconnected sectors, may experience reduced access to capital,
impairing their ability to cover operating costs.

\subsubsection{Transitions}

The existence of multiple and potentially unstable equilibria implies, in a
dynamic context, the possibility of transitions between collective states.
These transitions can occur from a non-default state to a default state, or
more generally between different non-default states.

\paragraph{Transition between non-default states ( \`{a} reprendre)}

We showed\footnote{%
See section 10.1 and Gosselin and Lotz (2025).} that each collective state
decomposes into a set of groups. Each of these groups is in a given phase,
and for each group, multiple phases exist, each with different sensitivities
to fluctuations. Once a phase becomes unstable, it may undergo a transition
to another phase, possibly leading to default. Even a stable phase, when
subjected to sufficiently large fluctuations, may be driven into another
phase. At the level of collective states, the transitions in the phases of
groups constitute a transition of the collective state itself. Moreover, if
groups are sufficiently interconnected, cascades of group transitions may
occur, inducing transitions of collective states.

Bank loans facilitate access to capital and partly modify the possible
phases of a group, but they do not reduce the multiplicity of collective
states and their associated transitions, as presented in Gosselin and Lotz
(2025). We showed that high level investors act as sources of large capital
fluctuations, which may induce phase transitions. The behavior of banks
impact these mechanismsby their behavior toward investors. Over-allocation
of capital to investors exacerbates these instability mechanisms and may
ultimately trigger phase transitions. Transitions driven by endogenous
factors primarily result in default states, whereas those triggered by
exogenous shocks tend to produce shifts into new phases.

\paragraph{Transition towards a default}

In the absence of banks, a mechanism leading high-return investors towards
default emerged\footnote{%
See Gosselin and Lotz (2025).}. Because these investors are less stable than
low-return investors, they depend heavily on the overall fluctuations of the
system. Negative fluctuations, coupled with investors' own instability, lead
to losses in capital and returns, which are directly transmitted to sector
firms, thereby reducing capital and potentially triggering defaults that may
propagate throughout the system.

When banks are added to the system, the mechanism of transition toward
default differs: provided that banks do not behave as high-return investors
and the banking system is sufficiently large, investors are bound to follow
the low-return solution, which is in general stable.\ However, when banks do
not behave as low-return investors, their disposable capital is primarily
(and excessively) allocated to investors.\ As a result, these investors will
come to control the majority of disposable capital in their sector firms.
Since investors are more sensitive to fluctuations in returns, a relative
increase in average financial returns compared to those of firms may induce
investors to withdraw their capital from firms, potentially driving the
firms into default.

\section{Discussion}

This paper illustrates the advantages of field models in economics. From a
modeling perspective, it shows how new types of agents can be added to a
basic field model by the introduction of additional fields.\ Moreover, any
variable or parameter could be endogeneized in the same way.

The flexibility provided by field models lies in that they encompass the
general, local, and modular aspects of any given model.\ At the global
level, the model averages describe aggregate results, whereas at the local
level, the sectoral quantities describe specific agents or sectors.\ These
two levels do not necessarily coincide, and their differences capture some
major features of the model. At the modular level, they provide a framework
for describing a wide variety of economic configurations: here for instance,
in a given sector, an investor may represent the financial activity of a
firm, a bank may be interpreted as the financial branch of a corporation,
and a firm may encompass all the productive activities of a broader
financial group. Because field models represent a fragmented reality made up
of interconnected sectors and agents organized within groups, each agent may
be conceived either as a subcomponent of a broader collective entity or as a
differentiated expression of an integrated, multidimensional agent.

The introduction of banks in our framework reveals robust collective
patterns across different specifications. Our results indicate that, unlike
conventional economic wisdom, no single representative equilibrium emerges
from a population of agents acting according to similar optimization
principles. There is no single equilibrium, and an equilibrium cannot be
described by the average of quantities, that is, by an aggregated
representative agent. This is a characteristic feature of field economics:
for a given type of agent, the same optimization behavior can lead to
multiple possible outcomes, depending on the initial conditions of each
agent. In the present context, two types of returns emerge for investors,
associated with their respective levels of capitalization in the collective
state.

The presence of banks does not, by itself, prevent the propagation of
defaults. In fact, monetary creation proves to be a double-edged mechanism.
When banks channel credit toward firms rather than investors, they mitigate
the risk of initial defaults. However, and particularly so when banks favor
investors, monetary creation systematically amplifies investors' behavior,
fueling financial intermediation at the expense of firms and potentially
amplifying the spread of defaults.

Banks loans can reduce the instability inherent to investors' capital
allocation. Even if investors underinvest in some sectors, bank loans
provide firms with the level of capital needed to produce and cover their
costs. Uncertainty and expected returns determine the level of capital
allocation, but overall, some stability is achieved within the sector space:
firms' disposable capital is provided by loans, and investors behave
similarly on average.

When banks allocate more capital to investors than to firms, they indirectly
alter the composition of firms' disposable capital. Investors are bound to
increase their exposure to shares or loans, while remaining sensitive to
firms' returns. Should an external shock modify these returns - or their
expectations - large capital flows from one sector to another would follow.
In this context, firms' dependence on investors' capital is a source of
risk, as investors' withdrawals may leave them insufficiently capitalized.
In the long run, banks' structural bias toward investors tends to shift
firms' ownership toward high-return investors, to the detriment of firms'
initial owners and low-return investors.

However, the allocation of capital by banks is not an exogenous input to the
model, but rather depends on the perceived risk associated with each type of
agent in each sector. In this regard, it should be noted that the situation
of investors and firms is not symmetric. Indeed, unlike firms, investors can
control the perceived riskiness of their investments through their
portfolios. Indeed, investors may appear safer by diversifying their
portfolios, thereby conveying a misleading appearance of lower risk. For
banks reluctant to bear firms' risks directly, high-return investors would
logically appear to be the most reliable borrowers: they are better
capitalized, more diversified, and capable of extracting higher yields from
their portfolios, and as such, seem less risky than low-return investors.
These characteristics are nonetheless misleading, since systemic risk is not
determined by investors' individual return profiles,\ but by the collective
sustainability of their exposures.

When banks are biased toward high-return investors, the challenge for these
investors becomes that of maintaining the very features that sustain banks'
confidence---extensive diversification, favorable yet often opaque
valuations, and attractive returns. In such an environment, the allocation
of capital to the real economy risks becoming merely a pretext for
speculative behavior. This equilibrium is inherently unstable over the
medium to long term: once the capital held by high-return investors
withdraws, the sector is left undercapitalized and structurally driven
toward default. The instability is further reinforced by circularity:
because high-return investors tend to be more interconnected than average
and to hold stakes in one another, feedback effects amplify disturbances and
can ultimately propagate throughout the entire system.

This equilibrium can only be altered by a macroprudential policy compelling
banks to take on firms' risks directly. By definition, any high-return
investor may offer apparent guarantees, and banks, by lending to them,
effectively finance risk indirectly. When extending credit, banks base their
decisions on capitalization and returns, which naturally leads them to favor
high-return investors. Yet a high-return profile is not an indicator of
lower risk; it is merely a characteristic that may be equally---or even
more---risky than a low-return profile.

Overall, the impact of monetary creation on the economy therefore depends on
banks' willingness to extend credit to firms and assume the associated
credit risks, the regulatory environment to which banks are subject, and
their accurate evaluation of the risks borne by both firms and investors.
Besides, even when banks regulate investors, they end up endorsing their
high- and low-return investment profiles, along with their intrinsic
instability. In the present framework, banks do not fundamentally depart
from investors in their optimization logic: both maximize their expected
returns.\ This result suggests that the system does not spontaneously
self-regulate, and that certain imbalances can only be corrected through
deliberate policy.

To fundamentally stabilize the collective state, a new type of agent must
exhibit optimization behaviors that differ from those prevailing in the
system, and a sufficiently large fraction of agents must exhibit these
distinct optimization patterns. When this condition is not met, the
inclusion of new agents merely reshapes the distribution of outcomes rather
than changing the nature of the equilibria.

From a financial stability perspective, this mechanism reveals the
endogenous nature of systemic risk. Stability cannot be achieved through the
mere addition of liquidity or diversification of balance sheets; it requires
a structural heterogeneity in optimization behaviors capable of dampening
the propagation of shocks. In its absence, the system remains characterized
by self-reinforcing dynamics in which profitability, capitalization, and
exposure evolve together---generating phases of apparent robustness followed
by abrupt transitions to fragility. Macroprudential tools can attenuate
local imbalances, but they cannot suppress the systemic mechanisms that
produce instability endogenously. In this sense, financial fragility emerges
as a structural property of interconnected capital systems rather than as a
failure of prudential oversight.

\section{Conclusion}

This paper has extended our analysis of the stability of collective states.
It shows that the presence of banks neither resolves the multiplicity of
such states nor prevents their potential instability. Instability manifests
itself through transitions between collective states, which represent
structural shifts within the system. These transitions may occur between
similar states---for instance, from one non-default collective state to
another. They can also involve paradigm shifts, such as transitions from a
non-default collective state to a default one. Moreover, such transitions
can unfold in cascades, with varying speeds and intervals.

Indeed, the possibility of transitions between collective states is the
hallmark of a more general feature.\ In general, collective states are not
fixed structures: they are subject to transitions driven either by internal
fluctuations or by exogenous shocks. However, until now, these transitions
have been derived under the simplifying assumption of independent---and
therefore rigid---groups. Allowing for interactions among groups may,
however, alter their composition, leading to aggregation, shrinkage, or
other structural changes without necessarily resulting in default. Even the
spatial localization of groups may evolve. These transitions---between
phases, but also potentially between groups---require a specific formalism,
which will be presented in the sequel of this work.

\section*{References}

%\begin{thebibliography}{Gehrkelager 2000}

\begin{description}
\item Abergel, F., Chakraborti, A., Muni Toke, I., \& Patriarca, M. (2011a).
Econophysics review: I. Empirical facts. Quantitative Finance, 11(7),
991--1012.

\item Abergel, F., Chakraborti, A., Muni Toke, I., \& Patriarca, M. (2011b).
Econophysics review: II. Agent-based models. Quantitative Finance, 11(7),
1013--1041.

\item Acemoglu, D., Ozdaglar, A., \& Tahbaz-Salehi, A. (2015). Systemic risk
and stability in financial networks. American Economic Review, 105(2),
564--608.

\item Acharya, V. V., Engle, R. \& Richardson, M. (2012). Capital shortfall:
A new approach to ranking and regulating systemic risks. (Quantifying
capital shortfalls under stress).

\item Acharya, V. V., Pedersen, L. H., Philippon, T., \& Richardson, M.
(2017). Measuring systemic risk. The Review of Financial Studies, 30(1),
2--47.

\item Adrian, T., \& Brunnermeier, M. K. (2016). CoVaR. American Economic
Review, 106(7), 1705--1741.

\item Allen, F., \& Gale, D. (2000). Financial contagion. Journal of
Political Economy, 108(1), 1--33.

\item Bardoscia, M., Livan, G., \& Marsili, M. (2017). Statistical mechanics
of complex economies. Journal of Statistical Mechanics: Theory and
Experiment, 2017, 1--21.

\item Bardoscia, M., Battiston, S., Caccioli, F., \& Caldarelli, G. (2019).
Pathways towards instability in financial networks. Nature Communications,
10(1), 1--9.

\item Battiston, S., Puliga, M., Kaushik, R., Tasca, P., \& Caldarelli, G.
(2012). Debtrank: Too central to fail? Financial networks, the FED and
systemic risk. Scientific Reports, 2, 541.

\item Battiston, S., Caldarelli, G., May, R. M., Roukny, T., \& Stiglitz, J.
E. (2020). The price of complexity in financial networks. Proceedings of the
National Academy of Sciences, 117(52), 32779--32786.

\item Bernanke, B., Gertler, M., \& Gilchrist, S. (1999). The financial
accelerator in a quantitative business cycle framework. In J. B. Taylor \&
M. Woodford (Eds.), Handbook of Macroeconomics (Vol. 1, Part C, pp.
1341--1393). Elsevier.

\item Bensoussan, A., Frehse, J., \& Yam, P. (2018). Mean Field Games and
Mean Field Type Control Theory. Springer, New York.

\item B\"{o}hm, V., Kikuchi, T., \& Vachadze, G. (2008). Asset pricing and
productivity growth: The role of consumption scenarios. Computational
Economics, 32, 163--181.

\item Brunnermeier, M. K. \& Sannikov, Y. (2014). A Macroeconomic Model with
a Financial Sector. (Seminal macro-financial modelling showing financial
amplification mechanisms)

\item Brunnermeier, M. K., Crockett, A., Goodhart, C., Persaud, A. D. \&
Shin, H. S. (2009). The fundamental principles of financial regulation.
Geneva Report on the World Economy (report outlining regulatory principles
post-crisis).

\item Borio, C. \& Drehmann, M. (2009). Assessing the risk of banking crises
--- revisited. BIS Quarterly Review.

\item Caggese, A., \& Orive, A. P. (2015). The interaction between household
and firm dynamics and the amplification of financial shocks. Barcelona GSE
Working Paper Series, Working Paper n%
%TCIMACRO{\U{ba} }%
%BeginExpansion
${{}^o}$
%EndExpansion
866.

\item Campello, M., Graham, J., \& Harvey, C. R. (2010). The real effects of
financial constraints: Evidence from a financial crisis. Journal of
Financial Economics, 97(3), 470--487.

\item Cerutti, E., Claessens, S. \& Laeven, L. (2017). The use and
effectiveness of macroprudential policies: New evidence. IMF Working Paper /
Journal version.

\item Cifuentes, R., Ferrucci, G., \& Shin, H. S. (2005). Liquidity risk and
contagion. Journal of the European Economic Association, 3(2--3), 556--566.

\item Claessens, S., Ghosh, S. R. \& Mihet, R. (2013). Macro-prudential
policy: What instruments and how effective? IMF Economic Review / background
reports on macroprudential toolkits.

\item Cochrane, J. H. (Ed.). (2006). Financial markets and the real economy.
Edward Elgar, International Library of Critical Writings in Financial
Economics, Vol. 18.

\item Diamond, D. W. \& Dybvig, P. H. (1983). Bank runs, deposit insurance,
and liquidity. Journal of Political Economy.

\item Diamond, D. W. \& Rajan, R. G. (2001). Liquidity risk, liquidity
creation and financial fragility: A theory of banking. Journal of Political
Economy.

\item Drehmann, M. \& Juselius, M. (2014). Evaluating early warning
indicators of banking crises: Sizing up the credit-to-GDP gap. International
Journal of Forecasting / BIS discussion.

\item Elliott, M., Golub, B., \& Jackson, M. O. (2014). Financial networks
and contagion. American Economic Review, 104(10), 3115--3153.

\item Freixas, X. \& Rochet, J.-C. (2008). Microeconomics of Banking.
(Second edition). MIT Press.

\item Freixas, X., Parigi, B. M. \& Rochet, J.-C. (2000). Systemic risk,
interbank relations, and liquidity provision by the central bank. Journal of
Money, Credit and Banking.

\item Gaffard, J.-L., \& Napoletano, M. (Eds.). (2012). Agent-based models
and economic policy. OFCE, Paris.

\item Gai, P., \& Kapadia, S. (2010). Contagion in financial networks.
Proceedings of the Royal Society A: Mathematical, Physical and Engineering
Sciences, 466(2120), 2401--2423.

\item Galati, G. \& Moessner, R. (2013). Macroprudential policy --- a
literature review. BIS Working Papers / BIS Quarterly Review articles.

\item Gennaioli, N., Shleifer, A., \& Vishny, R. W. (2012). Neglected risks,
financial innovation, and financial fragility. Journal of Financial
Economics, 104(3), 452--468.

\item Gertler, M. \& Kiyotaki, N. (2010). Financial intermediation and
credit policy in business cycle analysis. (Survey and modelling of the
financial accelerator and bank-mediated transmission).

\item Gomes, D. A., Nurbekyan, L., \& Pimentel, E. A. (2015). Economic
models and mean-field games theory. Publica\c{c}\~{o}es Matem\'{a}ticas do
IMPA, 30%
%TCIMACRO{\U{ba} }%
%BeginExpansion
${{}^o}$
%EndExpansion
Col\'{o}quio Brasileiro de Matem\'{a}tica, Rio de Janeiro.

\item Gosselin P., \& Lotz, A. (2024) Financial Interactions and Capital
Accumulation. hal-04573829.

\item Gosselin, P., \& Lotz, A. (2023a). A statistical field theory and
neural structures dynamics I: Action functionals, background states and
external perturbations. hal-04301171.

\item Gosselin, P., \& Lotz, A. (2023b). Statistical field theory and neural
structures dynamics II: Signals propagation, interferences, bound states.
hal-04301181v1.

\item Gosselin, P., \& Lotz, A. (2023c). A statistical field theory and
neural structures dynamics III: Effective action for connectivities,
interactions and emerging collective states. hal-04301185.

\item Gosselin, P., \& Lotz, A. (2023d). A statistical field theory and
neural structures dynamics IV: Field-theoretic formalism for interacting
collective states. hal-04301191.

\item Gosselin, P., Lotz, A., \& Wambst, M. (2017). A path integral approach
to interacting economic systems with multiple heterogeneous agents.
IF\_PREPUB, hal-01549586v2.

\item Gosselin, P., Lotz, A., \& Wambst, M. (2020). A path integral approach
to business cycle models with large number of agents. Journal of Economic
Interaction and Coordination, 15, 899--942.

\item Gosselin, P., Lotz, A., \& Wambst, M. (2021). A statistical field
approach to capital accumulation. Journal of Economic Interaction and
Coordination, 16, 817--908.

\item Grassetti, F., Mammana, C., \& Michetti, E. (2022). A dynamical model
for real economy and finance. Mathematical Finance and Economics.
https://doi.org/10.1007/s11579-021-00311-3

\item Greenwood, R., \& Hanson, S. G. (2013). Issuer quality and corporate
bond returns. The Review of Financial Studies, 26(6), 1483--1525.

\item Grosshans, D., \& Zeisberger, S. (2018). All's well that ends well? On
the importance of how returns are achieved. Journal of Banking \& Finance,
87, 397--410.

\item Haldane, A. G., \& May, R. M. (2011). Systemic risk in banking
ecosystems. Nature, 469(7330), 351--355.

\item Holmstrom, B., \& Tirole, J. (1997). Financial intermediation,
loanable funds, and the real sector. Quarterly Journal of Economics, 112(3),
663--691.

\item IMF (2014). Macroprudential policy: An organizing framework. IMF
Policy Paper (survey of policy objectives and instruments).

\item Jackson, M. O. (2010). Social and economic networks. Princeton
University Press, Princeton.

\item Jermann, U. J., \& Quadrini, V. (2012). Macroeconomic effects of
financial shocks. American Economic Review, 102(1), 238--271.

\item Khan, A., \& Thomas, J. K. (2013). Credit shocks and aggregate
fluctuations in an economy with production heterogeneity. Journal of
Political Economy, 121(6), 1055--1107.

\item Kaplan, G., \& Violante, L. (2018). Microeconomic heterogeneity and
macroeconomic shocks. Journal of Economic Perspectives, 32(3), 167--194.

\item Kleinert, H. (1989). Gauge fields in condensed matter (Vols. I--II).
World Scientific, Singapore.

\item Kleinert, H. (2009). Path integrals in quantum mechanics, statistics,
polymer physics, and financial markets (5th ed.). World Scientific,
Singapore.

\item Krugman, P. (1991). Increasing returns and economic geography. Journal
of Political Economy, 99(3), 483--499.

\item Langfield, S., Liu, Z., \& Ota, T. (2020). Mapping the network of
financial linkages: An industry-level analysis of the UK. Journal of Banking
\& Finance, 118, 105946.

\item Lasry, J.-M., Lions, P.-L., \& Gu\'{e}ant, O. (2010a). Application of
mean field games to growth theory. hal-00348376.

\item Lasry, J.-M., Lions, P.-L., \& Gu\'{e}ant, O. (2010b). Mean field
games and applications. In Paris-Princeton lectures on Mathematical Finance.
Springer. hal-01393103.

\item Lux, T. (2008). Applications of statistical physics in finance and
economics. Kiel Institute for the World Economy (IfW), Kiel.

\item Lux, T. (2016). Applications of statistical physics methods in
economics: Current state and perspectives. European Physical Journal Special
Topics, 225, 3255. https://doi.org/10.1140/epjst/e2016-60101-x

\item Mandel, A., Jaeger, C., F\"{u}rst, S., Lass, W., Lincke, D., Meissner,
F., Pablo-Marti, F., \& Wolf, S. (2010). Agent-based dynamics in
disaggregated growth models. Documents de travail du Centre d'Economie de la
Sorbonne, Paris.

\item Mandel, A. (2012). Agent-based dynamics in the general equilibrium
model. Complexity Economics, 1, 105--121.

\item Monacelli, T., Quadrini, V., \& Trigari, A. (2011). Financial markets
and unemployment. NBER Working Papers, 17389. National Bureau of Economic
Research.

\item Reinhart, C. M., \& Rogoff, K. S. (2009). This time is different:
Eight centuries of financial folly. Princeton University Press.

\item Sims, C. A. (2006). Rational inattention: Beyond the linear quadratic
case. American Economic Review, 96(2), 158--163.

\item Yang, J. (2018). Information theoretic approaches to economics.
Journal of Economic Surveys, 32(3), 940--960.

\pagebreak
\end{description}

\part*{Appendices}

\subsection*{A1.1 Returns equations}

In Gosselin and Lotz (2025), we obtained the return equations:%
\begin{eqnarray*}
&&\left( \Delta \left( X,X^{\prime }\right) -\frac{\hat{K}^{\prime }\hat{k}%
_{1}\left( X^{\prime },X\right) \left\vert \hat{\Psi}\left( \hat{K}^{\prime
},X^{\prime }\right) \right\vert ^{2}}{1+\underline{\hat{k}}\left( X^{\prime
}\right) +\underline{\hat{k}}_{1}^{B}\left( X^{\prime }\right) +\kappa \left[
\frac{\underline{\hat{k}}_{2}^{B}}{1+\bar{k}}\right] \left( X^{\prime
}\right) }\right) \frac{\hat{f}\left( \hat{K}^{\prime },X^{\prime }\right) -%
\bar{r}}{1+\underline{\hat{k}}_{2}\left( X^{\prime }\right) +\kappa \left[ 
\frac{\underline{\hat{k}}_{2}^{B}}{1+\bar{k}}\right] \left( X^{\prime
}\right) } \\
&&+\left\vert \hat{\Psi}\left( \hat{K}^{\prime },X^{\prime }\right)
\right\vert ^{2}\left( \frac{1+\hat{f}\left( X^{\prime }\right) }{\underline{%
\hat{k}}_{2}\left( X^{\prime }\right) +\kappa \left[ \frac{\underline{\hat{k}%
}_{2}^{B}}{1+\bar{k}}\right] \left( X^{\prime }\right) }H\left( -\frac{1+%
\hat{f}\left( X^{\prime }\right) }{\underline{\hat{k}}_{2}\left( X^{\prime
}\right) +\kappa \left[ \frac{\underline{\hat{k}}_{2}^{B}}{1+\bar{k}}\right]
\left( X^{\prime }\right) }\right) \right) \\
&&\times \frac{\hat{k}_{2}\left( X^{\prime },X\right) \hat{K}^{\prime }}{1+%
\underline{\hat{k}}\left( X^{\prime }\right) +\underline{\hat{k}}%
_{1}^{B}\left( X^{\prime }\right) +\kappa \left[ \frac{\underline{\hat{k}}%
_{2}^{B}}{1+\bar{k}}\right] \left( X^{\prime }\right) } \\
&&+\left\vert \Psi \left( K^{\prime },X^{\prime }\right) \right\vert
^{2}\left( \frac{1+f_{1}^{\prime }\left( X^{\prime }\right) }{\underline{k}%
_{2}\left( X^{\prime }\right) +\kappa \left[ \frac{\underline{k}_{2}^{B}}{1+%
\bar{k}}\right] \left( X^{\prime }\right) }H\left( -\frac{1+f_{1}^{\prime
}\left( X^{\prime }\right) }{\underline{k}_{2}\left( X^{\prime }\right)
+\kappa \left[ \frac{\underline{k}_{2}^{B}}{1+\bar{k}}\right] \left(
X^{\prime }\right) }\right) \right) \\
&&\times \frac{k_{2}\left( X^{\prime },X\right) \hat{K}^{\prime }}{1+%
\underline{k}\left( X^{\prime }\right) +\underline{k}_{1}^{\left( B\right)
}\left( X^{\prime }\right) +\kappa \left[ \frac{\underline{k}_{2}^{B}}{1+%
\bar{k}}\right] \left( X^{\prime }\right) } \\
&=&\frac{k_{1}\left( X^{\prime },X^{\prime }\right) K^{\prime }\left(
f_{1}\left( \bar{K},X,\Psi ,\hat{\Psi}\right) -\bar{r}\right) }{\left( 1+%
\underline{k}\left( X^{\prime }\right) +\underline{k}_{1}^{\left( B\right)
}\left( X^{\prime }\right) +\kappa \left[ \frac{\underline{k}_{2}^{B}}{1+%
\bar{k}}\right] \left( X^{\prime }\right) \right) \left( 1+\underline{k}%
_{2}\left( X^{\prime }\right) +\kappa \left[ \frac{\underline{k}_{2}^{B}}{1+%
\bar{k}}\right] \left( X^{\prime }\right) \right) }
\end{eqnarray*}%
for the investors, and:

\begin{eqnarray}
&&\left( 1-\frac{\bar{K}^{\prime }\bar{k}_{1}\left( X^{\prime },X\right)
\left\vert \bar{\Psi}\left( \bar{K}^{\prime },X^{\prime }\right) \right\vert
^{2}}{1+\underline{\bar{k}}\left( X^{\prime }\right) }\right) \frac{\bar{f}%
\left( X^{\prime }\right) -\left( 1+\kappa \right) \bar{r}}{1+\underline{%
\overline{\bar{k}}}_{2}\left( X^{\prime }\right) } \\
&&-\frac{\hat{K}^{\prime }\underline{\hat{k}}_{1}^{B}\left( X^{\prime
},X\right) \left\vert \hat{\Psi}\left( \hat{K}^{\prime },X^{\prime }\right)
\right\vert ^{2}}{1+\underline{\hat{k}}\left( X^{\prime }\right) +\underline{%
\hat{k}}_{1}^{B}\left( X^{\prime }\right) +\kappa \left[ \frac{\underline{%
\hat{k}}_{2}^{B}}{1+\bar{k}}\right] \left( X^{\prime }\right) }\frac{\hat{f}%
\left( X^{\prime }\right) -\bar{r}}{1+\underline{\hat{k}}_{2}\left(
X^{\prime }\right) +\kappa \frac{\underline{\hat{k}}_{2}^{B}\left( X^{\prime
}\right) }{1+\bar{k}\left( X\right) }}  \notag \\
&=&\left\vert \bar{\Psi}\left( \bar{K}^{\prime },X^{\prime }\right)
\right\vert ^{2}\left( \frac{1+\bar{f}\left( X^{\prime }\right) }{\underline{%
\overline{\bar{k}}}_{2}\left( X^{\prime }\right) }H\left( -\frac{1+\bar{f}%
\left( X^{\prime }\right) }{\underline{\overline{\bar{k}}}_{2}\left(
X^{\prime }\right) }\right) \right) \frac{\hat{K}^{\prime }\underline{\hat{k}%
}_{2}^{B}\left( X^{\prime },X\right) }{1+\underline{\overline{\bar{k}}}%
\left( X\right) }  \notag \\
&&+\left\vert \hat{\Psi}\left( \hat{K}^{\prime },X^{\prime }\right)
\right\vert ^{2}\left( \frac{1+\hat{f}\left( X^{\prime }\right) }{\underline{%
\hat{k}}_{2}\left( X^{\prime }\right) +\kappa \left[ \frac{\underline{\hat{k}%
}_{2}^{B}}{1+\bar{k}}\right] \left( X^{\prime }\right) }H\left( -\frac{1+%
\hat{f}\left( X^{\prime }\right) }{\underline{\hat{k}}_{2}\left( X^{\prime
}\right) +\kappa \left[ \frac{\underline{\hat{k}}_{2}^{B}}{1+\bar{k}}\right]
\left( X^{\prime }\right) }\right) \right)  \notag \\
&&\times \frac{\hat{K}^{\prime }\kappa \frac{\underline{\hat{k}}%
_{2}^{B}\left( X^{\prime },X\right) }{1+\bar{k}\left( X\right) }}{\left( 1+%
\underline{\hat{k}}\left( X^{\prime }\right) +\underline{\hat{k}}%
_{1}^{B}\left( X^{\prime }\right) +\kappa \left[ \frac{\underline{\hat{k}}%
_{2}^{B}}{1+\bar{k}}\right] \left( X^{\prime }\right) \right) }  \notag \\
&&+\left\vert \Psi \left( K^{\prime },X^{\prime }\right) \right\vert
^{2}\left( \frac{1+f_{1}^{\prime }\left( X^{\prime }\right) }{\underline{k}%
_{2}\left( X^{\prime }\right) +\kappa \left[ \frac{\underline{k}_{2}^{B}}{1+%
\bar{k}}\right] \left( X^{\prime }\right) }H\left( -\frac{1+f_{1}^{\prime
}\left( K^{\prime },X^{\prime }\right) }{\underline{k}_{2}\left( X^{\prime
}\right) +\kappa \left[ \frac{\underline{k}_{2}^{\left( B\right) }}{1+%
\underline{\bar{k}}}\right] \left( X^{\prime }\right) }\right) \right) 
\notag \\
&&\times \frac{K^{\prime }\kappa \frac{\underline{k}_{2}^{\left( B\right)
}\left( X^{\prime },X\right) }{1+\underline{\bar{k}}\left( X\right) }}{%
\left( 1+\underline{k}\left( X^{\prime }\right) +\underline{k}_{1}^{\left(
B\right) }\left( X^{\prime }\right) +\kappa \left[ \frac{\underline{k}%
_{2}^{\left( B\right) }}{1+\underline{\bar{k}}}\right] \left( X^{\prime
}\right) \right) }+\frac{K^{\prime }\underline{k}_{1}^{\left( B\right)
}\left( X^{\prime },X\right) }{\left( 1+\underline{k}\left( X^{\prime
}\right) +\underline{k}_{1}^{\left( B\right) }\left( X^{\prime }\right)
+\kappa \frac{\underline{k}_{2}^{\left( B\right) }\left( X^{\prime }\right) 
}{1+\underline{\bar{k}}}\right) }\left( f_{1}\left( \bar{K},X,\Psi ,\hat{\Psi%
}\right) -\bar{r}\right)  \notag
\end{eqnarray}%
for the banks.

\subsection*{A1.2 Average field and total capital per sector}

\subsubsection*{A1.2.1 Investors}

The return equation involve the average capital per sector and disposable
capital. The amount of capital for investors in sector $X_{1}$ is: 
\begin{eqnarray}
\hat{K}\left[ X_{1}\right] &=&\hat{K}_{X}\left\Vert \hat{\Psi}\left(
X_{1}\right) \right\Vert ^{2} \\
&\simeq &\frac{\hat{\mu}}{2\sigma _{\hat{K}}^{2}}\left( 6\frac{\sigma _{\hat{%
K}}^{2}}{\hat{\mu}}\frac{\left\Vert \hat{\Psi}_{0}\left( X_{1}\right)
\right\Vert ^{2}}{\left( 5+\frac{\left\langle \hat{g}^{ef}\right\rangle }{%
\left\langle \hat{g}\right\rangle }-\sqrt{\left( 1-\frac{\left\langle \hat{g}%
^{ef}\right\rangle }{\left\langle \hat{g}\right\rangle }\right) \left( 4-%
\frac{\left\langle \hat{g}^{ef}\right\rangle }{\left\langle \hat{g}%
\right\rangle }\right) }\right) }\right) ^{2}\left( \frac{1}{4\hat{g}%
^{2}\left( X_{1}\right) }-\frac{\left\langle \hat{g}\right\rangle \hat{g}%
^{ef}\left( X_{1}\right) }{3\hat{g}^{4}\left( X_{1}\right) }\right)  \notag
\end{eqnarray}%
and its associated field is:%
\begin{equation}
\left\Vert \hat{\Psi}\left( X_{1}\right) \right\Vert ^{2}=\frac{\hat{\mu}}{%
2\sigma _{\hat{K}}^{2}}\left( 6\frac{\sigma _{\hat{K}}^{2}}{\hat{\mu}}\frac{%
\left\Vert \hat{\Psi}_{0}\left( X_{1}\right) \right\Vert ^{2}}{\left( 5+%
\frac{\left\langle \hat{g}^{ef}\right\rangle }{\left\langle \hat{g}%
\right\rangle }-\sqrt{\left( 1-\frac{\left\langle \hat{g}^{ef}\right\rangle 
}{\left\langle \hat{g}\right\rangle }\right) \left( 4-\frac{\left\langle 
\hat{g}^{ef}\right\rangle }{\left\langle \hat{g}\right\rangle }\right) }%
\right) }\right) ^{\frac{3}{2}}\left( \frac{1}{3\hat{g}^{2}\left(
X_{1}\right) }-\frac{\left\langle \hat{g}\right\rangle \hat{g}^{ef}\left(
X_{1}\right) }{2\hat{g}^{4}\left( X_{1}\right) }\right)
\end{equation}%
\ In these formula, we used the modified returns:%
\begin{equation}
\bar{g}\left( \hat{K},X\right) =\left( 1-\bar{M}\left\vert \bar{\Psi}\left( 
\bar{K},X\right) \right\vert ^{2}\right) ^{-1}\bar{f}\left( \bar{K},X\right)
\end{equation}%
\begin{equation}
\hat{g}\left( \hat{K},X\right) =\left( 1-\hat{M}\left\vert \hat{\Psi}\left( 
\hat{K},X\right) \right\vert ^{2}\right) ^{-1}\hat{f}\left( \hat{K},X\right)
+\left( 1-\hat{M}\right) ^{-1}N\left( 1-\bar{M}\right) ^{-1}\bar{f}\left( 
\bar{K},X\right)
\end{equation}%
defined in part 1, and the brackets denote the average quantities. The field
definition of the matrices $\bar{M}$, $\hat{M}$ and $N$ are:%
\begin{equation*}
\bar{M}\left( \left( \bar{K},X\right) ,\left( \bar{K}^{\prime },X^{\prime
}\right) \right) =\frac{\overline{\bar{k}}\left( X,X^{\prime }\right) \bar{K}%
_{0}}{1+\int \overline{\bar{k}}\left( X,X^{\prime }\right) \bar{K}%
_{0}^{\prime }\left\vert \bar{\Psi}\left( \bar{K}_{0}^{\prime },X^{\prime
}\right) \right\vert ^{2}}
\end{equation*}%
\begin{eqnarray*}
&&\hat{M}\left( \left( \hat{K}^{\prime },X^{\prime }\right) ,\left( \hat{K}%
,X\right) \right) \\
&=&\frac{\hat{k}\left( X,X^{\prime }\right) \hat{K}}{1+\int \hat{k}\left(
X,X^{\prime }\right) \left\vert \hat{\Psi}\left( \hat{K}^{\prime },X^{\prime
}\right) \right\vert ^{2}+\int \hat{k}_{1}^{B}\left( X,X^{\prime }\right) 
\bar{K}_{0}^{\prime }\left\vert \bar{\Psi}\left( \bar{K}_{0}^{\prime
},X^{\prime }\right) \right\vert ^{2}+\int \hat{k}_{2}^{B}\left( X,X^{\prime
}\right) \frac{\bar{K}_{0}^{\prime }\left\vert \bar{\Psi}\left( \bar{K}%
_{0}^{\prime },X^{\prime }\right) \right\vert ^{2}}{1+\int \overline{\bar{k}}%
\left( X^{\prime },X^{\prime \prime }\right) \bar{K}_{0}^{\prime \prime
}\left\vert \bar{\Psi}\left( \bar{K}_{0}^{\prime \prime },X^{\prime \prime
}\right) \right\vert ^{2}}}
\end{eqnarray*}%
and ultimately:%
\begin{equation*}
\bar{N}=\frac{\left( \hat{k}_{1}^{B}\left( X,X^{\prime }\right) +\kappa 
\frac{\hat{k}_{2}^{B}\left( X,X^{\prime }\right) }{1+\int \overline{\bar{k}}%
\left( X^{\prime },X^{\prime \prime }\right) \bar{K}_{0}^{\prime \prime
}\left\vert \bar{\Psi}\left( \bar{K}_{0}^{\prime \prime },X^{\prime \prime
}\right) \right\vert ^{2}}-\kappa \int \frac{\hat{k}_{2}^{B}\left(
X,X^{\prime \prime }\right) \bar{K}_{0}^{\prime \prime }\overline{\bar{k}}%
\left( X^{\prime \prime },X^{\prime }\right) }{\left( 1+\int \overline{\bar{k%
}}\left( X^{\prime \prime },\bar{Y}\right) \bar{K}_{0}^{Y}\left\vert \bar{%
\Psi}\left( \bar{K}_{0}^{Y},\bar{Y}\right) \right\vert ^{2}\right) ^{2}}%
\right) \hat{K}}{1+\int \hat{k}\left( X,X^{\prime }\right) \left\vert \hat{%
\Psi}\left( \hat{K}^{\prime },X^{\prime }\right) \right\vert ^{2}+\int \hat{k%
}_{1}^{B}\left( X,X^{\prime }\right) \bar{K}_{0}^{\prime }\left\vert \bar{%
\Psi}\left( \bar{K}_{0}^{\prime },X^{\prime }\right) \right\vert ^{2}+\kappa
\int \hat{k}_{2}^{B}\left( X,X^{\prime }\right) \frac{\bar{K}_{0}^{\prime
}\left\vert \bar{\Psi}\left( \bar{K}_{0}^{\prime },X^{\prime }\right)
\right\vert ^{2}}{1+\int \overline{\bar{k}}\left( X^{\prime },X^{\prime
\prime }\right) \bar{K}_{0}^{\prime \prime }\left\vert \bar{\Psi}\left( \bar{%
K}_{0}^{\prime \prime },X^{\prime \prime }\right) \right\vert ^{2}}}
\end{equation*}

\subsubsection*{A1.2.2 Banks}

The amount of capital for banks in sector $X_{1}$ is given by: 
\begin{subequations}
\begin{equation}
\bar{K}\left[ X_{1}\right] \simeq 18\frac{\sigma _{\hat{K}}^{2}}{\hat{\mu}}%
\left( \frac{\sqrt{\left\langle \hat{g}\right\rangle ^{2}\left( 1+\frac{3}{4}%
\frac{\left\langle \hat{g}^{ef}\right\rangle }{\left\langle \hat{g}%
\right\rangle }\right) }\left\Vert \bar{\Psi}_{0}\left( X_{1}\right)
\right\Vert -\frac{3}{8}\left\langle \hat{g}\right\rangle \frac{\hat{g}%
^{Bef}\left( X_{1}\right) }{\bar{g}\left( X_{1}\right) }\left\Vert \hat{\Psi}%
_{0}\left( X_{1}\right) \right\Vert }{\sqrt{5+\frac{\left\langle \hat{g}%
^{ef}\right\rangle }{\left\langle \hat{g}\right\rangle }-\sqrt{\left( 1-%
\frac{\left\langle \hat{g}^{ef}\right\rangle }{\left\langle \hat{g}%
\right\rangle }\right) \left( 4-\frac{\left\langle \hat{g}^{ef}\right\rangle 
}{\left\langle \hat{g}\right\rangle }\right) }}}\right) ^{4}\left( \frac{1}{4%
\bar{g}^{2}\left( X_{1}\right) }-\frac{\left\langle \hat{g}\right\rangle 
\hat{g}^{Bef}\left( X_{1}\right) }{3\bar{g}^{4}\left( X_{1}\right) }\right)
\end{equation}%
along with its associated field: 
\end{subequations}
\begin{equation}
\left\Vert \bar{\Psi}\left( X_{1}\right) \right\Vert ^{2}\simeq 18\frac{%
\sigma _{\hat{K}}^{2}}{\hat{\mu}}\left( \frac{\sqrt{\left\langle \hat{g}%
\right\rangle ^{2}\left( 1+\frac{3}{4}\frac{\left\langle \hat{g}%
^{ef}\right\rangle }{\left\langle \hat{g}\right\rangle }\right) }\left\Vert 
\bar{\Psi}_{0}\left( X_{1}\right) \right\Vert -\frac{3}{8}\left\langle \hat{g%
}\right\rangle \frac{\hat{g}^{Bef}\left( X_{1}\right) }{\bar{g}\left(
X_{1}\right) }\left\Vert \hat{\Psi}_{0}\left( X_{1}\right) \right\Vert }{%
\sqrt{5+\frac{\left\langle \hat{g}^{ef}\right\rangle }{\left\langle \hat{g}%
\right\rangle }-\sqrt{\left( 1-\frac{\left\langle \hat{g}^{ef}\right\rangle 
}{\left\langle \hat{g}\right\rangle }\right) \left( 4-\frac{\left\langle 
\hat{g}^{ef}\right\rangle }{\left\langle \hat{g}\right\rangle }\right) }}}%
\right) ^{4}\left( \frac{1}{3\bar{g}^{2}\left( X_{1}\right) }-\frac{%
\left\langle \hat{g}\right\rangle \hat{g}^{Bef}\left( X_{1}\right) }{2\bar{g}%
^{4}\left( X_{1}\right) }\right)
\end{equation}

\section*{Appendix 2 Alternate formulation for return equation}

\subsection*{A2.1 Investors return equation}

We write investors'return equation in a modified form:%
\begin{eqnarray}
0 &=&\int \left( 1-\hat{S}_{E}\left( X^{\prime },\hat{K}^{\prime },X\right)
\right) \frac{\hat{f}\left( X^{\prime }\right) -\bar{r}}{1+\underline{\hat{k}%
}_{2}\left( X^{\prime }\right) +\kappa \frac{\underline{\hat{k}}%
_{2}^{B}\left( X^{\prime }\right) }{1+\bar{k}}}dX^{\prime }d\hat{K}^{\prime }
\label{Rt} \\
&&-\int \left( \frac{1+\hat{f}\left( X^{\prime }\right) }{\underline{\hat{k}}%
_{2}\left( X^{\prime }\right) +\kappa \left[ \frac{\underline{\hat{k}}%
_{2}^{B}}{1+\bar{k}}\right] \left( X^{\prime }\right) }H\left( -\frac{1+\hat{%
f}\left( X^{\prime }\right) }{\underline{\hat{k}}_{2}\left( X^{\prime
}\right) +\kappa \left[ \frac{\underline{\hat{k}}_{2}^{B}}{1+\bar{k}}\right]
\left( X^{\prime }\right) }\right) \right) \hat{S}_{L}\left( X^{\prime },%
\hat{K}^{\prime },X\right) dX^{\prime }d\hat{K}^{\prime }  \notag \\
&&-\int \frac{1+f_{1}^{\prime }\left( X^{\prime }\right) }{\underline{k}%
_{2}\left( X^{\prime }\right) +\kappa \left[ \frac{\underline{k}_{2}^{B}}{1+%
\bar{k}}\right] \left( X^{\prime }\right) }H\left( -\frac{1+f_{1}^{\prime
}\left( X^{\prime }\right) }{\underline{k}_{2}\left( X^{\prime }\right)
+\kappa \left[ \frac{\underline{k}_{2}^{B}}{1+\bar{k}}\right] \left(
X^{\prime }\right) }\right) S_{L}\left( X^{\prime },K^{\prime },X\right)
-\int S_{1}\left( X^{\prime },K^{\prime },X\right) \left( \hat{f}_{1}\left( 
\hat{K},X\right) -\bar{r}\right)  \notag
\end{eqnarray}%
where we define the shares:

\begin{eqnarray}
\hat{S}_{\eta }\left( X^{\prime },\hat{K}^{\prime },X\right) &=&\frac{\hat{K}%
^{\prime }\hat{k}_{\eta }\left( X^{\prime },X\right) \left\vert \hat{\Psi}%
\left( \hat{K}^{\prime },X^{\prime }\right) \right\vert ^{2}}{1+\underline{%
\hat{k}}\left( X^{\prime }\right) +\underline{\hat{k}}_{1}^{B}\left(
X^{\prime }\right) +\kappa \left[ \frac{\underline{\hat{k}}_{2}^{B}}{1+\bar{k%
}}\right] \left( X^{\prime }\right) }  \label{CF1} \\
\hat{S}_{\eta }\left( X^{\prime },X\right) &=&\int \frac{\hat{K}^{\prime }%
\hat{k}_{\eta }\left( X^{\prime },X\right) \left\vert \hat{\Psi}\left( \hat{K%
}^{\prime },X^{\prime }\right) \right\vert ^{2}}{1+\underline{\hat{k}}\left(
X^{\prime }\right) +\underline{\hat{k}}_{1}^{B}\left( X^{\prime }\right)
+\kappa \left[ \frac{\underline{\hat{k}}_{2}^{B}}{1+\bar{k}}\right] \left(
X^{\prime }\right) }d\hat{K}^{\prime }=\frac{\hat{K}_{X^{\prime }}\hat{k}%
_{\eta }\left( X^{\prime },X\right) \left\vert \hat{\Psi}\left( X^{\prime
}\right) \right\vert ^{2}}{1+\underline{\hat{k}}\left( X^{\prime }\right) +%
\underline{\hat{k}}_{1}^{B}\left( X^{\prime }\right) +\kappa \left[ \frac{%
\underline{\hat{k}}_{2}^{B}}{1+\bar{k}}\right] \left( X^{\prime }\right) } 
\notag
\end{eqnarray}%
and:%
\begin{eqnarray}
S_{\eta }\left( X^{\prime },K^{\prime },X\right) &=&\frac{k_{\eta }\left(
X^{\prime },X\right) K^{\prime }\left\vert \Psi \left( K^{\prime },X^{\prime
}\right) \right\vert ^{2}}{1+\underline{k}\left( X^{\prime }\right) +%
\underline{k}_{1}^{\left( B\right) }\left( X^{\prime }\right) +\kappa \left[ 
\frac{\underline{k}_{2}^{B}}{1+\bar{k}}\right] \left( X^{\prime }\right) }
\label{CF2} \\
S_{\eta }\left( X^{\prime },X\right) &=&\int \frac{k_{\eta }\left( X^{\prime
},X\right) K^{\prime }\left\vert \Psi \left( K^{\prime },X^{\prime }\right)
\right\vert ^{2}}{1+\underline{k}\left( X^{\prime }\right) +\underline{k}%
_{1}^{\left( B\right) }\left( X^{\prime }\right) +\kappa \left[ \frac{%
\underline{k}_{2}^{B}}{1+\bar{k}}\right] \left( X^{\prime }\right) }%
dK^{\prime }=\frac{k_{\eta }\left( X^{\prime },X\right) K_{X^{\prime
}}\left\vert \Psi \left( X^{\prime }\right) \right\vert ^{2}}{1+\underline{k}%
\left( X^{\prime }\right) +\underline{k}_{1}^{\left( B\right) }\left(
X^{\prime }\right) +\kappa \left[ \frac{\underline{k}_{2}^{B}}{1+\bar{k}}%
\right] \left( X^{\prime }\right) }  \notag
\end{eqnarray}%
After averaging over $\hat{K}^{\prime }$, equation (\ref{Rt}) writes:%
\begin{eqnarray}
0 &=&\int \left( 1-\hat{S}_{E}\left( X^{\prime },X\right) \right) \frac{\hat{%
f}\left( X^{\prime }\right) -\bar{r}}{1+\underline{\hat{k}}_{2}\left(
X^{\prime }\right) +\kappa \left[ \frac{\underline{\hat{k}}_{2}^{B}}{1+\bar{k%
}}\right] \left( X^{\prime }\right) }dX^{\prime } \\
&&-\int \left( \frac{1+\hat{f}\left( X^{\prime }\right) }{\underline{\hat{k}}%
_{2}\left( X^{\prime }\right) +\kappa \left[ \frac{\underline{\hat{k}}%
_{2}^{B}}{1+\bar{k}}\right] \left( X^{\prime }\right) }H\left( -\frac{1+\hat{%
f}\left( X^{\prime }\right) }{\underline{\hat{k}}_{2}\left( X^{\prime
}\right) +\kappa \left[ \frac{\underline{\hat{k}}_{2}^{B}}{1+\bar{k}}\right]
\left( X^{\prime }\right) }\right) \right) \hat{S}_{L}\left( X^{\prime
},X\right) dX^{\prime }  \notag \\
&&-\int \frac{1+f_{1}^{\prime }\left( X^{\prime }\right) }{\underline{k}%
_{2}\left( X^{\prime }\right) +\kappa \left[ \frac{\underline{k}_{2}^{B}}{1+%
\bar{k}}\right] \left( X^{\prime }\right) }H\left( -\frac{1+f_{1}^{\prime
}\left( X^{\prime }\right) }{\underline{k}_{2}\left( X^{\prime }\right)
+\kappa \left[ \frac{\underline{k}_{2}^{B}}{1+\bar{k}}\right] \left(
X^{\prime }\right) }\right) S_{L}\left( X^{\prime },X\right) -\int
S_{E}\left( X^{\prime },X\right) \left( \hat{f}_{1}\left( X\right) -\bar{r}%
\right)  \notag
\end{eqnarray}%
We show in Appendix 13. that:%
\begin{eqnarray*}
\underline{k}\left( X^{\prime }\right) &=&\frac{\int \frac{S\left( X^{\prime
},X\right) \hat{K}_{X}\left\vert \hat{\Psi}\left( X\right) \right\vert ^{2}}{%
K_{X^{\prime }}\left\vert \Psi \left( X^{\prime }\right) \right\vert ^{2}}dX%
}{1-\left( \int \frac{S\left( X^{\prime },X\right) \hat{K}_{X}\left\vert 
\hat{\Psi}\left( X\right) \right\vert ^{2}}{K_{X^{\prime }}\left\vert \Psi
\left( X^{\prime }\right) \right\vert ^{2}}dX+\int \frac{\left(
S_{E}^{B}\left( X^{\prime },X\right) +S_{L}^{B}\left( X^{\prime },X\right)
\right) \bar{K}_{X}\left\vert \bar{\Psi}\left( X\right) \right\vert ^{2}}{%
K_{X^{\prime }}\left\vert \Psi \left( X^{\prime }\right) \right\vert ^{2}}%
dX\right) } \\
\underline{k}_{1}^{\left( B\right) }\left( X^{\prime }\right) &=&\frac{\int
S_{E}^{B}\left( X^{\prime },X\right) \frac{\bar{K}_{X}\left\vert \bar{\Psi}%
\left( X\right) \right\vert ^{2}}{K_{X^{\prime }}\left\vert \Psi \left(
X^{\prime }\right) \right\vert ^{2}}dX}{1-\left( \int \frac{S\left(
X^{\prime },X\right) \hat{K}_{X}\left\vert \hat{\Psi}\left( X\right)
\right\vert ^{2}}{K_{X^{\prime }}\left\vert \Psi \left( X^{\prime }\right)
\right\vert ^{2}}dX+\int \frac{\left( S_{E}^{B}\left( X^{\prime },X\right)
+S_{L}^{B}\left( X^{\prime },X\right) \right) \bar{K}_{X}\left\vert \bar{\Psi%
}\left( X\right) \right\vert ^{2}}{K_{X^{\prime }}\left\vert \Psi \left(
X^{\prime }\right) \right\vert ^{2}}dX\right) } \\
\kappa \left[ \frac{\underline{k}_{2}^{B}}{1+\bar{k}}\right] \left(
X^{\prime }\right) &=&\frac{\int S_{L}^{B}\left( X^{\prime },X\right) \frac{%
\bar{K}_{X}\left\vert \bar{\Psi}\left( X\right) \right\vert ^{2}}{%
K_{X^{\prime }}\left\vert \Psi \left( X^{\prime }\right) \right\vert ^{2}}dX%
}{1-\left( \int \frac{S\left( X^{\prime },X\right) \hat{K}_{X}\left\vert 
\hat{\Psi}\left( X\right) \right\vert ^{2}}{K_{X^{\prime }}\left\vert \Psi
\left( X^{\prime }\right) \right\vert ^{2}}dX+\int \frac{\left(
S_{E}^{B}\left( X^{\prime },X\right) +S_{L}^{B}\left( X^{\prime },X\right)
\right) \bar{K}_{X}\left\vert \bar{\Psi}\left( X\right) \right\vert ^{2}}{%
K_{X^{\prime }}\left\vert \Psi \left( X^{\prime }\right) \right\vert ^{2}}%
dX\right) }
\end{eqnarray*}%
and that:%
\begin{equation*}
\frac{1}{1+\underline{\hat{k}}_{2}\left( X^{\prime }\right) +\kappa \left[ 
\frac{\underline{\hat{k}}_{2}^{B}}{1+\bar{k}}\right] \left( X^{\prime
}\right) }=\frac{1-\left( \int \frac{\hat{S}\left( X^{\prime },X\right) \hat{%
K}_{X}\left\vert \hat{\Psi}\left( X\right) \right\vert ^{2}}{\hat{K}%
_{X^{\prime }}\left\vert \hat{\Psi}\left( X^{\prime }\right) \right\vert ^{2}%
}dX+\int \frac{\left( \hat{S}_{E}^{B}\left( X^{\prime },X\right) +\hat{S}%
_{L}^{B}\left( X^{\prime },X\right) \right) \bar{K}_{X}\left\vert \bar{\Psi}%
\left( X\right) \right\vert ^{2}}{\hat{K}_{X^{\prime }}\left\vert \hat{\Psi}%
\left( X^{\prime }\right) \right\vert ^{2}}dX\right) }{1-\left( \int \frac{%
\hat{S}_{E}\left( X^{\prime },X\right) \hat{K}_{X}\left\vert \hat{\Psi}%
\left( X\right) \right\vert ^{2}}{\hat{K}_{X^{\prime }}\left\vert \hat{\Psi}%
\left( X^{\prime }\right) \right\vert ^{2}}dX+\int \frac{\hat{S}%
_{E}^{B}\left( X^{\prime },X\right) \bar{K}_{X}\left\vert \bar{\Psi}\left(
X\right) \right\vert ^{2}}{\hat{K}_{X^{\prime }}\left\vert \hat{\Psi}\left(
X^{\prime }\right) \right\vert ^{2}}dX\right) }
\end{equation*}

So that the investors equation, rewrites in the new variables:

\begin{eqnarray}
0 &=&\int \left( \delta \left( X-X^{\prime }\right) -\hat{S}_{E}\left(
X^{\prime },X\right) \right)  \label{nv} \\
&&\times \frac{1-\left( \int \frac{\hat{S}\left( X^{\prime },X\right) \hat{K}%
_{X}\left\vert \hat{\Psi}\left( X\right) \right\vert ^{2}}{\hat{K}%
_{X^{\prime }}\left\vert \hat{\Psi}\left( X^{\prime }\right) \right\vert ^{2}%
}dX+\int \frac{\left( \hat{S}_{E}^{B}\left( X^{\prime },X\right) +\hat{S}%
_{L}^{B}\left( X^{\prime },X\right) \right) \bar{K}_{X}\left\vert \bar{\Psi}%
\left( X\right) \right\vert ^{2}}{\hat{K}_{X^{\prime }}\left\vert \hat{\Psi}%
\left( X^{\prime }\right) \right\vert ^{2}}dX\right) }{1-\left( \int \frac{%
\hat{S}_{E}\left( X^{\prime },X\right) \hat{K}_{X}\left\vert \hat{\Psi}%
\left( X\right) \right\vert ^{2}}{\hat{K}_{X^{\prime }}\left\vert \hat{\Psi}%
\left( X^{\prime }\right) \right\vert ^{2}}dX+\int \frac{\hat{S}%
_{E}^{B}\left( X^{\prime },X\right) \bar{K}_{X}\left\vert \bar{\Psi}\left(
X\right) \right\vert ^{2}}{\hat{K}_{X^{\prime }}\left\vert \hat{\Psi}\left(
X^{\prime }\right) \right\vert ^{2}}dX\right) }\left( f\left( X^{\prime
}\right) -\bar{r}\right) dX^{\prime }  \notag \\
&&-\int S_{E}\left( X^{\prime },X\right) \left( \hat{f}_{1}\left( X\right) -%
\bar{r}\right)  \notag \\
&&-\int \frac{1+f\left( X^{\prime }\right) }{\underline{\hat{k}}_{2}\left(
X^{\prime }\right) }H\left( -\frac{1+f\left( X^{\prime }\right) }{\underline{%
\hat{k}}_{2}\left( X^{\prime }\right) }\right) \hat{S}_{L}\left( X^{\prime
},X\right) dX^{\prime }-\int \frac{1+f_{1}^{\prime }\left( X^{\prime
}\right) }{\underline{k}_{2}\left( X^{\prime }\right) }H\left( -\frac{%
1+f_{1}^{\prime }\left( X^{\prime }\right) }{\underline{k}_{2}\left(
X^{\prime }\right) }\right) S_{L}\left( X^{\prime },X\right) dX^{\prime } 
\notag
\end{eqnarray}%
The rates satisfy the following constraint:%
\begin{equation*}
\int \left( \hat{S}_{E}\left( X^{\prime },X\right) +\hat{S}_{L}\left(
X^{\prime },X\right) \right) dX^{\prime }+\int \left( S_{E}\left( X^{\prime
},X\right) +S_{L}\left( X^{\prime },X\right) \right) dX^{\prime }=1
\end{equation*}

\subsection*{A2.2 Banks returns equation}

We can express the return equations for banks in terms of shares by defining:%
\begin{eqnarray*}
\bar{S}_{\eta }\left( X^{\prime },\bar{K}^{\prime },X\right) &=&\frac{\bar{K}%
^{\prime }\bar{k}_{\eta }\left( X^{\prime },X\right) \left\vert \bar{\Psi}%
\left( \bar{K}^{\prime },X^{\prime }\right) \right\vert ^{2}}{1+\underline{%
\bar{k}}\left( X^{\prime }\right) } \\
\bar{S}_{\eta }\left( X^{\prime },X\right) &=&\int \frac{\bar{K}^{\prime }%
\bar{k}_{\eta }\left( X^{\prime },X\right) \left\vert \bar{\Psi}\left( \bar{K%
}^{\prime },X^{\prime }\right) \right\vert ^{2}}{1+\underline{\bar{k}}\left(
X^{\prime }\right) }d\bar{K}^{\prime }=\frac{\bar{K}_{X^{\prime }}\bar{k}%
_{\eta }\left( X^{\prime },X\right) \left\vert \bar{\Psi}\left( X^{\prime
}\right) \right\vert ^{2}}{1+\underline{\bar{k}}\left( X^{\prime }\right) }
\end{eqnarray*}%
\begin{eqnarray*}
\hat{S}_{E}^{B}\left( X^{\prime },\hat{K}^{\prime },X\right) &=&\frac{\hat{K}%
^{\prime }\underline{\hat{k}}_{1}^{B}\left( X^{\prime },X\right) \left\vert 
\hat{\Psi}\left( \hat{K}^{\prime },X^{\prime }\right) \right\vert ^{2}}{1+%
\underline{\hat{k}}\left( X^{\prime }\right) +\underline{\hat{k}}%
_{1}^{B}\left( X^{\prime }\right) +\kappa \left[ \frac{\underline{\hat{k}}%
_{2}^{B}}{1+\bar{k}}\right] \left( X^{\prime }\right) } \\
\hat{S}_{E}^{B}\left( X^{\prime },X\right) &=&\frac{\hat{K}_{X^{\prime }}%
\underline{\hat{k}}_{1}^{B}\left( X^{\prime },X\right) \left\vert \hat{\Psi}%
\left( X^{\prime }\right) \right\vert ^{2}}{1+\underline{\hat{k}}\left(
X^{\prime }\right) +\underline{\hat{k}}_{1}^{B}\left( X^{\prime }\right)
+\kappa \left[ \frac{\underline{\hat{k}}_{2}^{B}}{1+\bar{k}}\right] \left(
X^{\prime }\right) }
\end{eqnarray*}%
\begin{eqnarray*}
\hat{S}_{L}^{B}\left( X^{\prime },\hat{K}^{\prime },X\right) &=&\frac{\hat{K}%
^{\prime }\frac{\kappa \underline{\hat{k}}_{2}^{B}\left( X^{\prime
},X\right) }{1+\underline{\bar{k}}\left( X\right) }\left\vert \hat{\Psi}%
\left( \hat{K}^{\prime },X^{\prime }\right) \right\vert ^{2}}{1+\underline{%
\hat{k}}\left( X^{\prime }\right) +\underline{\hat{k}}_{1}^{B}\left(
X^{\prime }\right) +\kappa \left[ \frac{\underline{\hat{k}}_{2}^{B}}{1+\bar{k%
}}\right] \left( X^{\prime }\right) } \\
\hat{S}_{L}^{B}\left( X^{\prime },X\right) &=&\frac{\hat{K}_{X^{\prime }}%
\frac{\kappa \underline{\hat{k}}_{2}^{B}\left( X^{\prime },X\right) }{1+%
\underline{\bar{k}}\left( X\right) }\left\vert \hat{\Psi}\left( X^{\prime
}\right) \right\vert ^{2}}{1+\underline{\hat{k}}\left( X^{\prime }\right) +%
\underline{\hat{k}}_{1}^{B}\left( X^{\prime }\right) +\kappa \frac{%
\underline{\hat{k}}_{2}^{B}\left( X^{\prime }\right) }{1+\bar{k}}}
\end{eqnarray*}%
\begin{equation*}
\left[ \frac{\underline{k}_{2}^{B}}{1+\bar{k}}\right] \left( X^{\prime
}\right) =\int \frac{\underline{k}_{2}^{B}\left( X^{\prime },X\right) }{1+%
\bar{k}\left( X\right) }\bar{K}\left\vert \bar{\Psi}\left( \bar{K},X\right)
\right\vert ^{2}d\bar{K}dX
\end{equation*}%
\begin{eqnarray}
S_{E}^{B}\left( X^{\prime },K^{\prime },X\right) &=&\frac{K^{\prime }%
\underline{k}_{1}^{\left( B\right) }\left( X^{\prime },X\right) \left\vert
\Psi \left( K^{\prime },X^{\prime }\right) \right\vert ^{2}}{1+\underline{k}%
\left( X^{\prime }\right) +\underline{k}_{1}^{\left( B\right) }\left(
X^{\prime }\right) +\kappa \left[ \frac{\underline{k}_{2}^{\left( B\right) }%
}{1+\underline{\bar{k}}}\right] \left( X^{\prime }\right) }  \label{DFrApp}
\\
S_{L}^{B}\left( X^{\prime },K^{\prime },X\right) &=&\frac{\frac{\kappa 
\underline{k}_{2}^{\left( B\right) }\left( X^{\prime },X\right) }{1+%
\underline{\bar{k}}\left( X\right) }K^{\prime }\left\vert \Psi \left(
K^{\prime },X^{\prime }\right) \right\vert ^{2}}{1+\underline{k}\left(
X^{\prime }\right) +\underline{k}_{1}^{\left( B\right) }\left( X^{\prime
}\right) +\kappa \left[ \frac{\underline{k}_{2}^{\left( B\right) }}{1+%
\underline{\bar{k}}}\right] \left( X^{\prime }\right) }  \notag \\
S_{E}^{B}\left( X^{\prime },X\right) &=&\frac{K_{X^{\prime }}\underline{k}%
_{1}^{\left( B\right) }\left( X^{\prime },X\right) \left\vert \Psi \left(
K^{\prime },X^{\prime }\right) \right\vert ^{2}}{1+\underline{k}\left(
X^{\prime }\right) +\underline{k}_{1}^{\left( B\right) }\left( X^{\prime
}\right) +\kappa \left[ \frac{\underline{k}_{2}^{\left( B\right) }}{1+%
\underline{\bar{k}}}\right] \left( X^{\prime }\right) }\left\vert \Psi
\left( X^{\prime }\right) \right\vert ^{2}  \notag \\
S_{L}^{B}\left( X^{\prime },X\right) &=&\frac{\frac{\kappa \underline{k}%
_{2}^{\left( B\right) }\left( X^{\prime },X\right) }{1+\underline{\bar{k}}%
\left( X\right) }K_{X^{\prime }}\left\vert \Psi \left( X^{\prime }\right)
\right\vert ^{2}}{1+\underline{k}\left( X^{\prime }\right) +\underline{k}%
_{1}^{\left( B\right) }\left( X^{\prime }\right) +\kappa \left[ \frac{%
\underline{k}_{2}^{\left( B\right) }}{1+\underline{\bar{k}}}\right] \left(
X^{\prime }\right) }  \notag
\end{eqnarray}%
and the banks return equation writes with this parametrization: 
\begin{eqnarray}
0 &=&\left( 1-\bar{S}_{E}\left( X^{\prime },X\right) \right) \frac{\bar{f}%
\left( X^{\prime }\right) -\left( 1+\kappa \right) \bar{r}}{1+\underline{%
\overline{\bar{k}}}_{2}\left( X^{\prime }\right) }-\hat{S}_{E}^{B}\left(
X^{\prime },X\right) \left( \frac{\hat{f}\left( X^{\prime }\right) -\bar{r}}{%
1+\underline{\hat{k}}_{2}\left( X^{\prime }\right) +\kappa \frac{\underline{%
\hat{k}}_{2}^{B}\left( X^{\prime }\right) }{1+\bar{k}\left( X\right) }}%
\right)  \label{bt} \\
&&+\frac{\left( 1+\bar{f}\left( X^{\prime }\right) \right) H\left( -\left( 1+%
\bar{f}\left( X^{\prime }\right) \right) \right) }{\underline{\overline{\bar{%
k}}}_{2}\left( X^{\prime }\right) }\bar{S}_{L}\left( X^{\prime },X\right) +%
\frac{\left( 1+\hat{f}\left( X^{\prime }\right) \right) H\left( -\left( 1+%
\hat{f}\left( X^{\prime }\right) \right) \right) }{\underline{\hat{k}}%
_{2}\left( X^{\prime }\right) +\kappa \left[ \frac{\underline{\hat{k}}%
_{2}^{B}}{1+\bar{k}}\right] \left( X^{\prime }\right) }\hat{S}_{L}^{B}\left(
X^{\prime },X\right)  \notag \\
&&+\frac{\left( 1+f_{1}^{\prime }\left( X^{\prime }\right) \right) H\left(
1+f_{1}^{\prime }\left( K^{\prime },X^{\prime }\right) \right) }{\underline{k%
}_{2}\left( X^{\prime }\right) +\kappa \left[ \frac{\underline{k}%
_{2}^{\left( B\right) }}{1+\underline{\bar{k}}}\right] \left( X^{\prime
}\right) }S_{E}^{B}\left( X^{\prime },X\right) -S_{E}^{B}\left( X^{\prime
},X\right) \left( \frac{\left( f_{1}^{\prime }\left( \bar{K},X\right) -\bar{r%
}\right) }{1+\underline{k}_{2}\left( X^{\prime }\right) +\kappa \frac{%
\underline{k}_{2}^{\left( B\right) }\left( X^{\prime }\right) }{1+\underline{%
\bar{k}}}}+\Delta F_{\tau }\left( \bar{R}\left( K,X\right) \right) \right) 
\notag
\end{eqnarray}%
with constraint:%
\begin{equation*}
\int \left( \hat{S}_{E}\left( X^{\prime },X\right) +\hat{S}_{L}\left(
X^{\prime },X\right) \right) dX^{\prime }+\int \left( S_{E}\left( X^{\prime
},X\right) +S_{L}\left( X^{\prime },X\right) \right) dX^{\prime }=1
\end{equation*}

\begin{equation*}
\int \bar{S}\left( X^{\prime },X\right) \frac{\bar{K}_{X}\left\vert \bar{\Psi%
}\left( X\right) \right\vert ^{2}}{\bar{K}_{X^{\prime }}\left\vert \bar{\Psi}%
\left( X^{\prime }\right) \right\vert ^{2}}dX=\frac{\underline{\bar{k}}%
\left( X^{\prime }\right) }{1+\underline{\bar{k}}\left( X^{\prime }\right) }
\end{equation*}%
\begin{equation*}
\bar{S}\left( X^{\prime },X\right) =\bar{S}_{E}\left( X^{\prime },X\right) +%
\bar{S}_{L}\left( X^{\prime },X\right)
\end{equation*}%
We can replace all coefficients in (\ref{bt}) as function of the new
parametrization. Writing:%
\begin{equation*}
\frac{1}{1+\underline{\bar{k}}\left( X^{\prime }\right) }=1-\int \bar{S}%
\left( X^{\prime },X\right) \frac{\bar{K}_{X}\left\vert \bar{\Psi}\left(
X\right) \right\vert ^{2}}{\bar{K}_{X^{\prime }}\left\vert \bar{\Psi}\left(
X^{\prime }\right) \right\vert ^{2}}dX
\end{equation*}%
we find:%
\begin{eqnarray*}
\underline{\bar{k}}\left( X^{\prime }\right) &=&\frac{\int \bar{S}\left(
X^{\prime },X\right) \frac{\bar{K}_{X}\left\vert \bar{\Psi}\left( X\right)
\right\vert ^{2}}{\bar{K}_{X^{\prime }}\left\vert \bar{\Psi}\left( X^{\prime
}\right) \right\vert ^{2}}dX}{1-\int \bar{S}\left( X^{\prime },X\right) 
\frac{\bar{K}_{X}\left\vert \bar{\Psi}\left( X\right) \right\vert ^{2}}{\bar{%
K}_{X^{\prime }}\left\vert \bar{\Psi}\left( X^{\prime }\right) \right\vert
^{2}}dX} \\
\underline{\bar{k}}_{\eta }\left( X^{\prime }\right) &=&\frac{\int \bar{S}%
_{\eta }\left( X^{\prime },X\right) \frac{\bar{K}_{X}\left\vert \bar{\Psi}%
\left( X\right) \right\vert ^{2}}{\bar{K}_{X^{\prime }}\left\vert \bar{\Psi}%
\left( X^{\prime }\right) \right\vert ^{2}}dX}{1-\int \bar{S}\left(
X^{\prime },X\right) \frac{\bar{K}_{X}\left\vert \bar{\Psi}\left( X\right)
\right\vert ^{2}}{\bar{K}_{X^{\prime }}\left\vert \bar{\Psi}\left( X^{\prime
}\right) \right\vert ^{2}}dX}
\end{eqnarray*}%
\begin{equation*}
1+\underline{\bar{k}}_{2}\left( X^{\prime }\right) =\frac{1-\int \bar{S}%
_{E}\left( X^{\prime },X\right) \frac{\bar{K}_{X}\left\vert \bar{\Psi}\left(
X\right) \right\vert ^{2}}{\bar{K}_{X^{\prime }}\left\vert \bar{\Psi}\left(
X^{\prime }\right) \right\vert ^{2}}dX}{1-\int \bar{S}\left( X^{\prime
},X\right) \frac{\bar{K}_{X}\left\vert \bar{\Psi}\left( X\right) \right\vert
^{2}}{\bar{K}_{X^{\prime }}\left\vert \bar{\Psi}\left( X^{\prime }\right)
\right\vert ^{2}}dX}
\end{equation*}%
Ultimately, using:%
\begin{equation*}
\hat{S}_{\eta }\left( X^{\prime },X\right) =\frac{\hat{K}_{X^{\prime }}\hat{k%
}_{\eta }\left( X^{\prime },X\right) \left\vert \hat{\Psi}\left( X^{\prime
}\right) \right\vert ^{2}}{1+\underline{\hat{k}}\left( X^{\prime }\right) +%
\underline{\hat{k}}_{1}^{B}\left( X^{\prime }\right) +\kappa \left[ \frac{%
\underline{\hat{k}}_{2}^{B}}{1+\bar{k}}\right] \left( X^{\prime }\right) }
\end{equation*}%
\begin{equation*}
S_{\eta }\left( X^{\prime },X\right) =\frac{k_{\eta }\left( X^{\prime
},X\right) K_{X^{\prime }}\left\vert \Psi \left( X^{\prime }\right)
\right\vert ^{2}}{1+\underline{k}\left( X^{\prime }\right) +\underline{k}%
_{1}^{\left( B\right) }\left( X^{\prime }\right) +\kappa \left[ \frac{%
\underline{k}_{2}^{B}}{1+\bar{k}}\right] \left( X^{\prime }\right) }
\end{equation*}%
and the following relations:%
\begin{equation*}
\int \hat{S}\left( X^{\prime },X\right) \frac{\hat{K}_{X}\left\vert \hat{\Psi%
}\left( X\right) \right\vert ^{2}}{\hat{K}_{X^{\prime }}\left\vert \hat{\Psi}%
\left( X^{\prime }\right) \right\vert ^{2}}dX=\frac{\underline{\hat{k}}%
\left( X^{\prime }\right) }{1+\underline{\hat{k}}\left( X^{\prime }\right) +%
\underline{\hat{k}}_{1}^{B}\left( X^{\prime }\right) +\kappa \left[ \frac{%
\underline{\hat{k}}_{2}^{B}}{1+\bar{k}}\right] \left( X^{\prime }\right) }
\end{equation*}%
\begin{equation*}
\int \hat{S}_{E}^{B}\left( X^{\prime },X\right) \frac{\bar{K}_{X}\left\vert 
\bar{\Psi}\left( X\right) \right\vert ^{2}}{\hat{K}_{X^{\prime }}\left\vert 
\hat{\Psi}\left( X^{\prime }\right) \right\vert ^{2}}dX=\frac{\underline{%
\hat{k}}_{1}^{B}\left( X^{\prime }\right) }{1+\underline{\hat{k}}\left(
X^{\prime }\right) +\underline{\hat{k}}_{1}^{B}\left( X^{\prime }\right)
+\kappa \left[ \frac{\underline{\hat{k}}_{2}^{B}}{1+\bar{k}}\right] \left(
X^{\prime }\right) }
\end{equation*}%
\begin{equation*}
\int \hat{S}_{L}^{B}\left( X^{\prime },X\right) \frac{\bar{K}_{X}\left\vert 
\bar{\Psi}\left( X\right) \right\vert ^{2}}{\hat{K}_{X^{\prime }}\left\vert 
\hat{\Psi}\left( X^{\prime }\right) \right\vert ^{2}}=\frac{\kappa \left[ 
\frac{\underline{\hat{k}}_{2}^{B}}{1+\bar{k}}\right] \left( X^{\prime
}\right) }{1+\underline{\hat{k}}\left( X^{\prime }\right) +\underline{\hat{k}%
}_{1}^{B}\left( X^{\prime }\right) +\kappa \left[ \frac{\underline{\hat{k}}%
_{2}^{B}}{1+\bar{k}}\right] \left( X^{\prime }\right) }
\end{equation*}%
and defining the quantity $\left[ \frac{\underline{\hat{k}}_{2}^{B}}{1+\bar{k%
}}\right] \left( X^{\prime }\right) $:%
\begin{equation*}
\left[ \frac{\underline{\hat{k}}_{2}^{B}}{1+\bar{k}}\right] \left( X^{\prime
}\right) =\int \frac{\underline{\hat{k}}_{2}^{B}\left( X^{\prime },X\right) 
}{1+\bar{k}\left( X\right) }\bar{K}\left\vert \bar{\Psi}\left( \bar{K}%
,X\right) \right\vert ^{2}d\bar{K}dX
\end{equation*}%
leads to:%
\begin{eqnarray*}
&&\int \frac{\hat{S}\left( X^{\prime },X\right) \hat{K}_{X}\left\vert \hat{%
\Psi}\left( X\right) \right\vert ^{2}}{\hat{K}_{X^{\prime }}\left\vert \hat{%
\Psi}\left( X^{\prime }\right) \right\vert ^{2}}dX+\int \frac{\left( \hat{S}%
_{E}^{B}\left( X^{\prime },X\right) +\hat{S}_{L}^{B}\left( X^{\prime
},X\right) \right) \bar{K}_{X}\left\vert \bar{\Psi}\left( X\right)
\right\vert ^{2}}{\hat{K}_{X^{\prime }}\left\vert \hat{\Psi}\left( X^{\prime
}\right) \right\vert ^{2}}dX \\
&=&\frac{\underline{\hat{k}}\left( X^{\prime }\right) +\underline{\hat{k}}%
_{1}^{B}\left( X^{\prime }\right) +\kappa \left[ \frac{\underline{\hat{k}}%
_{2}^{B}}{1+\bar{k}}\right] \left( X^{\prime }\right) }{1+\underline{\hat{k}}%
\left( X^{\prime }\right) +\underline{\hat{k}}_{1}^{B}\left( X^{\prime
}\right) +\kappa \left[ \frac{\underline{\hat{k}}_{2}^{B}}{1+\bar{k}}\right]
\left( X^{\prime }\right) }
\end{eqnarray*}%
This allows to rewrite the parameters in terms of rates. First, for the
investors we have:%
\begin{eqnarray*}
&&\frac{1}{1+\underline{\hat{k}}\left( X^{\prime }\right) +\underline{\hat{k}%
}_{1}^{B}\left( X^{\prime }\right) +\kappa \left[ \frac{\underline{\hat{k}}%
_{2}^{B}}{1+\bar{k}}\right] \left( X^{\prime }\right) } \\
&=&1-\left( \int \frac{\hat{S}\left( X^{\prime },X\right) \hat{K}%
_{X}\left\vert \hat{\Psi}\left( X\right) \right\vert ^{2}}{\hat{K}%
_{X^{\prime }}\left\vert \hat{\Psi}\left( X^{\prime }\right) \right\vert ^{2}%
}dX+\int \frac{\left( \hat{S}_{E}^{B}\left( X^{\prime },X\right) +\hat{S}%
_{L}^{B}\left( X^{\prime },X\right) \right) \bar{K}_{X}\left\vert \bar{\Psi}%
\left( X\right) \right\vert ^{2}}{\hat{K}_{X^{\prime }}\left\vert \hat{\Psi}%
\left( X^{\prime }\right) \right\vert ^{2}}dX\right)
\end{eqnarray*}%
and:%
\begin{eqnarray*}
\underline{\hat{k}}\left( X^{\prime }\right) &=&\frac{\int \hat{S}\left(
X^{\prime },X\right) \frac{\hat{K}_{X}\left\vert \hat{\Psi}\left( X\right)
\right\vert ^{2}}{\hat{K}_{X^{\prime }}\left\vert \hat{\Psi}\left( X^{\prime
}\right) \right\vert ^{2}}dX}{1-\left( \int \frac{\hat{S}\left( X^{\prime
},X\right) \hat{K}_{X}\left\vert \hat{\Psi}\left( X\right) \right\vert ^{2}}{%
\hat{K}_{X^{\prime }}\left\vert \hat{\Psi}\left( X^{\prime }\right)
\right\vert ^{2}}dX+\int \frac{\left( \hat{S}_{E}^{B}\left( X^{\prime
},X\right) +\hat{S}_{L}^{B}\left( X^{\prime },X\right) \right) \bar{K}%
_{X}\left\vert \bar{\Psi}\left( X\right) \right\vert ^{2}}{\hat{K}%
_{X^{\prime }}\left\vert \hat{\Psi}\left( X^{\prime }\right) \right\vert ^{2}%
}dX\right) } \\
\underline{\hat{k}}_{1}^{B}\left( X^{\prime }\right) &=&\frac{\int \hat{S}%
_{E}^{B}\left( X^{\prime },X\right) \frac{\bar{K}_{X}\left\vert \bar{\Psi}%
\left( X\right) \right\vert ^{2}}{\hat{K}_{X^{\prime }}\left\vert \hat{\Psi}%
\left( X^{\prime }\right) \right\vert ^{2}}dX}{1-\left( \int \frac{\hat{S}%
\left( X^{\prime },X\right) \hat{K}_{X}\left\vert \hat{\Psi}\left( X\right)
\right\vert ^{2}}{\hat{K}_{X^{\prime }}\left\vert \hat{\Psi}\left( X^{\prime
}\right) \right\vert ^{2}}dX+\int \frac{\left( \hat{S}_{E}^{B}\left(
X^{\prime },X\right) +\hat{S}_{L}^{B}\left( X^{\prime },X\right) \right) 
\bar{K}_{X}\left\vert \bar{\Psi}\left( X\right) \right\vert ^{2}}{\hat{K}%
_{X^{\prime }}\left\vert \hat{\Psi}\left( X^{\prime }\right) \right\vert ^{2}%
}dX\right) } \\
\kappa \left[ \frac{\underline{\hat{k}}_{2}^{B}}{1+\bar{k}}\right] \left(
X^{\prime }\right) &=&\frac{\int \hat{S}_{L}^{B}\left( X^{\prime },X\right) 
\frac{\bar{K}_{X}\left\vert \bar{\Psi}\left( X\right) \right\vert ^{2}}{\hat{%
K}_{X^{\prime }}\left\vert \hat{\Psi}\left( X^{\prime }\right) \right\vert
^{2}}dX}{1-\left( \int \frac{\hat{S}\left( X^{\prime },X\right) \hat{K}%
_{X}\left\vert \hat{\Psi}\left( X\right) \right\vert ^{2}}{\hat{K}%
_{X^{\prime }}\left\vert \hat{\Psi}\left( X^{\prime }\right) \right\vert ^{2}%
}dX+\int \frac{\left( \hat{S}_{E}^{B}\left( X^{\prime },X\right) +\hat{S}%
_{L}^{B}\left( X^{\prime },X\right) \right) \bar{K}_{X}\left\vert \bar{\Psi}%
\left( X\right) \right\vert ^{2}}{\hat{K}_{X^{\prime }}\left\vert \hat{\Psi}%
\left( X^{\prime }\right) \right\vert ^{2}}dX\right) }
\end{eqnarray*}

Then for the firms, using:

\begin{equation*}
\int S\left( X^{\prime },X\right) \frac{K_{X}\left\vert \Psi \left( X\right)
\right\vert ^{2}}{K_{X^{\prime }}\left\vert \Psi \left( X^{\prime }\right)
\right\vert ^{2}}dX=\frac{\underline{\bar{k}}\left( X^{\prime }\right) }{1+%
\underline{k}\left( X^{\prime }\right) +\underline{k}_{1}^{\left( B\right)
}\left( X^{\prime }\right) +\kappa \left[ \frac{\underline{k}_{2}^{B}}{1+%
\bar{k}}\right] \left( X^{\prime }\right) }
\end{equation*}

\begin{equation*}
\left[ \frac{\underline{k}_{2}^{B}}{1+\bar{k}}\right] \left( X^{\prime
}\right) =\int \frac{\underline{k}_{2}^{B}\left( X^{\prime },X\right) }{1+%
\bar{k}\left( X\right) }\bar{K}\left\vert \bar{\Psi}\left( \bar{K},X\right)
\right\vert ^{2}d\bar{K}dX
\end{equation*}%
and (\ref{DFrApp}), we find:%
\begin{eqnarray*}
\underline{k}\left( X^{\prime }\right) &=&\frac{\int \frac{S\left( X^{\prime
},X\right) \hat{K}_{X}\left\vert \hat{\Psi}\left( X\right) \right\vert ^{2}}{%
K_{X^{\prime }}\left\vert \Psi \left( X^{\prime }\right) \right\vert ^{2}}dX%
}{1-\left( \int \frac{S\left( X^{\prime },X\right) \hat{K}_{X}\left\vert 
\hat{\Psi}\left( X\right) \right\vert ^{2}}{K_{X^{\prime }}\left\vert \Psi
\left( X^{\prime }\right) \right\vert ^{2}}dX+\int \frac{\left(
S_{E}^{B}\left( X^{\prime },X\right) +S_{L}^{B}\left( X^{\prime },X\right)
\right) \bar{K}_{X}\left\vert \bar{\Psi}\left( X\right) \right\vert ^{2}}{%
K_{X^{\prime }}\left\vert \Psi \left( X^{\prime }\right) \right\vert ^{2}}%
dX\right) } \\
\underline{k}_{1}^{\left( B\right) }\left( X^{\prime }\right) &=&\frac{\int
S_{E}^{B}\left( X^{\prime },X\right) \frac{\bar{K}_{X}\left\vert \bar{\Psi}%
\left( X\right) \right\vert ^{2}}{K_{X^{\prime }}\left\vert \Psi \left(
X^{\prime }\right) \right\vert ^{2}}dX}{1-\left( \int \frac{S\left(
X^{\prime },X\right) \hat{K}_{X}\left\vert \hat{\Psi}\left( X\right)
\right\vert ^{2}}{K_{X^{\prime }}\left\vert \Psi \left( X^{\prime }\right)
\right\vert ^{2}}dX+\int \frac{\left( S_{E}^{B}\left( X^{\prime },X\right)
+S_{L}^{B}\left( X^{\prime },X\right) \right) \bar{K}_{X}\left\vert \bar{\Psi%
}\left( X\right) \right\vert ^{2}}{K_{X^{\prime }}\left\vert \Psi \left(
X^{\prime }\right) \right\vert ^{2}}dX\right) } \\
\kappa \left[ \frac{\underline{k}_{2}^{B}}{1+\bar{k}}\right] \left(
X^{\prime }\right) &=&\frac{\int S_{L}^{B}\left( X^{\prime },X\right) \frac{%
\bar{K}_{X}\left\vert \bar{\Psi}\left( X\right) \right\vert ^{2}}{%
K_{X^{\prime }}\left\vert \Psi \left( X^{\prime }\right) \right\vert ^{2}}dX%
}{1-\left( \int \frac{S\left( X^{\prime },X\right) \hat{K}_{X}\left\vert 
\hat{\Psi}\left( X\right) \right\vert ^{2}}{K_{X^{\prime }}\left\vert \Psi
\left( X^{\prime }\right) \right\vert ^{2}}dX+\int \frac{\left(
S_{E}^{B}\left( X^{\prime },X\right) +S_{L}^{B}\left( X^{\prime },X\right)
\right) \bar{K}_{X}\left\vert \bar{\Psi}\left( X\right) \right\vert ^{2}}{%
K_{X^{\prime }}\left\vert \Psi \left( X^{\prime }\right) \right\vert ^{2}}%
dX\right) }
\end{eqnarray*}%
These formula allow to write (\ref{bt}) in terms of shares only:

\begin{eqnarray}
0 &=&\left( 1-\bar{S}_{E}\left( X^{\prime },X\right) \right) \left( \bar{f}%
\left( X^{\prime }\right) -\left( 1+\kappa \right) \bar{r}\right) \frac{%
1-\int \bar{S}\left( X^{\prime },X\right) \frac{\bar{K}_{X}\left\vert \bar{%
\Psi}\left( X\right) \right\vert ^{2}}{\bar{K}_{X^{\prime }}\left\vert \bar{%
\Psi}\left( X^{\prime }\right) \right\vert ^{2}}dX}{1-\int \bar{S}_{E}\left(
X^{\prime },X\right) \frac{\bar{K}_{X}\left\vert \bar{\Psi}\left( X\right)
\right\vert ^{2}}{\bar{K}_{X^{\prime }}\left\vert \bar{\Psi}\left( X^{\prime
}\right) \right\vert ^{2}}dX} \\
&&-\hat{S}_{E}^{B}\left( X^{\prime },X\right) \left( \hat{f}\left( X^{\prime
}\right) -\bar{r}\right)  \notag \\
&&\times \frac{1-\left( \int \frac{\hat{S}\left( X^{\prime },X\right) \hat{K}%
_{X}\left\vert \hat{\Psi}\left( X\right) \right\vert ^{2}}{\hat{K}%
_{X^{\prime }}\left\vert \hat{\Psi}\left( X^{\prime }\right) \right\vert ^{2}%
}dX+\int \frac{\left( \hat{S}_{E}^{B}\left( X^{\prime },X\right) +\hat{S}%
_{L}^{B}\left( X^{\prime },X\right) \right) \bar{K}_{X}\left\vert \bar{\Psi}%
\left( X\right) \right\vert ^{2}}{\hat{K}_{X^{\prime }}\left\vert \hat{\Psi}%
\left( X^{\prime }\right) \right\vert ^{2}}dX\right) }{1-\left( \int \frac{%
\hat{S}_{E}\left( X^{\prime },X\right) \hat{K}_{X}\left\vert \hat{\Psi}%
\left( X\right) \right\vert ^{2}}{\hat{K}_{X^{\prime }}\left\vert \hat{\Psi}%
\left( X^{\prime }\right) \right\vert ^{2}}dX+\int \frac{\hat{S}%
_{E}^{B}\left( X^{\prime },X\right) \bar{K}_{X}\left\vert \bar{\Psi}\left(
X\right) \right\vert ^{2}}{\hat{K}_{X^{\prime }}\left\vert \hat{\Psi}\left(
X^{\prime }\right) \right\vert ^{2}}dX\right) }  \notag \\
&&+\left( 1+\bar{f}\left( X^{\prime }\right) \right) H\left( -\left( 1+\bar{f%
}\left( X^{\prime }\right) \right) \right) \bar{S}_{L}\left( X^{\prime
},X\right) \frac{\left( 1-\int \bar{S}\left( X^{\prime },X\right) \frac{\bar{%
K}_{X}\left\vert \bar{\Psi}\left( X\right) \right\vert ^{2}}{\bar{K}%
_{X^{\prime }}\left\vert \bar{\Psi}\left( X^{\prime }\right) \right\vert ^{2}%
}dX\right) }{\int \bar{S}_{L}\left( X^{\prime },X\right) \frac{\bar{K}%
_{X}\left\vert \bar{\Psi}\left( X\right) \right\vert ^{2}}{\bar{K}%
_{X^{\prime }}\left\vert \bar{\Psi}\left( X^{\prime }\right) \right\vert ^{2}%
}dX}  \notag \\
&&+\left( 1+\hat{f}\left( X^{\prime }\right) \right) H\left( -\left( 1+\hat{f%
}\left( X^{\prime }\right) \right) \right) \hat{S}_{L}^{B}\left( X^{\prime
},X\right)  \notag \\
&&\times \frac{1-\left( \int \frac{S\left( X^{\prime },X\right) \hat{K}%
_{X}\left\vert \hat{\Psi}\left( X\right) \right\vert ^{2}}{K_{X^{\prime
}}\left\vert \Psi \left( X^{\prime }\right) \right\vert ^{2}}dX+\int \frac{%
\left( S_{E}^{B}\left( X^{\prime },X\right) +S_{L}^{B}\left( X^{\prime
},X\right) \right) \bar{K}_{X}\left\vert \bar{\Psi}\left( X\right)
\right\vert ^{2}}{K_{X^{\prime }}\left\vert \Psi \left( X^{\prime }\right)
\right\vert ^{2}}dX\right) }{\int \frac{S_{L}\left( X^{\prime },X\right) 
\hat{K}_{X}\left\vert \hat{\Psi}\left( X\right) \right\vert ^{2}}{%
K_{X^{\prime }}\left\vert \Psi \left( X^{\prime }\right) \right\vert ^{2}}%
dX+\int S_{L}^{B}\left( X^{\prime },X\right) \frac{\bar{K}_{X}\left\vert 
\bar{\Psi}\left( X\right) \right\vert ^{2}}{K_{X^{\prime }}\left\vert \Psi
\left( X^{\prime }\right) \right\vert ^{2}}dX}  \notag \\
&&+\left( 1+f_{1}^{\prime }\left( X^{\prime }\right) \right) H\left( -\left(
1+f_{1}^{\prime }\left( K^{\prime },X^{\prime }\right) \right) \right)
S_{L}^{B}\left( X^{\prime },X\right)  \notag \\
&&\times \frac{1-\left( \int \frac{\hat{S}\left( X^{\prime },X\right) \hat{K}%
_{X}\left\vert \hat{\Psi}\left( X\right) \right\vert ^{2}}{\hat{K}%
_{X^{\prime }}\left\vert \hat{\Psi}\left( X^{\prime }\right) \right\vert ^{2}%
}dX+\int \frac{\left( \hat{S}_{E}^{B}\left( X^{\prime },X\right) +\hat{S}%
_{L}^{B}\left( X^{\prime },X\right) \right) \bar{K}_{X}\left\vert \bar{\Psi}%
\left( X\right) \right\vert ^{2}}{\hat{K}_{X^{\prime }}\left\vert \hat{\Psi}%
\left( X^{\prime }\right) \right\vert ^{2}}dX\right) }{\int \hat{S}%
_{L}\left( X^{\prime },X\right) \frac{\hat{K}_{X}\left\vert \hat{\Psi}\left(
X\right) \right\vert ^{2}}{\hat{K}_{X^{\prime }}\left\vert \hat{\Psi}\left(
X^{\prime }\right) \right\vert ^{2}}dX+\int \hat{S}_{L}^{B}\left( X^{\prime
},X\right) \frac{\bar{K}_{X}\left\vert \bar{\Psi}\left( X\right) \right\vert
^{2}}{\hat{K}_{X^{\prime }}\left\vert \hat{\Psi}\left( X^{\prime }\right)
\right\vert ^{2}}dX} \\
&&-S_{E}^{B}\left( X^{\prime },X\right) \left( f_{1}^{\prime }\left(
X^{\prime }\right) -\bar{r}+\Delta F_{\tau }\left( \bar{R}\left( K,X\right)
\right) \right)  \notag \\
&&\times \frac{1-\left( \int \frac{S\left( X^{\prime },X\right) \hat{K}%
_{X}\left\vert \hat{\Psi}\left( X\right) \right\vert ^{2}}{K_{X^{\prime
}}\left\vert \Psi \left( X^{\prime }\right) \right\vert ^{2}}dX+\int \frac{%
\left( S_{E}^{B}\left( X^{\prime },X\right) +S_{L}^{B}\left( X^{\prime
},X\right) \right) \bar{K}_{X}\left\vert \bar{\Psi}\left( X\right)
\right\vert ^{2}}{K_{X^{\prime }}\left\vert \Psi \left( X^{\prime }\right)
\right\vert ^{2}}dX\right) }{1-\int \frac{S_{E}\left( X^{\prime },X\right) 
\hat{K}_{X}\left\vert \hat{\Psi}\left( X\right) \right\vert ^{2}}{%
K_{X^{\prime }}\left\vert \Psi \left( X^{\prime }\right) \right\vert ^{2}}%
dX-\int S_{E}^{B}\left( X^{\prime },X\right) \frac{\bar{K}_{X}\left\vert 
\bar{\Psi}\left( X\right) \right\vert ^{2}}{K_{X^{\prime }}\left\vert \Psi
\left( X^{\prime }\right) \right\vert ^{2}}dX}  \notag
\end{eqnarray}%
with the constraints on shares:%
\begin{equation*}
\int \left( \hat{S}_{E}\left( X^{\prime },X\right) +\hat{S}_{L}\left(
X^{\prime },X\right) \right) dX^{\prime }+\int \left( S_{E}\left( X^{\prime
},X\right) +S_{L}\left( X^{\prime },X\right) \right) dX^{\prime }=1
\end{equation*}%
\begin{equation*}
\int \left( \bar{S}_{E}\left( X^{\prime },X\right) +\bar{S}_{L}\left(
X^{\prime },X\right) \right) dX^{\prime }+\int \hat{S}_{E}^{B}\left(
X^{\prime },X\right) dX^{\prime }+\int S_{E}^{B}\left( X^{\prime },X\right)
dX^{\prime }=1
\end{equation*}%
\begin{eqnarray*}
\int \hat{S}_{L}^{B}\left( X^{\prime },X\right) dX^{\prime }+\int
S_{L}^{B}\left( X^{\prime },X\right) dX^{\prime } &=&\frac{\kappa }{1+\bar{k}%
\left( X\right) } \\
&=&\kappa \left( 1-\int \bar{S}\left( X,\bar{Y}\right) \frac{\bar{K}_{\bar{Y}%
}\left\vert \bar{\Psi}\left( \bar{Y}\right) \right\vert ^{2}}{\bar{K}%
_{X}\left\vert \bar{\Psi}\left( X\right) \right\vert ^{2}}d\bar{Y}\right) \\
&\rightarrow &\kappa \left( 1-\bar{S}\left( X\right) \right)
\end{eqnarray*}%
Note ultimatly that the various matrices defined in the first part and
arising in the return equation can be rewritten in terms of shares:%
\begin{eqnarray*}
\bar{M}\left( \left( \bar{K},X\right) ,\left( \bar{K}^{\prime },X^{\prime
}\right) \right) &=&\frac{\overline{\bar{k}}\left( X,X^{\prime }\right) \bar{%
K}}{1+\int \overline{\bar{k}}\left( X,X^{\prime }\right) \bar{K}_{0}^{\prime
}\left\vert \bar{\Psi}\left( \bar{K}_{0}^{\prime },X^{\prime }\right)
\right\vert ^{2}} \\
&=&\bar{S}\left( \left( \bar{K},X\right) ,\left( \bar{K}^{\prime },X^{\prime
}\right) \right)
\end{eqnarray*}%
\begin{eqnarray*}
&&\hat{M}\left( \left( \hat{K}^{\prime },X^{\prime }\right) ,\left( \hat{K}%
,X\right) \right) \\
&=&\frac{\hat{k}\left( X,X^{\prime }\right) \hat{K}}{1+\int \hat{k}\left(
X,X^{\prime }\right) \left\vert \hat{\Psi}\left( \hat{K}^{\prime },X^{\prime
}\right) \right\vert ^{2}+\int \hat{k}_{1}^{B}\left( X,X^{\prime }\right) 
\bar{K}_{0}^{\prime }\left\vert \bar{\Psi}\left( \bar{K}_{0}^{\prime
},X^{\prime }\right) \right\vert ^{2}+\int \hat{k}_{2}^{B}\left( X,X^{\prime
}\right) \frac{\bar{K}_{0}^{\prime }\left\vert \bar{\Psi}\left( \bar{K}%
_{0}^{\prime },X^{\prime }\right) \right\vert ^{2}}{1+\int \overline{\bar{k}}%
\left( X^{\prime },X^{\prime \prime }\right) \bar{K}_{0}^{\prime \prime
}\left\vert \bar{\Psi}\left( \bar{K}_{0}^{\prime \prime },X^{\prime \prime
}\right) \right\vert ^{2}}} \\
&=&\hat{S}\left( \left( \hat{K}^{\prime },X^{\prime }\right) ,\left( \hat{K}%
,X\right) \right)
\end{eqnarray*}%
and ultimately:%
\begin{eqnarray*}
\bar{N} &\rightarrow &\frac{\left( \hat{k}_{1}^{B}\left( X,X^{\prime
}\right) +\kappa \frac{\hat{k}_{2}^{B}\left( X,X^{\prime }\right) }{1+\int 
\overline{\bar{k}}\left( X^{\prime },X^{\prime \prime }\right) \bar{K}%
_{0}^{\prime \prime }\left\vert \bar{\Psi}\left( \bar{K}_{0}^{\prime \prime
},X^{\prime \prime }\right) \right\vert ^{2}}-\kappa \int \frac{\hat{k}%
_{2}^{B}\left( X,X^{\prime \prime }\right) \bar{K}_{0}^{\prime \prime }%
\overline{\bar{k}}\left( X^{\prime \prime },X^{\prime }\right) }{\left(
1+\int \overline{\bar{k}}\left( X^{\prime \prime },\bar{Y}\right) \bar{K}%
_{0}^{Y}\left\vert \bar{\Psi}\left( \bar{K}_{0}^{Y},\bar{Y}\right)
\right\vert ^{2}\right) ^{2}}\right) \hat{K}}{1+\int \hat{k}\left(
X,X^{\prime }\right) \left\vert \hat{\Psi}\left( \hat{K}^{\prime },X^{\prime
}\right) \right\vert ^{2}+\int \hat{k}_{1}^{B}\left( X,X^{\prime }\right) 
\bar{K}_{0}^{\prime }\left\vert \bar{\Psi}\left( \bar{K}_{0}^{\prime
},X^{\prime }\right) \right\vert ^{2}+\kappa \int \hat{k}_{2}^{B}\left(
X,X^{\prime }\right) \frac{\bar{K}_{0}^{\prime }\left\vert \bar{\Psi}\left( 
\bar{K}_{0}^{\prime },X^{\prime }\right) \right\vert ^{2}}{1+\int \overline{%
\bar{k}}\left( X^{\prime },X^{\prime \prime }\right) \bar{K}_{0}^{\prime
\prime }\left\vert \bar{\Psi}\left( \bar{K}_{0}^{\prime \prime },X^{\prime
\prime }\right) \right\vert ^{2}}} \\
&=&\hat{S}_{E}^{B}\left( \left( \hat{K},X\right) ,\left( \bar{K}^{\prime
},X^{\prime }\right) \right) +\hat{S}_{L}^{B}\left( \left( \hat{K},X\right)
,\left( \bar{K}^{\prime },X^{\prime }\right) \right)
\end{eqnarray*}

\subsection*{A2.3 Return equations in terms of partial averages of shares}

We will also need to define the total shares invested in one sector. They
are defined by summing all shares invested in this sector:

\begin{eqnarray*}
\hat{S}_{\eta }\left( X^{\prime }\right) &=&\int \hat{S}_{\eta }\left(
X^{\prime },X\right) \frac{\hat{K}_{X}\left\vert \hat{\Psi}\left( X\right)
\right\vert ^{2}}{\hat{K}_{X^{\prime }}\left\vert \hat{\Psi}\left( X^{\prime
}\right) \right\vert ^{2}}dX \\
\hat{S}\left( X^{\prime }\right) &=&\hat{S}_{E}\left( X^{\prime }\right) +%
\hat{S}_{L}\left( X^{\prime }\right)
\end{eqnarray*}%
\begin{eqnarray*}
\hat{S}_{\eta }^{B}\left( X^{\prime }\right) &=&\int \frac{\hat{S}_{\eta
}^{B}\left( X^{\prime },X\right) \bar{K}_{X}\left\vert \bar{\Psi}\left(
X\right) \right\vert ^{2}}{\hat{K}_{X^{\prime }}\left\vert \hat{\Psi}\left(
X^{\prime }\right) \right\vert ^{2}}dX \\
\hat{S}^{B}\left( X^{\prime }\right) &=&\hat{S}_{E}^{B}\left( X^{\prime
}\right) +\hat{S}_{L}^{B}\left( X^{\prime }\right)
\end{eqnarray*}

\begin{eqnarray*}
\bar{S}_{\eta }\left( X^{\prime }\right) &=&\int \bar{S}_{\eta }\left(
X^{\prime },X\right) \frac{\bar{K}_{X}\left\vert \bar{\Psi}\left( X\right)
\right\vert ^{2}}{\bar{K}_{X^{\prime }}\left\vert \bar{\Psi}\left( X^{\prime
}\right) \right\vert ^{2}}dX \\
\bar{S}\left( X^{\prime }\right) &=&\bar{S}_{E}\left( X^{\prime }\right) +%
\bar{S}_{L}\left( X^{\prime }\right)
\end{eqnarray*}%
\begin{eqnarray*}
S_{\eta }\left( X^{\prime }\right) &=&\int \frac{S_{\eta }\left( X^{\prime
},X\right) \hat{K}_{X}\left\vert \hat{\Psi}\left( X\right) \right\vert ^{2}}{%
K_{X^{\prime }}\left\vert \Psi \left( X^{\prime }\right) \right\vert ^{2}}dX
\\
S\left( X^{\prime }\right) &=&S_{E}\left( X^{\prime }\right) +S_{L}\left(
X^{\prime }\right)
\end{eqnarray*}%
\begin{eqnarray*}
S_{\eta }^{B}\left( X^{\prime }\right) &=&\int \frac{S_{\eta }^{B}\left(
X^{\prime },X\right) \bar{K}_{X}\left\vert \bar{\Psi}\left( X\right)
\right\vert ^{2}}{K_{X^{\prime }}\left\vert \Psi \left( X^{\prime }\right)
\right\vert ^{2}}dX \\
S^{B}\left( X^{\prime }\right) &=&S_{E}^{B}\left( X^{\prime }\right)
+S_{L}^{B}\left( X^{\prime }\right)
\end{eqnarray*}%
These formula allow to write (\ref{nv}) and (\ref{bt}) in the following form:

\begin{eqnarray*}
0 &=&\int \left( \Delta \left( X,X^{\prime }\right) -\hat{S}_{E}\left(
X^{\prime },X\right) \right) \frac{1-\hat{S}\left( X^{\prime }\right) }{1-%
\hat{S}_{E}\left( X^{\prime }\right) }\left( f\left( X^{\prime }\right) -%
\bar{r}\right) dX^{\prime } \\
&&-\int S_{E}\left( X^{\prime },X\right) \frac{1-\left( S\left( X^{\prime
}\right) +\left( S_{E}^{B}\left( X^{\prime }\right) +S_{L}^{B}\left(
X^{\prime }\right) \right) \right) }{1-S_{E}\left( X^{\prime }\right)
-S_{E}^{B}\left( X^{\prime }\right) }\left( \left( f_{1}^{\prime }\left(
X^{\prime }\right) -\bar{r}\right) +\Delta F_{\tau }\left( \bar{R}\left(
K,X\right) \right) \right) dX^{\prime } \\
&&-\int \frac{1-\left( \hat{S}\left( X^{\prime }\right) +\hat{S}%
_{E}^{B}\left( X^{\prime }\right) +\hat{S}_{L}^{B}\left( X^{\prime }\right)
\right) }{\hat{S}_{L}\left( X^{\prime }\right) }\left( 1+f\left( X^{\prime
}\right) \right) H\left( -\left( 1+f\left( X^{\prime }\right) \right)
\right) \hat{S}_{L}\left( X^{\prime },X\right) dX^{\prime } \\
&&-\int \frac{1-\left( S\left( X^{\prime }\right) +S_{E}^{B}\left( X^{\prime
}\right) +S_{L}^{B}\left( X^{\prime }\right) \right) }{S_{L}\left( X^{\prime
}\right) }\left( 1+f_{1}^{\prime }\left( X^{\prime }\right) \right) H\left(
-\left( 1+f_{1}^{\prime }\left( X^{\prime }\right) \right) \right)
S_{L}\left( X^{\prime },X\right) dX^{\prime }
\end{eqnarray*}%
for investors, and:%
\begin{eqnarray}
0 &=&\left( 1-\bar{S}_{E}\left( X^{\prime },X\right) \right) \left( \bar{f}%
\left( X^{\prime }\right) -\left( 1+\kappa \right) \bar{r}\right) \frac{1-%
\bar{S}\left( X^{\prime }\right) }{1-\bar{S}_{E}\left( X^{\prime }\right) }
\label{Rb} \\
&&-\hat{S}_{E}^{B}\left( X^{\prime },X\right) \left( \hat{f}\left( X^{\prime
}\right) -\bar{r}\right) \frac{1-\left( \hat{S}\left( X^{\prime }\right) +%
\hat{S}_{E}^{B}\left( X^{\prime }\right) +\hat{S}_{L}^{B}\left( X^{\prime
}\right) \right) }{1-\left( \hat{S}_{E}\left( X^{\prime }\right) +\hat{S}%
_{E}^{B}\left( X^{\prime }\right) \right) }  \notag \\
&&-\left( 1+\bar{f}\left( X^{\prime }\right) \right) H\left( -\left( 1+\bar{f%
}\left( X^{\prime }\right) \right) \right) \bar{S}_{L}\left( X^{\prime
},X\right) \frac{\left( 1-\bar{S}\left( X^{\prime }\right) \right) }{\bar{S}%
_{L}\left( X^{\prime }\right) }  \notag \\
&&-\left( 1+\hat{f}\left( X^{\prime }\right) \right) H\left( -\left( 1+\hat{f%
}\left( X^{\prime }\right) \right) \right) \hat{S}_{L}^{B}\left( X^{\prime
},X\right) \frac{1-\left( S\left( X^{\prime }\right) +\left( S_{E}^{B}\left(
X^{\prime }\right) +S_{L}^{B}\left( X^{\prime }\right) \right) \right) }{%
S_{L}\left( X^{\prime }\right) +S_{L}^{B}\left( X^{\prime }\right) }  \notag
\\
&&-\left( 1+f_{1}^{\prime }\left( X^{\prime }\right) \right) H\left( -\left(
1+f_{1}^{\prime }\left( X^{\prime }\right) \right) \right) S_{L}^{B}\left(
X^{\prime },X\right) \frac{1-\left( \hat{S}\left( X^{\prime },X\right)
+\left( \hat{S}_{E}^{B}\left( X^{\prime }\right) +\hat{S}_{L}^{B}\left(
X^{\prime }\right) \right) \right) }{\hat{S}_{L}\left( X^{\prime }\right) +%
\hat{S}_{L}^{B}\left( X^{\prime }\right) }  \notag \\
&&-S_{E}^{B}\left( X^{\prime },X\right) \left\{ \frac{1-\left( S\left(
X^{\prime }\right) +\left( S_{E}^{B}\left( X^{\prime }\right)
+S_{L}^{B}\left( X^{\prime }\right) \right) \right) }{1-S_{E}\left(
X^{\prime }\right) -S_{E}^{B}\left( X^{\prime }\right) }\left( \left(
f_{1}^{\prime }\left( X^{\prime }\right) -\bar{r}\right) +\Delta F_{\tau
}\left( \bar{R}\left( K,X\right) \right) \right) \right\}  \notag
\end{eqnarray}%
for banks.

In the sequel:%
\begin{equation*}
\frac{1-\left( S\left( X^{\prime }\right) +\left( S_{E}^{B}\left( X^{\prime
}\right) +S_{L}^{B}\left( X^{\prime }\right) \right) \right) }{1-S_{E}\left(
X^{\prime }\right) -S_{E}^{B}\left( X^{\prime }\right) }\left( \left(
f_{1}^{\prime }\left( X^{\prime }\right) -\bar{r}\right) +\Delta F_{\tau
}\left( \bar{R}\left( K,X\right) \right) \right)
\end{equation*}%
computes the firms returns. It will be given by:%
\begin{equation*}
\left( f_{1}\left( X^{\prime }\right) -\bar{r}\right) +\Delta F_{\tau
}\left( \bar{R}\left( K,X\right) \right)
\end{equation*}%
for constant return to scales, and by:

\begin{eqnarray*}
f_{1}\left( X^{\prime },\hat{K}\left[ X^{\prime }\right] ,\bar{K}\left[
X^{\prime }\right] \right) &\simeq &\frac{f_{1}\left( X\right) }{\left(
\left( 1+k\left( X\right) \hat{K}\left[ X\right] \right) +\left(
k_{1}^{B}\left( X\right) +\kappa \left[ \frac{\underline{k}_{2}^{B}\left(
X\right) }{1+\bar{k}}\right] \right) \bar{K}\left[ X\right] \right) ^{r}}%
-C_{0} \\
&=&\left( 1-\left( S\left( X^{\prime }\right) +\left( S_{E}^{B}\left(
X^{\prime }\right) +S_{L}^{B}\left( X^{\prime }\right) \right) \right)
\right) ^{r}\left( f_{1}\left( X^{\prime }\right) -\bar{r}\right) \\
&&+\Delta F_{\tau }\left( \bar{R}\left( K,X\right) \right) -C_{0}-\left(
1-\left( S\left( X^{\prime }\right) +\left( S_{E}^{B}\left( X^{\prime
}\right) +S_{L}^{B}\left( X^{\prime }\right) \right) \right) \right) C
\end{eqnarray*}%
for decreasing retrn.

In the sequel, as in part 1, we will perform the computations for constant
returns to scale and then include corrections due to decreasing rtrns.

\section*{Appendix 3 \ Field translation}

For investors, the field action has been derived in part I. It is:%
\begin{eqnarray}
&&-\sigma _{\hat{K}}^{2}\sum_{\eta }\int \Gamma ^{\dag }\left( \hat{S}%
^{\left( T\right) },X^{\prime },X\right) \nabla _{\hat{S}_{\eta }^{\left(
T\right) }}^{2}\Gamma \left( \hat{S}^{\left( T\right) },X^{\prime },X\right)
d\left( \hat{S}^{\left( T\right) },X^{\prime },X\right) -\int \beta
\left\vert \Gamma \left( \hat{S}^{\left( T\right) },X^{\prime },X\right)
\right\vert ^{2}d\left( \hat{S}^{\left( T\right) },X^{\prime },X\right) 
\notag \\
&&+\sum_{\eta }\int \left( \frac{\left( \hat{S}_{\eta }^{\left( T\right)
}\right) ^{2}}{2\hat{w}_{\eta }\left( X^{\prime },X\right) }-\hat{V}_{\eta }%
\hat{S}_{\eta }^{\left( T\right) }\right) \left\vert \Gamma \left( \hat{S}%
^{\left( T\right) },X^{\prime },X\right) \right\vert ^{2}d\left( \hat{S}%
^{\left( T\right) },X^{\prime },X\right)  \notag \\
+ &&\int \lambda \left( X\right) \left( \sum_{\eta }\int \hat{S}_{\eta
}^{\left( T\right) }\left\vert \Gamma \left( \hat{S}^{\left( T\right)
},X^{\prime },X\right) \right\vert ^{2}dX^{\prime }d\hat{S}^{\left( T\right)
}-1\right) \left\vert \Gamma \left( \hat{S}^{\left( T\right) },X^{\prime
},X\right) \right\vert ^{2}d\left( \hat{S}^{\left( T\right) },X^{\prime
},X\right)  \label{CFN}
\end{eqnarray}%
where:%
\begin{equation*}
\Gamma \left( \hat{S}^{\left( T\right) },X^{\prime },X\right) \equiv \Gamma
\left( S_{E},\hat{S}_{E},S_{L},\hat{S}_{L},X^{\prime },X\right)
\end{equation*}

Considerng banks, we start with the full objective function for the system
of bnks:%
\begin{eqnarray}
&&\int dt\sum_{i}\left( \sum_{j}\bar{S}_{Eij}^{B}\bar{f}_{j}+\sum_{j}\bar{S}%
_{Lij}^{B}\bar{r}_{j}-\frac{1}{2}\sum_{j}\frac{\left( \bar{S}_{\eta
ij}\right) ^{2}}{\bar{w}_{\eta _{i}}\left( X_{j}\right) }\right.  \label{MNF}
\\
&&+\left. \sum_{j}\hat{S}_{Eij}^{B}\hat{f}_{j}-\frac{1}{2}\sum_{j}\frac{%
\left( \hat{S}_{E,ij}^{B}\right) ^{2}}{\hat{w}_{1i}\left( X_{j}\right) }%
+\sum_{k}S_{Eik}^{B}f_{k}-\frac{1}{2}\sum_{k}\frac{\left( S_{Eik}^{B}\right)
^{2}}{w_{\eta ik}\left( X_{k}\right) }\right)  \notag
\end{eqnarray}%
and for loans:%
\begin{equation}
\int dt\sum_{i}\left( \sum_{j}\hat{S}_{Lij}^{B}\hat{r}_{j}-\frac{1}{2}%
\sum_{j}\frac{\left( \hat{S}_{Lij}^{B}\right) ^{2}}{\hat{w}_{\eta
_{Li}}\left( X_{j}\right) }+\sum_{k}S_{Lik}^{B}\bar{r}_{k}-\frac{1}{2}%
\sum_{k}\frac{\left( S_{Lik}^{B}\right) ^{2}}{w_{\eta ik}\left( X_{k}\right) 
}\right)  \label{MNf}
\end{equation}

Starting with banks shares flds:%
\begin{equation*}
\bar{\Gamma}\left( \bar{S}^{\left( T\right) },X^{\prime },X^{\prime
},X\right) =\bar{\Gamma}\left( \bar{S}_{E},S_{E},\hat{S}_{E},\bar{S}%
_{L},S_{L},\hat{S}_{L},X^{\prime },X^{\prime },X\right)
\end{equation*}%
The functions (\ref{MNF}) and (\ref{MNf}) are translated by the action
functionals:%
\begin{equation}
\sum_{\eta }\int \left( \frac{\left( \bar{S}_{\eta }^{T}\right) ^{2}}{2\bar{w%
}_{\eta }^{T}\left( X^{\prime },X^{\prime },X\right) }-\bar{V}_{\eta }\bar{S}%
_{\eta }^{T}\right) \left\vert \bar{\Gamma}\left( \bar{S}^{\left( T\right)
},X^{\prime },X^{\prime },X\right) \right\vert ^{2}d\left( \bar{S}^{\left(
T\right) },X^{\prime },X^{\prime },X\right)  \label{CTF}
\end{equation}%
where:%
\begin{equation*}
\bar{V}_{1}\left( X^{\prime },X^{\prime },X\right) =\bar{f}\left( X^{\prime
}\right) \text{, }\bar{V}_{2}\left( X^{\prime },X^{\prime },X\right) =\bar{r}%
\left( X^{\prime }\right)
\end{equation*}%
\begin{equation*}
\bar{V}_{3}\left( X^{\prime },X^{\prime },X\right) =\hat{f}\left( X^{\prime
}\right) \text{, }\bar{V}_{4}\left( X^{\prime },X^{\prime },X\right) =\hat{r}%
\left( X^{\prime }\right)
\end{equation*}%
\begin{equation*}
\bar{V}_{5}\left( X^{\prime },X^{\prime },X\right) =f\left( X\right) \text{, 
}\bar{V}_{6}\left( X^{\prime },X^{\prime },X\right) =r\left( X\right)
\end{equation*}%
and:%
\begin{equation*}
\bar{w}_{E}^{T}\left( X^{\prime },X^{\prime },X\right) =\bar{w}_{E}\left(
X^{\prime },X\right) \text{, }\bar{w}_{2}^{T}\left( X^{\prime },X^{\prime
},X\right) =\bar{w}_{L}\left( X^{\prime },X\right)
\end{equation*}%
\begin{equation*}
\hat{w}_{3}^{T}\left( X^{\prime },X^{\prime },X\right) =\hat{w}_{E}\left(
X^{\prime },X\right) \text{, }\bar{w}_{4}^{T}\left( X^{\prime },X^{\prime
},X\right) =\hat{w}_{L}\left( X^{\prime },X\right)
\end{equation*}%
\begin{equation*}
\hat{w}_{5}^{T}\left( X^{\prime },X^{\prime },X\right) =w_{E}\left(
X,X\right) \text{, }\bar{w}_{6}^{T}\left( X^{\prime },X^{\prime },X\right)
=w_{L}\left( X,X\right)
\end{equation*}%
As in part I, we add to the functional (\ref{CTF}) a term accounting for
inertia and a contribution characteristic of the time scale considered for
the collective states: 
\begin{eqnarray*}
&&-\sigma _{\hat{K}}^{2}\sum_{\eta }\int \bar{\Gamma}^{\dag }\left( \bar{S}%
^{\left( T\right) },X^{\prime },X^{\prime },X\right) \nabla _{\bar{S}_{\eta
}^{\left( T\right) }}^{2}\bar{\Gamma}\left( \bar{S}^{\left( T\right)
},X^{\prime },X^{\prime },X\right) d\left( \bar{S}^{\left( T\right)
},X^{\prime },X^{\prime },X\right) \\
&&-\int \beta \left\vert \bar{\Gamma}\left( \bar{S}^{\left( T\right)
},X^{\prime },X^{\prime },X\right) \right\vert ^{2}d\left( \bar{S}^{\left(
T\right) },X^{\prime },X^{\prime },X\right)
\end{eqnarray*}%
Translating the constraints:%
\begin{equation*}
\bar{S}_{Eij}+\bar{S}_{Lij}+\hat{S}_{Eij}^{B}+S_{Eik}^{B}=1
\end{equation*}%
and:%
\begin{equation*}
\hat{S}_{Lij}^{B}+S_{Lik}^{B}=\kappa \left( 1-\sum_{j}\bar{S}%
_{Lij}^{B}\right)
\end{equation*}%
leads to the two following contributions:%
\begin{eqnarray*}
&&\int \lambda \left( X\right) \left\vert \bar{\Gamma}\left( \bar{S}^{\left(
T\right) },X^{\prime },X^{\prime },X\right) \right\vert ^{2}d\left( \bar{S}%
^{\left( T\right) },X^{\prime },X^{\prime },X\right) \\
&&\times \left( \int \left\vert \bar{\Gamma}\left( \bar{S}^{\left( T\right)
},X^{\prime },X^{\prime },X\right) \right\vert ^{2}\left( \bar{S}_{E}d\left( 
\bar{S}_{E},X^{\prime },X^{\prime }\right) +\bar{S}_{L}d\left( \bar{S}%
_{L},X^{\prime },X^{\prime }\right) +\hat{S}_{E}^{B}d\left( \hat{S}%
_{E}^{B},X^{\prime },X^{\prime }\right) +S_{E}^{B}d\left(
S_{E}^{B},X^{\prime },X^{\prime }\right) \right) -1\right) \\
&&+\int \lambda ^{\prime }\left( X\right) \left\vert \bar{\Gamma}\left( \bar{%
S}^{\left( T\right) },X^{\prime },X^{\prime },X\right) \right\vert
^{2}d\left( \bar{S}^{\left( T\right) },X^{\prime },X^{\prime },X\right) \\
&&\times \left( \int \left\vert \bar{\Gamma}\left( \bar{S}^{\left( T\right)
},X^{\prime },X^{\prime },X\right) \right\vert ^{2}\left( \hat{S}%
_{L}^{B}d\left( \hat{S}_{E}^{B},X^{\prime },X^{\prime }\right)
+S_{L}^{B}d\left( S_{E}^{B},X^{\prime },X^{\prime }\right) \right) -\kappa
\left( 1-\bar{S}_{L}^{B}\right) d\left( S_{L}^{B},X^{\prime },X^{\prime
}\right) \right) \\
&&+\int \lambda \left( X\right) \left( \sum_{\eta }\int \bar{S}_{\eta
}\left( X^{\prime },X\right) dX^{\prime }+\int \hat{S}_{E}^{B}\left(
X^{\prime },X\right) dX^{\prime }+S_{E}^{B}\left( X,X\right) -1\right)
\left\vert \bar{\Gamma}\left( \bar{S}^{\left( T\right) },X^{\prime
},X^{\prime },X\right) \right\vert ^{2} \\
&&+\int \lambda ^{\prime }\left( X\right) \left( \int \hat{S}_{L}^{B}\left(
X^{\prime },X\right) dX^{\prime }+S_{L}^{B}\left( X,X\right) -\kappa \left(
1-\bar{S}_{\eta }^{B}\left( X\right) \right) \right) \left\vert \bar{\Gamma}%
\left( \bar{S}^{\left( T\right) },X^{\prime },X^{\prime },X\right)
\right\vert ^{2}
\end{eqnarray*}%
we define the various averages in the states defined by the fields:%
\begin{eqnarray*}
\bar{S}_{\eta }^{B}\left( X^{\prime },X\right) &=&\int \bar{S}_{\eta
}\left\vert \bar{\Gamma}\right\vert ^{2}d\left( \bar{S}_{E},S_{E},\hat{S}%
_{E},\bar{S}_{L},S_{L},\hat{S}_{L},X^{\prime }\right) \\
S_{\eta }^{B}\left( X,X\right) &=&\int S_{\eta }^{B}\left\vert \bar{\Gamma}%
\right\vert ^{2}d\left( \bar{S}_{E},S_{E},\hat{S}_{E},\bar{S}_{L},S_{L},\hat{%
S}_{L},X^{\prime },X^{\prime }\right) \\
\hat{S}_{\eta }^{B}\left( X^{\prime },X\right) &=&\int \hat{S}_{\eta
}^{B}\left\vert \bar{\Gamma}\right\vert ^{2}d\left( \bar{S}_{E},S_{E},\hat{S}%
_{E},\bar{S}_{L},S_{L},\hat{S}_{L},X^{\prime }\right)
\end{eqnarray*}%
We also define:%
\begin{equation*}
\bar{S}^{\left( T\right) }\left( X^{\prime },X\right) \left\vert \Gamma
\left( X^{\prime },X\right) \right\vert ^{2}=\int \bar{S}^{\left( T\right)
}\left\vert \Gamma \left( \bar{S},X^{\prime },X\right) \right\vert ^{2}
\end{equation*}%
\begin{eqnarray*}
\bar{S}_{\eta }^{B}\left( X\right) &=&\int \bar{S}_{\eta }^{B}\left(
X^{\prime },X\right) dX^{\prime } \\
\hat{S}_{\eta }^{B}\left( X\right) &=&\int \hat{S}_{\eta }^{B}\left(
X^{\prime },X\right) dX^{\prime }
\end{eqnarray*}%
and the constraints on allocation in this context writes:%
\begin{equation*}
\sum_{\eta =1}^{2}\int \bar{S}_{\eta }^{B}\left( X^{\prime },X\right)
dX^{\prime }+\int \hat{S}_{E}^{B}\left( X^{\prime },X\right) dX^{\prime
}+S_{E}^{B}\left( X,X\right) =1
\end{equation*}%
\begin{equation*}
\int \hat{S}_{L}^{B}\left( X^{\prime },X\right) dX^{\prime }+S_{L}^{B}\left(
X,X\right) =\kappa \left( 1-\bar{S}_{L}^{B}\left( X\right) \right)
\end{equation*}%
As a consequence, the contraint rewrites in terms of fields:

\begin{eqnarray*}
&&\int \lambda \left( X\right) \left( \sum_{\eta =1}^{2}\int \bar{S}_{\eta
}^{B}\left( X^{\prime },X\right) dX^{\prime }+\int \hat{S}_{E}^{B}\left(
X^{\prime },X\right) dX^{\prime }+S_{E}^{B}\left( X,X\right) -1\right)
\left\vert \bar{\Gamma}\left( \bar{S}^{\left( T\right) },X^{\prime
},X^{\prime },X\right) \right\vert ^{2} \\
&&+\int \lambda ^{\prime }\left( X\right) \left( \int \hat{S}_{L}^{B}\left(
X^{\prime },X\right) dX^{\prime }+S_{L}^{B}\left( X,X\right) -\kappa \left(
1-\bar{S}_{\eta }^{B}\left( X\right) \right) \right) \left\vert \bar{\Gamma}%
\left( \bar{S}^{\left( T\right) },X^{\prime },X^{\prime },X\right)
\right\vert ^{2}
\end{eqnarray*}%
The action functional for shares is thus defined by:%
\begin{eqnarray}
S\left( \bar{\Gamma}\right) &=&-\sigma _{\hat{K}}^{2}\sum_{\eta }\int \bar{%
\Gamma}^{\dag }\left( \bar{S}^{\left( T\right) },X^{\prime },X^{\prime
},X\right) \nabla _{\bar{S}_{\eta }^{\left( T\right) }}^{2}\bar{\Gamma}%
\left( \bar{S}^{\left( T\right) },X^{\prime },X^{\prime },X\right) d\left( 
\bar{S}^{\left( T\right) },X^{\prime },X^{\prime },X\right)  \label{CFB} \\
&&+\int \left( \sum_{\eta }\left( \frac{\left( \bar{S}_{\eta }^{\left(
T\right) }\right) ^{2}}{2\hat{w}_{\eta }\left( X^{\prime },X^{\prime
},X\right) }-\bar{V}_{\eta }\bar{S}_{\eta }^{\left( T\right) }\right) -\beta
\right) \left\vert \bar{\Gamma}\left( \bar{S}^{\left( T\right) },X^{\prime
},X^{\prime },X\right) \right\vert ^{2}d\left( \bar{S}^{\left( T\right)
},X^{\prime },X^{\prime },X\right)  \notag \\
&&+\int \lambda \left( X\right) \left( \sum_{\eta =1}^{2}\int \bar{S}_{\eta
}^{B}\left( X^{\prime },X\right) dX^{\prime }+\int \hat{S}_{E}^{B}\left(
X^{\prime },X\right) dX^{\prime }+S_{E}^{B}\left( X,X\right) -1\right)
\left\vert \bar{\Gamma}\left( \bar{S}^{\left( T\right) },X^{\prime
},X^{\prime },X\right) \right\vert ^{2}  \notag \\
&&+\int \lambda ^{\prime }\left( X\right) \left( \int \hat{S}_{L}^{B}\left(
X^{\prime },X\right) dX^{\prime }+S_{L}^{B}\left( X,X\right) -\kappa \left(
1-\bar{S}_{\eta }^{B}\left( X\right) \right) \right) \left\vert \bar{\Gamma}%
\left( \bar{S}^{\left( T\right) },X^{\prime },X^{\prime },X\right)
\right\vert ^{2}  \notag
\end{eqnarray}

\begin{equation*}
\bar{S}^{\left( T\right) }=\left[ \bar{S}_{E},S_{E},\hat{S}_{E},\bar{S}%
_{L},S_{L},\hat{S}_{L}\right]
\end{equation*}

The minimization equations for (\ref{CFN}) has been presented Gosselin and
Lotz (2025). The minimization equations for (\ref{CFB}) are similar. They
are obtained by using the derivatives with respect to $\Gamma \left( \hat{S}%
^{\left( T\right) },X^{\prime },X\right) $, $\lambda \left( X\right) $ and $%
\lambda ^{\prime }\left( X\right) $. These two derivatives implement the
constraints. We find:%
\begin{eqnarray*}
&&-\sigma _{\hat{K}}^{2}\sum_{\eta }\nabla _{\bar{S}_{\eta }^{\left(
T\right) }}^{2}\bar{\Gamma}\left( \bar{S}^{\left( T\right) },X^{\prime
},X^{\prime },X\right) \\
&&+\left( \sum_{\eta }\left( \frac{\left( \bar{S}_{\eta }^{\left( T\right)
}\right) ^{2}}{2\bar{w}_{\eta }\left( X^{\prime },X^{\prime },X\right) }-%
\bar{V}_{\eta }^{\left( T\right) }\bar{S}_{\eta }^{T}+\lambda _{\eta }\left(
X\right) \bar{S}_{\eta }^{T}\right) -\beta \right) \bar{\Gamma}\left( \bar{S}%
^{\left( T\right) },X^{\prime },X^{\prime },X\right)
\end{eqnarray*}%
with:%
\begin{eqnarray*}
\lambda _{\eta }\left( X\right) &=&\lambda \left( X\right) \text{ for }\eta
=1,2,3,5 \\
\lambda _{\eta }\left( X\right) &=&\lambda ^{\prime }\left( X\right) \text{
for }\eta =5,6
\end{eqnarray*}%
and where we rescaled:%
\begin{eqnarray*}
\lambda \left( X\right) \left\Vert \bar{\Gamma}\left( \bar{S}^{\left(
T\right) },X^{\prime },X^{\prime },X\right) \right\Vert _{X}^{2}
&\rightarrow &\lambda \left( X\right) \\
\lambda ^{\prime }\left( X\right) \left\Vert \bar{\Gamma}\left( \bar{S}%
^{\left( T\right) },X^{\prime },X^{\prime },X\right) \right\Vert _{X}^{2}
&\rightarrow &\lambda ^{\prime }\left( X\right)
\end{eqnarray*}%
with:%
\begin{equation*}
\left\Vert \bar{\Gamma}\left( \bar{S}^{\left( T\right) },X^{\prime
},X^{\prime },X\right) \right\Vert _{X}^{2}=\int \left\vert \bar{\Gamma}%
\left( \bar{S}^{\left( T\right) },X^{\prime },X^{\prime },X\right)
\right\vert ^{2}d\left( X^{\prime },X^{\prime }\right) d\bar{S}^{\left(
T\right) }
\end{equation*}%
As in Gosselin and Lotz (2025), we can adjust $\beta $, so that the solution
to the minimization equations have the form:%
\begin{equation*}
\Gamma _{0,X^{\prime },X^{\prime },X}\left( \bar{S}^{\left( T\right)
}\right) \Gamma \left( X^{\prime },X^{\prime },X\right)
\end{equation*}%
where:%
\begin{equation*}
\Gamma _{0,X^{\prime },X^{\prime },X}\left( \bar{S}^{\left( T\right)
}\right) =N\exp \left( -\sum_{\eta }\frac{\left( \bar{S}_{\eta }^{\left(
T\right) }-\overline{\bar{S}_{\eta }^{\left( T\right) }}\left( X^{\prime
},X^{\prime },X\right) \right) ^{2}}{2\sigma _{\hat{K}}^{2}}\right)
\end{equation*}%
and the average $\overline{\bar{S}_{\eta }^{\left( T\right) }}$ is given by:%
\begin{equation}
\overline{\bar{S}_{\eta }^{\left( T\right) }}\left( X^{\prime },X^{\prime
},X\right) =\bar{w}_{\eta }\left( X^{\prime },X^{\prime },X\right) \left( 
\bar{V}_{\eta }\left( X^{\prime },X^{\prime },X\right) +\lambda _{\eta
}\left( X\right) \right)  \label{MNS}
\end{equation}%
The average $\overline{\hat{S}^{\left( T\right) }}$ s also the average:%
\begin{eqnarray}
\overline{\bar{S}_{\eta }^{\left( T\right) }}\left( X^{\prime },X^{\prime
},X\right) &=&\frac{\int \bar{S}_{\eta }^{\left( T\right) }\left\vert \Gamma
_{0,X^{\prime },X^{\prime },X}\left( \bar{S}^{\left( T\right) }\right)
\right\vert ^{2}d\bar{S}^{\left( T\right) }}{\int \left\vert \Gamma
_{0,X^{\prime },X^{\prime },X}\left( \bar{S}^{\left( T\right) }\right)
\right\vert ^{2}d\bar{S}^{\left( T\right) }}  \label{VRD} \\
&=&\frac{\int \bar{S}_{\eta }^{\left( T\right) }\left\vert \Gamma \left( 
\bar{S}^{\left( T\right) },X^{\prime },X^{\prime },X\right) \right\vert ^{2}d%
\bar{S}^{\left( T\right) }}{\int \left\vert \Gamma \left( \bar{S}^{\left(
T\right) },X^{\prime },X^{\prime },X\right) \right\vert ^{2}d\bar{S}^{\left(
T\right) }}  \notag
\end{eqnarray}

Multiplying by $\Gamma \left( X^{\prime },X^{\prime },X\right) $, we
integrate with respect to $X^{\prime },X^{\prime }$:%
\begin{eqnarray*}
&&\int \bar{S}_{\eta }^{\left( T\right) }\left\vert \bar{\Gamma}\left( \bar{S%
}^{\left( T\right) },X^{\prime },X^{\prime },X\right) \right\vert
^{2}d\left( X^{\prime },X^{\prime }\right) d\bar{S}^{\left( T\right) } \\
&=&\int \bar{w}_{\eta }\left( X^{\prime },X^{\prime },X\right) \bar{V}_{\eta
}\left( \bar{S}^{\left( T\right) },X^{\prime },X^{\prime },X\right)
\left\vert \bar{\Gamma}\left( \bar{S}^{\left( T\right) },X^{\prime
},X^{\prime },X\right) \right\vert ^{2}d\left( X^{\prime },X^{\prime
}\right) d\bar{S}^{\left( T\right) } \\
&&+\lambda _{\eta }\left( X\right) \int \bar{w}_{\eta }\left( X^{\prime
},X^{\prime },X\right) \left\vert \bar{\Gamma}\left( X^{\prime },X^{\prime
},X\right) \right\vert ^{2}d\left( X^{\prime },X^{\prime }\right) d\bar{S}%
^{\left( T\right) }
\end{eqnarray*}%
Then the partial summation over $\eta $ and the constraints yield:%
\begin{eqnarray*}
1 &=&\sum_{\eta \in \left\{ 1,2,3,5,\right\} }\int \bar{w}_{\eta }\left(
X^{\prime },X^{\prime },X\right) \bar{V}_{\eta }\left( X^{\prime },X^{\prime
},X\right) \left\vert \bar{\Gamma}\left( \bar{S}^{\left( T\right)
},X^{\prime },X^{\prime },X\right) \right\vert ^{2}d\left( X^{\prime
},X^{\prime }\right) d\bar{S}^{\left( T\right) } \\
&&+\lambda \left( X\right) \sum_{\eta \in \left\{ 1,2,3,5,\right\} }\int 
\bar{w}_{\eta }\left( X^{\prime },X^{\prime },X\right) \left\vert \bar{\Gamma%
}\left( \bar{S}^{\left( T\right) },X^{\prime },X^{\prime },X\right)
\right\vert ^{2}d\left( X^{\prime },X^{\prime }\right) d\bar{S}^{\left(
T\right) }
\end{eqnarray*}%
\begin{eqnarray*}
&&\kappa \left( 1-\bar{S}_{L}^{B}\left( X\right) \right) \\
&=&\sum_{\eta \in \left\{ 4,6\right\} }\int \bar{w}_{\eta }\left( X^{\prime
},X^{\prime },X\right) \bar{V}_{\eta }\left( X^{\prime },X^{\prime
},X\right) \left\vert \bar{\Gamma}\left( \bar{S}^{\left( T\right)
},X^{\prime },X^{\prime },X\right) \right\vert ^{2}d\left( X^{\prime
},X^{\prime }\right) d\bar{S}^{\left( T\right) } \\
&&+\lambda ^{\prime }\left( X\right) \sum_{\eta \in \left\{ 4,6\right\}
}\int \bar{w}_{\eta }\left( X^{\prime },X^{\prime },X\right) \left\vert \bar{%
\Gamma}\left( \bar{S}^{\left( T\right) },X^{\prime },X^{\prime },X\right)
\right\vert ^{2}d\left( X^{\prime },X^{\prime }\right) d\bar{S}^{\left(
T\right) }
\end{eqnarray*}%
Using the normalization over $\hat{S}^{\left( T\right) }$, we find the
Lagrange multipliers:%
\begin{equation*}
\lambda \left( X\right) =\frac{1-\underset{\eta \in \left\{ 1,2,3,5,\right\} 
}{\sum }\int \bar{w}_{\eta }\left( X^{\prime },X^{\prime },X\right) \bar{V}%
_{\eta }\left( X^{\prime },X^{\prime },X\right) \left\vert \bar{\Gamma}%
\left( X^{\prime },X^{\prime },X\right) \right\vert ^{2}d\left( X^{\prime
},X^{\prime }\right) }{\underset{\eta \in \left\{ 1,2,3,5,\right\} }{\sum }%
\int \bar{w}_{\eta }\left( X^{\prime },X^{\prime },X\right) \bar{V}_{\eta
}\left( X^{\prime },X^{\prime },X\right) \left\vert \bar{\Gamma}\left(
X^{\prime },X^{\prime },X\right) \right\vert ^{2}d\left( X^{\prime
},X^{\prime }\right) }
\end{equation*}%
\begin{equation*}
\lambda ^{\prime }\left( X\right) =\frac{\kappa \left( 1-\bar{S}%
_{L}^{B}\left( X\right) \right) -\sum_{\eta \in \left\{ 4,6\right\} }\int 
\bar{w}_{\eta }\left( X^{\prime },X^{\prime },X\right) \bar{V}_{\eta }\left(
X^{\prime },X^{\prime },X\right) \left\vert \bar{\Gamma}\left( X^{\prime
},X^{\prime },X\right) \right\vert ^{2}d\left( X^{\prime },X^{\prime
}\right) d\bar{S}^{\left( T\right) }}{\sum_{\eta \in \left\{ 4,6\right\}
}\int \bar{w}_{\eta }\left( X^{\prime },X^{\prime },X\right) \left\vert \bar{%
\Gamma}\left( X^{\prime },X^{\prime },X\right) \right\vert ^{2}d\left(
X^{\prime },X^{\prime }\right) d\bar{S}^{\left( T\right) }}
\end{equation*}%
In the sequel we consider that the distribution $\left\vert \Gamma \left(
X^{\prime },X\right) \right\vert ^{2}$ varies slowly, so that, using the
explicit form for $\hat{V}_{\eta }\left( X^{\prime },X\right) $ leads to the
expanded form of $\lambda \left( X\right) $ and $\lambda ^{\prime }\left(
X\right) $:%
\begin{equation}
\lambda \left( X\right) =\frac{1-\int \bar{w}_{E}\left( X^{\prime },X\right) 
\bar{f}\left( X^{\prime }\right) -\int \bar{w}_{L}\left( X^{\prime
},X\right) \bar{r}\left( X^{\prime }\right) -\int \hat{w}_{E}^{B}\left(
X^{\prime },X\right) \hat{f}\left( X^{\prime }\right) -w_{E}^{B}\left(
X,X\right) f\left( X\right) }{\bar{w}_{E}\left( X\right) +\bar{w}_{E}\left(
X\right) +\hat{w}_{E}^{B}\left( X\right) +w_{E}^{B}\left( X\right) }
\label{LMN}
\end{equation}%
and:%
\begin{equation}
\lambda ^{\prime }\left( X\right) =\frac{\kappa \left( 1-\bar{S}%
_{L}^{B}\left( X\right) \right) -\int \hat{w}_{L}\left( X^{\prime },X\right) 
\hat{r}\left( X^{\prime }\right) -w_{L}\left( X,X\right) \bar{r}}{\hat{w}%
_{L}\left( X\right) +w_{L}\left( X\right) }
\end{equation}%
In the sequel, we will replace:%
\begin{equation*}
\overline{\bar{S}_{\eta }^{\left( T\right) }}\left( X^{\prime },X^{\prime
},X\right) \rightarrow \bar{S}_{\eta }^{\left( T\right) }\left( X^{\prime
},X^{\prime },X\right)
\end{equation*}%
Ultimately, remark that, as for investors, the same saddle point can be
obtained by considering the equivalent potential:%
\begin{eqnarray*}
&&V\left( \Gamma \right) \\
&\rightarrow &\int \left( \hat{S}_{E}\hat{f}\left( X^{\prime }\right) +\hat{S%
}_{L}\hat{r}\left( X^{\prime }\right) -\frac{1}{2}\frac{\left( \hat{S}_{\eta
}\right) ^{2}\left\vert \Gamma \left( \bar{S}^{\left( T\right) },X^{\prime
},X\right) \right\vert ^{2}}{\hat{w}_{\eta }\left( X^{\prime },X\right) }%
\right. \\
&&\left. +S_{E}f\left( X\right) +S_{L}\bar{r}\left( X\right) -\frac{1}{2}%
\frac{\left( S_{\eta }\right) ^{2}\left\vert \Gamma \left( \bar{S}^{\left(
T\right) },X^{\prime },X\right) \right\vert ^{2}}{w_{\eta }\left( X\right) }%
\right) \left\vert \Gamma \left( \bar{S}^{\left( T\right) },X^{\prime
},X\right) \right\vert ^{2} \\
&&+\int \left( \bar{S}_{E}\hat{f}\left( X^{\prime }\right) +\bar{S}_{L}\hat{r%
}\left( X^{\prime }\right) -\frac{1}{2}\frac{\left( \bar{S}_{\eta }\right)
^{2}\left\vert \bar{\Gamma}\left( \bar{S}^{\left( T\right) },X^{\prime
},X^{\prime },X\right) \right\vert ^{2}}{\hat{w}_{\eta }\left( X^{\prime
},X\right) }\right. \\
&&+\hat{S}_{E}^{B}\hat{f}\left( X^{\prime }\right) +\hat{S}_{L}^{B}\hat{r}%
\left( X^{\prime }\right) -\frac{1}{2}\frac{\left( \hat{S}_{\eta
}^{B}\right) ^{2}\left\vert \Gamma \left( \bar{S}^{\left( T\right)
},X^{\prime },X\right) \right\vert ^{2}}{\hat{w}_{\eta }^{B}\left( X^{\prime
},X\right) } \\
&&\left. +S_{E}^{B}f\left( X\right) +S_{L}^{B}\bar{r}\left( X\right) -\frac{1%
}{2}\frac{\left( S_{\eta }^{B}\right) ^{2}\left\vert \Gamma \left( \bar{S}%
^{\left( T\right) },X^{\prime },X\right) \right\vert ^{2}}{w_{\eta
}^{B}\left( X\right) }\right) \left\vert \bar{\Gamma}\left( \bar{S}^{\left(
T\right) },X^{\prime },X^{\prime },X\right) \right\vert ^{2}
\end{eqnarray*}

\section*{Appendix 4 Saddle point}

\subsection*{A4.1 Formula for shares between two sectors}

For investors' field, the saddle point equations are the same as in the
first part. For banks, they write:

\begin{eqnarray*}
\bar{S}_{E}\left( X^{\prime },X\right) &=&\bar{w}_{E}\left( X^{\prime
},X\right) \left( \bar{f}\left( X^{\prime }\right) +\lambda \left( X\right)
\right) \\
\bar{S}_{L}\left( X^{\prime },X\right) &=&\bar{w}_{L}\left( X^{\prime
},X\right) \left( \bar{r}\left( X^{\prime }\right) +\lambda \left( X\right)
\right)
\end{eqnarray*}%
\begin{eqnarray*}
\hat{S}_{E}^{B}\left( X^{\prime },X\right) &=&\hat{w}_{E}^{B}\left(
X^{\prime },X\right) \left( \hat{f}\left( X^{\prime }\right) +\lambda \left(
X\right) \right) \\
\hat{S}_{L}^{B}\left( X^{\prime },X\right) &=&\hat{w}_{L}^{B}\left(
X^{\prime },X\right) \left( \hat{r}\left( X^{\prime }\right) +\lambda
^{\prime }\left( X\right) \right)
\end{eqnarray*}%
\begin{eqnarray*}
S_{E}^{B}\left( X,X\right) &=&w_{E}^{B}\left( X\right) \left( f\left(
X\right) +\lambda \left( X\right) \right) \\
S_{L}^{B}\left( X,X\right) &=&w_{L}^{B}\left( X\right) \left( r\left(
X\right) +\lambda ^{\prime }\left( X\right) \right)
\end{eqnarray*}%
and the multiplers satisfy:%
\begin{eqnarray*}
&&\lambda \left( X\right) \\
&=&\frac{1-\int \bar{w}_{E}\left( X^{\prime },X\right) \hat{f}\left(
X^{\prime }\right) -\int \bar{w}_{L}\left( X^{\prime },X\right) \hat{r}%
\left( X^{\prime }\right) -\int \hat{w}_{E}^{B}\left( X^{\prime },X\right) 
\hat{f}\left( X^{\prime }\right) -w_{E}^{B}\left( X,X\right) f\left(
X\right) }{\bar{w}_{E}\left( X\right) +\bar{w}_{L}\left( X\right) +\hat{w}%
_{E}^{B}\left( X\right) +w_{E}^{B}\left( X\right) }
\end{eqnarray*}%
and:%
\begin{eqnarray*}
&&\lambda ^{\prime }\left( X\right) \\
&=&\frac{\kappa \left( 1-\bar{S}\left( X\right) \right) -\int \hat{w}%
_{L}^{B}\left( X^{\prime },X\right) \hat{r}\left( X^{\prime }\right)
-w_{L}^{B}\left( X,X\right) \bar{r}}{\hat{w}_{L}^{B}\left( X\right)
+w_{L}^{B}\left( X\right) }
\end{eqnarray*}%
we define the average coefficients:%
\begin{equation*}
\bar{w}_{\eta }\left( X\right) =\int \bar{w}_{\eta }\left( X^{\prime
},X\right)
\end{equation*}%
and:%
\begin{equation*}
\hat{w}_{\eta }^{B}\left( X\right) =\int \hat{w}_{\eta }^{B}\left( X^{\prime
},X\right)
\end{equation*}%
so that $\bar{S}_{E}\left( X^{\prime },X\right) $ and $\bar{S}_{L}\left(
X^{\prime },X\right) $ are given by:

\begin{eqnarray*}
&&\bar{S}_{E}\left( X^{\prime },X\right) \\
&=&\bar{S}_{E}\left( X^{\prime },X\right) +\frac{\bar{w}_{E}\left( X^{\prime
},X\right) }{\bar{w}_{E}\left( X\right) +\bar{w}_{L}\left( X\right) +\hat{w}%
_{E}^{B}\left( X\right) +w_{E}^{B}\left( X\right) } \\
&&\times \left\{ \bar{w}_{E}\left( X\right) \left( \bar{f}\left( X^{\prime
}\right) -\left\langle \bar{f}\left( X^{\prime }\right) \right\rangle _{\bar{%
w}_{E}}\right) +\bar{w}_{L}\left( X\right) \left( \bar{f}\left( X^{\prime
}\right) -\left\langle \bar{r}\left( X^{\prime }\right) \right\rangle _{\bar{%
w}_{L}}\right) \right. \\
&&\left. +\hat{w}_{E}^{B}\left( X\right) \left( \bar{f}\left( X^{\prime
}\right) -\left\langle \hat{f}\left( X^{\prime }\right) \right\rangle _{\hat{%
w}_{E}}\right) +w_{E}^{B}\left( X\right) \left( \bar{f}\left( X^{\prime
}\right) -f\left( X\right) \right) \right\}
\end{eqnarray*}%
\begin{eqnarray*}
&&\bar{S}_{L}\left( X^{\prime },X\right) \\
&=&\bar{S}_{L}\left( X^{\prime },X\right) +\frac{\bar{w}_{L}\left( X^{\prime
},X\right) }{\bar{w}_{E}\left( X\right) +\bar{w}_{L}\left( X\right) +\hat{w}%
_{E}^{B}\left( X\right) +w_{E}^{B}\left( X\right) } \\
&&\times \left\{ \bar{w}_{E}\left( X\right) \left( \bar{r}\left( X^{\prime
}\right) -\left\langle \bar{f}\left( X^{\prime }\right) \right\rangle _{\bar{%
w}_{E}}\right) +\bar{w}_{L}\left( X\right) \left( \bar{r}\left( X^{\prime
}\right) -\left\langle \bar{r}\left( X^{\prime }\right) \right\rangle _{\bar{%
w}_{L}}\right) \right. \\
&&\left. +\hat{w}_{E}^{B}\left( X\right) \left( \bar{r}\left( X^{\prime
}\right) -\left\langle \hat{f}\left( X^{\prime }\right) \right\rangle _{\hat{%
w}_{E}^{B}}\right) +w_{E}^{B}\left( X\right) \left( \bar{r}\left( X^{\prime
}\right) -f\left( X\right) \right) \right\}
\end{eqnarray*}%
Defining the weighted averages of a function by:%
\begin{equation*}
\left\langle F\left( X^{\prime }\right) \right\rangle _{\bar{w}_{\eta }}=%
\frac{\int F\left( X^{\prime }\right) \bar{w}_{\eta }\left( X^{\prime
},X\right) }{\bar{w}_{\eta }\left( X\right) }
\end{equation*}%
\begin{equation*}
\left\langle F\left( X^{\prime }\right) \right\rangle _{\hat{w}_{\eta }}=%
\frac{\int F\left( X^{\prime }\right) \hat{w}_{\eta }^{B}\left( X^{\prime
},X\right) }{\hat{w}_{\eta }^{B}\left( X\right) }
\end{equation*}%
for any functions $F\left( X^{\prime }\right) $, $F\left( X^{\prime }\right) 
$ and:%
\begin{equation*}
\underline{\bar{S}}_{\eta }\left( X^{\prime },X\right) =\frac{\bar{w}_{\eta
}\left( X^{\prime },X\right) }{\bar{w}_{E}\left( X\right) +\bar{w}_{L}\left(
X\right) +\hat{w}_{E}^{B}\left( X\right) +w_{E}^{B}\left( X\right) }
\end{equation*}%
the shares $\hat{S}_{E}^{B}\left( X^{\prime },X\right) $ and $\hat{S}%
_{L}^{B}\left( X^{\prime },X\right) $ are given by:

\begin{eqnarray*}
&&\hat{S}_{E}^{B}\left( X^{\prime },X\right) \\
&=&\underline{\hat{S}}_{E}^{B}\left( X^{\prime },X\right) +\frac{\hat{w}%
_{E}^{B}\left( X^{\prime },X\right) }{\bar{w}_{E}\left( X\right) +\bar{w}%
_{L}\left( X\right) +\hat{w}_{E}^{B}\left( X\right) +w_{E}^{B}\left(
X\right) } \\
&&\times \left\{ \bar{w}_{E}\left( X\right) \left( \hat{f}\left( X^{\prime
}\right) -\left\langle \bar{f}\left( X^{\prime }\right) \right\rangle _{\bar{%
w}_{E}}\right) +\bar{w}_{L}\left( X\right) \left( \hat{f}\left( X^{\prime
}\right) -\left\langle \bar{r}\left( X^{\prime }\right) \right\rangle _{\bar{%
w}_{L}}\right) \right. \\
&&\left. +\hat{w}_{E}^{B}\left( X\right) \left( \hat{f}\left( X^{\prime
}\right) -\left\langle \hat{f}\left( X^{\prime }\right) \right\rangle _{\hat{%
w}_{E}}\right) +w_{E}^{B}\left( X\right) \left( \hat{f}\left( X^{\prime
}\right) -f\left( X\right) \right) \right\}
\end{eqnarray*}%
and:%
\begin{eqnarray*}
&&\frac{\hat{S}_{L}^{B}\left( X^{\prime },X\right) }{\kappa \left( 1-\bar{S}%
\left( X\right) \right) }=\underline{\hat{S}}_{L}^{B}\left( X^{\prime
},X\right) \\
&&+\frac{\hat{w}_{L}^{B}\left( X^{\prime },X\right) }{\hat{w}_{L}^{B}\left(
X\right) +w_{L}^{B}\left( X\right) }\left[ \hat{w}_{L}^{B}\left( X\right)
\left( \hat{r}\left( X^{\prime }\right) -\left\langle \hat{r}\left(
X^{\prime }\right) \right\rangle _{\hat{w}_{E}}\right) +w_{L}^{B}\left(
X\right) \left( \hat{r}\left( X^{\prime }\right) -\left\langle r\left(
X^{\prime }\right) \right\rangle \right) \right] \\
&\simeq &\underline{\hat{S}}_{L}^{B}\left( X^{\prime },X\right) +\frac{\hat{w%
}_{L}^{B}\left( X^{\prime },X\right) }{\hat{w}_{L}^{B}\left( X\right)
+w_{L}^{B}\left( X\right) }w_{L}^{B}\left( X\right) \left( \hat{r}\left(
X^{\prime }\right) -\left\langle r\left( X^{\prime }\right) \right\rangle
\right)
\end{eqnarray*}%
where:%
\begin{equation*}
\underline{\hat{S}}_{E}\left( X^{\prime },X\right) =\frac{\hat{w}%
_{E}^{B}\left( X^{\prime },X\right) }{\bar{w}_{E}\left( X\right) +\bar{w}%
_{L}\left( X\right) +\hat{w}_{E}^{B}\left( X\right) +w_{E}^{B}\left(
X\right) }
\end{equation*}%
\begin{equation*}
\underline{\hat{S}}_{L}^{B}\left( X^{\prime },X\right) =\frac{\hat{w}%
_{L}^{B}\left( X^{\prime },X\right) }{\hat{w}_{L}^{B}\left( X\right)
+w_{L}^{B}\left( X\right) }
\end{equation*}%
The invested shares in firms are given by:%
\begin{eqnarray*}
&&S_{E}^{B}\left( X,X\right) \\
&=&\underline{S}_{E}^{B}\left( X,X\right) +\frac{w_{E}^{B}\left( X\right) }{%
\bar{w}_{E}\left( X\right) +\bar{w}_{L}\left( X\right) +\hat{w}%
_{E}^{B}\left( X\right) +w_{E}^{B}\left( X\right) } \\
&&\times \left\{ \bar{w}_{E}\left( X\right) \left( f\left( X\right)
-\left\langle \bar{f}\left( X^{\prime }\right) \right\rangle _{\bar{w}%
_{E}}\right) +\bar{w}_{L}\left( X\right) \left( f\left( X\right)
-\left\langle \bar{r}\left( X^{\prime }\right) \right\rangle _{\bar{w}%
_{L}}\right) \right. \\
&&\left. +\hat{w}_{E}^{B}\left( X\right) \left( f\left( X\right)
-\left\langle \hat{f}\left( X^{\prime }\right) \right\rangle _{\hat{w}%
_{E}}\right) \right\}
\end{eqnarray*}%
\begin{eqnarray*}
&&\frac{S_{L}^{B}\left( X,X\right) }{\kappa \left( 1-\bar{S}\left( X\right)
\right) } \\
&=&\underline{S}_{L}^{B}\left( X,X\right) +\frac{w_{L}^{B}\left( X\right) }{%
\hat{w}_{L}^{B}\left( X\right) +w_{L}^{B}\left( X\right) }\left( \hat{w}%
_{L}^{B}\left( X\right) \left( r\left( X\right) -\left\langle \hat{r}\left(
X^{\prime }\right) \right\rangle _{\hat{w}_{L}}\right) +w_{L}^{B}\left(
X\right) \left( r\left( X\right) -\left\langle r\left( X\right)
\right\rangle \right) \right) \\
&\simeq &\underline{S}_{L}^{B}\left( X,X\right) +\frac{w_{L}^{B}\left(
X\right) }{\hat{w}_{L}^{B}\left( X\right) +w_{L}^{B}\left( X\right) }\hat{w}%
_{L}^{B}\left( X\right) \left( r\left( X\right) -\left\langle \hat{r}\left(
X^{\prime }\right) \right\rangle _{\hat{w}_{L}}\right)
\end{eqnarray*}%
with:%
\begin{equation*}
\underline{S}_{E}^{B}\left( X,X\right) =\frac{w_{E}^{B}\left( X,X\right) }{%
\bar{w}_{E}\left( X\right) +\bar{w}_{L}\left( X\right) +\hat{w}%
_{E}^{B}\left( X\right) +w_{E}^{B}\left( X\right) }
\end{equation*}%
\begin{equation*}
\underline{S}_{L}^{B}\left( X,X\right) =\frac{w_{L}^{B}\left( X,X\right) }{%
\hat{w}_{L}^{B}\left( X\right) +w_{L}^{B}\left( X\right) }
\end{equation*}

The coefficients $\hat{w}_{\alpha }$, $w_{\alpha }$ are endogeneous since
they depend on the uncertainty on returns depending themselves on the $%
\underline{\hat{S}}_{\eta }$ $S_{\eta }$. Solving will be done by detailling
this uncertainties. Before doing so, we simplify slightly by imposing
several assumptions.

First, we normalize: 
\begin{equation*}
\bar{w}_{E}\left( X\right) +\bar{w}_{L}\left( X\right) +\hat{w}%
_{E}^{B}\left( X\right) +w_{E}^{B}\left( X\right) =1
\end{equation*}%
\begin{equation*}
\hat{w}_{L}^{B}\left( X\right) +w_{L}^{B}\left( X\right) =1
\end{equation*}%
and choose for the sake of simplicity:%
\begin{equation*}
\bar{w}_{E}\left( X^{\prime },X\right) =\bar{w}_{L}\left( X^{\prime
},X\right) =\frac{1}{2}\bar{w}\left( X^{\prime },X\right)
\end{equation*}%
which describes that the perceived uncertainty for participations and loans
are equal. This implies that:%
\begin{equation*}
\underline{\bar{S}}_{E}\left( X^{\prime },X\right) =\underline{\bar{S}}%
_{L}\left( X^{\prime },X\right) =\frac{\underline{\bar{S}}\left( X^{\prime
},X\right) }{2}=\frac{1}{2}\bar{w}\left( X^{\prime },X\right)
\end{equation*}%
Then defining: 
\begin{eqnarray*}
\bar{S}_{\eta }\left( X^{\prime }\right) &=&\int \bar{S}_{\eta }\left(
X^{\prime },X\right) \frac{\bar{K}_{X}\left\vert \bar{\Psi}\left( X\right)
\right\vert ^{2}}{\bar{K}_{X^{\prime }}\left\vert \bar{\Psi}\left( X^{\prime
}\right) \right\vert ^{2}}dX \\
&\simeq &\int \bar{S}_{\eta }\left( X^{\prime },X\right) dX\frac{%
\left\langle \bar{K}\right\rangle \left\Vert \bar{\Psi}\right\Vert ^{2}}{%
\bar{K}_{X^{\prime }}\left\vert \bar{\Psi}\left( X^{\prime }\right)
\right\vert ^{2}}
\end{eqnarray*}%
\begin{equation*}
\bar{w}\left( X^{\prime }\right) =\int \bar{w}\left( X^{\prime },X\right) dX
\end{equation*}%
\begin{eqnarray*}
\hat{w}\left( X^{\prime }\right) &=&\int \hat{w}\left( X^{\prime },X\right)
dX \\
\hat{w} &=&\int \hat{w}\left( X\right) dX
\end{eqnarray*}%
the shares rewrite as given in the text.

\subsection*{A4.2 Half average shares}

For the resulution, we need to define half averaged shares:%
\begin{equation*}
\bar{S}_{\eta }\left( X^{\prime }\right) =\int \bar{S}_{\eta }\left(
X^{\prime },X\right) \frac{\bar{K}_{X}\left\vert \bar{\Psi}\left( X\right)
\right\vert ^{2}}{\bar{K}_{X^{\prime }}\left\vert \bar{\Psi}\left( X^{\prime
}\right) \right\vert ^{2}}dX
\end{equation*}%
They satisfy the following formula:%
\begin{eqnarray*}
&&\bar{S}_{E}\left( X^{\prime }\right) \\
&=&\frac{\bar{w}\left( X^{\prime }\right) }{2}\left( 1+\left\{ \left\langle 
\bar{w}\left( X\right) \right\rangle \left( \bar{f}\left( X^{\prime }\right)
-\frac{\left\langle \bar{f}\left( X^{\prime }\right) \right\rangle _{\bar{w}%
_{E}}+\left\langle \bar{r}\left( X^{\prime }\right) \right\rangle _{\bar{w}%
_{L}}}{2}\right) \right. \right. \\
&&\left. \left. +\left\langle \hat{w}_{E}^{B}\left( X\right) \right\rangle
\left( \bar{f}\left( X^{\prime }\right) -\left\langle \hat{f}\left(
X^{\prime }\right) \right\rangle _{\hat{w}_{E}}\right) +\left\langle
w_{E}^{B}\left( X\right) \right\rangle \left( \bar{f}\left( X^{\prime
}\right) -\left\langle f\left( X\right) \right\rangle \right) \right\}
\right) \frac{\left\langle \bar{K}\right\rangle \left\Vert \bar{\Psi}%
\right\Vert ^{2}}{\bar{K}_{X^{\prime }}\left\vert \bar{\Psi}\left( X^{\prime
}\right) \right\vert ^{2}}
\end{eqnarray*}%
\begin{eqnarray*}
&&\bar{S}_{L}\left( X^{\prime }\right) \\
&=&\frac{\bar{w}\left( X^{\prime }\right) }{2}\left( 1+\left\{ \left\langle 
\bar{w}\left( X\right) \right\rangle \left( \bar{r}\left( X^{\prime }\right)
-\frac{\left\langle \bar{f}\left( X^{\prime }\right) \right\rangle _{\bar{w}%
_{E}}+\left\langle \bar{r}\left( X^{\prime }\right) \right\rangle _{\bar{w}%
_{L}}}{2}\right) \right. \right. \\
&&\left. \left. +\left\langle \hat{w}_{E}^{B}\left( X\right) \right\rangle
\left( \bar{r}\left( X^{\prime }\right) -\left\langle \hat{f}\left(
X^{\prime }\right) \right\rangle _{\hat{w}_{E}^{B}}\right) +\left\langle
w_{E}^{B}\left( X\right) \right\rangle \left( \bar{r}\left( X^{\prime
}\right) -\left\langle f\left( X\right) \right\rangle \right) \right\}
\right) \frac{\left\langle \bar{K}\right\rangle \left\Vert \bar{\Psi}%
\right\Vert ^{2}}{\bar{K}_{X^{\prime }}\left\vert \bar{\Psi}\left( X^{\prime
}\right) \right\vert ^{2}}
\end{eqnarray*}%
\begin{eqnarray*}
\bar{S}\left( X^{\prime }\right) &=&\bar{S}_{E}\left( X^{\prime }\right) +%
\bar{S}_{L}\left( X^{\prime }\right) \\
&=&\bar{w}\left( X^{\prime }\right) \left[ 1+\left\{ \left\langle \bar{w}%
\left( X\right) \right\rangle \left( \frac{\bar{f}\left( X^{\prime }\right) +%
\bar{r}\left( X^{\prime }\right) }{2}-\frac{\left\langle \bar{f}\left(
X^{\prime }\right) \right\rangle _{\bar{w}_{E}}+\left\langle \bar{r}\left(
X^{\prime }\right) \right\rangle _{\bar{w}_{L}}}{2}\right) \right. \right. \\
&&\left. \left. +\left\langle \hat{w}_{E}^{B}\left( X\right) \right\rangle
\left( \frac{\bar{f}\left( X^{\prime }\right) +\bar{r}\left( X^{\prime
}\right) }{2}-\left\langle \hat{f}\left( X^{\prime }\right) \right\rangle _{%
\hat{w}_{E}}\right) +\left\langle w_{E}^{B}\left( X\right) \right\rangle
\left( \frac{\bar{f}\left( X^{\prime }\right) +\bar{r}\left( X^{\prime
}\right) }{2}-\left\langle f\left( X\right) \right\rangle \right) \right\} %
\right] \frac{\left\langle \bar{K}\right\rangle \left\Vert \bar{\Psi}%
\right\Vert ^{2}}{\bar{K}_{X^{\prime }}\left\vert \bar{\Psi}\left( X^{\prime
}\right) \right\vert ^{2}}
\end{eqnarray*}%
with the partial averages:%
\begin{equation*}
\bar{w}_{\eta }\left( X\right) =\int \bar{w}_{\eta }\left( X^{\prime
},X\right) \text{, }\bar{w}\left( X^{\prime }\right) =\int \bar{w}\left(
X^{\prime },X\right) dX
\end{equation*}%
\begin{equation*}
\hat{w}_{\eta }^{B}\left( X\right) =\int \hat{w}_{\eta }^{B}\left( X^{\prime
},X\right) \text{, }\hat{w}\left( X^{\prime }\right) =\int \hat{w}\left(
X^{\prime },X\right) dX
\end{equation*}%
and the full average:%
\begin{equation*}
\hat{w}=\int \hat{w}\left( X\right) dX
\end{equation*}%
We will need also the coeficients $\hat{S}_{\eta }\left( X^{\prime }\right) $%
:%
\begin{equation*}
\hat{S}_{\eta }^{B}\left( X^{\prime }\right) =\int \hat{S}_{\eta }^{B}\left(
X^{\prime },X\right) \frac{\bar{K}_{X}\left\vert \bar{\Psi}\left( X\right)
\right\vert ^{2}}{\hat{K}_{X^{\prime }}\left\vert \hat{\Psi}\left( X^{\prime
}\right) \right\vert ^{2}}dX
\end{equation*}%
\begin{eqnarray*}
&&\hat{S}_{E}^{B}\left( X^{\prime }\right) \\
&=&\hat{w}_{E}^{B}\left( X^{\prime }\right) \left[ 1+\left\langle \bar{w}%
\left( X\right) \right\rangle \left( \hat{f}\left( X^{\prime }\right) -\frac{%
\left\langle \bar{f}\left( X^{\prime }\right) \right\rangle _{\bar{w}%
_{E}}+\left\langle \bar{r}\left( X^{\prime }\right) \right\rangle _{\bar{w}%
_{L}}}{2}\right) \right. \\
&&\left. +\left\langle \hat{w}_{E}^{B}\left( X\right) \right\rangle \left( 
\hat{f}\left( X^{\prime }\right) -\left\langle \hat{f}\left( X^{\prime
}\right) \right\rangle _{\hat{w}_{E}}\right) +\left\langle w_{E}^{B}\left(
X\right) \right\rangle \left( \hat{f}\left( X^{\prime }\right) -\left\langle
f\left( X\right) \right\rangle \right) \right] \frac{\left\langle \bar{K}%
\right\rangle \left\Vert \bar{\Psi}\right\Vert ^{2}}{\hat{K}_{X^{\prime
}}\left\vert \hat{\Psi}\left( X^{\prime }\right) \right\vert ^{2}}
\end{eqnarray*}%
\begin{eqnarray*}
&&\frac{\hat{S}_{L}^{B}\left( X^{\prime }\right) }{\kappa \left(
1-\left\langle \bar{S}\left( X\right) \right\rangle \right) }\hat{w}%
_{L}^{B}\left( X\right) \left( r\left( X\right) -\left\langle \hat{r}\left(
X^{\prime }\right) \right\rangle _{\hat{w}_{L}}\right) w_{L}^{B}\left(
X\right) \left( \hat{r}\left( X^{\prime }\right) -\left\langle r\left(
X^{\prime }\right) \right\rangle \right) \\
&=&\hat{w}_{L}^{B}\left( X^{\prime }\right) \left\{ 1+\left\langle
w_{L}^{B}\left( X\right) \right\rangle \left( \hat{r}\left( X^{\prime
}\right) -\left\langle r\left( X^{\prime }\right) \right\rangle \right)
\right\} \frac{\left\langle \bar{K}\right\rangle \left\Vert \bar{\Psi}%
\right\Vert ^{2}}{\hat{K}_{X^{\prime }}\left\vert \hat{\Psi}\left( X^{\prime
}\right) \right\vert ^{2}}
\end{eqnarray*}%
and the coefficient $S_{\eta }^{B}\left( X\right) $: 
\begin{equation*}
S_{\eta }^{B}\left( X\right) =S_{\eta }^{B}\left( X,X\right) \frac{\bar{K}%
_{X}\left\vert \bar{\Psi}\left( X\right) \right\vert ^{2}}{K_{X}\left\vert
\Psi \left( X\right) \right\vert ^{2}}
\end{equation*}%
\begin{equation*}
S^{B}\left( X\right) =\left( S_{E}^{B}\left( X,X\right) +S_{L}^{B}\left(
X,X\right) \right) \frac{\bar{K}_{X}\left\vert \bar{\Psi}\left( X\right)
\right\vert ^{2}}{K_{X}\left\vert \Psi \left( X\right) \right\vert ^{2}}%
=S^{B}\left( X,X\right) \frac{\bar{K}_{X}\left\vert \bar{\Psi}\left(
X\right) \right\vert ^{2}}{K_{X}\left\vert \Psi \left( X\right) \right\vert
^{2}}
\end{equation*}%
\begin{eqnarray*}
&&\frac{S_{L}^{B}\left( X\right) }{\kappa \left( 1-\left\langle \bar{S}%
\left( X\right) \right\rangle \right) } \\
&=&w_{L}^{B}\left( X^{\prime }\right) \left\{ 1+\hat{w}_{L}^{B}\left(
X\right) \left( r\left( X\right) -\left\langle \hat{r}\left( X^{\prime
}\right) \right\rangle _{\hat{w}_{L}}\right) \right\} \frac{\left\langle 
\bar{K}\right\rangle \left\Vert \bar{\Psi}\right\Vert ^{2}}{K_{X}\left\vert
\Psi \left( X\right) \right\vert ^{2}}
\end{eqnarray*}

\section*{Appendix 5 Uncertainty and coefficents}

Once the equations for the shares have been found, we can complete the
system by deriving formulas for uncertainty. For investors, formula are
similar to part 1. For banks, the derivation is the following.

\subsection*{A5.1 Form of uncertainty}

The principle is the same as for investors. The form of uncertainty is
derived by considering the return equations (\ref{Rb}) without default. We
consider participations only and use the return equation:%
\begin{eqnarray}
0 &=&\left( \delta \left( X-X^{\prime }\right) -\bar{S}_{E}\left( X^{\prime
},X\right) \right) \frac{1-\bar{S}\left( X^{\prime }\right) }{1-\bar{S}%
_{E}\left( X^{\prime }\right) }\left( \bar{f}\left( X^{\prime }\right)
-\left( 1+\kappa \right) \bar{r}\left( X^{\prime }\right) \right) \\
&&-\hat{S}_{E}^{B}\left( X^{\prime },X\right) \frac{1-\left( \hat{S}\left(
X^{\prime }\right) +\hat{S}_{E}^{B}\left( X^{\prime }\right) +\hat{S}%
_{L}^{B}\left( X^{\prime }\right) \right) }{1-\left( \hat{S}_{E}\left(
X^{\prime }\right) +\hat{S}_{E}^{B}\left( X^{\prime }\right) \right) }\left( 
\hat{f}\left( X^{\prime }\right) -\bar{r}\right)  \notag \\
&&-S_{E}^{B}\left( X,X\right) \left\{ \frac{1-\left( S\left( X^{\prime
}\right) +\left( S_{E}^{B}\left( X^{\prime }\right) +S_{L}^{B}\left(
X^{\prime }\right) \right) \right) }{1-S_{E}\left( X^{\prime }\right)
-S_{E}^{B}\left( X^{\prime }\right) }\left( \left( f_{1}^{\prime }\left(
X^{\prime }\right) -\bar{r}\right) +\Delta F_{\tau }\left( \bar{R}\left(
K,X\right) \right) \right) \right\}  \notag
\end{eqnarray}%
Expanding this equation in series leads to:%
\begin{eqnarray*}
&&\bar{f}\left( X^{\prime }\right) -\left( 1+\kappa \right) \bar{r}\left(
X^{\prime }\right) \\
&=&\frac{1-\bar{S}_{E}\left( X^{\prime }\right) }{1-\bar{S}\left( X^{\prime
}\right) }\hat{S}_{E}^{B}\left( X^{\prime },X^{\prime }\right) \frac{%
1-\left( \hat{S}\left( X^{\prime }\right) +\hat{S}_{E}^{B}\left( X^{\prime
}\right) +\hat{S}_{L}^{B}\left( X^{\prime }\right) \right) }{1-\left( \hat{S}%
_{E}\left( X^{\prime }\right) +\hat{S}_{E}^{B}\left( X^{\prime }\right)
\right) }\left( \hat{f}\left( X^{\prime }\right) -\bar{r}\right) \\
&&+\frac{1-\bar{S}_{E}\left( X^{\prime }\right) }{1-\bar{S}\left( X^{\prime
}\right) }S_{E}^{B}\left( X^{\prime },X^{\prime }\right) \left\{ \frac{%
1-\left( S\left( X^{\prime }\right) +\left( S_{E}^{B}\left( X^{\prime
}\right) +S_{L}^{B}\left( X^{\prime }\right) \right) \right) }{1-S_{E}\left(
X^{\prime }\right) -S_{E}^{B}\left( X^{\prime }\right) }\left( \left(
f_{1}^{\prime }\left( X^{\prime }\right) -\bar{r}\right) +\Delta F_{\tau
}\left( \bar{R}\left( K,X^{\prime }\right) \right) \right) \right\} \\
&&+\frac{1-\bar{S}_{E}\left( X^{\prime }\right) }{1-\bar{S}\left( X^{\prime
}\right) }\sum \bar{S}_{E}^{m}\left( \left( X^{\prime }\right) ^{\prime
},X^{\prime }\right) \\
&&\times \left[ \hat{S}_{E}^{B}\left( \left( X^{\prime }\right) ^{\prime
},\left( X^{\prime }\right) ^{\prime }\right) \frac{1-\left( \hat{S}\left(
\left( X^{\prime }\right) ^{\prime }\right) +\hat{S}_{E}^{B}\left( \left(
X^{\prime }\right) ^{\prime }\right) +\hat{S}_{L}^{B}\left( \left( X^{\prime
}\right) ^{\prime }\right) \right) }{1-\left( \hat{S}_{E}\left( \left(
X^{\prime }\right) ^{\prime }\right) +\hat{S}_{E}^{B}\left( \left( X^{\prime
}\right) ^{\prime }\right) \right) }\left( \hat{f}\left( \left( X^{\prime
}\right) ^{\prime }\right) -\bar{r}\right) \right. \\
&&\left. +S_{E}^{B}\left( \left( X^{\prime }\right) ^{\prime }\right)
\left\{ \frac{1-\left( S\left( \left( X^{\prime }\right) ^{\prime }\right)
+\left( S_{E}^{B}\left( \left( X^{\prime }\right) ^{\prime }\right)
+S_{L}^{B}\left( \left( X^{\prime }\right) ^{\prime }\right) \right) \right) 
}{1-S_{E}\left( \left( X^{\prime }\right) ^{\prime }\right) -S_{E}^{B}\left(
\left( X^{\prime }\right) ^{\prime }\right) }\left( \left( f_{1}^{\prime
}\left( \left( X^{\prime }\right) ^{\prime }\right) -\bar{r}\right) +\Delta
F_{\tau }\left( \bar{R}\left( K,\left( X^{\prime }\right) ^{\prime }\right)
\right) \right) \right\} \right]
\end{eqnarray*}%
This series shows that due to the diffusion, investing in one other
investors leads to a chain of far investments, which increases the
uncertainty, from the point of view of the investr located at $X$:

\begin{eqnarray*}
&&Un\left( X,\hat{f}\left( X^{\prime }\right) \right) \\
&=&Un\left[ \frac{1-\bar{S}_{E}\left( X^{\prime }\right) }{1-\bar{S}\left(
X^{\prime }\right) }\sum \bar{S}_{E}^{m}\left( \left( X^{\prime }\right)
^{\prime },X^{\prime }\right) \right. \\
&&\times \left[ \hat{S}_{E}^{B}\left( \left( X^{\prime }\right) ^{\prime
},\left( X^{\prime }\right) ^{\prime }\right) \frac{1-\left( \hat{S}\left(
\left( X^{\prime }\right) ^{\prime }\right) +\hat{S}_{E}^{B}\left( \left(
X^{\prime }\right) ^{\prime }\right) +\hat{S}_{L}^{B}\left( \left( X^{\prime
}\right) ^{\prime }\right) \right) }{1-\left( \hat{S}_{E}\left( \left(
X^{\prime }\right) ^{\prime }\right) +\hat{S}_{E}^{B}\left( \left( X^{\prime
}\right) ^{\prime }\right) \right) }\left( \hat{f}\left( \left( X^{\prime
}\right) ^{\prime }\right) -\bar{r}\right) \right. \\
&&\left. \left. +S_{E}^{B}\left( \left( X^{\prime }\right) ^{\prime }\right)
\left\{ \frac{1-\left( S\left( \left( X^{\prime }\right) ^{\prime }\right)
+\left( S_{E}^{B}\left( \left( X^{\prime }\right) ^{\prime }\right)
+S_{L}^{B}\left( \left( X^{\prime }\right) ^{\prime }\right) \right) \right)
\left( \left( f_{1}^{\prime }\left( \left( X^{\prime }\right) ^{\prime
}\right) -\bar{r}\right) +\Delta F_{\tau }\left( \bar{R}\left( K,\left(
X^{\prime }\right) ^{\prime }\right) \right) \right) }{1-S_{E}\left( \left(
X^{\prime }\right) ^{\prime }\right) -S_{E}^{B}\left( \left( X^{\prime
}\right) ^{\prime }\right) }\right\} \right] \right]
\end{eqnarray*}%
We consider as before that the uncertainty of the two terms is additive in
first approximation and that it is multiplicative for each term:%
\begin{eqnarray*}
&&Un\left( \bar{S}_{E}^{m}\left( \left( X^{\prime }\right) ^{\prime
},X^{\prime }\right) \left[ \hat{S}_{E}^{B}\left( \left( X^{\prime }\right)
^{\prime },\left( X^{\prime }\right) ^{\prime }\right) \frac{1-\left( \hat{S}%
\left( \left( X^{\prime }\right) ^{\prime }\right) +\hat{S}_{E}^{B}\left(
\left( X^{\prime }\right) ^{\prime }\right) +\hat{S}_{L}^{B}\left( \left(
X^{\prime }\right) ^{\prime }\right) \right) }{1-\left( \hat{S}_{1}\left(
\left( X^{\prime }\right) ^{\prime }\right) +\hat{S}_{1}^{B}\left( \left(
X^{\prime }\right) ^{\prime }\right) \right) }\left( \hat{f}\left( \left(
X^{\prime }\right) ^{\prime }\right) -\bar{r}\right) \right] \right) \\
&\rightarrow &\bar{\zeta}^{2}\left( \frac{1}{\bar{w}_{1}^{\left( 0\right)
}\left( \left( X^{\prime }\right) ^{\prime },X_{m-1}^{\prime }\right) ...%
\bar{w}_{1}^{\left( 0\right) }\left( X_{1}^{\prime },X^{\prime }\right) }%
\right) \bar{S}_{E}^{2m}\left( \left( X^{\prime }\right) ^{\prime
},X^{\prime }\right) Var\left( \hat{f}\left( \left( X^{\prime }\right)
^{\prime }\right) -\bar{r}\right)
\end{eqnarray*}%
$\bar{\zeta}^{2}$ is defined by:%
\begin{equation*}
\left\langle \frac{1-\bar{S}_{E}\left( X^{\prime }\right) }{1-\bar{S}\left(
X^{\prime }\right) }\right\rangle ^{2}\left\langle \hat{S}_{E}^{B}\left(
\left( X^{\prime }\right) ^{\prime },\left( X^{\prime }\right) ^{\prime
}\right) \right\rangle ^{2}\left\langle \frac{1-\left( \hat{S}\left( \left(
X^{\prime }\right) ^{\prime }\right) +\hat{S}_{E}^{B}\left( \left( X^{\prime
}\right) ^{\prime }\right) +\hat{S}_{L}^{B}\left( \left( X^{\prime }\right)
^{\prime }\right) \right) }{1-\left( \hat{S}_{E}\left( \left( X^{\prime
}\right) ^{\prime }\right) +\hat{S}_{E}^{B}\left( \left( X^{\prime }\right)
^{\prime }\right) \right) }\right\rangle ^{2}
\end{equation*}%
and is approximated by averages as before:

\begin{eqnarray*}
&&\bar{\zeta}^{2} \\
&\rightarrow &\left\langle \hat{S}_{E}^{B}\left( \left( X^{\prime }\right)
^{\prime },\left( X^{\prime }\right) ^{\prime }\right) \right\rangle
^{2}\left\langle \frac{1-\left( \hat{S}\left( \left( X^{\prime }\right)
^{\prime }\right) +\hat{S}_{E}^{B}\left( \left( X^{\prime }\right) ^{\prime
}\right) +\hat{S}_{L}^{B}\left( \left( X^{\prime }\right) ^{\prime }\right)
\right) }{1-\left( \hat{S}_{E}\left( \left( X^{\prime }\right) ^{\prime
}\right) +\hat{S}_{E}^{B}\left( \left( X^{\prime }\right) ^{\prime }\right)
\right) }\right\rangle ^{2}
\end{eqnarray*}%
that are considered as constant to focus rather on the diffusion of
uncertainty rather than on local uncertainties.

Similarly:%
\begin{eqnarray*}
&&Un\left( \bar{S}_{E}^{m}\left( X^{\prime },X\right) S_{E}^{B}\left( \left(
X^{\prime }\right) ^{\prime }\right) \right. \\
&&\times \left. \left\{ \frac{1-\left( S\left( \left( X^{\prime }\right)
^{\prime }\right) +\left( S_{E}^{B}\left( \left( X^{\prime }\right) ^{\prime
}\right) +S_{L}^{B}\left( \left( X^{\prime }\right) ^{\prime }\right)
\right) \right) \left( \left( f_{1}^{\prime }\left( \left( X^{\prime
}\right) ^{\prime }\right) -\bar{r}\right) +\Delta F_{\tau }\left( \bar{R}%
\left( K,\left( X^{\prime }\right) ^{\prime }\right) \right) \right) }{%
1-S_{E}\left( \left( X^{\prime }\right) ^{\prime }\right) -S_{E}^{B}\left(
\left( X^{\prime }\right) ^{\prime }\right) }\right\} \right) \\
&\rightarrow &\bar{\xi}^{2}\left( \frac{1}{\bar{w}_{1}^{\left( 0\right)
}\left( \left( X^{\prime }\right) ^{\prime },X_{m-1}^{\prime }\right) ...%
\bar{w}_{1}^{\left( 0\right) }\left( X_{1}^{\prime },X^{\prime }\right) }%
\right) \bar{S}_{E}^{2m}\left( X^{\prime },X\right)
\end{eqnarray*}%
$\bar{\xi}^{2}$ being the average variance:%
\begin{eqnarray*}
&&\left\langle S_{E}^{B}\left( \left( X^{\prime }\right) ^{\prime }\right)
\right\rangle ^{2}\left\langle \frac{1-\left( S\left( \left( X^{\prime
}\right) ^{\prime }\right) +\left( S_{E}^{B}\left( \left( X^{\prime }\right)
^{\prime }\right) +S_{L}^{B}\left( \left( X^{\prime }\right) ^{\prime
}\right) \right) \right) }{1-S_{E}\left( \left( X^{\prime }\right) ^{\prime
}\right) -S_{E}^{B}\left( \left( X^{\prime }\right) ^{\prime }\right) }%
\right\rangle ^{2} \\
&&\times Var\left( \left( f_{1}^{\prime }\left( \left( X^{\prime }\right)
^{\prime }\right) -\bar{r}\right) +\Delta F_{\tau }\left( \bar{R}\left(
K,\left( X^{\prime }\right) ^{\prime }\right) \right) \right)
\end{eqnarray*}%
We also assume that for paths that are disconnected the uncertainty is
additive, so that in first approximation:%
\begin{eqnarray*}
&&Un\left( X,\hat{f}\left( X^{\prime }\right) \right) \\
&\rightarrow &\sum \left( \bar{\zeta}^{2}Var\left( \hat{f}\left( \left(
X^{\prime }\right) ^{\prime }\right) -\bar{r}\right) +\bar{\xi}^{2}\right)
\left( \frac{1}{\bar{w}_{1}^{\left( 0\right) }\left( \left( X^{\prime
}\right) ^{\prime },X_{m-1}^{\prime }\right) ...\bar{w}_{1}^{\left( 0\right)
}\left( X_{1}^{\prime },X^{\prime }\right) }\right) \bar{S}_{E}^{2m}\left(
X^{\prime },X\right)
\end{eqnarray*}%
As before, this implies that connections are organized mainly in limited
zone of independent elements: if one element of some connected group
connects to an other group, it increases the uncertainty of the all groups.
As a consequence, it leads to an organization in several regions, or groups,
with limited connections between different groups. Due to inertia in stakes
dynamics, there are many possibilities of organized connexions between
investors, each of them presenting some persistence through time.

\subsection*{A5.2 Precise form of uncertainty and computation of $\hat{w}$}

The uncertainty can be evaluated more precisely. We consider that $\zeta
^{2} $ is the uncertainty for:$\ $%
\begin{equation*}
S_{E}\left( X^{\prime }\right) \frac{\left( 1-S\left( X^{\prime }\right)
\right) \left( f_{1}^{\prime }\left( X^{\prime }\right) -\bar{r}\right) }{%
1-S_{E}\left( X^{\prime }\right) }
\end{equation*}%
so that for choosing to take share in $\hat{f}\left( X^{\prime }\right) $ we
sum over diconnected paths:%
\begin{eqnarray}
&&\bar{f}\left( X^{\prime }\right) -\left( 1+\kappa \right) \bar{r}\left(
X^{\prime }\right)  \label{PRs} \\
&=&\frac{1-\bar{S}_{E}\left( X^{\prime }\right) }{1-\bar{S}\left( X^{\prime
}\right) }\hat{S}_{E}^{B}\left( X^{\prime },X^{\prime }\right) \frac{%
1-\left( \hat{S}\left( X^{\prime }\right) +\hat{S}_{E}^{B}\left( X^{\prime
}\right) +\hat{S}_{L}^{B}\left( X^{\prime }\right) \right) }{1-\left( \hat{S}%
_{E}\left( X^{\prime }\right) +\hat{S}_{E}^{B}\left( X^{\prime }\right)
\right) }\left( \hat{f}\left( X^{\prime }\right) -\bar{r}\right)  \notag \\
&&+\frac{1-\bar{S}_{E}\left( X^{\prime }\right) }{1-\bar{S}\left( X^{\prime
}\right) }S_{E}^{B}\left( X^{\prime },X^{\prime }\right) \left\{ \frac{%
1-\left( S\left( X^{\prime }\right) +\left( S_{E}^{B}\left( X^{\prime
}\right) +S_{L}^{B}\left( X^{\prime }\right) \right) \right) }{1-S_{E}\left(
X^{\prime }\right) -S_{E}^{B}\left( X^{\prime }\right) }\left( \left(
f_{1}^{\prime }\left( X^{\prime }\right) -\bar{r}\right) +\Delta F_{\tau
}\left( \bar{R}\left( K,X^{\prime }\right) \right) \right) \right\}  \notag
\\
&&+\frac{1-\bar{S}_{E}\left( X^{\prime }\right) }{1-\bar{S}\left( X^{\prime
}\right) }\sum \bar{S}_{E}^{m}\left( \left( X^{\prime }\right) ^{\prime
},X^{\prime }\right)  \notag \\
&&\times \left[ \hat{S}_{E}^{B}\left( \left( X^{\prime }\right) ^{\prime
},\left( X^{\prime }\right) ^{\prime }\right) \frac{1-\left( \hat{S}\left(
\left( X^{\prime }\right) ^{\prime }\right) +\hat{S}_{E}^{B}\left( \left(
X^{\prime }\right) ^{\prime }\right) +\hat{S}_{L}^{B}\left( \left( X^{\prime
}\right) ^{\prime }\right) \right) }{1-\left( \hat{S}_{E}\left( \left(
X^{\prime }\right) ^{\prime }\right) +\hat{S}_{E}^{B}\left( \left( X^{\prime
}\right) ^{\prime }\right) \right) }\left( \hat{f}\left( \left( X^{\prime
}\right) ^{\prime }\right) -\bar{r}\right) \right.  \notag \\
&&\left. +S_{E}^{B}\left( \left( X^{\prime }\right) ^{\prime }\right)
\left\{ \frac{1-\left( S\left( \left( X^{\prime }\right) ^{\prime }\right)
+\left( S_{E}^{B}\left( \left( X^{\prime }\right) ^{\prime }\right)
+S_{L}^{B}\left( \left( X^{\prime }\right) ^{\prime }\right) \right) \right) 
}{1-S_{E}\left( \left( X^{\prime }\right) ^{\prime }\right) -S_{E}^{B}\left(
\left( X^{\prime }\right) ^{\prime }\right) }\left( \left( f_{1}^{\prime
}\left( \left( X^{\prime }\right) ^{\prime }\right) -\bar{r}\right) +\Delta
F_{\tau }\left( \bar{R}\left( K,\left( X^{\prime }\right) ^{\prime }\right)
\right) \right) \right\} \right]  \notag
\end{eqnarray}%
and we assume as before that the uncertainty is multiplicative along paths
and additive for both component. Now we take into account the uncertainty
associated to:%
\begin{equation*}
\hat{f}\left( X^{\prime }\right) -\bar{r}
\end{equation*}%
so that we compute $Un\left( X,\bar{f}\left( X^{\prime }\right) \right) $:%
\begin{eqnarray}
&&Un\left[ \frac{1-\bar{S}_{E}\left( X^{\prime }\right) }{1-\bar{S}\left(
X^{\prime }\right) }\sum \bar{S}_{E}^{m}\left( \left( X^{\prime }\right)
^{\prime },X^{\prime }\right) \right.  \label{PR} \\
&&\left. \times \left[ \hat{S}_{E}^{B}\left( \left( X^{\prime }\right)
^{\prime },\left( X^{\prime }\right) ^{\prime }\right) \frac{1-\left( \hat{S}%
\left( \left( X^{\prime }\right) ^{\prime }\right) +\hat{S}_{E}^{B}\left(
\left( X^{\prime }\right) ^{\prime }\right) +\hat{S}_{L}^{B}\left( \left(
X^{\prime }\right) ^{\prime }\right) \right) }{1-\left( \hat{S}_{E}\left(
\left( X^{\prime }\right) ^{\prime }\right) +\hat{S}_{E}^{B}\left( \left(
X^{\prime }\right) ^{\prime }\right) \right) }\left( \hat{f}\left( \left(
X^{\prime }\right) ^{\prime }\right) -\bar{r}\right) \right. \right]  \notag
\\
&=&\left[ \frac{\bar{\zeta}^{2}}{\hat{w}_{1}^{\left( 0\right) B}\left(
\left( X^{\prime }\right) ^{\prime },X^{\prime }\right) }+\sum \frac{\bar{%
\zeta}^{2}}{\hat{w}_{1}^{\left( 0\right) B}\left( \left( X^{\prime }\right)
^{\prime },\left( X^{\prime }\right) ^{\prime }\right) }\left( \frac{1}{\bar{%
w}_{1}^{\left( 0\right) }\left( \left( X^{\prime }\right) ^{\prime
},X_{m-1}^{\prime }\right) ...\bar{w}_{1}^{\left( 0\right) }\left(
X_{1}^{\prime },X^{\prime }\right) }\right) \right.  \notag \\
&&\left. \times \bar{S}_{E}^{2}\left( \left( X^{\prime }\right) ^{\prime
},X_{m-1}^{\prime }\right) ...\bar{S}_{E}^{2}\left( X_{1}^{\prime
},X^{\prime }\right) \right] \times \left( \left\langle \hat{f}\left( \left(
X^{\prime }\right) ^{\prime }\right) -\bar{r}\right\rangle ^{2}+Un\left( 
\hat{f}\left( \left( X^{\prime }\right) ^{\prime }\right) -\bar{r}\right)
\right)  \notag \\
&&+\left[ \frac{1-\bar{S}_{E}\left( X^{\prime }\right) }{1-\bar{S}\left(
X^{\prime }\right) }\sum \bar{S}_{E}^{m}\left( \left( X^{\prime }\right)
^{\prime },X^{\prime }\right) \hat{S}_{E}^{B}\left( X^{\prime },X^{\prime
}\right) \hat{S}_{E}^{B}\left( \left( X^{\prime }\right) ^{\prime },\left(
X^{\prime }\right) ^{\prime }\right) \right.  \notag \\
&&\times \left. \frac{1-\left( \hat{S}\left( \left( X^{\prime }\right)
^{\prime }\right) +\hat{S}_{E}^{B}\left( \left( X^{\prime }\right) ^{\prime
}\right) +\hat{S}_{L}^{B}\left( \left( X^{\prime }\right) ^{\prime }\right)
\right) }{1-\left( \hat{S}_{E}\left( \left( X^{\prime }\right) ^{\prime
}\right) +\hat{S}_{E}^{B}\left( \left( X^{\prime }\right) ^{\prime }\right)
\right) }\right] ^{2}Un\left( \hat{f}\left( \left( X^{\prime }\right)
^{\prime }\right) -\bar{r}\right)  \notag
\end{eqnarray}%
where $\bar{\zeta}^{2}$ stands for: 
\begin{eqnarray*}
\bar{\zeta}^{2} &=&\left\langle \frac{1-\bar{S}_{E}\left( X^{\prime }\right) 
}{1-\bar{S}\left( X^{\prime }\right) }\right\rangle ^{2}\left\langle \hat{S}%
_{E}^{B}\left( \left( X^{\prime }\right) ^{\prime },\left( X^{\prime
}\right) \right) \right\rangle ^{2} \\
&&\times \left\langle \frac{1-\left( \hat{S}\left( \left( X^{\prime }\right)
^{\prime }\right) +\hat{S}_{E}^{B}\left( \left( X^{\prime }\right) ^{\prime
}\right) +\hat{S}_{L}^{B}\left( \left( X^{\prime }\right) ^{\prime }\right)
\right) }{1-\left( \hat{S}_{E}\left( \left( X^{\prime }\right) ^{\prime
}\right) +\hat{S}_{E}^{B}\left( \left( X^{\prime }\right) ^{\prime }\right)
\right) }\right\rangle ^{2}
\end{eqnarray*}%
Given the increasing character of uncertainty, we consider that:%
\begin{equation*}
\left\langle \hat{f}\left( X^{\prime }\right) -\bar{r}\right\rangle
^{2}<<Un\left( \hat{f}\left( X^{\prime }\right) -\bar{r}\right)
\end{equation*}%
and the last term is negligible compared to the composd uncertainty. As a
consequence, the uncertainty (\ref{PR}) writes:%
\begin{eqnarray}
&&\frac{1}{\bar{w}_{1}^{\left( 0\right) }\left( X^{\prime },X\right) }\left[ 
\frac{\bar{\zeta}^{2}}{\hat{w}_{1}^{\left( 0\right) B}\left( \left(
X^{\prime }\right) ^{\prime },X^{\prime }\right) }+\sum \frac{\bar{\zeta}^{2}%
}{\hat{w}_{1}^{\left( 0\right) B}\left( \left( X^{\prime }\right) ^{\prime
},\left( X^{\prime }\right) ^{\prime }\right) }\left( \frac{1}{\bar{w}%
_{1}^{\left( 0\right) }\left( \left( X^{\prime }\right) ^{\prime
},X_{m-1}^{\prime }\right) ...\bar{w}_{1}^{\left( 0\right) }\left(
X_{1}^{\prime },X^{\prime }\right) }\right) \right.  \notag \\
&&\left. \times \bar{S}_{E}^{2}\left( \left( X^{\prime }\right) ^{\prime
},X_{m-1}^{\prime }\right) ...\bar{S}_{1}^{2}\left( X_{1}^{\prime
},X^{\prime }\right) \right] \times Un\left( \hat{f}\left( X^{\prime
}\right) -\bar{r}\right)  \label{NTM} \\
&=&\frac{1}{\bar{w}_{1}^{\left( 0\right) }\left( X^{\prime },X\right) }\left[
\frac{\bar{\zeta}^{2}}{\hat{w}_{1}^{\left( 0\right) B}\left( \left(
X^{\prime }\right) ^{\prime },X^{\prime }\right) }+\sum \frac{\bar{\zeta}^{2}%
}{\hat{w}_{1}^{\left( 0\right) B}\left( \left( X^{\prime }\right) ^{\prime
},\left( X^{\prime }\right) ^{\prime }\right) }\left( \frac{1}{\bar{w}%
_{1}^{\left( 0\right) }\left( \left( X^{\prime }\right) ^{\prime
},X_{m-1}^{\prime }\right) ...\bar{w}_{1}^{\left( 0\right) }\left(
X_{1}^{\prime },X^{\prime }\right) }\right) \right.  \notag \\
&&\left. \times \bar{S}_{E}^{2}\left( \left( X^{\prime }\right) ^{\prime
},X_{m-1}^{\prime }\right) ...\bar{S}_{E}^{2}\left( X_{1}^{\prime
},X^{\prime }\right) \right] \times \frac{1}{\hat{w}_{1}^{\left( 0\right)
}\left( \left( X^{\prime }\right) ^{\prime },\left( X^{\prime }\right)
^{\prime }\right) }\left( \zeta ^{2}+\zeta ^{2}\frac{\left( \gamma
\left\langle \hat{S}_{E}\left( X_{1},\left( X^{\prime }\right) ^{\prime
}\right) \right\rangle _{X_{1}}\right) ^{2}}{1-\left( \gamma \left\langle 
\hat{S}_{E}\left( X^{\prime },\left( X^{\prime }\right) ^{\prime }\right)
\right\rangle \right) ^{2}}\right)  \notag \\
&=&\frac{1}{\bar{w}_{1}^{\left( 0\right) }\left( X^{\prime },X\right) }%
\left( \frac{\bar{\zeta}^{2}}{\hat{w}_{1}^{\left( 0\right) B}\left( \left(
X^{\prime }\right) ^{\prime },X^{\prime }\right) }+\frac{\bar{\zeta}^{2}}{%
\left\langle \hat{w}_{1}^{\left( 0\right) B}\left( \left( X^{\prime }\right)
^{\prime },X^{\prime }\right) \right\rangle _{\left( X^{\prime }\right)
^{\prime }}}\frac{\left( \bar{\gamma}\left\langle \bar{S}_{E}\left(
X_{1}^{\prime },X^{\prime }\right) \right\rangle _{X_{1}^{\prime }}\right)
^{2}}{1-\left( \bar{\gamma}\left\langle \bar{S}_{E}^{2}\left( X_{1}^{\prime
},X^{\prime }\right) \right\rangle \right) ^{2}}\right)  \notag \\
&&\times \left( \zeta ^{2}+\zeta ^{2}\frac{\left( \gamma \left\langle \hat{S}%
_{E}\left( X_{1},\left( X^{\prime }\right) ^{\prime }\right) \right\rangle
_{X_{1}}\right) ^{2}}{1-\left( \gamma \left\langle \hat{S}_{E}\left(
X^{\prime },\left( X^{\prime }\right) ^{\prime }\right) \right\rangle
\right) ^{2}}\right)  \notag
\end{eqnarray}%
with:%
\begin{eqnarray*}
&&\zeta ^{2}=Un\left( \frac{1-\hat{S}_{E}\left( X^{\prime }\right) }{1-\hat{S%
}\left( X^{\prime }\right) }S_{E}\left( X^{\prime }\right) \frac{\left(
1-S\left( X^{\prime }\right) \right) \left( f_{1}^{\prime }\left( X^{\prime
}\right) -\bar{r}\right) }{1-S_{E}\left( X^{\prime }\right) }\right) \\
&=&\left( \frac{1-\hat{S}_{E}\left( X^{\prime }\right) }{1-\hat{S}\left(
X^{\prime }\right) }S_{E}\left( X^{\prime }\right) \frac{\left( 1-S\left(
X^{\prime }\right) \right) }{1-S_{E}\left( X^{\prime }\right) }\right)
^{2}Var\left( f_{1}^{\prime }\left( X^{\prime }\right) -\bar{r}\right) \\
&\rightarrow &\left\langle \frac{1-\hat{S}_{E}\left( X^{\prime }\right) }{1-%
\hat{S}\left( X^{\prime }\right) }S_{E}\left( \left( X^{\prime }\right)
^{\prime }\right) \right\rangle ^{2}\left\langle \frac{\left( 1-S\left(
\left( X^{\prime }\right) ^{\prime }\right) \right) }{1-S_{E}\left( \left(
X^{\prime }\right) ^{\prime }\right) }\right\rangle ^{2}Var\left(
f_{1}^{\prime }\left( \left( X^{\prime }\right) ^{\prime }\right) -\bar{r}%
\right) \\
&\rightarrow &\left\langle \frac{1-\hat{S}_{E}\left( X^{\prime }\right) }{1-%
\hat{S}\left( X^{\prime }\right) }S_{E}\left( \left( X^{\prime }\right)
^{\prime }\right) \right\rangle ^{2}Var\left( \left( f_{1}\left( \left(
X^{\prime }\right) ^{\prime }\right) -\bar{r}\right) +\Delta F_{\tau }\left( 
\bar{R}\left( K,\left( X^{\prime }\right) ^{\prime }\right) \right) \right)
\end{eqnarray*}%
and:%
\begin{eqnarray*}
Un\left( \hat{f}\left( X^{\prime }\right) -\bar{r}\right) &=&Var\left( \hat{f%
}\left( \left( X^{\prime }\right) ^{\prime }\right) -\bar{r}\right) \\
&=&\zeta ^{2}+\zeta ^{2}\frac{\left( \gamma \left\langle \hat{S}_{E}\left(
X_{1},\left( X^{\prime }\right) ^{\prime }\right) \right\rangle
_{X_{1}}\right) ^{2}}{1-\left( \gamma \left\langle \hat{S}_{E}\left(
X^{\prime },\left( X^{\prime }\right) ^{\prime }\right) \right\rangle
\right) ^{2}}
\end{eqnarray*}%
Formula (\ref{NTM}) has to be summed over $\left( X^{\prime }\right)
^{\prime }$ leading to:%
\begin{equation*}
\frac{\bar{\zeta}^{2}\zeta ^{2}\left( 1+\frac{\left( \bar{\gamma}%
\left\langle \bar{S}_{E}\left( X_{1}^{\prime },X^{\prime }\right)
\right\rangle _{X_{1}^{\prime }}\right) ^{2}}{1-\left( \bar{\gamma}%
\left\langle \bar{S}_{E}^{2}\left( X_{1}^{\prime },X^{\prime }\right)
\right\rangle \right) ^{2}}\right) \left( 1+\frac{\left( \gamma \left\langle 
\hat{S}_{E}\left( X_{1},\left( X^{\prime }\right) ^{\prime }\right)
\right\rangle \right) ^{2}}{1-\left( \gamma \left\langle \hat{S}_{E}\left(
X^{\prime },\left( X^{\prime }\right) ^{\prime }\right) \right\rangle
\right) ^{2}}\right) }{\bar{w}_{1}^{\left( 0\right) }\left( X^{\prime
},X\right) \left\langle \hat{w}_{1}^{\left( 0\right) B}\left( \left(
X^{\prime }\right) ^{\prime },X^{\prime }\right) \right\rangle _{\left(
X^{\prime }\right) ^{\prime }}}
\end{equation*}%
The second uncertainty comes from the last contribution of (\ref{PRs}):%
\begin{eqnarray*}
&&Un\left\{ \frac{1-\bar{S}_{E}\left( X^{\prime }\right) }{1-\bar{S}\left(
X^{\prime }\right) }\sum \bar{S}_{E}^{m}\left( \left( X^{\prime }\right)
^{\prime },X^{\prime }\right) S_{E}^{B}\left( \left( X^{\prime }\right)
^{\prime },\left( X^{\prime }\right) ^{\prime }\right) \right. \\
&&\times \left. \left\{ \frac{1-\left( S\left( \left( X^{\prime }\right)
^{\prime }\right) +\left( S_{E}^{B}\left( \left( X^{\prime }\right) ^{\prime
}\right) +S_{L}^{B}\left( \left( X^{\prime }\right) ^{\prime }\right)
\right) \right) }{1-S_{E}\left( \left( X^{\prime }\right) ^{\prime }\right)
-S_{E}^{B}\left( \left( X^{\prime }\right) ^{\prime }\right) }\left( \left(
f_{1}^{\prime }\left( \left( X^{\prime }\right) ^{\prime }\right) -\bar{r}%
\right) +\Delta F_{\tau }\left( \bar{R}\left( K,\left( X^{\prime }\right)
^{\prime }\right) \right) \right) \right\} \right\} \\
&=&\frac{\xi ^{2}}{\bar{w}_{E}^{\left( 0\right) }\left( X^{\prime },X\right) 
} \\
&&+\sum \frac{\xi ^{2}}{\bar{w}_{E}^{\left( 0\right) }\left( X^{\prime
},X\right) }\left( \frac{1}{\bar{w}_{E}^{\left( 0\right) }\left( \left(
X^{\prime }\right) ^{\prime },X_{m-1}^{\prime }\right) ...\bar{w}%
_{E}^{\left( 0\right) }\left( X_{1}^{\prime },X^{\prime }\right) }\right) 
\bar{S}_{E}^{2}\left( \left( X^{\prime }\right) ^{\prime },X_{m-1}^{\prime
}\right) ...\bar{S}_{E}^{2}\left( X_{1}^{\prime },X^{\prime }\right) \\
&\simeq &\frac{1}{\bar{w}_{E}^{\left( 0\right) }\left( X^{\prime },X\right) }%
\left( \xi ^{2}+\left\langle \bar{S}_{E}\left( X_{1},X^{\prime }\right)
\right\rangle _{X_{1}}^{2}\sum \xi ^{2}\left( \bar{\gamma}^{2}\right)
^{m}\left\langle \bar{S}_{E}^{2}\left( \left( X^{\prime }\right) ^{\prime
},X^{\prime }\right) \right\rangle ^{m-1}\right) \\
&\simeq &\frac{\xi ^{2}}{\bar{w}_{E}^{\left( 0\right) }\left( X^{\prime
},X\right) }\left( 1+\frac{\left( \bar{\gamma}\left\langle \bar{S}_{E}\left(
X_{1},X^{\prime }\right) \right\rangle _{X_{1}}^{2}\right) ^{2}}{1-\left( 
\bar{\gamma}\left\langle \bar{S}_{E}^{2}\left( \left( X^{\prime }\right)
^{\prime },X^{\prime }\right) \right\rangle \right) ^{2}}\right)
\end{eqnarray*}%
where $\left\langle \bar{S}_{E}\left( X_{1},X^{\prime }\right) \right\rangle
_{X_{1}}^{2}$ is the average taken over $X_{1}$ and $\bar{\gamma}$
represents the average distance-dependent uncertainty 
\begin{equation*}
\bar{\gamma}^{2}\simeq \left( \frac{1}{\bar{w}_{1}^{\left( 0\right) }\left(
\left( X^{\prime }\right) ^{\prime },X_{m-1}^{\prime }\right) ...\bar{w}%
_{1}^{\left( 0\right) }\left( X_{1}^{\prime },X^{\prime }\right) }\right) ^{%
\frac{1}{m}}
\end{equation*}%
and $\xi ^{2}$ is the variance:%
\begin{eqnarray*}
&&\left\langle \frac{1-\bar{S}_{E}\left( X^{\prime }\right) }{1-\bar{S}%
\left( X^{\prime }\right) }\right\rangle ^{2}\left\langle S_{E}^{B}\left(
\left( X^{\prime }\right) ^{\prime },\left( X^{\prime }\right) ^{\prime
}\right) \right\rangle ^{2} \\
&&\left\langle \frac{1-\left( S\left( \left( X^{\prime }\right) ^{\prime
}\right) +\left( S_{E}^{B}\left( \left( X^{\prime }\right) ^{\prime }\right)
+S_{L}^{B}\left( \left( X^{\prime }\right) ^{\prime }\right) \right) \right) 
}{1-S_{E}\left( \left( X^{\prime }\right) ^{\prime }\right) -S_{E}^{B}\left(
\left( X^{\prime }\right) ^{\prime }\right) }\right\rangle ^{2}Var\left(
\left( f_{1}^{\prime }\left( \left( X^{\prime }\right) ^{\prime }\right) -%
\bar{r}\right) +\Delta F_{\tau }\left( \bar{R}\left( K,\left( X^{\prime
}\right) ^{\prime }\right) \right) \right) \\
&=&\left\langle \frac{1-\bar{S}_{E}\left( X^{\prime }\right) }{1-\bar{S}%
\left( X^{\prime }\right) }\right\rangle ^{2}\left\langle S_{E}^{B}\left(
\left( X^{\prime }\right) ^{\prime },\left( X^{\prime }\right) ^{\prime
}\right) \right\rangle ^{2}Var\left( \left( f_{1}\left( \left( X^{\prime
}\right) ^{\prime }\right) -\bar{r}\right) +\Delta F_{\tau }\left( \bar{R}%
\left( K,\left( X^{\prime }\right) ^{\prime }\right) \right) \right)
\end{eqnarray*}%
The uncertainty $Un\left( X,\bar{f}\left( X^{\prime }\right) \right) $ is
thus:%
\begin{eqnarray}
&&Un\left( X,\bar{f}\left( X^{\prime }\right) \right)  \label{NCR} \\
&=&\frac{1}{\bar{w}_{1}^{\left( 0\right) }\left( X^{\prime },X\right) }%
\left\{ \frac{\bar{\zeta}^{2}\zeta ^{2}\left( 1+\frac{\left( \gamma
\left\langle \hat{S}_{E}\left( X_{1},\left( X^{\prime }\right) ^{\prime
}\right) \right\rangle \right) ^{2}}{1-\left( \gamma \left\langle \hat{S}%
_{E}\left( X^{\prime },\left( X^{\prime }\right) ^{\prime }\right)
\right\rangle \right) ^{2}}\right) }{\left\langle \hat{w}_{1}^{\left(
0\right) B}\left( \left( X^{\prime }\right) ^{\prime },X^{\prime }\right)
\right\rangle _{\left( X^{\prime }\right) ^{\prime }}}+\xi ^{2}\right\}
\left( 1+\frac{\left( \bar{\gamma}\left\langle \bar{S}_{E}\left(
X_{1},X^{\prime }\right) \right\rangle _{X_{1}}\right) ^{2}}{1-\left( \bar{%
\gamma}\left\langle \bar{S}_{E}\left( \left( X^{\prime }\right) ^{\prime
},X^{\prime }\right) \right\rangle \right) ^{2}}\right)  \notag
\end{eqnarray}%
while the uncertainty $Un\left( X,\hat{f}\left( X^{\prime }\right) \right) $
becomes:%
\begin{equation}
Un\left( X,\hat{f}\left( X^{\prime }\right) \right) =\frac{\zeta ^{2}}{\hat{w%
}_{1}^{\left( 0\right) B}\left( X^{\prime },X\right) }\left( 1+\frac{\left(
\gamma \left\langle \hat{S}_{E}\left( X_{1},X^{\prime }\right) \right\rangle
_{X_{1}}\right) ^{2}}{1-\left( \gamma \left\langle \hat{S}_{E}\left(
X^{\prime },\left( X^{\prime }\right) ^{\prime }\right) \right\rangle
\right) ^{2}}\right)  \label{NCT}
\end{equation}%
and:%
\begin{equation}
Un\left( X,f^{\prime }\left( X\right) \right) \rightarrow \zeta ^{2}
\label{NCHApp}
\end{equation}%
To simplify the expressions, we also consider the same uncertainty for $\bar{%
S}_{E}$ and $\bar{S}_{L}$. as in Gosselin and Lotz (2025), so that: 
\begin{eqnarray*}
\frac{\hat{w}\left( X^{\prime },X\right) }{2} &\rightarrow &\frac{2\zeta ^{2}%
}{2\left( \zeta ^{2}+\frac{\zeta ^{2}}{\hat{w}_{1}^{\left( 0\right) }\left(
X^{\prime },X\right) }\left( 1+\frac{\left( \gamma \left\langle \hat{S}%
_{E}\left( X_{1},X^{\prime }\right) \right\rangle _{X_{1}}\right) ^{2}}{%
1-\left( \gamma \left\langle \hat{S}_{E}\left( X^{\prime },X\right)
\right\rangle \right) ^{2}}\right) \right) } \\
&=&\frac{\left( 1-\left( \gamma \left\langle \hat{S}_{E}\left( X\right)
\right\rangle \right) ^{2}\right) \hat{w}_{1}^{\left( 0\right) }\left(
X^{\prime },X\right) }{1+\hat{w}_{1}^{\left( 0\right) }\left( X^{\prime
},X\right) \left( 1-\left( \gamma \left\langle \hat{S}_{E}\left( X\right)
\right\rangle \right) ^{2}\right) +\left( \gamma \left\langle \hat{S}%
_{E}\left( X_{1},X^{\prime }\right) \right\rangle _{X_{1}}\right)
^{2}-\left( \gamma \left\langle \hat{S}_{E}\left( X\right) \right\rangle
\right) ^{2}}
\end{eqnarray*}%
We also include the uncertainty due to loans by replacing:%
\begin{equation*}
\gamma \left\langle \hat{S}_{E}\left( X_{1},X^{\prime }\right) \right\rangle
_{X_{1}}\rightarrow \gamma \left\langle \hat{S}\left( X_{1},X^{\prime
}\right) \right\rangle _{X_{1}}
\end{equation*}%
\begin{equation*}
\bar{\gamma}\left\langle \bar{S}_{E}\left( X_{1},X^{\prime }\right)
\right\rangle _{X_{1}}\rightarrow \bar{\gamma}\left\langle \bar{S}\left(
X_{1},X^{\prime }\right) \right\rangle _{X_{1}}
\end{equation*}%
and if we rescale $\bar{\gamma}\rightarrow \frac{\bar{\gamma}}{2}$:%
\begin{equation*}
\gamma \left\langle \hat{S}_{E}\left( X_{1},X^{\prime }\right) \right\rangle
_{X_{1}}\rightarrow \gamma \frac{\left\langle \hat{S}\left( X_{1},X^{\prime
}\right) \right\rangle _{X_{1}}}{2}\simeq \gamma \left\langle \hat{S}%
_{E}\left( X_{1},X^{\prime }\right) \right\rangle _{X_{1}}
\end{equation*}%
and:%
\begin{equation*}
\bar{\gamma}\left\langle \bar{S}_{E}\left( X_{1},X^{\prime }\right)
\right\rangle _{X_{1}}\rightarrow \bar{\gamma}\frac{\left\langle \bar{S}%
\left( X_{1},X^{\prime }\right) \right\rangle _{X_{1}}}{2}
\end{equation*}%
The others coefficients are computed by using both formula along wth the
fact that under our assumptions:%
\begin{equation*}
Un\left( X,\bar{f}\left( X^{\prime }\right) \right) +Un\left( X,\bar{r}%
\left( X^{\prime }\right) \right) =2Un\left( X,\bar{f}\left( X^{\prime
}\right) \right)
\end{equation*}%
The coefficient $\left( \bar{w}\left( X^{\prime },X\right) \right) ^{-1}$
measures this uncertainty:

\begin{eqnarray}
&&\left( \bar{w}\left( X^{\prime },X\right) \right) ^{-1}  \label{hbn} \\
&\rightarrow &1+\frac{1}{2\zeta ^{2}\bar{w}_{1}^{\left( 0\right) }\left(
X^{\prime },X\right) }\left\{ \frac{\bar{\zeta}^{2}\zeta ^{2}\left( 1+\frac{%
\left( \gamma \left\langle \hat{S}_{E}\left( X_{1},\left( X^{\prime }\right)
^{\prime }\right) \right\rangle \right) ^{2}}{1-\left( \gamma \left\langle 
\hat{S}_{E}\left( X^{\prime },\left( X^{\prime }\right) ^{\prime }\right)
\right\rangle \right) ^{2}}\right) }{\left\langle \hat{w}_{1}^{\left(
0\right) B}\left( \left( X^{\prime }\right) ^{\prime },X^{\prime }\right)
\right\rangle _{\left( X^{\prime }\right) ^{\prime }}}+\xi ^{2}\right\} 
\notag \\
&&\times \left( \frac{1+\frac{\left( \bar{\gamma}\frac{\left\langle \bar{S}%
\left( X_{E},X^{\prime }\right) \right\rangle _{X_{1}}}{2}\right) ^{2}}{%
1-\left( \bar{\gamma}\frac{\left\langle \bar{S}\left( X_{1},X^{\prime
}\right) \right\rangle _{X_{1}}}{2}\right) ^{2}}}{1+\frac{\left( \gamma
\left\langle \hat{S}_{E}\left( X_{1},X^{\prime }\right) \right\rangle
_{X_{1}}\right) ^{2}}{1-\left( \gamma \left\langle \hat{S}_{E}\left(
X^{\prime },\left( X^{\prime }\right) ^{\prime }\right) \right\rangle
\right) ^{2}}}\right) \left( w_{1}^{\left( 0\right) B}\left( X^{\prime
},X\right) +\frac{\zeta ^{2}}{\xi ^{2}}\left( 1+\frac{\left( \gamma
\left\langle \hat{S}_{E}\left( X_{1},X^{\prime }\right) \right\rangle
_{X_{1}}\right) ^{2}}{1-\left( \gamma \left\langle \hat{S}_{E}\left(
X^{\prime },\left( X^{\prime }\right) ^{\prime }\right) \right\rangle
\right) ^{2}}\right) \right)  \notag
\end{eqnarray}%
\begin{eqnarray}
&&\left( \hat{w}_{E}^{B}\left( X^{\prime },X\right) \right) ^{-1}
\label{hbt} \\
&\rightarrow &1+\frac{\frac{2\zeta ^{2}\bar{w}_{E}^{\left( 0\right) }\left(
X^{\prime },X\right) }{w_{E}^{\left( 0\right) B}\left( X^{\prime },X\right) }%
\left( \frac{1+\frac{\left( \gamma \left\langle \hat{S}_{E}\left(
X_{1},X^{\prime }\right) \right\rangle _{X_{1}}\right) ^{2}}{1-\left( \gamma
\left\langle \hat{S}_{E}\left( X^{\prime },\left( X^{\prime }\right)
^{\prime }\right) \right\rangle \right) ^{2}}}{1+\frac{\left( \bar{\gamma}%
\frac{\left\langle \bar{S}\left( X_{1},X^{\prime }\right) \right\rangle
_{X_{1}}}{2}\right) ^{2}}{1-\left( \bar{\gamma}\frac{\left\langle \bar{S}%
\left( X_{1},X^{\prime }\right) \right\rangle _{X_{1}}}{2}\right) ^{2}}}%
\right) }{\left\{ \frac{\bar{\zeta}^{2}\zeta ^{2}\left( 1+\frac{\left(
\gamma \left\langle \hat{S}_{E}\left( X_{1},\left( X^{\prime }\right)
^{\prime }\right) \right\rangle \right) ^{2}}{1-\left( \gamma \left\langle 
\hat{S}_{E}\left( X^{\prime },\left( X^{\prime }\right) ^{\prime }\right)
\right\rangle \right) ^{2}}\right) }{\left\langle \hat{w}_{E}^{\left(
0\right) B}\left( \left( X^{\prime }\right) ^{\prime },X^{\prime }\right)
\right\rangle _{\left( X^{\prime }\right) ^{\prime }}}+\xi ^{2}\right\} }+%
\frac{\zeta ^{2}}{\xi ^{2}w_{E}^{\left( 0\right) B}\left( X^{\prime
},X\right) }\left( 1+\frac{\left( \gamma \left\langle \hat{S}_{E}\left(
X_{1},X^{\prime }\right) \right\rangle _{X_{1}}\right) ^{2}}{1-\left( \gamma
\left\langle \hat{S}_{E}\left( X^{\prime },\left( X^{\prime }\right)
^{\prime }\right) \right\rangle \right) ^{2}}\right)  \notag \\
&\rightarrow &1+\frac{\left\langle \hat{w}_{E}^{\left( 0\right) B}\left(
\left( X^{\prime }\right) ^{\prime },X^{\prime }\right) \right\rangle
_{\left( X^{\prime }\right) ^{\prime }}\frac{2\zeta ^{2}\bar{w}_{E}^{\left(
0\right) }\left( X^{\prime },X\right) }{w_{E}^{\left( 0\right) B}\left(
X^{\prime },X\right) }\left( \frac{1+\frac{\left( \gamma \left\langle \hat{S}%
_{E}\left( X_{1},X^{\prime }\right) \right\rangle _{X_{1}}\right) ^{2}}{%
1-\left( \gamma \left\langle \hat{S}_{E}\left( X^{\prime },\left( X^{\prime
}\right) ^{\prime }\right) \right\rangle \right) ^{2}}}{1+\frac{\left( \bar{%
\gamma}\frac{\left\langle \bar{S}\left( X_{1},X^{\prime }\right)
\right\rangle _{X_{1}}}{2}\right) ^{2}}{1-\left( \bar{\gamma}\frac{%
\left\langle \bar{S}\left( X_{1},X^{\prime }\right) \right\rangle _{X_{1}}}{2%
}\right) ^{2}}}\right) }{\left( \bar{\zeta}^{2}\zeta ^{2}\left( 1+\frac{%
\left( \gamma \left\langle \hat{S}_{E}\left( X_{1},\left( X^{\prime }\right)
^{\prime }\right) \right\rangle \right) ^{2}}{1-\left( \gamma \left\langle 
\hat{S}_{E}\left( X^{\prime },\left( X^{\prime }\right) ^{\prime }\right)
\right\rangle \right) ^{2}}\right) +\xi ^{2}\left\langle \hat{w}_{E}^{\left(
0\right) B}\left( \left( X^{\prime }\right) ^{\prime },X^{\prime }\right)
\right\rangle _{\left( X^{\prime }\right) ^{\prime }}\right) }  \notag \\
&&+\frac{\zeta ^{2}\left\langle \hat{w}_{E}^{\left( 0\right) B}\left( \left(
X^{\prime }\right) ^{\prime },X^{\prime }\right) \right\rangle _{\left(
X^{\prime }\right) ^{\prime }}}{\xi ^{2}w_{E}^{\left( 0\right) B}\left(
X^{\prime },X\right) }\left( 1+\frac{\left( \gamma \left\langle \hat{S}%
_{E}\left( X_{1},X^{\prime }\right) \right\rangle _{X_{1}}\right) ^{2}}{%
1-\left( \gamma \left\langle \hat{S}_{E}\left( X^{\prime },\left( X^{\prime
}\right) ^{\prime }\right) \right\rangle \right) ^{2}}\right)  \notag
\end{eqnarray}

\section*{Appendix 6 Averages}

\subsection*{A6.1 Capital averages}

\subsubsection*{A6.1.1 Maximal capital}

The average return equations for shares depend on average capital ratios. We
recall the formula needed for the derivations expressd in terms of the
shares variables. We use the quantities:%
\begin{equation*}
\bar{f}\left( \bar{K}_{1},X_{1}\right) =\left( 1-\bar{M}\right) \bar{g}%
\left( \hat{K}^{\prime },X^{\prime }\right)
\end{equation*}%
\begin{eqnarray*}
&&\left\langle \bar{g}\right\rangle =\left( 1-\left\langle \bar{M}\left(
X^{\prime },X\right) \right\rangle \right) ^{-1}\left\langle \bar{f}%
\right\rangle \\
&=&\frac{1}{1-\left\langle \bar{S}\left( X^{\prime },X\right) \right\rangle }%
\left\langle \bar{f}\right\rangle
\end{eqnarray*}%
\begin{equation*}
\hat{f}\left( \hat{K}_{1},X_{1}\right) =\left( 1-\hat{M}\right) \hat{g}%
\left( \hat{K}^{\prime },X^{\prime }\right) +\bar{N}\bar{g}\left( \hat{K}%
^{\prime },X^{\prime }\right)
\end{equation*}%
The average maximal capital per sector for banks and investors are given in
first approximation by:%
\begin{equation*}
\left\langle \bar{K}_{0}\right\rangle ^{2}\simeq 2\frac{\sigma _{\hat{K}}^{2}%
}{\left\langle \bar{g}\right\rangle ^{2}}\left( \frac{\left\Vert \bar{\Psi}%
_{0}\right\Vert ^{2}}{\hat{\mu}}-\left( \frac{\left\langle \hat{K}%
\right\rangle ^{2}\left\langle \hat{g}\right\rangle }{\sigma _{\hat{K}}^{2}}%
\right) \left\langle \hat{g}^{Bef}\right\rangle \frac{\left\langle \bar{K}%
_{0}\right\rangle }{\left\langle \hat{K}\right\rangle }\right)
\end{equation*}%
and:%
\begin{equation}
\left\langle \hat{K}_{0}\right\rangle ^{2}\simeq 2\frac{\sigma _{\hat{K}}^{2}%
}{\left\langle \hat{g}\right\rangle ^{2}}\left( \frac{\left\Vert \hat{\Psi}%
_{0}\right\Vert ^{2}}{\hat{\mu}}-\left( \frac{\left\langle \hat{K}%
\right\rangle ^{2}\left\langle \hat{g}\right\rangle }{\sigma _{\hat{K}}^{2}}+%
\frac{1}{2}\right) \left\langle \hat{g}^{ef}\right\rangle \frac{\left\langle 
\hat{K}_{0}\right\rangle }{\left\langle \hat{K}\right\rangle }\right)
\end{equation}%
\begin{equation*}
\left\langle K_{0}\right\rangle ^{2}\simeq 2\frac{\sigma _{\hat{K}}^{2}}{%
\left\langle f_{1}\right\rangle ^{2}}\frac{\left\vert \Psi _{0}\left(
X\right) \right\vert ^{2}}{\epsilon }
\end{equation*}%
where:%
\begin{equation*}
\left\langle \hat{g}^{ef}\right\rangle =-\frac{\left( \kappa \left\langle %
\left[ \frac{\underline{\hat{k}}_{2}^{B}}{1+\bar{k}}\right] \right\rangle
\left( 1-\left\langle \underline{\hat{k}}\right\rangle \right) +\left\langle 
\hat{k}_{1}^{B}\right\rangle \left\langle \hat{k}_{2}\right\rangle \right)
\left( \left\langle \hat{g}\right\rangle +\frac{1}{1-\left\langle \underline{%
\hat{k}}\right\rangle }\bar{N}\left\langle \bar{g}\right\rangle \right) 
\frac{\left\Vert \bar{\Psi}_{0}\right\Vert ^{2}}{\left\Vert \hat{\Psi}%
_{0}\right\Vert ^{2}}}{\left( 1-\left( \left\langle \hat{k}_{1}\right\rangle
+\left\langle \hat{k}_{1}^{B}\right\rangle \frac{\left\Vert \bar{\Psi}%
_{0}\right\Vert ^{2}}{\left\Vert \hat{\Psi}_{0}\right\Vert ^{2}}\right)
\right) \left( 1-\left( \left\langle \underline{\hat{k}}\right\rangle
+\left( \left\langle \underline{\hat{k}}_{1}^{B}\right\rangle +\kappa
\left\langle \left[ \frac{\underline{\hat{k}}_{2}^{B}}{1+\bar{k}}\right]
\right\rangle \right) \frac{\left\Vert \bar{\Psi}_{0}\right\Vert ^{2}}{%
\left\Vert \hat{\Psi}_{0}\right\Vert ^{2}}\right) \right) }
\end{equation*}%
and: 
\begin{eqnarray*}
\left\langle \hat{g}^{Bef}\right\rangle &=&-\frac{\kappa \left\langle \left[ 
\frac{\underline{\hat{k}}_{2}^{B}}{1+\bar{k}}\right] \right\rangle \left(
1-\left\langle \hat{k}\right\rangle \right) +\left\langle \hat{k}%
_{1}^{B}\right\rangle \left\langle \hat{k}_{2}\right\rangle }{\left(
1-\left( \left\langle \hat{k}\right\rangle +\left( \left\langle \hat{k}%
_{1}^{B}\right\rangle +\kappa \left\langle \left[ \frac{\underline{\hat{k}}%
_{2}^{B}}{1+\bar{k}}\right] \right\rangle \right) \frac{\left\Vert \bar{\Psi}%
_{0}\right\Vert ^{2}}{\left\Vert \hat{\Psi}_{0}\right\Vert ^{2}}\right)
\right) \left( 1-\left( \left\langle \hat{k}_{1}\right\rangle +\left\langle 
\hat{k}_{1}^{B}\right\rangle \frac{\left\Vert \bar{\Psi}_{0}\right\Vert ^{2}%
}{\left\Vert \hat{\Psi}_{0}\right\Vert ^{2}}\right) \right) } \\
&&\times \left( \left\langle \hat{g}\right\rangle +\left( 1-\hat{M}\right)
^{-1}\bar{N}\left\langle \bar{g}\right\rangle \right)
\end{eqnarray*}

\subsubsection*{A6.1.2 Average disposable capital for investors}

The average disposable incomes are given by:%
\begin{eqnarray}
\left\langle \hat{K}\right\rangle \left\Vert \hat{\Psi}\right\Vert ^{2} &=&%
\hat{\mu}V\frac{\left\langle \hat{K}_{0}\right\rangle ^{4}}{2\sigma _{\hat{K}%
}^{2}}\left( \frac{1}{4}-\frac{\left\langle \hat{K}\right\rangle }{%
3\left\langle \hat{K}_{0}\right\rangle }\frac{\left\langle \hat{g}%
^{ef}\right\rangle }{\left\langle \hat{g}\right\rangle }\right) \left\langle 
\hat{g}\right\rangle ^{2}  \label{FRN} \\
&=&\frac{18\sigma _{\hat{K}}^{2}}{\hat{\mu}\left\langle \hat{g}\right\rangle
^{2}}V\left( \frac{\left\Vert \hat{\Psi}_{0}\right\Vert ^{2}}{\left( 5+\frac{%
\left\langle \hat{g}^{ef}\right\rangle }{\left\langle \hat{g}\right\rangle }-%
\sqrt{\left( 1-\frac{\left\langle \hat{g}^{ef}\right\rangle }{\left\langle 
\hat{g}\right\rangle }\right) \left( 4-\frac{\left\langle \hat{g}%
^{ef}\right\rangle }{\left\langle \hat{g}\right\rangle }\right) }\right) }%
\right) ^{2}  \notag \\
&&\times \left( \frac{1}{4}-\frac{1}{18}\left( 2+\frac{\left\langle \hat{g}%
^{ef}\right\rangle }{\left\langle \hat{g}\right\rangle }-\sqrt{\left( 1-%
\frac{\left\langle \hat{g}^{ef}\right\rangle }{\left\langle \hat{g}%
\right\rangle }\right) \left( 4-\frac{\left\langle \hat{g}^{ef}\right\rangle 
}{\left\langle \hat{g}\right\rangle }\right) }\right) \right)  \notag \\
&\simeq &\frac{9\sigma _{\hat{K}}^{2}}{2\hat{\mu}\left\langle \hat{g}%
\right\rangle ^{2}}V\left\Vert \hat{\Psi}_{0}\right\Vert ^{4}\simeq \frac{%
9\sigma _{\hat{K}}^{2}\left( 1-\hat{S}\right) ^{2}}{2\hat{\mu}\left(
\left\langle \hat{f}\right\rangle -\frac{\left\langle \hat{S}%
_{1}^{B}\right\rangle +\left\langle \hat{S}_{L}^{B}\right\rangle }{1-\bar{S}}%
\frac{\left\langle \bar{K}\right\rangle \left\Vert \bar{\Psi}\right\Vert ^{2}%
}{\left\langle \hat{K}\right\rangle \left\Vert \hat{\Psi}\right\Vert ^{2}}%
\left\langle \bar{f}\right\rangle \right) ^{2}}V\left\Vert \hat{\Psi}%
_{0}\right\Vert ^{4}  \notag
\end{eqnarray}

\subsubsection*{A6.1.3 Average disposable capital for banks}

The disposable capital for banks $\left\langle \bar{K}\right\rangle
\left\Vert \bar{\Psi}\right\Vert ^{2}$ is given by:%
\begin{eqnarray}
\left\langle \bar{K}\right\rangle \left\Vert \bar{\Psi}\right\Vert ^{2}
&\simeq &18\frac{\sigma _{\hat{K}}^{2}V}{\left\langle \bar{g}\right\rangle
^{2}\hat{\mu}}\left( \frac{\sqrt{\left( 1+\frac{3}{4}\frac{\left\langle \hat{%
g}^{ef}\right\rangle }{\left\langle \hat{g}\right\rangle }\right) }%
\left\Vert \bar{\Psi}_{0}\right\Vert -\frac{3}{8}\frac{\left\langle \hat{g}%
^{Bef}\right\rangle }{\left\langle \bar{g}\right\rangle }\left\Vert \hat{\Psi%
}_{0}\right\Vert }{\sqrt{5+\frac{\left\langle \hat{g}^{ef}\right\rangle }{%
\left\langle \hat{g}\right\rangle }-\sqrt{\left( 1-\frac{\left\langle \hat{g}%
^{ef}\right\rangle }{\left\langle \hat{g}\right\rangle }\right) \left( 4-%
\frac{\left\langle \hat{g}^{ef}\right\rangle }{\left\langle \hat{g}%
\right\rangle }\right) }}}\right) ^{4}\left( \frac{1}{4}-\frac{\left\langle 
\hat{g}\right\rangle \left\langle \hat{g}^{Bef}\right\rangle }{3\left\langle 
\bar{g}\right\rangle ^{2}}\right)  \label{FRD} \\
&\simeq &\frac{9}{2}\frac{\sigma _{\hat{K}}^{2}V}{\left\langle \bar{g}%
\right\rangle ^{2}\hat{\mu}}\left\Vert \bar{\Psi}_{0}\right\Vert ^{4}  \notag
\end{eqnarray}%
Combining (\ref{FRN}) and (\ref{FRD}) leads to the ratio:%
\begin{eqnarray*}
\frac{\left\langle \bar{K}\right\rangle \left\Vert \bar{\Psi}\right\Vert ^{2}%
}{\left\langle \hat{K}\right\rangle \left\Vert \hat{\Psi}\right\Vert ^{2}}
&\simeq &\frac{\left( \frac{\left\langle \hat{f}\right\rangle +\frac{%
\left\langle \hat{S}_{E}^{B}\right\rangle +\left\langle \hat{S}%
_{L}^{B}\right\rangle }{1-\bar{S}}\frac{\left\langle \bar{K}\right\rangle
\left\Vert \bar{\Psi}\right\Vert ^{2}}{\left\langle \hat{K}\right\rangle
\left\Vert \hat{\Psi}\right\Vert ^{2}}\left\langle \bar{f}\right\rangle }{1-%
\hat{S}}\right) ^{2}}{\left\langle \bar{g}\right\rangle ^{2}}\frac{%
\left\Vert \bar{\Psi}_{0}\right\Vert ^{4}}{\left\Vert \hat{\Psi}%
_{0}\right\Vert ^{4}} \\
&=&\frac{\frac{\left( 1-\bar{S}\right) ^{2}}{\left( 1-\hat{S}\right) ^{2}}%
\left( \left\langle \hat{f}\right\rangle +\frac{\left\langle \hat{S}%
_{E}^{B}\right\rangle +\left\langle \hat{S}_{L}^{B}\right\rangle }{1-\bar{S}}%
\frac{\left\langle \bar{K}\right\rangle \left\Vert \bar{\Psi}\right\Vert ^{2}%
}{\left\langle \hat{K}\right\rangle \left\Vert \hat{\Psi}\right\Vert ^{2}}%
\left\langle \bar{f}\right\rangle \right) ^{2}}{\left\langle \bar{f}%
\right\rangle ^{2}}\frac{\left\Vert \bar{\Psi}_{0}\right\Vert ^{4}}{%
\left\Vert \hat{\Psi}_{0}\right\Vert ^{4}}
\end{eqnarray*}%
with solution:%
\begin{eqnarray*}
&&\frac{\left\langle \bar{K}\right\rangle \left\Vert \bar{\Psi}\right\Vert
^{2}}{\left\langle \hat{K}\right\rangle \left\Vert \hat{\Psi}\right\Vert ^{2}%
} \\
&=&\frac{1+2\frac{\left( 1-\bar{S}\right) \left\langle \hat{f}\right\rangle 
}{\left( 1-\hat{S}\right) \left\langle \bar{f}\right\rangle }\frac{%
\left\Vert \bar{\Psi}_{0}\right\Vert ^{2}}{\left\Vert \hat{\Psi}%
_{0}\right\Vert ^{2}}\left( \left\langle \hat{S}_{E}^{B}\right\rangle
+\left\langle \hat{S}_{L}^{B}\right\rangle \right) -\sqrt{1+4\frac{\left( 1-%
\bar{S}\right) \left\langle \hat{f}\right\rangle }{\left( 1-\hat{S}\right)
\left\langle \bar{f}\right\rangle }\frac{\left\Vert \bar{\Psi}%
_{0}\right\Vert ^{2}}{\left\Vert \hat{\Psi}_{0}\right\Vert ^{2}}\left(
\left\langle \hat{S}_{E}^{B}\right\rangle +\left\langle \hat{S}%
_{L}^{B}\right\rangle \right) }}{2\left( \frac{\left\langle \hat{S}%
_{E}^{B}\right\rangle +\left\langle \hat{S}_{L}^{B}\right\rangle }{\left( 1-%
\hat{S}\right) }\right) ^{2}} \\
&=&\frac{2\left( 1-\bar{S}\right) ^{2}\left\Vert \bar{\Psi}_{0}\right\Vert
^{4}\left\langle \hat{f}\right\rangle ^{2}}{\left( 1-\hat{S}\right)
^{2}\left\Vert \hat{\Psi}_{0}\right\Vert ^{4}\left\langle \bar{f}%
\right\rangle ^{2}\left( 1+2\frac{\left( 1-\bar{S}\right) \left\langle \hat{f%
}\right\rangle }{\left( 1-\hat{S}\right) \left\langle \bar{f}\right\rangle }%
\frac{\left\Vert \bar{\Psi}_{0}\right\Vert ^{2}}{\left\Vert \hat{\Psi}%
_{0}\right\Vert ^{2}}\left( \left\langle \hat{S}_{E}^{B}\right\rangle
+\left\langle \hat{S}_{L}^{B}\right\rangle \right) +\sqrt{1+4\frac{\left( 1-%
\bar{S}\right) \left\langle \hat{f}\right\rangle }{\left( 1-\hat{S}\right)
\left\langle \bar{f}\right\rangle }\frac{\left\Vert \bar{\Psi}%
_{0}\right\Vert ^{2}}{\left\Vert \hat{\Psi}_{0}\right\Vert ^{2}}\left(
\left\langle \hat{S}_{E}^{B}\right\rangle +\left\langle \hat{S}%
_{L}^{B}\right\rangle \right) }\right) }
\end{eqnarray*}%
and this allows to rewrites:%
\begin{eqnarray}
&&\left\langle \hat{K}\right\rangle \left\Vert \hat{\Psi}\right\Vert ^{2}=%
\frac{\frac{9\sigma _{\hat{K}}^{2}V}{2\left\langle \bar{g}\right\rangle ^{2}%
\hat{\mu}}\left\Vert \bar{\Psi}_{0}\right\Vert ^{4}}{\frac{\left\langle \bar{%
K}\right\rangle \left\Vert \bar{\Psi}\right\Vert ^{2}}{\left\langle \hat{K}%
\right\rangle \left\Vert \hat{\Psi}\right\Vert ^{2}}}  \label{FRT} \\
&=&9\frac{\sigma _{\hat{K}}^{2}V\left( 1-\hat{S}\right) ^{2}\left\Vert \hat{%
\Psi}_{0}\right\Vert ^{4}}{4\hat{\mu}\left\langle \hat{f}\right\rangle ^{2}}
\notag \\
&&\times \left( 1+2\frac{\left( 1-\bar{S}\right) \left\langle \hat{f}%
\right\rangle \left\Vert \bar{\Psi}_{0}\right\Vert ^{2}\left( \left\langle 
\hat{S}_{E}^{B}\right\rangle +\left\langle \hat{S}_{K}^{B}\right\rangle
\right) }{\left( 1-\hat{S}\right) \left\langle \bar{f}\right\rangle
\left\Vert \hat{\Psi}_{0}\right\Vert ^{2}}+\sqrt{1+4\frac{\left( 1-\bar{S}%
\right) \left\langle \hat{f}\right\rangle \left\Vert \bar{\Psi}%
_{0}\right\Vert ^{2}\left( \left\langle \hat{S}_{E}^{B}\right\rangle
+\left\langle \hat{S}_{L}^{B}\right\rangle \right) }{\left( 1-\hat{S}\right)
\left\langle \bar{f}\right\rangle \left\Vert \hat{\Psi}_{0}\right\Vert ^{2}}}%
\right)  \notag
\end{eqnarray}%
At the lowest order:%
\begin{eqnarray}
&&\frac{\left\langle \bar{K}\right\rangle \left\Vert \bar{\Psi}\right\Vert
^{2}}{\left\langle \hat{K}\right\rangle \left\Vert \hat{\Psi}\right\Vert ^{2}%
}  \label{FRV} \\
&\simeq &\frac{2\left( 1-\bar{S}\right) ^{2}\left\Vert \bar{\Psi}%
_{0}\right\Vert ^{4}\left\langle \hat{f}\right\rangle ^{2}}{\left( 1-\hat{S}%
\right) ^{2}\left\Vert \hat{\Psi}_{0}\right\Vert ^{4}\left\langle \bar{f}%
\right\rangle ^{2}\left( 1+2\frac{\left( 1-\bar{S}\right) \left\langle \hat{f%
}\right\rangle }{\left( 1-\hat{S}\right) \left\langle \bar{f}\right\rangle }%
\frac{\left\Vert \bar{\Psi}_{0}\right\Vert ^{2}}{\left\Vert \hat{\Psi}%
_{0}\right\Vert ^{2}}\left( \left\langle \hat{S}_{E}^{B}\right\rangle
+\left\langle \hat{S}_{L}^{B}\right\rangle \right) +\sqrt{1+4\frac{\left( 1-%
\bar{S}\right) \left\langle \hat{f}\right\rangle }{\left( 1-\hat{S}\right)
\left\langle \bar{f}\right\rangle }\frac{\left\Vert \bar{\Psi}%
_{0}\right\Vert ^{2}}{\left\Vert \hat{\Psi}_{0}\right\Vert ^{2}}\left(
\left\langle \hat{S}_{E}^{B}\right\rangle +\left\langle \hat{S}%
_{L}^{B}\right\rangle \right) }\right) }  \notag \\
&\rightarrow &\frac{2\left\Vert \bar{\Psi}_{0}\right\Vert ^{4}}{\left\Vert 
\hat{\Psi}_{0}\right\Vert ^{4}\left( 1+2\frac{\left\Vert \bar{\Psi}%
_{0}\right\Vert ^{2}}{\left\Vert \hat{\Psi}_{0}\right\Vert ^{2}}\left(
\left\langle \hat{S}_{E}^{B}\right\rangle +\left\langle \hat{S}%
_{L}^{B}\right\rangle \right) +\sqrt{1+4\frac{\left\Vert \bar{\Psi}%
_{0}\right\Vert ^{2}}{\left\Vert \hat{\Psi}_{0}\right\Vert ^{2}}\left(
\left\langle \hat{S}_{E}^{B}\right\rangle +\left\langle \hat{S}%
_{L}^{B}\right\rangle \right) }\right) }\frac{\left( 1-\bar{S}\right)
^{2}\left\langle \hat{f}\right\rangle ^{2}}{\left( 1-\hat{S}\right)
^{2}\left\langle \bar{f}\right\rangle ^{2}}  \notag \\
&\simeq &\frac{\left\Vert \bar{\Psi}_{0}\right\Vert ^{4}}{\left\Vert \hat{%
\Psi}_{0}\right\Vert ^{4}\left( 1+\frac{\left\Vert \bar{\Psi}_{0}\right\Vert
^{2}}{\left\Vert \hat{\Psi}_{0}\right\Vert ^{2}}\left\langle \hat{S}%
_{L}^{B}\right\rangle \right) }\frac{\left( 1-\bar{S}\right)
^{2}\left\langle \hat{f}\right\rangle ^{2}}{\left( 1-\hat{S}\right)
^{2}\left\langle \bar{f}\right\rangle ^{2}}  \notag
\end{eqnarray}%
\begin{equation*}
\left\langle \bar{K}\right\rangle \left\Vert \bar{\Psi}\right\Vert ^{2}=9%
\frac{\sigma _{\hat{K}}^{2}V\left\Vert \bar{\Psi}_{0}\right\Vert ^{4}\left(
1-\bar{S}\right) ^{2}}{2\hat{\mu}\left\langle \bar{f}\right\rangle ^{2}}
\end{equation*}%
and:%
\begin{equation*}
\left\langle \hat{K}\right\rangle \left\Vert \hat{\Psi}\right\Vert ^{2}=%
\frac{9\sigma _{\hat{K}}^{2}\left\Vert \hat{\Psi}_{0}\right\Vert ^{4}\left(
1-\hat{S}\right) ^{2}\left( 1+\frac{\left\Vert \bar{\Psi}_{0}\right\Vert ^{2}%
}{\left\Vert \hat{\Psi}_{0}\right\Vert ^{2}}\left\langle \hat{S}%
_{L}^{B}\right\rangle \right) }{2\hat{\mu}\left\langle \hat{f}\right\rangle
^{2}}
\end{equation*}

\subsubsection*{A6.1.4 Average disposable capital for firms}

The formula for firms capital is: 
\begin{equation*}
\left\langle K\right\rangle \left\Vert \Psi \right\Vert ^{2}\simeq \frac{%
\epsilon K_{0}}{\sigma _{\hat{K}}^{2}f_{1}\left( X\right) }\frac{1}{12}%
\left( 3X-\bar{C}\right) \left( \bar{C}+X\right) ^{2}
\end{equation*}%
where:%
\begin{equation*}
X=\sqrt{\sigma _{\hat{K}}^{2}\left( \frac{\left\vert \Psi _{0}\left(
X\right) \right\vert ^{2}}{\epsilon }-\frac{f_{1}\left( X\right) }{2}\right) 
}
\end{equation*}%
Moreover, as in Part I:%
\begin{equation}
f_{1}^{\left( e\right) }\left( X\right) =\left( 1+\underline{k}_{2}\left(
X\right) +\kappa \left[ \frac{\underline{k}_{2}^{B}}{1+\bar{k}}\right]
\left( X\right) \right) f_{1}^{\prime }\left( X\right) -\left( \underline{k}%
_{2}\left( X\right) +\kappa \left[ \frac{\underline{k}_{2}^{B}}{1+\bar{k}}%
\right] \left( X\right) \right) \bar{r}
\end{equation}%
As a consequence, we replace:%
\begin{equation*}
f_{1}\left( X\right) \rightarrow \left( 1+\underline{k}_{2}\left( X^{\prime
}\right) +\kappa \left[ \frac{\underline{k}_{2}^{B}}{1+\bar{k}}\right]
\left( X^{\prime }\right) \right) \left( f_{1}\left( X\right) -r\left(
X\right) \right) +r\left( X\right)
\end{equation*}%
Given that:%
\begin{eqnarray*}
1+\underline{k}_{2}\left( X^{\prime }\right) +\kappa \left[ \frac{\underline{%
k}_{2}^{B}}{1+\bar{k}}\right] \left( X^{\prime }\right) &=&1+\frac{%
S_{L}\left( X^{\prime }\right) +S_{L}^{B}\left( X^{\prime }\right) }{%
1-\left( S\left( X^{\prime }\right) +S^{B}\left( X^{\prime }\right) \right) }
\\
&=&\frac{1-\left( S_{E}\left( X^{\prime }\right) +S_{E}^{B}\left( X^{\prime
}\right) \right) }{1-\left( S\left( X^{\prime }\right) +S^{B}\left(
X^{\prime }\right) \right) } \\
&\rightarrow &\frac{1-\left( \left\langle S_{E}\left( X^{\prime },X^{\prime
}\right) \right\rangle \frac{\left\langle \hat{K}\right\rangle \left\Vert 
\hat{\Psi}\right\Vert ^{2}}{\left\langle K\right\rangle \left\Vert \Psi
\right\Vert ^{2}}+\left\langle S_{E}^{B}\left( X^{\prime },X^{\prime
}\right) \right\rangle \frac{\left\langle \bar{K}\right\rangle \left\Vert 
\bar{\Psi}\right\Vert ^{2}}{\left\langle K\right\rangle \left\Vert \Psi
\right\Vert ^{2}}\right) }{1-\left( \left\langle S\left( X^{\prime
},X^{\prime }\right) \right\rangle \frac{\left\langle \hat{K}\right\rangle
\left\Vert \hat{\Psi}\right\Vert ^{2}}{\left\langle K\right\rangle
\left\Vert \Psi \right\Vert ^{2}}+\left\langle S^{B}\left( X^{\prime
},X^{\prime }\right) \right\rangle \frac{\left\langle \bar{K}\right\rangle
\left\Vert \bar{\Psi}\right\Vert ^{2}}{\left\langle K\right\rangle
\left\Vert \Psi \right\Vert ^{2}}\right) }
\end{eqnarray*}%
\begin{eqnarray*}
&&1+\underline{k}\left( X^{\prime }\right) +\underline{k}_{1}^{\left(
B\right) }\left( X^{\prime }\right) +\kappa \left[ \frac{\underline{k}%
_{2}^{B}}{1+\bar{k}}\right] \left( X^{\prime }\right) \\
&=&1+\frac{S\left( X^{\prime }\right) +S^{B}\left( X^{\prime }\right) }{%
1-\left( S\left( X^{\prime }\right) +S^{B}\left( X^{\prime }\right) \right) }%
=\frac{1}{1-\left( S\left( X^{\prime }\right) +S^{B}\left( X^{\prime
}\right) \right) }
\end{eqnarray*}%
we find that:%
\begin{eqnarray}
&&\frac{\epsilon K_{0}}{\sigma _{\hat{K}}^{2}f_{1}\left( X\right) }\frac{1}{%
12}\left( 3X-\bar{C}\right) \left( \bar{C}+X\right) ^{2}  \notag \\
&\rightarrow &\frac{\epsilon K_{0}}{\sigma _{\hat{K}}^{2}\frac{1-\left(
S_{E}\left( X\right) +S_{E}^{B}\left( X\right) \right) }{1-\left( S\left(
X\right) +S^{B}\left( X\right) \right) }\left( f_{1}\left( X\right) -r\left(
X\right) \right) +r\left( X\right) }  \notag \\
&&\times \frac{1}{12}\left( 3X-C\left( 1-\left( S_{E}\left( X\right)
+S_{E}^{B}\left( X\right) \right) \right) \right) \left( C\left( 1-\left(
S_{E}\left( X\right) +S_{E}^{B}\left( X\right) \right) \right) +X\right) ^{2}
\end{eqnarray}%
\begin{eqnarray*}
&&\left\langle K\right\rangle \left\Vert \Psi \right\Vert ^{2}\simeq \frac{%
\epsilon K_{0}}{\sigma _{\hat{K}}^{2}f_{1}\left( X\right) }\frac{1}{12}%
\left( 3X-\bar{C}\right) \left( \bar{C}+X\right) ^{2} \\
&\rightarrow &\frac{\epsilon \sqrt{\sigma _{\hat{K}}^{2}\frac{\left\vert
\Psi _{0}\left( X\right) \right\vert ^{2}}{\epsilon }}}{\sigma _{\hat{K}%
}^{2}\left( \frac{1-\left( S_{E}\left( X\right) +S_{E}^{B}\left( X\right)
\right) }{1-\left( S\left( X\right) +S^{B}\left( X\right) \right) }\left(
f_{1}\left( X\right) -r\left( X\right) \right) +r\left( X\right) \right) ^{2}%
} \\
&&\times \frac{1}{12}\left( 3X-C\left( 1-\left( S_{E}\left( X\right)
+S_{E}^{B}\left( X\right) \right) \right) \right) \left( C\left( 1-\left(
S_{E}\left( X\right) +S_{E}^{B}\left( X\right) \right) \right) +X\right) ^{2}
\end{eqnarray*}%
using in first approximation that:%
\begin{equation*}
\left( 3X-C\left( 1-S\right) \right) \left( C\left( 1-S\right) +X\right)
^{2}\simeq 3X^{3}
\end{equation*}%
we find:%
\begin{equation*}
\left\langle K\right\rangle \left\Vert \Psi \right\Vert ^{2}\rightarrow 
\frac{\epsilon \sqrt{\sigma _{\hat{K}}^{2}\frac{\left\vert \Psi _{0}\left(
X\right) \right\vert ^{2}}{\epsilon }}X^{3}}{4\sigma _{\hat{K}}^{2}\left( 
\frac{1-\left( S_{E}\left( X\right) +S_{E}^{B}\left( X\right) \right) }{%
1-\left( S\left( X\right) +S^{B}\left( X\right) \right) }\left( f_{1}\left(
X\right) -r\left( X\right) \right) +r\left( X\right) \right) ^{2}}
\end{equation*}

\subsection*{A6.2 Average capital ratio}

\subsubsection*{A6.2.1 Equations for capital ratio}

\begin{equation*}
\frac{\left\langle \hat{K}\right\rangle \left\Vert \hat{\Psi}\right\Vert ^{2}%
}{\left\langle K\right\rangle \left\Vert \Psi \right\Vert ^{2}}\simeq \frac{%
\frac{9\sigma _{\hat{K}}^{2}}{2\hat{\mu}\left\langle \hat{g}\right\rangle
^{2}}V\left\Vert \hat{\Psi}_{0}\right\Vert ^{4}}{\epsilon \sqrt{\sigma _{%
\hat{K}}^{2}\frac{\left\vert \Psi _{0}\left( X\right) \right\vert ^{2}}{%
\epsilon }}\frac{3}{12}X^{3}}\sigma _{\hat{K}}^{2}\left( \frac{1-\left(
\left\langle S_{E}\left( X\right) \right\rangle +\left\langle
S_{E}^{B}\left( X\right) \right\rangle \right) }{1-\left( \left\langle
S\left( X\right) \right\rangle +\left\langle S^{B}\left( X\right)
\right\rangle \right) }\left( f_{1}\left( X\right) -r\left( X\right) \right)
+r\left( X\right) \right) ^{2}
\end{equation*}%
\begin{equation*}
\frac{\left\langle \bar{K}\right\rangle \left\Vert \bar{\Psi}\right\Vert ^{2}%
}{\left\langle K\right\rangle \left\Vert \Psi \right\Vert ^{2}}\simeq \frac{%
18\frac{\sigma _{\hat{K}}^{2}V}{\left\langle \bar{g}\right\rangle ^{2}\hat{%
\mu}}\left( \left\Vert \bar{\Psi}_{0}\right\Vert \right) ^{4}}{\epsilon 
\sqrt{\sigma _{\hat{K}}^{2}\frac{\left\vert \Psi _{0}\left( X\right)
\right\vert ^{2}}{\epsilon }}\frac{3}{12}X^{3}}\sigma _{\hat{K}}^{2}\left( 
\frac{1-\left( \left\langle S_{E}\left( X\right) \right\rangle +\left\langle
S_{E}^{B}\left( X\right) \right\rangle \right) }{1-\left( \left\langle
S\left( X\right) \right\rangle +\left\langle S^{B}\left( X\right)
\right\rangle \right) }\left( f_{1}\left( X\right) -r\left( X\right) \right)
+r\left( X\right) \right) ^{2}
\end{equation*}%
These equations do not solve for capital ratios. Actually, the quantities in
the right hand side depend themselves on the capital ratios:%
\begin{equation*}
\frac{1-\left( \left\langle S_{E}\left( X\right) \right\rangle +\left\langle
S_{E}^{B}\left( X\right) \right\rangle \right) }{1-\left( \left\langle
S\left( X\right) \right\rangle +\left\langle S^{B}\left( X\right)
\right\rangle \right) }=\frac{1-\left( \left\langle S_{E}\left( X^{\prime
},X^{\prime }\right) \right\rangle \frac{\left\langle \hat{K}\right\rangle
\left\Vert \hat{\Psi}\right\Vert ^{2}}{\left\langle K\right\rangle
\left\Vert \Psi \right\Vert ^{2}}+\left\langle S_{E}^{B}\left( X^{\prime
},X^{\prime }\right) \right\rangle \frac{\left\langle \bar{K}\right\rangle
\left\Vert \bar{\Psi}\right\Vert ^{2}}{\left\langle K\right\rangle
\left\Vert \Psi \right\Vert ^{2}}\right) }{1-\left( \left\langle S\left(
X^{\prime },X^{\prime }\right) \right\rangle \frac{\left\langle \hat{K}%
\right\rangle \left\Vert \hat{\Psi}\right\Vert ^{2}}{\left\langle
K\right\rangle \left\Vert \Psi \right\Vert ^{2}}+\left\langle S^{B}\left(
X^{\prime },X^{\prime }\right) \right\rangle \frac{\left\langle \bar{K}%
\right\rangle \left\Vert \bar{\Psi}\right\Vert ^{2}}{\left\langle
K\right\rangle \left\Vert \Psi \right\Vert ^{2}}\right) }
\end{equation*}%
The term: 
\begin{equation*}
Z=\frac{1-\left( \left\langle S_{E}\left( X\right) \right\rangle
+\left\langle S_{E}^{B}\left( X\right) \right\rangle \right) }{1-\left(
\left\langle S\left( X\right) \right\rangle +\left\langle S^{B}\left(
X\right) \right\rangle \right) }
\end{equation*}%
can be computed as a function of external parameters solving the capital
rates.

\subsubsection*{A6.2.1 Equations for $Z$}

We write:%
\begin{eqnarray*}
&&\left\langle S_{E}\left( X^{\prime },X^{\prime }\right) \right\rangle 
\frac{\left\langle \hat{K}\right\rangle \left\Vert \hat{\Psi}\right\Vert ^{2}%
}{\left\langle K\right\rangle \left\Vert \Psi \right\Vert ^{2}}+\left\langle
S_{E}^{B}\left( X^{\prime },X^{\prime }\right) \right\rangle \frac{%
\left\langle \bar{K}\right\rangle \left\Vert \bar{\Psi}\right\Vert ^{2}}{%
\left\langle K\right\rangle \left\Vert \Psi \right\Vert ^{2}} \\
&\simeq &\frac{\left\langle S_{E}\left( X^{\prime },X^{\prime }\right)
\right\rangle \frac{9\sigma _{\hat{K}}^{2}}{2\hat{\mu}\left\langle \hat{g}%
\right\rangle ^{2}}V\left\Vert \hat{\Psi}_{0}\right\Vert ^{4}+\left\langle
S_{E}^{B}\left( X^{\prime },X^{\prime }\right) \right\rangle 18\frac{\sigma
_{\hat{K}}^{2}V}{\left\langle \bar{g}\right\rangle ^{2}\hat{\mu}}\left(
\left\Vert \bar{\Psi}_{0}\right\Vert \right) ^{4}}{\epsilon \sqrt{\sigma _{%
\hat{K}}^{2}\frac{\left\vert \Psi _{0}\left( X\right) \right\vert ^{2}}{%
\epsilon }}\frac{3}{12}X^{3}} \\
&&\times \sigma _{\hat{K}}^{2}\left( \frac{1-\left( \left\langle S_{E}\left(
X\right) \right\rangle +\left\langle S_{E}^{B}\left( X\right) \right\rangle
\right) }{1-\left( \left\langle S\left( X\right) \right\rangle +\left\langle
S^{B}\left( X\right) \right\rangle \right) }\left( f_{1}\left( X\right)
-r\left( X\right) \right) +r\left( X\right) \right) ^{2}
\end{eqnarray*}%
and:%
\begin{eqnarray*}
&&\left\langle S\left( X^{\prime },X^{\prime }\right) \right\rangle \frac{%
\left\langle \hat{K}\right\rangle \left\Vert \hat{\Psi}\right\Vert ^{2}}{%
\left\langle K\right\rangle \left\Vert \Psi \right\Vert ^{2}}+\left\langle
S^{B}\left( X^{\prime },X^{\prime }\right) \right\rangle \frac{\left\langle 
\bar{K}\right\rangle \left\Vert \bar{\Psi}\right\Vert ^{2}}{\left\langle
K\right\rangle \left\Vert \Psi \right\Vert ^{2}} \\
&\simeq &\frac{\left\langle S\left( X^{\prime },X^{\prime }\right)
\right\rangle \frac{9\sigma _{\hat{K}}^{2}}{2\hat{\mu}\left\langle \hat{g}%
\right\rangle ^{2}}V\left\Vert \hat{\Psi}_{0}\right\Vert ^{4}+\left\langle
S^{B}\left( X^{\prime },X^{\prime }\right) \right\rangle 18\frac{\sigma _{%
\hat{K}}^{2}V}{\left\langle \bar{g}\right\rangle ^{2}\hat{\mu}}\left(
\left\Vert \bar{\Psi}_{0}\right\Vert \right) ^{4}}{\epsilon \sqrt{\sigma _{%
\hat{K}}^{2}\frac{\left\vert \Psi _{0}\left( X\right) \right\vert ^{2}}{%
\epsilon }}\frac{3}{12}X^{3}} \\
&&\times \sigma _{\hat{K}}^{2}\left( \frac{1-\left( \left\langle S_{E}\left(
X\right) \right\rangle +\left\langle S_{E}^{B}\left( X\right) \right\rangle
\right) }{1-\left( S\left( X\right) +S^{B}\left( X\right) \right) }\left(
f_{1}\left( X\right) -r\left( X\right) \right) +r\left( X\right) \right) ^{2}
\end{eqnarray*}%
\begin{eqnarray*}
Z &=&\frac{1-\left( \left\langle S_{E}\left( X^{\prime },X^{\prime }\right)
\right\rangle \frac{\left\langle \hat{K}\right\rangle \left\Vert \hat{\Psi}%
\right\Vert ^{2}}{\left\langle K\right\rangle \left\Vert \Psi \right\Vert
^{2}}+\left\langle S_{E}^{B}\left( X^{\prime },X^{\prime }\right)
\right\rangle \frac{\left\langle \bar{K}\right\rangle \left\Vert \bar{\Psi}%
\right\Vert ^{2}}{\left\langle K\right\rangle \left\Vert \Psi \right\Vert
^{2}}\right) }{1-\left( \left\langle S\left( X^{\prime },X^{\prime }\right)
\right\rangle \frac{\left\langle \hat{K}\right\rangle \left\Vert \hat{\Psi}%
\right\Vert ^{2}}{\left\langle K\right\rangle \left\Vert \Psi \right\Vert
^{2}}+\left\langle S^{B}\left( X^{\prime },X^{\prime }\right) \right\rangle 
\frac{\left\langle \bar{K}\right\rangle \left\Vert \bar{\Psi}\right\Vert ^{2}%
}{\left\langle K\right\rangle \left\Vert \Psi \right\Vert ^{2}}\right) } \\
&=&\frac{1-\frac{\left\langle S_{E}\left( X^{\prime },X^{\prime }\right)
\right\rangle \frac{9\sigma _{\hat{K}}^{2}}{2\hat{\mu}\left\langle \hat{g}%
\right\rangle ^{2}}V\left\Vert \hat{\Psi}_{0}\right\Vert ^{4}+\left\langle
S_{E}^{B}\left( X^{\prime },X^{\prime }\right) \right\rangle 18\frac{\sigma
_{\hat{K}}^{2}V}{\left\langle \bar{g}\right\rangle ^{2}\hat{\mu}}\left(
\left\Vert \bar{\Psi}_{0}\right\Vert \right) ^{4}}{\epsilon \sqrt{\sigma _{%
\hat{K}}^{2}\frac{\left\vert \Psi _{0}\left( X\right) \right\vert ^{2}}{%
\epsilon }}\frac{3}{12}X^{3}}\sigma _{\hat{K}}^{2}\left( \frac{1-\left(
S_{E}\left( X\right) +S_{E}^{B}\left( X\right) \right) }{1-\left( S\left(
X\right) +S^{B}\left( X\right) \right) }\left( f_{1}\left( X\right) -r\left(
X\right) \right) +r\left( X\right) \right) ^{2}}{1-\frac{\left\langle
S\left( X^{\prime },X^{\prime }\right) \right\rangle \frac{9\sigma _{\hat{K}%
}^{2}}{2\hat{\mu}\left\langle \hat{g}\right\rangle ^{2}}V\left\Vert \hat{\Psi%
}_{0}\right\Vert ^{4}+\left\langle S^{B}\left( X^{\prime },X^{\prime
}\right) \right\rangle 18\frac{\sigma _{\hat{K}}^{2}V}{\left\langle \bar{g}%
\right\rangle ^{2}\hat{\mu}}\left( \left\Vert \bar{\Psi}_{0}\right\Vert
\right) ^{4}}{\epsilon \sqrt{\sigma _{\hat{K}}^{2}\frac{\left\vert \Psi
_{0}\left( X\right) \right\vert ^{2}}{\epsilon }}\frac{3}{12}X^{3}}\sigma _{%
\hat{K}}^{2}\left( \frac{1-\left( S_{E}\left( X\right) +S_{E}^{B}\left(
X\right) \right) }{1-\left( S\left( X\right) +S^{B}\left( X\right) \right) }%
\left( f_{1}\left( X\right) -r\left( X\right) \right) +r\left( X\right)
\right) ^{2}}
\end{eqnarray*}%
and this is an equation for $Z$:%
\begin{eqnarray}
Z &=&\frac{1-\frac{\left\langle S_{E}\left( X^{\prime },X^{\prime }\right)
\right\rangle \frac{9\sigma _{\hat{K}}^{2}}{2\hat{\mu}\left\langle \hat{g}%
\right\rangle ^{2}}V\left\Vert \hat{\Psi}_{0}\right\Vert ^{4}+\left\langle
S_{E}^{B}\left( X^{\prime },X^{\prime }\right) \right\rangle 18\frac{\sigma
_{\hat{K}}^{2}V}{\left\langle \bar{g}\right\rangle ^{2}\hat{\mu}}\left(
\left\Vert \bar{\Psi}_{0}\right\Vert \right) ^{4}}{\epsilon \sqrt{\sigma _{%
\hat{K}}^{2}\frac{\left\vert \Psi _{0}\left( X\right) \right\vert ^{2}}{%
\epsilon }}\frac{3}{12}X^{3}}\sigma _{\hat{K}}^{2}\left( Z\left( f_{1}\left(
X\right) -r\left( X\right) \right) +r\left( X\right) \right) ^{2}}{1-\frac{%
\left\langle S\left( X^{\prime },X^{\prime }\right) \right\rangle \frac{%
9\sigma _{\hat{K}}^{2}}{2\hat{\mu}\left\langle \hat{g}\right\rangle ^{2}}%
V\left\Vert \hat{\Psi}_{0}\right\Vert ^{4}+\left\langle S^{B}\left(
X^{\prime },X^{\prime }\right) \right\rangle 18\frac{\sigma _{\hat{K}}^{2}V}{%
\left\langle \bar{g}\right\rangle ^{2}\hat{\mu}}\left( \left\Vert \bar{\Psi}%
_{0}\right\Vert \right) ^{4}}{\epsilon \sqrt{\sigma _{\hat{K}}^{2}\frac{%
\left\vert \Psi _{0}\left( X\right) \right\vert ^{2}}{\epsilon }}\frac{3}{12}%
X^{3}}\sigma _{\hat{K}}^{2}\left( Z\left( f_{1}\left( X\right) -r\left(
X\right) \right) +r\left( X\right) \right) ^{2}}  \label{ZQN} \\
&=&\frac{1-A\left( Z\left( f_{1}\left( X\right) -r\left( X\right) \right)
+r\left( X\right) \right) ^{2}}{1-B\left( Z\left( f_{1}\left( X\right)
-r\left( X\right) \right) +r\left( X\right) \right) ^{2}}  \notag
\end{eqnarray}%
where:%
\begin{eqnarray*}
A &=&\frac{\left\langle S_{E}\left( X^{\prime },X^{\prime }\right)
\right\rangle \frac{9\sigma _{\hat{K}}^{2}}{2\hat{\mu}\left\langle \hat{g}%
\right\rangle ^{2}}V\left\Vert \hat{\Psi}_{0}\right\Vert ^{4}+\left\langle
S_{E}^{B}\left( X^{\prime },X^{\prime }\right) \right\rangle 18\frac{\sigma
_{\hat{K}}^{2}V}{\left\langle \bar{g}\right\rangle ^{2}\hat{\mu}}\left(
\left\Vert \bar{\Psi}_{0}\right\Vert \right) ^{4}}{\epsilon \sqrt{\sigma _{%
\hat{K}}^{2}\frac{\left\vert \Psi _{0}\left( X\right) \right\vert ^{2}}{%
\epsilon }}\frac{3}{12}X^{3}}\sigma _{\hat{K}}^{2} \\
B &=&\frac{\left\langle S\left( X^{\prime },X^{\prime }\right) \right\rangle 
\frac{9\sigma _{\hat{K}}^{2}}{2\hat{\mu}\left\langle \hat{g}\right\rangle
^{2}}V\left\Vert \hat{\Psi}_{0}\right\Vert ^{4}+\left\langle S^{B}\left(
X^{\prime },X^{\prime }\right) \right\rangle 18\frac{\sigma _{\hat{K}}^{2}V}{%
\left\langle \bar{g}\right\rangle ^{2}\hat{\mu}}\left( \left\Vert \bar{\Psi}%
_{0}\right\Vert \right) ^{4}}{\epsilon \sqrt{\sigma _{\hat{K}}^{2}\frac{%
\left\vert \Psi _{0}\left( X\right) \right\vert ^{2}}{\epsilon }}\frac{3}{12}%
X^{3}}\sigma _{\hat{K}}^{2}
\end{eqnarray*}%
Written in terms of $\left\langle \hat{f}\right\rangle $ and $\left\langle 
\bar{f}\right\rangle $ this becomes: 
\begin{eqnarray*}
A &=&\frac{\left\langle S_{E}\left( X^{\prime },X^{\prime }\right)
\right\rangle \frac{9\sigma _{\hat{K}}^{2}\left( 1-\left\langle \hat{S}%
\right\rangle \right) ^{2}V\left\Vert \hat{\Psi}_{0}\right\Vert ^{4}}{2\hat{%
\mu}\left( \left\langle \hat{f}\right\rangle +\frac{\left\langle \hat{S}%
^{B}\right\rangle +\left\langle \hat{S}_{L}^{B}\right\rangle }{%
1-\left\langle \bar{S}\right\rangle }\left\langle \bar{f}\right\rangle
\right) ^{2}}+\left\langle S_{E}^{B}\left( X^{\prime },X^{\prime }\right)
\right\rangle 18\frac{\sigma _{\hat{K}}^{2}V\left( 1-\bar{S}\right)
^{2}\left( \left\Vert \bar{\Psi}_{0}\right\Vert \right) ^{4}}{\left\langle 
\bar{f}\right\rangle ^{2}\hat{\mu}}}{\epsilon \sqrt{\sigma _{\hat{K}}^{2}%
\frac{\left\vert \Psi _{0}\left( X\right) \right\vert ^{2}}{\epsilon }}\frac{%
3}{12}X^{3}}\sigma _{\hat{K}}^{2} \\
B &=&\frac{\left\langle S\left( X^{\prime },X^{\prime }\right) \right\rangle 
\frac{9\sigma _{\hat{K}}^{2}\left( 1-\left\langle \hat{S}\right\rangle
\right) ^{2}V\left\Vert \hat{\Psi}_{0}\right\Vert ^{4}}{2\hat{\mu}\left(
\left\langle \hat{f}\right\rangle +\frac{\left\langle \hat{S}%
^{B}\right\rangle +\left\langle \hat{S}_{L}^{B}\right\rangle }{%
1-\left\langle \bar{S}\right\rangle }\left\langle \bar{f}\right\rangle
\right) ^{2}}+\left\langle S^{B}\left( X^{\prime },X^{\prime }\right)
\right\rangle 18\frac{\sigma _{\hat{K}}^{2}V\left( 1-\bar{S}\right)
^{2}\left( \left\Vert \bar{\Psi}_{0}\right\Vert \right) ^{4}}{\left\langle 
\bar{f}\right\rangle ^{2}\hat{\mu}}}{\epsilon \sqrt{\sigma _{\hat{K}}^{2}%
\frac{\left\vert \Psi _{0}\left( X\right) \right\vert ^{2}}{\epsilon }}\frac{%
3}{12}X^{3}}\sigma _{\hat{K}}^{2}
\end{eqnarray*}%
We set:%
\begin{eqnarray*}
f &=&\left( f_{1}\left( X\right) -r\left( X\right) \right) \\
r &=&r\left( X\right)
\end{eqnarray*}%
and the equation for $Z$ is: 
\begin{equation*}
Z^{3}-\frac{\left( Af^{2}-2Bfr\right) }{Bf^{2}}Z^{2}-\frac{\left(
-Br^{2}+2Afr+1\right) }{Bf^{2}}Z-\frac{\left( Ar^{2}-1\right) }{Bf^{2}}=0
\end{equation*}%
Written in a standard form, this equation writes:%
\begin{eqnarray*}
0 &=&\left( Z+\frac{\left( Af^{2}-2Bfr\right) }{3Bf^{2}}\right) ^{3}-\frac{%
\left( Af^{2}-2Bfr\right) }{Bf^{2}}\left( Z+\frac{\left( Af^{2}-2Bfr\right) 
}{3Bf^{2}}\right) ^{2} \\
&&-\frac{\left( -Br^{2}+2Afr+1\right) }{Bf^{2}}\left( Z+\frac{\left(
Af^{2}-2Bfr\right) }{3Bf^{2}}\right) -\frac{\left( Ar^{2}-1\right) }{Bf^{2}}
\end{eqnarray*}%
There is one solution:

\begin{eqnarray}
Z &\simeq &\sqrt{\frac{1}{3}\left( \frac{4\left\langle r\left( X\right)
\right\rangle ^{2}}{3\left( \left\langle f_{1}\left( X\right) \right\rangle
-\left\langle r\left( X\right) \right\rangle \right) ^{2}}+\frac{1+2A\left(
\left\langle f_{1}\left( X\right) \right\rangle -\left\langle r\left(
X\right) \right\rangle \right) \left\langle r\left( X\right) \right\rangle
-B\left\langle r\left( X\right) \right\rangle ^{2}}{B\left( \left\langle
f_{1}\left( X\right) \right\rangle -\left\langle r\left( X\right)
\right\rangle \right) ^{2}}\right) }  \label{ZSL} \\
&&\times \cosh \left( \frac{1}{3}ar\cosh \left( \frac{-\left( \frac{%
16\left\langle r\left( X\right) \right\rangle ^{3}}{27\left( \left\langle
f_{1}\left( X\right) \right\rangle -\left\langle r\left( X\right)
\right\rangle \right) ^{3}}+\frac{A\left\langle r\left( X\right)
\right\rangle ^{2}-1}{B\left( \left\langle f_{1}\left( X\right)
\right\rangle -\left\langle r\left( X\right) \right\rangle \right) ^{2}}-%
\frac{2r\left( 1+2A\left( \left\langle f_{1}\left( X\right) \right\rangle
-\left\langle r\left( X\right) \right\rangle \right) \left\langle r\left(
X\right) \right\rangle -B\left\langle r\left( X\right) \right\rangle
^{2}\right) }{3B\left( \left\langle f_{1}\left( X\right) \right\rangle
-\left\langle r\left( X\right) \right\rangle \right) ^{3}}\right) }{\left( 
\frac{1}{3}\left( \frac{4\left\langle r\left( X\right) \right\rangle ^{2}}{%
3\left( \left\langle f_{1}\left( X\right) \right\rangle -\left\langle
r\left( X\right) \right\rangle \right) ^{2}}+\frac{1+2A\left( \left\langle
f_{1}\left( X\right) \right\rangle -\left\langle r\left( X\right)
\right\rangle \right) \left\langle r\left( X\right) \right\rangle
-B\left\langle r\left( X\right) \right\rangle ^{2}}{B\left( \left\langle
f_{1}\left( X\right) \right\rangle -\left\langle r\left( X\right)
\right\rangle \right) ^{2}}\right) \right) ^{\frac{3}{2}}}\right) \right) 
\notag \\
&&-\frac{2\left\langle r\left( X\right) \right\rangle }{3\left( \left\langle
f_{1}\left( X\right) \right\rangle -\left\langle r\left( X\right)
\right\rangle \right) }  \notag
\end{eqnarray}%
For $r=\left\langle S_{E}\left( X,X\right) \right\rangle <<1$ and $%
s=\left\langle S\left( X,X\right) \right\rangle <<1$%
\begin{equation*}
Z\simeq \sqrt{\frac{1}{3}\left( \frac{4r^{2}}{3f^{2}}+\frac{-Br^{2}+2Afr+1}{%
Bf^{2}}\right) }\times \cosh \left( \frac{1}{3}ar\cosh \left( \frac{-\left( 
\frac{16r^{3}}{27f^{3}}+\frac{Ar^{2}-1}{Bf^{2}}-\frac{2r\left(
-Br^{2}+2Afr+1\right) }{3Bf^{3}}\right) }{\left( \frac{1}{3}\left( \frac{%
4r^{2}}{3f^{2}}+\frac{-Br^{2}+2Afr+1}{Bf^{2}}\right) \right) ^{\frac{3}{2}}}%
\right) \right) -\frac{2r}{3f}
\end{equation*}

\subsubsection*{A6.2.2 Expressions of capital ratios in functions of returns}

Inserting the expression for $Z$ in the expressions of capital ratios leads
to:%
\begin{eqnarray}
\frac{\left\langle \hat{K}\right\rangle \left\Vert \hat{\Psi}\right\Vert ^{2}%
}{\left\langle K\right\rangle \left\Vert \Psi \right\Vert ^{2}} &\simeq &%
\frac{\frac{18\sigma _{\hat{K}}^{2}}{\hat{\mu}\left\langle \hat{g}%
\right\rangle ^{2}}V\left\Vert \hat{\Psi}_{0}\right\Vert ^{4}}{\epsilon 
\sqrt{\sigma _{\hat{K}}^{2}\frac{\left\vert \Psi _{0}\left( X\right)
\right\vert ^{2}}{\epsilon }}X^{3}}\sigma _{\hat{K}}^{2}\left( Z\left(
f_{1}\left( X\right) -r\left( X\right) \right) +r\left( X\right) \right) ^{2}
\label{KRN} \\
&\simeq &\frac{\frac{18\sigma _{\hat{K}}^{2}}{\hat{\mu}}\frac{\left( 1-\hat{S%
}\right) ^{2}}{\left( \left\langle \hat{f}\right\rangle +\frac{\left\langle 
\hat{S}^{B}\right\rangle +\left\langle \hat{S}_{L}^{B}\right\rangle \left(
1-\left\langle \bar{S}\right\rangle \right) }{1-\bar{S}}\left\langle \bar{f}%
\right\rangle \right) ^{2}}V\left\Vert \hat{\Psi}_{0}\right\Vert ^{4}}{%
\epsilon \sqrt{\sigma _{\hat{K}}^{2}\frac{\left\vert \Psi _{0}\left(
X\right) \right\vert ^{2}}{\epsilon }}X^{3}}\sigma _{\hat{K}}^{2}\left(
Z\left( f_{1}\left( X\right) -r\left( X\right) \right) +r\left( X\right)
\right) ^{2}  \notag
\end{eqnarray}%
\begin{eqnarray}
\frac{\left\langle \bar{K}\right\rangle \left\Vert \bar{\Psi}\right\Vert ^{2}%
}{\left\langle K\right\rangle \left\Vert \Psi \right\Vert ^{2}} &\simeq &%
\frac{18\frac{\sigma _{\hat{K}}^{2}V}{\left\langle \bar{g}\right\rangle ^{2}%
\hat{\mu}}\left( \left\Vert \bar{\Psi}_{0}\right\Vert \right) ^{4}}{\epsilon 
\sqrt{\sigma _{\hat{K}}^{2}\frac{\left\vert \Psi _{0}\left( X\right)
\right\vert ^{2}}{\epsilon }}X^{3}}\sigma _{\hat{K}}^{2}\left( Z\left(
f_{1}\left( X\right) -r\left( X\right) \right) +r\left( X\right) \right) ^{2}
\label{KRT} \\
&\simeq &\frac{18\frac{\sigma _{\hat{K}}^{2}V\left( 1-\bar{S}\right) ^{2}}{%
\left\langle \bar{f}\right\rangle ^{2}\hat{\mu}}\left( \left\Vert \bar{\Psi}%
_{0}\right\Vert \right) ^{4}}{\epsilon \sqrt{\sigma _{\hat{K}}^{2}\frac{%
\left\vert \Psi _{0}\left( X\right) \right\vert ^{2}}{\epsilon }}X^{3}}%
\sigma _{\hat{K}}^{2}\left( Z\left( f_{1}\left( X\right) -r\left( X\right)
\right) +r\left( X\right) \right) ^{2}  \notag
\end{eqnarray}

\subsubsection*{A6.2.3 Corrections for decreasing returns}

For decrasing returns, the expression of firms capital is given by:%
\begin{equation*}
K_{X}\left\vert \Psi \left( X\right) \right\vert ^{2}\simeq \left( 1-S\left(
X\right) \right) \frac{2\epsilon }{3\sigma _{\hat{K}}^{2}}\left( \frac{%
f_{1}\left( X\right) }{C_{0}+\frac{S_{L}\left( X\right) }{1-S_{E}\left(
X\right) }\bar{r}}\right) ^{\frac{2}{r}}
\end{equation*}%
Here:%
\begin{equation*}
S\left( X\right) =\left\langle S\left( X,X\right) \right\rangle \frac{%
\left\langle \hat{K}\right\rangle \left\Vert \hat{\Psi}\right\Vert ^{2}}{%
\left\langle K\right\rangle \left\Vert \Psi \right\Vert ^{2}}+\left\langle
S^{B}\left( X,X\right) \right\rangle \frac{\left\langle \bar{K}\right\rangle
\left\Vert \bar{\Psi}\right\Vert ^{2}}{\left\langle K\right\rangle
\left\Vert \Psi \right\Vert ^{2}}
\end{equation*}%
and:%
\begin{equation*}
\frac{S_{L}\left( X\right) }{1-S_{E}\left( X\right) }\rightarrow \frac{%
\underline{k}_{2}\left( X\right) +\underline{k}_{2}^{B}\left( X\right) }{%
\left( 1+\underline{k}_{2}\left( X\right) +\underline{k}_{2}^{B}\left(
X\right) \right) }
\end{equation*}%
We replace:%
\begin{equation*}
\frac{S_{L}\left( X\right) }{1-S_{E}\left( X\right) }\bar{r}\rightarrow \bar{%
r}
\end{equation*}%
We proceed as in the first part. In first approximation:%
\begin{equation*}
\left\langle K\right\rangle \left\Vert \Psi \right\Vert ^{2}\simeq \left( 1-%
\frac{\left\langle S\left( X,X\right) \right\rangle \left\langle \hat{K}%
\right\rangle \left\Vert \hat{\Psi}\right\Vert ^{2}+\left\langle S^{B}\left(
X,X\right) \right\rangle \left\langle \bar{K}\right\rangle \left\Vert \bar{%
\Psi}\right\Vert ^{2}}{\frac{2\epsilon }{3\sigma _{\hat{K}}^{2}}\left( \frac{%
\left\langle f_{1}\right\rangle }{C_{0}+\bar{r}}\right) ^{\frac{2}{r}}}%
\right) \left( \left( \frac{2\epsilon }{3\sigma _{\hat{K}}^{2}}\right) ^{%
\frac{r}{2}}\frac{f_{1}\left( X\right) }{C_{0}+\frac{S_{L}\left( X\right) }{%
1-S_{E}\left( X\right) }\bar{r}}\right) ^{\frac{2}{r}}
\end{equation*}%
and given (\ref{FRn}) (\ref{Frb}):%
\begin{equation}
\left\langle \hat{K}\right\rangle \left\Vert \hat{\Psi}\right\Vert ^{2}=%
\frac{9\sigma _{\hat{K}}^{2}\left\Vert \hat{\Psi}_{0}\right\Vert ^{4}\left(
1-\hat{S}\right) ^{2}\left( 1+\frac{\left\Vert \bar{\Psi}_{0}\right\Vert ^{2}%
}{\left\Vert \hat{\Psi}_{0}\right\Vert ^{2}}\left\langle \hat{S}%
_{L}^{B}\right\rangle \right) }{2\hat{\mu}\left\langle \hat{f}\right\rangle
^{2}}  \label{FRn}
\end{equation}%
\begin{equation}
\left\langle \bar{K}\right\rangle \left\Vert \bar{\Psi}\right\Vert ^{2}=9%
\frac{\sigma _{\hat{K}}^{2}V\left\Vert \bar{\Psi}_{0}\right\Vert ^{4}\left(
1-\bar{S}\right) ^{2}}{2\hat{\mu}\left\langle \bar{f}\right\rangle ^{2}}
\label{Frb}
\end{equation}%
the capital ratios becomes:%
\begin{equation}
\frac{\left\langle \hat{K}\right\rangle \left\Vert \hat{\Psi}\right\Vert ^{2}%
}{\left\langle K\right\rangle \left\Vert \Psi \right\Vert ^{2}}\simeq \frac{%
\left\langle \hat{K}\right\rangle \left\Vert \hat{\Psi}\right\Vert ^{2}}{%
\left( 1-\frac{\left\langle S\left( X,X\right) \right\rangle \left\langle 
\hat{K}\right\rangle \left\Vert \hat{\Psi}\right\Vert ^{2}+\left\langle
S^{B}\left( X,X\right) \right\rangle \left\langle \bar{K}\right\rangle
\left\Vert \bar{\Psi}\right\Vert ^{2}}{\frac{2\epsilon }{3\sigma _{\hat{K}%
}^{2}}\left( \frac{\left\langle f_{1}\left( X\right) \right\rangle }{C_{0}+%
\bar{r}}\right) ^{\frac{2}{r}}}\right) \left( \left( \frac{2\epsilon }{%
3\sigma _{\hat{K}}^{2}}\right) ^{\frac{r}{2}}\frac{\left\langle f_{1}\left(
X\right) \right\rangle }{C_{0}+\left\langle \frac{S_{L}\left( X\right) }{%
1-S_{E}\left( X\right) }\right\rangle \bar{r}}\right) ^{\frac{2}{r}}}
\label{ora}
\end{equation}%
and:%
\begin{equation}
\frac{\left\langle \bar{K}\right\rangle \left\Vert \bar{\Psi}\right\Vert ^{2}%
}{\left\langle K\right\rangle \left\Vert \Psi \right\Vert ^{2}}\simeq \frac{%
\left\langle \bar{K}\right\rangle \left\Vert \bar{\Psi}\right\Vert ^{2}}{%
\left( 1-\frac{\left\langle S\left( X,X\right) \right\rangle \left\langle 
\hat{K}\right\rangle \left\Vert \hat{\Psi}\right\Vert ^{2}+\left\langle
S^{B}\left( X,X\right) \right\rangle \left\langle \bar{K}\right\rangle
\left\Vert \bar{\Psi}\right\Vert ^{2}}{\frac{2\epsilon }{3\sigma _{\hat{K}%
}^{2}}\left( \frac{\left\langle f_{1}\left( X\right) \right\rangle }{C_{0}+%
\bar{r}}\right) ^{\frac{2}{r}}}\right) \left( \left( \frac{2\epsilon }{%
3\sigma _{\hat{K}}^{2}}\right) ^{\frac{r}{2}}\frac{\left\langle f_{1}\left(
X\right) \right\rangle }{C_{0}+\left\langle \frac{S_{L}\left( X\right) }{%
1-S_{E}\left( X\right) }\right\rangle \bar{r}}\right) ^{\frac{2}{r}}}
\label{ore}
\end{equation}

\subsection*{A6.3 Coefficients averages $\left\langle \bar{w}\left(
X^{\prime },X\right) \right\rangle $, $\left\langle \hat{w}_{1}^{B}\left(
X^{\prime },X\right) \right\rangle $, $\left\langle \hat{w}_{1}^{B}\left(
X^{\prime },X\right) \right\rangle $ as functions of shares}

Averages of coefficients (\ref{hbn}), (\ref{hbt}) can be computed. We find:%
\begin{eqnarray}
&&\left\langle \bar{w}\left( X^{\prime },X\right) \right\rangle  \label{Mhb}
\\
&\rightarrow &\frac{1}{1+\frac{\left\{ \frac{\bar{\zeta}^{2}\zeta ^{2}}{%
\left\langle \hat{w}_{1}^{\left( 0\right) B}\left( \left( X^{\prime }\right)
^{\prime },X^{\prime }\right) \right\rangle }+\xi ^{2}\left( 1-\left( \gamma
\left\langle \hat{S}_{E}\right\rangle \right) ^{2}\right) \right\} }{2\zeta
^{2}\left\langle \bar{w}_{1}^{\left( 0\right) }\left( X^{\prime },X\right)
\right\rangle \left( 1-\left( \bar{\gamma}\frac{\left\langle \bar{S}%
\right\rangle }{2}\right) ^{2}\right) }\left( \left\langle w_{1}^{\left(
0\right) B}\left( X^{\prime },X\right) \right\rangle +\frac{\zeta ^{2}}{\xi
^{2}}\frac{1}{1-\left( \gamma \left\langle \hat{S}_{E}\right\rangle \right)
^{2}}\right) }  \notag \\
&\rightarrow &\frac{1}{1+\frac{\left\{ \frac{\bar{\zeta}^{2}}{\left\langle 
\hat{w}_{1}^{\left( 0\right) B}\left( \left( X^{\prime }\right) ^{\prime
},X^{\prime }\right) \right\rangle }+\frac{\xi ^{2}}{\zeta ^{2}}\left(
1-\left( \gamma \left\langle \hat{S}_{E}\right\rangle \right) ^{2}\right)
\right\} }{2\left\langle \bar{w}_{1}^{\left( 0\right) }\left( X^{\prime
},X\right) \right\rangle \left( 1-\left( \bar{\gamma}\frac{\left\langle \bar{%
S}\right\rangle }{2}\right) ^{2}\right) }\left( \left\langle w_{1}^{\left(
0\right) B}\left( X^{\prime },X\right) \right\rangle +\frac{\zeta ^{2}}{\xi
^{2}}\frac{1}{1-\left( \gamma \left\langle \hat{S}_{E}\right\rangle \right)
^{2}}\right) }  \notag
\end{eqnarray}%
\begin{eqnarray}
&&\left\langle \hat{w}_{1}^{B}\left( X^{\prime },X\right) \right\rangle
\label{Mc} \\
&\rightarrow &\frac{1}{1+\frac{\left\langle \hat{w}_{1}^{\left( 0\right)
B}\left( \left( X^{\prime }\right) ^{\prime },X^{\prime }\right)
\right\rangle \frac{2\zeta ^{2}\left\langle \bar{w}_{1}^{\left( 0\right)
}\left( X^{\prime },X\right) \right\rangle }{\left\langle w_{1}^{\left(
0\right) B}\left( X^{\prime },X\right) \right\rangle }\left( \frac{1-\left( 
\bar{\gamma}\frac{\left\langle \bar{S}\right\rangle }{2}\right) ^{2}}{%
1-\left( \gamma \left\langle \hat{S}_{E}\right\rangle \right) ^{2}}\right) }{%
\left( \bar{\zeta}^{2}\zeta ^{2}\left( \frac{1}{1-\left( \gamma \left\langle 
\hat{S}_{E}\right\rangle \right) ^{2}}\right) +\xi ^{2}\left\langle \hat{w}%
_{1}^{\left( 0\right) B}\left( \left( X^{\prime }\right) ^{\prime
},X^{\prime }\right) \right\rangle \right) }+\frac{\zeta ^{2}\left\langle 
\hat{w}_{1}^{\left( 0\right) B}\left( \left( X^{\prime }\right) ^{\prime
},X^{\prime }\right) \right\rangle _{\left( X^{\prime }\right) ^{\prime }}}{%
\xi ^{2}w_{1}^{\left( 0\right) B}\left( X^{\prime },X\right) }\frac{1}{%
1-\left( \gamma \left\langle \hat{S}_{E}\right\rangle \right) ^{2}}}  \notag
\\
&\rightarrow &\frac{1}{1+\frac{\left\langle \hat{w}_{1}^{\left( 0\right)
B}\left( \left( X^{\prime }\right) ^{\prime },X^{\prime }\right)
\right\rangle \frac{2\left\langle \bar{w}_{1}^{\left( 0\right) }\left(
X^{\prime },X\right) \right\rangle }{\left\langle w_{1}^{\left( 0\right)
B}\left( X^{\prime },X\right) \right\rangle }\left( 1-\left( \bar{\gamma}%
\frac{\left\langle \bar{S}\right\rangle }{2}\right) ^{2}\right) }{\left( 
\bar{\zeta}^{2}+\frac{\xi ^{2}}{\zeta ^{2}}\left\langle \hat{w}_{1}^{\left(
0\right) B}\left( \left( X^{\prime }\right) ^{\prime },X^{\prime }\right)
\right\rangle \left( \left( 1-\left( \gamma \left\langle \hat{S}%
_{E}\right\rangle \right) ^{2}\right) \right) \right) }+\frac{\left\langle 
\hat{w}_{1}^{\left( 0\right) B}\left( \left( X^{\prime }\right) ^{\prime
},X^{\prime }\right) \right\rangle _{\left( X^{\prime }\right) ^{\prime }}}{%
\frac{\xi ^{2}}{\zeta ^{2}}w_{1}^{\left( 0\right) B}\left( X^{\prime
},X\right) \left( 1-\left( \gamma \left\langle \hat{S}_{E}\right\rangle
\right) ^{2}\right) }}  \notag
\end{eqnarray}%
The coefficient $\left\langle \hat{w}_{1}^{B}\left( X^{\prime },X\right)
\right\rangle $ also wrtes:%
\begin{eqnarray}
\left\langle \hat{w}_{E}^{B}\left( X^{\prime },X\right) \right\rangle
&\rightarrow &1+\frac{\left\langle \hat{w}_{1}^{\left( 0\right) B}\left(
\left( X^{\prime }\right) ^{\prime },X^{\prime }\right) \right\rangle
_{\left( X^{\prime }\right) }\frac{2\zeta ^{2}\bar{w}_{1}^{\left( 0\right)
}\left( X^{\prime },X\right) }{w_{1}^{\left( 0\right) B}\left( X^{\prime
},X\right) }\left( \frac{1+\frac{\left( \gamma \left\langle \hat{S}%
_{E}\left( X_{1},X^{\prime }\right) \right\rangle _{X_{1}}\right) ^{2}}{%
1-\left( \gamma \left\langle \hat{S}_{E}\left( X^{\prime },\left( X^{\prime
}\right) ^{\prime }\right) \right\rangle \right) ^{2}}}{1+\frac{\left( \bar{%
\gamma}\frac{\left\langle \bar{S}\left( X_{1},X^{\prime }\right)
\right\rangle _{X_{1}}}{2}\right) ^{2}}{1-\left( \bar{\gamma}\frac{%
\left\langle \bar{S}\left( X_{1},X^{\prime }\right) \right\rangle _{X_{1}}}{2%
}\right) ^{2}}}\right) }{\left( \bar{\zeta}^{2}\zeta ^{2}\left( 1+\frac{%
\left( \gamma \left\langle \hat{S}_{E}\left( X_{1},\left( X^{\prime }\right)
^{\prime }\right) \right\rangle \right) ^{2}}{1-\left( \gamma \left\langle 
\hat{S}_{E}\left( X^{\prime },\left( X^{\prime }\right) ^{\prime }\right)
\right\rangle \right) ^{2}}\right) +\xi ^{2}\left\langle \hat{w}_{1}^{\left(
0\right) B}\left( \left( X^{\prime }\right) ^{\prime },X^{\prime }\right)
\right\rangle _{\left( X^{\prime }\right) ^{\prime }}\right) }  \notag \\
&&+\frac{\zeta ^{2}\left\langle \hat{w}_{1}^{\left( 0\right) B}\left( \left(
X^{\prime }\right) ^{\prime },X^{\prime }\right) \right\rangle _{\left(
X^{\prime }\right) ^{\prime }}}{\xi ^{2}w_{1}^{\left( 0\right) B}\left(
X^{\prime },X\right) }\left( 1+\frac{\left( \gamma \left\langle \hat{S}%
_{E}\left( X_{1},X^{\prime }\right) \right\rangle _{X_{1}}\right) ^{2}}{%
1-\left( \gamma \left\langle \hat{S}_{E}\left( X^{\prime },\left( X^{\prime
}\right) ^{\prime }\right) \right\rangle \right) ^{2}}\right)  \notag
\end{eqnarray}%
which is also in first approximation:%
\begin{eqnarray}
&&\left\langle \hat{w}_{E}^{B}\left( X^{\prime },X\right) \right\rangle \\
&\simeq &1+\frac{2\left( 1-\left( \bar{\gamma}\left\langle \frac{\bar{S}%
\left( \left( X^{\prime }\right) ^{\prime },X^{\prime }\right) }{2}%
\right\rangle \right) ^{2}\right) }{\bar{\zeta}^{2}+\frac{\xi ^{2}}{\zeta
^{2}}\left( 1-\left( \gamma \left\langle \hat{S}_{E}\left( X^{\prime
},\left( X^{\prime }\right) ^{\prime }\right) \right\rangle \right)
^{2}\right) }+\frac{\zeta ^{2}}{\xi ^{2}}\left( \frac{1}{1-\left( \gamma
\left\langle \hat{S}_{E}\left( X^{\prime },\left( X^{\prime }\right)
^{\prime }\right) \right\rangle \right) ^{2}}\right)  \notag
\end{eqnarray}%
where the variances $\zeta ^{2}$ and $\zeta ^{2}$ and $\bar{\xi}^{2}$are
given by: 
\begin{eqnarray*}
&&\zeta ^{2} \\
&\rightarrow &\left\langle \frac{1-\hat{S}_{E}\left( X^{\prime }\right) }{1-%
\hat{S}\left( X^{\prime }\right) }S_{E}\left( \left( X^{\prime }\right)
^{\prime }\right) \right\rangle ^{2}Var\left( f_{1}\left( \left( X^{\prime
}\right) ^{\prime }\right) -\bar{r}\right)
\end{eqnarray*}%
and:%
\begin{eqnarray*}
&&\bar{\xi}^{2} \\
&\rightarrow &\left\langle \frac{1-\bar{S}_{E}\left( X^{\prime }\right) }{1-%
\bar{S}\left( X^{\prime }\right) }\right\rangle ^{2}\left\langle
S_{E}^{B}\left( \left( X^{\prime }\right) ^{\prime },\left( X^{\prime
}\right) ^{\prime }\right) \right\rangle ^{2}Var\left( f_{1}\left( \left(
X^{\prime }\right) ^{\prime }\right) -\bar{r}\right)
\end{eqnarray*}%
\begin{eqnarray*}
&&\bar{\zeta}^{2} \\
&\rightarrow &\left\langle \frac{1-\bar{S}_{E}\left( X^{\prime }\right) }{1-%
\bar{S}\left( X^{\prime }\right) }\right\rangle ^{2}\left\langle \hat{S}%
_{E}^{B}\left( \left( X^{\prime }\right) ^{\prime },\left( X^{\prime
}\right) ^{\prime }\right) \right\rangle ^{2}\left\langle \frac{1-\left( 
\hat{S}\left( \left( X^{\prime }\right) ^{\prime }\right) +\hat{S}%
_{E}^{B}\left( \left( X^{\prime }\right) ^{\prime }\right) +\hat{S}%
_{L}^{B}\left( \left( X^{\prime }\right) ^{\prime }\right) \right) }{%
1-\left( \hat{S}_{1}\left( \left( X^{\prime }\right) ^{\prime }\right) +\hat{%
S}_{E}^{B}\left( \left( X^{\prime }\right) ^{\prime }\right) \right) }%
\right\rangle ^{2}
\end{eqnarray*}%
Their ratios arising in (\ref{Mhb}), (\ref{Mc}) are:%
\begin{eqnarray}
&&\frac{\xi ^{2}}{\zeta ^{2}}  \label{ZTR} \\
&\rightarrow &\frac{\left\langle S_{E}^{B}\right\rangle ^{2}\left\langle 
\frac{1-\bar{S}_{E}}{1-\bar{S}}\right\rangle ^{2}}{\left\langle \frac{1-\hat{%
S}_{E}}{1-\hat{S}}S_{E}\right\rangle ^{2}}=\frac{\left\langle
S_{E}^{B}\right\rangle ^{2}\left\langle \frac{1-\bar{S}_{E}}{1-\bar{S}}%
\right\rangle ^{2}}{\left\langle \frac{1-\hat{S}_{E}}{1-\hat{S}}%
S_{E}\right\rangle ^{2}}  \notag
\end{eqnarray}%
and $\bar{\zeta}^{2}$ writes also:%
\begin{equation}
\bar{\zeta}^{2}\rightarrow \left\langle \hat{S}_{E}^{B}\right\rangle
^{2}\left\langle \frac{1-\bar{S}_{E}\left( X^{\prime }\right) }{1-\bar{S}%
\left( X^{\prime }\right) }\right\rangle ^{2}  \label{ZTB}
\end{equation}%
As before $\left\langle \bar{S}_{E}\right\rangle <<\bar{\gamma}\left\langle 
\bar{S}_{E}\right\rangle $, and $\left\langle \hat{S}_{E}\right\rangle
<<\gamma \left\langle \hat{S}_{E}\right\rangle $ and we approximate:%
\begin{equation*}
\frac{\left\langle \frac{1-\bar{S}_{E}}{1-\bar{S}}\right\rangle ^{2}}{%
\left\langle \frac{1-\hat{S}_{E}}{1-\hat{S}}\right\rangle ^{2}}\simeq c
\end{equation*}%
so that:%
\begin{equation*}
\frac{\xi ^{2}}{\zeta ^{2}}\simeq c\frac{\left\langle S_{E}^{B}\right\rangle
^{2}}{\left\langle S_{E}\right\rangle ^{2}}
\end{equation*}%
The constant $c$ is chosen so that for $\gamma =0$, $\frac{\xi ^{2}}{\zeta
^{2}}\rightarrow 1$, $c\frac{\left\langle \hat{S}_{E}^{B}\right\rangle ^{2}}{%
\left\langle S_{E}\right\rangle ^{2}}=1$, and this implies that $c=1$. We
will thus replace:%
\begin{equation*}
\frac{\xi ^{2}}{\zeta ^{2}}\rightarrow \frac{\left\langle
S_{E}^{B}\right\rangle ^{2}}{\left\langle S_{E}\right\rangle ^{2}}=\left[ 
\frac{\left\langle S_{E}^{B}\right\rangle }{\left\langle S_{E}\right\rangle }%
\right] ^{2}
\end{equation*}%
Moreover:%
\begin{equation*}
\bar{\zeta}^{2}\rightarrow \left\langle \frac{1-\bar{S}_{E}}{1-\bar{S}}%
\right\rangle ^{2}\left\langle \hat{S}_{E}^{B}\right\rangle ^{2}
\end{equation*}%
The coefficients corresponding to participations are given by:%
\begin{eqnarray*}
&&\left\langle \bar{w}\left( X^{\prime },X\right) \right\rangle \\
&\rightarrow &\frac{1}{1+\frac{\left\{ \frac{\left\langle \frac{1-\bar{S}_{E}%
}{1-\bar{S}}\right\rangle ^{2}\left\langle \hat{S}_{E}^{B}\right\rangle ^{2}%
}{\left\langle \hat{w}_{1}^{\left( 0\right) B}\left( \left( X^{\prime
}\right) ^{\prime },X^{\prime }\right) \right\rangle }+\left[ \frac{%
\left\langle S_{E}^{B}\right\rangle }{\left\langle S_{E}\right\rangle }%
\right] ^{2}\left( 1-\left( \gamma \left\langle \hat{S}_{E}\right\rangle
\right) ^{2}\right) \right\} }{2\left\langle \bar{w}_{1}^{\left( 0\right)
}\left( X^{\prime },X\right) \right\rangle \left( 1-\left( \bar{\gamma}\frac{%
\left\langle \bar{S}\right\rangle }{2}\right) ^{2}\right) }\left(
\left\langle w_{1}^{\left( 0\right) B}\left( X^{\prime },X\right)
\right\rangle +\frac{1}{\left[ \frac{\left\langle S_{E}^{B}\right\rangle }{%
\left\langle S_{E}\right\rangle }\right] ^{2}}\frac{1}{1-\left( \gamma
\left\langle \hat{S}_{E}\right\rangle \right) ^{2}}\right) } \\
&\rightarrow &\frac{2\left\langle \bar{w}_{1}^{\left( 0\right) }\left(
X^{\prime },X\right) \right\rangle \left( 1-\left( \bar{\gamma}\frac{%
\left\langle \bar{S}\right\rangle }{2}\right) ^{2}\right) }{2\left\langle 
\bar{w}_{1}^{\left( 0\right) }\left( X^{\prime },X\right) \right\rangle
\left( 1-\left( \bar{\gamma}\frac{\left\langle \bar{S}\right\rangle }{2}%
\right) ^{2}\right) +D}
\end{eqnarray*}%
with:%
\begin{eqnarray*}
D &=&\left\{ \frac{\left\langle \frac{1-\bar{S}_{E}}{1-\bar{S}}\right\rangle
^{2}\left\langle \hat{S}_{E}^{B}\right\rangle ^{2}}{\left\langle \hat{w}%
_{1}^{\left( 0\right) B}\left( \left( X^{\prime }\right) ^{\prime
},X^{\prime }\right) \right\rangle }+\left[ \frac{\left\langle
S_{E}^{B}\right\rangle }{\left\langle S_{E}\right\rangle }\right] ^{2}\left(
1-\left( \gamma \left\langle \hat{S}_{E}\right\rangle \right) ^{2}\right)
\right\} \\
&&\left( \left\langle w_{1}^{\left( 0\right) B}\left( X^{\prime },X\right)
\right\rangle +\frac{1}{\left[ \frac{\left\langle S_{E}^{B}\right\rangle }{%
\left\langle S_{E}\right\rangle }\right] ^{2}\left( 1-\left( \gamma
\left\langle \hat{S}_{E}\right\rangle \right) ^{2}\right) }\right)
\end{eqnarray*}%
\begin{eqnarray}
&&\left\langle \hat{w}_{E}^{B}\left( X^{\prime },X\right) \right\rangle
\label{HN} \\
&\rightarrow &\frac{1}{1+\frac{\left\langle \hat{w}_{1}^{\left( 0\right)
B}\left( \left( X^{\prime }\right) ^{\prime },X^{\prime }\right)
\right\rangle \frac{2\left\langle \bar{w}_{1}^{\left( 0\right) }\left(
X^{\prime },X\right) \right\rangle }{\left\langle w_{1}^{\left( 0\right)
B}\left( X^{\prime },X\right) \right\rangle }\left( 1-\left( \bar{\gamma}%
\frac{\left\langle \bar{S}\right\rangle }{2}\right) ^{2}\right) }{\left(
\left\langle \frac{1-\bar{S}_{E}}{1-\bar{S}}\right\rangle ^{2}\left\langle 
\hat{S}_{E}^{B}\right\rangle ^{2}+\left[ \frac{\left\langle
S_{E}^{B}\right\rangle }{\left\langle S_{E}\right\rangle }\right]
^{2}\left\langle \hat{w}_{1}^{\left( 0\right) B}\left( \left( X^{\prime
}\right) ^{\prime },X^{\prime }\right) \right\rangle \left( \left( 1-\left(
\gamma \left\langle \hat{S}_{E}\right\rangle \right) ^{2}\right) \right)
\right) }+\frac{\left\langle \hat{w}_{1}^{\left( 0\right) B}\left( \left(
X^{\prime }\right) ^{\prime },X^{\prime }\right) \right\rangle }{\left[ 
\frac{\left\langle S_{E}^{B}\right\rangle }{\left\langle S_{E}\right\rangle }%
\right] ^{2}w_{1}^{\left( 0\right) B}\left( X^{\prime },X\right) \left(
1-\left( \gamma \left\langle \hat{S}_{E}\right\rangle \right) ^{2}\right) }}
\notag \\
&\rightarrow &\frac{\left[ \frac{\left\langle S_{E}^{B}\right\rangle }{%
\left\langle S_{E}\right\rangle }\right] ^{2}\left\langle w_{1}^{\left(
0\right) B}\right\rangle \left( 1-\left( \gamma \left\langle \hat{S}%
_{E}\right\rangle \right) ^{2}\right) +\left\langle \hat{S}%
_{E}^{B}\right\rangle ^{2}}{\left\langle \hat{w}_{1}^{\left( 0\right)
B}\left( \left( X^{\prime }\right) ^{\prime },X^{\prime }\right)
\right\rangle \frac{2\left\langle \bar{w}_{1}^{\left( 0\right) }\left(
X^{\prime },X\right) \right\rangle }{\left\langle w_{1}^{\left( 0\right)
B}\left( X^{\prime },X\right) \right\rangle }\left( 1-\left( \bar{\gamma}%
\frac{\left\langle \bar{S}\right\rangle }{2}\right) ^{2}\right) +D^{\prime }}
\notag
\end{eqnarray}%
where:%
\begin{eqnarray*}
D^{\prime } &=&\left\langle \hat{w}_{1}^{\left( 0\right) B}\left( \left(
X^{\prime }\right) ^{\prime },X^{\prime }\right) \right\rangle \\
&&+\frac{\left\langle \frac{1-\bar{S}_{E}}{1-\bar{S}}\right\rangle
^{2}\left\langle \hat{S}_{E}^{B}\right\rangle ^{2}\left\langle \hat{w}%
_{1}^{\left( 0\right) B}\left( \left( X^{\prime }\right) ^{\prime
},X^{\prime }\right) \right\rangle }{\left[ \frac{\left\langle
S_{E}^{B}\right\rangle }{\left\langle S_{E}\right\rangle }\right]
^{2}w_{1}^{\left( 0\right) B}\left( X^{\prime },X\right) \left( 1-\left(
\gamma \left\langle \hat{S}_{E}\right\rangle \right) ^{2}\right) }+\left[ 
\frac{\left\langle S_{E}^{B}\right\rangle }{\left\langle S_{E}\right\rangle }%
\right] ^{2}\left\langle \hat{w}_{1}^{\left( 0\right) B}\left( \left(
X^{\prime }\right) ^{\prime },X^{\prime }\right) \right\rangle \left( \left(
1-\left( \gamma \left\langle \hat{S}_{E}\right\rangle \right) ^{2}\right)
\right) +\left\langle \hat{S}_{E}^{B}\right\rangle ^{2}
\end{eqnarray*}%
That is, in first approximtion:%
\begin{eqnarray}
&&\left\langle \bar{w}\left( X^{\prime },X\right) \right\rangle  \label{HBr}
\\
&\rightarrow &\frac{2\left( 1-\left( \bar{\gamma}\frac{\left\langle \bar{S}%
\right\rangle }{2}\right) ^{2}\right) }{1+\left[ \frac{\left\langle
S_{E}^{B}\right\rangle }{\left\langle S_{E}\right\rangle }\right] ^{2}\left(
1-\left( \gamma \left\langle \hat{S}_{E}\right\rangle \right) ^{2}\right)
+2\left( 1-\left( \bar{\gamma}\frac{\left\langle \bar{S}\right\rangle }{2}%
\right) ^{2}\right) +\frac{\left\langle \frac{1-\bar{S}_{E}}{1-\bar{S}}%
\right\rangle ^{2}\left\langle \hat{S}_{E}^{B}\right\rangle ^{2}}{\left[ 
\frac{\left\langle S_{E}^{B}\right\rangle }{\left\langle S_{E}\right\rangle }%
\right] ^{2}}\frac{1}{1-\left( \gamma \left\langle \hat{S}_{E}\right\rangle
\right) ^{2}}+\left\langle \frac{1-\bar{S}_{E}}{1-\bar{S}}\right\rangle
^{2}\left\langle \hat{S}_{E}^{B}\right\rangle ^{2}}  \notag
\end{eqnarray}%
\begin{equation}
\left\langle \hat{w}_{E}^{B}\left( X^{\prime },X\right) \right\rangle
\rightarrow \frac{\left\langle \frac{1-\bar{S}_{E}}{1-\bar{S}}\right\rangle
^{2}\left\langle \hat{S}_{E}^{B}\right\rangle ^{2}+\left[ \frac{\left\langle
S_{E}^{B}\right\rangle }{\left\langle S_{E}\right\rangle }\right] ^{2}\left(
\left( 1-\left( \gamma \left\langle \hat{S}_{E}\right\rangle \right)
^{2}\right) \right) }{1+2\left( 1-\left( \bar{\gamma}\frac{\left\langle \bar{%
S}\right\rangle }{2}\right) ^{2}\right) +\left[ \frac{\left\langle
S_{E}^{B}\right\rangle }{\left\langle S_{E}\right\rangle }\right] ^{2}\left(
\left( 1-\left( \gamma \left\langle \hat{S}_{E}\right\rangle \right)
^{2}\right) \right) +\frac{\left\langle \frac{1-\bar{S}_{E}}{1-\bar{S}}%
\right\rangle ^{2}\left\langle \hat{S}_{E}^{B}\right\rangle ^{2}}{\left[ 
\frac{\left\langle S_{E}^{B}\right\rangle }{\left\langle S_{E}\right\rangle }%
\right] ^{2}\left( 1-\left( \gamma \left\langle \hat{S}_{E}\right\rangle
\right) ^{2}\right) }+\left\langle \frac{1-\bar{S}_{E}}{1-\bar{S}}%
\right\rangle ^{2}\left\langle \hat{S}_{E}^{B}\right\rangle ^{2}}
\label{HBh}
\end{equation}%
Moreover:%
\begin{eqnarray}
&&\left\langle w_{E}^{B}\left( X,X\right) \right\rangle  \label{Hn} \\
&=&1-\left\langle \bar{w}\left( X^{\prime },X\right) \right\rangle
-\left\langle \hat{w}_{E}^{B}\left( X^{\prime },X\right) \right\rangle 
\notag \\
&\rightarrow &\frac{1+\frac{\left\langle \frac{1-\bar{S}_{E}}{1-\bar{S}}%
\right\rangle ^{2}\left\langle \hat{S}_{E}^{B}\right\rangle ^{2}}{\left[ 
\frac{\left\langle S_{E}^{B}\right\rangle }{\left\langle S_{E}\right\rangle }%
\right] ^{2}\left( 1-\left( \gamma \left\langle \hat{S}_{E}\right\rangle
\right) ^{2}\right) }}{1+2\left( 1-\left( \bar{\gamma}\frac{\left\langle 
\bar{S}\right\rangle }{2}\right) ^{2}\right) +\left[ \frac{\left\langle
S_{E}^{B}\right\rangle }{\left\langle S_{E}\right\rangle }\right] ^{2}\left(
\left( 1-\left( \gamma \left\langle \hat{S}_{E}\right\rangle \right)
^{2}\right) \right) +\frac{\left\langle \frac{1-\bar{S}_{E}}{1-\bar{S}}%
\right\rangle ^{2}\left\langle \hat{S}_{E}^{B}\right\rangle ^{2}}{\left[ 
\frac{\left\langle S_{E}^{B}\right\rangle }{\left\langle S_{E}\right\rangle }%
\right] ^{2}\left( 1-\left( \gamma \left\langle \hat{S}_{E}\right\rangle
\right) ^{2}\right) }+\left\langle \frac{1-\bar{S}_{E}}{1-\bar{S}}%
\right\rangle ^{2}\left\langle \hat{S}_{E}^{B}\right\rangle ^{2}}  \notag
\end{eqnarray}%
For coefficients related to loans, we assume that the uncertainties to
compute are:%
\begin{equation}
Un\left( X,\hat{f}\left( X^{\prime }\right) \right) =\frac{\zeta ^{2}}{%
w_{1}^{\left( 0\right) B}\left( X^{\prime },X\right) }\left( 1+\frac{\left(
\gamma \left\langle \hat{S}_{E}\left( X_{1},X^{\prime }\right) \right\rangle
_{X_{1}}\right) ^{2}}{1-\left( \gamma \left\langle \hat{S}_{E}\left(
X^{\prime },\left( X^{\prime }\right) ^{\prime }\right) \right\rangle
\right) ^{2}}\right)
\end{equation}%
and that the uncertainty of firms' returns are:%
\begin{equation}
Un\left( X,f^{\prime }\left( X\right) \right) \rightarrow \xi ^{2}
\end{equation}%
Ultimately, as in part one, we consider that uncertainty for loans is
identical that uncertainty for participations. As consequence:%
\begin{equation}
\left\langle \hat{w}_{2}^{B}\left( X^{\prime },X\right) \right\rangle =\frac{%
1-\left( \gamma \left\langle \hat{S}_{E}\left( X^{\prime },X\right)
\right\rangle \right) ^{2}}{2-\left( \gamma \left\langle \hat{S}_{E}\left(
X^{\prime },X\right) \right\rangle \right) ^{2}}=\left\langle \hat{w}%
_{2}\left( X^{\prime },X\right) \right\rangle =\left\langle \hat{w}%
_{E}\left( X^{\prime },X\right) \right\rangle  \label{Stn}
\end{equation}%
and:%
\begin{equation}
\left\langle w_{2}^{B}\left( X\right) \right\rangle =\frac{1}{2-\left(
\gamma \left\langle \hat{S}_{E}\left( X^{\prime },X\right) \right\rangle
\right) ^{2}}=\left\langle w_{2}\left( X^{\prime },X\right) \right\rangle
=\left\langle w_{E}\left( X^{\prime },X\right) \right\rangle  \label{ST}
\end{equation}%
which in turn leads to the shares of loans:%
\begin{eqnarray*}
&&\frac{\hat{S}_{L}^{B}\left( X^{\prime }\right) }{\kappa \left(
1-\left\langle \bar{S}\left( X\right) \right\rangle \right) } \\
&=&\hat{w}_{2}^{B}\left( X^{\prime }\right) \left\{ 1+w_{2}^{B}\left(
X\right) \left( \hat{r}\left( X^{\prime }\right) -\left\langle r\left(
X^{\prime }\right) \right\rangle \right) \right\} \frac{\left\langle \bar{K}%
\right\rangle \left\Vert \bar{\Psi}\right\Vert ^{2}}{\hat{K}_{X^{\prime
}}\left\vert \hat{\Psi}\left( X^{\prime }\right) \right\vert ^{2}}
\end{eqnarray*}%
\begin{equation}
\frac{S_{L}^{B}\left( X,X\right) }{\kappa \left( 1-\bar{S}\left( X\right)
\right) }=w_{2}^{B}\left( X\right) \left[ 1+\hat{w}_{2}^{B}\left( X\right)
\left( r\left( X\right) -\left\langle \hat{r}\left( X^{\prime }\right)
\right\rangle _{\hat{w}_{2}}\right) \right]
\end{equation}

\subsection*{A6.4. Averaged shares}

Average capital ratios and coefficients $w$ have been found as function of
return and shares. We can thus derive average shares equations.

\subsubsection*{A6.4.1 Investors average shares}

As in part I:%
\begin{eqnarray*}
\left\langle S_{E}\left( X\right) \right\rangle &=&\left\langle S_{E}\left(
X,X\right) \right\rangle \frac{\left\langle \hat{K}\right\rangle \left\Vert 
\hat{\Psi}\right\Vert ^{2}}{\left\langle K\right\rangle \left\Vert \Psi
\right\Vert ^{2}}=\left\langle S_{E}\right\rangle \frac{\left\langle \hat{K}%
\right\rangle \left\Vert \hat{\Psi}\right\Vert ^{2}}{\left\langle
K\right\rangle \left\Vert \Psi \right\Vert ^{2}} \\
\left\langle S\left( X\right) \right\rangle &=&\left\langle S\left(
X,X\right) \right\rangle \frac{\left\langle \hat{K}\right\rangle \left\Vert 
\hat{\Psi}\right\Vert ^{2}}{\left\langle K\right\rangle \left\Vert \Psi
\right\Vert ^{2}}=\left\langle S\right\rangle \frac{\left\langle \hat{K}%
\right\rangle \left\Vert \hat{\Psi}\right\Vert ^{2}}{\left\langle
K\right\rangle \left\Vert \Psi \right\Vert ^{2}}
\end{eqnarray*}%
where the average shares are given by:%
\begin{eqnarray*}
\left\langle S_{E}\right\rangle &=&\left\langle S_{E}\left( X,X\right)
\right\rangle \\
&=&\frac{w\left( X\right) }{2}\left( 1+\left( \left\langle \hat{w}\left(
X\right) \right\rangle \left( \left\langle f\left( X\right) \right\rangle -%
\frac{\left\langle \hat{f}\left( X^{\prime }\right) \right\rangle _{\hat{w}%
_{E}}+\left\langle \hat{r}\left( X^{\prime }\right) \right\rangle _{\hat{w}%
_{2}}}{2}\right) +\frac{w\left( X\right) }{2}\left( f\left( X\right) -\bar{r}%
\left( X\right) \right) \right) \right)
\end{eqnarray*}%
\begin{equation*}
\left\langle S\left( X,X\right) \right\rangle =w\left( X\right) \left(
1+\left( \hat{w}\left( X\right) \left( \frac{\left\langle f\left( X\right)
\right\rangle +\left\langle \bar{r}\left( X\right) \right\rangle }{2}-\frac{%
\left\langle \hat{f}\left( X^{\prime }\right) \right\rangle _{\hat{w}%
_{E}}+\left\langle \hat{r}\left( X^{\prime }\right) \right\rangle _{\hat{w}%
_{2}}}{2}\right) \right) \right)
\end{equation*}

\subsubsection*{A6.4.2 Average coeffcients}

The average coeffcients are:%
\begin{equation*}
\left\langle \hat{w}\right\rangle \rightarrow \frac{\zeta ^{2}}{\zeta
^{2}+\zeta ^{2}\frac{1}{1-\left( \gamma \left\langle \hat{S}_{E}\left(
X\right) \right\rangle \right) ^{2}}}=\frac{1-\left( \gamma \left\langle 
\hat{S}_{E}\left( X\right) \right\rangle \right) ^{2}}{2-\left( \gamma
\left\langle \hat{S}_{E}\left( X\right) \right\rangle \right) ^{2}}
\end{equation*}%
\begin{equation*}
\left\langle w\right\rangle \rightarrow \frac{1}{2-\left( \gamma
\left\langle \hat{S}_{E}\left( X\right) \right\rangle \right) ^{2}}
\end{equation*}

\subsubsection{A6.4.3 Bank average shares: formula for participtions}

Averaging the shares formula leads to:%
\begin{eqnarray}
&&\left\langle \bar{S}_{E}\right\rangle =\frac{\left\langle \bar{w}\left(
X^{\prime }\right) \right\rangle }{2}\left( 1+\left\{ \left\langle \bar{w}%
\left( X\right) \right\rangle \left( \frac{\left\langle \bar{f}\left(
X^{\prime }\right) \right\rangle -\left\langle \bar{r}\left( X^{\prime
}\right) \right\rangle }{2}\right) \right. \right.  \label{SNB} \\
&&\left. \left. +\left\langle \hat{w}_{E}^{B}\left( X\right) \right\rangle
\left( \left\langle \bar{f}\left( X^{\prime }\right) \right\rangle
-\left\langle \hat{f}\left( X^{\prime }\right) \right\rangle _{\hat{w}%
_{E}}\right) +\left\langle w_{E}^{B}\left( X\right) \right\rangle \left(
\left\langle \bar{f}\left( X^{\prime }\right) \right\rangle -\left\langle
f\left( X\right) \right\rangle \right) \right\} \right)  \notag
\end{eqnarray}%
where:%
\begin{equation*}
\left\langle \bar{w}\left( X\right) \right\rangle \simeq \frac{2\left(
1-\left( \bar{\gamma}\frac{\left\langle \bar{S}\right\rangle }{2}\right)
^{2}\right) }{1+\left( \left[ \frac{\left\langle S_{E}^{B}\right\rangle }{%
\left\langle S_{E}\right\rangle }\right] ^{2}+2\right) \left( 1-\left(
\gamma \left\langle \hat{S}_{E}\right\rangle \right) ^{2}\right) +\frac{%
\left\langle \frac{1-\bar{S}_{E}}{1-\bar{S}}\right\rangle ^{2}\left\langle 
\hat{S}_{E}^{B}\right\rangle ^{2}}{\left[ \frac{\left\langle
S_{E}^{B}\right\rangle }{\left\langle S_{E}\right\rangle }\right] ^{2}\left(
1-\left( \gamma \left\langle \hat{S}_{E}\right\rangle \right) ^{2}\right) }%
+\left\langle \frac{1-\bar{S}_{E}}{1-\bar{S}}\right\rangle ^{2}\left\langle 
\hat{S}_{E}^{B}\right\rangle ^{2}}
\end{equation*}%
\begin{equation}
\left\langle \bar{S}\right\rangle =\left\langle \bar{w}\left( X^{\prime
}\right) \right\rangle \left[ 1+\left\langle \hat{w}_{E}^{B}\left( X\right)
\right\rangle \left( \frac{\left\langle \bar{f}\left( X^{\prime }\right)
\right\rangle +\left\langle \bar{r}\left( X^{\prime }\right) \right\rangle }{%
2}-\left\langle \hat{f}\left( X^{\prime }\right) \right\rangle _{\hat{w}%
_{E}}\right) +\left\langle w_{E}^{B}\left( X\right) \right\rangle \left( 
\frac{\left\langle \bar{f}\left( X^{\prime }\right) \right\rangle
+\left\langle \bar{r}\left( X^{\prime }\right) \right\rangle }{2}%
-\left\langle f\left( X\right) \right\rangle \right) \right]  \label{Sb}
\end{equation}%
The shares $\left\langle \bar{S}\right\rangle $ and $\left\langle \bar{S}%
_{E}\right\rangle $ satisfy the relation:%
\begin{equation}
\left\langle \bar{S}\right\rangle =2\left\langle \bar{S}_{E}\right\rangle
-\left\langle \bar{w}\left( X^{\prime }\right) \right\rangle \left( \frac{%
\left\langle \bar{f}\left( X^{\prime }\right) \right\rangle -\left\langle 
\bar{r}\left( X^{\prime }\right) \right\rangle }{2}\right)  \label{Sbn}
\end{equation}%
The shares of participations invested in investrs and firms are given by:%
\begin{eqnarray}
\left\langle \hat{S}_{E}^{B}\right\rangle &=&\left\langle \hat{S}%
_{E}^{B}\left( X^{\prime },X\right) \right\rangle  \label{VRH} \\
&=&\left\langle \hat{w}_{E}^{B}\left( X^{\prime }\right) \right\rangle \left[
1+\left\langle \bar{w}\left( X\right) \right\rangle \left( \left\langle \hat{%
f}\left( X^{\prime }\right) \right\rangle -\frac{\left\langle \bar{f}\left(
X^{\prime }\right) \right\rangle _{\bar{w}_{E}}+\left\langle \bar{r}\left(
X^{\prime }\right) \right\rangle _{\bar{w}_{2}}}{2}\right) \right.  \notag \\
&&\left. +\left\langle w_{E}^{B}\left( X\right) \right\rangle \left(
\left\langle \hat{f}\left( X^{\prime }\right) \right\rangle -\left\langle
f\left( X\right) \right\rangle \right) \right]  \notag
\end{eqnarray}%
and:%
\begin{eqnarray}
\left\langle S_{E}^{B}\right\rangle &=&\left\langle S_{E}^{B}\left(
X,X\right) \right\rangle  \label{VRB} \\
&=&\left\langle w_{E}^{B}\left( X\right) \right\rangle \left\{
1+\left\langle \bar{w}\left( X\right) \right\rangle \left( \left\langle
f\left( X\right) \right\rangle -\frac{\left\langle \bar{f}\left( X^{\prime
}\right) \right\rangle _{\bar{w}_{E}}+\left\langle \bar{r}\left( X^{\prime
}\right) \right\rangle _{\bar{w}_{2}}}{2}\right) +\left\langle \hat{w}%
_{E}^{B}\left( X\right) \right\rangle \left( \left\langle f\left( X\right)
\right\rangle -\left\langle \hat{f}\left( X^{\prime }\right) \right\rangle _{%
\hat{w}_{E}}\right) \right\}  \notag
\end{eqnarray}

with $\left\langle \hat{w}_{E}^{B}\left( X^{\prime }\right) \right\rangle $%
and $\left\langle w_{E}^{B}\left( X\right) \right\rangle $ given by (\ref{HN}%
) and (\ref{Hn}). Ultimately, we have:%
\begin{equation*}
\left\langle \hat{S}_{E}^{B}\left( X^{\prime }\right) \right\rangle
=\left\langle \hat{S}_{E}^{B}\right\rangle \frac{\left\langle \bar{K}%
\right\rangle \left\Vert \bar{\Psi}\right\Vert ^{2}}{\left\langle \hat{K}%
\right\rangle \left\Vert \hat{\Psi}\right\Vert ^{2}}
\end{equation*}%
\begin{eqnarray*}
\left\langle S_{E}^{B}\left( X\right) \right\rangle &=&\left\langle
S_{E}^{B}\left( X,X\right) \right\rangle \frac{\left\langle \bar{K}%
\right\rangle \left\Vert \bar{\Psi}\right\Vert ^{2}}{\left\langle
K\right\rangle \left\Vert \Psi \right\Vert ^{2}} \\
&=&\left\langle S_{E}^{B}\right\rangle \frac{\left\langle \bar{K}%
\right\rangle \left\Vert \bar{\Psi}\right\Vert ^{2}}{\left\langle
K\right\rangle \left\Vert \Psi \right\Vert ^{2}}
\end{eqnarray*}

\subsubsection{A6.4.4 Bank average shares: formula for loans}

Usng (\ref{Stn}) and (\ref{ST}), the coefficients associatd to loans are:%
\begin{equation}
\left\langle \hat{w}_{2}^{B}\left( X^{\prime },X\right) \right\rangle =\frac{%
1-\left( \gamma \left\langle \hat{S}_{E}\left( X^{\prime },X\right)
\right\rangle \right) ^{2}}{2-\left( \gamma \left\langle \hat{S}_{E}\left(
X^{\prime },X\right) \right\rangle \right) ^{2}}=\left\langle \hat{w}%
_{2}\left( X^{\prime },X\right) \right\rangle =\left\langle \hat{w}%
_{E}\left( X^{\prime },X\right) \right\rangle
\end{equation}%
and:%
\begin{equation}
\left\langle w_{2}^{B}\left( X\right) \right\rangle =\frac{1}{2-\left(
\gamma \left\langle \hat{S}_{E}\left( X^{\prime },X\right) \right\rangle
\right) ^{2}}=\left\langle w_{2}\left( X^{\prime },X\right) \right\rangle
=\left\langle w_{E}\left( X^{\prime },X\right) \right\rangle
\end{equation}%
so that the shares for loans are:%
\begin{equation}
\frac{\left\langle \hat{S}_{L}^{B}\left( X^{\prime },X\right) \right\rangle 
}{\kappa \left( 1-\left\langle \bar{S}\left( X\right) \right\rangle \right) }%
=\frac{1-\left( \gamma \left\langle \hat{S}_{E}\left( X^{\prime },X\right)
\right\rangle \right) ^{2}}{2-\left( \gamma \left\langle \hat{S}_{E}\left(
X^{\prime },X\right) \right\rangle \right) ^{2}}\left\{ 1+\frac{\left(
\left\langle \hat{r}\left( X^{\prime }\right) \right\rangle -\left\langle
r\left( X^{\prime }\right) \right\rangle \right) }{2-\left( \gamma
\left\langle \hat{S}_{E}\left( X^{\prime },X\right) \right\rangle \right)
^{2}}\right\}  \label{frmn}
\end{equation}%
\begin{equation}
\frac{\left\langle S_{L}^{B}\left( X,X\right) \right\rangle }{\kappa \left(
1-\left\langle \bar{S}\left( X\right) \right\rangle \right) }=\frac{1}{%
2-\left( \gamma \left\langle \hat{S}_{E}\left( X^{\prime },X\right)
\right\rangle \right) ^{2}}\left[ 1+\frac{1-\left( \gamma \left\langle \hat{S%
}_{E}\left( X^{\prime },X\right) \right\rangle \right) ^{2}}{2-\left( \gamma
\left\langle \hat{S}_{E}\left( X^{\prime },X\right) \right\rangle \right)
^{2}}\left( \left\langle r\left( X\right) \right\rangle -\left\langle \hat{r}%
\left( X^{\prime }\right) \right\rangle _{\hat{w}_{2}}\right) \right]
\label{frmd}
\end{equation}

\subsection*{A6.5 Equations for $\left\langle \hat{S}_{E}\left( X\right)
\right\rangle $, $\left\langle S_{E}^{B}\right\rangle $, $\left\langle \hat{S%
}_{E}^{B}\right\rangle $}

Equations (\ref{HBr}), (\ref{HBh}), (\ref{Hn}) and the constraint along with
the constrnt:%
\begin{equation*}
\left\langle \hat{S}_{E}^{B}\right\rangle +\left\langle
S_{E}^{B}\right\rangle +2\left\langle \bar{S}_{E}\right\rangle =1
\end{equation*}%
yields a system of equations:%
\begin{eqnarray*}
\frac{\left\langle w_{E}^{B}\left( X,X\right) \right\rangle }{\left\langle 
\bar{w}\left( X^{\prime },X\right) \right\rangle } &=&\frac{1+\frac{%
\left\langle \frac{1-\bar{S}_{E}}{1-\bar{S}}\right\rangle ^{2}\left\langle 
\hat{S}_{E}^{B}\right\rangle ^{2}}{\left[ \frac{\left\langle
S_{E}^{B}\right\rangle }{\left\langle S_{E}\right\rangle }\right] ^{2}\left(
1-\left( \gamma \left\langle \hat{S}_{E}\right\rangle \right) ^{2}\right) }}{%
2\left( 1-\left( \bar{\gamma}\frac{\left\langle \bar{S}\right\rangle }{2}%
\right) ^{2}\right) } \\
\frac{\left\langle \hat{w}_{E}^{B}\left( X^{\prime },X\right) \right\rangle 
}{\left\langle \bar{w}\left( X^{\prime },X\right) \right\rangle } &=&\frac{%
\left\langle \frac{1-\bar{S}_{E}}{1-\bar{S}}\right\rangle ^{2}\left\langle 
\hat{S}_{E}^{B}\right\rangle ^{2}+\left[ \frac{\left\langle
S_{1}^{B}\right\rangle }{\left\langle S_{E}\right\rangle }\right] ^{2}\left(
\left( 1-\left( \gamma \left\langle \hat{S}_{E}\right\rangle \right)
^{2}\right) \right) }{2\left( 1-\left( \bar{\gamma}\frac{\left\langle \bar{S}%
\right\rangle }{2}\right) ^{2}\right) } \\
\left\langle \hat{S}_{E}^{B}\right\rangle +\left\langle
S_{E}^{B}\right\rangle +2\left\langle \bar{S}_{E}\right\rangle &=&1
\end{eqnarray*}%
The coefficients are themselves expressed in terms of shares by using (\ref%
{SNB}), (\ref{Sbn}), (\ref{VRH}):%
\begin{eqnarray}
&&\left\langle \bar{S}_{E}\right\rangle =\frac{\left\langle \bar{w}\left(
X^{\prime }\right) \right\rangle }{2}\left( 1+\left\{ \left\langle \bar{w}%
\left( X\right) \right\rangle \left( \frac{\left\langle \bar{f}\left(
X^{\prime }\right) \right\rangle -\left\langle \bar{r}\left( X^{\prime
}\right) \right\rangle }{2}\right) \right. \right. \\
&&\left. \left. +\left\langle \hat{w}_{E}^{B}\left( X\right) \right\rangle
\left( \left\langle \bar{f}\left( X^{\prime }\right) \right\rangle
-\left\langle \hat{f}\left( X^{\prime }\right) \right\rangle _{\hat{w}%
_{E}}\right) +\left\langle w_{E}^{B}\left( X\right) \right\rangle \left(
\left\langle \bar{f}\left( X^{\prime }\right) \right\rangle -\left\langle
f\left( X\right) \right\rangle \right) \right\} \right)  \notag
\end{eqnarray}%
\begin{eqnarray}
\left\langle \hat{S}_{E}^{B}\right\rangle &=&\left\langle \hat{S}%
_{E}^{B}\left( X^{\prime },X\right) \right\rangle \\
&=&\left\langle \hat{w}_{E}^{B}\left( X^{\prime }\right) \right\rangle \left[
1+\left\langle \bar{w}\left( X\right) \right\rangle \left( \left\langle \hat{%
f}\left( X^{\prime }\right) \right\rangle -\frac{\left\langle \bar{f}\left(
X^{\prime }\right) \right\rangle _{\bar{w}_{E}}+\left\langle \bar{r}\left(
X^{\prime }\right) \right\rangle _{\bar{w}_{2}}}{2}\right) \right.  \notag \\
&&\left. +\left\langle w_{E}^{B}\left( X\right) \right\rangle \left(
\left\langle \hat{f}\left( X^{\prime }\right) \right\rangle -\left\langle
f\left( X\right) \right\rangle \right) \right]  \notag
\end{eqnarray}%
and:%
\begin{eqnarray}
\left\langle S_{E}^{B}\right\rangle &=&\left\langle S_{E}^{B}\left(
X,X\right) \right\rangle \\
&=&\left\langle w_{E}^{B}\left( X\right) \right\rangle \left\{
1+\left\langle \bar{w}\left( X\right) \right\rangle \left( \left\langle
f\left( X\right) \right\rangle -\frac{\left\langle \bar{f}\left( X^{\prime
}\right) \right\rangle _{\bar{w}_{E}}+\left\langle \bar{r}\left( X^{\prime
}\right) \right\rangle _{\bar{w}_{2}}}{2}\right) +\left\langle \hat{w}%
_{E}^{B}\left( X\right) \right\rangle \left( \left\langle f\left( X\right)
\right\rangle -\left\langle \hat{f}\left( X^{\prime }\right) \right\rangle _{%
\hat{w}_{E}}\right) \right\}  \notag
\end{eqnarray}%
These equations allow to express $\left\langle \bar{S}_{E}\right\rangle $ $%
\left\langle \hat{S}_{E}\right\rangle $, $\left\langle
S_{E}^{B}\right\rangle $, $\left\langle \hat{S}_{E}^{B}\right\rangle $ as
implicit functions of $\left\langle \bar{f}\left( X^{\prime }\right)
\right\rangle -\left\langle \bar{r}\left( X^{\prime }\right) \right\rangle $%
, $\left\langle \hat{f}\left( X^{\prime }\right) \right\rangle -\left\langle 
\bar{r}\left( X^{\prime }\right) \right\rangle $, $\left\langle f\left(
X\right) \right\rangle -\left\langle \bar{r}\left( X^{\prime }\right)
\right\rangle $:%
\begin{eqnarray}
&&\left\langle \bar{S}_{E}\right\rangle \left( \left\langle \bar{f}\left(
X^{\prime }\right) \right\rangle ,\left\langle \hat{f}\left( X^{\prime
}\right) \right\rangle ,\left\langle f\left( X\right) \right\rangle
,\left\langle \bar{r}\left( X^{\prime }\right) \right\rangle \right)
\label{DPNC} \\
&&\left\langle \hat{S}_{E}\right\rangle \left( \left\langle \bar{f}\left(
X^{\prime }\right) \right\rangle ,\left\langle \hat{f}\left( X^{\prime
}\right) \right\rangle ,\left\langle f\left( X\right) \right\rangle
,\left\langle \bar{r}\left( X^{\prime }\right) \right\rangle \right)  \notag
\\
&&\left\langle S_{E}^{B}\right\rangle \left( \left\langle \bar{f}\left(
X^{\prime }\right) \right\rangle ,\left\langle \hat{f}\left( X^{\prime
}\right) \right\rangle ,\left\langle f\left( X\right) \right\rangle
,\left\langle \bar{r}\left( X^{\prime }\right) \right\rangle \right)  \notag
\\
&&\left\langle \hat{S}_{E}^{B}\right\rangle \left( \left\langle \bar{f}%
\left( X^{\prime }\right) \right\rangle ,\left\langle \hat{f}\left(
X^{\prime }\right) \right\rangle ,\left\langle f\left( X\right)
\right\rangle ,\left\langle \bar{r}\left( X^{\prime }\right) \right\rangle
\right)  \notag
\end{eqnarray}%
We present in this section approximate solutions at the zeroth and first
order in these variables.

Alternatively, we will express below $\left\langle \bar{f}\left( X^{\prime
}\right) \right\rangle -\left\langle \bar{r}\left( X^{\prime }\right)
\right\rangle $ and $\left\langle \hat{f}\left( X^{\prime }\right)
\right\rangle -\left\langle \bar{r}\left( X^{\prime }\right) \right\rangle $
as functions of the shares $\left\langle \bar{S}_{E}\right\rangle $ and $%
\left\langle \hat{S}_{E}\left( X\right) \right\rangle $ and the retrn
equations will be expressd in these variables.

\subsubsection*{A6.5.1 Lowest order formula}

\paragraph*{A6.5.1.1 \ Computation of $\left\langle w_{E}^{B}\left( X\right)
\right\rangle $ and $\left\langle S_{E}^{B}\right\rangle $}

At the lowest order, given (\ref{Sb}), (\ref{VRB}), (\ref{VRH}) we have:%
\begin{equation*}
\left\langle \bar{S}\right\rangle =\bar{w}\left( X^{\prime }\right)
\end{equation*}%
\begin{equation*}
\frac{\left\langle \bar{S}\right\rangle }{2}=\left\langle \bar{S}%
_{E}\right\rangle
\end{equation*}%
\begin{equation*}
\left\langle S_{E}^{B}\right\rangle =\left\langle w_{E}^{B}\left( X\right)
\right\rangle
\end{equation*}%
\begin{equation*}
\left\langle \hat{S}_{E}^{B}\right\rangle =\left\langle \hat{w}%
_{E}^{B}\left( X^{\prime }\right) \right\rangle
\end{equation*}%
so that the ratios of stakes are given by:%
\begin{equation}
\frac{\left\langle S_{E}^{B}\right\rangle }{\left\langle \bar{S}%
_{E}\right\rangle }=\frac{1+\frac{\left( \frac{1-\bar{S}_{E}}{1-\bar{S}}%
\right) ^{2}\left\langle \hat{S}_{E}^{B}\right\rangle ^{2}}{\left[ \frac{%
\left\langle S_{E}^{B}\right\rangle }{\left\langle S_{E}\right\rangle }%
\right] ^{2}\left( 1-\left( \gamma \left\langle \hat{S}_{E}\right\rangle
\right) ^{2}\right) }}{\left( 1-\left( \bar{\gamma}\left\langle \bar{S}%
_{E}\right\rangle \right) ^{2}\right) }  \label{Ptn}
\end{equation}%
\begin{equation}
\left( \frac{1-\bar{S}_{E}}{1-\bar{S}}\right) ^{2}\left\langle \hat{S}%
_{E}^{B}\right\rangle ^{2}+\left[ \frac{\left\langle S_{E}^{B}\right\rangle 
}{\left\langle S_{E}\right\rangle }\right] ^{2}\left( 1-\left( \gamma
\left\langle \hat{S}_{E}\right\rangle \right) ^{2}\right) =\left( 1-\left( 
\bar{\gamma}\left\langle \bar{S}_{E}\right\rangle \right) ^{2}\right) \frac{%
\left\langle \hat{S}_{E}^{B}\right\rangle }{\left\langle \bar{S}%
_{E}\right\rangle }  \label{Pts}
\end{equation}

This is solved by replacing (\ref{Pts})\bigskip\ in (\ref{Ptn}) which
becomes:%
\begin{equation*}
\frac{\left\langle S_{E}^{B}\right\rangle }{\left\langle \bar{S}%
_{E}\right\rangle }=\frac{1+\frac{\left( \frac{1-\bar{S}_{E}}{1-\bar{S}}%
\right) ^{2}\left\langle \hat{S}_{E}^{B}\right\rangle ^{2}}{\left( 1-\left( 
\bar{\gamma}\left\langle \bar{S}_{E}\right\rangle \right) ^{2}\right) \frac{%
\left\langle \hat{S}_{E}^{B}\right\rangle }{\left\langle \bar{S}%
_{E}\right\rangle }-\left( \frac{1-\bar{S}_{E}}{1-\bar{S}}\right)
^{2}\left\langle \hat{S}_{E}^{B}\right\rangle ^{2}}}{\left( 1-\left( \bar{%
\gamma}\left\langle \bar{S}_{E}\right\rangle \right) ^{2}\right) }=\frac{1+%
\frac{1}{\left( 1-\left( \bar{\gamma}\left\langle \bar{S}_{E}\right\rangle
\right) ^{2}\right) \frac{1}{\left( \frac{1-\bar{S}_{E}}{1-\bar{S}}\right)
^{2}\bar{S}_{E}\left\langle \hat{S}_{E}^{B}\right\rangle }-1}}{\left(
1-\left( \bar{\gamma}\left\langle \bar{S}_{E}\right\rangle \right)
^{2}\right) }
\end{equation*}%
Rewriting this equation as:%
\begin{equation*}
\frac{\left\langle S_{E}^{B}\right\rangle }{\left\langle \bar{S}%
_{E}\right\rangle }\left( 1-\left( \bar{\gamma}\left\langle \bar{S}%
_{E}\right\rangle \right) ^{2}\right) =1+\frac{\left( \frac{1-\bar{S}_{E}}{1-%
\bar{S}}\right) ^{2}\bar{S}_{E}\left\langle \hat{S}_{E}^{B}\right\rangle }{%
\left( 1-\left( \bar{\gamma}\left\langle \bar{S}_{E}\right\rangle \right)
^{2}\right) -\left( \frac{1-\bar{S}_{E}}{1-\bar{S}}\right) ^{2}\bar{S}%
_{E}\left\langle \hat{S}_{E}^{B}\right\rangle }
\end{equation*}%
yields the following relation:%
\begin{equation*}
\left\langle S_{E}^{B}\right\rangle \left( \left( 1-\left( \bar{\gamma}%
\left\langle \bar{S}_{E}\right\rangle \right) ^{2}\right) -\left( \frac{1-%
\bar{S}_{E}}{1-\bar{S}}\right) ^{2}\bar{S}_{E}\left\langle \hat{S}%
_{E}^{B}\right\rangle \right) =\left\langle \bar{S}_{E}\right\rangle
\end{equation*}%
Then, using that:%
\begin{equation*}
\left\langle \hat{S}_{E}^{B}\right\rangle =1-2\left\langle \bar{S}%
_{E}\right\rangle -\left\langle S_{E}^{B}\right\rangle
\end{equation*}%
this becomes an equation for $\left\langle S_{E}^{B}\right\rangle $ as a
function of $\left\langle \bar{S}_{E}\right\rangle $:%
\begin{equation}
0=\left\langle S_{E}^{B}\right\rangle \left( \left( 1-\left( \bar{\gamma}%
\left\langle \bar{S}_{E}\right\rangle \right) ^{2}\right) -\left( \frac{1-%
\bar{S}_{E}}{1-\bar{S}}\right) ^{2}\left\langle \bar{S}_{E}\right\rangle
\left( 1-2\bar{S}_{E}-\left\langle S_{E}^{B}\right\rangle \right) \right) -%
\bar{S}_{E}  \label{QNSn}
\end{equation}%
with solution:%
\begin{eqnarray}
\left\langle S_{E}^{B}\right\rangle &=&\left\langle w_{1}^{B}\left( X\right)
\right\rangle _{0}  \label{hbz} \\
&=&\frac{\sqrt{\left( 1-\left( \bar{\gamma}\left\langle \bar{S}%
_{E}\right\rangle \right) ^{2}-\left( \frac{1-\bar{S}_{E}}{1-2\bar{S}_{E}}%
\right) ^{2}\left( \left\langle \bar{S}_{E}\right\rangle -2\left\langle \bar{%
S}_{E}\right\rangle ^{2}\right) \right) ^{2}+4\left( \frac{1-\bar{S}_{E}}{1-2%
\bar{S}_{E}}\left\langle \bar{S}_{E}\right\rangle \right) ^{2}}}{2\left( 
\frac{1-\bar{S}_{E}}{1-2\bar{S}_{E}}\right) ^{2}\left\langle \bar{S}%
_{E}\right\rangle }  \notag \\
&&-\frac{\left( 1-\left( \bar{\gamma}\left\langle \bar{S}_{E}\right\rangle
\right) ^{2}-\left( \frac{1-\bar{S}_{E}}{1-2\bar{S}_{E}}\right) ^{2}\left(
\left\langle \bar{S}_{E}\right\rangle -2\left\langle \bar{S}%
_{E}\right\rangle ^{2}\right) \right) }{2\left( \frac{1-\bar{S}_{E}}{1-2\bar{%
S}_{E}}\right) ^{2}\left\langle \bar{S}_{E}\right\rangle }  \notag \\
&=&\frac{1}{2}\left( \sqrt{\left( \frac{1-\left( \bar{\gamma}\left\langle 
\bar{S}_{E}\right\rangle \right) ^{2}}{\left( \frac{1-\bar{S}_{E}}{1-2\bar{S}%
_{E}}\right) ^{2}\left\langle \bar{S}_{E}\right\rangle }-\left(
1-2\left\langle \bar{S}_{E}\right\rangle \right) \right) ^{2}+\frac{4}{%
\left( \frac{1-\bar{S}_{E}}{1-2\bar{S}_{E}}\right) ^{2}}}-\left( \frac{%
1-\left( \bar{\gamma}\left\langle \bar{S}_{E}\right\rangle \right) ^{2}}{%
\left( \frac{1-\bar{S}_{E}}{1-2\bar{S}_{E}}\right) ^{2}\left\langle \bar{S}%
_{E}\right\rangle }-\left( 1-2\left\langle \bar{S}_{E}\right\rangle \right)
\right) \right)  \notag
\end{eqnarray}

\paragraph*{A6.5.1.2 \ Computation of $\left\langle \hat{w}_{E}^{B}\left(
X^{\prime }\right) \right\rangle $ and $\left\langle \hat{S}%
_{E}^{B}\right\rangle $}

Average $\left\langle \hat{S}_{E}^{B}\right\rangle $ is computed at the
lowest order using:%
\begin{eqnarray}
\left\langle \hat{S}_{E}^{B}\right\rangle &=&\left\langle \hat{w}%
_{1}^{B}\left( X^{\prime }\right) \right\rangle _{0}  \label{hbzt} \\
&=&1-2\bar{S}_{E}-\left\langle S_{E}^{B}\right\rangle  \notag \\
&=&\frac{1}{2}\left( \left( \frac{1-\left( \bar{\gamma}\left\langle \bar{S}%
_{E}\right\rangle \right) ^{2}}{\left( \frac{1-\bar{S}_{E}}{1-2\bar{S}_{E}}%
\right) ^{2}\left\langle \bar{S}_{E}\right\rangle }+\left( 1-2\left\langle 
\bar{S}_{E}\right\rangle \right) \right) -\sqrt{\left( \frac{1-\left( \bar{%
\gamma}\left\langle \bar{S}_{E}\right\rangle \right) ^{2}}{\left( \frac{1-%
\bar{S}_{E}}{1-2\bar{S}_{E}}\right) ^{2}\left\langle \bar{S}%
_{E}\right\rangle }-\left( 1-2\left\langle \bar{S}_{E}\right\rangle \right)
\right) ^{2}+\frac{4}{\left( \frac{1-\bar{S}_{E}}{1-2\bar{S}_{E}}\right) ^{2}%
}}\right)  \notag
\end{eqnarray}

\paragraph*{A6.5.1.2 \ Equation for $\bar{S}_{E}$}

We use both equation for $\left\langle S_{E}^{B}\right\rangle $ and $%
\left\langle \hat{S}_{E}^{B}\right\rangle $ to find the zeroth order
equation for $\bar{S}_{E}$: 
\begin{equation*}
\frac{\left\langle S_{E}^{B}\right\rangle }{\left\langle \bar{S}%
_{E}\right\rangle }=\frac{1+\frac{\left( \frac{1-\bar{S}_{E}}{1-\bar{S}}%
\right) ^{2}\left\langle \hat{S}_{E}^{B}\right\rangle ^{2}}{\left[ \frac{%
\left\langle S_{E}^{B}\right\rangle }{\left\langle S_{E}\right\rangle }%
\right] ^{2}\left( 1-\left( \gamma \left\langle \hat{S}_{E}\right\rangle
\right) ^{2}\right) }}{\left( 1-\left( \bar{\gamma}\left\langle \bar{S}%
_{E}\right\rangle \right) ^{2}\right) }
\end{equation*}%
writes:%
\begin{equation*}
\frac{\left\langle S_{E}^{B}\right\rangle }{\left\langle \bar{S}%
_{E}\right\rangle }\left( 1-\left( \bar{\gamma}\left\langle \bar{S}%
_{E}\right\rangle \right) ^{2}\right) =\frac{\left[ \frac{\left\langle
S_{E}^{B}\right\rangle }{\left\langle S_{E}\right\rangle }\right] ^{2}\left(
1-\left( \gamma \left\langle \hat{S}_{E}\right\rangle \right) ^{2}\right)
+\left( \frac{1-\bar{S}_{E}}{1-\bar{S}}\right) ^{2}\left\langle \hat{S}%
_{E}^{B}\right\rangle ^{2}}{\left[ \frac{\left\langle S_{E}^{B}\right\rangle 
}{\left\langle S_{E}\right\rangle }\right] ^{2}\left( 1-\left( \gamma
\left\langle \hat{S}_{E}\right\rangle \right) ^{2}\right) }
\end{equation*}%
Compared to:%
\begin{equation*}
\left( \frac{1-\bar{S}_{E}}{1-\bar{S}}\right) ^{2}\left\langle \hat{S}%
_{E}^{B}\right\rangle ^{2}+\left[ \frac{\left\langle S_{E}^{B}\right\rangle 
}{\left\langle S_{E}\right\rangle }\right] ^{2}\left( 1-\left( \gamma
\left\langle \hat{S}_{E}\right\rangle \right) ^{2}\right) =\left( 1-\left( 
\bar{\gamma}\left\langle \bar{S}_{E}\right\rangle \right) ^{2}\right) \frac{%
\left\langle \hat{S}_{E}^{B}\right\rangle }{\left\langle \bar{S}%
_{E}\right\rangle }
\end{equation*}%
We find the ratio $\frac{\left\langle \hat{S}_{E}^{B}\right\rangle }{%
\left\langle S_{E}^{B}\right\rangle }$:%
\begin{equation*}
\frac{\left\langle \hat{S}_{E}^{B}\right\rangle }{\left\langle
S_{E}^{B}\right\rangle }=\left[ \frac{\left\langle S_{E}^{B}\right\rangle }{%
\left\langle S_{E}\right\rangle }\right] ^{2}\left( 1-\left( \gamma
\left\langle \hat{S}_{E}\right\rangle \right) ^{2}\right)
\end{equation*}%
\begin{equation*}
\left\langle \hat{S}_{E}^{B}\right\rangle =\left[ \frac{\left\langle
S_{E}^{B}\right\rangle }{\left\langle S_{E}\right\rangle }\right] ^{2}\left(
1-\left( \gamma \left\langle \hat{S}_{E}\right\rangle \right) ^{2}\right)
\left\langle S_{E}^{B}\right\rangle
\end{equation*}%
and the equation for $\bar{S}_{E}$ is given b th constraint: 
\begin{equation*}
1-2\bar{S}_{E}=\left\langle \hat{S}_{E}^{B}\right\rangle +\left\langle
S_{E}^{B}\right\rangle =\left( \left[ \frac{\left\langle
S_{E}^{B}\right\rangle }{\left\langle S_{E}\right\rangle }\right] ^{2}\left(
1-\left( \gamma \left\langle \hat{S}_{E}\right\rangle \right) ^{2}\right)
+1\right) \left\langle S_{E}^{B}\right\rangle
\end{equation*}

\paragraph*{A6.5.1.3 \ Approximate solution for $\bar{S}_{E}$}

We use that at the zeroth order:%
\begin{equation*}
\left( 1-\left( \gamma \left\langle \hat{S}_{E}\right\rangle \right)
^{2}\right) =\frac{2z}{\left( 1-2z\right) }
\end{equation*}%
and:%
\begin{equation*}
\left\langle S_{E}\right\rangle =\frac{1-2z}{2}
\end{equation*}%
we write the equation for $\bar{S}_{E}$ written $x$: 
\begin{equation*}
2\left( 1-2x\right) =\left( \left( \sqrt{F^{2}+\frac{4}{\left( \frac{1-x}{%
1-2x}\right) ^{2}}}-F\right) ^{2}\frac{2z}{\left( 1-2z\right) ^{3}}+1\right)
\left( \sqrt{F^{2}+\frac{4}{\left( \frac{1-x}{1-2x}\right) ^{2}}}-F\right)
\end{equation*}%
with:%
\begin{equation*}
F=\left( \frac{1-\left( \bar{\gamma}x\right) ^{2}}{\left( \frac{1-x}{1-2x}%
\right) ^{2}x}-\left( 1-2x\right) \right)
\end{equation*}%
We look for solutions with relatively small values of $z$. At lowest order:%
\begin{equation*}
2\left( 1-2x\right) =\left( \sqrt{F^{2}+\frac{4}{\left( \frac{1-\bar{S}_{E}}{%
1-2\bar{S}_{E}}\right) ^{2}}}-F\right)
\end{equation*}%
and the equation can be rewritten by including the correction: 
\begin{equation*}
2\left( 1-2x\right) =\left( 2\left( 1-2x\right) \frac{2z}{\left( 1-2z\right)
^{3}}+1\right) \left( \sqrt{F^{2}+\frac{4}{\left( \frac{1-\bar{S}_{E}}{1-2%
\bar{S}_{E}}\right) ^{2}}}-F\right)
\end{equation*}%
tht becomes:%
\begin{equation*}
\left( \frac{2\left( 1-2x\right) }{2\left( 1-2x\right) \frac{2z}{\left(
1-2z\right) ^{3}}+1}+F\right) ^{2}=F^{2}+\frac{4}{\left( \frac{1-\bar{S}_{E}%
}{1-2\bar{S}_{E}}\right) ^{2}}
\end{equation*}%
or:%
\begin{equation*}
\left( \frac{1-\left( \bar{\gamma}x\right) ^{2}}{\left( \frac{1-x}{1-2x}%
\right) ^{2}x}-\left( 1-2x\right) \right) \frac{1-2x}{2\left( 1-2x\right) 
\frac{2z}{\left( 1-2z\right) ^{3}}+1}+\left( \frac{1-2x}{2\left( 1-2x\right) 
\frac{2z}{\left( 1-2z\right) ^{3}}+1}\right) ^{2}=\frac{1}{\left( \frac{1-x}{%
1-2x}\right) ^{2}}
\end{equation*}%
and this allows to obtain $1-\left( \bar{\gamma}x\right) ^{2}$ as a function
of $x$:%
\begin{equation*}
\left( \frac{1-\left( \bar{\gamma}x\right) ^{2}}{\left( \frac{1-x}{1-2x}%
\right) ^{2}x}-\left( 1-2x\right) \right) =\left( \frac{1}{\left( \frac{1-x}{%
1-2x}\right) ^{2}}-\left( \frac{1-2x}{2\left( 1-2x\right) \frac{2z}{\left(
1-2z\right) ^{3}}+1}\right) ^{2}\right) \frac{2\left( 1-2x\right) \frac{2z}{%
\left( 1-2z\right) ^{3}}+1}{1-2x}
\end{equation*}%
with solution:%
\begin{equation*}
1-\left( \bar{\gamma}x\right) ^{2}=\left( \left( \frac{1}{\left( \frac{1-x}{%
1-2x}\right) ^{2}}-\left( \frac{1-2x}{2\left( 1-2x\right) \frac{2z}{\left(
1-2z\right) ^{3}}+1}\right) ^{2}\right) \frac{2\left( 1-2x\right) \frac{2z}{%
\left( 1-2z\right) ^{3}}+1}{1-2x}+\left( 1-2x\right) \right) \left( \frac{1-x%
}{1-2x}\right) ^{2}x
\end{equation*}%
Recall that $z_{0}=\left\langle \hat{S}_{1}\right\rangle _{0}$ is solution
of:

\begin{equation}
\frac{2\left( 2-\left( \gamma z\right) ^{2}\right) }{1-\left( \gamma
z\right) ^{2}}z-1=0
\end{equation}%
and we can write: 
\begin{equation*}
\gamma =\frac{1}{z}\sqrt{\frac{1-4z}{1-2z}}
\end{equation*}%
As a consequence, the equation for $x$ becoms:%
\begin{equation*}
0=1-\frac{1}{z^{2}}\frac{1-4z}{1-2z}x^{2}-\left( \left( \frac{1}{\left( 
\frac{1-x}{1-2x}\right) ^{2}}-\left( \frac{1-2x}{2\left( 1-2x\right) \frac{2z%
}{\left( 1-2z\right) ^{3}}+1}\right) ^{2}\right) \frac{2\left( 1-2x\right) 
\frac{2z}{\left( 1-2z\right) ^{3}}+1}{1-2x}+\left( 1-2x\right) \right)
\left( \frac{1-x}{1-2x}\right) ^{2}x
\end{equation*}%
and at lowest order we find:%
\begin{equation*}
x\simeq z
\end{equation*}%
and:%
\begin{equation*}
1-\left( \bar{\gamma}x\right) ^{2}=\left( \left( \frac{1}{\left( \frac{1-x}{%
1-2x}\right) ^{2}}-\left( \frac{1-2x}{2\left( 1-2x\right) \frac{2z}{\left(
1-2z\right) ^{3}}+1}\right) ^{2}\right) \frac{2\left( 1-2x\right) \frac{2z}{%
\left( 1-2z\right) ^{3}}+1}{1-2x}+\left( 1-2x\right) \right) \left( \frac{1-x%
}{1-2x}\right) ^{2}x
\end{equation*}%
a first order expansion yields:%
\begin{equation*}
1-\left( \bar{\gamma}x\right) ^{2}\simeq x
\end{equation*}%
\begin{equation*}
0=1-\frac{1}{z^{2}}\frac{1-4z}{1-2z}x^{2}-\left( \left( \frac{1}{\left( 
\frac{1-x}{1-2x}\right) ^{2}}-\left( \frac{1-2x}{2\left( 1-2x\right) \frac{2z%
}{\left( 1-2z\right) ^{3}}+1}\right) ^{2}\right) \frac{2\left( 1-2x\right) 
\frac{2z}{\left( 1-2z\right) ^{3}}+1}{1-2x}+\left( 1-2x\right) \right)
\left( \frac{1-x}{1-2x}\right) ^{2}x
\end{equation*}

\paragraph*{A6.5.1.3 \ Full solution for $1-\left( \bar{\protect\gamma}%
x\right) ^{2}$ and general equation for $\bar{S}_{E}$}

We first define:%
\begin{equation*}
Y=\sqrt{F^{2}+\frac{4}{\left( \frac{1-\bar{S}_{E}}{1-2\bar{S}_{E}}\right)
^{2}}}-F
\end{equation*}%
and write the equation for $x$ as:%
\begin{equation*}
2\left( 1-2x\right) =\left( Y^{2}\frac{2z}{\left( 1-2z\right) ^{3}}+1\right)
Y
\end{equation*}%
\begin{equation*}
Y^{3}+\frac{\left( 1-2z\right) ^{3}}{2z}Y-\frac{\left( 1-2z\right) ^{3}}{z}%
\left( 1-2x\right) =0
\end{equation*}%
The solution is given by:%
\begin{eqnarray*}
Y &=&\left( \frac{\left( 1-2z\right) ^{3}}{2z}\left( 1-2x\right) +\sqrt{%
\frac{\left( \frac{\left( 1-2z\right) ^{3}\left( 1-2x\right) }{z}\right) ^{2}%
}{4}+\frac{\left( \frac{\left( 1-2z\right) ^{3}}{2z}\right) ^{3}}{27}}%
\right) ^{\frac{1}{3}} \\
&&+\left( \frac{\left( 1-2z\right) ^{3}}{2z}\left( 1-2x\right) -\sqrt{\frac{%
\left( \frac{\left( 1-2z\right) ^{3}\left( 1-2x\right) }{z}\right) ^{2}}{4}+%
\frac{\left( \frac{\left( 1-2z\right) ^{3}}{2z}\right) ^{3}}{27}}\right) ^{%
\frac{1}{3}}
\end{eqnarray*}%
which allows to obtain an equation for $1-\left( \bar{\gamma}x\right) ^{2}$
by writing: 
\begin{equation*}
\sqrt{F^{2}+\frac{4}{\left( \frac{1-\bar{S}_{E}}{1-2\bar{S}_{E}}\right) ^{2}}%
}=F+Y
\end{equation*}%
taking the square yields:%
\begin{equation*}
\frac{4}{\left( \frac{1-\bar{S}_{E}}{1-2\bar{S}_{E}}\right) ^{2}}=Y^{2}+2FY
\end{equation*}%
with solution for $F$:%
\begin{equation*}
F=\frac{1}{\left( \frac{Y}{2}\right) \left( \frac{1-x}{1-2x}\right) ^{2}}%
-\left( \frac{Y}{2}\right)
\end{equation*}%
Replacing $F$ with its expression allows to derive $1-\left( \bar{\gamma}%
x\right) ^{2}$: 
\begin{equation*}
\frac{1-\left( \bar{\gamma}x\right) ^{2}}{\left( \frac{1-x}{1-2x}\right)
^{2}x}-\left( 1-2x\right) =\frac{1}{\left( \frac{Y}{2}\right) \left( \frac{%
1-x}{1-2x}\right) ^{2}}-\left( \frac{Y}{2}\right)
\end{equation*}%
with solution:%
\begin{equation*}
1-\left( \bar{\gamma}x\right) ^{2}=\left( \left( \left( 1-2x\right) -\left( 
\frac{Y}{2}\right) \right) \left( \frac{1-x}{1-2x}\right) ^{2}+\frac{1}{%
\left( \frac{Y}{2}\right) }\right) x
\end{equation*}%
As before replcing 
\begin{equation*}
\left( \bar{\gamma}x\right) ^{2}=\frac{1}{z^{2}}\frac{1-4z}{1-2z}x^{2}
\end{equation*}%
yields the equation for $x$:%
\begin{equation*}
1-\frac{1}{z^{2}}\frac{1-4z}{1-2z}x^{2}=\left( \left( \left( 1-2x\right)
-\left( \frac{Y}{2}\right) \right) \left( \frac{1-x}{1-2x}\right) ^{2}+\frac{%
1}{\left( \frac{Y}{2}\right) }\right) x
\end{equation*}%
At the lowest order:%
\begin{equation*}
Y\simeq 2\left( 1-x\right)
\end{equation*}%
for $z<<1$, and the solution is:%
\begin{equation*}
x=z
\end{equation*}

\paragraph{A6.5.1.4 Interpolation formula}

Given that for $x=\frac{1}{4}$, there is no uncertainty and we have:%
\begin{equation*}
\bar{S}_{E}=\bar{S}_{B}=\left\langle S_{E}^{B}\right\rangle =\left\langle 
\hat{S}_{E}^{B}\right\rangle =\frac{1}{4}
\end{equation*}%
we can interpolate the solution by adding a quadratic correction to $%
\left\langle \hat{S}_{E}^{B}\right\rangle $ and $\left\langle
S_{E}^{B}\right\rangle $:%
\begin{eqnarray*}
\bar{S} &=&2x \\
\left\langle \hat{S}_{E}^{B}\right\rangle &=&4x-12x^{2} \\
\left\langle S_{E}^{B}\right\rangle &=&1-6x+12x^{2}
\end{eqnarray*}

\subsubsection*{A6.5.2 First order formula}

\paragraph*{A6.5.2.1 \ First order formula for $\left\langle
S_{E}^{B}\right\rangle $ and $\left\langle \hat{S}_{E}^{B}\right\rangle $}

Given (\ref{SNB}) (\ref{Sb}), (\ref{Sbn}), (\ref{VRB}), (\ref{VRH}) we can
compute $\left\langle S_{E}^{B}\right\rangle $ and $\left\langle \hat{S}%
_{E}^{B}\right\rangle $ to the first ordr. To do so we use $\frac{%
\left\langle \bar{S}\right\rangle }{2}$ at this order:%
\begin{equation}
\frac{\left\langle \bar{S}\right\rangle }{2}=\frac{\left\langle \bar{w}%
\left( X^{\prime }\right) \right\rangle }{2}+\Delta \frac{\left\langle \bar{S%
}\right\rangle }{2}
\end{equation}%
\begin{equation}
\left\langle \hat{S}_{E}^{B}\right\rangle =\left\langle \hat{w}%
_{E}^{B}\left( X^{\prime }\right) \right\rangle +\Delta \left\langle \hat{S}%
_{E}^{B}\right\rangle
\end{equation}%
and:%
\begin{equation}
\left\langle S_{E}^{B}\right\rangle =\left\langle w_{E}^{B}\left( X\right)
\right\rangle +\Delta \left\langle S_{E}^{B}\right\rangle
\end{equation}%
with:%
\begin{equation*}
\Delta \frac{\left\langle \bar{S}\right\rangle }{2}=\frac{\left\langle \bar{w%
}\left( X^{\prime }\right) \right\rangle _{0}}{2}\left( \left\langle \hat{w}%
_{E}^{B}\left( X\right) \right\rangle _{0}\left( \frac{\left\langle \bar{f}%
\left( X^{\prime }\right) \right\rangle +\left\langle \bar{r}\left(
X^{\prime }\right) \right\rangle }{2}-\left\langle \hat{f}\left( X^{\prime
}\right) \right\rangle _{\hat{w}_{E}}\right) +\left\langle w_{E}^{B}\left(
X\right) \right\rangle _{0}\left( \frac{\left\langle \bar{f}\left( X^{\prime
}\right) \right\rangle +\left\langle \bar{r}\left( X^{\prime }\right)
\right\rangle }{2}-\left\langle f\left( X\right) \right\rangle \right)
\right)
\end{equation*}%
\begin{eqnarray*}
\Delta \left\langle \hat{S}_{E}^{B}\right\rangle &=&\left\langle \hat{w}%
_{E}^{B}\left( X^{\prime }\right) \right\rangle _{0}\left[ \left\langle \bar{%
w}\left( X\right) \right\rangle _{0}\left( \left\langle \hat{f}\left(
X^{\prime }\right) \right\rangle -\frac{\left\langle \bar{f}\left( X^{\prime
}\right) \right\rangle _{\bar{w}_{E}}+\left\langle \bar{r}\left( X^{\prime
}\right) \right\rangle _{\bar{w}_{2}}}{2}\right) \right. \\
&&\left. +\left\langle w_{E}^{B}\left( X\right) \right\rangle _{0}\left(
\left\langle \hat{f}\left( X^{\prime }\right) \right\rangle -\left\langle
f\left( X\right) \right\rangle \right) \right]
\end{eqnarray*}%
\begin{eqnarray*}
&&\Delta \left\langle S_{E}^{B}\right\rangle \\
&=&\left\langle w_{E}^{B}\left( X\right) \right\rangle _{0}\left\{
\left\langle \bar{w}\left( X\right) \right\rangle _{0}\left( \left\langle
f\left( X\right) \right\rangle -\frac{\left\langle \bar{f}\left( X^{\prime
}\right) \right\rangle _{\bar{w}_{E}}+\left\langle \bar{r}\left( X^{\prime
}\right) \right\rangle _{\bar{w}_{2}}}{2}\right) +\left\langle \hat{w}%
_{E}^{B}\left( X\right) \right\rangle _{0}\left( \left\langle f\left(
X\right) \right\rangle -\left\langle \hat{f}\left( X^{\prime }\right)
\right\rangle _{\hat{w}_{E}}\right) \right\}
\end{eqnarray*}%
and we have the fllwng frm fr the equations (\ref{VRB}), (\ref{VRH}):%
\begin{equation*}
\frac{\left\langle S_{E}^{B}\right\rangle -\Delta \left\langle
S_{E}^{B}\right\rangle }{\frac{\left\langle \bar{S}\right\rangle }{2}-\Delta 
\frac{\left\langle \bar{S}\right\rangle }{2}}=\frac{1+\frac{\left( \frac{1-%
\bar{S}_{E}}{1-\bar{S}}\right) ^{2}\left\langle \hat{S}_{E}^{B}\right\rangle
^{2}}{\left[ \frac{\left\langle S_{E}^{B}\right\rangle }{\left\langle
S_{E}\right\rangle }\right] ^{2}\left( 1-\left( \gamma \left\langle \hat{S}%
_{E}\right\rangle \right) ^{2}\right) }}{\left( 1-\left( \bar{\gamma}\frac{%
\left\langle \bar{S}\right\rangle }{2}\right) ^{2}\right) }
\end{equation*}%
\begin{equation*}
\frac{\left\langle \hat{S}_{E}^{B}\right\rangle -\Delta \left\langle \hat{S}%
_{E}^{B}\right\rangle }{\frac{\left\langle \bar{S}\right\rangle }{2}-\Delta 
\frac{\left\langle \bar{S}\right\rangle }{2}}=\frac{\left( \frac{1-\bar{S}%
_{E}}{1-\bar{S}}\right) ^{2}\left\langle \hat{S}_{E}^{B}\right\rangle ^{2}+%
\left[ \frac{\left\langle S_{E}^{B}\right\rangle }{\left\langle
S_{E}\right\rangle }\right] ^{2}\left( 1-\left( \gamma \left\langle \hat{S}%
_{E}\right\rangle \right) ^{2}\right) }{\left( 1-\left( \bar{\gamma}\frac{%
\left\langle \bar{S}\right\rangle }{2}\right) ^{2}\right) }
\end{equation*}%
These two equations factr as:%
\begin{equation*}
\frac{\left\langle S_{E}^{B}\right\rangle }{\frac{\left\langle \bar{S}%
\right\rangle }{2}}\frac{1-\frac{\Delta \left\langle S_{E}^{B}\right\rangle 
}{\left\langle S_{E}^{B}\right\rangle }}{1-\frac{\Delta \left\langle \bar{S}%
\right\rangle }{\left\langle \bar{S}\right\rangle }}=\frac{1+\frac{\left( 
\frac{1-\bar{S}_{E}}{1-\bar{S}}\right) ^{2}\left\langle \hat{S}%
_{E}^{B}\right\rangle ^{2}}{\left[ \frac{\left\langle S_{E}^{B}\right\rangle 
}{\left\langle S_{E}\right\rangle }\right] ^{2}\left( 1-\left( \gamma
\left\langle \hat{S}_{E}\right\rangle \right) ^{2}\right) }}{\left( 1-\left( 
\bar{\gamma}\frac{\left\langle \bar{S}\right\rangle }{2}\right) ^{2}\right) }
\end{equation*}%
\begin{equation*}
\frac{\left\langle \hat{S}_{E}^{B}\right\rangle }{\frac{\left\langle \bar{S}%
\right\rangle }{2}}\left( \frac{1-\frac{\Delta \left\langle \hat{S}%
_{E}^{B}\right\rangle }{\left\langle \hat{S}_{E}^{B}\right\rangle }}{1-\frac{%
\Delta \left\langle \bar{S}\right\rangle }{\left\langle \bar{S}\right\rangle 
}}\right) =\frac{\left( \frac{1-\bar{S}_{E}}{1-\bar{S}}\right)
^{2}\left\langle \hat{S}_{E}^{B}\right\rangle ^{2}+\left[ \frac{\left\langle
S_{E}^{B}\right\rangle }{\left\langle S_{E}\right\rangle }\right] ^{2}\left(
1-\left( \gamma \left\langle \hat{S}_{E}\right\rangle \right) ^{2}\right) }{%
\left( 1-\left( \bar{\gamma}\frac{\left\langle \bar{S}\right\rangle }{2}%
\right) ^{2}\right) }
\end{equation*}%
We use that:%
\begin{eqnarray*}
&&\frac{\Delta \left\langle S_{E}^{B}\right\rangle }{\left\langle
S_{E}^{B}\right\rangle } \\
&=&\left\langle \bar{w}\left( X\right) \right\rangle _{0}\left( \left\langle
f\left( X\right) \right\rangle -\frac{\left\langle \bar{f}\left( X^{\prime
}\right) \right\rangle _{\bar{w}_{E}}+\left\langle \bar{r}\left( X^{\prime
}\right) \right\rangle _{\bar{w}_{2}}}{2}\right) +\left\langle \hat{w}%
_{E}^{B}\left( X\right) \right\rangle _{0}\left( \left\langle f\left(
X\right) \right\rangle -\left\langle \hat{f}\left( X^{\prime }\right)
\right\rangle _{\hat{w}_{E}}\right)
\end{eqnarray*}%
to compute the corrections as:%
\begin{eqnarray*}
&&\frac{\Delta \left\langle \bar{S}\right\rangle }{\left\langle \bar{S}%
\right\rangle }-\frac{\Delta \left\langle S_{E}^{B}\right\rangle }{%
\left\langle S_{E}^{B}\right\rangle }=\left\langle \hat{w}_{E}^{B}\left(
X\right) \right\rangle _{0}\left( \frac{\left\langle \bar{f}\left( X^{\prime
}\right) \right\rangle +\left\langle \bar{r}\left( X^{\prime }\right)
\right\rangle }{2}-\left\langle \hat{f}\left( X^{\prime }\right)
\right\rangle _{\hat{w}_{E}}\right) \\
&&+\left\langle w_{E}^{B}\left( X\right) \right\rangle _{0}\left( \frac{%
\left\langle \bar{f}\left( X^{\prime }\right) \right\rangle +\left\langle 
\bar{r}\left( X^{\prime }\right) \right\rangle }{2}-\left\langle f\left(
X\right) \right\rangle \right) \\
&&-\left\langle \bar{w}\left( X\right) \right\rangle _{0}\left( \left\langle
f\left( X\right) \right\rangle -\frac{\left\langle \bar{f}\left( X^{\prime
}\right) \right\rangle _{\bar{w}_{E}}+\left\langle \bar{r}\left( X^{\prime
}\right) \right\rangle _{\bar{w}_{2}}}{2}\right) -\left\langle \hat{w}%
_{E}^{B}\left( X\right) \right\rangle _{0}\left( \left\langle f\left(
X\right) \right\rangle -\left\langle \hat{f}\left( X^{\prime }\right)
\right\rangle _{\hat{w}_{E}}\right) \\
&=&\left\langle \hat{w}_{E}^{B}\left( X\right) \right\rangle _{0}\left( 
\frac{\left\langle \bar{f}\left( X^{\prime }\right) \right\rangle
+\left\langle \bar{r}\left( X^{\prime }\right) \right\rangle }{2}%
-\left\langle f\left( X\right) \right\rangle \right) +\left\langle
w_{E}^{B}\left( X\right) \right\rangle _{0}\left( \frac{\left\langle \bar{f}%
\left( X^{\prime }\right) \right\rangle +\left\langle \bar{r}\left(
X^{\prime }\right) \right\rangle }{2}-\left\langle f\left( X\right)
\right\rangle \right) \\
&&-\left\langle \bar{w}\left( X\right) \right\rangle _{0}\left( \left\langle
f\left( X\right) \right\rangle -\frac{\left\langle \bar{f}\left( X^{\prime
}\right) \right\rangle _{\bar{w}_{E}}+\left\langle \bar{r}\left( X^{\prime
}\right) \right\rangle _{\bar{w}_{2}}}{2}\right) \\
&=&\left( \frac{\left\langle \bar{f}\left( X^{\prime }\right) \right\rangle
+\left\langle \bar{r}\left( X^{\prime }\right) \right\rangle }{2}%
-\left\langle f\left( X\right) \right\rangle \right)
\end{eqnarray*}%
and similrly:%
\begin{equation*}
\frac{\Delta \left\langle \bar{S}\right\rangle }{\left\langle \bar{S}%
\right\rangle }-\frac{\Delta \left\langle \hat{S}_{E}^{B}\right\rangle }{%
\left\langle \hat{S}_{E}^{B}\right\rangle }=\frac{\left\langle \bar{f}\left(
X^{\prime }\right) \right\rangle +\left\langle \bar{r}\left( X^{\prime
}\right) \right\rangle }{2}-\left\langle \hat{f}\left( X^{\prime }\right)
\right\rangle
\end{equation*}%
As a consequence, (\ref{VRB}), (\ref{VRH}) rewrite:%
\begin{equation*}
\frac{\left\langle S_{E}^{B}\right\rangle }{\frac{\left\langle \bar{S}%
\right\rangle }{2}}\left( 1+\frac{\left\langle \bar{f}\left( X^{\prime
}\right) \right\rangle +\left\langle \bar{r}\left( X^{\prime }\right)
\right\rangle }{2}-\left\langle f\left( X\right) \right\rangle \right) =%
\frac{1+\frac{\left( \frac{1-\bar{S}_{E}}{1-\bar{S}}\right) ^{2}\left\langle 
\hat{S}_{E}^{B}\right\rangle ^{2}}{\left[ \frac{\left\langle
S_{E}^{B}\right\rangle }{\left\langle S_{E}\right\rangle }\right] ^{2}\left(
1-\left( \gamma \left\langle \hat{S}_{E}\right\rangle \right) ^{2}\right) }}{%
\left( 1-\left( \bar{\gamma}\frac{\left\langle \bar{S}\right\rangle }{2}%
\right) ^{2}\right) }
\end{equation*}%
and:%
\begin{equation*}
\frac{\left\langle \hat{S}_{E}^{B}\right\rangle }{\frac{\left\langle \bar{S}%
\right\rangle }{2}}\left( 1+\frac{\left\langle \bar{f}\left( X^{\prime
}\right) \right\rangle +\left\langle \bar{r}\left( X^{\prime }\right)
\right\rangle }{2}-\left\langle \hat{f}\left( X^{\prime }\right)
\right\rangle \right) =\frac{\left( \frac{1-\bar{S}_{E}}{1-\bar{S}}\right)
^{2}\left\langle \hat{S}_{E}^{B}\right\rangle ^{2}+\left[ \frac{\left\langle
S_{E}^{B}\right\rangle }{\left\langle S_{E}\right\rangle }\right] ^{2}\left(
1-\left( \gamma \left\langle \hat{S}_{E}\right\rangle \right) ^{2}\right) }{%
\left( 1-\left( \bar{\gamma}\frac{\left\langle \bar{S}\right\rangle }{2}%
\right) ^{2}\right) }
\end{equation*}%
or, equivalently:%
\begin{equation*}
\frac{\left\langle S_{E}^{B}\right\rangle }{\frac{\left\langle \bar{S}%
\right\rangle }{2}}=\frac{1+\frac{\left( \frac{1-\bar{S}_{E}}{1-\bar{S}}%
\right) ^{2}\left\langle \hat{S}_{E}^{B}\right\rangle ^{2}}{\left[ \frac{%
\left\langle S_{E}^{B}\right\rangle }{\left\langle S_{E}\right\rangle }%
\right] ^{2}\left( 1-\left( \gamma \left\langle \hat{S}_{E}\right\rangle
\right) ^{2}\right) }}{\left( 1-\left( \bar{\gamma}\frac{\left\langle \bar{S}%
\right\rangle }{2}\right) ^{2}\right) \left( 1+\frac{\left\langle \bar{f}%
\left( X^{\prime }\right) \right\rangle +\left\langle \bar{r}\left(
X^{\prime }\right) \right\rangle }{2}-\left\langle f\left( X\right)
\right\rangle \right) }
\end{equation*}%
\begin{eqnarray*}
\frac{\left\langle \hat{S}_{E}^{B}\right\rangle }{\frac{\left\langle \bar{S}%
\right\rangle }{2}} &=&\frac{\left( \frac{1-\bar{S}_{E}}{1-\bar{S}}\right)
^{2}\left\langle \hat{S}_{E}^{B}\right\rangle ^{2}+\left[ \frac{\left\langle
S_{E}^{B}\right\rangle }{\left\langle S_{E}\right\rangle }\right] ^{2}\left(
1-\left( \gamma \left\langle \hat{S}_{E}\right\rangle \right) ^{2}\right) }{%
\left( 1-\left( \bar{\gamma}\frac{\left\langle \bar{S}\right\rangle }{2}%
\right) ^{2}\right) \left( 1+\frac{\left\langle \bar{f}\left( X^{\prime
}\right) \right\rangle +\left\langle \bar{r}\left( X^{\prime }\right)
\right\rangle }{2}-\left\langle \hat{f}\left( X^{\prime }\right)
\right\rangle \right) } \\
&=&\frac{\left( \frac{1-\bar{S}_{E}}{1-\bar{S}}\right) ^{2}\frac{1+\frac{%
\left\langle \bar{f}\left( X^{\prime }\right) \right\rangle +\left\langle 
\bar{r}\left( X^{\prime }\right) \right\rangle }{2}-\left\langle f\left(
X\right) \right\rangle }{1+\frac{\left\langle \bar{f}\left( X^{\prime
}\right) \right\rangle +\left\langle \bar{r}\left( X^{\prime }\right)
\right\rangle }{2}-\left\langle \hat{f}\left( X^{\prime }\right)
\right\rangle }\left\langle \hat{S}_{E}^{B}\right\rangle ^{2}+\left[ \frac{%
\left\langle S_{E}^{B}\right\rangle }{\left\langle S_{E}\right\rangle }%
\right] ^{2}\left( 1-\left( \gamma \left\langle \hat{S}_{E}\right\rangle
\right) ^{2}\right) \frac{1+\frac{\left\langle \bar{f}\left( X^{\prime
}\right) \right\rangle +\left\langle \bar{r}\left( X^{\prime }\right)
\right\rangle }{2}-\left\langle f\left( X\right) \right\rangle }{\left( 1+%
\frac{\left\langle \bar{f}\left( X^{\prime }\right) \right\rangle
+\left\langle \bar{r}\left( X^{\prime }\right) \right\rangle }{2}%
-\left\langle \hat{f}\left( X^{\prime }\right) \right\rangle \right) }}{%
\left( 1-\left( \bar{\gamma}\frac{\left\langle \bar{S}\right\rangle }{2}%
\right) ^{2}\right) \left( 1+\frac{\left\langle \bar{f}\left( X^{\prime
}\right) \right\rangle +\left\langle \bar{r}\left( X^{\prime }\right)
\right\rangle }{2}-\left\langle f\left( X\right) \right\rangle \right) } \\
&=&\frac{\left( \frac{1-\bar{S}_{E}}{1-\bar{S}}\right) ^{2}\left(
1+\left\langle \hat{f}\left( X^{\prime }\right) \right\rangle -\left\langle
f\left( X\right) \right\rangle \right) \left\langle \hat{S}%
_{E}^{B}\right\rangle ^{2}+\left[ \frac{\left\langle S_{E}^{B}\right\rangle 
}{\left\langle S_{E}\right\rangle }\right] ^{2}\left( 1-\left( \gamma
\left\langle \hat{S}_{E}\right\rangle \right) ^{2}\right) \left(
1+\left\langle \hat{f}\left( X^{\prime }\right) \right\rangle -\left\langle
f\left( X\right) \right\rangle \right) }{\left( 1-\left( \bar{\gamma}\frac{%
\left\langle \bar{S}\right\rangle }{2}\right) ^{2}\right) \left( 1+\frac{%
\left\langle \bar{f}\left( X^{\prime }\right) \right\rangle +\left\langle 
\bar{r}\left( X^{\prime }\right) \right\rangle }{2}-\left\langle f\left(
X\right) \right\rangle \right) }
\end{eqnarray*}%
This allows to obtain directly the solution as corrections to the zeroth
order cas by subtituting:%
\begin{equation*}
\left( 1+\frac{\left\langle \bar{f}\left( X^{\prime }\right) \right\rangle
+\left\langle \bar{r}\left( X^{\prime }\right) \right\rangle }{2}%
-\left\langle f\left( X\right) \right\rangle \right) \rightarrow \left(
1-\left( \bar{\gamma}\left\langle \bar{S}_{E}\right\rangle \right)
^{2}\right) \left( 1+\frac{\left\langle \bar{f}\left( X^{\prime }\right)
\right\rangle +\left\langle \bar{r}\left( X^{\prime }\right) \right\rangle }{%
2}-\left\langle f\left( X\right) \right\rangle \right)
\end{equation*}%
\begin{equation*}
\left( \frac{1-\bar{S}_{E}}{1-\bar{S}}\right) ^{2}\rightarrow \left( \frac{1-%
\bar{S}_{E}}{1-\bar{S}}\right) ^{2}\left( 1+\left\langle \hat{f}\left(
X^{\prime }\right) \right\rangle -\left\langle f\left( X\right)
\right\rangle \right)
\end{equation*}%
\begin{equation*}
\left( 1-\left( \gamma \left\langle \hat{S}_{E}\right\rangle \right)
^{2}\right) \rightarrow \left( 1-\left( \gamma \left\langle \hat{S}%
_{E}\right\rangle \right) ^{2}\right) \left( 1+\left\langle \hat{f}\left(
X^{\prime }\right) \right\rangle -\left\langle f\left( X\right)
\right\rangle \right)
\end{equation*}%
and the solutions are directly:%
\begin{eqnarray}
\frac{\left\langle S_{E}^{B}\right\rangle }{\frac{\left\langle \bar{S}%
\right\rangle }{2}} &=&\frac{1}{2}\left\{ \left( \frac{\left( 1-\left( \bar{%
\gamma}\left\langle \bar{S}_{E}\right\rangle \right) ^{2}\right) \left( 1+%
\frac{\left\langle \bar{f}\left( X^{\prime }\right) \right\rangle
+\left\langle \bar{r}\left( X^{\prime }\right) \right\rangle }{2}%
-\left\langle f\left( X\right) \right\rangle \right) }{\left( 1+\left\langle 
\hat{f}\left( X^{\prime }\right) \right\rangle -\left\langle f\left(
X\right) \right\rangle \right) \left( \frac{1-\bar{S}_{E}}{1-2\bar{S}_{E}}%
\right) ^{2}\left\langle \bar{S}_{E}\right\rangle }-\left( 1-2\left\langle 
\bar{S}_{E}\right\rangle \right) \right) ^{2}\right.  \label{SBNl} \\
&&\left. +\frac{4}{\left( 1+\left\langle \hat{f}\left( X^{\prime }\right)
\right\rangle -\left\langle f\left( X\right) \right\rangle \right) \left( 
\frac{1-\bar{S}_{E}}{1-2\bar{S}_{E}}\right) ^{2}}\right\}  \notag \\
&&-\frac{1}{2}\left( \frac{\left( 1-\left( \bar{\gamma}\left\langle \bar{S}%
_{E}\right\rangle \right) ^{2}\right) \left( 1+\frac{\left\langle \bar{f}%
\left( X^{\prime }\right) \right\rangle +\left\langle \bar{r}\left(
X^{\prime }\right) \right\rangle }{2}-\left\langle f\left( X\right)
\right\rangle \right) }{\left( 1+\left\langle \hat{f}\left( X^{\prime
}\right) \right\rangle -\left\langle f\left( X\right) \right\rangle \right)
\left( \frac{1-\bar{S}_{E}}{1-2\bar{S}_{E}}\right) ^{2}\left\langle \bar{S}%
_{E}\right\rangle }-\left( 1-2\frac{\left\langle \bar{S}\right\rangle }{2}%
\right) \right)  \notag
\end{eqnarray}

\begin{eqnarray}
&&\frac{\left\langle \hat{S}_{E}^{B}\right\rangle }{\frac{\left\langle \bar{S%
}\right\rangle }{2}}  \label{SBHl} \\
&=&\frac{1}{2}\left( \frac{\left( 1-\left( \bar{\gamma}\left\langle \bar{S}%
_{E}\right\rangle \right) ^{2}\right) \left( 1+\frac{\left\langle \bar{f}%
\left( X^{\prime }\right) \right\rangle +\left\langle \bar{r}\left(
X^{\prime }\right) \right\rangle }{2}-\left\langle \hat{f}\left( X^{\prime
}\right) \right\rangle \right) }{\left( \frac{1-\bar{S}_{E}}{1-2\bar{S}_{E}}%
\right) ^{2}\left\langle \bar{S}_{E}\right\rangle }+\left( 1-2\frac{%
\left\langle \bar{S}\right\rangle }{2}\right) \right)  \notag \\
&&-\frac{\sqrt{\left( \frac{1-\left( \bar{\gamma}\frac{\left\langle \bar{S}%
\right\rangle }{2}\right) ^{2}\left( 1+\frac{\left\langle \bar{f}\left(
X^{\prime }\right) \right\rangle +\left\langle \bar{r}\left( X^{\prime
}\right) \right\rangle }{2}-\left\langle \hat{f}\left( X^{\prime }\right)
\right\rangle \right) }{\left( \frac{1-\bar{S}_{E}}{1-2\bar{S}_{E}}\right)
^{2}\frac{\left\langle \bar{S}\right\rangle }{2}}-\left( 1-2\frac{%
\left\langle \bar{S}\right\rangle }{2}\right) \right) ^{2}+\frac{4}{\left(
1+\left\langle \hat{f}\left( X^{\prime }\right) \right\rangle -\left\langle
f\left( X\right) \right\rangle \right) \left( \frac{1-\bar{S}_{E}}{1-2\bar{S}%
_{E}}\right) ^{2}}}}{2}  \notag
\end{eqnarray}

\paragraph*{A6.5.2.2 \ First order corrections}

A first order expansion of (\ref{SBNl}), (\ref{SBHl}) yields the first order
correction for $\left\langle S_{E}^{B}\right\rangle $:%
\begin{eqnarray*}
&&\frac{\left\langle S_{E}^{B}\right\rangle -\left\langle
S_{E}^{B}\right\rangle _{0}}{\frac{\left\langle \bar{S}\right\rangle }{2}} \\
&=&\frac{\left( \frac{1-\left( \bar{\gamma}\left\langle \bar{S}%
_{E}\right\rangle \right) ^{2}}{\left( \frac{1-\bar{S}_{E}}{1-2\bar{S}_{E}}%
\right) ^{2}\left\langle \bar{S}_{E}\right\rangle }-\left( 1-2\left\langle 
\bar{S}_{E}\right\rangle \right) \right) \frac{1-\left( \bar{\gamma}%
\left\langle \bar{S}_{E}\right\rangle \right) ^{2}}{\left( \frac{1-\bar{S}%
_{E}}{1-2\bar{S}_{E}}\right) ^{2}\left\langle \bar{S}_{E}\right\rangle }%
\left( \frac{\left\langle \bar{f}\left( X^{\prime }\right) \right\rangle
+\left\langle \bar{r}\left( X^{\prime }\right) \right\rangle }{2}%
-\left\langle \hat{f}\left( X^{\prime }\right) \right\rangle \right) -\frac{2%
}{\left( \frac{1-\bar{S}_{E}}{1-2\bar{S}_{E}}\right) ^{2}}\left(
\left\langle \hat{f}\left( X^{\prime }\right) \right\rangle -\left\langle
f\left( X\right) \right\rangle \right) }{2\sqrt{\left( \frac{1-\left( \bar{%
\gamma}\left\langle \bar{S}_{E}\right\rangle \right) ^{2}}{\left( \frac{1-%
\bar{S}_{E}}{1-2\bar{S}_{E}}\right) ^{2}\left\langle \bar{S}%
_{E}\right\rangle }-\left( 1-2\left\langle \bar{S}_{E}\right\rangle \right)
\right) ^{2}+\frac{4}{\left( \frac{1-\bar{S}_{E}}{1-2\bar{S}_{E}}\right) ^{2}%
}}} \\
&&-\frac{1}{2}\frac{1-\left( \bar{\gamma}\left\langle \bar{S}%
_{E}\right\rangle \right) ^{2}}{\left( \frac{1-\bar{S}_{E}}{1-2\bar{S}_{E}}%
\right) ^{2}\left\langle \bar{S}_{E}\right\rangle }\left( \frac{\left\langle 
\bar{f}\left( X^{\prime }\right) \right\rangle +\left\langle \bar{r}\left(
X^{\prime }\right) \right\rangle }{2}-\left\langle \hat{f}\left( X^{\prime
}\right) \right\rangle \right)
\end{eqnarray*}%
that is:%
\begin{eqnarray}
&&\frac{\left\langle S_{E}^{B}\right\rangle -\left\langle
S_{E}^{B}\right\rangle _{0}}{\frac{\left\langle \bar{S}\right\rangle }{2}}
\label{FPr} \\
&=&\frac{1-\left( \bar{\gamma}\left\langle \bar{S}_{E}\right\rangle \right)
^{2}}{\left( \frac{1-\bar{S}_{E}}{1-2\bar{S}_{E}}\right) ^{2}\left\langle 
\bar{S}_{E}\right\rangle }\frac{\left( \frac{1-\left( \bar{\gamma}%
\left\langle \bar{S}_{E}\right\rangle \right) ^{2}}{\left( \frac{1-\bar{S}%
_{E}}{1-2\bar{S}_{E}}\right) ^{2}\left\langle \bar{S}_{E}\right\rangle }%
-\left( 1-2\left\langle \bar{S}_{E}\right\rangle \right) \right) -\sqrt{%
\left( \frac{1-\left( \bar{\gamma}\left\langle \bar{S}_{E}\right\rangle
\right) ^{2}}{\left( \frac{1-\bar{S}_{E}}{1-2\bar{S}_{E}}\right)
^{2}\left\langle \bar{S}_{E}\right\rangle }-\left( 1-2\left\langle \bar{S}%
_{E}\right\rangle \right) \right) ^{2}+\frac{4}{\left( \frac{1-\bar{S}_{E}}{%
1-2\bar{S}_{E}}\right) ^{2}}}}{2\sqrt{\left( \frac{1-\left( \bar{\gamma}%
\left\langle \bar{S}_{E}\right\rangle \right) ^{2}}{\left( \frac{1-\bar{S}%
_{E}}{1-2\bar{S}_{E}}\right) ^{2}\left\langle \bar{S}_{E}\right\rangle }%
-\left( 1-2\left\langle \bar{S}_{E}\right\rangle \right) \right) ^{2}+\frac{4%
}{\left( \frac{1-\bar{S}_{E}}{1-2\bar{S}_{E}}\right) ^{2}}}}  \notag \\
&&\times \left( \frac{\left\langle \bar{f}\left( X^{\prime }\right)
\right\rangle +\left\langle \bar{r}\left( X^{\prime }\right) \right\rangle }{%
2}-\left\langle \hat{f}\left( X^{\prime }\right) \right\rangle \right) 
\notag \\
&&-\frac{\frac{2}{\left( \frac{1-\bar{S}_{E}}{1-2\bar{S}_{E}}\right) ^{2}}%
\left( \left\langle \hat{f}\left( X^{\prime }\right) \right\rangle
-\left\langle f\left( X\right) \right\rangle \right) }{2\sqrt{\left( \frac{%
1-\left( \bar{\gamma}\left\langle \bar{S}_{E}\right\rangle \right) ^{2}}{%
\left( \frac{1-\bar{S}_{E}}{1-2\bar{S}_{E}}\right) ^{2}\left\langle \bar{S}%
_{E}\right\rangle }-\left( 1-2\left\langle \bar{S}_{E}\right\rangle \right)
\right) ^{2}+\frac{4}{\left( \frac{1-\bar{S}_{E}}{1-2\bar{S}_{E}}\right) ^{2}%
}}}  \notag
\end{eqnarray}%
and this also expressed as:%
\begin{eqnarray}
&&\frac{\left\langle S_{E}^{B}\right\rangle -\left\langle
S_{E}^{B}\right\rangle _{0}}{\frac{\left\langle \bar{S}\right\rangle }{2}}
\label{Fr} \\
&=&-\frac{1-\left( \bar{\gamma}\frac{\left\langle \bar{S}\right\rangle }{2}%
\right) ^{2}}{\left( \frac{1-\bar{S}_{E}}{1-2\bar{S}_{E}}\right) ^{2}\frac{%
\left\langle \bar{S}\right\rangle }{2}}\frac{\left\langle
S_{E}^{B}\right\rangle \left( \left\langle \bar{f}\left( X^{\prime }\right)
\right\rangle -\left\langle \hat{f}\left( X^{\prime }\right) \right\rangle
\right) +\frac{2\left\langle \bar{S}_{E}\right\rangle }{1-\left( \bar{\gamma}%
\left\langle \bar{S}_{E}\right\rangle \right) ^{2}}\left( \left\langle \hat{f%
}\left( X^{\prime }\right) \right\rangle -\left\langle f\left( X\right)
\right\rangle \right) }{2\left( \left\langle S_{E}^{B}\right\rangle
-\left\langle \hat{S}_{E}^{B}\right\rangle +\frac{1-\left( \bar{\gamma}%
\left\langle \bar{S}_{E}\right\rangle \right) ^{2}}{\left( \frac{1-\bar{S}%
_{E}}{1-2\bar{S}_{E}}\right) ^{2}\left\langle \bar{S}_{E}\right\rangle }%
\right) }  \notag \\
&\rightarrow &-\frac{\left\langle S_{E}^{B}\right\rangle \left( \left\langle 
\bar{f}\left( X^{\prime }\right) \right\rangle -\left\langle \hat{f}\left(
X^{\prime }\right) \right\rangle \right) +\frac{2\left\langle \bar{S}%
_{E}\right\rangle }{1-\left( \bar{\gamma}\left\langle \bar{S}%
_{E}\right\rangle \right) ^{2}}\left( \left\langle \hat{f}\left( X^{\prime
}\right) \right\rangle -\left\langle f\left( X\right) \right\rangle \right) 
}{2\left( 1+\left( \left\langle S_{E}^{B}\right\rangle -\left\langle \hat{S}%
_{E}^{B}\right\rangle \right) \frac{\left( \frac{1-\bar{S}_{E}}{1-2\bar{S}%
_{E}}\right) ^{2}}{1-\left( \bar{\gamma}\left\langle \bar{S}%
_{E}\right\rangle \right) ^{2}}\right) }  \notag
\end{eqnarray}%
Moreover:%
\begin{equation*}
\frac{\left\langle \hat{S}_{E}^{B}\right\rangle -\left\langle \hat{S}%
_{E}^{B}\right\rangle _{0}}{\frac{\left\langle \bar{S}\right\rangle }{2}}=-%
\frac{\left\langle S_{E}^{B}\right\rangle -\left\langle
S_{E}^{B}\right\rangle _{0}}{\frac{\left\langle \bar{S}\right\rangle }{2}}
\end{equation*}

\paragraph*{A6.5.2.3 \ Computation of $\frac{\left\langle \bar{w}\left(
X\right) \right\rangle }{2}$ and $\frac{\left\langle \bar{S}\right\rangle }{2%
}$}

We can compute $\frac{\left\langle \bar{w}\left( X\right) \right\rangle }{2}$%
:

\begin{eqnarray*}
\frac{1}{\frac{\left\langle \bar{w}\left( X\right) \right\rangle }{2}} &=&2+%
\frac{\left\langle \hat{w}_{E}^{B}\left( X^{\prime }\right) \right\rangle }{%
\frac{\left\langle \bar{w}\left( X\right) \right\rangle }{2}}+\frac{%
\left\langle w_{E}^{B}\left( X\right) \right\rangle }{\frac{\left\langle 
\bar{w}\left( X\right) \right\rangle }{2}} \\
&&\left( 2+\frac{1+\frac{\left( \frac{1-\bar{S}_{E}}{1-\bar{S}}\right)
^{2}\left\langle \hat{S}_{E}^{B}\right\rangle ^{2}}{\left[ \frac{%
\left\langle S_{E}^{B}\right\rangle }{\left\langle S_{E}\right\rangle }%
\right] ^{2}\left( 1-\left( \gamma \left\langle \hat{S}_{E}\right\rangle
\right) ^{2}\right) }}{\left( 1-\left( \bar{\gamma}\frac{\left\langle \bar{S}%
\right\rangle }{2}\right) ^{2}\right) }+\frac{\left( \frac{1-\bar{S}_{E}}{1-%
\bar{S}}\right) ^{2}\left\langle \hat{S}_{E}^{B}\right\rangle ^{2}+\left[ 
\frac{\left\langle S_{E}^{B}\right\rangle }{\left\langle S_{E}\right\rangle }%
\right] ^{2}\left( 1-\left( \gamma \left\langle \hat{S}_{E}\right\rangle
\right) ^{2}\right) }{\left( 1-\left( \bar{\gamma}\frac{\left\langle \bar{S}%
\right\rangle }{2}\right) ^{2}\right) }\right) \\
&\rightarrow &\left( 2+\frac{\left\langle S_{E}^{B}\right\rangle -\Delta
\left\langle S_{E}^{B}\right\rangle }{\frac{\left\langle \bar{S}%
\right\rangle }{2}-\Delta \frac{\left\langle \bar{S}\right\rangle }{2}}+%
\frac{\left\langle \hat{S}_{E}^{B}\right\rangle -\Delta \left\langle \hat{S}%
_{E}^{B}\right\rangle }{\frac{\left\langle \bar{S}\right\rangle }{2}-\Delta 
\frac{\left\langle \bar{S}\right\rangle }{2}}\right) \\
&=&\left( 2+\frac{\left\langle S_{E}^{B}\right\rangle }{\frac{\left\langle 
\bar{S}\right\rangle }{2}}\left( 1+\frac{\left\langle \bar{f}\left(
X^{\prime }\right) \right\rangle +\left\langle \bar{r}\left( X^{\prime
}\right) \right\rangle }{2}-\left\langle f\left( X\right) \right\rangle
\right) +\frac{\left\langle \hat{S}_{E}^{B}\right\rangle }{\frac{%
\left\langle \bar{S}\right\rangle }{2}}\left( 1+\frac{\left\langle \bar{f}%
\left( X^{\prime }\right) \right\rangle +\left\langle \bar{r}\left(
X^{\prime }\right) \right\rangle }{2}-\left\langle \hat{f}\left( X^{\prime
}\right) \right\rangle \right) \right) \\
&=&2+\frac{\left\langle S_{E}^{B}\right\rangle _{0}+\delta \left\langle
S_{E}^{B}\right\rangle }{\frac{\left\langle \bar{S}\right\rangle }{2}}\left(
1+\frac{\left\langle \bar{f}\left( X^{\prime }\right) \right\rangle
+\left\langle \bar{r}\left( X^{\prime }\right) \right\rangle }{2}%
-\left\langle f\left( X\right) \right\rangle \right) \\
&&+\frac{\left\langle \hat{S}_{E}^{B}\right\rangle _{0}-\delta \left\langle
S_{E}^{B}\right\rangle }{\frac{\left\langle \bar{S}\right\rangle }{2}}\left(
1+\frac{\left\langle \bar{f}\left( X^{\prime }\right) \right\rangle
+\left\langle \bar{r}\left( X^{\prime }\right) \right\rangle }{2}%
-\left\langle \hat{f}\left( X^{\prime }\right) \right\rangle \right) \\
&=&\left( 2+\frac{\left\langle S_{E}^{B}\right\rangle _{0}}{\frac{%
\left\langle \bar{S}\right\rangle }{2}}\left( 1+\frac{\left\langle \bar{f}%
\left( X^{\prime }\right) \right\rangle +\left\langle \bar{r}\left(
X^{\prime }\right) \right\rangle }{2}-\left\langle f\left( X\right)
\right\rangle \right) +\frac{\left\langle \hat{S}_{E}^{B}\right\rangle _{0}}{%
\frac{\left\langle \bar{S}\right\rangle }{2}}\left( 1+\frac{\left\langle 
\bar{f}\left( X^{\prime }\right) \right\rangle +\left\langle \bar{r}\left(
X^{\prime }\right) \right\rangle }{2}-\left\langle \hat{f}\left( X^{\prime
}\right) \right\rangle \right) \right)
\end{eqnarray*}%
Thus, we obtain for$\frac{\left\langle \bar{w}\left( X\right) \right\rangle 
}{2}$:%
\begin{eqnarray}
&&\frac{\left\langle \bar{w}\left( X\right) \right\rangle }{2}  \label{Hb} \\
&\rightarrow &\frac{1}{\left( 2+\frac{\left\langle S_{E}^{B}\right\rangle
_{0}}{\frac{\left\langle \bar{S}\right\rangle }{2}}+\frac{\left\langle \hat{S%
}_{E}^{B}\right\rangle _{0}}{\frac{\left\langle \bar{S}\right\rangle }{2}}+%
\frac{\left\langle S_{E}^{B}\right\rangle _{0}}{\frac{\left\langle \bar{S}%
\right\rangle }{2}}\left( \frac{\left\langle \bar{f}\left( X^{\prime
}\right) \right\rangle +\left\langle \bar{r}\left( X^{\prime }\right)
\right\rangle }{2}-\left\langle f\left( X\right) \right\rangle \right) +%
\frac{\left\langle \hat{S}_{E}^{B}\right\rangle _{0}}{\frac{\left\langle 
\bar{S}\right\rangle }{2}}\left( \frac{\left\langle \bar{f}\left( X^{\prime
}\right) \right\rangle +\left\langle \bar{r}\left( X^{\prime }\right)
\right\rangle }{2}-\left\langle \hat{f}\left( X^{\prime }\right)
\right\rangle \right) \right) }  \notag \\
&=&\frac{\left\langle \bar{w}\left( X\right) \right\rangle _{0}}{2}\frac{1}{%
\left( 1+\frac{\frac{\left\langle S_{E}^{B}\right\rangle _{0}}{\frac{%
\left\langle \bar{S}\right\rangle }{2}}\left( \frac{\left\langle \bar{f}%
\left( X^{\prime }\right) \right\rangle +\left\langle \bar{r}\left(
X^{\prime }\right) \right\rangle }{2}-\left\langle f\left( X\right)
\right\rangle \right) +\frac{\left\langle \hat{S}_{E}^{B}\right\rangle _{0}}{%
\frac{\left\langle \bar{S}\right\rangle }{2}}\left( \frac{\left\langle \bar{f%
}\left( X^{\prime }\right) \right\rangle +\left\langle \bar{r}\left(
X^{\prime }\right) \right\rangle }{2}-\left\langle \hat{f}\left( X^{\prime
}\right) \right\rangle \right) }{2+\frac{\left\langle S_{E}^{B}\right\rangle
_{0}}{\frac{\left\langle \bar{S}\right\rangle }{2}}+\frac{\left\langle \hat{S%
}_{E}^{B}\right\rangle _{0}}{\frac{\left\langle \bar{S}\right\rangle }{2}}}%
\right) }  \notag \\
&=&\frac{\left\langle \bar{w}\left( X\right) \right\rangle _{0}}{2}  \notag
\\
&&\times \left( 1-\frac{\left\langle \bar{w}\left( X\right) \right\rangle }{2%
}\left( \frac{\left\langle S_{E}^{B}\right\rangle _{0}}{\frac{\left\langle 
\bar{S}\right\rangle }{2}}\left( \frac{\left\langle \bar{f}\left( X^{\prime
}\right) \right\rangle +\left\langle \bar{r}\left( X^{\prime }\right)
\right\rangle }{2}-\left\langle f\left( X\right) \right\rangle \right) +%
\frac{\left\langle \hat{S}_{E}^{B}\right\rangle _{0}}{\frac{\left\langle 
\bar{S}\right\rangle }{2}}\left( \frac{\left\langle \bar{f}\left( X^{\prime
}\right) \right\rangle +\left\langle \bar{r}\left( X^{\prime }\right)
\right\rangle }{2}-\left\langle \hat{f}\left( X^{\prime }\right)
\right\rangle \right) \right) \right)  \notag \\
&=&\frac{\left\langle \bar{w}\left( X\right) \right\rangle _{0}}{2}\left(
1-\left( \left\langle w_{E}^{B}\left( X\right) \right\rangle _{0}\left(
\left\langle \bar{f}\left( X^{\prime }\right) \right\rangle -\left\langle
f\left( X\right) \right\rangle \right) +\left\langle \hat{w}_{E}^{B}\left(
X\right) \right\rangle _{0}\left( \left\langle \bar{f}\left( X^{\prime
}\right) \right\rangle -\left\langle \hat{f}\left( X^{\prime }\right)
\right\rangle \right) \right) \right)  \notag
\end{eqnarray}%
As a consequence:%
\begin{eqnarray}
\frac{\left\langle \bar{S}\right\rangle }{2} &=&\frac{\left\langle \bar{w}%
\left( X\right) \right\rangle _{0}}{2}+\frac{\left\langle \bar{w}\left(
X^{\prime }\right) \right\rangle _{0}}{2}  \label{SBh} \\
&&\times \left\{ \left\langle \hat{w}_{E}^{B}\left( X\right) \right\rangle
_{0}\left( \frac{\left\langle \bar{f}\left( X^{\prime }\right) \right\rangle
+\left\langle \bar{r}\left( X^{\prime }\right) \right\rangle }{2}%
-\left\langle \hat{f}\left( X^{\prime }\right) \right\rangle _{\hat{w}%
_{E}}\right) +\left\langle w_{E}^{B}\left( X\right) \right\rangle _{0}\left( 
\frac{\left\langle \bar{f}\left( X^{\prime }\right) \right\rangle
+\left\langle \bar{r}\left( X^{\prime }\right) \right\rangle }{2}%
-\left\langle f\left( X\right) \right\rangle \right) \right\}  \notag \\
&&-\frac{\left\langle \bar{w}\left( X\right) \right\rangle _{0}}{2}\left(
\left\langle w_{E}^{B}\left( X\right) \right\rangle _{0}\left( \frac{%
\left\langle \bar{f}\left( X^{\prime }\right) \right\rangle +\left\langle 
\bar{r}\left( X^{\prime }\right) \right\rangle }{2}-\left\langle f\left(
X\right) \right\rangle \right) +\left\langle \hat{w}_{E}^{B}\left( X\right)
\right\rangle _{0}\left( \frac{\left\langle \bar{f}\left( X^{\prime }\right)
\right\rangle +\left\langle \bar{r}\left( X^{\prime }\right) \right\rangle }{%
2}-\left\langle \hat{f}\left( X^{\prime }\right) \right\rangle \right)
\right)  \notag \\
&=&\frac{\left\langle \bar{w}\left( X\right) \right\rangle _{0}}{2}  \notag
\end{eqnarray}

\paragraph*{A6.5.2.4 \ Computation of $\left\langle \bar{S}_{E}\right\rangle 
$ to the first order}

Given (\ref{SBh}) we can derive directly $\left\langle \bar{S}%
_{E}\right\rangle $:%
\begin{eqnarray}
&&\left\langle \bar{S}_{E}\right\rangle =\frac{\left\langle \bar{w}\left(
X^{\prime }\right) \right\rangle _{0}}{2}\left( 1+\left\{ \left\langle \bar{w%
}\left( X\right) \right\rangle _{0}\left( \frac{\left\langle \bar{f}\left(
X^{\prime }\right) \right\rangle -\left\langle \bar{r}\left( X^{\prime
}\right) \right\rangle }{2}\right) \right. \right.  \label{SNf} \\
&&\left. \left. +\left\langle \hat{w}_{E}^{B}\left( X\right) \right\rangle
_{0}\left( \left\langle \bar{f}\left( X^{\prime }\right) \right\rangle
-\left\langle \hat{f}\left( X^{\prime }\right) \right\rangle _{\hat{w}%
_{E}}\right) +\left\langle w_{E}^{B}\left( X\right) \right\rangle _{0}\left(
\left\langle \bar{f}\left( X^{\prime }\right) \right\rangle -\left\langle
f\left( X\right) \right\rangle \right) \right\} \right)  \notag
\end{eqnarray}%
The shares $\left\langle \bar{S}\right\rangle $ and $\left\langle \bar{S}%
_{E}\right\rangle $ satisfy the relation:%
\begin{equation}
\left\langle \bar{S}\right\rangle =2\left\langle \bar{S}_{E}\right\rangle
-\left\langle \bar{w}\left( X^{\prime }\right) \right\rangle _{0}\left( 
\frac{\left\langle \bar{f}\left( X^{\prime }\right) \right\rangle
-\left\langle \bar{r}\left( X^{\prime }\right) \right\rangle }{2}\right)
\end{equation}

\paragraph*{A6.5.2.5 \ Full expansion of $\left\langle
S_{E}^{B}\right\rangle $ and $\left\langle \hat{S}_{E}^{B}\right\rangle $ at
first order}

To obtain the first order formla for $\left\langle S_{E}^{B}\right\rangle $
and $\left\langle \hat{S}_{E}^{B}\right\rangle $ we first expand\ (\ref{hbz}%
) and (\ref{hbzt}) using (\ref{QNSn}). Considering a firs order variation $%
\delta \bar{S}_{E}$ and the corresponding variation $\delta \left\langle
S_{E}^{B}\right\rangle $, the variation of (\ref{QNSn}) becomes:%
\begin{eqnarray*}
0 &=&\delta \left\langle S_{E}^{B}\right\rangle \left( \left( 1-\left( \bar{%
\gamma}\left\langle \bar{S}_{E}\right\rangle \right) ^{2}\right) -\left( 
\frac{1-\bar{S}_{E}}{1-\bar{S}}\right) ^{2}\left\langle \bar{S}%
_{E}\right\rangle \left( 1-2\bar{S}_{E}-\left\langle S_{E}^{B}\right\rangle
\right) \right) +\left( \frac{1-\bar{S}_{E}}{1-\bar{S}}\right)
^{2}\left\langle \bar{S}_{E}\right\rangle \left\langle
S_{E}^{B}\right\rangle \delta \left\langle S_{E}^{B}\right\rangle \\
&&-\delta \bar{S}_{E}\left( 2\frac{\left( \bar{\gamma}x\right) ^{2}}{x}%
+\left( 1-x\right) \frac{-3x+4x^{2}+1}{\left( 1-2x\right) ^{2}}\right)
\end{eqnarray*}%
Given (\ref{QNSn}), this also writes:%
\begin{equation}
0=\left( \frac{\bar{S}_{E}}{\left\langle S_{E}^{B}\right\rangle }+\left( 
\frac{1-\bar{S}_{E}}{1-\bar{S}}\right) ^{2}\left\langle \bar{S}%
_{E}\right\rangle \left\langle S_{E}^{B}\right\rangle \right) \delta
\left\langle S_{E}^{B}\right\rangle -\delta \bar{S}_{E}\left( 2\frac{\left( 
\bar{\gamma}x\right) ^{2}}{x}+\left( 1-x\right) \frac{-3x+4x^{2}+1}{\left(
1-2x\right) ^{2}}\right)
\end{equation}%
with solution:%
\begin{equation*}
\delta \left\langle S_{E}^{B}\right\rangle =\frac{2\frac{\left( \bar{\gamma}%
x\right) ^{2}}{x}+\left( 1-x\right) \frac{-3x+4x^{2}+1}{\left( 1-2x\right)
^{2}}}{\left( \frac{\bar{S}_{E}}{\left\langle S_{E}^{B}\right\rangle }%
+\left( \frac{1-\bar{S}_{E}}{1-\bar{S}}\right) ^{2}\left\langle \bar{S}%
_{E}\right\rangle \left\langle S_{E}^{B}\right\rangle \right) }\delta \bar{S}%
_{E}
\end{equation*}%
Using (\ref{SNf}), this leads to:%
\begin{eqnarray}
&&\delta \left\langle S_{E}^{B}\right\rangle =\frac{2\frac{\left( \bar{\gamma%
}x\right) ^{2}}{x}+\left( 1-x\right) \frac{-3x+4x^{2}+1}{\left( 1-2x\right)
^{2}}}{\left( 1+\left( \frac{1-\bar{S}_{E}}{1-\bar{S}}\right)
^{2}\left\langle S_{E}^{B}\right\rangle ^{2}\right) }^{2}\left\langle
S_{E}^{B}\right\rangle _{0}\left\{ 2\left\langle \bar{S}_{E}\right\rangle
_{0}\left( \frac{\left\langle \bar{f}\left( X^{\prime }\right) \right\rangle
-\left\langle \bar{r}\left( X^{\prime }\right) \right\rangle }{2}\right)
\right. \\
&&\left. \left. +\left\langle \hat{S}_{E}^{B}\left( X\right) \right\rangle
_{0}\left( \left\langle \bar{f}\left( X^{\prime }\right) \right\rangle
-\left\langle \hat{f}\left( X^{\prime }\right) \right\rangle _{\hat{w}%
_{E}}\right) +\left\langle S_{E}^{B}\left( X\right) \right\rangle _{0}\left(
\left\langle \bar{f}\left( X^{\prime }\right) \right\rangle -\left\langle
f\left( X\right) \right\rangle \right) \right\} \right)  \notag
\end{eqnarray}%
Considerng the additional contribution (\ref{Fr}) yields:%
\begin{eqnarray}
\left\langle S_{E}^{B}\right\rangle &=&\left\langle S_{E}^{B}\right\rangle
_{0}+\frac{2\frac{\left( \bar{\gamma}x\right) ^{2}}{x}+\left( 1-x\right) 
\frac{-3x+4x^{2}+1}{\left( 1-2x\right) ^{2}}}{\left( 1+\left( \frac{1-\bar{S}%
_{E}}{1-\bar{S}}\right) ^{2}\left\langle S_{E}^{B}\right\rangle ^{2}\right) }%
^{2}\left\langle S_{E}^{B}\right\rangle _{0}\left\{ 2\left\langle \bar{S}%
_{E}\right\rangle _{0}\left( \frac{\left\langle \bar{f}\left( X^{\prime
}\right) \right\rangle -\left\langle \bar{r}\left( X^{\prime }\right)
\right\rangle }{2}\right) \right.  \label{FBN} \\
&&\left. \left. +\left\langle \hat{S}_{E}^{B}\left( X\right) \right\rangle
_{0}\left( \left\langle \bar{f}\left( X^{\prime }\right) \right\rangle
-\left\langle \hat{f}\left( X^{\prime }\right) \right\rangle _{\hat{w}%
_{E}}\right) +\left\langle S_{E}^{B}\left( X\right) \right\rangle _{0}\left(
\left\langle \bar{f}\left( X^{\prime }\right) \right\rangle -\left\langle
f\left( X\right) \right\rangle \right) \right\} \right)  \notag \\
&&-\frac{\left\langle S_{E}^{B}\right\rangle \left( \left\langle \bar{f}%
\left( X^{\prime }\right) \right\rangle -\left\langle \hat{f}\left(
X^{\prime }\right) \right\rangle \right) +\frac{2\left\langle \bar{S}%
_{E}\right\rangle }{1-\left( \bar{\gamma}\left\langle \bar{S}%
_{E}\right\rangle \right) ^{2}}\left( \left\langle \hat{f}\left( X^{\prime
}\right) \right\rangle -\left\langle f\left( X\right) \right\rangle \right) 
}{2\left( 1+\left( \left\langle S_{E}^{B}\right\rangle -\left\langle \hat{S}%
_{E}^{B}\right\rangle \right) \frac{\left( \frac{1-\bar{S}_{E}}{1-2\bar{S}%
_{E}}\right) ^{2}}{1-\left( \bar{\gamma}\left\langle \bar{S}%
_{E}\right\rangle \right) ^{2}}\right) }\left\langle S_{E}^{B}\right\rangle
_{0}  \notag
\end{eqnarray}%
Moreover since:%
\begin{eqnarray*}
\delta \left\langle S_{E}^{B}\right\rangle +\delta \left\langle \hat{S}%
_{E}^{B}\right\rangle &=&-2\delta \left\langle \bar{S}_{E}\right\rangle _{0}-%
\frac{2\frac{\left( \bar{\gamma}x\right) ^{2}}{x}+\left( 1-x\right) \frac{%
-3x+4x^{2}+1}{\left( 1-2x\right) ^{2}}}{\left( 1+\left( \frac{1-\bar{S}_{E}}{%
1-\bar{S}}\right) ^{2}\left\langle S_{E}^{B}\right\rangle ^{2}\right) }%
^{2}\left\langle S_{E}^{B}\right\rangle _{0}\left\{ 2\left\langle \bar{S}%
_{E}\right\rangle _{0}\left( \frac{\left\langle \bar{f}\left( X^{\prime
}\right) \right\rangle -\left\langle \bar{r}\left( X^{\prime }\right)
\right\rangle }{2}\right) \right. \\
&&\left. \left. +\left\langle \hat{S}_{E}^{B}\left( X\right) \right\rangle
_{0}\left( \left\langle \bar{f}\left( X^{\prime }\right) \right\rangle
-\left\langle \hat{f}\left( X^{\prime }\right) \right\rangle _{\hat{w}%
_{E}}\right) +\left\langle S_{E}^{B}\left( X\right) \right\rangle _{0}\left(
\left\langle \bar{f}\left( X^{\prime }\right) \right\rangle -\left\langle
f\left( X\right) \right\rangle \right) \right\} \right) \\
&&+\frac{\left\langle S_{E}^{B}\right\rangle \left( \left\langle \bar{f}%
\left( X^{\prime }\right) \right\rangle -\left\langle \hat{f}\left(
X^{\prime }\right) \right\rangle \right) +\frac{2\left\langle \bar{S}%
_{E}\right\rangle }{1-\left( \bar{\gamma}\left\langle \bar{S}%
_{E}\right\rangle \right) ^{2}}\left( \left\langle \hat{f}\left( X^{\prime
}\right) \right\rangle -\left\langle f\left( X\right) \right\rangle \right) 
}{2\left( 1+\left( \left\langle S_{E}^{B}\right\rangle -\left\langle \hat{S}%
_{E}^{B}\right\rangle \right) \frac{\left( \frac{1-\bar{S}_{E}}{1-2\bar{S}%
_{E}}\right) ^{2}}{1-\left( \bar{\gamma}\left\langle \bar{S}%
_{E}\right\rangle \right) ^{2}}\right) }\left\langle S_{E}^{B}\right\rangle
_{0}
\end{eqnarray*}%
we obtain:%
\begin{eqnarray}
\left\langle \hat{S}_{E}^{B}\right\rangle &=&\left\langle \hat{S}%
_{E}^{B}\right\rangle _{0}-2\delta \left\langle \bar{S}_{E}\right\rangle
_{0}-\frac{2\frac{\left( \bar{\gamma}x\right) ^{2}}{x}+\left( 1-x\right) 
\frac{-3x+4x^{2}+1}{\left( 1-2x\right) ^{2}}}{\left( 1+\left( \frac{1-\bar{S}%
_{E}}{1-\bar{S}}\right) ^{2}\left\langle S_{E}^{B}\right\rangle ^{2}\right) }%
^{2}\left\langle S_{E}^{B}\right\rangle _{0}\left\{ 2\left\langle \bar{S}%
_{E}\right\rangle _{0}\left( \frac{\left\langle \bar{f}\left( X^{\prime
}\right) \right\rangle -\left\langle \bar{r}\left( X^{\prime }\right)
\right\rangle }{2}\right) \right.  \notag \\
&&\left. \left. +\left\langle \hat{S}_{E}^{B}\left( X\right) \right\rangle
_{0}\left( \left\langle \bar{f}\left( X^{\prime }\right) \right\rangle
-\left\langle \hat{f}\left( X^{\prime }\right) \right\rangle _{\hat{w}%
_{E}}\right) +\left\langle S_{E}^{B}\left( X\right) \right\rangle _{0}\left(
\left\langle \bar{f}\left( X^{\prime }\right) \right\rangle -\left\langle
f\left( X\right) \right\rangle \right) \right\} \right)  \notag \\
&&+\frac{\left\langle S_{E}^{B}\right\rangle \left( \left\langle \bar{f}%
\left( X^{\prime }\right) \right\rangle -\left\langle \hat{f}\left(
X^{\prime }\right) \right\rangle \right) +\frac{2\left\langle \bar{S}%
_{E}\right\rangle }{1-\left( \bar{\gamma}\left\langle \bar{S}%
_{E}\right\rangle \right) ^{2}}\left( \left\langle \hat{f}\left( X^{\prime
}\right) \right\rangle -\left\langle f\left( X\right) \right\rangle \right) 
}{2\left( 1+\left( \left\langle S_{E}^{B}\right\rangle -\left\langle \hat{S}%
_{E}^{B}\right\rangle \right) \frac{\left( \frac{1-\bar{S}_{E}}{1-2\bar{S}%
_{E}}\right) ^{2}}{1-\left( \bar{\gamma}\left\langle \bar{S}%
_{E}\right\rangle \right) ^{2}}\right) }\left\langle S_{E}^{B}\right\rangle
_{0}  \label{FHN}
\end{eqnarray}

\subsection*{A6.6 Average return $\left\langle \hat{f}\left( X^{\prime
}\right) \right\rangle $ as a function of $\left\langle \hat{S}_{E}\left(
X\right) \right\rangle $}

As in part 1, (\ref{hb}) and (\ref{Shn}) allow to write $\left\langle \hat{f}%
\left( X^{\prime }\right) \right\rangle $ as function of $\left\langle \hat{S%
}_{E}\left( X^{\prime },X\right) \right\rangle $ and external parameters $%
\left\langle f\left( X^{\prime }\right) \right\rangle $ and $\left\langle
r\left( X\right) \right\rangle $:

\begin{equation}
\left\langle \hat{f}\left( X^{\prime }\right) \right\rangle =\frac{\frac{%
2\left( 2-\left( \gamma \left\langle \hat{S}_{E}\left( X\right)
\right\rangle \right) ^{2}\right) }{1-\left( \gamma \left\langle \hat{S}%
_{E}\left( X\right) \right\rangle \right) ^{2}}\left\langle \hat{S}%
_{E}\left( X\right) \right\rangle -1+\frac{\left\langle \hat{r}\left(
X^{\prime }\right) \right\rangle _{\hat{w}_{2}}}{2}+\frac{\left\langle
f\left( X\right) \right\rangle }{2\left( 2-\left( \gamma \left\langle \hat{S}%
_{E}\left( X\right) \right\rangle \right) ^{2}\right) }}{\left( 1-\frac{1}{2}%
\frac{1-\left( \gamma \left\langle \hat{S}_{E}\left( X\right) \right\rangle
\right) ^{2}}{2-\left( \gamma \left\langle \hat{S}_{E}\left( X\right)
\right\rangle \right) ^{2}}\right) }  \label{Fsh}
\end{equation}%
Note for the sequel that:%
\begin{equation}
\left\langle \hat{f}\left( X^{\prime }\right) \right\rangle -\left\langle 
\hat{r}\left( X^{\prime }\right) \right\rangle =\frac{\frac{2\left( 2-\left(
\gamma \left\langle \hat{S}_{E}\left( X\right) \right\rangle \right)
^{2}\right) }{1-\left( \gamma \left\langle \hat{S}_{E}\left( X\right)
\right\rangle \right) ^{2}}\left\langle \hat{S}_{E}\left( X\right)
\right\rangle -1+\frac{\left\langle f\left( X\right) \right\rangle
-\left\langle \hat{r}\left( X^{\prime }\right) \right\rangle _{\hat{w}_{2}}}{%
2\left( 2-\left( \gamma \left\langle \hat{S}_{E}\left( X\right)
\right\rangle \right) ^{2}\right) }}{\left( 1-\frac{1}{2}\frac{1-\left(
\gamma \left\langle \hat{S}_{E}\left( X\right) \right\rangle \right) ^{2}}{%
2-\left( \gamma \left\langle \hat{S}_{E}\left( X\right) \right\rangle
\right) ^{2}}\right) }  \label{Fsk}
\end{equation}

\subsection*{A6.7 Computing $\left\langle \bar{f}\left( X^{\prime }\right)
\right\rangle $, $\left\langle S_{E}^{B}\right\rangle $ and $\left\langle
S_{E}^{B}\left( X\right) \right\rangle $, $\left\langle \hat{S}%
_{E}^{B}\right\rangle $ and $\left\langle \hat{S}_{E}^{B}\left( X\right)
\right\rangle $\ as functions of $\left\langle \bar{S}_{E}\right\rangle $}

To solve return equation for banks, we will need\ to express returns and
shares as functions of $\left\langle \bar{S}_{E}\right\rangle $ and $%
\left\langle \hat{S}_{E}\right\rangle $. This is already done for $%
\left\langle \hat{S}_{L}^{B}\left( X^{\prime },X\right) \right\rangle $ and $%
\left\langle S_{L}^{B}\left( X,X\right) \right\rangle $ ((\ref{frmn}) nd (%
\ref{frmd})), but similar expression have to found for the other shares.

\subsubsection*{A6.7.1 Finding $\left\langle \bar{f}\left( X^{\prime
}\right) \right\rangle $, $\left\langle \bar{f}\left( X^{\prime }\right)
\right\rangle -\left\langle \bar{r}\left( X^{\prime }\right) \right\rangle $
as function of $\left\langle \bar{S}_{E}\right\rangle $}

Starting with equation of shares (\ref{SNB}), we can express $\left\langle 
\bar{f}\left( X^{\prime }\right) \right\rangle $ as:%
\begin{equation*}
\left\langle \bar{f}\left( X^{\prime }\right) \right\rangle =\bar{F}\left(
\left\langle \bar{S}_{E}\right\rangle ,\left\langle \bar{r}\left( X^{\prime
}\right) \right\rangle ,\left\langle \hat{f}\left( X^{\prime }\right)
\right\rangle _{\hat{w}_{E}}\right)
\end{equation*}%
This function is found by regrouping the terms proportional to $\left\langle 
\bar{f}\left( X^{\prime }\right) \right\rangle $, with solution: 
\begin{equation}
\left\langle \bar{f}\left( X^{\prime }\right) \right\rangle =\frac{\frac{%
\left\langle \bar{S}_{E}\right\rangle }{\frac{\left\langle \bar{w}\left(
X^{\prime }\right) \right\rangle }{2}}-1}{1-\frac{\left\langle \bar{w}\left(
X\right) \right\rangle }{2}}+\frac{\frac{\left\langle \bar{w}\left( X\right)
\right\rangle \left\langle \bar{r}\left( X^{\prime }\right) \right\rangle }{2%
}+\left\langle \hat{w}_{E}^{B}\left( X\right) \right\rangle \left\langle 
\hat{f}\left( X^{\prime }\right) \right\rangle _{\hat{w}_{E}}+\left\langle
w_{E}^{B}\left( X\right) \right\rangle \left\langle f\left( X\right)
\right\rangle }{1-\frac{\left\langle \bar{w}\left( X\right) \right\rangle }{2%
}}  \label{Fb}
\end{equation}%
and:%
\begin{eqnarray}
&&\left\langle \bar{f}\left( X^{\prime }\right) \right\rangle -\left\langle 
\bar{r}\left( X^{\prime }\right) \right\rangle  \label{Fd} \\
&=&\frac{\frac{\left\langle \bar{S}_{E}\right\rangle }{\frac{\left\langle 
\bar{w}\left( X^{\prime }\right) \right\rangle }{2}}-1+\left\langle \hat{w}%
_{E}^{B}\left( X\right) \right\rangle \left( \left\langle \hat{f}\left(
X^{\prime }\right) \right\rangle _{\hat{w}_{E}}-\left\langle \bar{r}\left(
X^{\prime }\right) \right\rangle \right) +\left\langle w_{E}^{B}\left(
X\right) \right\rangle \left( \left\langle f\left( X\right) \right\rangle
-\left\langle \bar{r}\left( X^{\prime }\right) \right\rangle \right) }{1-%
\frac{\left\langle \bar{w}\left( X\right) \right\rangle }{2}}  \notag
\end{eqnarray}%
As in (\ref{Sb}) the coefficients $w$, $\bar{w}$ can be expressed as
functions of $\left\langle \bar{S}_{E}\right\rangle $ with $\left\langle 
\bar{w}\left( X^{\prime }\right) \right\rangle $, $\left\langle \hat{w}%
_{E}^{B}\left( X\right) \right\rangle $, $\left\langle w_{E}^{B}\left(
X\right) \right\rangle $ replaced with their zeroth order approximations: 
\begin{equation*}
\left\langle \bar{w}\left( X\right) \right\rangle _{0}=\frac{\left( 1-\left( 
\bar{\gamma}\left\langle \bar{S}_{E}\right\rangle \right) ^{2}\right) }{1+%
\left[ \frac{\left\langle S_{E}^{B}\right\rangle }{\left\langle
S_{E}\right\rangle }\right] ^{2}\left( 1-\left( \gamma \left\langle \hat{S}%
_{E}\right\rangle \right) ^{2}\right) +2\left( 1-\left( \bar{\gamma}%
\left\langle \bar{S}_{E}\right\rangle \right) ^{2}\right) +\frac{%
\left\langle \frac{1-\bar{S}_{E}}{1-\bar{S}}\right\rangle ^{2}\left\langle 
\hat{S}_{E}^{B}\right\rangle ^{2}}{\left[ \frac{\left\langle
S_{E}^{B}\right\rangle }{\left\langle S_{E}\right\rangle }\right] ^{2}}\frac{%
1}{1-\left( \gamma \left\langle \hat{S}_{E}\right\rangle \right) ^{2}}%
+\left\langle \frac{1-\bar{S}_{E}}{1-\bar{S}}\right\rangle ^{2}\left\langle 
\hat{S}_{E}^{B}\right\rangle ^{2}}
\end{equation*}%
and $\left\langle \hat{w}_{E}^{B}\left( X\right) \right\rangle $, $%
\left\langle w_{E}^{B}\left( X\right) \right\rangle $ defined in (\ref{hbz})
and (\ref{hbzt}) and:%
\begin{eqnarray*}
&&1-\frac{\left\langle \bar{w}\left( X\right) \right\rangle _{0}}{2} \\
&=&\frac{1+\left[ \frac{\left\langle S_{E}^{B}\right\rangle }{\left\langle
S_{E}\right\rangle }\right] ^{2}\left( 1-\left( \gamma \left\langle \hat{S}%
_{E}\right\rangle \right) ^{2}\right) +\left( 1-\left( \bar{\gamma}%
\left\langle \bar{S}_{E}\right\rangle \right) ^{2}\right) +\frac{%
\left\langle \frac{1-\bar{S}_{E}}{1-\bar{S}}\right\rangle ^{2}\left\langle 
\hat{S}_{E}^{B}\right\rangle ^{2}}{\left[ \frac{\left\langle
S_{E}^{B}\right\rangle }{\left\langle S_{E}\right\rangle }\right] ^{2}}\frac{%
1}{1-\left( \gamma \left\langle \hat{S}_{E}\right\rangle \right) ^{2}}%
+\left\langle \frac{1-\bar{S}_{E}}{1-\bar{S}}\right\rangle ^{2}\left\langle 
\hat{S}_{E}^{B}\right\rangle ^{2}}{1+\left[ \frac{\left\langle
S_{E}^{B}\right\rangle }{\left\langle S_{E}\right\rangle }\right] ^{2}\left(
1-\left( \gamma \left\langle \hat{S}_{E}\right\rangle \right) ^{2}\right)
+2\left( 1-\left( \bar{\gamma}\left\langle \bar{S}_{E}\right\rangle \right)
^{2}\right) +\frac{\left\langle \frac{1-\bar{S}_{E}}{1-\bar{S}}\right\rangle
^{2}\left\langle \hat{S}_{E}^{B}\right\rangle ^{2}}{\left[ \frac{%
\left\langle S_{E}^{B}\right\rangle }{\left\langle S_{E}\right\rangle }%
\right] ^{2}}\frac{1}{1-\left( \gamma \left\langle \hat{S}_{E}\right\rangle
\right) ^{2}}+\left\langle \frac{1-\bar{S}_{E}}{1-\bar{S}}\right\rangle
^{2}\left\langle \hat{S}_{E}^{B}\right\rangle ^{2}}
\end{eqnarray*}

\subsubsection*{A6.7.2 Computation of $\left\langle S_{E}^{B}\right\rangle $%
, $\left\langle S_{E}^{B}\left( X^{\prime }\right) \right\rangle $, $%
\left\langle \hat{S}_{E}^{B}\right\rangle $ and $\left\langle \hat{S}%
_{E}^{B}\left( X^{\prime }\right) \right\rangle $}

Using formula (\ref{Fsk}) and (\ref{Fd}) and (\ref{DPNC}) allow to rewrite $%
\left\langle S_{E}^{B}\right\rangle $ and $\left\langle \hat{S}%
_{E}^{B}\right\rangle $ as functions of $\left\langle \bar{S}%
_{E}\right\rangle $ and $\left\langle \hat{S}_{E}\right\rangle $: 
\begin{eqnarray}
&&\left\langle S_{E}^{B}\right\rangle \left( \left\langle \bar{f}\left(
X^{\prime }\right) \right\rangle ,\left\langle \hat{f}\left( X^{\prime
}\right) \right\rangle ,\left\langle f\left( X\right) \right\rangle
,\left\langle \bar{r}\left( X^{\prime }\right) \right\rangle \right)  \notag
\\
&&\left\langle \hat{S}_{E}^{B}\right\rangle \left( \left\langle \bar{f}%
\left( X^{\prime }\right) \right\rangle ,\left\langle \hat{f}\left(
X^{\prime }\right) \right\rangle ,\left\langle f\left( X\right)
\right\rangle ,\left\langle \bar{r}\left( X^{\prime }\right) \right\rangle
\right)  \notag
\end{eqnarray}%
and:%
\begin{equation*}
\left\langle S_{E}^{B}\left( X^{\prime }\right) \right\rangle =\left\langle
S_{E}^{B}\right\rangle \frac{\left\langle \bar{K}\right\rangle \left\Vert 
\bar{\Psi}\right\Vert ^{2}}{\left\langle K\right\rangle \left\Vert \Psi
\right\Vert ^{2}}
\end{equation*}%
\begin{equation*}
\left\langle \hat{S}_{E}^{B}\left( X^{\prime }\right) \right\rangle
=\left\langle \hat{S}_{E}^{B}\right\rangle \frac{\left\langle \bar{K}%
\right\rangle \left\Vert \bar{\Psi}\right\Vert ^{2}}{\left\langle \hat{K}%
\right\rangle \left\Vert \hat{\Psi}\right\Vert ^{2}}
\end{equation*}%
Equation (\ref{FBN}), (\ref{FHN}) allow to find the first order expansion
for $\left\langle S_{E}^{B}\right\rangle $ and $\left\langle \hat{S}%
_{E}^{B}\right\rangle $ by using (\ref{Fsk}) and (\ref{Fd}) to replace $%
\left\langle \bar{f}\left( X^{\prime }\right) \right\rangle -\left\langle 
\bar{r}\left( X^{\prime }\right) \right\rangle $ and $\left\langle \hat{f}%
\left( X^{\prime }\right) \right\rangle -\left\langle \hat{r}\left(
X^{\prime }\right) \right\rangle $. This can be written using the zeroth
order solutions (\ref{hbz}), (\ref{hbzt}) and first order correction (\ref%
{FPr}):

\begin{eqnarray}
\left\langle S_{E}^{B}\right\rangle &=&\frac{1}{2}\left( \sqrt{\left( \frac{%
1-\left( \bar{\gamma}\left\langle \bar{S}_{E}\right\rangle \right) ^{2}}{%
\left( \frac{1-\bar{S}_{E}}{1-2\bar{S}_{E}}\right) ^{2}\left\langle \bar{S}%
_{E}\right\rangle }-\left( 1-2\left\langle \bar{S}_{E}\right\rangle \right)
\right) ^{2}+\frac{4}{\left( \frac{1-\bar{S}_{E}}{1-2\bar{S}_{E}}\right) ^{2}%
}}-\left( \frac{1-\left( \bar{\gamma}\left\langle \bar{S}_{E}\right\rangle
\right) ^{2}}{\left( \frac{1-\bar{S}_{E}}{1-2\bar{S}_{E}}\right)
^{2}\left\langle \bar{S}_{E}\right\rangle }-\left( 1-2\left\langle \bar{S}%
_{E}\right\rangle \right) \right) \right)  \notag \\
&&-\frac{\left\langle S_{E}^{B}\right\rangle \left( \left\langle \bar{f}%
\left( X^{\prime }\right) \right\rangle -\left\langle \hat{f}\left(
X^{\prime }\right) \right\rangle \right) +\frac{2\left\langle \bar{S}%
_{E}\right\rangle }{1-\left( \bar{\gamma}\left\langle \bar{S}%
_{E}\right\rangle \right) ^{2}}\left( \left\langle \hat{f}\left( X^{\prime
}\right) \right\rangle -\left\langle f\left( X\right) \right\rangle \right) 
}{2\left( 1+\left( \left\langle S_{E}^{B}\right\rangle -\left\langle \hat{S}%
_{E}^{B}\right\rangle \right) \frac{\left( \frac{1-\bar{S}_{E}}{1-2\bar{S}%
_{E}}\right) ^{2}}{1-\left( \bar{\gamma}\left\langle \bar{S}%
_{E}\right\rangle \right) ^{2}}\right) }  \label{Snbc}
\end{eqnarray}%
\begin{eqnarray}
\left\langle \hat{S}_{E}^{B}\right\rangle &=&\frac{1}{2}\left( \left( \frac{%
1-\left( \bar{\gamma}\left\langle \bar{S}_{E}\right\rangle \right) ^{2}}{%
\left( \frac{1-\bar{S}_{E}}{1-2\bar{S}_{E}}\right) ^{2}\left\langle \bar{S}%
_{E}\right\rangle }+\left( 1-2\left\langle \bar{S}_{E}\right\rangle \right)
\right) -\sqrt{\left( \frac{1-\left( \bar{\gamma}\left\langle \bar{S}%
_{E}\right\rangle \right) ^{2}}{\left( \frac{1-\bar{S}_{E}}{1-2\bar{S}_{E}}%
\right) ^{2}\left\langle \bar{S}_{E}\right\rangle }-\left( 1-2\left\langle 
\bar{S}_{E}\right\rangle \right) \right) ^{2}+\frac{4}{\left( \frac{1-\bar{S}%
_{E}}{1-2\bar{S}_{E}}\right) ^{2}}}\right)  \notag \\
&&+\frac{-\left\langle S_{E}^{B}\right\rangle \left( \left\langle \bar{f}%
\left( X^{\prime }\right) \right\rangle -\left\langle \hat{f}\left(
X^{\prime }\right) \right\rangle \right) +\frac{2\left\langle \bar{S}%
_{E}\right\rangle }{1-\left( \bar{\gamma}\left\langle \bar{S}%
_{E}\right\rangle \right) ^{2}}\left( \left\langle \hat{f}\left( X^{\prime
}\right) \right\rangle -\left\langle f\left( X\right) \right\rangle \right) 
}{2\left( 1+\left( \left\langle S_{E}^{B}\right\rangle -\left\langle \hat{S}%
_{E}^{B}\right\rangle \right) \frac{\left( \frac{1-\bar{S}_{E}}{1-2\bar{S}%
_{E}}\right) ^{2}}{1-\left( \bar{\gamma}\left\langle \bar{S}%
_{E}\right\rangle \right) ^{2}}\right) }  \label{Snbd}
\end{eqnarray}%
where:%
\begin{eqnarray*}
&&-\frac{\left\langle S_{E}^{B}\right\rangle \left( \left\langle \bar{f}%
\left( X^{\prime }\right) \right\rangle -\left\langle \hat{f}\left(
X^{\prime }\right) \right\rangle \right) +\frac{2\left\langle \bar{S}%
_{E}\right\rangle }{1-\left( \bar{\gamma}\left\langle \bar{S}%
_{E}\right\rangle \right) ^{2}}\left( \left\langle \hat{f}\left( X^{\prime
}\right) \right\rangle -\left\langle f\left( X\right) \right\rangle \right) 
}{2\left( 1+\left( \left\langle S_{E}^{B}\right\rangle -\left\langle \hat{S}%
_{E}^{B}\right\rangle \right) \frac{\left( \frac{1-\bar{S}_{E}}{1-2\bar{S}%
_{E}}\right) ^{2}}{1-\left( \bar{\gamma}\left\langle \bar{S}%
_{E}\right\rangle \right) ^{2}}\right) } \\
&=&\frac{1-\left( \bar{\gamma}\left\langle \bar{S}_{E}\right\rangle \right)
^{2}}{\left( \frac{1-\bar{S}_{E}}{1-2\bar{S}_{E}}\right) ^{2}\left\langle 
\bar{S}_{E}\right\rangle }\frac{\left( \frac{1-\left( \bar{\gamma}%
\left\langle \bar{S}_{E}\right\rangle \right) ^{2}}{\left( \frac{1-\bar{S}%
_{E}}{1-2\bar{S}_{E}}\right) ^{2}\left\langle \bar{S}_{E}\right\rangle }%
-\left( 1-2\left\langle \bar{S}_{E}\right\rangle \right) \right) -\sqrt{%
\left( \frac{1-\left( \bar{\gamma}\left\langle \bar{S}_{E}\right\rangle
\right) ^{2}}{\left( \frac{1-\bar{S}_{E}}{1-2\bar{S}_{E}}\right)
^{2}\left\langle \bar{S}_{E}\right\rangle }-\left( 1-2\left\langle \bar{S}%
_{E}\right\rangle \right) \right) ^{2}+\frac{4}{\left( \frac{1-\bar{S}_{E}}{%
1-2\bar{S}_{E}}\right) ^{2}}}}{2\sqrt{\left( \frac{1-\left( \bar{\gamma}%
\left\langle \bar{S}_{E}\right\rangle \right) ^{2}}{\left( \frac{1-\bar{S}%
_{E}}{1-2\bar{S}_{E}}\right) ^{2}\left\langle \bar{S}_{E}\right\rangle }%
-\left( 1-2\left\langle \bar{S}_{E}\right\rangle \right) \right) ^{2}+\frac{4%
}{\left( \frac{1-\bar{S}_{E}}{1-2\bar{S}_{E}}\right) ^{2}}}} \\
&&\times \left( \frac{\left\langle \bar{f}\left( X^{\prime }\right)
\right\rangle +\left\langle \bar{r}\left( X^{\prime }\right) \right\rangle }{%
2}-\left\langle \hat{f}\left( X^{\prime }\right) \right\rangle \right) \\
&&-\frac{1-\left( \bar{\gamma}\left\langle \bar{S}_{E}\right\rangle \right)
^{2}}{\left( \frac{1-\bar{S}_{E}}{1-2\bar{S}_{E}}\right) ^{2}\left\langle 
\bar{S}_{E}\right\rangle }\frac{\left\langle S_{E}^{B}\right\rangle \left( 
\frac{\left\langle \bar{f}\left( X^{\prime }\right) \right\rangle
+\left\langle \bar{r}\left( X^{\prime }\right) \right\rangle }{2}%
-\left\langle \hat{f}\left( X^{\prime }\right) \right\rangle \right) }{2%
\sqrt{\left( \frac{1-\left( \bar{\gamma}\left\langle \bar{S}%
_{E}\right\rangle \right) ^{2}}{\left( \frac{1-\bar{S}_{E}}{1-2\bar{S}_{E}}%
\right) ^{2}\left\langle \bar{S}_{E}\right\rangle }-\left( 1-2\left\langle 
\bar{S}_{E}\right\rangle \right) \right) ^{2}+\frac{4}{\left( \frac{1-\bar{S}%
_{E}}{1-2\bar{S}_{E}}\right) ^{2}}}}
\end{eqnarray*}%
and the first order corrctions are obtaind as functions f $\left\langle \bar{%
S}_{E}\right\rangle $ by using (\ref{Fsk}) and (\ref{Fd}).

\section*{Appendix 7 Average equations for investors and banks in terms of $%
\left\langle \hat{S}_{E}\right\rangle $ and $\left\langle \bar{S}%
_{E}\right\rangle $ Case $C=0$ and constant return to scale}

\subsection*{A7.1 Average equations for investors}

Once the averages have been expressed in functions of $\left\langle \bar{S}%
_{E}\right\rangle $ and $\left\langle \hat{S}_{E}\right\rangle $, the return
equations (\ref{QDM}) can be considered in average. The investors equation
without default writes: 
\begin{eqnarray*}
0 &=&\int \left( \delta \left( X-X^{\prime }\right) -\hat{S}_{E}\left(
X^{\prime },X\right) \right) \frac{1-\left\langle \hat{S}\right\rangle
-\left( \left\langle \hat{S}_{E}^{B}\right\rangle +\left\langle \hat{S}%
_{L}^{B}\right\rangle \right) \frac{\left\langle \bar{K}\right\rangle
\left\Vert \bar{\Psi}\right\Vert ^{2}}{\left\langle \hat{K}\right\rangle
\left\Vert \hat{\Psi}\right\Vert ^{2}}}{1-\left\langle \hat{S}%
_{E}\right\rangle -\left( \left\langle \hat{S}_{E}^{B}\right\rangle \right) 
\frac{\left\langle \bar{K}\right\rangle \left\Vert \bar{\Psi}\right\Vert ^{2}%
}{\left\langle \hat{K}\right\rangle \left\Vert \hat{\Psi}\right\Vert ^{2}}}%
\left( f\left( X^{\prime }\right) -\bar{r}\right) dX^{\prime } \\
&&-\int S_{E}\left( X^{\prime },X\right) \frac{1-\left( S\left( X^{\prime
}\right) +\left( S_{E}^{B}\left( X^{\prime }\right) +S_{L}^{B}\left(
X^{\prime }\right) \right) \right) }{1-S_{E}\left( X^{\prime }\right)
-S_{E}^{B}\left( X^{\prime }\right) }\left( \left( f_{1}^{\prime }\left(
X^{\prime }\right) -\bar{r}\right) +\Delta F_{\tau }\left( \bar{R}\left(
K,X\right) \right) \right) dX^{\prime }
\end{eqnarray*}%
Given that $\kappa >>1$, in first approximation:%
\begin{equation*}
\frac{\left\langle \bar{K}\right\rangle \left\Vert \bar{\Psi}\right\Vert ^{2}%
}{\left\langle \hat{K}\right\rangle \left\Vert \hat{\Psi}\right\Vert ^{2}}<<1
\end{equation*}%
and:%
\begin{equation*}
\frac{1-\left\langle \hat{S}\right\rangle -\left( \left\langle \hat{S}%
_{E}^{B}\right\rangle +\left\langle \hat{S}_{L}^{B}\right\rangle \right) 
\frac{\left\langle \bar{K}\right\rangle \left\Vert \bar{\Psi}\right\Vert ^{2}%
}{\left\langle \hat{K}\right\rangle \left\Vert \hat{\Psi}\right\Vert ^{2}}}{%
1-\left\langle \hat{S}_{E}\right\rangle -\left( \left\langle \hat{S}%
_{E}^{B}\right\rangle \right) \frac{\left\langle \bar{K}\right\rangle
\left\Vert \bar{\Psi}\right\Vert ^{2}}{\left\langle \hat{K}\right\rangle
\left\Vert \hat{\Psi}\right\Vert ^{2}}}\simeq \frac{1-\left\langle \hat{S}%
\right\rangle -\left\langle \hat{S}_{L}^{B}\right\rangle \frac{\left\langle 
\bar{K}\right\rangle \left\Vert \bar{\Psi}\right\Vert ^{2}}{\left\langle 
\hat{K}\right\rangle \left\Vert \hat{\Psi}\right\Vert ^{2}}}{1-\left\langle 
\hat{S}_{E}\right\rangle }
\end{equation*}%
We will show below in Appendix 20.2.2.3 that $\left\langle \hat{S}%
_{L}^{B}\right\rangle $ is of order $\kappa $ and $\frac{\left\langle \bar{K}%
\right\rangle \left\Vert \bar{\Psi}\right\Vert ^{2}}{\left\langle \hat{K}%
\right\rangle \left\Vert \hat{\Psi}\right\Vert ^{2}}$ of order $\frac{1}{%
\kappa ^{2}}$ so that:%
\begin{equation*}
\frac{1-\left\langle \hat{S}\right\rangle -\left( \left\langle \hat{S}%
_{E}^{B}\right\rangle +\left\langle \hat{S}_{L}^{B}\right\rangle \right) 
\frac{\left\langle \bar{K}\right\rangle \left\Vert \bar{\Psi}\right\Vert ^{2}%
}{\left\langle \hat{K}\right\rangle \left\Vert \hat{\Psi}\right\Vert ^{2}}}{%
1-\left\langle \hat{S}_{E}\right\rangle -\left( \left\langle \hat{S}%
_{E}^{B}\right\rangle \right) \frac{\left\langle \bar{K}\right\rangle
\left\Vert \bar{\Psi}\right\Vert ^{2}}{\left\langle \hat{K}\right\rangle
\left\Vert \hat{\Psi}\right\Vert ^{2}}}\simeq \frac{1-\left\langle \hat{S}%
\right\rangle }{1-\left\langle \hat{S}_{E}\right\rangle }
\end{equation*}%
so that the return equation is in average:%
\begin{equation*}
\left( 1-\left\langle \hat{S}\left( X^{\prime },X\right) \right\rangle
\right) \left( \left\langle \hat{f}\left( X^{\prime }\right) \right\rangle -%
\bar{r}\right) =\left\langle S_{E}\left( X,X\right) \right\rangle \left(
\left\langle f_{1}\left( X\right) \right\rangle -\bar{r}\right)
\end{equation*}

where:%
\begin{equation*}
\left\langle f\left( X\right) \right\rangle =\left\langle f_{1}\left(
X\right) \right\rangle -\bar{r}
\end{equation*}

\subsection*{A7.2 Average equations for banks}

The average equation for banks is obtained similarly:%
\begin{eqnarray*}
0 &=&\left( 1-\left\langle \bar{S}\right\rangle \right) \left( \left\langle 
\bar{f}\left( X^{\prime }\right) \right\rangle -\left( 1+\kappa \right) \bar{%
r}\right) -\frac{1-\left\langle \hat{S}\right\rangle -\left( \left\langle 
\hat{S}_{E}^{B}\right\rangle +\left\langle \hat{S}_{L}^{B}\right\rangle
\right) \frac{\left\langle \bar{K}\right\rangle \left\Vert \bar{\Psi}%
\right\Vert ^{2}}{\left\langle \hat{K}\right\rangle \left\Vert \hat{\Psi}%
\right\Vert ^{2}}}{1-\left\langle \hat{S}_{E}\right\rangle -\left(
\left\langle \hat{S}_{E}^{B}\right\rangle \right) \frac{\left\langle \bar{K}%
\right\rangle \left\Vert \bar{\Psi}\right\Vert ^{2}}{\left\langle \hat{K}%
\right\rangle \left\Vert \hat{\Psi}\right\Vert ^{2}}}\left\langle \hat{S}%
_{E}^{B}\right\rangle \frac{\left\langle \bar{K}\right\rangle \left\Vert 
\bar{\Psi}\right\Vert ^{2}}{\left\langle \hat{K}\right\rangle \left\Vert 
\hat{\Psi}\right\Vert ^{2}}\left( \left\langle \hat{f}\left( X^{\prime
}\right) \right\rangle -\bar{r}\right) \\
&&-\left\langle S_{E}^{B}\left( X^{\prime },X^{\prime }\right) \right\rangle
\left( \left\langle f_{1}\left( X^{\prime }\right) \right\rangle -\bar{r}%
\right)
\end{eqnarray*}%
Given the solution for investors return equation, this writes also:%
\begin{eqnarray}
0 &=&\left( 1-\left\langle \bar{S}\right\rangle \right) \left( \left\langle 
\bar{f}\left( X^{\prime }\right) \right\rangle -\left( 1+\kappa \right) \bar{%
r}\right)  \label{LT} \\
&&-\left( \frac{1-\left\langle \hat{S}\right\rangle -\left( \left\langle 
\hat{S}_{E}^{B}\right\rangle +\left\langle \hat{S}_{L}^{B}\right\rangle
\right) \frac{\left\langle \bar{K}\right\rangle \left\Vert \bar{\Psi}%
\right\Vert ^{2}}{\left\langle \hat{K}\right\rangle \left\Vert \hat{\Psi}%
\right\Vert ^{2}}}{1-\left\langle \hat{S}_{E}\right\rangle -\left(
\left\langle \hat{S}_{E}^{B}\right\rangle \right) \frac{\left\langle \bar{K}%
\right\rangle \left\Vert \bar{\Psi}\right\Vert ^{2}}{\left\langle \hat{K}%
\right\rangle \left\Vert \hat{\Psi}\right\Vert ^{2}}}\frac{\left\langle \hat{%
S}_{E}^{B}\right\rangle \left\langle \bar{K}\right\rangle \left\Vert \bar{%
\Psi}\right\Vert ^{2}}{\left\langle \hat{K}\right\rangle \left\Vert \hat{\Psi%
}\right\Vert ^{2}}\frac{\left\langle S_{E}\left( X,X\right) \right\rangle }{%
\left( 1-\left\langle \hat{S}\left( X^{\prime },X\right) \right\rangle
\right) }+\left\langle S_{E}^{B}\left( X^{\prime },X^{\prime }\right)
\right\rangle \right) \left( \left\langle f_{1}\left( X\right) \right\rangle
-\bar{r}\right)  \notag
\end{eqnarray}

\subsection*{A7.3 Solving returns for investors in terms of $\left\langle 
\hat{S}_{E}\right\rangle $}

The resolution of:%
\begin{equation}
\left( 1-\left\langle \hat{S}\left( X^{\prime },X\right) \right\rangle
\right) \left( \left\langle \hat{f}\left( X^{\prime }\right) \right\rangle -%
\bar{r}\right) =\left\langle S_{E}\left( X,X\right) \right\rangle \left(
\left\langle f_{1}\left( X\right) \right\rangle -\bar{r}\right)  \label{rtqw}
\end{equation}%
is similar to part one. Given that $\left\langle \hat{S}_{E}\left( X\right)
\right\rangle =\left\langle \hat{S}_{E}\right\rangle $, setting:

\begin{equation*}
\left\langle \hat{S}\left( X^{\prime },X\right) \right\rangle =2\left\langle 
\hat{S}_{E}\left( X^{\prime },X\right) \right\rangle +\left\langle \hat{w}%
\left( X^{\prime }\right) \right\rangle \frac{\left\langle \hat{r}\left(
X^{\prime },X\right) \right\rangle -\left\langle \hat{f}\left( X^{\prime
},X\right) \right\rangle }{2}
\end{equation*}%
with:%
\begin{equation*}
\left\langle \hat{w}\left( X^{\prime }\right) \right\rangle =\frac{\left(
1-\left( \gamma \left\langle \hat{S}_{E}\left( X\right) \right\rangle
\right) ^{2}\right) }{2-\left( \gamma \left\langle \hat{S}_{E}\left(
X\right) \right\rangle \right) ^{2}}
\end{equation*}%
We write the return in terms of share, using equation (\ref{Fsk}):

\begin{equation}
\left\langle \hat{f}\left( X^{\prime }\right) \right\rangle -\left\langle 
\hat{r}\left( X^{\prime }\right) \right\rangle _{\hat{w}_{2}}=\frac{\frac{%
2\left( 2-\left( \gamma \left\langle \hat{S}_{E}\left( X\right)
\right\rangle \right) ^{2}\right) }{1-\left( \gamma \left\langle \hat{S}%
_{E}\left( X\right) \right\rangle \right) ^{2}}\left\langle \hat{S}%
_{E}\left( X\right) \right\rangle -1+\frac{\left\langle f\left( X\right)
\right\rangle -\left\langle \hat{r}\left( X^{\prime }\right) \right\rangle _{%
\hat{w}_{2}}}{2\left( 2-\left( \gamma \left\langle \hat{S}_{E}\left(
X\right) \right\rangle \right) ^{2}\right) }}{\left( 1-\frac{1}{2}\frac{%
1-\left( \gamma \left\langle \hat{S}_{E}\left( X\right) \right\rangle
\right) ^{2}}{2-\left( \gamma \left\langle \hat{S}_{E}\left( X\right)
\right\rangle \right) ^{2}}\right) }
\end{equation}%
\begin{equation*}
\left\langle f\left( X\right) \right\rangle =\left\langle f_{1}\left(
X\right) \right\rangle
\end{equation*}%
As in part one, we set $z=\left\langle \hat{S}_{E}\left( X\right)
\right\rangle $, so that the equation (\ref{rtqw}) becomes:%
\begin{eqnarray}
&&\left( 1-2z+\frac{1-\left( \gamma z\right) ^{2}}{3-\left( \gamma z\right)
^{2}}\left( \frac{2\left( 2-\left( \gamma z\right) ^{2}\right) }{1-\left(
\gamma z\right) ^{2}}z-1+\frac{\left\langle f\left( X\right) \right\rangle
-\left\langle \hat{r}\left( X^{\prime }\right) \right\rangle _{\hat{w}_{2}}}{%
2\left( 2-\left( \gamma z\right) ^{2}\right) }\right) \right)  \label{DQ} \\
&&\times \frac{\frac{2\left( 2-\left( \gamma z\right) ^{2}\right) }{1-\left(
\gamma z\right) ^{2}}z-1+\frac{\left\langle f\left( X\right) \right\rangle
-\left\langle \hat{r}\left( X^{\prime }\right) \right\rangle _{\hat{w}_{2}}}{%
2\left( 2-\left( \gamma z\right) ^{2}\right) }}{\left( 1-\frac{1}{2}\frac{%
1-\left( \gamma z\right) ^{2}}{2-\left( \gamma z\right) ^{2}}\right) } 
\notag \\
&=&\frac{1}{2}\frac{1}{2-\left( \gamma z\right) ^{2}}\left( 1+\left( \frac{%
3-\left( \gamma z\right) ^{2}}{2\left( 2-\left( \gamma z\right) ^{2}\right) }%
\right) \left( \left\langle f_{1}\left( X\right) \right\rangle -\left\langle 
\hat{r}\left( X^{\prime }\right) \right\rangle _{\hat{w}_{2}}\right) \right.
\notag \\
&&\left. -\frac{\frac{2\left( 2-\left( \gamma z\right) ^{2}\right) }{%
1-\left( \gamma z\right) ^{2}}z-1+\frac{\left\langle f\left( X\right)
\right\rangle -\left\langle \hat{r}\left( X^{\prime }\right) \right\rangle _{%
\hat{w}_{2}}}{2\left( 2-\left( \gamma z\right) ^{2}\right) }}{2\left( 1-%
\frac{1}{2}\frac{1-\left( \gamma z\right) ^{2}}{2-\left( \gamma z\right) ^{2}%
}\right) }\right) \left[ \left\langle f_{1}\left( X\right) \right\rangle
-\left\langle \bar{r}\left( X\right) \right\rangle \right]  \notag
\end{eqnarray}%
The solutions are similar to that of part one, with the same properties.

We showed in part I that the approximate solutions at the first order in $%
\left\langle f\left( X\right) \right\rangle -\left\langle \hat{r}\left(
X^{\prime }\right) \right\rangle _{\hat{w}_{2}}$ are:%
\begin{equation*}
\left\langle \hat{S}_{E}\left( X^{\prime }\right) \right\rangle =z_{0}\left(
\gamma \right) \left( 1+\frac{1}{2}\frac{z_{0}^{2}}{1-5z_{0}+8z_{0}^{2}}%
\left( \left\langle f_{1}\left( X\right) \right\rangle -\left\langle \bar{r}%
\left( X\right) \right\rangle \right) \right)
\end{equation*}%
where the share $z$ satisfies:%
\begin{equation}
\frac{2\left( 2-\left( \gamma z\right) ^{2}\right) }{1-\left( \gamma
z\right) ^{2}}z-1=0
\end{equation}%
that is:%
\begin{equation*}
1-4z-z^{2}\gamma ^{2}+2z^{3}\gamma ^{2}=0
\end{equation*}%
under the condition that $z\left( \gamma \right) \rightarrow \frac{1}{4}$
when $\gamma \rightarrow 0$.

\subsection*{A7.4 Solving equation for banks' returns}

The equation for banks average returns is (\ref{LT}):%
\begin{eqnarray}
0 &=&\left( 1-\left\langle \bar{S}\right\rangle \right) \left( \left\langle 
\bar{f}\left( X^{\prime }\right) \right\rangle -\left( 1+\kappa \right) \bar{%
r}\right)  \label{PRV} \\
&&-\left( \frac{1-\left\langle \hat{S}\right\rangle -\left( \left\langle 
\hat{S}_{E}^{B}\right\rangle +\left\langle \hat{S}_{L}^{B}\right\rangle
\right) \frac{\left\langle \bar{K}\right\rangle \left\Vert \bar{\Psi}%
\right\Vert ^{2}}{\left\langle \hat{K}\right\rangle \left\Vert \hat{\Psi}%
\right\Vert ^{2}}}{1-\left\langle \hat{S}_{E}\right\rangle -\left(
\left\langle \hat{S}_{E}^{B}\right\rangle \right) \frac{\left\langle \bar{K}%
\right\rangle \left\Vert \bar{\Psi}\right\Vert ^{2}}{\left\langle \hat{K}%
\right\rangle \left\Vert \hat{\Psi}\right\Vert ^{2}}}\frac{\left\langle \hat{%
S}_{E}^{B}\right\rangle \left\langle \bar{K}\right\rangle \left\Vert \bar{%
\Psi}\right\Vert ^{2}}{\left\langle \hat{K}\right\rangle \left\Vert \hat{\Psi%
}\right\Vert ^{2}}\frac{\left\langle S_{E}\left( X,X\right) \right\rangle }{%
\left( 1-\left\langle \hat{S}\left( X^{\prime },X\right) \right\rangle
\right) }+\left\langle S_{E}^{B}\left( X^{\prime },X^{\prime }\right)
\right\rangle \right) \left( \left\langle f_{1}\left( X\right) \right\rangle
-\bar{r}\right)  \notag
\end{eqnarray}%
where the various terms have been obtained previously in terms of $%
\left\langle \bar{S}_{E}\right\rangle $, $\left\langle \hat{S}%
_{E}\right\rangle $, $\left\langle \left( f_{1}\left( X^{\prime }\right)
\right) \right\rangle $: 
\begin{equation}
\left\langle \bar{S}\right\rangle =2\left\langle \bar{S}_{E}\right\rangle
-\left\langle \bar{w}\left( X^{\prime }\right) \right\rangle \left( \frac{%
\left\langle \bar{f}\left( X^{\prime }\right) \right\rangle -\left\langle 
\bar{r}\left( X^{\prime }\right) \right\rangle }{2}\right)  \label{Frs}
\end{equation}%
with:%
\begin{eqnarray*}
\left\langle \bar{w}\left( X^{\prime }\right) \right\rangle &=&\frac{2\left(
1-\left( \bar{\gamma}\frac{\left\langle \bar{S}\right\rangle }{2}\right)
^{2}\right) }{1+\left[ \frac{\left\langle S_{E}^{B}\right\rangle }{%
\left\langle S_{E}\right\rangle }\right] ^{2}\left( 1-\left( \gamma
\left\langle \hat{S}_{E}\right\rangle \right) ^{2}\right) +2\left( 1-\left( 
\bar{\gamma}\frac{\left\langle \bar{S}\right\rangle }{2}\right) ^{2}\right) +%
\frac{\left\langle \frac{1-\bar{S}_{E}}{1-\bar{S}}\right\rangle
^{2}\left\langle \hat{S}_{E}^{B}\right\rangle ^{2}}{\left[ \frac{%
\left\langle S_{E}^{B}\right\rangle }{\left\langle S_{E}\right\rangle }%
\right] ^{2}}\frac{1}{1-\left( \gamma \left\langle \hat{S}_{E}\right\rangle
\right) ^{2}}+\left\langle \frac{1-\bar{S}_{E}}{1-\bar{S}}\right\rangle
^{2}\left\langle \hat{S}_{E}^{B}\right\rangle ^{2}} \\
&\simeq &\frac{2\left( 1-\left( \bar{\gamma}\left\langle \bar{S}%
_{E}\right\rangle \right) ^{2}\right) }{1+\left[ \frac{\left\langle
S_{E}^{B}\right\rangle }{\left\langle S_{E}\right\rangle }\right] ^{2}\left(
1-\left( \gamma \left\langle \hat{S}_{E}\right\rangle \right) ^{2}\right)
+2\left( 1-\left( \bar{\gamma}\left\langle \bar{S}_{E}\right\rangle \right)
^{2}\right) +\frac{\left\langle \frac{1-\bar{S}_{E}}{1-\bar{S}}\right\rangle
^{2}\left\langle \hat{S}_{E}^{B}\right\rangle ^{2}}{\left[ \frac{%
\left\langle S_{E}^{B}\right\rangle }{\left\langle S_{E}\right\rangle }%
\right] ^{2}}\frac{1}{1-\left( \gamma \left\langle \hat{S}_{E}\right\rangle
\right) ^{2}}+\left\langle \frac{1-\bar{S}_{E}}{1-\bar{S}}\right\rangle
^{2}\left\langle \hat{S}_{E}^{B}\right\rangle ^{2}}
\end{eqnarray*}%
where at lowest ordr the equatn is satisfd:%
\begin{equation*}
\left\langle \bar{w}\left( X^{\prime }\right) \right\rangle \simeq
\left\langle \bar{w}\left( X^{\prime }\right) \right\rangle
_{0}=\left\langle \bar{S}_{E}\right\rangle
\end{equation*}%
and:

\begin{eqnarray*}
\frac{\left\langle \hat{S}_{L}^{B}\right\rangle }{\kappa \left(
1-\left\langle \bar{S}\left( X\right) \right\rangle \right) } &=&\frac{%
1-\left( \gamma \left\langle \hat{S}_{E}\right\rangle \right) ^{2}}{2-\left(
\gamma \left\langle \hat{S}_{E}\right\rangle \right) ^{2}} \\
&&\times \left\{ 1+\frac{1-\left( \gamma \left\langle \hat{S}%
_{E}\right\rangle \right) ^{2}}{2-\left( \gamma \left\langle \hat{S}%
_{E}\right\rangle \right) ^{2}}\left( \left\langle \hat{r}\left( X^{\prime
}\right) \right\rangle -\left\langle \hat{f}\left( X^{\prime }\right)
\right\rangle _{\hat{w}_{1}}\right) +\frac{\left\langle \hat{r}\left(
X^{\prime }\right) \right\rangle -\left\langle f\left( X\right)
\right\rangle }{2-\left( \gamma \left\langle \hat{S}_{E}\left( X^{\prime
},X\right) \right\rangle \right) ^{2}}\right\}
\end{eqnarray*}%
The formula for $\left\langle S_{E}^{B}\right\rangle $ and $\left\langle 
\hat{S}_{E}^{B}\right\rangle $ are given in (\ref{Snbc}) and (\ref{Snbd}).

From (\ref{Fd}), we find $\left\langle \bar{f}\left( X^{\prime }\right)
\right\rangle -\left\langle \bar{r}\left( X^{\prime }\right) \right\rangle $:%
\begin{equation}
\left\langle \bar{f}\left( X^{\prime }\right) \right\rangle -\left( 1+\kappa
\right) \left\langle \bar{r}\left( X^{\prime }\right) \right\rangle =\frac{%
\frac{\left\langle \bar{S}_{E}\right\rangle }{\frac{\left\langle \bar{w}%
\left( X^{\prime }\right) \right\rangle }{2}}-1+\left\langle \hat{w}%
_{1}^{B}\left( X\right) \right\rangle \left( \left\langle \hat{f}\left(
X^{\prime }\right) \right\rangle _{\hat{w}_{1}}-\left\langle \bar{r}\left(
X^{\prime }\right) \right\rangle \right) +\left\langle w_{1}^{B}\left(
X\right) \right\rangle \left( \left\langle f\left( X\right) \right\rangle
-\left\langle \bar{r}\left( X^{\prime }\right) \right\rangle \right) }{1-%
\frac{\left\langle \bar{w}\left( X\right) \right\rangle }{2}}  \label{DFR}
\end{equation}%
To the first order we can replace: 
\begin{eqnarray*}
\left\langle \hat{w}_{1}^{B}\left( X\right) \right\rangle &\rightarrow
&\left\langle \hat{S}_{E}^{B}\right\rangle _{0} \\
\left\langle w_{1}^{B}\left( X\right) \right\rangle &\rightarrow
&\left\langle S_{E}^{B}\right\rangle _{0}
\end{eqnarray*}%
The ratio:%
\begin{equation*}
Z=\frac{1-\left( \left\langle S_{E}\left( X^{\prime },X^{\prime }\right)
\right\rangle \frac{\left\langle \hat{K}\right\rangle \left\Vert \hat{\Psi}%
\right\Vert ^{2}}{\left\langle K\right\rangle \left\Vert \Psi \right\Vert
^{2}}+\left\langle S_{E}^{B}\left( X^{\prime },X^{\prime }\right)
\right\rangle \frac{\left\langle \bar{K}\right\rangle \left\Vert \bar{\Psi}%
\right\Vert ^{2}}{\left\langle K\right\rangle \left\Vert \Psi \right\Vert
^{2}}\right) }{1-\left( \left\langle S\left( X^{\prime },X^{\prime }\right)
\right\rangle \frac{\left\langle \hat{K}\right\rangle \left\Vert \hat{\Psi}%
\right\Vert ^{2}}{\left\langle K\right\rangle \left\Vert \Psi \right\Vert
^{2}}+\left\langle S^{B}\left( X^{\prime },X^{\prime }\right) \right\rangle 
\frac{\left\langle \bar{K}\right\rangle \left\Vert \bar{\Psi}\right\Vert ^{2}%
}{\left\langle K\right\rangle \left\Vert \Psi \right\Vert ^{2}}\right) }
\end{equation*}%
was defined in (\ref{ZQN}) and capital ratios derived in (\ref{KRN}) and (%
\ref{KRT}).

Along with the return equation for $\left\langle \hat{f}\right\rangle $ we
can check numericlly that $\left\langle \hat{f}\right\rangle $ and $%
\left\langle \bar{f}\right\rangle $ increase with $\left\langle f_{1}\left(
X\right) \right\rangle $, and that capital ratis $\frac{\left\langle \hat{K}%
\right\rangle \left\Vert \hat{\Psi}\right\Vert ^{2}}{\left\langle
K\right\rangle \left\Vert \Psi \right\Vert ^{2}}$ and $\frac{\left\langle 
\bar{K}\right\rangle \left\Vert \bar{\Psi}\right\Vert ^{2}}{\left\langle
K\right\rangle \left\Vert \Psi \right\Vert ^{2}}$ decrease. Disposable
capital decrease, circulation is reduced and capital is invested directly in
firms.As for investors, we solve this equation at first order in $\left(
\left\langle f_{1}\left( X\right) \right\rangle -\bar{r}\right) $. This
equation (\ref{PRV}) will be solved below in an approximated form for
decreasing return.

\section*{Appendix 8 Including decreasing return and $C \neq 0$}

\subsection*{A8. 1 Modification of return equations under decreasing return
to scale}

Including decreasing returns modify the formulas for return equations. As in
the first part we replace in (\ref{DQ}) and (\ref{PRV}): 
\begin{equation*}
\left\langle f_{1}\left( X^{\prime }\right) \right\rangle -\bar{r}%
\rightarrow \left\langle f_{1}\left( X\right) \right\rangle _{dr}-\bar{r}
\end{equation*}%
where:%
\begin{equation}
\left\langle f_{1}\left( X\right) \right\rangle _{dr}=\frac{\left\langle
f_{1}\left( X^{\prime }\right) \right\rangle }{\left( \left\langle
K\right\rangle \left\Vert \Psi \right\Vert ^{2}\right) ^{r}}-\frac{C}{%
\left\langle K\right\rangle \left\Vert \Psi \right\Vert ^{2}}  \label{FDN}
\end{equation}%
For example: 
\begin{equation*}
Z=\frac{1-\left( \left\langle S_{E}\left( X^{\prime },X^{\prime }\right)
\right\rangle \frac{\left\langle \hat{K}\right\rangle \left\Vert \hat{\Psi}%
\right\Vert ^{2}}{\left\langle K\right\rangle \left\Vert \Psi \right\Vert
^{2}}+\left\langle S_{E}^{B}\left( X^{\prime },X^{\prime }\right)
\right\rangle \frac{\left\langle \bar{K}\right\rangle \left\Vert \bar{\Psi}%
\right\Vert ^{2}}{\left\langle K\right\rangle \left\Vert \Psi \right\Vert
^{2}}\right) }{1-\left( \left\langle S\left( X^{\prime },X^{\prime }\right)
\right\rangle \frac{\left\langle \hat{K}\right\rangle \left\Vert \hat{\Psi}%
\right\Vert ^{2}}{\left\langle K\right\rangle \left\Vert \Psi \right\Vert
^{2}}+\left\langle S^{B}\left( X^{\prime },X^{\prime }\right) \right\rangle 
\frac{\left\langle \bar{K}\right\rangle \left\Vert \bar{\Psi}\right\Vert ^{2}%
}{\left\langle K\right\rangle \left\Vert \Psi \right\Vert ^{2}}\right) }
\end{equation*}%
is computed by using capital ratios (\ref{ora}), (\ref{ore}) and
modification in $\left\langle S_{E}\left( X^{\prime },X^{\prime }\right)
\right\rangle $, $\left\langle S_{E}^{B}\left( X^{\prime },X^{\prime
}\right) \right\rangle $... This will be developped in the approximate
solutions of Appendix 8.2.6.

\subsection*{A8.2 Approximate solution}

\subsubsection*{A8.2.1 Investors equations}

Approximate solutions of equation (\ref{DQ}) have been derived in part I. We
recall the results for later purposes:%
\begin{eqnarray}
&&\left\langle \hat{S}_{E}\left( X^{\prime },X\right) \right\rangle \\
&=&\left\langle \hat{S}_{E}\left( X^{\prime }\right) \right\rangle
=z_{0}\left( \gamma \right) \left( 1+\left( \frac{z_{0}\left( 1-\gamma
^{2}z_{0}^{2}\right) }{2\left( \left( \gamma ^{2}z_{0}^{2}\right)
^{2}-\gamma ^{2}z_{0}^{2}+2\right) \left( 1-2z_{0}\right) }\right) \left(
\left\langle f_{1}\left( X\right) \right\rangle -\left\langle \bar{r}\left(
X\right) \right\rangle \right) \right)  \notag \\
&=&z_{0}\left( \gamma \right) \left( 1+\frac{1}{2}\frac{z_{0}^{2}}{%
1-5z_{0}+8z_{0}^{2}}\left( \left\langle f_{1}\left( X\right) \right\rangle
-\left\langle \bar{r}\left( X\right) \right\rangle \right) \right)  \notag
\end{eqnarray}%
where the share $\left\langle \hat{S}_{E}\left( X^{\prime },X\right)
\right\rangle _{0}=z_{0}$ satisfies:%
\begin{equation}
\frac{2\left( 2-\left( \gamma z_{0}\right) ^{2}\right) }{1-\left( \gamma
z_{0}\right) ^{2}}z_{0}-1=0
\end{equation}%
Average investors return satifies:%
\begin{equation}
\left\langle \hat{f}\left( X^{\prime }\right) \right\rangle -\left\langle 
\hat{r}\left( X^{\prime }\right) \right\rangle _{\hat{w}_{2}}=\frac{\left(
\left\langle f_{1}\left( X\right) \right\rangle ^{\left( dr\right)
}-\left\langle \hat{r}\left( X^{\prime }\right) \right\rangle _{\hat{w}%
_{2}}\right) }{2}  \label{Rtp}
\end{equation}%
The shares of participations and loans are given by:%
\begin{eqnarray}
&&\left\langle \hat{S}\left( X^{\prime },X\right) \right\rangle  \label{SHT}
\\
&=&2z_{0}\left( \gamma \right) \left( 1-\frac{1}{4}\frac{\left(
1-3z_{0}\right) \left( 1-2z_{0}\right) }{-5z_{0}+8z_{0}^{2}+1}\left(
\left\langle f_{1}\left( X\right) \right\rangle ^{\left( dr\right)
}-\left\langle \hat{r}\left( X^{\prime }\right) \right\rangle _{\hat{w}%
_{2}}\right) \right)  \notag
\end{eqnarray}%
The participations invested in real sector are:%
\begin{eqnarray}
&&\left\langle S_{E}\left( X,X\right) \right\rangle  \label{SNF} \\
&=&\frac{1-2z_{0}}{2}\left( 1+\left( \frac{z_{0}^{3}\left( 1-4z_{0}\right) }{%
\left( 1-5z_{0}+8z_{0}^{2}\right) }+\left( \frac{3}{4}-z_{0}\right) \right)
\left( \left\langle f_{1}\left( X\right) \right\rangle ^{\left( dr\right)
}-\left\langle \hat{r}\left( X^{\prime }\right) \right\rangle _{\hat{w}%
_{2}}\right) \right)  \notag
\end{eqnarray}%
and the total share $\left\langle S\left( X,X\right) \right\rangle $ to the
first order are given order:

\begin{equation}
\left\langle S\left( X,X\right) \right\rangle =\left( 1-2z_{0}\right) \left(
1+\left( \frac{1}{2}+\frac{z_{0}^{2}\left( 1-4z_{0}\right) }{\left(
1-5z_{0}+8z_{0}^{2}\right) }\right) z_{0}\left( \left\langle f_{1}\left(
X\right) \right\rangle ^{\left( dr\right) }-\left\langle \hat{r}\left(
X^{\prime }\right) \right\rangle _{\hat{w}_{2}}\right) \right)  \label{STF}
\end{equation}

\subsubsection*{A8.2.2 Banks equations}

\paragraph*{A8.2.2.1 Zeroth order solutions}

At zeroth order in $\left\langle f_{1}\left( X\right) \right\rangle
_{dr}-\left\langle r\left( X^{\prime }\right) \right\rangle $, equation (\ref%
{PRV}) writes:%
\begin{equation*}
\left( 1-\left\langle \bar{S}\right\rangle \right) \left( \left\langle \bar{f%
}\left( X^{\prime }\right) \right\rangle -\bar{r}\right) =0
\end{equation*}%
As for investors, the only possible solutions arise from:%
\begin{equation}
\left( \left\langle \bar{f}\left( X^{\prime }\right) \right\rangle -\bar{r}%
\right) =0  \label{RTn}
\end{equation}%
and givn (\ref{DFR}) the solutions are:%
\begin{equation*}
\left\langle \bar{S}_{E}\right\rangle =\frac{\left\langle \bar{w}\left(
X^{\prime }\right) \right\rangle _{0}}{2}
\end{equation*}%
These are the solutns obtained in Appendix 6.5. The variable $x=\left\langle 
\bar{S}_{E}\right\rangle $ satisifies: 
\begin{equation*}
0=1-\frac{1}{z^{2}}\frac{1-4z}{1-2z}x^{2}-\left( \left( \frac{1}{\left( 
\frac{1-x}{1-2x}\right) ^{2}}-\left( \frac{1-2x}{2\left( 1-2x\right) \frac{2z%
}{\left( 1-2z\right) ^{3}}+1}\right) ^{2}\right) \frac{2\left( 1-2x\right) 
\frac{2z}{\left( 1-2z\right) ^{3}}+1}{1-2x}+\left( 1-2x\right) \right)
\left( \frac{1-x}{1-2x}\right) ^{2}x
\end{equation*}%
and at lowest order for $z<<1$, we find:%
\begin{equation*}
x\simeq z
\end{equation*}%
\begin{equation*}
\left( 1-\left( \bar{\gamma}\left\langle \bar{S}_{E}\right\rangle \right)
^{2}\right) \simeq \frac{x}{\left( 1-2x\right) \left( 1-8z\right) }
\end{equation*}%
\begin{eqnarray*}
\left\langle S_{E}^{B}\right\rangle &\simeq &\frac{1}{2}\left( \sqrt{\left( 
\frac{1-\left( \bar{\gamma}\left\langle \bar{S}_{E}\right\rangle \right) ^{2}%
}{\left( \frac{1-\bar{S}_{E}}{1-2\bar{S}_{E}}\right) ^{2}\left\langle \bar{S}%
_{E}\right\rangle }-\left( 1-2\left\langle \bar{S}_{E}\right\rangle \right)
\right) ^{2}+\frac{4}{\left( \frac{1-\bar{S}_{E}}{1-2\bar{S}_{E}}\right) ^{2}%
}}-\left( \frac{1-\left( \bar{\gamma}\left\langle \bar{S}_{E}\right\rangle
\right) ^{2}}{\left( \frac{1-\bar{S}_{E}}{1-2\bar{S}_{E}}\right)
^{2}\left\langle \bar{S}_{E}\right\rangle }-\left( 1-2\left\langle \bar{S}%
_{E}\right\rangle \right) \right) \right) \\
&\simeq &1-2x-4z\simeq 1-6x
\end{eqnarray*}%
\begin{eqnarray*}
\left\langle \hat{S}_{E}^{B}\right\rangle &=&1-2\bar{S}_{E}-2\left\langle
S_{E}^{B}\right\rangle \\
&\simeq &4z\simeq 4x
\end{eqnarray*}%
Moreover, the coefficients for banks loans are given by:%
\begin{eqnarray*}
\left\langle \hat{S}_{L}^{B}\left( X^{\prime }\right) \right\rangle _{0}
&=&\kappa \left( 1-\left\langle \bar{S}\left( X\right) \right\rangle
_{0}\right) \frac{1-\left( \gamma \left\langle \hat{S}_{E}\left( X^{\prime
},X\right) \right\rangle \right) ^{2}}{2-\left( \gamma \left\langle \hat{S}%
_{E}\left( X^{\prime },X\right) \right\rangle \right) ^{2}}\simeq \kappa
\left( 1-2z\right) \frac{1-\left( \gamma \left\langle \hat{S}_{E}\left(
X^{\prime },X\right) \right\rangle \right) ^{2}}{2-\left( \gamma
\left\langle \hat{S}_{E}\left( X^{\prime },X\right) \right\rangle \right)
^{2}} \\
&\simeq &\kappa \left( 1-2z\right) 2z
\end{eqnarray*}%
\begin{eqnarray*}
\left\langle S_{L}^{B}\left( X^{\prime }\right) \right\rangle _{0} &=&\kappa
\left( 1-\left\langle \bar{S}\left( X\right) \right\rangle _{0}\right) \frac{%
1}{2-\left( \gamma \left\langle \hat{S}_{E}\left( X^{\prime },X\right)
\right\rangle \right) ^{2}} \\
&=&\kappa \left( 1-2z\right) \left( 1-2z\right)
\end{eqnarray*}

\paragraph{A8.2.2.3 First order solutions for constant returns}

To find approximate solutions to the frst order, we define:%
\begin{equation*}
\left\langle \bar{S}_{E}\right\rangle =\left\langle \bar{S}_{E}\right\rangle
_{0}+s_{E}
\end{equation*}%
To find $s_{E}$, we expand the LHS of (\ref{PRV}) to the first order by
insrtn with (\ref{Fd}):%
\begin{eqnarray}
&&\left\langle \bar{f}\left( X^{\prime }\right) \right\rangle -\left(
1+\kappa \right) \left\langle \bar{r}\left( X^{\prime }\right) \right\rangle
\\
&=&\frac{\frac{\left\langle \bar{S}_{E}\right\rangle }{\frac{\left\langle 
\bar{w}\left( X^{\prime }\right) \right\rangle _{0}}{2}}-1+\left\langle \hat{%
w}_{1}^{B}\left( X\right) \right\rangle _{0}\left( \left\langle \hat{f}%
\left( X^{\prime }\right) \right\rangle _{\hat{w}_{1}}-\left\langle \bar{r}%
\left( X^{\prime }\right) \right\rangle \right) +\left\langle
w_{1}^{B}\left( X\right) \right\rangle _{0}\left( \left\langle f\left(
X\right) \right\rangle -\left\langle \bar{r}\left( X^{\prime }\right)
\right\rangle \right) }{1-\frac{\left\langle \bar{w}\left( X\right)
\right\rangle _{0}}{2}}  \notag
\end{eqnarray}%
\begin{equation*}
\frac{\frac{\left\langle \bar{S}_{E}\right\rangle _{0}+s_{E}}{\left\langle 
\bar{S}_{E}\right\rangle _{0}}-1+\left\langle \hat{w}_{1}^{B}\left( X\right)
\right\rangle _{0}\left( \left\langle \hat{f}\left( X^{\prime }\right)
\right\rangle _{\hat{w}_{1}}-\left\langle \bar{r}\left( X^{\prime }\right)
\right\rangle \right) +\left\langle w_{1}^{B}\left( X\right) \right\rangle
_{0}\left( \left\langle f\left( X\right) \right\rangle -\left\langle \bar{r}%
\left( X^{\prime }\right) \right\rangle \right) }{1-\left\langle \bar{S}%
_{E}\right\rangle _{0}}
\end{equation*}%
\begin{eqnarray}
0 &=&\left( 1-\left\langle \bar{S}\right\rangle \right) \left( \left\langle 
\bar{f}\left( X^{\prime }\right) \right\rangle -\left( 1+\kappa \right) \bar{%
r}\right)  \label{Rtns} \\
&&-\left( \frac{1-\left\langle \hat{S}\right\rangle -\left( \left\langle 
\hat{S}_{E}^{B}\right\rangle +\left\langle \hat{S}_{L}^{B}\right\rangle
\right) \frac{\left\langle \bar{K}\right\rangle \left\Vert \bar{\Psi}%
\right\Vert ^{2}}{\left\langle \hat{K}\right\rangle \left\Vert \hat{\Psi}%
\right\Vert ^{2}}}{1-\left\langle \hat{S}_{E}\right\rangle -\left(
\left\langle \hat{S}_{E}^{B}\right\rangle \right) \frac{\left\langle \bar{K}%
\right\rangle \left\Vert \bar{\Psi}\right\Vert ^{2}}{\left\langle \hat{K}%
\right\rangle \left\Vert \hat{\Psi}\right\Vert ^{2}}}\frac{\left\langle \hat{%
S}_{E}^{B}\right\rangle \left\langle \bar{K}\right\rangle \left\Vert \bar{%
\Psi}\right\Vert ^{2}}{\left\langle \hat{K}\right\rangle \left\Vert \hat{\Psi%
}\right\Vert ^{2}}\frac{\left\langle S_{E}\left( X,X\right) \right\rangle }{%
\left( 1-\left\langle \hat{S}\left( X^{\prime },X\right) \right\rangle
\right) }+\left\langle S_{E}^{B}\left( X^{\prime },X^{\prime }\right)
\right\rangle \right) \left( \left\langle f_{1}\left( X\right) \right\rangle
-\bar{r}\right)  \notag
\end{eqnarray}

\begin{eqnarray}
0 &=&\left( 1-\left\langle \bar{S}\right\rangle _{0}\right) \frac{\frac{s_{E}%
}{\left\langle \bar{S}_{E}\right\rangle _{0}}+\left\langle \hat{w}%
_{1}^{B}\left( X\right) \right\rangle _{0}\left( \left\langle \hat{f}\left(
X^{\prime }\right) \right\rangle _{\hat{w}_{1}}-\left\langle \bar{r}\left(
X^{\prime }\right) \right\rangle \right) +\left\langle w_{1}^{B}\left(
X\right) \right\rangle _{0}\left( \left\langle f\left( X\right)
\right\rangle -\left\langle \bar{r}\left( X^{\prime }\right) \right\rangle
\right) }{1-\left\langle \bar{S}_{E}\right\rangle _{0}} \\
&&-\left( \frac{1-\left\langle \hat{S}\right\rangle _{0}-\left( \left\langle 
\hat{S}_{E}^{B}\right\rangle _{0}+\left\langle \hat{S}_{L}^{B}\right\rangle
_{0}\right) \frac{\left\langle \bar{K}\right\rangle \left\Vert \bar{\Psi}%
\right\Vert ^{2}}{\left\langle \hat{K}\right\rangle \left\Vert \hat{\Psi}%
\right\Vert ^{2}}}{1-\left\langle \hat{S}_{E}\right\rangle _{0}-\left\langle 
\hat{S}_{E}^{B}\right\rangle \frac{\left\langle \bar{K}\right\rangle
\left\Vert \bar{\Psi}\right\Vert ^{2}}{\left\langle \hat{K}\right\rangle
\left\Vert \hat{\Psi}\right\Vert ^{2}}}\frac{\left\langle \hat{S}%
_{E}^{B}\right\rangle _{0}\left\langle \bar{K}\right\rangle \left\Vert \bar{%
\Psi}\right\Vert ^{2}}{\left\langle \hat{K}\right\rangle \left\Vert \hat{\Psi%
}\right\Vert ^{2}}\frac{\left\langle S_{E}\right\rangle _{0}}{\left(
1-\left\langle \hat{S}\right\rangle _{0}\right) }+\left\langle
S_{E}^{B}\right\rangle _{0}\right) \left( \left\langle f_{1}\left( X\right)
\right\rangle -\bar{r}\right)  \notag
\end{eqnarray}%
\begin{eqnarray*}
&&s_{E}=\left\langle \bar{S}_{E}\right\rangle _{0}\frac{1-\left\langle \bar{S%
}_{E}\right\rangle _{0}}{\left( 1-2\left\langle \bar{S}_{E}\right\rangle
\right) }\left( \left\langle S_{E}^{\left( e\right) }\right\rangle
_{0}\right) \left( \left\langle f_{1}\left( X\right) \right\rangle -\bar{r}%
\right) \\
&&-\left\langle \bar{S}_{E}\right\rangle _{0}\left( \left\langle \hat{w}%
_{1}^{B}\left( X\right) \right\rangle _{0}\left( \left\langle \hat{f}\left(
X^{\prime }\right) \right\rangle _{\hat{w}_{1}}-\left\langle \bar{r}\left(
X^{\prime }\right) \right\rangle \right) +\left\langle w_{1}^{B}\left(
X\right) \right\rangle _{0}\left( \left\langle f\left( X\right)
\right\rangle -\left\langle \bar{r}\left( X^{\prime }\right) \right\rangle
\right) \right)
\end{eqnarray*}%
where the effective coefficient $\left\langle S_{E}^{\left( e\right)
}\right\rangle _{0}$ is computed at zeroth order:%
\begin{eqnarray}
&&\left\langle S_{E}^{\left( e\right) }\right\rangle _{0} \\
&=&\frac{1-\left\langle \hat{S}\right\rangle _{0}-\left( \left\langle \hat{S}%
_{E}^{B}\right\rangle _{0}+\left\langle \hat{S}_{L}^{B}\right\rangle
_{0}\right) \frac{\left\langle \bar{K}\right\rangle \left\Vert \bar{\Psi}%
\right\Vert ^{2}}{\left\langle \hat{K}\right\rangle \left\Vert \hat{\Psi}%
\right\Vert ^{2}}}{1-\left\langle \hat{S}_{E}\right\rangle _{0}-\left(
\left\langle \hat{S}_{E}^{B}\right\rangle _{0}\right) \frac{\left\langle 
\bar{K}\right\rangle \left\Vert \bar{\Psi}\right\Vert ^{2}}{\left\langle 
\hat{K}\right\rangle \left\Vert \hat{\Psi}\right\Vert ^{2}}}\frac{%
\left\langle \hat{S}_{E}^{B}\right\rangle _{0}\left\langle \bar{K}%
\right\rangle \left\Vert \bar{\Psi}\right\Vert ^{2}}{\left\langle \hat{K}%
\right\rangle \left\Vert \hat{\Psi}\right\Vert ^{2}}\frac{\left\langle
S_{E}\right\rangle _{0}}{\left( 1-\left\langle \hat{S}\left( X^{\prime
},X\right) \right\rangle _{0}\right) }+\left\langle S_{E}^{B}\left(
X^{\prime },X^{\prime }\right) \right\rangle _{0}  \notag
\end{eqnarray}%
This approximation is suficient to solve (\ref{PRV}) at first order, since
this coefficient appears in muliplied by $\left( \left\langle f_{1}\left(
X\right) \right\rangle -\bar{r}\right) $. The coefficients have been
computed in the previous prgrphs.

We thus replace:%
\begin{equation*}
\left\langle \bar{K}\right\rangle \left\Vert \bar{\Psi}\right\Vert
^{2}\simeq 18\frac{\sigma _{\hat{K}}^{2}V\left( 1-\bar{S}\right) ^{2}}{\bar{r%
}^{2}\left( 1+\kappa \right) ^{2}\hat{\mu}}\left\Vert \bar{\Psi}%
_{0}\right\Vert ^{4}
\end{equation*}%
that is:%
\begin{eqnarray*}
&&\frac{\left\langle \bar{K}\right\rangle \left\Vert \bar{\Psi}\right\Vert
^{2}}{\left\langle \hat{K}\right\rangle \left\Vert \hat{\Psi}\right\Vert ^{2}%
} \\
&\simeq &\frac{2\left( 1-\bar{S}\right) ^{2}\left\Vert \bar{\Psi}%
_{0}\right\Vert ^{4}}{\left( 1-\hat{S}\right) ^{2}\left\Vert \hat{\Psi}%
_{0}\right\Vert ^{4}\left( 1+\kappa \right) ^{2}\left( 1+2\frac{\left( 1-%
\bar{S}\right) \left\langle \hat{S}_{L}^{B}\right\rangle }{\left( 1-\hat{S}%
\right) \left( 1+\kappa \right) }\frac{\left\Vert \bar{\Psi}_{0}\right\Vert
^{2}}{\left\Vert \hat{\Psi}_{0}\right\Vert ^{2}}+\sqrt{1+4\frac{\left( 1-%
\bar{S}\right) \left\langle \hat{S}_{L}^{B}\right\rangle }{\left( 1-\hat{S}%
\right) \left( 1+\kappa \right) }\frac{\left\Vert \bar{\Psi}_{0}\right\Vert
^{2}}{\left\Vert \hat{\Psi}_{0}\right\Vert ^{2}}}\right) } \\
&\simeq &\frac{2\left( 1-2x\right) ^{2}\left\Vert \bar{\Psi}_{0}\right\Vert
^{4}}{\left( 1-2z\right) ^{2}\left\Vert \hat{\Psi}_{0}\right\Vert ^{4}\left(
1+\kappa \right) ^{2}\left( 1+2\frac{\left( 1-2x\right) \kappa \left(
1-2z\right) 2z}{\left( 1-2z\right) \left( 1+\kappa \right) }\frac{\left\Vert 
\bar{\Psi}_{0}\right\Vert ^{2}}{\left\Vert \hat{\Psi}_{0}\right\Vert ^{2}}+%
\sqrt{1+4\frac{\left( 1-2x\right) \kappa \left( 1-2z\right) 2z}{\left(
1-2z\right) \left( 1+\kappa \right) }\frac{\left\Vert \bar{\Psi}%
_{0}\right\Vert ^{2}}{\left\Vert \hat{\Psi}_{0}\right\Vert ^{2}}}\right) } \\
&\simeq &\frac{2\left\Vert \bar{\Psi}_{0}\right\Vert ^{4}}{\left\Vert \hat{%
\Psi}_{0}\right\Vert ^{4}\left( 1+\kappa \right) ^{2}\left( 1+2\frac{\kappa
\left( 1-2z\right) 2z}{\left( 1+\kappa \right) }\frac{\left\Vert \bar{\Psi}%
_{0}\right\Vert ^{2}}{\left\Vert \hat{\Psi}_{0}\right\Vert ^{2}}+\sqrt{1+4%
\frac{\kappa \left( 1-2z\right) 2z}{\left( 1+\kappa \right) }\frac{%
\left\Vert \bar{\Psi}_{0}\right\Vert ^{2}}{\left\Vert \hat{\Psi}%
_{0}\right\Vert ^{2}}}\right) }
\end{eqnarray*}%
This justifies our assumption in Appendix 7.1 in that $\frac{\left\langle 
\bar{K}\right\rangle \left\Vert \bar{\Psi}\right\Vert ^{2}}{\left\langle 
\hat{K}\right\rangle \left\Vert \hat{\Psi}\right\Vert ^{2}}$ is of order $%
\frac{1}{\kappa ^{2}}$.

Moreover, given our previous results:%
\begin{eqnarray*}
&&\frac{1-\left\langle \hat{S}\right\rangle _{0}-\left( \left\langle \hat{S}%
_{E}^{B}\right\rangle _{0}+\left\langle \hat{S}_{L}^{B}\right\rangle
_{0}\right) \frac{\left\langle \bar{K}\right\rangle \left\Vert \bar{\Psi}%
\right\Vert ^{2}}{\left\langle \hat{K}\right\rangle \left\Vert \hat{\Psi}%
\right\Vert ^{2}}}{1-\left\langle \hat{S}_{E}\right\rangle _{0}-\left(
\left\langle \hat{S}_{E}^{B}\right\rangle _{0}\right) \frac{\left\langle 
\bar{K}\right\rangle \left\Vert \bar{\Psi}\right\Vert ^{2}}{\left\langle 
\hat{K}\right\rangle \left\Vert \hat{\Psi}\right\Vert ^{2}}} \\
&\simeq &\frac{1-2z-\left( 4z+\kappa \left( 1-2z\right) 2z\right) \frac{%
2\left\Vert \bar{\Psi}_{0}\right\Vert ^{4}}{\left\Vert \hat{\Psi}%
_{0}\right\Vert ^{4}\left( 1+\kappa \right) ^{2}\left( 1+2\frac{\kappa
\left( 1-2z\right) 2z}{\left( 1+\kappa \right) }\frac{\left\Vert \bar{\Psi}%
_{0}\right\Vert ^{2}}{\left\Vert \hat{\Psi}_{0}\right\Vert ^{2}}+\sqrt{1+4%
\frac{\kappa \left( 1-2z\right) 2z}{\left( 1+\kappa \right) }\frac{%
\left\Vert \bar{\Psi}_{0}\right\Vert ^{2}}{\left\Vert \hat{\Psi}%
_{0}\right\Vert ^{2}}}\right) }}{1-z-4z\frac{2\left\Vert \bar{\Psi}%
_{0}\right\Vert ^{4}}{\left\Vert \hat{\Psi}_{0}\right\Vert ^{4}\left(
1+\kappa \right) ^{2}\left( 1+2\frac{\kappa \left( 1-2z\right) 2z}{\left(
1+\kappa \right) }\frac{\left\Vert \bar{\Psi}_{0}\right\Vert ^{2}}{%
\left\Vert \hat{\Psi}_{0}\right\Vert ^{2}}+\sqrt{1+4\frac{\kappa \left(
1-2z\right) 2z}{\left( 1+\kappa \right) }\frac{\left\Vert \bar{\Psi}%
_{0}\right\Vert ^{2}}{\left\Vert \hat{\Psi}_{0}\right\Vert ^{2}}}\right) }}
\end{eqnarray*}%
and:%
\begin{eqnarray}
&&\left\langle S_{E}^{\left( e\right) }\right\rangle _{0} \\
&=&1-2x-4z  \notag \\
&&+\frac{1-2z-\left( 4z+\kappa \left( 1-2z\right) 2z\right) \frac{%
2\left\Vert \bar{\Psi}_{0}\right\Vert ^{4}}{\left\Vert \hat{\Psi}%
_{0}\right\Vert ^{4}\left( 1+\kappa \right) ^{2}\left( 1+2\frac{\kappa
\left( 1-2z\right) 2z}{\left( 1+\kappa \right) }\frac{\left\Vert \bar{\Psi}%
_{0}\right\Vert ^{2}}{\left\Vert \hat{\Psi}_{0}\right\Vert ^{2}}+\sqrt{1+4%
\frac{\kappa \left( 1-2z\right) 2z}{\left( 1+\kappa \right) }\frac{%
\left\Vert \bar{\Psi}_{0}\right\Vert ^{2}}{\left\Vert \hat{\Psi}%
_{0}\right\Vert ^{2}}}\right) }}{1-z-4z\frac{2\left\Vert \bar{\Psi}%
_{0}\right\Vert ^{4}}{\left\Vert \hat{\Psi}_{0}\right\Vert ^{4}\left(
1+\kappa \right) ^{2}\left( 1+2\frac{\kappa \left( 1-2z\right) 2z}{\left(
1+\kappa \right) }\frac{\left\Vert \bar{\Psi}_{0}\right\Vert ^{2}}{%
\left\Vert \hat{\Psi}_{0}\right\Vert ^{2}}+\sqrt{1+4\frac{\kappa \left(
1-2z\right) 2z}{\left( 1+\kappa \right) }\frac{\left\Vert \bar{\Psi}%
_{0}\right\Vert ^{2}}{\left\Vert \hat{\Psi}_{0}\right\Vert ^{2}}}\right) }} 
\notag \\
&&\times \frac{4z\left( 1-\bar{S}\right) ^{2}}{2\left( 1-\hat{S}\right) ^{2}}%
\frac{2\left\Vert \bar{\Psi}_{0}\right\Vert ^{4}}{\left\Vert \hat{\Psi}%
_{0}\right\Vert ^{4}\left( 1+\kappa \right) ^{2}\left( 1+2\frac{\kappa
\left( 1-2z\right) 2z}{\left( 1+\kappa \right) }\frac{\left\Vert \bar{\Psi}%
_{0}\right\Vert ^{2}}{\left\Vert \hat{\Psi}_{0}\right\Vert ^{2}}+\sqrt{1+4%
\frac{\kappa \left( 1-2z\right) 2z}{\left( 1+\kappa \right) }\frac{%
\left\Vert \bar{\Psi}_{0}\right\Vert ^{2}}{\left\Vert \hat{\Psi}%
_{0}\right\Vert ^{2}}}\right) }  \notag
\end{eqnarray}

The first order terms $v$ and $s_{E}$ have been obtained in part I:%
\begin{equation*}
v=\frac{1}{2}\frac{z_{0}^{3}}{1-5z_{0}+8z_{0}^{2}}\left( \left\langle
f_{1}\left( X\right) \right\rangle _{dr}-\left\langle r\left( X^{\prime
}\right) \right\rangle \right)
\end{equation*}%
Given that:%
\begin{equation*}
\left\langle S_{E}\right\rangle _{0}=\frac{1-2z_{0}}{2}
\end{equation*}

\begin{eqnarray*}
&&s_{E}=\left\langle \bar{S}_{E}\right\rangle _{0}\frac{1-\left\langle \bar{S%
}_{E}\right\rangle _{0}}{\left( 1-2\left\langle \bar{S}_{E}\right\rangle
\right) }\left\langle S_{E}^{\left( e\right) }\right\rangle _{0}\left(
\left\langle f_{1}\left( X\right) \right\rangle -\bar{r}\right) \\
&&-\left\langle \bar{S}_{E}\right\rangle _{0}\left( \left\langle \hat{w}%
_{1}^{B}\left( X\right) \right\rangle _{0}\left( \left\langle \hat{f}\left(
X^{\prime }\right) \right\rangle _{\hat{w}_{1}}-\left\langle \bar{r}\left(
X^{\prime }\right) \right\rangle \right) +\left\langle w_{1}^{B}\left(
X\right) \right\rangle _{0}\left( \left\langle f\left( X\right)
\right\rangle -\left\langle \bar{r}\left( X^{\prime }\right) \right\rangle
\right) \right)
\end{eqnarray*}%
\begin{equation*}
s_{E}=\left\langle \bar{S}_{E}\right\rangle _{0}\left( \frac{1-\left\langle 
\bar{S}_{E}\right\rangle _{0}}{\left( 1-2\left\langle \bar{S}%
_{E}\right\rangle \right) }\left\langle S_{E}^{\left( e\right)
}\right\rangle _{0}-\left\langle S_{E}^{B}\right\rangle _{0}\right) \left(
\left\langle f_{1}\left( X\right) \right\rangle -\bar{r}\right)
-\left\langle \bar{S}_{E}\right\rangle _{0}\left\langle \hat{S}%
_{E}^{B}\right\rangle _{0}\left( \left\langle \hat{f}\left( X^{\prime
}\right) \right\rangle _{\hat{w}_{1}}-\left\langle \bar{r}\left( X^{\prime
}\right) \right\rangle \right)
\end{equation*}%
Given (\ref{Rtp}) this is:%
\begin{equation*}
s_{E}=\left\langle \bar{S}_{E}\right\rangle _{0}\left( \frac{1-\left\langle 
\bar{S}_{E}\right\rangle _{0}}{\left( 1-2\left\langle \bar{S}%
_{E}\right\rangle \right) }\left\langle S_{E}^{\left( e\right)
}\right\rangle _{0}-\left\langle S_{E}^{B}\right\rangle _{0}-\frac{%
\left\langle \hat{S}_{E}^{B}\right\rangle _{0}}{2}\right) \left(
\left\langle f_{1}\left( X\right) \right\rangle -\bar{r}\right)
\end{equation*}%
\begin{eqnarray*}
\left\langle \bar{S}_{E}\right\rangle &=&\left( 1+\left( \frac{%
1-\left\langle \bar{S}_{E}\right\rangle _{0}}{\left( 1-2\left\langle \bar{S}%
_{E}\right\rangle \right) }\left\langle S_{E}^{\left( e\right)
}\right\rangle _{0}-\left\langle S_{E}^{B}\right\rangle _{0}-\frac{%
\left\langle \hat{S}_{E}^{B}\right\rangle _{0}}{2}\right) \left(
\left\langle f_{1}\left( X\right) \right\rangle -\bar{r}\right) \right)
\left\langle \bar{S}_{E}\right\rangle _{0} \\
&\simeq &\left( 1+\left( \frac{1-\left\langle \bar{S}_{E}\right\rangle _{0}}{%
\left( 1-2\left\langle \bar{S}_{E}\right\rangle \right) }\left\langle \bar{S}%
_{E}\right\rangle _{0}-\left\langle S_{E}^{B}\right\rangle _{0}-\frac{%
\left\langle \hat{S}_{E}^{B}\right\rangle _{0}}{2}\right) \left(
\left\langle f_{1}\left( X\right) \right\rangle -\bar{r}\right) \right)
\left\langle \bar{S}_{E}\right\rangle _{0} \\
&=&\left( 1+\frac{1}{2}\left( \frac{\left\langle \bar{S}_{E}\right\rangle
_{0}}{\left( 1-2\left\langle \bar{S}_{E}\right\rangle \right) }-\left(
1-2\left\langle \bar{S}_{E}\right\rangle \right) \right) \left( \left\langle
f_{1}\left( X\right) \right\rangle -\bar{r}\right) \right) \left\langle \bar{%
S}_{E}\right\rangle _{0}
\end{eqnarray*}%
and we obtain $\left\langle \bar{S}\right\rangle $ by: 
\begin{equation*}
\left\langle \bar{S}\right\rangle =2\left\langle \bar{S}_{E}\right\rangle
-\left\langle \bar{w}\left( X^{\prime }\right) \right\rangle \left( \frac{%
\left\langle \bar{f}\left( X^{\prime }\right) \right\rangle -\left\langle 
\bar{r}\left( X^{\prime }\right) \right\rangle }{2}\right)
\end{equation*}%
That is:%
\begin{eqnarray*}
\left\langle \bar{S}\right\rangle &\simeq &2\left\langle \bar{S}%
_{E}\right\rangle -\frac{\left\langle \bar{S}_{E}\right\rangle }{\left(
1-2\left\langle \bar{S}_{E}\right\rangle \right) }\left\langle S_{E}^{\left(
e\right) }\right\rangle _{0}\left( \left\langle f_{1}\left( X\right)
\right\rangle -\bar{r}\right) \\
&\simeq &2\left( 1+\left( \frac{1-2\left\langle \bar{S}_{E}\right\rangle _{0}%
}{2\left( 1-2\left\langle \bar{S}_{E}\right\rangle _{0}\right) }\left\langle
S_{E}^{\left( e\right) }\right\rangle _{0}-\left\langle
S_{E}^{B}\right\rangle _{0}-\frac{\left\langle \hat{S}_{E}^{B}\right\rangle
_{0}}{2}\right) \left( \left\langle f_{1}\left( X\right) \right\rangle -\bar{%
r}\right) \right) \left\langle \bar{S}_{E}\right\rangle _{0}
\end{eqnarray*}%
Given that in first approximation:%
\begin{equation*}
\left\langle S_{E}^{\left( e\right) }\right\rangle _{0}\simeq \left\langle
S_{E}^{B}\right\rangle _{0}
\end{equation*}%
This also writes:%
\begin{eqnarray*}
\left\langle \bar{S}\right\rangle &\simeq &2\left( 1-\frac{1}{2}\left(
\left\langle S_{E}^{B}\right\rangle _{0}+\left\langle \hat{S}%
_{E}^{B}\right\rangle _{0}\right) \left( \left\langle f_{1}\left( X\right)
\right\rangle -\bar{r}\right) \right) \left\langle \bar{S}_{E}\right\rangle
_{0} \\
&\simeq &2\left( 1-\frac{1-2z}{2}\left( \left\langle f_{1}\left( X\right)
\right\rangle -\bar{r}\right) \right) z
\end{eqnarray*}%
\begin{equation*}
1-\left\langle \bar{S}\right\rangle \rightarrow \left( 1-2z\right) \left(
1+z\left( \left\langle f_{1}\left( X\right) \right\rangle -\bar{r}\right)
\right)
\end{equation*}%
To derive $\left\langle \bar{K}\right\rangle \left\Vert \bar{\Psi}%
\right\Vert ^{2}$, we use: 
\begin{equation*}
\left\langle \bar{K}\right\rangle \left\Vert \bar{\Psi}\right\Vert
^{2}\simeq 9\frac{\sigma _{\hat{K}}^{2}V\left( 1-\bar{S}\right) ^{2}}{%
2\left\langle \bar{f}\right\rangle ^{2}\hat{\mu}}\left( \left\Vert \bar{\Psi}%
_{0}\right\Vert \right) ^{4}
\end{equation*}%
which is in first approximation:%
\begin{equation*}
\frac{\left( 1-2\left\langle \bar{S}_{E}\right\rangle _{0}\right) ^{2}}{%
\left( \left\langle \bar{r}\left( X\right) \right\rangle \left( 1+\kappa
\right) \right) ^{2}\hat{\mu}}\left( 1-\frac{2}{1-2\left\langle \bar{S}%
_{E}\right\rangle _{0}}\left( 2\left( \left\langle \bar{S}_{E}\right\rangle
-\left\langle \bar{S}_{E}\right\rangle _{0}\right) -\frac{\left\langle \bar{S%
}_{E}\right\rangle }{\left( 1-2\left\langle \bar{S}_{E}\right\rangle \right) 
}\left\langle S_{E}^{\left( e\right) }\right\rangle _{0}+\frac{\left\langle
S_{E}^{\left( e\right) }\right\rangle _{0}}{\left( 1+\kappa \right) }\right)
\left( \left\langle f_{1}\left( X\right) \right\rangle -\bar{r}\right)
\right)
\end{equation*}%
or in expanded form:%
\begin{eqnarray*}
&&\frac{\left( 1-2\left\langle \bar{S}_{E}\right\rangle _{0}\right) ^{2}}{%
\left( \left\langle \bar{r}\left( X\right) \right\rangle \left( 1+\kappa
\right) \right) ^{2}\hat{\mu}}\left( 1-\frac{2}{1-2\left\langle \bar{S}%
_{E}\right\rangle _{0}}\right. \\
&&\times \left. \left( 2\left( \frac{1-\left\langle \bar{S}_{E}\right\rangle
_{0}}{\left( 1-2\left\langle \bar{S}_{E}\right\rangle \right) }\left\langle
S_{E}^{\left( e\right) }\right\rangle _{0}-\left\langle
S_{E}^{B}\right\rangle _{0}-\frac{\left\langle \hat{S}_{E}^{B}\right\rangle
_{0}}{2}\right) \left\langle \bar{S}_{E}\right\rangle _{0}-\frac{%
\left\langle \bar{S}_{E}\right\rangle _{0}\left\langle S_{E}^{\left(
e\right) }\right\rangle _{0}}{\left( 1-2\left\langle \bar{S}%
_{E}\right\rangle \right) }+\frac{\left\langle S_{E}^{\left( e\right)
}\right\rangle _{0}}{\left( 1+\kappa \right) }\right) \left( \left\langle
f_{1}\left( X\right) \right\rangle -\bar{r}\right) \right)
\end{eqnarray*}

\subsubsection*{A8.2.3 Capital ratios for decreasing returns}

We compute the sectors' disposable capital and capital ratios for decreasing
return to scale.

Given (\ref{FRT}) and (\ref{FRV}):%
\begin{eqnarray}
&&\left\langle \hat{K}\right\rangle \left\Vert \hat{\Psi}\right\Vert ^{2}=%
\frac{\frac{9\sigma _{\hat{K}}^{2}V}{2\left\langle \bar{g}\right\rangle ^{2}%
\hat{\mu}}\left\Vert \bar{\Psi}_{0}\right\Vert ^{4}}{\frac{\left\langle \bar{%
K}\right\rangle \left\Vert \bar{\Psi}\right\Vert ^{2}}{\left\langle \hat{K}%
\right\rangle \left\Vert \hat{\Psi}\right\Vert ^{2}}} \\
&=&9\frac{\sigma _{\hat{K}}^{2}V\left( 1-\hat{S}\right) ^{2}\left\Vert \hat{%
\Psi}_{0}\right\Vert ^{4}}{4\hat{\mu}\left\langle \hat{f}\right\rangle ^{2}}
\notag \\
&&\times \left( 1+2\frac{\left( 1-\bar{S}\right) \left\langle \hat{f}%
\right\rangle \left\Vert \bar{\Psi}_{0}\right\Vert ^{2}\left( \left\langle 
\hat{S}_{E}^{B}\right\rangle +\left\langle \hat{S}_{L}^{B}\right\rangle
\right) }{\left( 1-\hat{S}\right) \left\langle \bar{f}\right\rangle
\left\Vert \hat{\Psi}_{0}\right\Vert ^{2}}+\sqrt{1+4\frac{\left( 1-\bar{S}%
\right) \left\langle \hat{f}\right\rangle \left\Vert \bar{\Psi}%
_{0}\right\Vert ^{2}\left( \left\langle \hat{S}_{E}^{B}\right\rangle
+\left\langle \hat{S}_{L}^{B}\right\rangle \right) }{\left( 1-\hat{S}\right)
\left\langle \bar{f}\right\rangle \left\Vert \hat{\Psi}_{0}\right\Vert ^{2}}}%
\right)  \notag
\end{eqnarray}%
\begin{equation*}
\left\langle \bar{K}\right\rangle \left\Vert \bar{\Psi}\right\Vert ^{2}=18%
\frac{\sigma _{\hat{K}}^{2}V\left\Vert \bar{\Psi}_{0}\right\Vert ^{4}\left(
1-\bar{S}\right) ^{2}}{2\hat{\mu}\left\langle \bar{f}\right\rangle ^{2}}
\end{equation*}

\begin{eqnarray*}
&&\frac{\left\langle \bar{K}\right\rangle \left\Vert \bar{\Psi}\right\Vert
^{2}}{\left\langle \hat{K}\right\rangle \left\Vert \hat{\Psi}\right\Vert ^{2}%
} \\
&=&\frac{2\left( 1-\bar{S}\right) ^{2}\left\Vert \bar{\Psi}_{0}\right\Vert
^{4}\left\langle \hat{f}\right\rangle ^{2}}{\left( 1-\hat{S}\right)
^{2}\left\Vert \hat{\Psi}_{0}\right\Vert ^{4}\left\langle \bar{f}%
\right\rangle ^{2}\left( 1+2\frac{\left( 1-\bar{S}\right) \left\langle \hat{f%
}\right\rangle }{\left( 1-\hat{S}\right) \left\langle \bar{f}\right\rangle }%
\frac{\left\Vert \bar{\Psi}_{0}\right\Vert ^{2}}{\left\Vert \hat{\Psi}%
_{0}\right\Vert ^{2}}\left( \left\langle \hat{S}_{E}^{B}\right\rangle
+\left\langle \hat{S}_{L}^{B}\right\rangle \right) +\sqrt{1+4\frac{\left( 1-%
\bar{S}\right) \left\langle \hat{f}\right\rangle }{\left( 1-\hat{S}\right)
\left\langle \bar{f}\right\rangle }\frac{\left\Vert \bar{\Psi}%
_{0}\right\Vert ^{2}}{\left\Vert \hat{\Psi}_{0}\right\Vert ^{2}}\left(
\left\langle \hat{S}_{E}^{B}\right\rangle +\left\langle \hat{S}%
_{L}^{B}\right\rangle \right) }\right) }
\end{eqnarray*}

As in part one, we first find $\left\langle \hat{f}\right\rangle $:%
\begin{eqnarray*}
\left\langle \hat{f}\right\rangle &=&\left\langle \hat{r}\left( X^{\prime
}\right) \right\rangle _{\hat{w}_{2}}+\frac{\left\langle f\right\rangle
-\left\langle \hat{r}\right\rangle }{2} \\
&\rightarrow &\left\langle \hat{r}\left( X^{\prime }\right) \right\rangle _{%
\hat{w}_{2}}+\frac{\left( \left[ K_{X}\right] ^{-r}\left\langle f_{1}\left(
X\right) \right\rangle -\frac{C}{\left( \frac{f_{1}\left( X\right) }{C_{0}+%
\bar{r}}\right) ^{\frac{1}{r}}}-\left\langle \hat{r}\left( X^{\prime
}\right) \right\rangle _{\hat{w}_{2}}\right) }{2}
\end{eqnarray*}%
that writes:%
\begin{equation*}
\left\langle \hat{f}\right\rangle =\frac{\left\langle \hat{r}\left(
X^{\prime }\right) \right\rangle _{\hat{w}_{2}}}{2}+\frac{\left\langle
f_{1}\left( X\right) \right\rangle -2C\left( \frac{f_{1}\left( X\right) }{%
C_{0}+\bar{r}}\right) ^{-\frac{1}{r}}}{2\left( 1-\frac{\left\langle S\left(
X,X\right) \right\rangle \left\langle \hat{K}\right\rangle \left\Vert \hat{%
\Psi}\right\Vert ^{2}+\left\langle S^{B}\left( X,X\right) \right\rangle
\left\langle \bar{K}\right\rangle \left\Vert \bar{\Psi}\right\Vert ^{2}}{%
\frac{2\epsilon }{3\sigma _{\hat{K}}^{2}}\left( \frac{\left\langle
f_{1}\right\rangle }{C_{0}+\bar{r}}\right) ^{\frac{2}{r}}}\right)
^{2r}\left( \left( \frac{2\epsilon }{3\sigma _{\hat{K}}^{2}}\right) ^{\frac{r%
}{2}}\frac{\left\langle f_{1}\right\rangle }{C_{0}+\bar{r}}\right) ^{2}}
\end{equation*}%
Then we obtain $\left\langle \bar{f}\right\rangle $: 
\begin{eqnarray}
\left\langle \bar{f}\right\rangle &=&\left( 1+\kappa \right) \bar{r}+\frac{%
\left\langle S_{E}^{\left( e\right) }\right\rangle _{0}}{1-\left\langle \bar{%
S}\right\rangle _{0}}\left( \left\langle f_{1}\left( X\right) \right\rangle -%
\bar{r}\right)  \label{Bt} \\
&\rightarrow &\left( 1+\kappa \right) \bar{r}+\frac{\left\langle
S_{E}^{\left( e\right) }\right\rangle _{0}\left( \left[ K_{X}\right]
^{-r}\left\langle f_{1}\left( X\right) \right\rangle -\frac{C}{\left( \frac{%
f_{1}\left( X\right) }{C_{0}+\bar{r}}\right) ^{\frac{1}{r}}}-\left\langle 
\hat{r}\left( X^{\prime }\right) \right\rangle _{\hat{w}_{2}}\right) }{%
1-\left\langle \bar{S}\right\rangle _{0}}  \notag
\end{eqnarray}%
In first approximation, for:%
\begin{equation*}
\left\langle \hat{f}\right\rangle \simeq \left\langle \hat{r}\left(
X^{\prime }\right) \right\rangle _{\hat{w}_{2}}
\end{equation*}%
we can replace:%
\begin{equation*}
\left\langle \bar{f}\right\rangle \simeq \left( 1+\kappa \right) \bar{r}
\end{equation*}%
and $\left\langle \bar{K}\right\rangle \left\Vert \bar{\Psi}\right\Vert ^{2}$
is given by:%
\begin{equation*}
\left\langle \bar{K}\right\rangle \left\Vert \bar{\Psi}\right\Vert
^{2}\simeq 9\frac{\sigma _{\hat{K}}^{2}V\left\Vert \bar{\Psi}_{0}\right\Vert
^{4}\left( 1-\bar{S}\right) ^{2}}{\hat{\mu}\left( \left( 1+\kappa \right) 
\bar{r}\right) ^{2}}
\end{equation*}%
while $\left\langle \hat{K}\right\rangle \left\Vert \hat{\Psi}\right\Vert
^{2}$ expresses as: 
\begin{eqnarray*}
\left\langle \hat{K}\right\rangle \left\Vert \hat{\Psi}\right\Vert ^{2}
&\simeq &9\frac{\sigma _{\hat{K}}^{2}V\left( 1-\hat{S}\right) ^{2}\left\Vert 
\hat{\Psi}_{0}\right\Vert ^{4}}{4\hat{\mu}\left\langle \hat{f}\right\rangle
^{2}} \\
&&\times \left( 1+2\frac{\left( 1-\bar{S}\right) \left\langle \hat{f}%
\right\rangle \left\Vert \bar{\Psi}_{0}\right\Vert ^{2}\left\langle \hat{S}%
_{L}^{B}\right\rangle }{\left( 1-\hat{S}\right) \left( 1-\frac{\left\langle
S_{E}^{\left( e\right) }\right\rangle _{0}}{1-\left\langle \bar{S}%
\right\rangle _{0}}+\kappa \right) \bar{r}\left\Vert \hat{\Psi}%
_{0}\right\Vert ^{2}}+\sqrt{1+4\frac{\left( 1-\bar{S}\right) \left\langle 
\hat{f}\right\rangle \left\Vert \bar{\Psi}_{0}\right\Vert ^{2}\left\langle 
\hat{S}_{L}^{B}\right\rangle }{\left( 1-\hat{S}\right) \left( 1-\frac{%
\left\langle S_{E}^{\left( e\right) }\right\rangle _{0}}{1-\left\langle \bar{%
S}\right\rangle _{0}}+\kappa \right) \bar{r}\left\Vert \hat{\Psi}%
_{0}\right\Vert ^{2}}}\right) \\
&\simeq &9\frac{\sigma _{\hat{K}}^{2}V\left( 1-\hat{S}\right) ^{2}\left\Vert 
\hat{\Psi}_{0}\right\Vert ^{4}}{2\hat{\mu}\bar{r}^{2}}\left( 1+2\frac{\left(
1-\bar{S}\right) \left\Vert \bar{\Psi}_{0}\right\Vert ^{2}\left\langle \hat{S%
}_{L}^{B}\right\rangle }{\left( 1-\hat{S}\right) \left( 1-\frac{\left\langle
S_{E}^{\left( e\right) }\right\rangle _{0}}{1-\left\langle \bar{S}%
\right\rangle _{0}}+\kappa \right) \left\Vert \hat{\Psi}_{0}\right\Vert ^{2}}%
\right)
\end{eqnarray*}

\begin{eqnarray*}
&&\left\langle S\left( X,X\right) \right\rangle \left\langle \hat{K}%
\right\rangle \left\Vert \hat{\Psi}\right\Vert ^{2}+\left\langle S^{B}\left(
X,X\right) \right\rangle \left\langle \bar{K}\right\rangle \left\Vert \bar{%
\Psi}\right\Vert ^{2} \\
&\simeq &9\frac{\left\langle S\left( X,X\right) \right\rangle \sigma _{\hat{K%
}}^{2}V\left( 1-\hat{S}\right) ^{2}\left\Vert \hat{\Psi}_{0}\right\Vert ^{4}%
}{2\hat{\mu}\bar{r}^{2}}\left( 1+2\frac{\left( 1-\bar{S}\right) \left\Vert 
\bar{\Psi}_{0}\right\Vert ^{2}\left\langle \hat{S}_{L}^{B}\right\rangle }{%
\left( 1-\hat{S}\right) \left( 1-\frac{\left\langle S_{E}^{\left( e\right)
}\right\rangle _{0}}{1-\left\langle \bar{S}\right\rangle _{0}}+\kappa
\right) \left\Vert \hat{\Psi}_{0}\right\Vert ^{2}}\right) \\
&&+9\frac{\left\langle S_{L}^{B}\left( X,X\right) \right\rangle \sigma _{%
\hat{K}}^{2}V\left\Vert \bar{\Psi}_{0}\right\Vert ^{4}\left( 1-\bar{S}%
\right) ^{2}}{\hat{\mu}\left( \left( 1+\kappa \right) \bar{r}\right) ^{2}}
\end{eqnarray*}%
\begin{equation*}
\left\langle \hat{f}\right\rangle \simeq \frac{\left\langle \hat{r}\left(
X^{\prime }\right) \right\rangle _{\hat{w}_{2}}}{2}+\frac{\left\langle
f_{1}\left( X\right) \right\rangle -2\left( \frac{f_{1}\left( X\right) }{%
C_{0}+\bar{r}}\right) ^{-\frac{1}{r}}}{2\left( 1-\frac{9\sigma _{\hat{K}%
}^{2}V}{\hat{\mu}\bar{r}^{2}\frac{2\epsilon }{3\sigma _{\hat{K}}^{2}}\left( 
\frac{\left\langle f_{1}\right\rangle }{C_{0}+\bar{r}}\right) ^{\frac{2}{r}}}%
T\right) ^{2r}\left( \left( \frac{2\epsilon }{3\sigma _{\hat{K}}^{2}}\right)
^{\frac{r}{2}}\frac{\left\langle f_{1}\right\rangle }{C_{0}+\bar{r}}\right)
^{2}}
\end{equation*}%
where:%
\begin{equation*}
T=\frac{\left\langle S\left( X,X\right) \right\rangle \left( 1-\hat{S}%
\right) ^{2}\left\Vert \hat{\Psi}_{0}\right\Vert ^{4}}{2\bar{r}^{2}}\left(
1+2\frac{\left( 1-\bar{S}\right) \left\Vert \bar{\Psi}_{0}\right\Vert
^{2}\left\langle \hat{S}_{L}^{B}\right\rangle }{\left( 1-\hat{S}\right)
\left( 1-\frac{\left\langle S_{E}^{\left( e\right) }\right\rangle _{0}}{%
1-\left\langle \bar{S}\right\rangle _{0}}+\kappa \right) \left\Vert \hat{\Psi%
}_{0}\right\Vert ^{2}}\right) +\frac{\left\langle S_{L}^{B}\left( X,X\right)
\right\rangle \left\Vert \bar{\Psi}_{0}\right\Vert ^{4}\left( 1-\bar{S}%
\right) ^{2}}{\left( \left( 1+\kappa \right) \bar{r}\right) ^{2}}
\end{equation*}%
\begin{eqnarray*}
\left\langle \bar{f}\right\rangle &\simeq &\left( 1-\frac{\left\langle
S_{E}^{\left( e\right) }\right\rangle _{0}}{1-\left\langle \bar{S}%
\right\rangle _{0}}+\kappa \right) \bar{r} \\
&&+\frac{\left\langle S_{E}^{\left( e\right) }\right\rangle _{0}}{%
1-\left\langle \bar{S}\right\rangle _{0}}\frac{\left\langle f_{1}\left(
X\right) \right\rangle -2\left( \frac{f_{1}\left( X\right) }{C_{0}+\bar{r}}%
\right) ^{-\frac{1}{r}}}{\left( 1-\frac{9\sigma _{\hat{K}}^{2}V}{\hat{\mu}%
\bar{r}^{2}\frac{2\epsilon }{3\sigma _{\hat{K}}^{2}}\left( \frac{%
\left\langle f_{1}\right\rangle }{C_{0}+\bar{r}}\right) ^{\frac{2}{r}}}%
T\right) ^{2r}\left( \left( \frac{2\epsilon }{3\sigma _{\hat{K}}^{2}}\right)
^{\frac{r}{2}}\frac{\left\langle f_{1}\right\rangle }{C_{0}+\bar{r}}\right)
^{2}}
\end{eqnarray*}%
\begin{eqnarray*}
&&\left\langle \bar{K}\right\rangle \left\Vert \bar{\Psi}\right\Vert ^{2} \\
&\simeq &\frac{18\sigma _{\hat{K}}^{2}V\left( 1-\bar{S}\right) ^{2}\left( 1-%
\frac{9\sigma _{\hat{K}}^{2}V}{\hat{\mu}\bar{r}^{2}\frac{2\epsilon }{3\sigma
_{\hat{K}}^{2}}\left( \frac{\left\langle f_{1}\right\rangle }{C_{0}+\bar{r}}%
\right) ^{\frac{2}{r}}}T\right) ^{4r}\left( \left( \frac{2\epsilon }{3\sigma
_{\hat{K}}^{2}}\right) ^{\frac{r}{2}}\frac{\left\langle f_{1}\right\rangle }{%
C_{0}+\bar{r}}\right) ^{4}\left\Vert \bar{\Psi}_{0}\right\Vert ^{4}}{2\left(
\left( 1-\frac{\left\langle S_{E}^{\left( e\right) }\right\rangle _{0}}{%
1-\left\langle \bar{S}\right\rangle _{0}}+\kappa \right) \bar{r}\left( 1-%
\frac{9\sigma _{\hat{K}}^{2}V}{\hat{\mu}\bar{r}^{2}\frac{2\epsilon }{3\sigma
_{\hat{K}}^{2}}\left( \frac{\left\langle f_{1}\right\rangle }{C_{0}+\bar{r}}%
\right) ^{\frac{2}{r}}}T\right) ^{2r}\left( \left( \frac{2\epsilon }{3\sigma
_{\hat{K}}^{2}}\right) ^{\frac{r}{2}}\frac{\left\langle f_{1}\right\rangle }{%
C_{0}+\bar{r}}\right) ^{2}+\frac{\left\langle S_{E}^{\left( e\right)
}\right\rangle _{0}}{1-\left\langle \bar{S}\right\rangle _{0}}\left(
\left\langle f_{1}\left( X\right) \right\rangle -2\left( \frac{f_{1}\left(
X\right) }{C_{0}+\bar{r}}\right) ^{-\frac{1}{r}}\right) \right) ^{2}\hat{\mu}%
}
\end{eqnarray*}

\begin{eqnarray*}
\left\langle K\right\rangle \left\Vert \Psi \right\Vert ^{2} &\simeq &\left(
1-\frac{\left\langle S\left( X,X\right) \right\rangle \left\langle \hat{K}%
\right\rangle \left\Vert \hat{\Psi}\right\Vert ^{2}+\left\langle S^{B}\left(
X,X\right) \right\rangle \left\langle \bar{K}\right\rangle \left\Vert \bar{%
\Psi}\right\Vert ^{2}}{\frac{2\epsilon }{3\sigma _{\hat{K}}^{2}}\left( \frac{%
\left\langle f_{1}\right\rangle }{C_{0}+\bar{r}}\right) ^{\frac{2}{r}}}%
\right) ^{2}\left( \left( \frac{2\epsilon }{3\sigma _{\hat{K}}^{2}}\right) ^{%
\frac{r}{2}}\frac{f_{1}\left( X\right) }{C_{0}+\frac{S_{L}\left( X\right) }{%
1-S_{E}\left( X\right) }\bar{r}}\right) ^{\frac{2}{r}} \\
&=&\left( 1-\frac{9\sigma _{\hat{K}}^{2}V}{\hat{\mu}\bar{r}^{2}\frac{%
2\epsilon }{3\sigma _{\hat{K}}^{2}}\left( \frac{\left\langle
f_{1}\right\rangle }{C_{0}+\bar{r}}\right) ^{\frac{2}{r}}}T\right)
^{2}\left( \left( \frac{2\epsilon }{3\sigma _{\hat{K}}^{2}}\right) ^{\frac{r%
}{2}}\frac{f_{1}\left( X\right) }{C_{0}+\frac{S_{L}\left( X\right) }{%
1-S_{E}\left( X\right) }\bar{r}}\right) ^{\frac{2}{r}}
\end{eqnarray*}%
and the capital ratios becomes:%
\begin{equation}
\frac{\left\langle \hat{K}\right\rangle \left\Vert \hat{\Psi}\right\Vert ^{2}%
}{\left\langle K\right\rangle \left\Vert \Psi \right\Vert ^{2}}\simeq \frac{%
\left\langle \hat{K}\right\rangle \left\Vert \hat{\Psi}\right\Vert ^{2}}{%
\left( 1-\frac{\left\langle S\left( X,X\right) \right\rangle \left\langle 
\hat{K}\right\rangle \left\Vert \hat{\Psi}\right\Vert ^{2}+\left\langle
S^{B}\left( X,X\right) \right\rangle \left\langle \bar{K}\right\rangle
\left\Vert \bar{\Psi}\right\Vert ^{2}}{\frac{2\epsilon }{3\sigma _{\hat{K}%
}^{2}}\left( \frac{\left\langle f_{1}\right\rangle }{C_{0}+\bar{r}}\right) ^{%
\frac{2}{r}}}\right) ^{2}\left( \left( \frac{2\epsilon }{3\sigma _{\hat{K}%
}^{2}}\right) ^{\frac{r}{2}}\frac{f_{1}\left( X\right) }{C_{0}+\frac{%
S_{L}\left( X\right) }{1-S_{E}\left( X\right) }\bar{r}}\right) ^{\frac{2}{r}}%
}  \label{oraA}
\end{equation}%
and:%
\begin{equation}
\frac{\left\langle \bar{K}\right\rangle \left\Vert \bar{\Psi}\right\Vert ^{2}%
}{\left\langle K\right\rangle \left\Vert \Psi \right\Vert ^{2}}\simeq \frac{%
\left\langle \bar{K}\right\rangle \left\Vert \bar{\Psi}\right\Vert ^{2}}{%
\left( 1-\frac{\left\langle S\left( X,X\right) \right\rangle \left\langle 
\hat{K}\right\rangle \left\Vert \hat{\Psi}\right\Vert ^{2}+\left\langle
S^{B}\left( X,X\right) \right\rangle \left\langle \bar{K}\right\rangle
\left\Vert \bar{\Psi}\right\Vert ^{2}}{\frac{2\epsilon }{3\sigma _{\hat{K}%
}^{2}}\left( \frac{\left\langle f_{1}\right\rangle }{C_{0}+\bar{r}}\right) ^{%
\frac{2}{r}}}\right) ^{2}\left( \left( \frac{2\epsilon }{3\sigma _{\hat{K}%
}^{2}}\right) ^{\frac{r}{2}}\frac{f_{1}\left( X\right) }{C_{0}+\frac{%
S_{L}\left( X\right) }{1-S_{E}\left( X\right) }\bar{r}}\right) ^{\frac{2}{r}}%
}  \label{oreA}
\end{equation}

\subsubsection*{A8.2.4 First order values of $\left\langle
S_{E}^{B}\right\rangle $ and $\left\langle \hat{S}_{E}^{B}\right\rangle $}

We start with the formula for $\left\langle S_{E}^{B}\right\rangle $:%
\begin{eqnarray*}
\left\langle S_{E}^{B}\right\rangle &=&\frac{1}{2}\left( \sqrt{\left( \frac{%
1-\left( \bar{\gamma}\left\langle \bar{S}_{E}\right\rangle \right) ^{2}}{%
\left( \frac{1-\bar{S}_{E}}{1-2\bar{S}_{E}}\right) ^{2}\left\langle \bar{S}%
_{E}\right\rangle }-\left( 1-2\left\langle \bar{S}_{E}\right\rangle \right)
\right) ^{2}+\frac{4}{\left( \frac{1-\bar{S}_{E}}{1-2\bar{S}_{E}}\right) ^{2}%
}}-\left( \frac{1-\left( \bar{\gamma}\left\langle \bar{S}_{E}\right\rangle
\right) ^{2}}{\left( \frac{1-\bar{S}_{E}}{1-2\bar{S}_{E}}\right)
^{2}\left\langle \bar{S}_{E}\right\rangle }-\left( 1-2\left\langle \bar{S}%
_{E}\right\rangle \right) \right) \right) \\
&&-\frac{\left\langle S_{E}^{B}\right\rangle \left( \left\langle \bar{f}%
\left( X^{\prime }\right) \right\rangle -\left\langle \hat{f}\left(
X^{\prime }\right) \right\rangle \right) +\frac{2\left\langle \bar{S}%
_{E}\right\rangle }{1-\left( \bar{\gamma}\left\langle \bar{S}%
_{E}\right\rangle \right) ^{2}}\left( \left\langle \hat{f}\left( X^{\prime
}\right) \right\rangle -\left\langle f\left( X\right) \right\rangle \right) 
}{2\left( 1+\left( \left\langle S_{E}^{B}\right\rangle -\left\langle \hat{S}%
_{E}^{B}\right\rangle \right) \frac{\left( \frac{1-\bar{S}_{E}}{1-2\bar{S}%
_{E}}\right) ^{2}}{1-\left( \bar{\gamma}\left\langle \bar{S}%
_{E}\right\rangle \right) ^{2}}\right) }
\end{eqnarray*}%
and expand to the first order in $x=\left\langle \bar{S}_{E}\right\rangle $:%
\begin{equation*}
\frac{1}{2}\left( \sqrt{\left( \frac{x}{\left( \frac{1-x}{1-2x}\right) ^{2}x}%
-\left( 1-2x\right) \right) ^{2}+\frac{4}{\left( \frac{1-x}{1-2x}\right) ^{2}%
}}-\left( \frac{x}{\left( \frac{1-x}{1-2\bar{S}_{E}}\right) ^{2}x}-\left(
1-2x\right) \right) \right)
\end{equation*}%
whose derivative writes:%
\begin{eqnarray*}
&&\frac{\left( \left( \frac{x}{\left( \frac{1-x}{1-2x}\right) ^{2}x}-\left(
1-2x\right) \right) -\sqrt{\left( \frac{x}{\left( \frac{1-x}{1-2x}\right)
^{2}x}-\left( 1-2x\right) \right) ^{2}+\frac{4}{\left( \frac{1-x}{1-2x}%
\right) ^{2}}}\right) \frac{d}{dx}\left( \frac{x}{\left( \frac{1-x}{1-2\bar{S%
}_{E}}\right) ^{2}x}-\left( 1-2x\right) \right) +\frac{d}{dx}\frac{2}{\left( 
\frac{1-x}{1-2x}\right) ^{2}}}{2\sqrt{\left( \frac{x}{\left( \frac{1-x}{1-2x}%
\right) ^{2}x}-\left( 1-2x\right) \right) ^{2}+\frac{4}{\left( \frac{1-x}{%
1-2x}\right) ^{2}}}} \\
&=&\frac{-\left\langle S_{E}^{B}\right\rangle \frac{d}{dx}\left( \frac{x}{%
\left( \frac{1-x}{1-2\bar{S}_{E}}\right) ^{2}x}-\left( 1-2x\right) \right) +%
\frac{d}{dx}\frac{2}{\left( \frac{1-x}{1-2x}\right) ^{2}}}{2\sqrt{\left( 
\frac{x}{\left( \frac{1-x}{1-2x}\right) ^{2}x}-\left( 1-2x\right) \right)
^{2}+\frac{4}{\left( \frac{1-x}{1-2x}\right) ^{2}}}}
\end{eqnarray*}%
Using that:%
\begin{equation*}
\frac{d}{dx}\left( \frac{x}{\left( \frac{1-x}{1-2x}\right) ^{2}x}-\left(
1-2x\right) \right) =2x\frac{\left( 1-x\right) ^{2}-x}{\left( 1-x\right) ^{3}%
}
\end{equation*}%
\begin{equation*}
\frac{d}{dx}\frac{2}{\left( \frac{1-x}{1-2x}\right) ^{2}}=4\frac{1-2x}{%
\left( x-1\right) ^{3}}
\end{equation*}%
This induces a first order contribution:%
\begin{equation*}
\left( \frac{-\left\langle S_{E}^{B}\right\rangle x\frac{\left( 1-x\right)
^{2}-x}{\left( 1-x\right) ^{3}}+2\frac{1-2x}{\left( x-1\right) ^{3}}}{\sqrt{%
\left( \frac{x}{\left( \frac{1-x}{1-2x}\right) ^{2}x}-\left( 1-2x\right)
\right) ^{2}+\frac{4}{\left( \frac{1-x}{1-2x}\right) ^{2}}}}\right) \left(
\left( \frac{1-\left\langle \bar{S}_{E}\right\rangle _{0}}{\left(
1-2\left\langle \bar{S}_{E}\right\rangle \right) }\left\langle S_{E}^{\left(
e\right) }\right\rangle _{0}-\left\langle S_{E}^{B}\right\rangle _{0}-\frac{%
\left\langle \hat{S}_{E}^{B}\right\rangle _{0}}{2}\right) \left(
\left\langle f_{1}\left( X\right) \right\rangle -\bar{r}\right) \left\langle 
\bar{S}_{E}\right\rangle _{0}\right)
\end{equation*}%
and this leads to the total variation:%
\begin{eqnarray*}
&&\frac{\left\langle S_{E}^{B}\right\rangle -\left\langle
S_{E}^{B}\right\rangle _{0}}{\frac{\left\langle \bar{S}\right\rangle }{2}} \\
&=&\frac{\left( \frac{1-\left( \bar{\gamma}\left\langle \bar{S}%
_{E}\right\rangle \right) ^{2}}{\left( \frac{1-\bar{S}_{E}}{1-2\bar{S}_{E}}%
\right) ^{2}\left\langle \bar{S}_{E}\right\rangle }-\left( 1-2\left\langle 
\bar{S}_{E}\right\rangle \right) \right) \frac{1-\left( \bar{\gamma}%
\left\langle \bar{S}_{E}\right\rangle \right) ^{2}\left( \frac{\left\langle 
\bar{f}\left( X^{\prime }\right) \right\rangle +\left\langle \bar{r}\left(
X^{\prime }\right) \right\rangle }{2}-\left\langle \hat{f}\left( X^{\prime
}\right) \right\rangle \right) }{\left\langle \bar{S}_{E}\right\rangle }%
-2\left( \left\langle \hat{f}\left( X^{\prime }\right) \right\rangle
-\left\langle f\left( X\right) \right\rangle \right) }{2\sqrt{\left( \frac{%
1-\left( \bar{\gamma}\left\langle \bar{S}_{E}\right\rangle \right) ^{2}}{%
\left( \frac{1-\bar{S}_{E}}{1-2\bar{S}_{E}}\right) ^{2}\left\langle \bar{S}%
_{E}\right\rangle }-\left( 1-2\left\langle \bar{S}_{E}\right\rangle \right)
\right) ^{2}+\frac{4}{\left( \frac{1-\bar{S}_{E}}{1-2\bar{S}_{E}}\right) ^{2}%
}}\left( \frac{1-\bar{S}_{E}}{1-2\bar{S}_{E}}\right) ^{2}} \\
&&-\frac{1}{2}\frac{1-\left( \bar{\gamma}\left\langle \bar{S}%
_{E}\right\rangle \right) ^{2}}{\left( \frac{1-\bar{S}_{E}}{1-2\bar{S}_{E}}%
\right) ^{2}\left\langle \bar{S}_{E}\right\rangle }\left( \frac{\left\langle 
\bar{f}\left( X^{\prime }\right) \right\rangle +\left\langle \bar{r}\left(
X^{\prime }\right) \right\rangle }{2}-\left\langle \hat{f}\left( X^{\prime
}\right) \right\rangle \right) \\
&=&\frac{-\left\langle S_{E}^{B}\right\rangle \frac{1-\left( \bar{\gamma}%
\left\langle \bar{S}_{E}\right\rangle \right) ^{2}}{\left( \frac{1-\bar{S}%
_{E}}{1-2\bar{S}_{E}}\right) ^{2}\left\langle \bar{S}_{E}\right\rangle }%
\left( \frac{\left\langle \bar{f}\left( X^{\prime }\right) \right\rangle
+\left\langle \bar{r}\left( X^{\prime }\right) \right\rangle }{2}%
-\left\langle \hat{f}\left( X^{\prime }\right) \right\rangle \right) -\frac{1%
}{\left( \frac{1-\bar{S}_{E}}{1-2\bar{S}_{E}}\right) ^{2}}\left(
\left\langle \hat{f}\left( X^{\prime }\right) \right\rangle -\left\langle
f\left( X\right) \right\rangle \right) }{\sqrt{\left( \frac{1-\left( \bar{%
\gamma}\left\langle \bar{S}_{E}\right\rangle \right) ^{2}}{\left( \frac{1-%
\bar{S}_{E}}{1-2\bar{S}_{E}}\right) ^{2}\left\langle \bar{S}%
_{E}\right\rangle }-\left( 1-2\left\langle \bar{S}_{E}\right\rangle \right)
\right) ^{2}+\frac{4}{\left( \frac{1-\bar{S}_{E}}{1-2\bar{S}_{E}}\right) ^{2}%
}}}
\end{eqnarray*}%
Given that in first approximation:%
\begin{equation*}
\frac{-\left\langle S_{E}^{B}\right\rangle \frac{1-\left( \bar{\gamma}%
\left\langle \bar{S}_{E}\right\rangle \right) ^{2}}{\left( \frac{1-\bar{S}%
_{E}}{1-2\bar{S}_{E}}\right) ^{2}}\left( \frac{\left\langle \bar{f}\left(
X^{\prime }\right) \right\rangle +\left\langle \bar{r}\left( X^{\prime
}\right) \right\rangle }{2}-\left\langle \hat{f}\left( X^{\prime }\right)
\right\rangle \right) -\frac{\left\langle \bar{S}_{E}\right\rangle }{\left( 
\frac{1-\bar{S}_{E}}{1-2\bar{S}_{E}}\right) ^{2}}\left( \left\langle \hat{f}%
\left( X^{\prime }\right) \right\rangle -\left\langle f\left( X\right)
\right\rangle \right) }{\sqrt{\left( \frac{1-\left( \bar{\gamma}\left\langle 
\bar{S}_{E}\right\rangle \right) ^{2}}{\left( \frac{1-\bar{S}_{E}}{1-2\bar{S}%
_{E}}\right) ^{2}\left\langle \bar{S}_{E}\right\rangle }-\left(
1-2\left\langle \bar{S}_{E}\right\rangle \right) \right) ^{2}+\frac{4}{%
\left( \frac{1-\bar{S}_{E}}{1-2\bar{S}_{E}}\right) ^{2}}}}
\end{equation*}%
can be replaced by:%
\begin{equation*}
\frac{-\left\langle S_{E}^{B}\right\rangle \frac{\left\langle \bar{S}%
_{E}\right\rangle }{\left( \frac{1-\bar{S}_{E}}{1-2\bar{S}_{E}}\right) ^{2}}%
\left( \frac{\left\langle \bar{f}\left( X^{\prime }\right) \right\rangle
+\left\langle \bar{r}\left( X^{\prime }\right) \right\rangle }{2}%
-\left\langle \hat{f}\left( X^{\prime }\right) \right\rangle \right) -\frac{%
\left\langle \bar{S}_{E}\right\rangle }{\left( \frac{1-\bar{S}_{E}}{1-2\bar{S%
}_{E}}\right) ^{2}}\left( \left\langle \hat{f}\left( X^{\prime }\right)
\right\rangle -\left\langle f\left( X\right) \right\rangle \right) }{\sqrt{%
\left( \frac{1-\left( \bar{\gamma}\left\langle \bar{S}_{E}\right\rangle
\right) ^{2}}{\left( \frac{1-\bar{S}_{E}}{1-2\bar{S}_{E}}\right)
^{2}\left\langle \bar{S}_{E}\right\rangle }-\left( 1-2\left\langle \bar{S}%
_{E}\right\rangle \right) \right) ^{2}+\frac{4}{\left( \frac{1-\bar{S}_{E}}{%
1-2\bar{S}_{E}}\right) ^{2}}}}
\end{equation*}%
we obtain to the first order:%
\begin{equation*}
\left\langle \bar{S}_{E}\right\rangle =\left( 1+\left( \frac{1-\left\langle 
\bar{S}_{E}\right\rangle _{0}}{\left( 1-2\left\langle \bar{S}%
_{E}\right\rangle \right) }\left\langle S_{E}^{\left( e\right)
}\right\rangle _{0}-\left\langle S_{E}^{B}\right\rangle _{0}-\frac{%
\left\langle \hat{S}_{E}^{B}\right\rangle _{0}}{2}\right) \left(
\left\langle f_{1}\left( X\right) \right\rangle -\bar{r}\right) \right)
\left\langle \bar{S}_{E}\right\rangle _{0}
\end{equation*}%
\begin{equation*}
\left\langle \bar{S}\right\rangle \rightarrow 2\left\langle \bar{S}%
_{E}\right\rangle -\frac{\left\langle \bar{S}_{E}\right\rangle }{\left(
1-2\left\langle \bar{S}_{E}\right\rangle \right) }\left\langle S_{E}^{\left(
e\right) }\right\rangle _{0}\left( \left\langle f_{1}\left( X\right)
\right\rangle -\bar{r}\right)
\end{equation*}%
\begin{eqnarray*}
\left\langle \bar{S}\right\rangle &\rightarrow &2\left( 1+\left( \frac{%
1-2\left\langle \bar{S}_{E}\right\rangle _{0}}{2\left( 1-\left\langle \bar{S}%
_{E}\right\rangle _{0}\right) }\left\langle S_{E}^{\left( e\right)
}\right\rangle _{0}-\left\langle S_{E}^{B}\right\rangle _{0}-\frac{%
\left\langle \hat{S}_{E}^{B}\right\rangle _{0}}{2}\right) \left(
\left\langle f_{1}\left( X\right) \right\rangle -\bar{r}\right) \right)
\left\langle \bar{S}_{E}\right\rangle _{0} \\
&\rightarrow &2x-2x\left( \frac{\left\langle \hat{S}_{E}^{B}\right\rangle
_{0}}{2}+\frac{\left\langle S_{E}^{B}\right\rangle _{0}}{2\left(
1-\left\langle \bar{S}_{E}\right\rangle _{0}\right) }\right) \left(
\left\langle f_{1}\left( X\right) \right\rangle -\bar{r}\right)
\end{eqnarray*}%
\begin{eqnarray*}
\left\langle S_{E}^{B}\right\rangle &\rightarrow &\left\langle
S_{E}^{B}\right\rangle _{0}+\left( \frac{-\left\langle
S_{E}^{B}\right\rangle x\frac{\left( 1-x\right) ^{2}-x}{\left( 1-x\right)
^{3}}+2\frac{1-2x}{\left( x-1\right) ^{3}}}{\sqrt{\left( \frac{x}{\left( 
\frac{1-x}{1-2x}\right) ^{2}x}-\left( 1-2x\right) \right) ^{2}+\frac{4}{%
\left( \frac{1-x}{1-2x}\right) ^{2}}}}\right) \\
&&\times \left( \frac{1-\left\langle \bar{S}_{E}\right\rangle _{0}}{\left(
1-2\left\langle \bar{S}_{E}\right\rangle \right) }\left\langle S_{E}^{\left(
e\right) }\right\rangle _{0}-\left\langle S_{E}^{B}\right\rangle _{0}-\frac{%
\left\langle \hat{S}_{E}^{B}\right\rangle _{0}}{2}\right) \left(
\left\langle f_{1}\left( X\right) \right\rangle -\bar{r}\right) \left\langle 
\bar{S}_{E}\right\rangle _{0} \\
&&-x\frac{\left( 1-2x\right) ^{2}}{\left( 1-x\right) ^{2}}\frac{\left\langle 
\hat{f}\left( X^{\prime }\right) \right\rangle +\left\langle f\left(
X\right) \right\rangle +\left\langle S_{E}^{B}\right\rangle \left( \frac{%
\left\langle \bar{f}\left( X^{\prime }\right) \right\rangle +\left\langle 
\bar{r}\left( X^{\prime }\right) \right\rangle }{2}-\left\langle \hat{f}%
\left( X^{\prime }\right) \right\rangle \right) }{\sqrt{\left( \frac{x}{%
\left( \frac{1-x}{1-2x}\right) ^{2}x}-\left( 1-2x\right) \right) ^{2}+\frac{4%
}{\left( \frac{1-\bar{S}_{E}}{1-2\bar{S}_{E}}\right) ^{2}}}}
\end{eqnarray*}%
\begin{eqnarray}
\left\langle S_{E}^{B}\right\rangle &\rightarrow &\left( 1-6x-\frac{x}{2}%
\left( 1+\frac{5}{2}x\right) \left( 1-7x\right) \left( \left\langle
f_{1}\left( X\right) \right\rangle -\bar{r}\right) \right)  \label{SBn} \\
&&-\frac{1}{2}\left( 1-x\right) x\left( \left\langle \hat{f}\left( X^{\prime
}\right) \right\rangle -\left\langle f_{1}\left( X\right) \right\rangle
+x\left( \frac{\left\langle \bar{f}\left( X^{\prime }\right) \right\rangle
+\left\langle \bar{r}\left( X^{\prime }\right) \right\rangle }{2}%
-\left\langle \hat{f}\left( X^{\prime }\right) \right\rangle \right) \right)
\notag \\
&=&1-6x+\frac{7}{4}x^{2}\left( 1+5x\right) \left( \left\langle f_{1}\left(
X\right) \right\rangle -\bar{r}\right) -\frac{1}{2}\left( 1-x\right)
^{2}x\left( \left\langle \hat{f}\left( X^{\prime }\right) \right\rangle -%
\bar{r}\right) -\frac{1}{4}\left( 1-x\right) x^{2}\left( \left\langle \bar{f}%
\left( X^{\prime }\right) \right\rangle -\left\langle \bar{r}\left(
X^{\prime }\right) \right\rangle \right)  \notag
\end{eqnarray}

and similarly:%
\begin{eqnarray*}
\left\langle \hat{S}_{E}^{B}\right\rangle &\rightarrow &\left\langle \hat{S}%
_{E}^{B}\right\rangle _{0}+\left( \frac{\left\langle S_{E}^{B}\right\rangle x%
\frac{\left( 1-x\right) ^{2}-x}{\left( 1-x\right) ^{3}}+2\frac{1-2x}{\left(
1-x\right) ^{3}}}{\sqrt{\left( \frac{x}{\left( \frac{1-x}{1-2x}\right) ^{2}x}%
-\left( 1-2x\right) \right) ^{2}+\frac{4}{\left( \frac{1-x}{1-2x}\right) ^{2}%
}}}-2\right) \\
&&\times \left( \frac{1-\left\langle \bar{S}_{E}\right\rangle _{0}}{\left(
1-2\left\langle \bar{S}_{E}\right\rangle \right) }\left\langle S_{E}^{\left(
e\right) }\right\rangle _{0}-\left\langle S_{E}^{B}\right\rangle _{0}-\frac{%
\left\langle \hat{S}_{E}^{B}\right\rangle _{0}}{2}\right) \left(
\left\langle f_{1}\left( X\right) \right\rangle -\bar{r}\right) \left\langle 
\bar{S}_{E}\right\rangle _{0} \\
&&+x\frac{\left( 1-2x\right) ^{2}}{\left( 1-x\right) ^{2}}\frac{\left\langle 
\hat{f}\left( X^{\prime }\right) \right\rangle -\left\langle f\left(
X\right) \right\rangle -\left\langle S_{E}^{B}\right\rangle \left( \frac{%
\left\langle \bar{f}\left( X^{\prime }\right) \right\rangle +\left\langle 
\bar{r}\left( X^{\prime }\right) \right\rangle }{2}-\left\langle \hat{f}%
\left( X^{\prime }\right) \right\rangle \right) }{\sqrt{\left( \frac{%
1-\left( \bar{\gamma}\left\langle \bar{S}_{E}\right\rangle \right) ^{2}}{%
\left( \frac{1-\bar{S}_{E}}{1-2\bar{S}_{E}}\right) ^{2}\left\langle \bar{S}%
_{E}\right\rangle }-\left( 1-2\left\langle \bar{S}_{E}\right\rangle \right)
\right) ^{2}+\frac{4}{\left( \frac{1-\bar{S}_{E}}{1-2\bar{S}_{E}}\right) ^{2}%
}}}
\end{eqnarray*}

\subsubsection*{A8.2.5 First order solutions for banks loans}

\paragraph*{A8.2.5.1 Relative shares}

Ultimately, we derive the loans of banks to investors and firms. We statr
with loans to investors. We use that, given (\ref{Stn}) and (\ref{ST}):%
\begin{equation}
\left\langle \hat{w}_{2}^{B}\left( X^{\prime },X\right) \right\rangle =\frac{%
1-\left( \gamma \left\langle \hat{S}_{E}\left( X^{\prime },X\right)
\right\rangle \right) ^{2}}{2-\left( \gamma \left\langle \hat{S}_{E}\left(
X^{\prime },X\right) \right\rangle \right) ^{2}}=\left\langle \hat{w}%
_{2}\left( X^{\prime },X\right) \right\rangle =\left\langle \hat{w}%
_{1}\left( X^{\prime },X\right) \right\rangle
\end{equation}%
and:%
\begin{equation}
\left\langle w_{2}^{B}\left( X\right) \right\rangle =\frac{1}{2-\left(
\gamma \left\langle \hat{S}_{E}\left( X^{\prime },X\right) \right\rangle
\right) ^{2}}=\left\langle w_{2}\left( X^{\prime },X\right) \right\rangle
=\left\langle w_{1}\left( X^{\prime },X\right) \right\rangle
\end{equation}%
which in turn leads to the relative shares of loans $\frac{\left\langle \hat{%
S}_{L}^{B}\left( X^{\prime }\right) \right\rangle }{\kappa \left(
1-\left\langle \bar{S}\left( X\right) \right\rangle \right) }$ and $\frac{%
\left\langle S_{L}^{B}\left( X,X\right) \right\rangle }{\kappa \left(
1-\left\langle \bar{S}\left( X\right) \right\rangle \right) }$ with $%
\left\langle \bar{S}\left( X\right) \right\rangle =\left\langle \bar{S}%
\right\rangle $:%
\begin{eqnarray}
&&\frac{\left\langle \hat{S}_{L}^{B}\left( X^{\prime }\right) \right\rangle 
}{\kappa \left( 1-\left\langle \bar{S}\right\rangle \right) }  \label{STb} \\
&=&\frac{1-\left( \gamma \left\langle \hat{S}_{E}\left( X^{\prime },X\right)
\right\rangle \right) ^{2}}{2-\left( \gamma \left\langle \hat{S}_{E}\left(
X^{\prime },X\right) \right\rangle \right) ^{2}}\left( 1+\frac{\left( \hat{r}%
\left( X^{\prime }\right) -\left\langle r\left( X\right) \right\rangle
\right) }{2-\left( \gamma \left\langle \hat{S}_{E}\left( X^{\prime
},X\right) \right\rangle \right) ^{2}}\right)  \notag
\end{eqnarray}%
Writing $z_{0}=\left\langle \hat{S}_{E}\left( X^{\prime },X\right)
\right\rangle _{0}$ the zeroth order participation of investors in firms, we
showed in part I:%
\begin{equation*}
\gamma ^{2}z_{0}^{2}=\frac{1-4z_{0}}{1-2z_{0}},1-\left( \gamma z_{0}\right)
^{2}=1-\frac{1-4z_{0}}{1-2z_{0}}=2\frac{z_{0}}{1-2z_{0}}
\end{equation*}%
\begin{equation*}
2-\left( \gamma z_{0}\right) ^{2}=\frac{1}{1-2z_{0}},3-\frac{1-4z_{0}}{%
1-2z_{0}}=2\frac{1-z_{0}}{1-2z_{0}}
\end{equation*}%
and the first order correction to $\left\langle \hat{S}_{E}\left( X^{\prime
},X\right) \right\rangle $ is: 
\begin{equation*}
v=\frac{z_{0}^{3}\left( \left\langle f_{1}\left( X\right) \right\rangle
-\left\langle \bar{r}\left( X\right) \right\rangle \right) }{2\left(
1-5z_{0}+8z_{0}^{2}\right) }
\end{equation*}%
Then developping to the first order the factor:%
\begin{eqnarray*}
&&\frac{1-\left( \gamma \left( z+v\right) \right) ^{2}}{2-\left( \gamma
\left( z+v\right) \right) ^{2}} \\
&=&\frac{1-z^{2}\gamma ^{2}}{2-z^{2}\gamma ^{2}}-v2z\gamma ^{2}\left( \frac{1%
}{2-z^{2}\gamma ^{2}}-\frac{1-z^{2}\gamma ^{2}}{\left( 2-z^{2}\gamma
^{2}\right) ^{2}}\right) \\
&=&\frac{2\frac{z_{0}}{1-2z_{0}}}{\frac{1}{1-2z_{0}}}-v2\frac{1-4z_{0}}{%
z_{0}\left( 1-2z_{0}\right) }\left( \left( 1-2z_{0}\right) -2\frac{z_{0}}{%
1-2z_{0}}\left( 1-2z_{0}\right) ^{2}\right) \\
&=&2z_{0}-2\left( 1-4z_{0}\right) \frac{1-2z_{0}}{z_{0}}v
\end{eqnarray*}%
equation (\ref{STb}) becomes:%
\begin{eqnarray*}
\frac{\left\langle \hat{S}_{L}^{B}\left( X^{\prime }\right) \right\rangle }{%
\kappa \left( 1-\left\langle \bar{S}\right\rangle \right) } &=&\left(
2z_{0}-2\frac{\left( 1-4z_{0}\right) \left( 1-2z_{0}\right) }{z_{0}}v\right)
\left( 1+\left( 1-2z_{0}\right) \left( \left\langle \hat{r}\left( X^{\prime
}\right) \right\rangle -\left\langle r\left( X\right) \right\rangle \right)
\right) \\
&=&\left( 2z_{0}-\left( 1-4z_{0}\right) \left( 1-2z_{0}\right) \frac{%
z_{0}^{2}\left( \left\langle f_{1}\left( X\right) \right\rangle ^{\left(
dr\right) }-\left\langle \bar{r}\left( X\right) \right\rangle \right) }{%
\left( 1-5z_{0}+8z_{0}^{2}\right) }\right) \\
&&\times \left( 1+\left( 1-2z_{0}\right) \left( \left\langle \hat{r}\left(
X^{\prime }\right) \right\rangle -\left\langle r\left( X\right)
\right\rangle \right) \right) \\
&=&2z_{0}\left( 1-\frac{\left( 1-4z_{0}\right) \left( 1-2z_{0}\right)
z_{0}\left( \left\langle f_{1}\left( X\right) \right\rangle ^{\left(
dr\right) }-\left\langle \bar{r}\left( X\right) \right\rangle \right) }{%
2\left( 1-5z_{0}+8z_{0}^{2}\right) }\right. \\
&&\left. +\left( 1-2z_{0}\right) \left( \left\langle \hat{r}\left( X^{\prime
}\right) \right\rangle -\left\langle r\left( X\right) \right\rangle \right)
\right)
\end{eqnarray*}%
Similarly, we obtain the relative loans of banks to firms:

\begin{eqnarray}
&&\frac{S_{L}^{B}\left( X,X\right) }{\kappa \left( 1-\left\langle \bar{S}%
\right\rangle \right) } \\
&=&w_{2}^{B}\left( X\right) \left[ 1+\hat{w}_{2}^{B}\left( X\right) \left(
r\left( X\right) -\left\langle \hat{r}\left( X^{\prime }\right)
\right\rangle _{\hat{w}_{2}}\right) \right]  \notag \\
&=&\frac{1}{2-\left( \gamma \left\langle \hat{S}_{E}\left( X^{\prime
},X\right) \right\rangle \right) ^{2}}\left[ 1+\frac{1-\left( \gamma
\left\langle \hat{S}_{E}\left( X^{\prime },X\right) \right\rangle \right)
^{2}}{2-\left( \gamma \left\langle \hat{S}_{E}\left( X^{\prime },X\right)
\right\rangle \right) ^{2}}\left( r\left( X\right) -\left\langle \hat{r}%
\left( X^{\prime }\right) \right\rangle _{\hat{w}_{2}}\right) \right]  \notag
\end{eqnarray}%
The expansion of the first factor of the right hand side:%
\begin{eqnarray*}
\frac{1}{2-\left( \gamma \left( z+v\right) \right) ^{2}} &=&\frac{1}{%
2-z^{2}\gamma ^{2}}+2vz\frac{\gamma ^{2}}{\left( z^{2}\gamma ^{2}-2\right)
^{2}} \\
&=&1-2z_{0}+\frac{2\left( 1-2z_{0}\right) \left( 1-4z_{0}\right) }{z_{0}}v
\end{eqnarray*}%
leads to the following formula:%
\begin{eqnarray}
\frac{\left\langle S_{L}^{B}\left( X,X\right) \right\rangle }{\kappa \left(
1-\left\langle \bar{S}\right\rangle \right) } &=&\left( 1-2z_{0}+\frac{%
2\left( 1-2z_{0}\right) \left( 1-4z_{0}\right) }{z_{0}}v\right)
\label{SBdme} \\
&&\times \left( 1+2z_{0}\left( \left\langle r\left( X\right) \right\rangle
-\left\langle \hat{r}\left( X^{\prime }\right) \right\rangle _{\hat{w}%
_{2}}\right) \right)  \notag \\
&=&\left( 1-2z_{0}\right) \left( 1+\frac{2\left( 1-4z_{0}\right) }{z_{0}}%
v+2z_{0}\left( \left\langle r\left( X\right) \right\rangle -\left\langle 
\hat{r}\left( X^{\prime }\right) \right\rangle _{\hat{w}_{2}}\right) \right)
\notag \\
&=&\left( 1-2z_{0}\right)  \notag \\
&&\times \left( 1+\frac{\left( 1-4z_{0}\right) z_{0}^{2}\left( \left\langle
f_{1}\left( X\right) \right\rangle ^{\left( dr\right) }-\left\langle \bar{r}%
\left( X\right) \right\rangle \right) }{\left( 1-5z_{0}+8z_{0}^{2}\right) }%
+2z_{0}\left( \left\langle r\left( X\right) \right\rangle -\left\langle \hat{%
r}\left( X^{\prime }\right) \right\rangle _{\hat{w}_{2}}\right) \right) 
\notag
\end{eqnarray}

\paragraph*{A8.2.5.2 Loans shares}

To derive $\left\langle \hat{S}_{L}^{B}\left( X^{\prime }\right)
\right\rangle $ and $\left\langle S_{L}^{B}\left( X,X\right) \right\rangle $%
, we have to include the factor:%
\begin{eqnarray*}
&&\kappa \left( 1-\left\langle \bar{S}\right\rangle \right) \\
&\rightarrow &\kappa \left( 1-2z\right) \left( 1+z\left( \left\langle
f_{1}\left( X\right) \right\rangle -\bar{r}\right) \right)
\end{eqnarray*}%
whose value can be found using (\ref{frmn}). Neglecting the variations in
interest rates, we thus have:%
\begin{eqnarray*}
&&\left\langle \hat{S}_{L}^{B}\left( X^{\prime }\right) \right\rangle \\
&=&\kappa \left( 1-2z_{0}\right) 2z_{0}\left( 1+z_{0}\left( 1-\frac{\left(
1-4z_{0}\right) \left( 1-2z_{0}\right) }{2\left( 1-5z_{0}+8z_{0}^{2}\right) }%
\right) \left( \left\langle f_{1}\left( X\right) \right\rangle ^{\left(
dr\right) }-\left\langle \bar{r}\left( X\right) \right\rangle \right) \right.
\\
&&\left. +\left( 1-2z_{0}\right) \left( \left\langle \hat{r}\left( X^{\prime
}\right) \right\rangle -\left\langle r\left( X\right) \right\rangle \right)
\right)
\end{eqnarray*}%
\begin{eqnarray*}
&&\left\langle S_{L}^{B}\left( X,X\right) \right\rangle \\
&=&\kappa \left( 1-2z_{0}\right) ^{2} \\
&&\times \left( 1+z_{0}\left( 1+\frac{\left( 1-4z_{0}\right) z_{0}}{\left(
1-5z_{0}+8z_{0}^{2}\right) }\right) \left( \left\langle f_{1}\left( X\right)
\right\rangle ^{\left( dr\right) }-\left\langle \bar{r}\left( X\right)
\right\rangle \right) +2z_{0}\left( \left\langle r\left( X\right)
\right\rangle -\left\langle \hat{r}\left( X^{\prime }\right) \right\rangle _{%
\hat{w}_{2}}\right) \right)
\end{eqnarray*}

\bigskip

\subsubsection*{A8.2.6 Returns and capital ratios under decreasing returns}

As in Part one, formulas (\ref{Frn}) and (\ref{Frb}) are developed in the
following way.

The returns served to investors, without banks was:

\begin{eqnarray*}
\frac{f_{1}^{\left( e\right) }\left( X\right) -r}{\left( 1+\underline{k}%
_{2}\left( X\right) \right) } &\rightarrow &rC_{0}+\frac{\left( 1+\frac{%
\underline{k}\left( X\right) }{\left\langle K\right\rangle }\hat{K}%
_{X}\left\vert \hat{\Psi}\left( X\right) \right\vert ^{2}\right) \left(
3X^{\left( e\right) }-C^{\left( e\right) }\right) \left( C^{\left( e\right)
}+X^{\left( e\right) }\right) }{\left( 1+\frac{\underline{k}\left( X\right) 
}{\left\langle K\right\rangle }\hat{K}_{X}\left\vert \hat{\Psi}\left(
X\right) \right\vert ^{2}\right) K_{X}\left( 1+\underline{k}_{2}\left(
X\right) \right) \left( 2X^{\left( e\right) }-C^{\left( e\right) }\right) }
\\
&&-\frac{C}{\left( 1+\underline{k}_{2}\left( X\right) \right) \left( 1+\frac{%
\underline{k}\left( X\right) }{\left\langle K\right\rangle }\hat{K}%
_{X}\left\vert \hat{\Psi}\left( X\right) \right\vert ^{2}\right) K_{X}}
\end{eqnarray*}%
rewritten in terms of stakes, it writes:%
\begin{equation*}
\frac{f_{1}^{\left( e\right) }\left( X\right) -r}{\left( 1+\underline{k}%
_{2}\left( X\right) \right) }\rightarrow rC_{0}+\left( \frac{C_{0}+\frac{%
S_{L}\left( X^{\prime }\right) }{1-S_{E}\left( X^{\prime }\right) }\bar{r}}{%
f_{1}\left( X\right) }\right) ^{\frac{1}{r}}\left( 1-S_{L}\left( X^{\prime
}\right) \right) \left( \frac{\left( 3X^{\left( e\right) }-C^{\left(
e\right) }\right) \left( C^{\left( e\right) }+X^{\left( e\right) }\right) }{%
1-S\left( X^{\prime }\right) }-C\right)
\end{equation*}

Including banks leads to replace:%
\begin{equation*}
S_{L}\left( X^{\prime }\right) \rightarrow S_{L}\left( X^{\prime }\right)
+S_{L}^{B}\left( X^{\prime }\right)
\end{equation*}%
and:%
\begin{equation*}
S\left( X^{\prime }\right) \rightarrow S\left( X^{\prime }\right)
+S^{B}\left( X^{\prime }\right)
\end{equation*}%
with:%
\begin{equation*}
S_{L}\left( X^{\prime }\right) =\frac{\left( 1-2z_{0}\right) }{2}\frac{%
\left\langle \hat{K}\right\rangle \left\Vert \hat{\Psi}\right\Vert ^{2}}{%
\left\langle K\right\rangle \left\Vert \Psi \right\Vert ^{2}}-e\frac{%
f_{1}^{\left( e\right) }\left( X\right) -r}{\left( 1+\underline{k}_{2}\left(
X\right) \right) }
\end{equation*}%
\begin{equation*}
e=\frac{1-2z_{0}}{2}\left( \frac{3}{8}-z_{0}-\frac{1}{2}\frac{%
z_{0}^{3}\left( 1-4z_{0}\right) }{\left( 1-5z_{0}+8z_{0}^{2}\right) }\right)
\end{equation*}%
Using (\ref{STF}) and (\ref{SHT}):%
\begin{equation*}
S\left( X^{\prime }\right) \simeq \left( 1-2z_{0}\right) \frac{\left\langle 
\hat{K}\right\rangle \left\Vert \hat{\Psi}\right\Vert ^{2}}{\left\langle
K\right\rangle \left\Vert \Psi \right\Vert ^{2}}+d\frac{f_{1}^{\left(
e\right) }\left( X\right) -r}{\left( 1+\underline{k}_{2}\left( X\right)
\right) }
\end{equation*}%
with:%
\begin{equation*}
d=\left( 1-2z_{0}\right) \left( \frac{1}{2}+\frac{z_{0}^{2}\left(
1-4z_{0}\right) }{\left( 1-5z_{0}+8z_{0}^{2}\right) }\right) z_{0}
\end{equation*}

(\ref{SBn}) and (\ref{SBdme}):%
\begin{eqnarray}
\left\langle S_{E}^{B}\right\rangle &\rightarrow &x\left( 1-6x-\frac{1}{2}%
\left( 1+\frac{5}{2}x\right) \left( 1-7x\right) \left( \left\langle
f_{1}\left( X\right) \right\rangle -\bar{r}\right) \right) \\
&&-\frac{1}{2}\left( 1-x\right) x\left( \left\langle \hat{f}\left( X^{\prime
}\right) \right\rangle -\left\langle f\left( X\right) \right\rangle -x\left( 
\frac{\left\langle \bar{f}\left( X^{\prime }\right) \right\rangle
+\left\langle \bar{r}\left( X^{\prime }\right) \right\rangle }{2}%
-\left\langle \hat{f}\left( X^{\prime }\right) \right\rangle \right) \right)
\notag
\end{eqnarray}%
\begin{eqnarray*}
\left\langle S_{L}^{B}\left( X,X\right) \right\rangle &\simeq &\kappa \left(
1-2x\right) \left( 1-2z_{0}\right) \\
&&\times \left( 1+\frac{\left( 1-4z_{0}\right) z_{0}^{2}\left( \left\langle
f_{1}\left( X\right) \right\rangle ^{\left( dr\right) }-\left\langle \bar{r}%
\left( X\right) \right\rangle \right) }{\left( 1-5z_{0}+8z_{0}^{2}\right) }%
+2z_{0}\left( \left\langle r\left( X\right) \right\rangle -\left\langle \hat{%
r}\left( X^{\prime }\right) \right\rangle _{\hat{w}_{2}}\right) \right) \\
&\simeq &\kappa \left( 1-2z_{0}\right) ^{2}\left( 1+\frac{\left(
1-4z_{0}\right) z_{0}^{2}\left( \left\langle f_{1}\left( X\right)
\right\rangle ^{\left( dr\right) }-\left\langle \bar{r}\left( X\right)
\right\rangle \right) }{\left( 1-5z_{0}+8z_{0}^{2}\right) }\right)
\end{eqnarray*}%
\bigskip

\begin{eqnarray*}
&&1-\left( S_{L}\left( X^{\prime }\right) +S_{L}^{B}\left( X^{\prime
}\right) \right) \\
&=&1-\frac{\left( 1-2z_{0}\right) }{2}\frac{\left\langle \hat{K}%
\right\rangle \left\Vert \hat{\Psi}\right\Vert ^{2}}{\left\langle
K\right\rangle \left\Vert \Psi \right\Vert ^{2}}-\kappa \left(
1-2z_{0}\right) ^{2}\frac{\left\langle \bar{K}\right\rangle \left\Vert \bar{%
\Psi}\right\Vert ^{2}}{\left\langle K\right\rangle \left\Vert \Psi
\right\Vert ^{2}}-e\frac{f_{1}^{\left( e\right) }\left( X\right) -r}{\left(
1+\underline{k}_{2}\left( X\right) \right) }
\end{eqnarray*}%
\begin{eqnarray*}
e &=&\kappa \left( 1-2z_{0}\right) ^{2}z_{0}\left( 1+\frac{\left(
1-4z_{0}\right) z_{0}}{\left( 1-5z_{0}+8z_{0}^{2}\right) }\right) \frac{%
\left\langle \bar{K}\right\rangle \left\Vert \bar{\Psi}\right\Vert ^{2}}{%
\left\langle K\right\rangle \left\Vert \Psi \right\Vert ^{2}} \\
&&-\frac{1-2z_{0}}{2}\left( \frac{3}{8}-z_{0}-\frac{1}{2}\frac{%
z_{0}^{3}\left( 1-4z_{0}\right) }{\left( 1-5z_{0}+8z_{0}^{2}\right) }\right) 
\frac{\left\langle \hat{K}\right\rangle \left\Vert \hat{\Psi}\right\Vert ^{2}%
}{\left\langle K\right\rangle \left\Vert \Psi \right\Vert ^{2}} \\
&\simeq &\kappa \left( 1-2z_{0}\right) ^{2}z_{0}\frac{\left\langle \bar{K}%
\right\rangle \left\Vert \bar{\Psi}\right\Vert ^{2}}{\left\langle
K\right\rangle \left\Vert \Psi \right\Vert ^{2}}-\frac{1-2z_{0}}{2}\left( 
\frac{3}{8}-z_{0}\right) \frac{\left\langle \hat{K}\right\rangle \left\Vert 
\hat{\Psi}\right\Vert ^{2}}{\left\langle K\right\rangle \left\Vert \Psi
\right\Vert ^{2}}
\end{eqnarray*}%
\begin{eqnarray*}
&&1-S\left( X^{\prime }\right) \\
&=&1-\left( 1-2z_{0}\right) \frac{\left\langle \hat{K}\right\rangle
\left\Vert \hat{\Psi}\right\Vert ^{2}}{\left\langle K\right\rangle
\left\Vert \Psi \right\Vert ^{2}} \\
&&-x\left( \left( 1-6x\right) -\frac{1}{2}\left( 1-x\right) x\left(
\left\langle \hat{f}\left( X^{\prime }\right) \right\rangle -\left\langle
r\left( X\right) \right\rangle -\left( \frac{\left\langle \bar{f}\left(
X^{\prime }\right) \right\rangle +\left\langle \bar{r}\left( X^{\prime
}\right) \right\rangle }{2}-\left\langle \hat{f}\left( X^{\prime }\right)
\right\rangle \right) \right) \right) \frac{\left\langle \bar{K}%
\right\rangle \left\Vert \bar{\Psi}\right\Vert ^{2}}{\left\langle
K\right\rangle \left\Vert \Psi \right\Vert ^{2}} \\
&&-d_{n}\frac{f_{1}^{\left( e\right) }\left( X\right) -r}{\left( 1+%
\underline{k}_{2}\left( X\right) \right) }\frac{\left\langle \hat{K}%
\right\rangle \left\Vert \hat{\Psi}\right\Vert ^{2}}{\left\langle
K\right\rangle \left\Vert \Psi \right\Vert ^{2}}-d_{b}\frac{f_{1}^{\left(
e\right) }\left( X\right) -r}{\left( 1+\underline{k}_{2}\left( X\right)
\right) }\frac{\left\langle \bar{K}\right\rangle \left\Vert \bar{\Psi}%
\right\Vert ^{2}}{\left\langle K\right\rangle \left\Vert \Psi \right\Vert
^{2}}
\end{eqnarray*}%
\begin{equation*}
d_{n}=\left( 1-2z_{0}\right) \left( \frac{1}{2}+\frac{z_{0}^{2}\left(
1-4z_{0}\right) }{\left( 1-5z_{0}+8z_{0}^{2}\right) }\right) z_{0}
\end{equation*}%
\begin{eqnarray*}
d_{b}^{\prime } &=&-\frac{1}{2}\left( 1+\frac{5}{2}x\right) \left(
1-7x\right) +\kappa \left( 1-2z_{0}\right) ^{2}\left( 1+\frac{\left(
1-4z_{0}\right) z_{0}^{2}}{\left( 1-5z_{0}+8z_{0}^{2}\right) }\right) +\frac{%
1}{2}\left( 1-x\right) x \\
&\rightarrow &\allowbreak \frac{1}{2}\left( 1-\frac{11x+33x^{2}}{2}\right)
+\kappa \left( 1-2z_{0}\right) ^{2}\left( 1+\frac{\left( 1-4z_{0}\right)
z_{0}^{2}}{\left( 1-5z_{0}+8z_{0}^{2}\right) }\right)
\end{eqnarray*}%
\begin{eqnarray*}
&&1-S\left( X^{\prime }\right) \\
&=&1-\left( 1-2z_{0}\right) \frac{\left\langle \hat{K}\right\rangle
\left\Vert \hat{\Psi}\right\Vert ^{2}}{\left\langle K\right\rangle
\left\Vert \Psi \right\Vert ^{2}} \\
&&-x\left( \left( 1-6x\right) +\frac{1}{2}\left( 1-x\right) x\left( \frac{%
\left\langle \bar{f}\left( X^{\prime }\right) \right\rangle -\left\langle 
\bar{r}\left( X^{\prime }\right) \right\rangle }{2}\right) \right) \frac{%
\left\langle \bar{K}\right\rangle \left\Vert \bar{\Psi}\right\Vert ^{2}}{%
\left\langle K\right\rangle \left\Vert \Psi \right\Vert ^{2}} \\
&&-\kappa \left( 1-2z_{0}\right) ^{2}\frac{\left\langle \bar{K}\right\rangle
\left\Vert \bar{\Psi}\right\Vert ^{2}}{\left\langle K\right\rangle
\left\Vert \Psi \right\Vert ^{2}} \\
&&-d_{n}\frac{f_{1}^{\left( e\right) }\left( X\right) -r}{\left( 1+%
\underline{k}_{2}\left( X\right) \right) }\frac{\left\langle \hat{K}%
\right\rangle \left\Vert \hat{\Psi}\right\Vert ^{2}}{\left\langle
K\right\rangle \left\Vert \Psi \right\Vert ^{2}}-d_{b}^{\prime }\frac{%
f_{1}^{\left( e\right) }\left( X\right) -r}{\left( 1+\underline{k}_{2}\left(
X\right) \right) }\frac{\left\langle \bar{K}\right\rangle \left\Vert \bar{%
\Psi}\right\Vert ^{2}}{\left\langle K\right\rangle \left\Vert \Psi
\right\Vert ^{2}}
\end{eqnarray*}%
Given (\ref{Rtns}), we have in first approximation:%
\begin{eqnarray}
&&\left( \left\langle \bar{f}\left( X^{\prime }\right) \right\rangle -\bar{r}%
\right) \\
&=&\frac{\left( \frac{1-\left\langle \hat{S}\right\rangle -\left(
\left\langle \hat{S}_{E}^{B}\right\rangle +\left\langle \hat{S}%
_{L}^{B}\right\rangle \right) \frac{\left\langle \bar{K}\right\rangle
\left\Vert \bar{\Psi}\right\Vert ^{2}}{\left\langle \hat{K}\right\rangle
\left\Vert \hat{\Psi}\right\Vert ^{2}}}{1-\left\langle \hat{S}%
_{E}\right\rangle -\left( \left\langle \hat{S}_{E}^{B}\right\rangle \right) 
\frac{\left\langle \bar{K}\right\rangle \left\Vert \bar{\Psi}\right\Vert ^{2}%
}{\left\langle \hat{K}\right\rangle \left\Vert \hat{\Psi}\right\Vert ^{2}}}%
\frac{\left\langle \hat{S}_{E}^{B}\right\rangle \left\langle \bar{K}%
\right\rangle \left\Vert \bar{\Psi}\right\Vert ^{2}}{\left\langle \hat{K}%
\right\rangle \left\Vert \hat{\Psi}\right\Vert ^{2}}\frac{\left\langle
S_{E}\left( X,X\right) \right\rangle }{\left( 1-\left\langle \hat{S}\left(
X^{\prime },X\right) \right\rangle \right) }+\left\langle S_{E}^{B}\left(
X^{\prime },X^{\prime }\right) \right\rangle \right) }{\left( 1-\left\langle 
\bar{S}\right\rangle \right) }\left( \left\langle f_{1}\left( X\right)
\right\rangle -\bar{r}\right)  \notag \\
&\simeq &\frac{\left\langle S_{E}^{B}\left( X^{\prime },X^{\prime }\right)
\right\rangle }{\left( 1-\left\langle \bar{S}\right\rangle \right) }\left(
\left\langle f_{1}\left( X\right) \right\rangle -\bar{r}\right)  \notag \\
&\simeq &\frac{1-6x}{1-2x}\left( \left\langle f_{1}\left( X\right)
\right\rangle -\bar{r}\right)  \notag
\end{eqnarray}%
\begin{eqnarray*}
&&1-S\left( X^{\prime }\right) \\
&=&1-\left( 1-2z_{0}\right) \frac{\left\langle \hat{K}\right\rangle
\left\Vert \hat{\Psi}\right\Vert ^{2}}{\left\langle K\right\rangle
\left\Vert \Psi \right\Vert ^{2}}-\left( x\left( \left( 1-6x\right) \right)
+\kappa \left( 1-2z_{0}\right) ^{2}\right) \frac{\left\langle \bar{K}%
\right\rangle \left\Vert \bar{\Psi}\right\Vert ^{2}}{\left\langle
K\right\rangle \left\Vert \Psi \right\Vert ^{2}} \\
&&-d_{n}\frac{f_{1}^{\left( e\right) }\left( X\right) -r}{\left( 1+%
\underline{k}_{2}\left( X\right) \right) }\frac{\left\langle \hat{K}%
\right\rangle \left\Vert \hat{\Psi}\right\Vert ^{2}}{\left\langle
K\right\rangle \left\Vert \Psi \right\Vert ^{2}}-d_{b}\frac{f_{1}^{\left(
e\right) }\left( X\right) -r}{\left( 1+\underline{k}_{2}\left( X\right)
\right) }\frac{\left\langle \bar{K}\right\rangle \left\Vert \bar{\Psi}%
\right\Vert ^{2}}{\left\langle K\right\rangle \left\Vert \Psi \right\Vert
^{2}}
\end{eqnarray*}%
\begin{equation*}
d_{b}\rightarrow \allowbreak \frac{1}{4}\left( 2x+1\right) \frac{%
2-19x+28x^{2}+3x^{3}}{1-2x}+\kappa \left( 1-2z_{0}\right) ^{2}\left( 1+\frac{%
\left( 1-4z_{0}\right) z_{0}^{2}}{\left( 1-5z_{0}+8z_{0}^{2}\right) }\right)
\end{equation*}%
$\allowbreak $

and $V=\frac{f_{1}^{\left( e\right) }\left( X\right) -r}{\left( 1+\underline{%
k}_{2}\left( X\right) \right) }$ satisfies:%
\begin{equation*}
V=rC_{0}+A\left( 1-EV\right) \left( \frac{G}{1-DV}-BC\right)
\end{equation*}

\bigskip with:%
\begin{equation*}
A=\left( \frac{C_{0}+\frac{S_{L}\left( X^{\prime }\right) }{1-S_{E}\left(
X^{\prime }\right) }\bar{r}}{f_{1}\left( X\right) }\right) ^{\frac{1}{r}}%
\frac{1-\left( \frac{\left( 1-2z_{0}\right) }{2}\right) \frac{\left\langle 
\hat{K}\right\rangle \left\Vert \hat{\Psi}\right\Vert ^{2}}{\left\langle
K\right\rangle \left\Vert \Psi \right\Vert ^{2}}-\kappa \left(
1-2z_{0}\right) ^{2}\frac{\left\langle \bar{K}\right\rangle \left\Vert \bar{%
\Psi}\right\Vert ^{2}}{\left\langle K\right\rangle \left\Vert \Psi
\right\Vert ^{2}}}{1-\left( 1-2z_{0}\right) \frac{\left\langle \hat{K}%
\right\rangle \left\Vert \hat{\Psi}\right\Vert ^{2}}{\left\langle
K\right\rangle \left\Vert \Psi \right\Vert ^{2}}-x\left( \left( 1-6x\right)
\right) \frac{\left\langle \bar{K}\right\rangle \left\Vert \bar{\Psi}%
\right\Vert ^{2}}{\left\langle K\right\rangle \left\Vert \Psi \right\Vert
^{2}}}
\end{equation*}%
\begin{equation*}
D=\frac{d_{n}\frac{\left\langle \hat{K}\right\rangle \left\Vert \hat{\Psi}%
\right\Vert ^{2}}{\left\langle K\right\rangle \left\Vert \Psi \right\Vert
^{2}}+d_{b}\frac{\left\langle \bar{K}\right\rangle \left\Vert \bar{\Psi}%
\right\Vert ^{2}}{\left\langle K\right\rangle \left\Vert \Psi \right\Vert
^{2}}}{1-\left( 1-2z_{0}\right) \frac{\left\langle \hat{K}\right\rangle
\left\Vert \hat{\Psi}\right\Vert ^{2}}{\left\langle K\right\rangle
\left\Vert \Psi \right\Vert ^{2}}-x\left( \left( 1-6x\right) \right) \frac{%
\left\langle \bar{K}\right\rangle \left\Vert \bar{\Psi}\right\Vert ^{2}}{%
\left\langle K\right\rangle \left\Vert \Psi \right\Vert ^{2}}}
\end{equation*}%
\begin{equation*}
E=\frac{e}{1-\left( \frac{\left( 1-2z_{0}\right) }{2}\right) \frac{%
\left\langle \hat{K}\right\rangle \left\Vert \hat{\Psi}\right\Vert ^{2}}{%
\left\langle K\right\rangle \left\Vert \Psi \right\Vert ^{2}}-\kappa \left(
1-2z_{0}\right) ^{2}\frac{\left\langle \bar{K}\right\rangle \left\Vert \bar{%
\Psi}\right\Vert ^{2}}{\left\langle K\right\rangle \left\Vert \Psi
\right\Vert ^{2}}}
\end{equation*}%
\begin{equation*}
G=\frac{\left( 3X^{\left( e\right) }-C^{\left( e\right) }\right) \left(
C^{\left( e\right) }+X^{\left( e\right) }\right) }{1-\left( 1-2z_{0}\right) 
\frac{\left\langle \hat{K}\right\rangle \left\Vert \hat{\Psi}\right\Vert ^{2}%
}{\left\langle K\right\rangle \left\Vert \Psi \right\Vert ^{2}}}
\end{equation*}%
\begin{equation*}
F=\frac{d}{1-\left( 1-2z_{0}\right) \frac{\left\langle \hat{K}\right\rangle
\left\Vert \hat{\Psi}\right\Vert ^{2}}{\left\langle K\right\rangle
\left\Vert \Psi \right\Vert ^{2}}}
\end{equation*}%
The solution:%
\begin{eqnarray}
&&\frac{f_{1}^{\left( e\right) }\left( X\right) -r}{\left( 1+\underline{k}%
_{2}\left( X\right) \right) } \\
&=&\frac{1-ABC\left( E+D\right) +AGE+rC_{0}D}{2D\left( 1-ABCE\right) } 
\notag \\
&&+\sqrt{\left( \frac{1-ABC\left( E+D\right) +AGE+rC_{0}D}{2D\left(
1-ABCE\right) }\right) ^{2}-\frac{A\left( G-BC\right) +rC_{0}}{\allowbreak
D\left( 1-ABCE\right) }}  \notag
\end{eqnarray}%
is at the lowest order:%
\begin{eqnarray*}
\frac{f_{1}^{\left( e\right) }\left( X\right) -r}{\left( 1+\underline{k}%
_{2}\left( X\right) \right) } &\simeq &rC_{0}+A\left( G-BC\right) \\
&=&rC_{0}-\left( \frac{C_{0}+\frac{S_{L}\left( X^{\prime }\right) }{%
1-S_{E}\left( X^{\prime }\right) }\bar{r}}{f_{1}\left( X\right) }\right) ^{%
\frac{1}{r}}\frac{1-\left( \frac{\left( 1-2z_{0}\right) }{2}\right) \frac{%
\left\langle \hat{K}\right\rangle \left\Vert \hat{\Psi}\right\Vert ^{2}}{%
\left\langle K\right\rangle \left\Vert \Psi \right\Vert ^{2}}-\kappa \left(
1-2z_{0}\right) ^{2}\frac{\left\langle \bar{K}\right\rangle \left\Vert \bar{%
\Psi}\right\Vert ^{2}}{\left\langle K\right\rangle \left\Vert \Psi
\right\Vert ^{2}}}{1-\left( 1-2z_{0}\right) \frac{\left\langle \hat{K}%
\right\rangle \left\Vert \hat{\Psi}\right\Vert ^{2}}{\left\langle
K\right\rangle \left\Vert \Psi \right\Vert ^{2}}-x\left( \left( 1-6x\right)
\right) \frac{\left\langle \bar{K}\right\rangle \left\Vert \bar{\Psi}%
\right\Vert ^{2}}{\left\langle K\right\rangle \left\Vert \Psi \right\Vert
^{2}}} \\
&&\times \left( C\left( 1-\left( 1-2z_{0}\right) \frac{\left\langle \hat{K}%
\right\rangle \left\Vert \hat{\Psi}\right\Vert ^{2}}{\left\langle
K\right\rangle \left\Vert \Psi \right\Vert ^{2}}-x\left( \left( 1-6x\right)
\right) \frac{\left\langle \bar{K}\right\rangle \left\Vert \bar{\Psi}%
\right\Vert ^{2}}{\left\langle K\right\rangle \left\Vert \Psi \right\Vert
^{2}}\right) -3X^{2}\right)
\end{eqnarray*}%
We have the investors and banks returns:%
\begin{equation*}
\left\langle \hat{f}\right\rangle =\left\langle r\right\rangle +\frac{1}{2}%
\frac{f_{1}^{\left( e\right) }\left( X\right) -r}{\left( 1+\underline{k}%
_{2}\left( X\right) \right) }=\left\langle r\right\rangle +\frac{%
f_{a}-f_{b}\left( \frac{C_{0}+\frac{S_{L}\left( X^{\prime }\right) }{%
1-S_{E}\left( X^{\prime }\right) }\bar{r}}{f_{1}\left( X\right) }\right) ^{%
\frac{1}{r}}}{2}
\end{equation*}%
with:%
\begin{eqnarray*}
f_{a} &=&rC_{0} \\
f_{b} &\simeq &\frac{1-\left( \frac{\left( 1-2z_{0}\right) }{2}\right) \frac{%
\left\langle \hat{K}\right\rangle \left\Vert \hat{\Psi}\right\Vert ^{2}}{%
\left\langle K\right\rangle \left\Vert \Psi \right\Vert ^{2}}-\kappa \left(
1-2z_{0}\right) ^{2}\frac{\left\langle \bar{K}\right\rangle \left\Vert \bar{%
\Psi}\right\Vert ^{2}}{\left\langle K\right\rangle \left\Vert \Psi
\right\Vert ^{2}}}{1-\left( 1-2z_{0}\right) \frac{\left\langle \hat{K}%
\right\rangle \left\Vert \hat{\Psi}\right\Vert ^{2}}{\left\langle
K\right\rangle \left\Vert \Psi \right\Vert ^{2}}-x\left( \left( 1-6x\right)
\right) \frac{\left\langle \bar{K}\right\rangle \left\Vert \bar{\Psi}%
\right\Vert ^{2}}{\left\langle K\right\rangle \left\Vert \Psi \right\Vert
^{2}}}C
\end{eqnarray*}%
and given (\ref{Bt}):%
\begin{equation*}
\left\langle \bar{f}\right\rangle \simeq \left( 1+\kappa \right) \bar{r}
\end{equation*}%
Using:%
\begin{equation*}
\left\langle K\right\rangle \left\Vert \Psi \right\Vert ^{2}\simeq \left( 1-%
\frac{\left\langle S\left( X,X\right) \right\rangle \left\langle \hat{K}%
\right\rangle \left\Vert \hat{\Psi}\right\Vert ^{2}+\left\langle S^{B}\left(
X,X\right) \right\rangle \left\langle \bar{K}\right\rangle \left\Vert \bar{%
\Psi}\right\Vert ^{2}}{\frac{2\epsilon }{3\sigma _{\hat{K}}^{2}}\left( \frac{%
\left\langle f_{1}\right\rangle }{C_{0}+\bar{r}}\right) ^{\frac{2}{r}}}%
\right) \left( \left( \frac{2\epsilon }{3\sigma _{\hat{K}}^{2}}\right) ^{%
\frac{r}{2}}\frac{f_{1}\left( X\right) }{C_{0}+\frac{S_{L}\left( X\right) }{%
1-S_{E}\left( X\right) }\bar{r}}\right) ^{\frac{2}{r}}
\end{equation*}%
Given (\ref{FRn}), (\ref{Frb}):%
\begin{equation*}
\left\langle \hat{K}\right\rangle \left\Vert \hat{\Psi}\right\Vert ^{2}=%
\frac{9\sigma _{\hat{K}}^{2}\left\Vert \hat{\Psi}_{0}\right\Vert ^{4}\left(
1-\hat{S}\right) ^{2}\left( 1+\frac{\left\Vert \bar{\Psi}_{0}\right\Vert ^{2}%
}{\left\Vert \hat{\Psi}_{0}\right\Vert ^{2}}\left\langle \hat{S}%
_{L}^{B}\right\rangle \right) }{2\hat{\mu}\left\langle \hat{f}\right\rangle
^{2}}
\end{equation*}%
\begin{equation*}
\left\langle \bar{K}\right\rangle \left\Vert \bar{\Psi}\right\Vert ^{2}=9%
\frac{\sigma _{\hat{K}}^{2}V\left\Vert \bar{\Psi}_{0}\right\Vert ^{4}\left(
1-\bar{S}\right) ^{2}}{2\hat{\mu}\left\langle \bar{f}\right\rangle ^{2}}
\end{equation*}%
are at the lowest order:%
\begin{equation*}
\left\langle \hat{K}\right\rangle \left\Vert \hat{\Psi}\right\Vert ^{2}=%
\frac{9\sigma _{\hat{K}}^{2}\left\Vert \hat{\Psi}_{0}\right\Vert ^{4}\left(
1-\hat{S}\right) ^{2}\left( 1+\frac{\left\Vert \bar{\Psi}_{0}\right\Vert ^{2}%
}{\left\Vert \hat{\Psi}_{0}\right\Vert ^{2}}\left\langle \hat{S}%
_{L}^{B}\right\rangle \right) }{2\hat{\mu}\left( \left\langle r\right\rangle
+\frac{rC_{0}}{2}\right) ^{2}}
\end{equation*}%
\begin{equation*}
\left\langle \bar{K}\right\rangle \left\Vert \bar{\Psi}\right\Vert ^{2}=9%
\frac{\sigma _{\hat{K}}^{2}V\left\Vert \bar{\Psi}_{0}\right\Vert ^{4}\left(
1-\bar{S}\right) ^{2}}{2\hat{\mu}\left( \left( 1+\kappa \right) \bar{r}%
\right) ^{2}}
\end{equation*}

\bigskip Given (\ref{ora}), (\ref{ore}) the capital ratios becomes:%
\begin{eqnarray*}
&&\frac{\left\langle \hat{K}\right\rangle \left\Vert \hat{\Psi}\right\Vert
^{2}}{\left\langle K\right\rangle \left\Vert \Psi \right\Vert ^{2}} \\
&\simeq &\frac{\frac{9\sigma _{\hat{K}}^{2}\left\Vert \hat{\Psi}%
_{0}\right\Vert ^{4}\left( 1-\hat{S}\right) ^{2}\left( 1+\frac{\left\Vert 
\bar{\Psi}_{0}\right\Vert ^{2}}{\left\Vert \hat{\Psi}_{0}\right\Vert ^{2}}%
\left\langle \hat{S}_{L}^{B}\right\rangle \right) }{2\hat{\mu}\left(
\left\langle r\right\rangle +\frac{rC_{0}}{2}\right) ^{2}}}{\left( 1-\frac{%
\left\langle S\left( X,X\right) \right\rangle \frac{9\sigma _{\hat{K}%
}^{2}\left\Vert \hat{\Psi}_{0}\right\Vert ^{4}\left( 1-\hat{S}\right)
^{2}\left( 1+\frac{\left\Vert \bar{\Psi}_{0}\right\Vert ^{2}}{\left\Vert 
\hat{\Psi}_{0}\right\Vert ^{2}}\left\langle \hat{S}_{L}^{B}\right\rangle
\right) }{2\hat{\mu}\left( \left\langle r\right\rangle +\frac{rC_{0}}{2}%
\right) ^{2}}+\left\langle S^{B}\left( X,X\right) \right\rangle 9\frac{%
\sigma _{\hat{K}}^{2}V\left\Vert \bar{\Psi}_{0}\right\Vert ^{4}\left( 1-\bar{%
S}\right) ^{2}}{2\hat{\mu}\left( \left( 1+\kappa \right) \bar{r}\right) ^{2}}%
}{\frac{2\epsilon }{3\sigma _{\hat{K}}^{2}}\left( \frac{\left\langle
f_{1}\left( X\right) \right\rangle }{C_{0}+\bar{r}}\right) ^{\frac{2}{r}}}%
\right) \left( \left( \frac{2\epsilon }{3\sigma _{\hat{K}}^{2}}\right) ^{%
\frac{r}{2}}\frac{\left\langle f_{1}\left( X\right) \right\rangle }{%
C_{0}+\left\langle \frac{S_{L}\left( X\right) }{1-S_{E}\left( X\right) }%
\right\rangle \bar{r}}\right) ^{\frac{2}{r}}}
\end{eqnarray*}%
and:%
\begin{eqnarray*}
&&\frac{\left\langle \bar{K}\right\rangle \left\Vert \bar{\Psi}\right\Vert
^{2}}{\left\langle K\right\rangle \left\Vert \Psi \right\Vert ^{2}} \\
&\simeq &\frac{9\frac{\sigma _{\hat{K}}^{2}V\left\Vert \bar{\Psi}%
_{0}\right\Vert ^{4}\left( 1-\bar{S}\right) ^{2}}{2\hat{\mu}\left( \left(
1+\kappa \right) \bar{r}\right) ^{2}}}{\left( 1-\frac{\left\langle S\left(
X,X\right) \right\rangle \frac{9\sigma _{\hat{K}}^{2}\left\Vert \hat{\Psi}%
_{0}\right\Vert ^{4}\left( 1-\hat{S}\right) ^{2}\left( 1+\frac{\left\Vert 
\bar{\Psi}_{0}\right\Vert ^{2}}{\left\Vert \hat{\Psi}_{0}\right\Vert ^{2}}%
\left\langle \hat{S}_{L}^{B}\right\rangle \right) }{2\hat{\mu}\left(
\left\langle r\right\rangle +\frac{rC_{0}}{2}\right) ^{2}}+\left\langle
S^{B}\left( X,X\right) \right\rangle 9\frac{\sigma _{\hat{K}}^{2}V\left\Vert 
\bar{\Psi}_{0}\right\Vert ^{4}\left( 1-\bar{S}\right) ^{2}}{2\hat{\mu}\left(
\left( 1+\kappa \right) \bar{r}\right) ^{2}}}{\frac{2\epsilon }{3\sigma _{%
\hat{K}}^{2}}\left( \frac{\left\langle f_{1}\left( X\right) \right\rangle }{%
C_{0}+\bar{r}}\right) ^{\frac{2}{r}}}\right) \left( \left( \frac{2\epsilon }{%
3\sigma _{\hat{K}}^{2}}\right) ^{\frac{r}{2}}\frac{\left\langle f_{1}\left(
X\right) \right\rangle }{C_{0}+\left\langle \frac{S_{L}\left( X\right) }{%
1-S_{E}\left( X\right) }\right\rangle \bar{r}}\right) ^{\frac{2}{r}}}
\end{eqnarray*}

\section*{Appendix 9. Computation of outward and inward aggregate stakes}

As in part 1 we include decreasing return to scale so that we replace the
return by the sector version of (\ref{FDN}):%
\begin{equation}
f_{1}\left( X\right) _{dr}=\frac{f_{1}\left( X\right) }{\left( K_{X}\right)
^{r}}-\frac{C}{K_{X}}
\end{equation}%
or, if we normalize the private capital, we can use in first approximation:%
\begin{eqnarray}
&&f_{1}\left( X\right) -C  \label{FDNr} \\
&\rightarrow &\left( 1-\left( \left\langle S\left( X^{\prime },X^{\prime
}\right) \right\rangle \frac{\hat{K}_{X^{\prime }}\left\Vert \hat{\Psi}%
\right\Vert ^{2}}{K_{X^{\prime }}\left\vert \Psi \left( X^{\prime }\right)
\right\vert ^{2}}+\left\langle S^{B}\left( X^{\prime },X^{\prime }\right)
\right\rangle \frac{\bar{K}_{X^{\prime }}\left\vert \bar{\Psi}\left(
X^{\prime }\right) \right\vert ^{2}}{K_{X^{\prime }}\left\vert \Psi \left(
X^{\prime }\right) \right\vert ^{2}}\right) \right) ^{r}f_{1}\left( X\right)
\notag \\
&&-\left( 1-\left( \left\langle S\left( X^{\prime },X^{\prime }\right)
\right\rangle \frac{\hat{K}_{X^{\prime }}\left\Vert \hat{\Psi}\right\Vert
^{2}}{K_{X^{\prime }}\left\vert \Psi \left( X^{\prime }\right) \right\vert
^{2}}+\left\langle S^{B}\left( X^{\prime },X^{\prime }\right) \right\rangle 
\frac{\hat{K}_{X^{\prime }}\left\Vert \hat{\Psi}\right\Vert ^{2}}{%
K_{X^{\prime }}\left\vert \Psi \left( X^{\prime }\right) \right\vert ^{2}}%
\right) \right) C  \notag
\end{eqnarray}%
\begin{equation*}
\Delta F_{\tau }\left( \bar{R}\left( K,X\right) \right) =\tau \left(
\left\langle f_{1}\left( X\right) \right\rangle -\left\langle f_{1}\left(
X^{\prime }\right) \right\rangle \right)
\end{equation*}%
We give the formula for partial, that is, half, averages. Then we have to
express them as functions of few of them to write the return equations for
each sector.

\subsection*{A9.1 Formula for $\left\langle \hat{S}_{E}\left( X^{\prime
},X\right) \right\rangle _{X^{\prime }}$}

As in part 1:

\begin{eqnarray*}
\left\langle \hat{S}_{E}\left( X^{\prime },X\right) \right\rangle
_{X^{\prime }} &\simeq &\frac{\left\langle \underline{\hat{S}}\left(
X^{\prime },X\right) \right\rangle }{2} \\
&&+\frac{\left\langle \hat{w}\left( X^{\prime },X\right) \right\rangle }{2}%
\left( \hat{w}\left( X\right) \left( \frac{\left\langle \hat{f}\left(
X^{\prime }\right) \right\rangle -\left\langle \hat{r}\left( X^{\prime
}\right) \right\rangle _{\hat{w}_{2}}}{2}\right) +w\left( X\right) \left(
\left\langle \hat{f}\left( X^{\prime }\right) \right\rangle -\frac{f\left(
X\right) +r\left( X\right) }{2}\right) \right) \\
&\simeq &\left\langle \hat{w}\left( X^{\prime },X\right) \right\rangle
\left( 1-\left\langle w\left( X\right) \right\rangle \Delta \left( \frac{%
f\left( X\right) +r\left( X\right) }{2}\right) +\frac{\left\langle \hat{f}%
\left( X^{\prime }\right) \right\rangle -\left\langle \hat{r}\left(
X^{\prime }\right) \right\rangle _{\hat{w}_{2}}}{2}\right)
\end{eqnarray*}%
\begin{equation*}
\Delta \left( \frac{f\left( X\right) +r\left( X\right) }{2}\right) =\left( 
\frac{f\left( X\right) +r\left( X\right) }{2}-\frac{\left\langle f\left(
X\right) \right\rangle +\left\langle r\left( X\right) \right\rangle }{2}%
\right)
\end{equation*}

\subsection*{A9.2 Formula for $\left\langle \bar{S}_{E}\left( X^{\prime
},X\right) \right\rangle _{X^{\prime }}$, $\left\langle \bar{S}_{L}\left(
X^{\prime },X\right) \right\rangle _{X^{\prime }}$ and $\left\langle \bar{S}%
\left( X^{\prime },X\right) \right\rangle _{X^{\prime }}$}

The computations of averages for variable $X^{\prime }$ or straightforward.
We find:%
\begin{eqnarray}
&&\left\langle \bar{S}_{E}\left( X^{\prime },X\right) \right\rangle
_{X^{\prime }}  \label{Snmm} \\
&=&\frac{\left\langle \bar{w}\left( X^{\prime },X\right) \right\rangle }{2} 
\notag \\
&&\times \left( 1+\bar{w}\left( X\right) \left( \frac{\left\langle \bar{f}%
\left( X^{\prime }\right) \right\rangle _{\bar{w}_{1}}-\left\langle \bar{r}%
\left( X^{\prime }\right) \right\rangle _{\bar{w}_{2}}}{2}\right) +\hat{w}%
_{1}^{B}\left( X\right) \left( \left\langle \bar{f}\left( X^{\prime }\right)
\right\rangle -\left\langle \hat{f}\left( X^{\prime }\right) \right\rangle _{%
\hat{w}_{1}}\right) +w_{1}^{B}\left( X\right) \left( \left\langle \bar{f}%
\left( X^{\prime }\right) \right\rangle -f\left( X\right) \right) \right) 
\notag \\
&\simeq &\frac{\left\langle \bar{w}\left( X^{\prime },X\right) \right\rangle 
}{2}  \notag \\
&&\times \left( 1+\left\langle \bar{w}\left( X\right) \right\rangle \left( 
\frac{\left\langle \bar{f}\left( X^{\prime }\right) \right\rangle _{\bar{w}%
_{1}}-\left\langle \bar{r}\left( X^{\prime }\right) \right\rangle _{\bar{w}%
_{2}}}{2}\right) +\left\langle \hat{w}_{1}^{B}\left( X\right) \right\rangle
\left( \left\langle \bar{f}\left( X^{\prime }\right) \right\rangle
-\left\langle \hat{f}\left( X^{\prime }\right) \right\rangle _{\hat{w}%
_{1}}\right) +\left\langle w_{1}^{B}\left( X\right) \right\rangle \left(
\left\langle \bar{f}\left( X^{\prime }\right) \right\rangle -f\left(
X\right) \right) \right)  \notag
\end{eqnarray}%
\begin{eqnarray*}
&&\left\langle \bar{S}_{L}\left( X^{\prime },X\right) \right\rangle
_{X^{\prime }} \\
&\simeq &\frac{\left\langle \bar{w}\left( X^{\prime },X\right) \right\rangle 
}{2}\left( 1+\left\{ \left\langle \bar{w}\left( X\right) \right\rangle
\left( \frac{\left\langle \bar{r}\left( X^{\prime }\right) \right\rangle _{%
\bar{w}_{2}}-\left\langle \bar{f}\left( X^{\prime }\right) \right\rangle _{%
\bar{w}_{1}}}{2}\right) \right. \right. \\
&&\left. \left. +\hat{w}_{1}^{B}\left( X\right) \left( \left\langle \bar{r}%
\left( X^{\prime }\right) \right\rangle -\left\langle \hat{f}\left(
X^{\prime }\right) \right\rangle _{\hat{w}_{1}^{B}}\right) +\left\langle
w_{1}^{B}\left( X\right) \right\rangle \left( \left\langle \bar{r}\left(
X^{\prime }\right) \right\rangle -f\left( X\right) \right) \right\} \right)
\end{eqnarray*}%
and as a consequence:%
\begin{eqnarray*}
\left\langle \bar{S}\left( X^{\prime },X\right) \right\rangle _{X^{\prime }}
&=&\left\langle \bar{S}_{E}\left( X^{\prime },X\right) \right\rangle
_{X^{\prime }}+\left\langle \bar{S}_{L}\left( X^{\prime },X\right)
\right\rangle _{X^{\prime }} \\
&\simeq &\left\langle \bar{w}\left( X^{\prime },X\right) \right\rangle \left[
1+\left\langle \hat{w}_{1}^{B}\left( X\right) \right\rangle \left( \frac{%
\left\langle \bar{f}\left( X^{\prime }\right) \right\rangle +\left\langle 
\bar{r}\left( X^{\prime }\right) \right\rangle }{2}-\left\langle \hat{f}%
\left( X^{\prime }\right) \right\rangle _{\hat{w}_{1}}\right) \right. \\
&&\left. +\left\langle w_{1}^{B}\left( X\right) \right\rangle \left( \frac{%
\left\langle \bar{f}\left( X^{\prime }\right) \right\rangle +\left\langle 
\bar{r}\left( X^{\prime }\right) \right\rangle }{2}-f\left( X\right) \right) %
\right]
\end{eqnarray*}

\subsection*{A9.3 Formula for $\left\langle \hat{S}_{E}^{B}\left( X^{\prime
},X\right) \right\rangle _{X^{\prime }}$ and $\left\langle \hat{S}%
_{L}^{B}\left( X^{\prime },X\right) \right\rangle _{X^{\prime }}$}

\begin{eqnarray*}
&&\left\langle \hat{S}_{E}^{B}\left( X^{\prime },X\right) \right\rangle
_{X^{\prime }} \\
&\simeq &\left\langle \hat{w}_{1}^{B}\left( X^{\prime },X\right)
\right\rangle \left[ 1+\left\langle \bar{w}\left( X\right) \right\rangle
\left( \left\langle \hat{f}\left( X^{\prime }\right) \right\rangle -\frac{%
\left\langle \bar{f}\left( X^{\prime }\right) \right\rangle _{\bar{w}%
_{1}}+\left\langle \bar{r}\left( X^{\prime }\right) \right\rangle _{\bar{w}%
_{2}}}{2}\right) +\left\langle w_{1}^{B}\left( X\right) \right\rangle \left(
\left\langle \hat{f}\left( X^{\prime }\right) \right\rangle -f\left(
X\right) \right) \right]
\end{eqnarray*}%
\begin{eqnarray*}
\frac{\left\langle \hat{S}_{L}^{B}\left( X^{\prime },X\right) \right\rangle
_{X^{\prime }}}{\kappa \left( 1-\bar{S}\left( X\right) \right) } &\simeq
&\left\langle \hat{w}_{2}^{B}\left( X^{\prime },X\right) \right\rangle
\left\{ 1+\left\langle \hat{w}_{2}^{B}\left( X\right) \right\rangle \left(
\left\langle \hat{r}\left( X^{\prime }\right) \right\rangle -\left\langle 
\hat{f}\left( X^{\prime }\right) \right\rangle _{\hat{w}_{1}}\right)
+\left\langle w_{2}^{B}\left( X\right) \right\rangle \left( \left\langle 
\hat{r}\left( X^{\prime }\right) \right\rangle -f\left( X\right) \right)
\right\} \\
&=&\left\langle \hat{w}_{1}\left( X^{\prime },X\right) \right\rangle \left\{
1+\left\langle \hat{w}_{1}\left( X\right) \right\rangle \left( \left\langle 
\hat{r}\left( X^{\prime }\right) \right\rangle -\left\langle \hat{f}\left(
X^{\prime }\right) \right\rangle _{\hat{w}_{1}}\right) +\left\langle
w_{1}\left( X\right) \right\rangle \left( \left\langle \hat{r}\left(
X^{\prime }\right) \right\rangle -f\left( X\right) \right) \right\}
\end{eqnarray*}

\subsection*{A9.4 Formula for $\frac{\bar{K}_{X}\left\vert \bar{\Psi}\left(
X\right) \right\vert ^{2}}{K_{X}\left\vert \Psi \left( X\right) \right\vert
^{2}}$, $\frac{\left\langle \bar{K}\right\rangle \left\Vert \bar{\Psi}%
\right\Vert ^{2}}{\hat{K}_{X}\left\vert \hat{\Psi}\left( X\right)
\right\vert ^{2}}$, $\frac{\hat{K}_{X}\left\vert \hat{\Psi}\left( X\right)
\right\vert ^{2}}{K_{X}\left\vert \Psi \left( X\right) \right\vert ^{2}}$\
and $\frac{\left\langle \hat{K}\right\rangle \left\Vert \hat{\Psi}%
\right\Vert ^{2}}{\hat{K}_{X}\left\vert \hat{\Psi}\left( X\right)
\right\vert ^{2}}$}

To compute the ratios, we will use that:%
\begin{eqnarray*}
\bar{g}\left( X\right) &=&\int \left( 1-\bar{S}\left( X^{\prime },X\right)
\right) ^{-1}\bar{f}\left( X^{\prime }\right) dX^{\prime } \\
&\rightarrow &\bar{f}\left( X\right) +\int \bar{S}\left( X^{\prime
},X\right) \left( 1-\bar{S}\right) ^{-1}\bar{f}\left( X^{\prime }\right) \\
&\rightarrow &\bar{f}\left( X\right) +\frac{\left\langle \bar{S}\left(
X^{\prime },X\right) \right\rangle _{X^{\prime }}}{\left( 1-\left\langle 
\bar{S}\right\rangle \right) }\left\langle \bar{f}\right\rangle
\end{eqnarray*}%
and that:%
\begin{eqnarray*}
\hat{g}\left( X\right) &=&\left( \frac{1}{1-\hat{S}}\hat{f}\right) \left(
X\right) +\frac{1}{1-\hat{S}}\left( \hat{S}_{E}^{B}+\hat{S}_{L}^{B}\right) 
\frac{1}{1-\bar{S}}\frac{\left\langle \bar{K}\right\rangle \left\Vert \bar{%
\Psi}\right\Vert ^{2}}{\hat{K}_{X}\left\vert \hat{\Psi}\left( X\right)
\right\vert ^{2}}\bar{f}\left( X\right) \\
&\rightarrow &\hat{f}\left( X\right) +\frac{\left\langle \hat{S}\left(
X^{\prime },X\right) \right\rangle _{X^{\prime }}\left\langle \hat{f}%
\right\rangle }{1-\left\langle \hat{S}\right\rangle } \\
&&+\frac{\left( \left\langle \hat{S}_{E}^{B}\left( X,X^{\prime }\right)
\right\rangle _{X^{\prime }}+\left\langle \hat{S}_{L}^{B}\left( X,X^{\prime
}\right) \right\rangle _{X^{\prime }}+\left\langle \hat{S}\left( X^{\prime
},X\right) \right\rangle _{X^{\prime }}\frac{\left\langle \hat{S}%
_{E}^{B}\right\rangle +\left\langle \hat{S}_{L}^{B}\right\rangle }{%
1-\left\langle \hat{S}\right\rangle }\right) \frac{\left\langle \bar{K}%
\right\rangle \left\Vert \bar{\Psi}\right\Vert ^{2}}{\left\langle \hat{K}%
\right\rangle \left\Vert \hat{\Psi}\right\Vert ^{2}}\left\langle \bar{f}%
\right\rangle }{1-\left\langle \bar{S}\right\rangle } \\
&\rightarrow &\hat{f}\left( X\right) +\frac{\left( \left\langle \hat{S}%
_{E}^{B}\left( X,X^{\prime }\right) \right\rangle _{X^{\prime
}}+\left\langle \hat{S}_{L}^{B}\left( X,X^{\prime }\right) \right\rangle
_{X^{\prime }}\right) \frac{\left\langle \bar{K}\right\rangle \left\Vert 
\bar{\Psi}\right\Vert ^{2}}{\left\langle \hat{K}\right\rangle \left\Vert 
\hat{\Psi}\right\Vert ^{2}}\left\langle \bar{f}\right\rangle }{%
1-\left\langle \bar{S}\right\rangle } \\
&&+\frac{\left\langle \hat{S}\left( X^{\prime },X\right) \right\rangle
_{X^{\prime }}\left\langle \hat{f}\right\rangle }{1-\left\langle \hat{S}%
\right\rangle }+\frac{\left\langle \hat{S}\left( X^{\prime },X\right)
\right\rangle _{X^{\prime }}\frac{\left\langle \hat{S}_{E}^{B}\right\rangle
+\left\langle \hat{S}_{L}^{B}\right\rangle }{1-\left\langle \hat{S}%
\right\rangle }\frac{\left\langle \bar{K}\right\rangle \left\Vert \bar{\Psi}%
\right\Vert ^{2}}{\left\langle \hat{K}\right\rangle \left\Vert \hat{\Psi}%
\right\Vert ^{2}}\left\langle \bar{f}\right\rangle }{1-\left\langle \bar{S}%
\right\rangle }
\end{eqnarray*}%
\begin{equation*}
\frac{\left\langle \hat{g}\right\rangle }{\hat{g}\left( X\right) }%
\rightarrow \frac{\left\langle \hat{f}\right\rangle +\frac{\left( \hat{S}%
_{E}^{B}+\hat{S}_{L}^{B}\right) }{1-\bar{S}}\frac{\left\langle \bar{K}%
\right\rangle \left\Vert \bar{\Psi}\right\Vert ^{2}}{\left\langle \hat{K}%
\right\rangle \left\Vert \hat{\Psi}\right\Vert ^{2}}\left\langle \bar{f}%
\right\rangle }{\left( \hat{f}\left( X\right) +\frac{\left( \left\langle 
\hat{S}_{E}^{B}\left( X,X^{\prime }\right) \right\rangle _{X^{\prime
}}+\left\langle \hat{S}_{L}^{B}\left( X,X^{\prime }\right) \right\rangle
_{X^{\prime }}\right) \frac{\left\langle \bar{K}\right\rangle \left\Vert 
\bar{\Psi}\right\Vert ^{2}}{\left\langle \hat{K}\right\rangle \left\Vert 
\hat{\Psi}\right\Vert ^{2}}\left\langle \bar{f}\right\rangle }{%
1-\left\langle \bar{S}\right\rangle }\right) +\left\langle \hat{S}\left(
X^{\prime },X\right) \right\rangle _{X^{\prime }}\left( \left\langle \hat{f}%
\right\rangle +\frac{\left( \hat{S}_{E}^{B}+\hat{S}_{L}^{B}\right) }{1-\bar{S%
}}\frac{\left\langle \bar{K}\right\rangle \left\Vert \bar{\Psi}\right\Vert
^{2}}{\left\langle \hat{K}\right\rangle \left\Vert \hat{\Psi}\right\Vert ^{2}%
}\left\langle \bar{f}\right\rangle \right) }
\end{equation*}%
\bigskip

In (Gosselin and Lotz 24), we obtained:%
\begin{equation*}
\hat{K}\left[ X^{\prime }\right] \simeq \left\langle \hat{K}\right\rangle
\left\Vert \hat{\Psi}\right\Vert ^{2}\left( \frac{\left\langle \hat{g}%
\right\rangle }{\hat{g}\left( X^{\prime }\right) }\frac{\hat{k}\left(
X^{\prime },\left\langle X\right\rangle \right) }{\hat{k}}\right) ^{2}
\end{equation*}%
\begin{equation*}
\bar{K}\left[ X^{\prime }\right] \simeq \left\langle \bar{K}\right\rangle
\left\Vert \bar{\Psi}\right\Vert ^{2}\left( \frac{\left\langle \bar{g}%
\right\rangle }{\bar{g}\left( X^{\prime }\right) }\frac{\bar{k}\left(
X^{\prime },\left\langle X\right\rangle \right) }{\bar{k}}\right) ^{2}
\end{equation*}%
We use tht:%
\begin{equation*}
\hat{k}=\frac{\left\langle \hat{S}\left( X^{\prime }\right) \right\rangle }{%
1-\left\langle \hat{S}\left( X^{\prime }\right) \right\rangle }=\frac{%
\left\langle \hat{S}\left( X^{\prime },X\right) \right\rangle }{%
1-\left\langle \hat{S}\left( X^{\prime },X\right) \right\rangle }
\end{equation*}%
and:%
\begin{equation*}
\hat{S}\left( X^{\prime }\right) =\frac{\hat{k}\left( X^{\prime
},\left\langle X\right\rangle \right) \frac{\left\langle \hat{K}%
\right\rangle \left\Vert \hat{\Psi}\right\Vert ^{2}}{\hat{K}\left[ X^{\prime
}\right] }}{1+\hat{k}\left( X^{\prime },\left\langle X\right\rangle \right) 
\frac{\left\langle \hat{K}\right\rangle \left\Vert \hat{\Psi}\right\Vert ^{2}%
}{\hat{K}\left[ X^{\prime }\right] }}
\end{equation*}%
so that the ratio $\frac{\hat{k}\left( X^{\prime },\left\langle
X\right\rangle \right) }{\hat{k}}$ becomes: 
\begin{equation*}
\frac{\hat{k}\left( X^{\prime },\left\langle X\right\rangle \right) }{\hat{k}%
}=\frac{\frac{\left\langle \hat{S}\left( X^{\prime },X\right) \right\rangle
_{X}}{1-\hat{S}\left( X^{\prime }\right) }}{\frac{\left\langle \hat{S}\left(
X^{\prime },X\right) \right\rangle }{1-\left\langle \hat{S}\left( X^{\prime
},X\right) \right\rangle }}
\end{equation*}%
and:%
\begin{equation*}
\frac{\bar{k}\left( X^{\prime },\left\langle X\right\rangle \right) }{\bar{k}%
}=\frac{\frac{\left\langle \bar{S}\left( X^{\prime },X\right) \right\rangle
_{X}}{1-\bar{S}\left( X^{\prime }\right) }}{\frac{\left\langle \bar{S}\left(
X^{\prime },X\right) \right\rangle }{1-\left\langle \bar{S}\left( X^{\prime
},X\right) \right\rangle }}
\end{equation*}

\bigskip For disposable capital that follow the equations:%
\begin{equation}
\hat{K}\left[ X^{\prime }\right] \simeq \left\langle \hat{K}\right\rangle
\left\Vert \hat{\Psi}\right\Vert ^{2}\left( \frac{\left\langle \hat{g}%
\right\rangle }{\hat{g}\left( X^{\prime }\right) }\frac{\frac{\left\langle 
\hat{S}\left( X^{\prime },X\right) \right\rangle _{X}}{1-\left\langle \hat{S}%
\left( X^{\prime },X\right) \right\rangle _{X}\frac{\left\langle \hat{K}%
\right\rangle \left\Vert \hat{\Psi}\right\Vert ^{2}}{\hat{K}\left[ X^{\prime
}\right] }}}{\frac{\left\langle \hat{S}\left( X^{\prime },X\right)
\right\rangle }{1-\left\langle \hat{S}\left( X^{\prime },X\right)
\right\rangle }}\right) ^{2}  \label{Gnr}
\end{equation}%
\begin{equation}
\bar{K}\left[ X^{\prime }\right] \simeq \left\langle \bar{K}\right\rangle
\left\Vert \bar{\Psi}\right\Vert ^{2}\left( \frac{\left\langle \bar{g}%
\right\rangle }{\bar{g}\left( X^{\prime }\right) }\frac{\frac{\left\langle 
\bar{S}\left( X^{\prime },X\right) \right\rangle _{X}}{1-\bar{S}\left(
X^{\prime },X\right) \frac{\left\langle \bar{K}\right\rangle \left\Vert \bar{%
\Psi}\right\Vert ^{2}}{\bar{K}\left[ X^{\prime }\right] }}}{\frac{%
\left\langle \bar{S}\left( X^{\prime },X\right) \right\rangle }{%
1-\left\langle \bar{S}\left( X^{\prime },X\right) \right\rangle }}\right)
^{2}\simeq \left\langle \bar{K}\right\rangle \left\Vert \bar{\Psi}%
\right\Vert ^{2}\left( \frac{\left\langle \bar{g}\right\rangle }{\bar{g}%
\left( X^{\prime }\right) }\frac{\frac{\left\langle \bar{S}\left( X^{\prime
},X\right) \right\rangle _{X}}{1-\bar{S}\left( X^{\prime },X\right) \left( 
\frac{\bar{g}\left( X^{\prime }\right) }{\left\langle \bar{g}\right\rangle }%
\right) ^{2}}}{\frac{\left\langle \bar{S}\left( X^{\prime },X\right)
\right\rangle }{1-\left\langle \bar{S}\left( X^{\prime },X\right)
\right\rangle }}\right) ^{2}  \label{Gnt}
\end{equation}%
with solution depending on average disposable capital:%
\begin{equation*}
\frac{\left\langle \hat{K}\right\rangle \left\Vert \hat{\Psi}\right\Vert ^{2}%
}{\hat{K}\left[ X^{\prime }\right] }=\frac{1}{2}\frac{\left( \frac{%
\left\langle \hat{g}\right\rangle }{\hat{g}\left( X^{\prime }\right) }\frac{%
\left\langle \hat{S}\left( X^{\prime },X\right) \right\rangle _{X}\left(
1-\left\langle \hat{S}\left( X^{\prime },X\right) \right\rangle \right) }{%
\left\langle \hat{S}\left( X^{\prime },X\right) \right\rangle }\right)
^{2}\left( 1-\sqrt{1+\frac{4\left\langle \hat{S}\left( X^{\prime },X\right)
\right\rangle _{X}}{\left( \frac{\left\langle \hat{g}\right\rangle }{\hat{g}%
\left( X^{\prime }\right) }\frac{\left\langle \hat{S}\left( X^{\prime
},X\right) \right\rangle _{X}\left( 1-\left\langle \hat{S}\left( X^{\prime
},X\right) \right\rangle \right) }{\left\langle \hat{S}\left( X^{\prime
},X\right) \right\rangle }\right) ^{2}}}\right) +2\left\langle \hat{S}\left(
X^{\prime },X\right) \right\rangle _{X}}{\left( \left\langle \hat{S}\left(
X^{\prime },X\right) \right\rangle _{X}\right) ^{2}}
\end{equation*}%
\begin{equation*}
\frac{\left\langle \bar{K}\right\rangle \left\Vert \bar{\Psi}\right\Vert ^{2}%
}{\bar{K}\left[ X^{\prime }\right] }=\frac{1}{2}\frac{\left( \frac{%
\left\langle \bar{g}\right\rangle }{\bar{g}\left( X^{\prime }\right) }\frac{%
\left\langle \bar{S}\left( X^{\prime },X\right) \right\rangle _{X}\left(
1-\left\langle \bar{S}\left( X^{\prime },X\right) \right\rangle \right) }{%
\left\langle \bar{S}\left( X^{\prime },X\right) \right\rangle }\right)
^{2}\left( 1-\sqrt{1+\frac{4\left\langle \bar{S}\left( X^{\prime },X\right)
\right\rangle _{X}}{\left( \frac{\left\langle \hat{g}\right\rangle }{\hat{g}%
\left( X^{\prime }\right) }\frac{\left\langle \bar{S}\left( X^{\prime
},X\right) \right\rangle _{X}\left( 1-\left\langle \bar{S}\left( X^{\prime
},X\right) \right\rangle \right) }{\left\langle \hat{S}\left( X^{\prime
},X\right) \right\rangle }\right) ^{2}}}\right) +2\left\langle \bar{S}\left(
X^{\prime },X\right) \right\rangle _{X}}{\left( \left\langle \bar{S}\left(
X^{\prime },X\right) \right\rangle _{X}\right) ^{2}}
\end{equation*}%
Thus, at lowest order, the ratio average-sector for investor becomes: 
\begin{eqnarray*}
&&\frac{\left\langle \hat{K}\right\rangle \left\Vert \hat{\Psi}\right\Vert
^{2}}{\hat{K}\left[ X\right] } \\
&\simeq &\left( \frac{\hat{g}\left( X\right) }{\left\langle \hat{g}%
\right\rangle }\frac{\frac{\left\langle \hat{S}\left( X^{\prime },X\right)
\right\rangle }{1-\left\langle \hat{S}\left( X^{\prime },X\right)
\right\rangle }}{\frac{\left\langle \hat{S}\left( X^{\prime },X\right)
\right\rangle _{X}}{1-\left\langle \hat{S}\left( X^{\prime },X\right)
\right\rangle _{X}\left( \frac{\hat{g}\left( X^{\prime }\right) }{%
\left\langle \hat{g}\right\rangle }\right) ^{2}}}\right) ^{2} \\
&\simeq &\left( \frac{\left( \hat{f}\left( X\right) +\frac{\left(
\left\langle \hat{S}_{E}^{B}\left( X,X^{\prime }\right) \right\rangle
_{X^{\prime }}+\left\langle \hat{S}_{L}^{B}\left( X,X^{\prime }\right)
\right\rangle _{X^{\prime }}\right) \frac{\left\langle \bar{K}\right\rangle
\left\Vert \bar{\Psi}\right\Vert ^{2}}{\left\langle \hat{K}\right\rangle
\left\Vert \hat{\Psi}\right\Vert ^{2}}\left\langle \bar{f}\right\rangle }{%
1-\left\langle \bar{S}\right\rangle }\right) +\left\langle \hat{S}\left(
X^{\prime },X\right) \right\rangle _{X^{\prime }}\left( \left\langle \hat{f}%
\right\rangle +\frac{\left( \hat{S}_{E}^{B}+\hat{S}_{L}^{B}\right) }{1-\bar{S%
}}\frac{\left\langle \bar{K}\right\rangle \left\Vert \bar{\Psi}\right\Vert
^{2}}{\left\langle \hat{K}\right\rangle \left\Vert \hat{\Psi}\right\Vert ^{2}%
}\left\langle \bar{f}\right\rangle \right) }{\left\langle \hat{f}%
\right\rangle +\frac{\left( \hat{S}_{E}^{B}+\hat{S}_{L}^{B}\right) }{1-\bar{S%
}}\frac{\left\langle \bar{K}\right\rangle \left\Vert \bar{\Psi}\right\Vert
^{2}}{\left\langle \hat{K}\right\rangle \left\Vert \hat{\Psi}\right\Vert ^{2}%
}\left\langle \bar{f}\right\rangle }\right) ^{2} \\
&&\times \left( \frac{\frac{\left\langle \hat{S}\left( X^{\prime },X\right)
\right\rangle }{1-\left\langle \hat{S}\left( X^{\prime },X\right)
\right\rangle }}{\frac{\left\langle \hat{S}\left( X^{\prime },X\right)
\right\rangle _{X}}{1-\left\langle \hat{S}\left( X^{\prime },X\right)
\right\rangle _{X}}}\right) ^{2}
\end{eqnarray*}%
and the ratio average-sector for banks is:%
\begin{equation}
\frac{\left\langle \bar{K}\right\rangle \left\Vert \bar{\Psi}\right\Vert ^{2}%
}{\bar{K}_{X}\left\vert \bar{\Psi}\left( X\right) \right\vert ^{2}}\simeq
\left( \frac{\bar{g}\left( X^{\prime }\right) \frac{\left\langle \bar{S}%
\left( X^{\prime },X\right) \right\rangle }{1-\left\langle \bar{S}\left(
X^{\prime },X\right) \right\rangle }}{\left\langle \bar{g}\right\rangle 
\frac{\left\langle \bar{S}\left( X^{\prime },X\right) \right\rangle _{X}}{1-%
\bar{S}\left( X^{\prime },X\right) \left( \frac{\bar{g}\left( X^{\prime
}\right) }{\left\langle \bar{g}\right\rangle }\right) ^{2}}}\right)
^{2}\simeq \left( \frac{\left( \left( 1-\left\langle \bar{S}\right\rangle
\right) \bar{f}\left( X\right) +\left\langle \bar{S}\left( X^{\prime
},X\right) \right\rangle _{X^{\prime }}\left\langle \bar{f}\right\rangle
\right) \frac{\left\langle \bar{S}\left( X^{\prime },X\right) \right\rangle 
}{1-\left\langle \bar{S}\left( X^{\prime },X\right) \right\rangle }}{%
\left\langle \bar{f}\right\rangle \frac{\left\langle \bar{S}\left( X^{\prime
},X\right) \right\rangle _{X}}{1-\bar{S}\left( X^{\prime },X\right) }}%
\right) ^{2}  \label{Rtb}
\end{equation}%
and using (\ref{FRV}), the ratio bank's average-investors' sector is: 
\begin{eqnarray}
&&\frac{\left\langle \bar{K}\right\rangle \left\Vert \bar{\Psi}\right\Vert
^{2}}{\hat{K}_{X}\left\vert \hat{\Psi}\left( X\right) \right\vert ^{2}}
\label{Rcb} \\
&\simeq &\frac{\left\langle \bar{K}\right\rangle \left\Vert \bar{\Psi}%
\right\Vert ^{2}}{\left\langle \hat{K}\right\rangle \left\Vert \hat{\Psi}%
\right\Vert ^{2}}\left( \frac{\hat{g}\left( X^{\prime }\right) }{%
\left\langle \hat{g}\right\rangle }\frac{\frac{\left\langle \hat{S}\left(
X^{\prime },X\right) \right\rangle }{1-\left\langle \hat{S}\left( X^{\prime
},X\right) \right\rangle }}{\frac{\left\langle \hat{S}\left( X^{\prime
},X\right) \right\rangle _{X}}{1-\left\langle \hat{S}\left( X^{\prime
},X\right) \right\rangle _{X}\frac{\left\langle \hat{K}\right\rangle
\left\Vert \hat{\Psi}\right\Vert ^{2}}{\hat{K}\left[ X^{\prime }\right] }}}%
\right) ^{2}  \notag \\
&\simeq &\frac{\left( 1-\bar{S}\right) ^{2}\left( \hat{f}\left( X\right) +%
\frac{\left\langle \hat{S}\left( X^{\prime },X\right) \right\rangle
_{X^{\prime }}\left\langle \hat{f}\right\rangle }{1-\left\langle \hat{S}%
\right\rangle }+\frac{\left( \left\langle \hat{S}_{E}^{B}\left( X,X^{\prime
}\right) \right\rangle _{X^{\prime }}+\left\langle \hat{S}_{L}^{B}\left(
X,X^{\prime }\right) \right\rangle _{X^{\prime }}+\frac{\left\langle \hat{S}%
_{E}^{B}\right\rangle +\left\langle \hat{S}_{L}^{B}\right\rangle }{%
1-\left\langle \hat{S}\right\rangle }\right) \frac{\left\langle \bar{K}%
\right\rangle \left\Vert \bar{\Psi}\right\Vert ^{2}}{\left\langle \hat{K}%
\right\rangle \left\Vert \hat{\Psi}\right\Vert ^{2}}\left\langle \bar{f}%
\right\rangle }{1-\left\langle \bar{S}\right\rangle }\right) ^{2}}{%
\left\langle \bar{f}\right\rangle ^{2}}  \notag \\
&&\times \frac{\left\Vert \bar{\Psi}_{0}\right\Vert ^{4}}{\left\Vert \hat{%
\Psi}_{0}\right\Vert ^{4}\left( 1+\frac{\left\Vert \bar{\Psi}_{0}\right\Vert
^{2}}{\left\Vert \hat{\Psi}_{0}\right\Vert ^{2}}\left\langle \hat{S}%
_{L}^{B}\right\rangle \right) }  \notag
\end{eqnarray}%
with:%
\begin{eqnarray*}
\frac{\left\langle \bar{K}\right\rangle \left\Vert \bar{\Psi}\right\Vert ^{2}%
}{\left\langle \hat{K}\right\rangle \left\Vert \hat{\Psi}\right\Vert ^{2}}
&\simeq &\frac{\left( \frac{\left\langle \hat{f}\right\rangle +\frac{%
\left\langle \hat{S}_{E}^{B}\right\rangle +\left\langle \hat{S}%
_{L}^{B}\right\rangle }{1-\bar{S}}\frac{\left\langle \bar{K}\right\rangle
\left\Vert \bar{\Psi}\right\Vert ^{2}}{\left\langle \hat{K}\right\rangle
\left\Vert \hat{\Psi}\right\Vert ^{2}}\left\langle \bar{f}\right\rangle }{1-%
\hat{S}}\right) ^{2}}{\left\langle \bar{g}\right\rangle ^{2}}\frac{%
\left\Vert \bar{\Psi}_{0}\right\Vert ^{4}}{\left\Vert \hat{\Psi}%
_{0}\right\Vert ^{4}\left( 1+\frac{\left\Vert \bar{\Psi}_{0}\right\Vert ^{2}%
}{\left\Vert \hat{\Psi}_{0}\right\Vert ^{2}}\left\langle \hat{S}%
_{L}^{B}\right\rangle \right) } \\
&=&\frac{\frac{\left( 1-\bar{S}\right) ^{2}}{\left( 1-\hat{S}\right) ^{2}}%
\left( \left\langle \hat{f}\right\rangle +\frac{\left\langle \hat{S}%
_{E}^{B}\right\rangle +\left\langle \hat{S}_{L}^{B}\right\rangle }{1-\bar{S}}%
\frac{\left\langle \bar{K}\right\rangle \left\Vert \bar{\Psi}\right\Vert ^{2}%
}{\left\langle \hat{K}\right\rangle \left\Vert \hat{\Psi}\right\Vert ^{2}}%
\left\langle \bar{f}\right\rangle \right) ^{2}}{\left\langle \bar{f}%
\right\rangle ^{2}}\frac{\left\Vert \bar{\Psi}_{0}\right\Vert ^{4}}{%
\left\Vert \hat{\Psi}_{0}\right\Vert ^{4}\left( 1+\frac{\left\Vert \bar{\Psi}%
_{0}\right\Vert ^{2}}{\left\Vert \hat{\Psi}_{0}\right\Vert ^{2}}\left\langle 
\hat{S}_{L}^{B}\right\rangle \right) }
\end{eqnarray*}%
The ratio banks-firms and investors-firms per sector are:%
\begin{eqnarray*}
\frac{\bar{K}\left[ X\right] }{\left\langle K_{X}\right\rangle \left\Vert
\Psi \left( X\right) \right\Vert ^{2}} &\simeq &\frac{18\frac{\sigma _{\hat{K%
}}^{2}V}{\hat{\mu}}\left( \left\Vert \bar{\Psi}_{0}\right\Vert \right)
^{4}\left( \frac{\frac{\left\langle \bar{S}\left( X^{\prime },X\right)
\right\rangle _{X}}{1-\bar{S}\left( X^{\prime },X\right) }}{\bar{g}\left(
X^{\prime }\right) \frac{\left\langle \bar{S}\left( X^{\prime },X\right)
\right\rangle }{1-\left\langle \bar{S}\left( X^{\prime },X\right)
\right\rangle }}\right) ^{2}}{\epsilon \sqrt{\sigma _{\hat{K}}^{2}\frac{%
\left\vert \Psi _{0}\left( X\right) \right\vert ^{2}}{\epsilon }}\frac{3}{12}%
X^{3}}\sigma _{\hat{K}}^{2}\left( Z\left( X\right) \left( f_{1}\left(
X\right) -r\left( X\right) \right) +r\left( X\right) \right) ^{2} \\
&\simeq &\frac{18\frac{\sigma _{\hat{K}}^{2}V}{\hat{\mu}}\left( \left\Vert 
\bar{\Psi}_{0}\right\Vert \right) ^{4}\left( \frac{\frac{\left\langle \bar{S}%
\left( X^{\prime },X\right) \right\rangle _{X}}{1-\bar{S}\left( X^{\prime
},X\right) }}{\left( \bar{f}\left( X\right) +\frac{\left\langle \bar{S}%
\left( X^{\prime },X\right) \right\rangle _{X^{\prime }}}{\left(
1-\left\langle \bar{S}\right\rangle \right) }\left\langle \bar{f}%
\right\rangle \right) ^{2}\frac{\left\langle \bar{S}\left( X^{\prime
},X\right) \right\rangle }{1-\left\langle \bar{S}\left( X^{\prime },X\right)
\right\rangle }}\right) ^{2}}{\epsilon \sqrt{\sigma _{\hat{K}}^{2}\frac{%
\left\vert \Psi _{0}\left( X\right) \right\vert ^{2}}{\epsilon }}\frac{3}{12}%
X^{3}}\sigma _{\hat{K}}^{2}\left( Z\left( X\right) \left( f_{1}\left(
X\right) -r\left( X\right) \right) +r\left( X\right) \right) ^{2}
\end{eqnarray*}

\begin{equation}
\frac{\hat{K}\left[ X\right] }{\left\langle K_{X}\right\rangle \left\Vert
\Psi \left( X\right) \right\Vert ^{2}}\simeq \frac{\frac{9\sigma _{\hat{K}%
}^{2}}{2\hat{\mu}}V\left\Vert \hat{\Psi}_{0}\right\Vert ^{4}\left( \frac{%
\frac{\left\langle \hat{S}\left( X^{\prime },X\right) \right\rangle _{X}}{%
1-\left\langle \hat{S}\left( X^{\prime },X\right) \right\rangle _{X}\frac{%
\left\langle \hat{K}\right\rangle \left\Vert \hat{\Psi}\right\Vert ^{2}}{%
\hat{K}\left[ X^{\prime }\right] }}}{\hat{g}\left( X^{\prime }\right) \frac{%
\left\langle \hat{S}\left( X^{\prime },X\right) \right\rangle }{%
1-\left\langle \hat{S}\left( X^{\prime },X\right) \right\rangle }}\right)
^{2}}{\epsilon \sqrt{\sigma _{\hat{K}}^{2}\frac{\left\vert \Psi _{0}\left(
X\right) \right\vert ^{2}}{\epsilon }}\frac{3}{12}X^{3}}\sigma _{\hat{K}%
}^{2}\left( Z\left( X\right) \left( f_{1}\left( X\right) -r\left( X\right)
\right) +r\left( X\right) \right) ^{2}  \label{Rtcl}
\end{equation}%
with:%
\begin{equation*}
Z\left( X\right) =\frac{1-\left( \left\langle S_{E}\left( X\right)
\right\rangle +\left\langle S_{E}^{B}\left( X\right) \right\rangle \right) }{%
1-\left( \left\langle S\left( X\right) \right\rangle +\left\langle
S^{B}\left( X\right) \right\rangle \right) }
\end{equation*}%
\begin{eqnarray*}
&&\frac{\hat{K}_{X}\left\vert \hat{\Psi}\left( X\right) \right\vert ^{2}}{%
K_{X}\left\vert \Psi \left( X\right) \right\vert ^{2}} \\
&\simeq &\frac{\frac{18\sigma _{\hat{K}}^{2}}{\hat{\mu}\hat{g}^{2}\left(
X\right) }V\left\Vert \hat{\Psi}_{0}\left( X\right) \right\Vert ^{4}}{%
\epsilon \sqrt{\sigma _{\hat{K}}^{2}\frac{\left\vert \Psi _{0}\left(
X\right) \right\vert ^{2}}{\epsilon }}X^{3}}\left( Z\left( X\right) \left(
f_{1}\left( X\right) -r\left( X\right) \right) +r\left( X\right) \right) ^{2}
\\
&\simeq &\frac{\frac{18\sigma _{\hat{K}}^{2}}{\hat{\mu}}V\left\Vert \hat{\Psi%
}_{0}\right\Vert ^{4}\sigma _{\hat{K}}^{2}\left( Z\left( X\right) \left(
f_{1}\left( X\right) -r\left( X\right) \right) +r\left( X\right) \right) ^{2}%
}{\left( \hat{f}\left( X\right) +\frac{\left\langle \hat{S}\left( X^{\prime
},X\right) \right\rangle _{X^{\prime }}\left\langle \hat{f}\right\rangle }{%
1-\left\langle \hat{S}\right\rangle }+\frac{\left( \left\langle \hat{S}%
_{E}^{B}\left( X,X^{\prime }\right) \right\rangle _{X^{\prime
}}+\left\langle \hat{S}_{L}^{B}\left( X,X^{\prime }\right) \right\rangle
_{X^{\prime }}+\left\langle \hat{S}\left( X^{\prime },X\right) \right\rangle
_{X^{\prime }}\frac{\left\langle \hat{S}_{E}^{B}\right\rangle +\left\langle 
\hat{S}_{L}^{B}\right\rangle }{1-\left\langle \hat{S}\right\rangle }\right) 
\frac{\left\langle \bar{K}\right\rangle \left\Vert \bar{\Psi}\right\Vert ^{2}%
}{\left\langle \hat{K}\right\rangle \left\Vert \hat{\Psi}\right\Vert ^{2}}%
\left\langle \bar{f}\right\rangle }{1-\left\langle \bar{S}\right\rangle }%
\right) ^{2}\epsilon \sqrt{\sigma _{\hat{K}}^{2}\frac{\left\vert \Psi
_{0}\left( X\right) \right\vert ^{2}}{\epsilon }}X^{3}}
\end{eqnarray*}%
where:%
\begin{eqnarray*}
Z\left( X\right) &=&\frac{1-\left( S_{E}\left( X,X\right) \frac{\left\langle 
\hat{K}\right\rangle \left\Vert \hat{\Psi}\right\Vert ^{2}}{K_{X}\left\vert
\Psi \left( X\right) \right\vert ^{2}}+S_{E}^{B}\left( X,X\right) \frac{%
\left\langle \bar{K}\right\rangle \left\Vert \bar{\Psi}\right\Vert ^{2}}{%
K_{X}\left\vert \Psi \left( X\right) \right\vert ^{2}}\right) }{1-\left(
S\left( X,X\right) \frac{\left\langle \hat{K}\right\rangle \left\Vert \hat{%
\Psi}\right\Vert ^{2}}{K_{X}\left\vert \Psi \left( X\right) \right\vert ^{2}}%
+S^{B}\left( X,X\right) \frac{\left\langle \bar{K}\right\rangle \left\Vert 
\bar{\Psi}\right\Vert ^{2}}{K_{X}\left\vert \Psi \left( X\right) \right\vert
^{2}}\right) } \\
&\rightarrow &\sqrt{\frac{1}{3}\left( \frac{4r^{2}\left( X\right) }{3\left(
f_{1}\left( X\right) -r\left( X\right) \right) ^{2}}+\frac{1+2A\left(
f_{1}\left( X\right) -r\left( X\right) \right) r\left( X\right)
-Br^{2}\left( X\right) }{B\left( f_{1}\left( X\right) -r\left( X\right)
\right) ^{2}}\right) } \\
&&\times \cosh \left( \frac{1}{3}ar\cosh \left( \frac{-\left( \frac{%
16r^{3}\left( X\right) }{27\left( f_{1}\left( X\right) -r\left( X\right)
\right) ^{3}}+\frac{Ar^{2}\left( X\right) -1}{B\left( f_{1}\left( X\right)
-r\left( X\right) \right) ^{2}}-\frac{2r\left( 1+2A\left( f_{1}\left(
X\right) -r\left( X\right) \right) r\left( X\right) -Br^{2}\left( X\right)
\right) }{3B\left( f_{1}\left( X\right) -r\left( X\right) \right) ^{3}}%
\right) }{\left( \frac{1}{3}\left( \frac{4\left\langle r\left( X\right)
\right\rangle ^{2}}{3\left( \left\langle f_{1}\left( X\right) \right\rangle
-\left\langle r\left( X\right) \right\rangle \right) ^{2}}+\frac{1+2A\left(
f_{1}\left( X\right) -r\left( X\right) \right) r\left( X\right)
-Br^{2}\left( X\right) }{B\left( \left\langle f_{1}\left( X\right)
\right\rangle -\left\langle r\left( X\right) \right\rangle \right) ^{2}}%
\right) \right) ^{\frac{3}{2}}}\right) \right) \\
&&-\frac{2r\left( X\right) }{3\left( f_{1}\left( X\right) -r\left( X\right)
\right) }
\end{eqnarray*}%
and:%
\begin{eqnarray*}
&&A\left( \epsilon \sqrt{\sigma _{\hat{K}}^{2}\frac{\left\vert \Psi
_{0}\left( X\right) \right\vert ^{2}}{\epsilon }}\frac{3}{12}X^{3}\right) \\
&=&\frac{S_{E}\left( X,X\right) \frac{18\sigma _{\hat{K}}^{2}}{\hat{\mu}}%
V\left\Vert \hat{\Psi}_{0}\right\Vert ^{4}}{\left( \hat{f}\left( X\right) +%
\frac{\left\langle \hat{S}\left( X^{\prime },X\right) \right\rangle
_{X^{\prime }}\left\langle \hat{f}\right\rangle }{1-\left\langle \hat{S}%
\right\rangle }+\frac{\left( \left\langle \hat{S}_{E}^{B}\left( X,X^{\prime
}\right) \right\rangle _{X^{\prime }}+\left\langle \hat{S}_{L}^{B}\left(
X,X^{\prime }\right) \right\rangle _{X^{\prime }}+\left\langle \hat{S}\left(
X^{\prime },X\right) \right\rangle _{X^{\prime }}\frac{\left\langle \hat{S}%
_{E}^{B}\right\rangle +\left\langle \hat{S}_{L}^{B}\right\rangle }{%
1-\left\langle \hat{S}\right\rangle }\right) \frac{\left\langle \bar{K}%
\right\rangle \left\Vert \bar{\Psi}\right\Vert ^{2}}{\left\langle \hat{K}%
\right\rangle \left\Vert \hat{\Psi}\right\Vert ^{2}}\left\langle \bar{f}%
\right\rangle }{1-\left\langle \bar{S}\right\rangle }\right) ^{2}} \\
&&+\frac{\left\langle S_{E}^{B}\left( X^{\prime },X^{\prime }\right)
\right\rangle _{X^{\prime }}18\sigma _{\hat{K}}^{2}V}{\left( \bar{f}\left(
X\right) +\frac{\left\langle \bar{S}\left( X^{\prime },X\right)
\right\rangle _{X^{\prime }}}{\left( 1-\left\langle \bar{S}\right\rangle
\right) }\left\langle \bar{f}\right\rangle \right) ^{2}\hat{\mu}}\left(
\left\Vert \bar{\Psi}_{0}\right\Vert \right) ^{4}
\end{eqnarray*}%
\begin{eqnarray*}
&&B\epsilon \sqrt{\sigma _{\hat{K}}^{2}\frac{\left\vert \Psi _{0}\left(
X\right) \right\vert ^{2}}{\epsilon }}\frac{3}{12}X^{3} \\
&=&\frac{S\left( X,X\right) \frac{18\sigma _{\hat{K}}^{2}}{\hat{\mu}}%
V\left\Vert \hat{\Psi}_{0}\right\Vert ^{4}}{\left( \hat{f}\left( X\right) +%
\frac{\left\langle \hat{S}\left( X^{\prime },X\right) \right\rangle
_{X^{\prime }}\left\langle \hat{f}\right\rangle }{1-\left\langle \hat{S}%
\right\rangle }+\frac{\left( \left\langle \hat{S}_{E}^{B}\left( X,X^{\prime
}\right) \right\rangle _{X^{\prime }}+\left\langle \hat{S}_{L}^{B}\left(
X,X^{\prime }\right) \right\rangle _{X^{\prime }}+\left\langle \hat{S}\left(
X^{\prime },X\right) \right\rangle _{X^{\prime }}\frac{\left\langle \hat{S}%
_{E}^{B}\right\rangle +\left\langle \hat{S}_{L}^{B}\right\rangle }{%
1-\left\langle \hat{S}\right\rangle }\right) \frac{\left\langle \bar{K}%
\right\rangle \left\Vert \bar{\Psi}\right\Vert ^{2}}{\left\langle \hat{K}%
\right\rangle \left\Vert \hat{\Psi}\right\Vert ^{2}}\left\langle \bar{f}%
\right\rangle }{1-\left\langle \bar{S}\right\rangle }\right) ^{2}} \\
&&+\left\langle S^{B}\left( X^{\prime },X^{\prime }\right) \right\rangle
_{X^{\prime }}\frac{18\sigma _{\hat{K}}^{2}V}{\left( \bar{f}\left( X\right) +%
\frac{\left\langle \bar{S}\left( X^{\prime },X\right) \right\rangle
_{X^{\prime }}}{\left( 1-\left\langle \bar{S}\right\rangle \right) }%
\left\langle \bar{f}\right\rangle \right) ^{2}\hat{\mu}}\left( \left\Vert 
\bar{\Psi}_{0}\right\Vert \right) ^{4}
\end{eqnarray*}

\subsection*{A9.5 Formula for $\hat{S}_{E}\left( X\right) $ and $\hat{S}%
\left( X\right) $}

Similar derivations to those performed in Part 1 lead to the following
results:%
\begin{equation*}
\left\langle \frac{\hat{w}\left( X^{\prime },X\right) }{2}\right\rangle
_{X}\simeq \frac{\left( 1-\left( \gamma \left\langle \hat{S}_{E}\left(
X\right) \right\rangle \right) ^{2}\right) \left\langle \hat{w}_{1}^{\left(
0\right) }\left( X^{\prime },X\right) \right\rangle _{X}}{1+\left\langle 
\hat{w}_{1}^{\left( 0\right) }\left( X^{\prime },X\right) \right\rangle
_{X}\left( 1-\left( \gamma \left\langle \hat{S}_{E}\left( X\right)
\right\rangle \right) ^{2}\right) +\left( \gamma \left\langle \hat{S}%
_{E}\left( X_{1},X^{\prime }\right) \right\rangle _{X_{1}}\right)
^{2}-\left( \gamma \left\langle \hat{S}_{E}\left( X\right) \right\rangle
\right) ^{2}}
\end{equation*}%
\begin{equation*}
\left\langle \hat{w}_{1}^{\left( 0\right) }\left( X^{\prime },X\right)
\right\rangle _{X}\simeq 1
\end{equation*}%
which allows to rewrite the average over $X$: 
\begin{eqnarray}
&&\left\langle \hat{S}_{E}\left( X^{\prime },X\right) \right\rangle _{X}
\label{SVb} \\
&\simeq &\frac{\left\langle \hat{w}\left( X^{\prime },X\right) \right\rangle
_{X}}{2}\left( 1+\left( \left\langle \hat{w}\left( X\right) \right\rangle
\left( \hat{f}\left( X^{\prime }\right) -\frac{\left\langle \hat{f}\left(
X^{\prime }\right) \right\rangle _{\hat{w}_{1}}+\left\langle \hat{r}\left(
X^{\prime }\right) \right\rangle _{\hat{w}_{2}}}{2}\right) \right. \right. 
\notag \\
&&\left. \left. +\left\langle w\left( X\right) \right\rangle \left( \hat{f}%
\left( X^{\prime }\right) -\frac{\left\langle f\left( X\right) \right\rangle
+\left\langle r\left( X\right) \right\rangle }{2}\right) \right) \right) 
\notag \\
&=&\frac{\left\langle \hat{w}\left( X^{\prime },X\right) \right\rangle _{X}}{%
2}\left( 1+\hat{f}\left( X^{\prime }\right) -\left( \left\langle \hat{w}%
\left( X\right) \right\rangle \frac{\left\langle \hat{f}\left( X^{\prime
}\right) \right\rangle _{\hat{w}_{1}}+\left\langle \hat{r}\left( X^{\prime
}\right) \right\rangle _{\hat{w}_{2}}}{2}+\left\langle w\left( X\right)
\right\rangle \frac{\left\langle f\left( X\right) \right\rangle
+\left\langle r\left( X\right) \right\rangle }{2}\right) \right)  \notag \\
&=&\frac{\left\langle \hat{w}\left( X^{\prime },X\right) \right\rangle _{X}}{%
2}\left( 1+\Delta \hat{f}\left( X^{\prime }\right) \right)  \notag
\end{eqnarray}%
where:%
\begin{equation*}
\left\langle \frac{\hat{w}\left( X^{\prime },X\right) }{2}\right\rangle
_{X}\simeq \frac{\left( 1-\left( \gamma \left\langle \hat{S}_{E}\left(
X\right) \right\rangle \right) ^{2}\right) }{2-\left( \gamma \left\langle 
\hat{S}_{E}\left( X\right) \right\rangle \right) ^{2}+\left( \gamma
\left\langle \hat{S}_{E}\left( X_{1},X^{\prime }\right) \right\rangle
_{X_{1}}\right) ^{2}-\left( \gamma \left\langle \hat{S}_{E}\left( X\right)
\right\rangle \right) ^{2}}
\end{equation*}%
\begin{eqnarray*}
\left\langle \hat{w}\left( X\right) \right\rangle &=&\frac{\left( 1-\left(
\gamma \left\langle \hat{S}_{E}\left( X\right) \right\rangle \right)
^{2}\right) }{2-\left( \gamma \left\langle \hat{S}_{E}\left( X\right)
\right\rangle \right) ^{2}} \\
\left\langle w\left( X\right) \right\rangle &=&\frac{1}{2-\left( \gamma
\left\langle \hat{S}_{E}\left( X\right) \right\rangle \right) ^{2}}
\end{eqnarray*}%
and:%
\begin{equation*}
\Delta \hat{f}\left( X^{\prime }\right) =\hat{f}\left( X^{\prime }\right)
-\left( \left\langle \hat{w}\left( X\right) \right\rangle \frac{\left\langle 
\hat{f}\left( X^{\prime }\right) \right\rangle _{\hat{w}_{1}}+\left\langle 
\hat{r}\left( X^{\prime }\right) \right\rangle _{\hat{w}_{2}}}{2}%
+\left\langle w\left( X\right) \right\rangle \frac{\left\langle f\left(
X\right) \right\rangle +\left\langle r\left( X\right) \right\rangle }{2}%
\right)
\end{equation*}%
Similarly, we find:%
\begin{equation*}
\left\langle \hat{S}\left( X^{\prime },X\right) \right\rangle _{X}\simeq
\left\langle \hat{w}\left( X^{\prime },X\right) \right\rangle _{X}\left( 1+%
\frac{\Delta \hat{f}\left( X^{\prime }\right) +\Delta \hat{r}\left(
X^{\prime }\right) }{2}\right)
\end{equation*}%
\begin{eqnarray}
\hat{S}_{E}\left( X\right) &=&\left\langle \hat{S}_{E}\left( X,X^{\prime
}\right) \right\rangle _{X^{\prime }}\frac{\left\langle \hat{K}\right\rangle
\left\Vert \hat{\Psi}\right\Vert ^{2}}{\hat{K}_{X}\left\vert \hat{\Psi}%
\left( X\right) \right\vert ^{2}}  \label{seh} \\
&=&\left\langle \hat{S}_{E}\left( X,X^{\prime }\right) \right\rangle
_{X^{\prime }}\frac{\left\Vert \hat{\Psi}_{0}\right\Vert ^{4}\hat{g}%
^{2}\left( X\right) \left( \left\langle Z\left( X\right) \right\rangle
\left( \left\langle f_{1}\left( X\right) \right\rangle -\left\langle r\left(
X\right) \right\rangle \right) +\left\langle r\left( X\right) \right\rangle
\right) ^{2}}{\left\vert \hat{\Psi}_{0}\left( X\right) \right\vert
^{4}\left\langle \hat{g}\left( X\right) \right\rangle ^{2}\left( Z\left(
X\right) \left( f_{1}\left( X\right) -r\left( X\right) \right) +r\left(
X\right) \right) ^{2}}  \notag
\end{eqnarray}%
and:%
\begin{eqnarray*}
\hat{S}\left( X\right) &=&\left\langle \hat{S}\left( X,X^{\prime }\right)
\right\rangle _{X^{\prime }}\frac{\left\langle \hat{K}\right\rangle
\left\Vert \hat{\Psi}\right\Vert ^{2}}{\hat{K}_{X}\left\vert \hat{\Psi}%
\left( X\right) \right\vert ^{2}} \\
&=&\left\langle \hat{S}\left( X,X^{\prime }\right) \right\rangle _{X^{\prime
}}\frac{\left\Vert \hat{\Psi}_{0}\right\Vert ^{4}\hat{g}^{2}\left( X\right)
\left( \left\langle Z\left( X\right) \right\rangle \left( \left\langle
f_{1}\left( X\right) \right\rangle -\left\langle r\left( X\right)
\right\rangle \right) +r\left( X\right) \right) ^{2}}{\left\vert \hat{\Psi}%
_{0}\left( X\right) \right\vert ^{4}\left\langle \hat{g}\left( X\right)
\right\rangle ^{2}\left( Z\left( X\right) \left( f_{1}\left( X\right)
-r\left( X\right) \right) +r\left( X\right) \right) ^{2}}
\end{eqnarray*}

\subsection*{A9.5 Formula for $S_{E}\left( X,X\right) $, $S\left( X,X\right) 
$ and $S_{E}\left( X\right) $, $S\left( X\right) $}

\begin{eqnarray}
&&S_{E}\left( X,X\right) \\
&=&\frac{w\left( X\right) }{2}\left( 1+\left( \hat{w}\left( X\right) \left(
f\left( X\right) -\frac{\left\langle \hat{f}\left( X^{\prime }\right)
\right\rangle _{\hat{w}_{1}}+\left\langle \hat{r}\left( X^{\prime }\right)
\right\rangle _{\hat{w}_{2}}}{2}\right) +\frac{w\left( X\right) }{2}\left(
f\left( X\right) -\bar{r}\left( X\right) \right) \right) \right)  \notag
\end{eqnarray}%
\begin{eqnarray*}
\hat{w}\left( X\right) &=&\frac{1-\left( \gamma \left\langle \hat{S}%
_{E}\left( X^{\prime },X\right) \right\rangle _{X^{\prime }}\right) ^{2}}{%
2-\left( \gamma \left\langle \hat{S}_{E}\left( X^{\prime },X\right)
\right\rangle _{X^{\prime }}\right) ^{2}} \\
w\left( X\right) &=&\frac{1}{2-\left( \gamma \left\langle \hat{S}_{E}\left(
X^{\prime },X\right) \right\rangle _{X^{\prime }}\right) ^{2}}
\end{eqnarray*}%
\begin{equation}
S\left( X,X\right) =w\left( X\right) \left( 1+\left( \hat{w}\left( X\right)
\left( \frac{f\left( X\right) +\bar{r}\left( X\right) }{2}-\frac{%
\left\langle \hat{f}\left( X^{\prime }\right) \right\rangle _{\hat{w}%
_{1}}+\left\langle \hat{r}\left( X^{\prime }\right) \right\rangle _{\hat{w}%
_{2}}}{2}\right) \right) \right)
\end{equation}%
\begin{eqnarray}
S_{E}\left( X\right) &=&S_{E}\left( X,X\right) \frac{\hat{K}_{X}\left\vert 
\hat{\Psi}\left( X\right) \right\vert ^{2}}{K_{X}\left\vert \Psi \left(
X\right) \right\vert ^{2}}  \label{Vp} \\
S\left( X\right) &=&S\left( X,X\right) \frac{\hat{K}_{X}\left\vert \hat{\Psi}%
\left( X\right) \right\vert ^{2}}{K_{X}\left\vert \Psi \left( X\right)
\right\vert ^{2}}  \notag
\end{eqnarray}%
with $\frac{\hat{K}_{X}\left\vert \hat{\Psi}\left( X\right) \right\vert ^{2}%
}{K_{X}\left\vert \Psi \left( X\right) \right\vert ^{2}}$ given by (\ref%
{Rtcl}).

\subsection*{A9.6 Formula for $\bar{S}_{E}\left( X^{\prime }\right) $, $\bar{%
S}_{L}\left( X^{\prime }\right) $ and $\bar{S}\left( X^{\prime }\right) $}

These averages are given by:%
\begin{eqnarray}
\left\langle \bar{S}_{E}\left( X^{\prime },X\right) \right\rangle _{X}
&=&\left\langle \frac{\bar{w}\left( X^{\prime },X\right) }{2}\right\rangle
_{X}\left( 1+\left\{ \left\langle \bar{w}\left( X\right) \right\rangle
\left( \bar{f}\left( X^{\prime }\right) -\frac{\left\langle \bar{f}\left(
X^{\prime }\right) \right\rangle _{\bar{w}_{1}}+\left\langle \bar{r}\left(
X^{\prime }\right) \right\rangle _{\bar{w}_{2}}}{2}\right) \right. \right.
\label{seb} \\
&&\left. \left. +\left\langle \hat{w}_{1}^{B}\left( X\right) \right\rangle
\left( \bar{f}\left( X^{\prime }\right) -\left\langle \hat{f}\left(
X^{\prime }\right) \right\rangle _{\hat{w}_{1}}\right) +\left\langle
w_{1}^{B}\left( X\right) \right\rangle \left( \bar{f}\left( X^{\prime
}\right) -\left\langle f\left( X\right) \right\rangle \right) \right\}
\right)  \notag
\end{eqnarray}

\begin{eqnarray}
&&\left\langle \bar{S}_{L}\left( X^{\prime },X\right) \right\rangle _{X}
\label{sbl} \\
&=&\left\langle \frac{\bar{w}\left( X^{\prime },X\right) }{2}\right\rangle
\left( 1+\left\{ \bar{w}\left( X\right) \left( \bar{r}\left( X^{\prime
}\right) -\frac{\left\langle \bar{f}\left( X^{\prime }\right) \right\rangle
_{\bar{w}_{1}}+\left\langle \bar{r}\left( X^{\prime }\right) \right\rangle _{%
\bar{w}_{2}}}{2}\right) \right. \right.  \notag \\
&&\left. \left. +\left\langle \hat{w}_{1}^{B}\left( X\right) \right\rangle
\left( \bar{r}\left( X^{\prime }\right) -\left\langle \hat{f}\left(
X^{\prime }\right) \right\rangle _{\hat{w}_{1}^{B}}\right) +\left\langle
w_{1}^{B}\left( X\right) \right\rangle \left( \bar{r}\left( X^{\prime
}\right) -\left\langle f\left( X\right) \right\rangle \right) \right\}
\right)  \notag
\end{eqnarray}

\begin{eqnarray}
\left\langle \bar{S}\left( X^{\prime },X\right) \right\rangle _{X}
&=&\left\langle \bar{S}_{E}\left( X^{\prime },X\right) \right\rangle
_{X}+\left\langle \bar{S}_{L}\left( X^{\prime },X\right) \right\rangle _{X}
\label{SB} \\
&=&\left\langle \bar{w}\left( X^{\prime },X\right) \right\rangle \left[
1+\left\{ \bar{w}\left( X\right) \left( \frac{\bar{f}\left( X^{\prime
}\right) +\bar{r}\left( X^{\prime }\right) }{2}-\frac{\left\langle \bar{f}%
\left( X^{\prime }\right) \right\rangle _{\bar{w}_{1}}+\left\langle \bar{r}%
\left( X^{\prime }\right) \right\rangle _{\bar{w}_{2}}}{2}\right) \right.
\right.  \notag \\
&&\left. \left. +\hat{w}_{1}^{B}\left( X\right) \left( \frac{\bar{f}\left(
X^{\prime }\right) +\bar{r}\left( X^{\prime }\right) }{2}-\left\langle \hat{f%
}\left( X^{\prime }\right) \right\rangle _{\hat{w}_{1}}\right)
+w_{1}^{B}\left( X\right) \left( \frac{\bar{f}\left( X^{\prime }\right) +%
\bar{r}\left( X^{\prime }\right) }{2}-\left\langle f\left( X\right)
\right\rangle \right) \right\} \right]  \notag
\end{eqnarray}%
while the uncertainty are obtained by the formulas: 
\begin{eqnarray*}
&&\left( \left\langle \bar{w}\left( X^{\prime },X\right) \right\rangle
_{X}\right) ^{-1} \\
&=&1+\frac{4}{\zeta ^{2}\left\langle \bar{w}_{1}^{\left( 0\right) }\left(
X^{\prime },X\right) \right\rangle _{X}}\left\{ \frac{\bar{\zeta}^{2}\zeta
^{2}\left( 1+\frac{\left( \gamma \left\langle \hat{S}_{E}\left( X_{1},\left(
X^{\prime }\right) ^{\prime }\right) \right\rangle \right) ^{2}}{1-\left(
\gamma \left\langle \hat{S}_{E}\left( X^{\prime },\left( X^{\prime }\right)
^{\prime }\right) \right\rangle \right) ^{2}}\right) }{\left\langle \hat{w}%
_{1}^{\left( 0\right) B}\left( \left( X^{\prime }\right) ^{\prime
},X^{\prime }\right) \right\rangle _{\left( X^{\prime }\right) ^{\prime }}}%
+\xi ^{2}\right\} \\
&&\times \left( \frac{1+\frac{\left( \bar{\gamma}\left\langle \bar{S}%
_{E}\left( X_{1},X^{\prime }\right) \right\rangle _{X_{1}}\right) ^{2}}{%
1-\left( \bar{\gamma}\left\langle \bar{S}_{E}\left( \left( X^{\prime
}\right) ^{\prime },X^{\prime }\right) \right\rangle \right) ^{2}}}{1+\frac{%
\left( \gamma \left\langle \hat{S}_{E}\left( X_{1},X^{\prime }\right)
\right\rangle _{X_{1}}\right) ^{2}}{1-\left( \gamma \left\langle \hat{S}%
_{E}\left( X^{\prime },\left( X^{\prime }\right) ^{\prime }\right)
\right\rangle \right) ^{2}}}\right) \left( w_{1}^{\left( 0\right) B}\left(
X^{\prime },X\right) +\frac{\zeta ^{2}}{\xi ^{2}}\left( 1+\frac{\left(
\gamma \left\langle \hat{S}_{E}\left( X_{1},X^{\prime }\right) \right\rangle
_{X_{1}}\right) ^{2}}{1-\left( \gamma \left\langle \hat{S}_{E}\left(
X^{\prime },\left( X^{\prime }\right) ^{\prime }\right) \right\rangle
\right) ^{2}}\right) \right) \\
&\simeq &1+\frac{4}{\zeta ^{2}}\left\{ \bar{\zeta}^{2}\zeta ^{2}\left( 1+%
\frac{\left( \gamma \left\langle \hat{S}_{E}\right\rangle \right) ^{2}}{%
1-\left( \gamma \left\langle \hat{S}_{E}\right\rangle \right) ^{2}}\right)
+\xi ^{2}\right\} \left( \frac{1+\frac{\left( \bar{\gamma}\left\langle \bar{S%
}_{E}\left( X_{1},X^{\prime }\right) \right\rangle _{X_{1}}\right) ^{2}}{%
1-\left( \bar{\gamma}\left\langle \bar{S}_{E}\right\rangle \right) ^{2}}}{1+%
\frac{\left( \gamma \left\langle \hat{S}_{E}\left( X_{1},X^{\prime }\right)
\right\rangle _{X_{1}}\right) ^{2}}{1-\left( \gamma \left\langle \hat{S}%
_{E}\right\rangle \right) ^{2}}}\right) \\
&&\times \left( 1+\frac{\zeta ^{2}}{\xi ^{2}}\left( 1+\frac{\left( \gamma
\left\langle \hat{S}_{E}\left( X_{1},X^{\prime }\right) \right\rangle
_{X_{1}}\right) ^{2}}{1-\left( \gamma \left\langle \hat{S}_{E}\right\rangle
\right) ^{2}}\right) \right)
\end{eqnarray*}%
\begin{eqnarray*}
\left( \left\langle \hat{w}_{1}^{B}\left( X^{\prime },X\right) \right\rangle
_{X}\right) ^{-1} &=&1+\frac{\left\langle \hat{w}_{1}^{\left( 0\right)
B}\left( \left( X^{\prime }\right) ^{\prime },X^{\prime }\right)
\right\rangle _{\left( X^{\prime }\right) ^{\prime }}\frac{\zeta
^{2}\left\langle \bar{w}_{1}^{\left( 0\right) }\left( X^{\prime },X\right)
\right\rangle _{X}}{\left\langle w_{1}^{\left( 0\right) B}\left( X^{\prime
},X\right) \right\rangle _{X}}\left( \frac{1+\frac{\left( \gamma
\left\langle \hat{S}_{E}\left( X_{1},X^{\prime }\right) \right\rangle
_{X_{1}}\right) ^{2}}{1-\left( \gamma \left\langle \hat{S}_{E}\left(
X^{\prime },\left( X^{\prime }\right) ^{\prime }\right) \right\rangle
\right) ^{2}}}{1+\frac{\left( \bar{\gamma}\left\langle \bar{S}_{E}\left(
X_{1},X^{\prime }\right) \right\rangle _{X_{1}}\right) ^{2}}{1-\left( \bar{%
\gamma}\left\langle \bar{S}_{E}\left( \left( X^{\prime }\right) ^{\prime
},X^{\prime }\right) \right\rangle \right) ^{2}}}\right) }{4\left( \bar{\zeta%
}^{2}\zeta ^{2}\left( 1+\frac{\left( \gamma \left\langle \hat{S}_{E}\left(
X_{1},\left( X^{\prime }\right) ^{\prime }\right) \right\rangle \right) ^{2}%
}{1-\left( \gamma \left\langle \hat{S}_{E}\left( X^{\prime },\left(
X^{\prime }\right) ^{\prime }\right) \right\rangle \right) ^{2}}\right) +\xi
^{2}\left\langle \hat{w}_{1}^{\left( 0\right) B}\left( \left( X^{\prime
}\right) ^{\prime },X^{\prime }\right) \right\rangle _{\left( X^{\prime
}\right) ^{\prime }}\right) } \\
&&+\frac{\zeta ^{2}\left\langle \hat{w}_{1}^{\left( 0\right) B}\left( \left(
X^{\prime }\right) ^{\prime },X^{\prime }\right) \right\rangle _{\left(
X^{\prime }\right) ^{\prime }}}{\xi ^{2}\left\langle w_{1}^{\left( 0\right)
B}\left( X^{\prime },X\right) \right\rangle _{X}}\left( 1+\frac{\left(
\gamma \left\langle \hat{S}_{E}\left( X_{1},X^{\prime }\right) \right\rangle
_{X_{1}}\right) ^{2}}{1-\left( \gamma \left\langle \hat{S}_{E}\left(
X^{\prime },\left( X^{\prime }\right) ^{\prime }\right) \right\rangle
\right) ^{2}}\right) \\
&\simeq &1+\frac{\zeta ^{2}\left( \frac{1+\frac{\left( \gamma \left\langle 
\hat{S}_{E}\left( X_{1},X^{\prime }\right) \right\rangle _{X_{1}}\right) ^{2}%
}{1-\left( \gamma \left\langle \hat{S}_{E}\right\rangle \right) ^{2}}}{1+%
\frac{\left( \bar{\gamma}\left\langle \bar{S}_{E}\left( X_{1},X^{\prime
}\right) \right\rangle _{X_{1}}\right) ^{2}}{1-\left( \bar{\gamma}%
\left\langle \bar{S}_{E}\right\rangle \right) ^{2}}}\right) }{4\left( \bar{%
\zeta}^{2}\zeta ^{2}\left( 1+\frac{\left( \gamma \left\langle \hat{S}%
_{E}\right\rangle \right) ^{2}}{1-\left( \gamma \left\langle \hat{S}%
_{E}\right\rangle \right) ^{2}}\right) +\xi ^{2}\right) }+\frac{\zeta ^{2}}{%
\xi ^{2}}\left( 1+\frac{\left( \gamma \left\langle \hat{S}_{E}\left(
X_{1},X^{\prime }\right) \right\rangle _{X_{1}}\right) ^{2}}{1-\left( \gamma
\left\langle \hat{S}_{E}\right\rangle \right) ^{2}}\right)
\end{eqnarray*}%
where the various parameters are:%
\begin{eqnarray*}
&&\frac{\xi ^{2}}{\zeta ^{2}} \\
&\rightarrow &\frac{\left\langle \hat{S}_{E}^{B}\right\rangle
^{2}\left\langle \frac{1-\left( \hat{S}+\hat{S}_{E}^{B}+\hat{S}%
_{L}^{B}\right) }{1-\left( \hat{S}_{E}+\hat{S}_{E}^{B}\right) }\right\rangle
^{2}\left\langle \frac{1-\bar{S}_{E}}{1-\bar{S}}\right\rangle ^{2}}{%
\left\langle \frac{1-\hat{S}_{E}}{1-\hat{S}}S_{E}\right\rangle
^{2}\left\langle \frac{\left( 1-S\right) }{1-S_{E}}\right\rangle ^{2}} \\
&\rightarrow &\frac{\left\langle \hat{S}_{E}^{B}\right\rangle
^{2}\left\langle \frac{1-2\left( \hat{S}_{E}+\hat{S}_{E}^{B}\right) }{%
1-\left( \hat{S}_{E}+\hat{S}_{E}^{B}\right) }\right\rangle ^{2}\left\langle 
\frac{1-\bar{S}_{E}}{1-\bar{S}}\right\rangle ^{2}}{\left\langle \frac{1-\hat{%
S}_{E}}{1-\hat{S}}S_{E}\right\rangle ^{2}\left\langle \frac{\left(
1-S\right) }{1-S_{E}}\right\rangle ^{2}}
\end{eqnarray*}%
\begin{eqnarray*}
&&\bar{\zeta}^{2} \\
&\rightarrow &\left\langle S_{E}^{B}\right\rangle ^{2}\left\langle \frac{%
1-2\left( S_{E}+S_{E}^{B}\right) }{1-S_{E}-S_{E}^{B}}\right\rangle ^{2}
\end{eqnarray*}%
As before $\left\langle \bar{S}_{E}\right\rangle <<\bar{\gamma}\left\langle 
\bar{S}_{E}\right\rangle $, and $\left\langle \hat{S}_{E}\right\rangle
<<\gamma \left\langle \hat{S}_{E}\right\rangle $.%
\begin{eqnarray*}
\frac{\xi ^{2}}{\zeta ^{2}} &\rightarrow &\frac{\left\langle \hat{S}%
_{E}^{B}\right\rangle ^{2}}{\left\langle S_{E}\right\rangle ^{2}} \\
\bar{\zeta}^{2} &\rightarrow &\left\langle S_{E}^{B}\right\rangle ^{2}
\end{eqnarray*}%
\begin{eqnarray}
\bar{S}_{E}\left( X^{\prime }\right) &=&\left\langle \bar{S}_{E}\left(
X^{\prime },X\right) \right\rangle _{X}\frac{\left\langle \bar{K}%
\right\rangle \left\Vert \bar{\Psi}\right\Vert ^{2}}{\left\langle \bar{K}%
_{X^{\prime }}\right\rangle \left\vert \bar{\Psi}\left( X^{\prime }\right)
\right\vert ^{2}}  \label{SG} \\
\bar{S}_{L}\left( X\right) &=&\left\langle \bar{S}_{L}\left( X^{\prime
},X\right) \right\rangle _{X}\frac{\left\langle \bar{K}\right\rangle
\left\Vert \bar{\Psi}\right\Vert ^{2}}{\left\langle \bar{K}_{X^{\prime
}}\right\rangle \left\vert \bar{\Psi}\left( X^{\prime }\right) \right\vert
^{2}}  \notag \\
\bar{S}\left( X^{\prime },X\right) &=&\left\langle \bar{S}\left( X^{\prime
}\right) \right\rangle _{X}\frac{\left\langle \bar{K}\right\rangle
\left\Vert \bar{\Psi}\right\Vert ^{2}}{\left\langle \bar{K}_{X^{\prime
}}\right\rangle \left\vert \bar{\Psi}\left( X^{\prime }\right) \right\vert
^{2}}  \notag
\end{eqnarray}%
where $\frac{\left\langle \bar{K}\right\rangle \left\Vert \bar{\Psi}%
\right\Vert ^{2}}{\bar{K}_{X}\left\vert \bar{\Psi}\left( X\right)
\right\vert ^{2}}$ is given by (\ref{Rtb}).

Formula for $\left\langle \hat{w}_{E}^{B}\left( X\right) \right\rangle $ and 
$\left\langle w_{E}^{B}\left( X\right) \right\rangle $ given in (\ref{hbz})
and (\ref{hbzt}):%
\begin{eqnarray*}
&&\left\langle w_{E}^{B}\left( X\right) \right\rangle \\
&\rightarrow &\frac{1+\left( 1-\left( \bar{\gamma}\left\langle \bar{S}%
_{E}\right\rangle \right) ^{2}\right) \left( 4+3\left( 1-\left( \bar{\gamma}%
\left\langle \bar{S}_{E}\right\rangle \right) ^{2}\right) -\frac{%
\left\langle \hat{S}_{E}^{B}\right\rangle ^{2}}{\left\langle
S_{E}\right\rangle ^{2}}\left( 1-\left( \gamma \left\langle \hat{S}%
_{E}\right\rangle \right) ^{2}\right) \right) }{\left( 1+4\left( 1-\left( 
\bar{\gamma}\left\langle \bar{S}_{E}\right\rangle \right) ^{2}\right)
\right) \left( \frac{\left\langle \hat{S}_{E}^{B}\right\rangle ^{2}}{%
\left\langle S_{E}\right\rangle ^{2}}\left( 1-\left( \gamma \left\langle 
\hat{S}_{E}\right\rangle \right) ^{2}\right) +\left( 1-\left( \bar{\gamma}%
\left\langle \bar{S}_{E}\right\rangle \right) ^{2}\right) +1\right) }
\end{eqnarray*}

\subsection*{A9.8 Formula for $\left\langle \hat{S}_{E}^{B}\left( X^{\prime
},X\right) \right\rangle _{X}$, $\hat{S}_{E}^{B}\left( X^{\prime }\right) $
and $S_{E}^{B}\left( X,X\right) $, $S_{E}^{B}\left( X\right) $}

The (non-normalized) aggregate stakes are given by:%
\begin{eqnarray}
&&\left\langle \hat{S}_{E}^{B}\left( X^{\prime },X\right) \right\rangle _{X}
\label{snB} \\
&=&\left\langle \hat{w}_{1}^{B}\left( X^{\prime },X\right) \right\rangle _{X}%
\left[ 1+\left\langle \bar{w}\left( X\right) \right\rangle \left( \hat{f}%
\left( X^{\prime }\right) -\frac{\left\langle \bar{f}\left( X^{\prime
}\right) \right\rangle _{\bar{w}_{1}}+\left\langle \bar{r}\left( X^{\prime
}\right) \right\rangle _{\bar{w}_{2}}}{2}\right) \right.  \notag \\
&&\left. +\left\langle \hat{w}_{1}^{B}\left( X\right) \right\rangle \left( 
\hat{f}\left( X^{\prime }\right) -\left\langle \hat{f}\left( X^{\prime
}\right) \right\rangle _{\hat{w}_{1}}\right) +\left\langle w_{1}^{B}\left(
X\right) \right\rangle \left( \hat{f}\left( X^{\prime }\right) -f\left(
X\right) \right) \right]  \notag
\end{eqnarray}

\begin{eqnarray*}
&&S_{E}^{B}\left( X,X\right) \\
&=&w_{1}^{B}\left( X\right) \left\{ 1+\left\langle \bar{w}\left( X\right)
\right\rangle \left( f\left( X\right) -\frac{\left\langle \bar{f}\left(
X^{\prime }\right) \right\rangle _{\bar{w}_{1}}+\left\langle \bar{r}\left(
X^{\prime }\right) \right\rangle _{\bar{w}_{2}}}{2}\right) +\left\langle 
\hat{w}_{1}^{B}\left( X\right) \right\rangle \left( f\left( X\right)
-\left\langle \hat{f}\left( X^{\prime }\right) \right\rangle _{\hat{w}%
_{1}}\right) \right\}
\end{eqnarray*}%
where the coefficients are:%
\begin{eqnarray*}
w_{1}^{B}\left( X\right) &=&\left\langle w_{1}^{B}\left( X\right)
\right\rangle \\
&\rightarrow &\frac{1+\left( 1-\left( \bar{\gamma}\left\langle \bar{S}%
_{E}\right\rangle \right) ^{2}\right) \left( 4+3\left( 1-\left( \bar{\gamma}%
\left\langle \bar{S}_{E}\right\rangle \right) ^{2}\right) -\frac{%
\left\langle \hat{S}_{E}^{B}\right\rangle ^{2}}{\left\langle
S_{E}\right\rangle ^{2}}\left( 1-\left( \gamma \left\langle \hat{S}%
_{E}\right\rangle \right) ^{2}\right) \right) }{\left( 1+4\left( 1-\left( 
\bar{\gamma}\left\langle \bar{S}_{E}\right\rangle \right) ^{2}\right)
\right) \left( \frac{\left\langle \hat{S}_{E}^{B}\right\rangle ^{2}}{%
\left\langle S_{E}\right\rangle ^{2}}\left( 1-\left( \gamma \left\langle 
\hat{S}_{E}\right\rangle \right) ^{2}\right) +\left( 1-\left( \bar{\gamma}%
\left\langle \bar{S}_{E}\right\rangle \right) ^{2}\right) +1\right) }
\end{eqnarray*}%
The normalized agregate stakes write:%
\begin{eqnarray*}
\hat{S}_{E}^{B}\left( X^{\prime }\right) &=&\hat{S}_{E}^{B}\left( X^{\prime
},X\right) \frac{\left\langle \bar{K}\right\rangle \left\Vert \bar{\Psi}%
\right\Vert ^{2}}{\hat{K}_{X^{\prime }}\left\vert \hat{\Psi}\left( X^{\prime
}\right) \right\vert ^{2}} \\
S_{E}^{B}\left( X\right) &=&S_{E}^{B}\left( X,X\right) \frac{\left\langle 
\bar{K}_{X}\right\rangle \left\vert \bar{\Psi}\left( X\right) \right\vert
^{2}}{K_{X}\left\vert \Psi \left( X\right) \right\vert ^{2}}
\end{eqnarray*}

\subsection*{A9.9 Formula for $\hat{S}_{L}^{B}\left( X^{\prime }\right) $
and $S_{L}^{B}\left( X^{\prime }\right) $}

\begin{equation}
\frac{\left\langle \hat{S}_{L}^{B}\left( X^{\prime },X\right) \right\rangle
_{X}}{\kappa \left( 1-\bar{S}\left( X\right) \right) }=\left\langle \hat{w}%
_{E}^{B}\left( X^{\prime },X\right) \right\rangle _{X}\left\{ 1+\left\langle 
\hat{w}_{2}^{B}\left( X\right) \right\rangle \left( \hat{r}\left( X^{\prime
}\right) -\left\langle \hat{f}\left( X^{\prime }\right) \right\rangle _{\hat{%
w}_{1}}\right) +\left\langle w_{E}^{B}\left( X\right) \right\rangle \left( 
\hat{r}\left( X^{\prime }\right) -f\left( X\right) \right) \right\}
\label{frl}
\end{equation}%
\begin{eqnarray*}
\left\langle \hat{w}_{2}^{B}\left( X^{\prime },X\right) \right\rangle _{X}
&\simeq &\left\langle \hat{w}_{2}\left( X^{\prime },X\right) \right\rangle
_{X} \\
&\simeq &\left\langle \hat{w}_{1}\left( X^{\prime },X\right) \right\rangle
_{X} \\
&\rightarrow &\frac{\left( 1-\left( \gamma \left\langle \hat{S}_{E}\left(
X\right) \right\rangle \right) ^{2}\right) }{2-\left( \gamma \left\langle 
\hat{S}_{E}\left( X\right) \right\rangle \right) ^{2}+\left( \gamma
\left\langle \hat{S}_{E}\left( X_{1},X^{\prime }\right) \right\rangle
_{X_{1}}\right) ^{2}-\left( \gamma \left\langle \hat{S}_{E}\left( X\right)
\right\rangle \right) ^{2}}
\end{eqnarray*}%
\begin{equation*}
\frac{S_{L}^{B}\left( X,X\right) }{\kappa \left( 1-\bar{S}\left( X\right)
\right) }=w_{2}^{B}\left( X\right) \left[ 1+\hat{w}_{2}^{B}\left( X\right)
\left( r\left( X\right) -\left\langle \hat{r}\left( X^{\prime }\right)
\right\rangle _{\hat{w}_{2}}\right) \right]
\end{equation*}%
\begin{eqnarray*}
w_{2}^{B}\left( X\right) &\simeq &\left\langle w_{2}^{B}\left( X\right)
\right\rangle \\
\hat{w}_{2}^{B}\left( X\right) &\simeq &1-w_{2}^{B}\left( X\right)
=1-\left\langle w_{2}^{B}\left( X\right) \right\rangle
\end{eqnarray*}%
\begin{equation}
\frac{S_{L}^{B}\left( X,X\right) }{\kappa \left( 1-\left\langle \bar{S}%
\left( X\right) \right\rangle \right) }=\frac{1}{2-\left( \gamma
\left\langle \hat{S}_{E}\left( X^{\prime },X\right) \right\rangle \right)
^{2}}\left[ 1+\frac{1-\left( \gamma \left\langle \hat{S}_{E}\left( X^{\prime
},X\right) \right\rangle \right) ^{2}}{2-\left( \gamma \left\langle \hat{S}%
_{E}\left( X^{\prime },X\right) \right\rangle \right) ^{2}}\left( r\left(
X\right) -\left\langle \hat{r}\left( X^{\prime }\right) \right\rangle _{\hat{%
w}_{2}}\right) \right]  \label{sfB}
\end{equation}%
\begin{eqnarray*}
\hat{S}_{L}^{B}\left( X^{\prime }\right) &=&\left\langle \hat{S}%
_{L}^{B}\left( X^{\prime },X\right) \right\rangle _{X}\frac{\left\langle 
\bar{K}\right\rangle \left\Vert \bar{\Psi}\right\Vert ^{2}}{\hat{K}%
_{X^{\prime }}\left\vert \hat{\Psi}\left( X^{\prime }\right) \right\vert ^{2}%
} \\
S_{L}^{B}\left( X\right) &=&S_{L}^{B}\left( X,X\right) \frac{\left\langle 
\bar{K}_{X}\right\rangle \left\vert \bar{\Psi}\left( X\right) \right\vert
^{2}}{K_{X}\left\vert \Psi \left( X\right) \right\vert ^{2}}
\end{eqnarray*}%
\bigskip

\section*{Appendix 10 Equations for returns in terms of $\hat{f}\left(
X^{\prime }\right) $, $\bar{f}\left( X^{\prime }\right) $}

As before we can rewrite the equations for returns in term of $\hat{f}\left(
X^{\prime }\right) $, $\bar{f}\left( X^{\prime }\right) $ only. We can
retrieve $\hat{S}_{E}\left( X\right) $ and $\bar{S}_{E}\left( X\right) $\
using (\ref{seh}) (\ref{SB}), (\ref{SG}) and (\ref{Sn}). Starting from these
equations:%
\begin{eqnarray*}
0 &=&\int \left( \Delta \left( X,X^{\prime }\right) -\hat{S}_{E}\left(
X^{\prime },X\right) \right) \frac{1-\hat{S}\left( X^{\prime }\right) }{1-%
\hat{S}_{E}\left( X^{\prime }\right) }\left( \hat{f}\left( X^{\prime
}\right) -\bar{r}\right) dX^{\prime } \\
&&-\int S_{E}\left( X^{\prime },X\right) \frac{1-\left( S\left( X^{\prime
}\right) +\left( S_{E}^{B}\left( X^{\prime }\right) +S_{L}^{B}\left(
X^{\prime }\right) \right) \right) }{1-S_{E}\left( X^{\prime }\right)
-S_{E}^{B}\left( X^{\prime }\right) }\left( \left( f_{1}^{\prime }\left(
X^{\prime }\right) -\bar{r}\right) +\Delta F_{\tau }\left( \bar{R}\left(
K,X\right) \right) \right) dX^{\prime }
\end{eqnarray*}

\begin{eqnarray}
0 &=&\left( 1-\bar{S}_{E}\left( X^{\prime },X\right) \right) \left( \bar{f}%
\left( X^{\prime }\right) -\left( 1+\kappa \right) \bar{r}\right) \frac{1-%
\bar{S}\left( X^{\prime }\right) }{1-\bar{S}_{E}\left( X^{\prime }\right) }
\\
&&-\hat{S}_{E}^{B}\left( X^{\prime },X\right) \left( \hat{f}\left( X^{\prime
}\right) -\bar{r}\right) \frac{1-\left( \hat{S}\left( X^{\prime }\right) +%
\hat{S}_{E}^{B}\left( X^{\prime }\right) +\hat{S}_{L}^{B}\left( X^{\prime
}\right) \right) }{1-\left( \hat{S}_{E}\left( X^{\prime }\right) +\hat{S}%
_{E}^{B}\left( X^{\prime }\right) \right) }  \notag \\
&&-S_{E}^{B}\left( X^{\prime },X\right) \left\{ \frac{1-\left( S\left(
X^{\prime }\right) +\left( S_{E}^{B}\left( X^{\prime }\right)
+S_{L}^{B}\left( X^{\prime }\right) \right) \right) }{1-S_{E}\left(
X^{\prime }\right) -S_{E}^{B}\left( X^{\prime }\right) }\left( \left(
f_{1}^{\prime }\left( X^{\prime }\right) -\bar{r}\right) +\Delta F_{\tau
}\left( \bar{R}\left( K,X\right) \right) \right) \right\}  \notag
\end{eqnarray}%
averaged over $X^{\prime }$ and $X^{\prime }$:%
\begin{eqnarray*}
0 &=&\frac{1-\hat{S}\left( X\right) }{1-\hat{S}_{E}\left( X\right) }\left( 
\hat{f}\left( X\right) -\bar{r}\right) -\left\langle \hat{S}_{E}\left(
X^{\prime },X\right) \right\rangle _{X^{\prime }}\frac{1-\left\langle \hat{S}%
\left( X^{\prime }\right) \right\rangle }{1-\left\langle \hat{S}_{E}\left(
X^{\prime }\right) \right\rangle }\left( \left\langle \hat{f}\left(
X^{\prime }\right) \right\rangle -\left\langle \bar{r}\right\rangle \right)
\\
&&-S_{E}\left( X,X\right) \frac{1-\left( S\left( X\right) +\left(
S_{E}^{B}\left( X\right) +S_{L}^{B}\left( X\right) \right) \right) }{%
1-S_{E}\left( X\right) -S_{E}^{B}\left( X\right) }\left( \left(
f_{1}^{\prime }\left( X\right) -\bar{r}\right) +\Delta F_{\tau }\left( \bar{R%
}\left( K,X\right) \right) \right)
\end{eqnarray*}%
\begin{eqnarray*}
0 &=&\frac{1-\hat{S}\left( X\right) }{1-\hat{S}_{E}\left( X\right) }\left( 
\hat{f}\left( X^{\prime }\right) -\bar{r}\right) -\left\langle \hat{S}%
_{E}\left( X^{\prime },X\right) \right\rangle _{X^{\prime }}\frac{%
1-\left\langle \hat{S}\left( X^{\prime }\right) \right\rangle }{%
1-\left\langle \hat{S}_{E}\left( X^{\prime }\right) \right\rangle }\left(
\left\langle \hat{f}\left( X^{\prime }\right) \right\rangle -\left\langle 
\bar{r}\right\rangle \right) \\
&&-S_{E}\left( X,X\right) \left( f_{1}\left( X\right) -\bar{r}\right)
\end{eqnarray*}%
\begin{eqnarray}
0 &=&\frac{1-\bar{S}\left( X\right) }{1-\bar{S}_{E}\left( X\right) }\left( 
\bar{f}\left( X\right) -\left( 1+\kappa \right) \bar{r}\right) -\left\langle 
\bar{S}_{E}\left( X^{\prime },X\right) \right\rangle _{X^{\prime }}\frac{%
1-\left\langle \bar{S}\left( X^{\prime }\right) \right\rangle }{%
1-\left\langle \bar{S}_{E}\left( X^{\prime }\right) \right\rangle }\left(
\left\langle \bar{f}\left( X^{\prime }\right) \right\rangle -\left( 1+\kappa
\right) \bar{r}\right) \\
&&-\left\langle \hat{S}_{E}^{B}\left( X^{\prime },X\right) \right\rangle
_{X^{\prime }}\frac{1-\left\langle \hat{S}\left( X^{\prime }\right)
\right\rangle -\left\langle \hat{S}_{E}^{B}\left( X^{\prime }\right)
\right\rangle -\left\langle \hat{S}_{L}^{B}\left( X^{\prime }\right)
\right\rangle }{1-\left\langle \hat{S}_{E}\left( X^{\prime }\right)
\right\rangle -\left\langle \hat{S}_{E}^{B}\left( X^{\prime }\right)
\right\rangle }\left( \left\langle \hat{f}\left( X^{\prime }\right)
\right\rangle -\bar{r}\right) -S_{E}^{B}\left( X,X\right) \left( f_{1}\left(
X\right) -\bar{r}\right)  \notag
\end{eqnarray}%
These are equations for $\left\langle \hat{S}_{E}\left( X^{\prime },X\right)
\right\rangle _{X^{\prime }}$ and $\left\langle \bar{S}_{E}\left( X^{\prime
},X\right) \right\rangle _{X^{\prime }}$. Once solved, this leads to find $%
\bar{S}_{E}\left( X\right) $.%
\begin{equation*}
\Delta F_{\tau }\left( \bar{R}\left( K,X\right) \right) =\tau \left(
f_{1}\left( X\right) -\left\langle f_{1}\left( X^{\prime }\right)
\right\rangle \right)
\end{equation*}

\subsection*{A10.1 Solving for investors}

The equations are similar to part one, the expression involved being
different due to the presence of banks. We have:

\begin{eqnarray*}
&&\left( 1-\left\langle \hat{S}\right\rangle \right) \left( \hat{f}\left(
X\right) +\frac{\left\langle \hat{S}_{E}^{B}\left( X,X^{\prime }\right)
\right\rangle _{X^{\prime }}+\left\langle \hat{S}_{L}^{B}\left( X,X^{\prime
}\right) \right\rangle _{X^{\prime }}\left\langle \bar{S}\right\rangle }{%
1-\left\langle \bar{S}_{E}\right\rangle }\left\langle \bar{f}\right\rangle
\right) \\
&&+\left\langle \hat{S}\left( X^{\prime },X\right) \right\rangle _{X^{\prime
}}\left( \left\langle \hat{f}\right\rangle +\frac{\left\langle \hat{S}%
_{E}^{B}\right\rangle +\left\langle \hat{S}_{L}^{B}\right\rangle
\left\langle \bar{S}\right\rangle }{1-\left\langle \bar{S}_{E}\right\rangle }%
\left\langle \bar{f}\right\rangle \right) \\
&=&\left( 1-\left\langle \hat{S}\right\rangle \right) \left( \hat{f}\left(
X\right) \left( 1+\frac{\left\langle \hat{w}_{1}^{B}\right\rangle }{%
1-\left\langle \bar{S}_{E}\right\rangle }\left\langle \bar{f}\right\rangle
\right) +\frac{\left\langle \hat{S}_{E,0}^{B}\left( X,X\right) \right\rangle
+\left\langle \hat{S}_{L}^{B}\left( X,X\right) \right\rangle
_{X}\left\langle \bar{S}\right\rangle }{1-\left\langle \bar{S}%
_{E}\right\rangle }\left\langle \bar{f}\right\rangle \right) \\
&&+\left\langle \hat{S}\left( X^{\prime },X\right) \right\rangle _{X^{\prime
}}\left( \left\langle \hat{f}\right\rangle +\frac{\left\langle \hat{S}%
_{E}^{B}\right\rangle +\left\langle \hat{S}_{L}^{B}\right\rangle
\left\langle \bar{S}\right\rangle }{1-\left\langle \bar{S}_{E}\right\rangle }%
\left\langle \bar{f}\right\rangle \right)
\end{eqnarray*}%
with:%
\begin{eqnarray}
&&\left\langle \hat{S}_{E}^{B}\left( X,X\right) \right\rangle _{X} \\
&=&\left\langle \hat{w}_{1}^{B}\left( X,X\right) \right\rangle _{X}\left[
1+\left\langle \bar{w}\left( X\right) \right\rangle \left( \hat{f}\left(
X\right) -\frac{\left\langle \bar{f}\left( X^{\prime }\right) \right\rangle
_{\bar{w}_{1}}+\left\langle \bar{r}\left( X^{\prime }\right) \right\rangle _{%
\bar{w}_{2}}}{2}\right) \right.  \notag \\
&&\left. +\left\langle \hat{w}_{1}^{B}\left( X\right) \right\rangle \left( 
\hat{f}\left( X\right) -\left\langle \hat{f}\left( X^{\prime }\right)
\right\rangle _{\hat{w}_{1}}\right) +\left\langle w_{1}^{B}\left( X\right)
\right\rangle \left( \hat{f}\left( X\right) -f\left( X\right) \right) \right]
\notag \\
&=&\left\langle \hat{w}_{1}^{B}\left( X,X\right) \right\rangle _{X}\hat{f}%
\left( X\right) +\left\langle \hat{S}_{E,0}^{B}\left( X,X\right)
\right\rangle  \notag
\end{eqnarray}%
and:%
\begin{eqnarray}
&&\left\langle \hat{S}_{E,0}^{B}\left( X,X\right) \right\rangle _{X} \\
&=&\left\langle \hat{w}_{1}^{B}\left( X,X\right) \right\rangle _{X}\left[
1-\left\langle \bar{w}\left( X\right) \right\rangle \left( \frac{%
\left\langle \bar{f}\left( X^{\prime }\right) \right\rangle _{\bar{w}%
_{1}}+\left\langle \bar{r}\left( X^{\prime }\right) \right\rangle _{\bar{w}%
_{2}}}{2}\right) -\left\langle \hat{w}_{1}^{B}\left( X\right) \right\rangle
\left\langle \hat{f}\left( X^{\prime }\right) \right\rangle _{\hat{w}%
_{1}}-\left\langle w_{1}^{B}\left( X\right) \right\rangle f\left( X\right) %
\right]  \notag
\end{eqnarray}%
with (\ref{snB}) and (\ref{frl}):%
\begin{equation*}
\frac{\left\langle \hat{S}_{L}^{B}\left( X^{\prime },X\right) \right\rangle
_{X}}{\kappa \left( 1-\bar{S}\left( X\right) \right) }=\left\langle \hat{w}%
_{E}^{B}\left( X^{\prime },X\right) \right\rangle _{X}\left\{ 1+\left\langle 
\hat{w}_{E}^{B}\left( X\right) \right\rangle \left( \hat{r}\left( X^{\prime
}\right) -\left\langle \hat{f}\left( X^{\prime }\right) \right\rangle _{\hat{%
w}_{E}}\right) +\left\langle w_{E}^{B}\left( X\right) \right\rangle \left( 
\hat{r}\left( X^{\prime }\right) -f\left( X\right) \right) \right\}
\end{equation*}%
\begin{eqnarray*}
\left\langle \hat{w}_{2}^{B}\left( X^{\prime },X\right) \right\rangle _{X}
&\simeq &\left\langle \hat{w}_{E}\left( X^{\prime },X\right) \right\rangle
_{X} \\
&\simeq &\left\langle \hat{w}_{E}\left( X^{\prime },X\right) \right\rangle
_{X} \\
&\rightarrow &\frac{\left( 1-\left( \gamma \left\langle \hat{S}_{E}\left(
X\right) \right\rangle \right) ^{2}\right) }{2-\left( \gamma \left\langle 
\hat{S}_{E}\left( X\right) \right\rangle \right) ^{2}+\left( \gamma
\left\langle \hat{S}_{E}\left( X_{1},X^{\prime }\right) \right\rangle
_{X_{1}}\right) ^{2}-\left( \gamma \left\langle \hat{S}_{E}\left( X\right)
\right\rangle \right) ^{2}}
\end{eqnarray*}%
The equation for investors return is the same as in part 1:%
\begin{eqnarray}
&&0=\frac{1-\frac{\left( 1-\left( \gamma \left\langle \hat{S}_{E}\left(
X\right) \right\rangle \right) ^{2}\right) \left( 1+\frac{\Delta \hat{f}%
\left( X\right) +\Delta \hat{r}\left( X\right) }{2}\right) }{2-\left( \gamma
\left\langle \hat{S}_{E}\left( X\right) \right\rangle \right) ^{2}+\left(
\gamma \left\langle \hat{S}_{E}\left( X_{1},X\right) \right\rangle
_{X_{1}}\right) ^{2}-\left( \gamma \left\langle \hat{S}_{E}\left( X\right)
\right\rangle \right) ^{2}}\frac{\left\langle \hat{K}\right\rangle
\left\Vert \hat{\Psi}\right\Vert ^{2}}{\hat{K}_{X}\left\vert \hat{\Psi}%
\left( X\right) \right\vert ^{2}}}{1-\frac{1}{2}\frac{\left( 1-\left( \gamma
\left\langle \hat{S}_{E}\left( X\right) \right\rangle \right) ^{2}\right)
\left( 1+\Delta \hat{f}\left( X\right) \right) }{2-\left( \gamma
\left\langle \hat{S}_{E}\left( X\right) \right\rangle \right) ^{2}+\left(
\gamma \left\langle \hat{S}_{E}\left( X_{1},X\right) \right\rangle
_{X_{1}}\right) ^{2}-\left( \gamma \left\langle \hat{S}_{E}\left( X\right)
\right\rangle \right) ^{2}}\frac{\left\langle \hat{K}\right\rangle
\left\Vert \hat{\Psi}\right\Vert ^{2}}{\hat{K}_{X}\left\vert \hat{\Psi}%
\left( X\right) \right\vert ^{2}}}\left( \hat{f}\left( X\right) -\bar{r}%
\right)  \label{NVT} \\
&&-\frac{\left( 1-\left( \gamma \left\langle \hat{S}_{E}\left( X\right)
\right\rangle \right) ^{2}\right) }{2-\left( \gamma \left\langle \hat{S}%
_{E}\left( X\right) \right\rangle \right) ^{2}}\left( 1-\frac{\Delta \left( 
\frac{f\left( X\right) +r\left( X\right) }{2}\right) }{2-\left( \gamma
\left\langle \hat{S}_{E}\left( X\right) \right\rangle \right) ^{2}}+\frac{%
\left\langle \hat{f}\left( X^{\prime }\right) \right\rangle -\left\langle 
\hat{r}\left( X^{\prime }\right) \right\rangle _{\hat{w}_{2}}}{2}\right) 
\notag \\
&&\times \frac{1-\left\langle \hat{S}\left( X^{\prime }\right) \right\rangle 
}{1-\left\langle \hat{S}_{E}\left( X^{\prime }\right) \right\rangle }\left(
\left\langle \hat{f}\left( X^{\prime }\right) \right\rangle -\left\langle 
\bar{r}\right\rangle \right) -S_{E}\left( X,X\right) \left( f\left( X\right)
-r\right)  \notag
\end{eqnarray}%
but with different formulas for capital ratio. We use the first order proxy: 
\begin{eqnarray}
&&\frac{\left\langle \hat{K}\right\rangle \left\Vert \hat{\Psi}\right\Vert
^{2}}{\hat{K}_{X}\left\vert \hat{\Psi}\left( X\right) \right\vert ^{2}}%
\simeq \left( \frac{\hat{g}\left( X^{\prime }\right) \frac{\left\langle \hat{%
S}\left( X^{\prime },X\right) \right\rangle }{1-\left\langle \hat{S}\left(
X^{\prime },X\right) \right\rangle }}{\left\langle \hat{g}\right\rangle 
\frac{\left\langle \hat{S}\left( X^{\prime },X\right) \right\rangle _{X}}{%
1-\left\langle \hat{S}\left( X^{\prime },X\right) \right\rangle _{X}\frac{%
\left\langle \hat{K}\right\rangle \left\Vert \hat{\Psi}\right\Vert ^{2}}{%
\hat{K}_{X}\left\vert \hat{\Psi}\left( X\right) \right\vert ^{2}}}}\right)
^{2}  \label{RTC} \\
&\simeq &\frac{\left( \left( \hat{f}\left( X\right) +\frac{\left\langle \hat{%
S}\left( X^{\prime },X\right) \right\rangle _{X^{\prime }}\left\langle \hat{f%
}\right\rangle }{1-\left\langle \hat{S}\right\rangle }+\frac{\left(
\left\langle \hat{S}_{E}^{B}\left( X,X^{\prime }\right) \right\rangle
_{X^{\prime }}+\left\langle \hat{S}_{L}^{B}\left( X,X^{\prime }\right)
\right\rangle _{X^{\prime }}+\left\langle \hat{S}\left( X^{\prime },X\right)
\right\rangle _{X^{\prime }}\frac{\left\langle \hat{S}_{E}^{B}\right\rangle
+\left\langle \hat{S}_{L}^{B}\right\rangle }{1-\left\langle \hat{S}%
\right\rangle }\right) \frac{\left\langle \bar{K}\right\rangle \left\Vert 
\bar{\Psi}\right\Vert ^{2}}{\left\langle \hat{K}\right\rangle \left\Vert 
\hat{\Psi}\right\Vert ^{2}}\left\langle \bar{f}\right\rangle }{%
1-\left\langle \bar{S}\right\rangle }\right) \frac{\left\langle \hat{S}%
\left( X^{\prime },X\right) \right\rangle }{1-\left\langle \hat{S}\left(
X^{\prime },X\right) \right\rangle }\right) ^{2}}{\left( \left( \left\langle 
\hat{f}\right\rangle +\frac{\left\langle \hat{S}_{E}^{B}\right\rangle
+\left\langle \hat{S}_{L}^{B}\right\rangle }{1-\bar{S}}\frac{\left\langle 
\bar{K}\right\rangle \left\Vert \bar{\Psi}\right\Vert ^{2}}{\left\langle 
\hat{K}\right\rangle \left\Vert \hat{\Psi}\right\Vert ^{2}}\left\langle \bar{%
f}\right\rangle \right) \frac{\left\langle \hat{S}\left( X^{\prime
},X\right) \right\rangle _{X}}{1-\left\langle \hat{S}\left( X^{\prime
},X\right) \right\rangle _{X}\frac{\left\langle \hat{K}\right\rangle
\left\Vert \hat{\Psi}\right\Vert ^{2}}{\hat{K}_{X}\left\vert \hat{\Psi}%
\left( X\right) \right\vert ^{2}}}\right) ^{2}}  \notag
\end{eqnarray}%
where $\frac{\left\langle \hat{K}\right\rangle \left\Vert \hat{\Psi}%
\right\Vert ^{2}}{\hat{K}_{X}\left\vert \hat{\Psi}\left( X\right)
\right\vert ^{2}}$ is replaced by its zeroth order approximation:%
\begin{equation*}
\frac{\left( \left( \hat{f}\left( X\right) +\frac{\left\langle \hat{S}\left(
X^{\prime },X\right) \right\rangle _{X^{\prime }}\left\langle \hat{f}%
\right\rangle }{1-\left\langle \hat{S}\right\rangle }+\frac{\left(
\left\langle \hat{S}_{E}^{B}\left( X,X^{\prime }\right) \right\rangle
_{X^{\prime }}+\left\langle \hat{S}_{L}^{B}\left( X,X^{\prime }\right)
\right\rangle _{X^{\prime }}+\left\langle \hat{S}\left( X^{\prime },X\right)
\right\rangle _{X^{\prime }}\frac{\left\langle \hat{S}_{E}^{B}\right\rangle
+\left\langle \hat{S}_{L}^{B}\right\rangle }{1-\left\langle \hat{S}%
\right\rangle }\right) \frac{\left\langle \bar{K}\right\rangle \left\Vert 
\bar{\Psi}\right\Vert ^{2}}{\left\langle \hat{K}\right\rangle \left\Vert 
\hat{\Psi}\right\Vert ^{2}}\left\langle \bar{f}\right\rangle }{%
1-\left\langle \bar{S}\right\rangle }\right) \frac{\left\langle \hat{S}%
\left( X^{\prime },X\right) \right\rangle }{1-\left\langle \hat{S}\left(
X^{\prime },X\right) \right\rangle }\right) ^{2}}{\left( \left( \left\langle 
\hat{f}\right\rangle +\frac{\left\langle \hat{S}_{E}^{B}\right\rangle
+\left\langle \hat{S}_{L}^{B}\right\rangle }{1-\bar{S}}\frac{\left\langle 
\bar{K}\right\rangle \left\Vert \bar{\Psi}\right\Vert ^{2}}{\left\langle 
\hat{K}\right\rangle \left\Vert \hat{\Psi}\right\Vert ^{2}}\left\langle \bar{%
f}\right\rangle \right) \frac{\left\langle \hat{S}\left( X^{\prime
},X\right) \right\rangle _{X}}{1-\left\langle \hat{S}\left( X^{\prime
},X\right) \right\rangle _{X}}\right) ^{2}}
\end{equation*}

\bigskip

and this leads to the equation:

\begin{equation*}
\hat{G}=\frac{\left( \left\langle \hat{f}\right\rangle +\frac{\left\langle 
\hat{S}_{E}^{B}\right\rangle +\left\langle \hat{S}_{L}^{B}\right\rangle }{1-%
\bar{S}}\frac{\left\langle \bar{K}\right\rangle \left\Vert \bar{\Psi}%
\right\Vert ^{2}}{\left\langle \hat{K}\right\rangle \left\Vert \hat{\Psi}%
\right\Vert ^{2}}\left\langle \bar{f}\right\rangle \right) ^{2}\left( \frac{%
\frac{\left\langle \hat{S}\left( X^{\prime },X\right) \right\rangle _{X}}{%
1-\left\langle \hat{S}\left( X^{\prime },X\right) \right\rangle _{X}\frac{%
\left\langle \hat{K}\right\rangle \left\Vert \hat{\Psi}\right\Vert ^{2}}{%
\hat{K}_{X}\left\vert \hat{\Psi}\left( X\right) \right\vert ^{2}}}}{\frac{%
\left\langle \hat{S}\left( X^{\prime },X\right) \right\rangle }{%
1-\left\langle \hat{S}\left( X^{\prime },X\right) \right\rangle }}\right)
^{2}-A\left( 1+\frac{\Delta \hat{f}\left( X\right) +\Delta \hat{r}\left(
X\right) }{2}\right) F_{1}^{2}}{\left( \left\langle \hat{f}\right\rangle +%
\frac{\left\langle \hat{S}_{E}^{B}\right\rangle +\left\langle \hat{S}%
_{L}^{B}\right\rangle }{1-\bar{S}}\frac{\left\langle \bar{K}\right\rangle
\left\Vert \bar{\Psi}\right\Vert ^{2}}{\left\langle \hat{K}\right\rangle
\left\Vert \hat{\Psi}\right\Vert ^{2}}\left\langle \bar{f}\right\rangle
\right) ^{2}\left( \frac{\frac{\left\langle \hat{S}\left( X^{\prime
},X\right) \right\rangle _{X}}{1-\left\langle \hat{S}\left( X^{\prime
},X\right) \right\rangle _{X}\frac{\left\langle \hat{K}\right\rangle
\left\Vert \hat{\Psi}\right\Vert ^{2}}{\hat{K}_{X}\left\vert \hat{\Psi}%
\left( X\right) \right\vert ^{2}}}}{\frac{\left\langle \hat{S}\left(
X^{\prime },X\right) \right\rangle }{1-\left\langle \hat{S}\left( X^{\prime
},X\right) \right\rangle }}\right) ^{2}-\frac{1}{2}A\left( 1+\Delta \hat{f}%
\left( X\right) \right) F_{1}^{2}}\left( \hat{f}\left( X\right) -\bar{r}%
\right)
\end{equation*}%
with:%
\begin{eqnarray}
F_{1} &=&\left( 1-\hat{S}\right) \hat{f}\left( X\right) +\left\langle \hat{S}%
\left( X^{\prime },X\right) \right\rangle _{X^{\prime }}\left\langle \hat{f}%
\right\rangle  \label{Dfn} \\
&&+\left( \left\langle \hat{S}_{E}^{B}\left( X,X^{\prime }\right)
\right\rangle _{X^{\prime }}+\left\langle \hat{S}_{L}^{B}\left( X,X^{\prime
}\right) \right\rangle _{X^{\prime }}+\left\langle \hat{S}\left( X^{\prime
},X\right) \right\rangle _{X^{\prime }}\frac{\left\langle \hat{S}%
_{E}^{B}\right\rangle +\left\langle \hat{S}_{L}^{B}\right\rangle }{%
1-\left\langle \hat{S}\right\rangle }\right) \frac{\left\langle \bar{K}%
\right\rangle \left\Vert \bar{\Psi}\right\Vert ^{2}}{\left\langle \hat{K}%
\right\rangle \left\Vert \hat{\Psi}\right\Vert ^{2}}\left\langle \bar{f}%
\right\rangle  \notag
\end{eqnarray}%
\begin{equation*}
A=\frac{\left( 1-\left( \gamma \left\langle \hat{S}_{E}\left( X\right)
\right\rangle \right) ^{2}\right) }{2-\left( \gamma \left\langle \hat{S}%
_{E}\left( X\right) \right\rangle \right) ^{2}+\left( \gamma \left\langle 
\hat{S}_{E}\left( X_{1},X\right) \right\rangle _{X_{1}}\right) ^{2}-\left(
\gamma \left\langle \hat{S}_{E}\left( X\right) \right\rangle \right) ^{2}}
\end{equation*}%
and:%
\begin{eqnarray*}
\hat{G} &=&\frac{\left( 1-\left( \gamma \left\langle \hat{S}_{E}\left(
X\right) \right\rangle \right) ^{2}\right) }{2-\left( \gamma \left\langle 
\hat{S}_{E}\left( X\right) \right\rangle \right) ^{2}}\left( 1-\frac{\Delta
\left( \frac{f\left( X\right) +r\left( X\right) }{2}\right) }{2-\left(
\gamma \left\langle \hat{S}_{E}\left( X\right) \right\rangle \right) ^{2}}+%
\frac{\left\langle \hat{f}\left( X^{\prime }\right) \right\rangle
-\left\langle \hat{r}\left( X^{\prime }\right) \right\rangle _{\hat{w}_{2}}}{%
2}\right) \\
&&\times \frac{1-\left\langle \hat{S}\left( X^{\prime }\right) \right\rangle 
}{1-\left\langle \hat{S}_{E}\left( X^{\prime }\right) \right\rangle }\left(
\left\langle \hat{f}\left( X^{\prime }\right) \right\rangle -\left\langle 
\bar{r}\right\rangle \right) -S_{E}\left( X,X\right) \left( f\left( X\right)
-r\right)
\end{eqnarray*}%
where:%
\begin{equation*}
\left\langle \tilde{f}\right\rangle +\left\langle \tilde{r}\right\rangle
=\left\langle \hat{w}\left( X\right) \right\rangle \frac{\left\langle \hat{f}%
\left( X^{\prime }\right) \right\rangle _{\hat{w}_{1}}+\left\langle \hat{r}%
\left( X^{\prime }\right) \right\rangle _{\hat{w}_{2}}}{2}+\left\langle
w\left( X\right) \right\rangle \frac{\left\langle f\left( X\right)
\right\rangle +\left\langle r\left( X\right) \right\rangle }{2}
\end{equation*}

\subsection*{A10.2 Solving for banks}

\begin{eqnarray}
0 &=&\frac{1-\bar{S}\left( X\right) }{1-\bar{S}_{E}\left( X\right) }\left( 
\bar{f}\left( X\right) -\bar{r}\right) -\left\langle \bar{S}_{E}\left(
X^{\prime },X\right) \right\rangle _{X^{\prime }}\frac{1-\left\langle \bar{S}%
\left( X^{\prime }\right) \right\rangle }{1-\left\langle \bar{S}_{E}\left(
X^{\prime }\right) \right\rangle }\left( \left\langle \bar{f}\left(
X^{\prime }\right) \right\rangle -\bar{r}\right) \\
&&-\left\langle \hat{S}_{E}^{B}\left( X^{\prime },X\right) \right\rangle
_{X^{\prime }}\frac{1-\left\langle \hat{S}\left( X^{\prime }\right)
\right\rangle +\left\langle \hat{S}_{E}^{B}\left( X^{\prime }\right)
\right\rangle +\left\langle \hat{S}_{L}^{B}\left( X^{\prime }\right)
\right\rangle }{1-\left\langle \hat{S}_{E}\left( X^{\prime }\right)
\right\rangle +\left\langle \hat{S}_{E}^{B}\left( X^{\prime }\right)
\right\rangle }\left( \left\langle \hat{f}\left( X^{\prime }\right)
\right\rangle -\bar{r}\right) -S_{E}^{B}\left( X,X\right) \left( f_{1}\left(
X\right) -\bar{r}\right)  \notag
\end{eqnarray}%
Using the previous results for $\left\langle \bar{S}_{E}\left( X^{\prime
},X\right) \right\rangle _{X}$, $\left\langle \bar{S}\left( X^{\prime
},X\right) \right\rangle _{X}$: 
\begin{eqnarray*}
\left\langle \bar{S}_{E}\left( X^{\prime },X\right) \right\rangle _{X}
&=&\left\langle \frac{\bar{w}\left( X^{\prime },X\right) }{2}\right\rangle
_{X}\left( 1+\left\{ \left\langle \bar{w}\left( X\right) \right\rangle
\left( \bar{f}\left( X^{\prime }\right) -\frac{\left\langle \bar{f}\left(
X^{\prime }\right) \right\rangle _{\bar{w}_{1}}+\left\langle \bar{r}\left(
X^{\prime }\right) \right\rangle _{\bar{w}_{2}}}{2}\right) \right. \right. \\
&&\left. \left. +\left\langle \hat{w}_{1}^{B}\left( X\right) \right\rangle
\left( \bar{f}\left( X^{\prime }\right) -\left\langle \hat{f}\left(
X^{\prime }\right) \right\rangle _{\hat{w}_{1}}\right) +\left\langle
w_{1}^{B}\left( X\right) \right\rangle \left( \bar{f}\left( X^{\prime
}\right) -f\left( X\right) \right) \right\} \right)
\end{eqnarray*}%
\begin{eqnarray*}
\left\langle \bar{S}\left( X^{\prime },X\right) \right\rangle _{X}
&=&\left\langle \bar{S}_{E}\left( X^{\prime },X\right) \right\rangle
_{X}+\left\langle \bar{S}_{L}\left( X^{\prime },X\right) \right\rangle _{X}
\\
&=&\left\langle \bar{w}\left( X^{\prime },X\right) \right\rangle \left[
1+\left\{ \bar{w}\left( X\right) \left( \frac{\bar{f}\left( X^{\prime
}\right) +\bar{r}\left( X^{\prime }\right) }{2}-\frac{\left\langle \bar{f}%
\left( X^{\prime }\right) \right\rangle _{\bar{w}_{1}}+\left\langle \bar{r}%
\left( X^{\prime }\right) \right\rangle _{\bar{w}_{2}}}{2}\right) \right.
\right. \\
&&\left. \left. +\hat{w}_{1}^{B}\left( X\right) \left( \frac{\bar{f}\left(
X^{\prime }\right) +\bar{r}\left( X^{\prime }\right) }{2}-\left\langle \hat{f%
}\left( X^{\prime }\right) \right\rangle _{\hat{w}_{1}}\right)
+w_{1}^{B}\left( X\right) \left( \frac{\bar{f}\left( X^{\prime }\right) +%
\bar{r}\left( X^{\prime }\right) }{2}-f\left( X\right) \right) \right\} %
\right]
\end{eqnarray*}%
and the ratio:%
\begin{equation}
\frac{\left\langle \bar{K}\right\rangle \left\Vert \bar{\Psi}\right\Vert ^{2}%
}{\bar{K}_{X}\left\vert \bar{\Psi}\left( X\right) \right\vert ^{2}}=\left( 
\frac{\left( \left( 1-\left\langle \bar{S}\right\rangle \right) \bar{f}%
\left( X\right) +\left\langle \bar{S}\left( X^{\prime },X\right)
\right\rangle _{X^{\prime }}\left\langle \bar{f}\right\rangle \right) }{%
\left\langle \bar{f}\right\rangle }\frac{\frac{\left\langle \bar{S}\left(
X^{\prime },X\right) \right\rangle }{1-\left\langle \bar{S}\left( X^{\prime
},X\right) \right\rangle }}{\frac{\left\langle \bar{S}\left( X^{\prime
},X\right) \right\rangle _{X^{\prime }}}{1-\left\langle \bar{S}\left(
X^{\prime },X\right) \right\rangle _{X^{\prime }}\left( \frac{\left( \left(
1-\left\langle \bar{S}\right\rangle \right) \bar{f}\left( X\right)
+\left\langle \bar{S}\left( X^{\prime },X\right) \right\rangle _{X^{\prime
}}\left\langle \bar{f}\right\rangle \right) \frac{\left\langle \bar{S}\left(
X^{\prime },X\right) \right\rangle }{1-\left\langle \bar{S}\left( X^{\prime
},X\right) \right\rangle }}{\left\langle \bar{f}\right\rangle \frac{%
\left\langle \bar{S}\left( X^{\prime },X\right) \right\rangle _{X^{\prime }}%
}{1-\left\langle \bar{S}\left( X^{\prime },X\right) \right\rangle
_{X^{\prime }}}}\right) ^{2}}}\right) ^{2}  \label{rtc}
\end{equation}%
with the half average:%
\begin{eqnarray*}
&&\left\langle \bar{S}_{E}\left( X^{\prime },X\right) \right\rangle
_{X^{\prime }} \\
&\rightarrow &\frac{\left\langle \bar{w}\left( X^{\prime },X\right)
\right\rangle }{2}\left( 1+\left\langle \bar{w}\left( X\right) \right\rangle
\left( \frac{\left\langle \bar{f}\left( X^{\prime }\right) \right\rangle _{%
\bar{w}_{1}}-\left\langle \bar{r}\left( X^{\prime }\right) \right\rangle _{%
\bar{w}_{2}}}{2}\right) \right. \\
&&\left. +\left\langle \hat{w}_{1}^{B}\left( X\right) \right\rangle \left(
\left\langle \bar{f}\left( X^{\prime }\right) \right\rangle -\left\langle 
\hat{f}\left( X^{\prime }\right) \right\rangle _{\hat{w}_{1}}\right)
+\left\langle w_{1}^{B}\left( X\right) \right\rangle \left( \left\langle 
\bar{f}\left( X^{\prime }\right) \right\rangle -f\left( X\right) \right)
\right)
\end{eqnarray*}%
the equation for $\bar{f}\left( X^{\prime }\right) $ becomes: 
\begin{equation*}
0=\frac{1-\bar{A}\left( 1+\frac{\Delta \bar{f}\left( X^{\prime }\right)
+\Delta \bar{r}\left( X^{\prime }\right) }{2}\right) \frac{\left\langle \bar{%
K}\right\rangle \left\Vert \bar{\Psi}\right\Vert ^{2}}{\bar{K}_{X}\left\vert 
\bar{\Psi}\left( X\right) \right\vert ^{2}}}{1-\frac{1}{2}\bar{A}\left(
1+\Delta \bar{f}\left( X^{\prime }\right) \right) \frac{\left\langle \bar{K}%
\right\rangle \left\Vert \bar{\Psi}\right\Vert ^{2}}{\bar{K}_{X}\left\vert 
\bar{\Psi}\left( X\right) \right\vert ^{2}}}\left( \bar{f}\left( X\right) -%
\bar{r}\right) -\bar{G}
\end{equation*}%
where:%
\begin{eqnarray}
\bar{G} &=&\left\langle \bar{S}_{E}\left( X^{\prime },X\right) \right\rangle
_{X^{\prime }}\frac{1-\left\langle \bar{S}\left( X^{\prime }\right)
\right\rangle }{1-\left\langle \bar{S}_{E}\left( X^{\prime }\right)
\right\rangle }\left( \left\langle \bar{f}\left( X^{\prime }\right)
\right\rangle -\bar{r}\right)  \label{Gbr} \\
&&+\left\langle \hat{S}_{E}^{B}\left( X^{\prime },X\right) \right\rangle
_{X^{\prime }}\frac{1-\left( \left\langle \hat{S}\left( X^{\prime }\right)
\right\rangle +\left\langle \hat{S}_{E}^{B}\left( X^{\prime }\right)
\right\rangle +\left\langle \hat{S}_{L}^{B}\left( X^{\prime }\right)
\right\rangle \right) }{1-\left( \left\langle \hat{S}_{E}\left( X^{\prime
}\right) \right\rangle +\left\langle \hat{S}_{E}^{B}\left( X^{\prime
}\right) \right\rangle \right) }\left( \left\langle \hat{f}\left( X^{\prime
}\right) \right\rangle -\bar{r}\right) +S_{E}^{B}\left( X,X\right) \left(
f_{1}\left( X\right) -\bar{r}\right)  \notag
\end{eqnarray}%
and:%
\begin{eqnarray*}
\bar{A} &=&\left( 1-\left( \bar{\gamma}\left\langle \bar{S}_{E}\right\rangle
\right) ^{2}\right) \left( 1-\left( \gamma \left\langle \hat{S}%
_{E}\right\rangle \right) ^{2}+\left( \gamma \left\langle \hat{S}_{E}\left(
X_{1},X^{\prime }\right) \right\rangle _{X_{1}}\right) ^{2}\right) \\
&&\times \left\{ 1-\left( \bar{\gamma}\left\langle \bar{S}_{E}\right\rangle
\right) ^{2}+\frac{4}{\zeta ^{2}}\left\{ \bar{\zeta}^{2}\zeta ^{2}\left( 1+%
\frac{\left( \gamma \left\langle \hat{S}_{E}\right\rangle \right) ^{2}}{%
1-\left( \gamma \left\langle \hat{S}_{E}\right\rangle \right) ^{2}}\right)
+\xi ^{2}\right\} \right. \\
&&\times \left. \left( 1-\left( \bar{\gamma}\left\langle \bar{S}%
_{E}\right\rangle \right) ^{2}+\left( \bar{\gamma}\left\langle \bar{S}%
_{E}\left( X_{1},X^{\prime }\right) \right\rangle _{X_{1}}\right)
^{2}\right) \left( 1-\left( \gamma \left\langle \hat{S}_{E}\right\rangle
\right) ^{2}\right) \left( 1+\frac{\zeta ^{2}}{\xi ^{2}}\left( 1+\frac{%
\left( \gamma \left\langle \hat{S}_{E}\left( X_{1},X^{\prime }\right)
\right\rangle _{X_{1}}\right) ^{2}}{1-\left( \gamma \left\langle \hat{S}%
_{E}\right\rangle \right) ^{2}}\right) \right) \right\} ^{-1}
\end{eqnarray*}%
\begin{equation*}
\Delta \bar{f}\left( X^{\prime }\right) =\bar{f}\left( X^{\prime }\right)
-\left( \left\langle \bar{w}\left( X\right) \right\rangle \frac{\left\langle 
\bar{f}\left( X^{\prime }\right) \right\rangle _{\bar{w}_{1}}+\left\langle 
\bar{r}\left( X^{\prime }\right) \right\rangle _{\bar{w}_{2}}}{2}%
+\left\langle \hat{w}_{1}^{B}\left( X\right) \right\rangle \left\langle \hat{%
f}\left( X^{\prime }\right) \right\rangle _{\hat{w}_{1}}+\left\langle
w_{1}^{B}\left( X\right) \right\rangle \left\langle f\left( X\right)
\right\rangle \right)
\end{equation*}%
\begin{equation*}
\Delta \hat{f}\left( X^{\prime }\right) =\hat{f}\left( X^{\prime }\right)
-\left( \left\langle \hat{w}\left( X\right) \right\rangle \frac{\left\langle 
\hat{f}\left( X^{\prime }\right) \right\rangle _{\hat{w}_{1}}+\left\langle 
\hat{r}\left( X^{\prime }\right) \right\rangle _{\hat{w}_{2}}}{2}%
+\left\langle w\left( X\right) \right\rangle \frac{\left\langle f\left(
X\right) \right\rangle +\left\langle r\left( X\right) \right\rangle }{2}%
\right)
\end{equation*}%
Using (\ref{rtc}) this leads to: 
\begin{subequations}
\begin{equation}
0=\frac{1-\bar{A}\left( 1+\frac{\Delta \bar{f}\left( X\right) +\Delta \bar{r}%
\left( X\right) }{2}\right) \frac{\left( \bar{f}\left( X\right) \left(
1-\left\langle \bar{S}\right\rangle \right) +\left\langle \bar{S}\left(
X^{\prime },X\right) \right\rangle _{X^{\prime }}\left\langle \bar{f}%
\right\rangle \right) ^{2}}{\left\langle \bar{f}\right\rangle ^{2}D}}{1-%
\frac{1}{2}\bar{A}\left( 1+\Delta \bar{f}\left( X\right) \right) \frac{%
\left( \bar{f}\left( X\right) \left( 1-\left\langle \bar{S}\right\rangle
\right) +\left\langle \bar{S}\left( X^{\prime },X\right) \right\rangle
_{X^{\prime }}\left\langle \bar{f}\right\rangle \right) ^{2}}{\left\langle 
\bar{f}\right\rangle ^{2}D}}\left( \bar{f}\left( X\right) -\bar{r}\right) -%
\bar{G}  \label{Brs}
\end{equation}%
with: 
\end{subequations}
\begin{equation*}
D=\frac{\frac{\left\langle \bar{S}\left( X^{\prime },X\right) \right\rangle
_{X^{\prime }}}{1-\left\langle \bar{S}\left( X^{\prime },X\right)
\right\rangle _{X^{\prime }}\left( \frac{\left( \left( 1-\left\langle \bar{S}%
\right\rangle \right) \bar{f}\left( X\right) +\left\langle \bar{S}\left(
X^{\prime },X\right) \right\rangle _{X^{\prime }}\left\langle \bar{f}%
\right\rangle \right) \frac{\left\langle \bar{S}\left( X^{\prime },X\right)
\right\rangle }{1-\left\langle \bar{S}\left( X^{\prime },X\right)
\right\rangle }}{\left\langle \bar{f}\right\rangle \frac{\left\langle \bar{S}%
\left( X^{\prime },X\right) \right\rangle _{X^{\prime }}}{1-\left\langle 
\bar{S}\left( X^{\prime },X\right) \right\rangle _{X^{\prime }}}}\right) ^{2}%
}}{\frac{\left\langle \bar{S}\left( X^{\prime },X\right) \right\rangle }{%
1-\left\langle \bar{S}\left( X^{\prime },X\right) \right\rangle }}
\end{equation*}%
or: 
\begin{equation}
0=\frac{\left\langle \bar{f}\right\rangle ^{2}D-\bar{A}\left( 1+\frac{\bar{f}%
\left( X\right) +\Delta \bar{r}\left( X\right) -\tilde{f}}{2}\right) \left( 
\bar{f}\left( X\right) \left( 1-\left\langle \bar{S}\right\rangle \right)
+\left\langle \bar{S}\left( X^{\prime },X\right) \right\rangle _{X^{\prime
}}\left\langle \bar{f}\right\rangle \right) ^{2}}{\left\langle \bar{f}%
\right\rangle ^{2}D-\frac{1}{2}\bar{A}\left( 1+\bar{f}\left( X\right) -%
\tilde{f}\right) \left( \bar{f}\left( X\right) \left( 1-\left\langle \bar{S}%
\right\rangle \right) +\left\langle \bar{S}\left( X^{\prime },X\right)
\right\rangle _{X^{\prime }}\left\langle \bar{f}\right\rangle \right) ^{2}}-%
\bar{G}  \label{Brv}
\end{equation}%
This is an equation of the form:%
\begin{equation*}
f^{2}D-A\left( 1+\frac{X+c-b}{2}\right) \left( X\left( 1-s\right) +sf\right)
^{2}-\bar{G}\left( f^{2}-\frac{1}{2}A\left( 1+X-b\right) \left( X\left(
1-s\right) +sf\right) ^{2}\right)
\end{equation*}%
and this equation has several solutions.

\subsection*{A10.3 Returns $\Delta \bar{f}\left( X^{\prime }\right) $ and $%
\Delta \hat{f}\left( X^{\prime }\right) $ as function of $\left\langle \bar{S%
}_{E}\left( X^{\prime },X\right) \right\rangle _{X}$ and $\left\langle \hat{S%
}_{E}\left( X^{\prime },X\right) \right\rangle _{X}$}

Given the formula for returns $\Delta \bar{f}\left( X^{\prime }\right) $ and 
$\Delta \hat{f}\left( X^{\prime }\right) $ 
\begin{equation*}
\Delta \bar{f}\left( X^{\prime }\right) =\bar{f}\left( X^{\prime }\right)
-\left( \left\langle \bar{w}\left( X\right) \right\rangle \frac{\left\langle 
\bar{f}\left( X^{\prime }\right) \right\rangle _{\bar{w}_{1}}+\left\langle 
\bar{r}\left( X^{\prime }\right) \right\rangle _{\bar{w}_{2}}}{2}%
+\left\langle \hat{w}_{1}^{B}\left( X\right) \right\rangle \left\langle \hat{%
f}\left( X^{\prime }\right) \right\rangle _{\hat{w}_{1}}+\left\langle
w_{1}^{B}\left( X\right) \right\rangle \left\langle f\left( X\right)
\right\rangle \right)
\end{equation*}%
\begin{equation*}
\Delta \hat{f}\left( X^{\prime }\right) =\hat{f}\left( X^{\prime }\right)
-\left( \left\langle \hat{w}\left( X\right) \right\rangle \frac{\left\langle 
\hat{f}\left( X^{\prime }\right) \right\rangle _{\hat{w}_{1}}+\left\langle 
\hat{r}\left( X^{\prime }\right) \right\rangle _{\hat{w}_{2}}}{2}%
+\left\langle w\left( X\right) \right\rangle \frac{\left\langle f\left(
X\right) \right\rangle +\left\langle r\left( X\right) \right\rangle }{2}%
\right)
\end{equation*}%
and the relations:%
\begin{equation*}
\left\langle \bar{S}_{E}\left( X^{\prime },X\right) \right\rangle
_{X}=\left\langle \frac{\bar{w}\left( X^{\prime },X\right) }{2}\right\rangle
_{X}\Delta \bar{f}\left( X^{\prime }\right) +1
\end{equation*}%
\begin{equation}
\left\langle \hat{S}_{E}\left( X^{\prime },X\right) \right\rangle _{X}=\frac{%
\left\langle \hat{w}\left( X^{\prime },X\right) \right\rangle _{X}}{2}\left(
1+\Delta \hat{f}\left( X^{\prime }\right) \right)
\end{equation}%
We can express returns in terms of shares:%
\begin{equation*}
\Delta \bar{f}\left( X^{\prime }\right) =\frac{\left\langle \bar{S}%
_{E}\left( X^{\prime },X\right) \right\rangle _{X}}{\left\langle \frac{\bar{w%
}\left( X^{\prime },X\right) }{2}\right\rangle _{X}}-1
\end{equation*}%
with:%
\begin{equation*}
\left\langle \frac{\hat{w}\left( X^{\prime },X\right) }{2}\right\rangle
_{X}\simeq \frac{\left( 1-\left( \gamma \left\langle \hat{S}_{E}\left(
X\right) \right\rangle \right) ^{2}\right) }{2-\left( \gamma \left\langle 
\hat{S}_{E}\left( X\right) \right\rangle \right) ^{2}+\left( \gamma
\left\langle \hat{S}_{E}\left( X_{1},X^{\prime }\right) \right\rangle
_{X_{1}}\right) ^{2}-\left( \gamma \left\langle \hat{S}_{E}\left( X\right)
\right\rangle \right) ^{2}}
\end{equation*}%
and:%
\begin{eqnarray}
&&\left\langle \bar{S}_{E}\left( X^{\prime },X\right) \right\rangle
_{X^{\prime }} \\
&\simeq &\frac{\left\langle \bar{w}\left( X^{\prime },X\right) \right\rangle 
}{2}\left( 1+\left\langle \bar{w}\left( X\right) \right\rangle \left( \frac{%
\left\langle \bar{f}\left( X^{\prime }\right) \right\rangle _{\bar{w}%
_{1}}-\left\langle \bar{r}\left( X^{\prime }\right) \right\rangle _{\bar{w}%
_{2}}}{2}\right) \right.  \notag \\
&&\left. +\left\langle \hat{w}_{1}^{B}\left( X\right) \right\rangle \left(
\left\langle \bar{f}\left( X^{\prime }\right) \right\rangle -\left\langle 
\hat{f}\left( X^{\prime }\right) \right\rangle _{\hat{w}_{1}}\right)
+\left\langle w_{1}^{B}\left( X\right) \right\rangle \left( \left\langle 
\bar{f}\left( X^{\prime }\right) \right\rangle -f\left( X\right) \right)
\right)  \notag
\end{eqnarray}%
for banks, and:%
\begin{equation*}
\Delta \hat{f}\left( X^{\prime }\right) =\frac{\left\langle \hat{S}%
_{E}\left( X^{\prime },X\right) \right\rangle _{X}}{\frac{\left\langle \hat{w%
}\left( X^{\prime },X\right) \right\rangle _{X}}{2}}-1
\end{equation*}%
\begin{equation*}
\left\langle \frac{\hat{w}\left( X^{\prime },X\right) }{2}\right\rangle
_{X}\simeq \frac{\left( 1-\left( \gamma \left\langle \hat{S}_{E}\left(
X\right) \right\rangle \right) ^{2}\right) }{2-\left( \gamma \left\langle 
\hat{S}_{E}\left( X\right) \right\rangle \right) ^{2}+\left( \gamma
\left\langle \hat{S}_{E}\left( X_{1},X^{\prime }\right) \right\rangle
_{X_{1}}\right) ^{2}-\left( \gamma \left\langle \hat{S}_{E}\left( X\right)
\right\rangle \right) ^{2}}
\end{equation*}

\begin{eqnarray*}
&&\left\langle \hat{S}_{E}\left( X^{\prime },X\right) \right\rangle
_{X^{\prime }} \\
&\simeq &\frac{\left\langle \underline{\hat{S}}\left( X^{\prime },X\right)
\right\rangle }{2}+\frac{\left\langle \hat{w}\left( X^{\prime },X\right)
\right\rangle }{2} \\
&&\left( \hat{w}\left( X\right) \left( \frac{\left\langle \hat{f}\left(
X^{\prime }\right) \right\rangle -\left\langle \hat{r}\left( X^{\prime
}\right) \right\rangle _{\hat{w}_{2}}}{2}\right) +w\left( X\right) \left(
\left\langle \hat{f}\left( X^{\prime }\right) \right\rangle -\frac{f\left(
X\right) +r\left( X\right) }{2}\right) \right) \\
&\simeq &\left\langle \hat{w}\left( X^{\prime },X\right) \right\rangle
\left( 1-\left\langle w\left( X\right) \right\rangle \Delta \left( \frac{%
f\left( X\right) +r\left( X\right) }{2}\right) +\frac{\left\langle \hat{f}%
\left( X^{\prime }\right) \right\rangle -\left\langle \hat{r}\left(
X^{\prime }\right) \right\rangle _{\hat{w}_{2}}}{2}\right)
\end{eqnarray*}%
\begin{equation*}
\Delta \left( \frac{f\left( X\right) +r\left( X\right) }{2}\right) =\left( 
\frac{f\left( X\right) +r\left( X\right) }{2}-\frac{\left\langle f\left(
X\right) \right\rangle +\left\langle r\left( X\right) \right\rangle }{2}%
\right)
\end{equation*}%
As explained in first part, this change of variable allows to consider
representation in terms of shares, the returns being expressed in terms of
these shares.

\subsection*{A10.4 Corrections for decreasing return to scale}

As for averages, we replace the constant returns with decreasing retrns:%
\begin{equation*}
f_{1}\left( X\right) \rightarrow f_{1}^{dr}\left( X\right)
\end{equation*}%
\begin{equation*}
f_{1}^{dr}\left( X\right) =\frac{f_{1}\left( X\right) }{K_{X}^{r}}-\frac{C}{%
K_{X}}
\end{equation*}%
To zeroth order, we replace in the derivation of the results:%
\begin{equation*}
\left( S\left( X,X\right) ,S^{B}\left( X,X\right) \right) \rightarrow \left(
S\left( X,X\right) _{cr},S^{B}\left( X,X\right) _{cr}\right)
\end{equation*}%
where $S\left( X,X\right) _{cr}$ and $S^{B}\left( X,X\right) _{cr}$ are
constant return solutions. To zeroth order the firms' capital writes:%
\begin{eqnarray}
K_{X}\left\vert \Psi \left( X\right) \right\vert ^{2} &\simeq &\left( \left(
1-S\left( X\right) \right) \right) ^{2}\frac{2\epsilon }{3\sigma _{\hat{K}%
}^{2}}\left( \left( \frac{2\epsilon }{3\sigma _{\hat{K}}^{2}}\right) ^{\frac{%
r}{2}}\frac{f_{1}\left( X\right) }{C_{0}+\frac{S_{L}\left( X\right) }{%
1-S_{E}\left( X\right) }\bar{r}}\right) ^{\frac{2}{r}}  \label{FCt} \\
&\simeq &\left( 1-\left( S\left( X,X\right) \frac{\hat{K}_{X}\left\vert \hat{%
\Psi}\left( X\right) \right\vert ^{2}}{K_{X}\left\vert \Psi \left( X\right)
\right\vert ^{2}}+S^{B}\left( X,X\right) \frac{\bar{K}_{X}\left\vert \bar{%
\Psi}\left( X\right) \right\vert ^{2}}{K_{X}\left\vert \Psi \left( X\right)
\right\vert ^{2}}\right) \right) \left( \left( \frac{2\epsilon }{3\sigma _{%
\hat{K}}^{2}}\right) ^{\frac{r}{2}}\frac{f_{1}\left( X\right) }{C_{0}+\frac{%
S_{L}\left( X\right) }{1-S_{E}\left( X\right) }\bar{r}}\right) ^{\frac{2}{r}}
\notag
\end{eqnarray}%
In first approximation, we have the following capital ratios:%
\begin{equation*}
\frac{\hat{K}_{X}\left\vert \hat{\Psi}\left( X\right) \right\vert ^{2}}{%
K_{X}\left\vert \Psi \left( X\right) \right\vert ^{2}}\simeq \frac{18\sigma
_{\hat{K}}^{2}V\left\Vert \hat{\Psi}_{0}\left( X\right) \right\Vert ^{4}}{%
\hat{\mu}F_{1}^{2}\left( \left( \frac{2\epsilon }{3\sigma _{\hat{K}}^{2}}%
\right) ^{\frac{r}{2}}\frac{f_{1}\left( X\right) }{C_{0}+\frac{S_{L}\left(
X\right) }{1-S_{E}\left( X\right) }\bar{r}}\right) ^{\frac{2}{r}}}
\end{equation*}%
where $F_{1}$ is defined in (\ref{Dfn}), and:%
\begin{equation*}
\frac{\bar{K}_{X}\left\vert \bar{\Psi}\left( X\right) \right\vert ^{2}}{%
K_{X}\left\vert \Psi \left( X\right) \right\vert ^{2}}\simeq \frac{18\sigma
_{\hat{K}}^{2}V\left\Vert \bar{\Psi}_{0}\left( X\right) \right\Vert ^{4}}{%
\left( \bar{f}\left( X\right) +\frac{\left\langle \bar{S}\left( X^{\prime
},X\right) \right\rangle _{X^{\prime }}}{\left( 1-\left\langle \bar{S}%
\right\rangle \right) }\left\langle \bar{f}\right\rangle \right) ^{2}\hat{\mu%
}\left( \left( \frac{2\epsilon }{3\sigma _{\hat{K}}^{2}}\right) ^{\frac{r}{2}%
}\frac{f_{1}\left( X\right) }{C_{0}+\frac{S_{L}\left( X\right) }{%
1-S_{E}\left( X\right) }\bar{r}}\right) ^{\frac{2}{r}}}
\end{equation*}%
These ratios, combined with (\ref{FCt}), leads to the zeroth order formla:%
\begin{equation*}
K_{X}\left\vert \Psi \left( X\right) \right\vert ^{2}\simeq \left( 1-\frac{%
18\sigma _{\hat{K}}^{2}V}{\hat{\mu}}\frac{\frac{S\left( X,X\right)
_{cr}\left( \left\Vert \bar{\Psi}_{0}\left( X\right) \right\Vert \right) ^{4}%
}{F_{1}^{2}}+\frac{S^{B}\left( X,X\right) _{cr}\left( \left\Vert \bar{\Psi}%
_{0}\left( X\right) \right\Vert \right) ^{4}}{\left( \bar{f}\left( X\right) +%
\frac{\left\langle \bar{S}\left( X^{\prime },X\right) \right\rangle
_{X^{\prime }}}{\left( 1-\left\langle \bar{S}\right\rangle \right) }%
\left\langle \bar{f}\right\rangle \right) ^{2}}}{\left( \left( \frac{%
2\epsilon }{3\sigma _{\hat{K}}^{2}}\right) ^{\frac{r}{2}}\frac{f_{1}\left(
X\right) }{C_{0}+\frac{S_{L}\left( X\right) }{1-S_{E}\left( X\right) }\bar{r}%
}\right) ^{\frac{2}{r}}}\right) \left( \left( \frac{2\epsilon }{3\sigma _{%
\hat{K}}^{2}}\right) ^{\frac{r}{2}}\frac{f_{1}\left( X\right) }{C_{0}+\frac{%
S_{L}\left( X\right) }{1-S_{E}\left( X\right) }\bar{r}}\right) ^{\frac{2}{r}}
\end{equation*}%
that yields the expanded form of capital ratios:%
\begin{eqnarray}
&&\frac{\hat{K}_{X}\left\vert \hat{\Psi}\left( X\right) \right\vert ^{2}}{%
K_{X}\left\vert \Psi \left( X\right) \right\vert ^{2}}  \label{cpn} \\
&\simeq &\frac{18\sigma _{\hat{K}}^{2}V\left( \left\Vert \hat{\Psi}%
_{0}\right\Vert \right) ^{4}}{\hat{\mu}F_{1}^{2}\left( 1-\frac{18\sigma _{%
\hat{K}}^{2}V}{\hat{\mu}}\frac{\frac{S\left( X,X\right) _{cr}\left(
\left\Vert \bar{\Psi}_{0}\left( X\right) \right\Vert \right) ^{4}}{F_{1}^{2}}%
+\frac{S^{B}\left( X,X\right) _{cr}\left( \left\Vert \bar{\Psi}_{0}\left(
X\right) \right\Vert \right) ^{4}}{\left( \bar{f}\left( X\right) +\frac{%
\left\langle \bar{S}\left( X^{\prime },X\right) \right\rangle _{X^{\prime }}%
}{\left( 1-\left\langle \bar{S}\right\rangle \right) }\left\langle \bar{f}%
\right\rangle \right) ^{2}}}{\left( \left( \frac{2\epsilon }{3\sigma _{\hat{K%
}}^{2}}\right) ^{\frac{r}{2}}\frac{f_{1}\left( X\right) }{C_{0}+\frac{%
S_{L}\left( X\right) }{1-S_{E}\left( X\right) }\bar{r}}\right) ^{\frac{2}{r}}%
}\right) ^{2}\left( \left( \frac{2\epsilon }{3\sigma _{\hat{K}}^{2}}\right)
^{\frac{r}{2}}\frac{f_{1}\left( X\right) }{C_{0}+\frac{S_{L}\left( X\right) 
}{1-S_{E}\left( X\right) }\bar{r}}\right) ^{\frac{2}{r}}}  \notag
\end{eqnarray}%
and:%
\begin{eqnarray}
&&\frac{\bar{K}_{X}\left\vert \bar{\Psi}\left( X\right) \right\vert ^{2}}{%
K_{X}\left\vert \Psi \left( X\right) \right\vert ^{2}}  \label{cpt} \\
&\simeq &\frac{18\frac{\sigma _{\hat{K}}^{2}}{\hat{\mu}}V\left( \left\Vert 
\bar{\Psi}_{0}\right\Vert \right) ^{4}}{\left( \bar{f}\left( X\right) +\frac{%
\left\langle \bar{S}\left( X^{\prime },X\right) \right\rangle _{X^{\prime }}%
}{\left( 1-\left\langle \bar{S}\right\rangle \right) }\left\langle \bar{f}%
\right\rangle \right) ^{2}\left( \left( \left( \frac{2\epsilon }{3\sigma _{%
\hat{K}}^{2}}\right) ^{\frac{r}{2}}\frac{f_{1}\left( X\right) }{C_{0}+\frac{%
S_{L}\left( X\right) }{1-S_{E}\left( X\right) }\bar{r}}\right) ^{\frac{1}{r}%
}-\frac{18\sigma _{\hat{K}}^{2}V}{\hat{\mu}}\frac{\frac{S\left( X,X\right)
_{cr}\left( \left\Vert \bar{\Psi}_{0}\left( X\right) \right\Vert \right) ^{4}%
}{F_{1}^{2}}+\frac{S^{B}\left( X,X\right) _{cr}\left( \left\Vert \bar{\Psi}%
_{0}\left( X\right) \right\Vert \right) ^{4}}{\left( \bar{f}\left( X\right) +%
\frac{\left\langle \bar{S}\left( X^{\prime },X\right) \right\rangle
_{X^{\prime }}}{\left( 1-\left\langle \bar{S}\right\rangle \right) }%
\left\langle \bar{f}\right\rangle \right) ^{2}}}{\left( \left( \frac{%
2\epsilon }{3\sigma _{\hat{K}}^{2}}\right) ^{\frac{r}{2}}\frac{f_{1}\left(
X\right) }{C_{0}+\frac{S_{L}\left( X\right) }{1-S_{E}\left( X\right) }\bar{r}%
}\right) ^{\frac{1}{r}}}\right) ^{2}}  \notag
\end{eqnarray}%
Formula for shares $S_{E}^{B}\left( X,X\right) $, $S^{B}\left( X,X\right) $, 
$S_{E}\left( X,X\right) $, $S\left( X,X\right) $ are thus modified at frst
order by: 
\begin{equation*}
f_{1}\left( X\right) \rightarrow f_{1}\left( X\right) _{dr}
\end{equation*}%
and the investors' equation writes:%
\begin{equation*}
\hat{G}=\frac{\left( \left\langle \hat{f}\right\rangle +\frac{\left\langle 
\hat{S}_{E}^{B}\right\rangle +\left\langle \hat{S}_{L}^{B}\right\rangle
\left\langle \bar{S}\right\rangle }{1-\left\langle \bar{S}_{E}\right\rangle }%
\left\langle \bar{f}\right\rangle \right) ^{2}\left( \frac{\frac{%
\left\langle \hat{S}\left( X^{\prime },X\right) \right\rangle _{X}}{%
1-\left\langle \hat{S}\left( X^{\prime },X\right) \right\rangle _{X}\frac{%
\left\langle \hat{K}\right\rangle \left\Vert \hat{\Psi}\right\Vert ^{2}}{%
\hat{K}_{X}\left\vert \hat{\Psi}\left( X\right) \right\vert ^{2}}}}{\frac{%
\left\langle \hat{S}\left( X^{\prime },X\right) \right\rangle }{%
1-\left\langle \hat{S}\left( X^{\prime },X\right) \right\rangle }}\right)
^{2}-A\left( 1+\frac{\hat{f}\left( X^{\prime }\right) +\left\langle \tilde{f}%
+\tilde{r}\right\rangle +\Delta \hat{r}\left( X\right) }{2}\right) F_{1}^{2}%
}{\left( \left\langle \hat{f}\right\rangle +\frac{\left\langle \hat{S}%
_{E}^{B}\right\rangle +\left\langle \hat{S}_{L}^{B}\right\rangle
\left\langle \bar{S}\right\rangle }{1-\left\langle \bar{S}_{E}\right\rangle }%
\left\langle \bar{f}\right\rangle \right) ^{2}\left( \frac{\frac{%
\left\langle \hat{S}\left( X^{\prime },X\right) \right\rangle _{X}}{%
1-\left\langle \hat{S}\left( X^{\prime },X\right) \right\rangle _{X}\frac{%
\left\langle \hat{K}\right\rangle \left\Vert \hat{\Psi}\right\Vert ^{2}}{%
\hat{K}_{X}\left\vert \hat{\Psi}\left( X\right) \right\vert ^{2}}}}{\frac{%
\left\langle \hat{S}\left( X^{\prime },X\right) \right\rangle }{%
1-\left\langle \hat{S}\left( X^{\prime },X\right) \right\rangle }}\right)
^{2}-\frac{1}{2}A\left( 1+\hat{f}\left( X^{\prime }\right) +\left\langle 
\tilde{f}+\tilde{r}\right\rangle \right) F_{1}^{2}}\left( f_{1}^{dr}\left(
X\right) -\bar{r}\right)
\end{equation*}%
with $F_{1}$ defined in (\ref{Dfn}) and $\hat{G}$ given by:%
\begin{eqnarray*}
\hat{G} &=&\frac{\left( 1-\left( \gamma \left\langle \hat{S}_{E}\left(
X\right) \right\rangle \right) ^{2}\right) }{2-\left( \gamma \left\langle 
\hat{S}_{E}\left( X\right) \right\rangle \right) ^{2}}\left( 1-\frac{\Delta
\left( \frac{f\left( X\right) +r\left( X\right) }{2}\right) }{2-\left(
\gamma \left\langle \hat{S}_{E}\left( X\right) \right\rangle \right) ^{2}}+%
\frac{\left\langle \hat{f}\left( X^{\prime }\right) \right\rangle
-\left\langle \hat{r}\left( X^{\prime }\right) \right\rangle _{\hat{w}_{2}}}{%
2}\right) \\
&&\times \frac{1-\left\langle \hat{S}\left( X^{\prime }\right) \right\rangle 
}{1-\left\langle \hat{S}_{E}\left( X^{\prime }\right) \right\rangle }\left(
\left\langle \hat{f}\left( X^{\prime }\right) \right\rangle -\left\langle 
\bar{r}\right\rangle \right) -S_{E}\left( X,X\right) \left( f_{1}^{dr}\left(
X\right) -r\right)
\end{eqnarray*}%
The banks retrn equations writes;%
\begin{equation*}
0=\frac{\left\langle \bar{f}\right\rangle ^{2}D-\bar{A}\left( 1+\frac{\bar{f}%
\left( X\right) +\Delta \bar{r}\left( X\right) -\tilde{f}}{2}\right) \left( 
\bar{f}\left( X\right) \left( 1-\left\langle \bar{S}\right\rangle \right)
+\left\langle \bar{S}\left( X^{\prime },X\right) \right\rangle _{X^{\prime
}}\left\langle \bar{f}\right\rangle \right) ^{2}}{\left\langle \bar{f}%
\right\rangle ^{2}D-\frac{1}{2}\bar{A}\left( 1+\bar{f}\left( X\right) -%
\tilde{f}\right) \left( \bar{f}\left( X\right) \left( 1-\left\langle \bar{S}%
\right\rangle \right) +\left\langle \bar{S}\left( X^{\prime },X\right)
\right\rangle _{X^{\prime }}\left\langle \bar{f}\right\rangle \right) ^{2}}%
\frac{1-\bar{A}\left( 1+\frac{\Delta \bar{f}\left( X^{\prime }\right)
+\Delta \bar{r}\left( X^{\prime }\right) }{2}\right) }{1-\frac{1}{2}\bar{A}%
\left( 1+\Delta \bar{f}\left( X^{\prime }\right) \right) }-\bar{G}
\end{equation*}%
where:%
\begin{eqnarray}
\bar{G} &=&\left\langle \bar{S}_{E}\left( X^{\prime },X\right) \right\rangle
_{X^{\prime }}\frac{1-\left\langle \bar{S}\left( X^{\prime }\right)
\right\rangle }{1-\left\langle \bar{S}_{E}\left( X^{\prime }\right)
\right\rangle }\left( \left\langle \bar{f}\left( X^{\prime }\right)
\right\rangle -\bar{r}\right) \\
&&+\left\langle \hat{S}_{E}^{B}\left( X^{\prime },X\right) \right\rangle
_{X^{\prime }}\frac{1-\left\langle \hat{S}\left( X^{\prime }\right)
\right\rangle +\left\langle \hat{S}_{E}^{B}\left( X^{\prime }\right)
\right\rangle +\left\langle \hat{S}_{L}^{B}\left( X^{\prime }\right)
\right\rangle }{1-\left\langle \hat{S}_{E}\left( X^{\prime }\right)
\right\rangle +\left\langle \hat{S}_{E}^{B}\left( X^{\prime }\right)
\right\rangle }\left( \left\langle \hat{f}\left( X^{\prime }\right)
\right\rangle -\bar{r}\right) +S_{E}^{B}\left( X,X\right) \left( f_{1}\left(
X\right) _{dr}-\bar{r}\right)  \notag
\end{eqnarray}%
These equations are similar to those obtained for investors. The main shift
is a direct effect of the modificatn of $\hat{G}$ and $\bar{G}$. This
reduces $\hat{G}$ and $\bar{G}$ which leads to a lower return and higher
capital solution. The effect is simplified by banks.

\subsection*{A10.5 Approximate solutions}

\subsubsection*{A10.5.1 Investors}

Solution of (\ref{NVT}) to the first order of approximation is similar to
the one obtained in Gosselin and Lotz (2025) under decreasing returns with $%
\frac{\left\langle \hat{K}\right\rangle \left\Vert \hat{\Psi}\right\Vert ^{2}%
}{\hat{K}_{X}\left\vert \hat{\Psi}\left( X\right) \right\vert ^{2}}$ given
by (\ref{RTC}). The equation writes:%
\begin{equation*}
\hat{G}=\frac{\left( \left\langle \hat{f}\right\rangle +\frac{\left\langle 
\hat{S}_{E}^{B}\right\rangle +\left\langle \hat{S}_{L}^{B}\right\rangle }{1-%
\bar{S}}\frac{\left\langle \bar{K}\right\rangle \left\Vert \bar{\Psi}%
\right\Vert ^{2}}{\left\langle \hat{K}\right\rangle \left\Vert \hat{\Psi}%
\right\Vert ^{2}}\left\langle \bar{f}\right\rangle \right) ^{2}\left( \frac{%
\frac{\left\langle \hat{S}\left( X^{\prime },X\right) \right\rangle _{X}}{%
1-\left\langle \hat{S}\left( X^{\prime },X\right) \right\rangle _{X}\frac{%
\left\langle \hat{K}\right\rangle \left\Vert \hat{\Psi}\right\Vert ^{2}}{%
\hat{K}_{X}\left\vert \hat{\Psi}\left( X\right) \right\vert ^{2}}}}{\frac{%
\left\langle \hat{S}\left( X^{\prime },X\right) \right\rangle }{%
1-\left\langle \hat{S}\left( X^{\prime },X\right) \right\rangle }}\right)
^{2}-A\left( 1+\frac{\Delta \hat{f}\left( X\right) +\Delta \hat{r}\left(
X\right) }{2}\right) F_{1}^{2}}{\left( \left\langle \hat{f}\right\rangle +%
\frac{\left\langle \hat{S}_{E}^{B}\right\rangle +\left\langle \hat{S}%
_{L}^{B}\right\rangle }{1-\bar{S}}\frac{\left\langle \bar{K}\right\rangle
\left\Vert \bar{\Psi}\right\Vert ^{2}}{\left\langle \hat{K}\right\rangle
\left\Vert \hat{\Psi}\right\Vert ^{2}}\left\langle \bar{f}\right\rangle
\right) ^{2}\left( \frac{\frac{\left\langle \hat{S}\left( X^{\prime
},X\right) \right\rangle _{X}}{1-\left\langle \hat{S}\left( X^{\prime
},X\right) \right\rangle _{X}\frac{\left\langle \hat{K}\right\rangle
\left\Vert \hat{\Psi}\right\Vert ^{2}}{\hat{K}_{X}\left\vert \hat{\Psi}%
\left( X\right) \right\vert ^{2}}}}{\frac{\left\langle \hat{S}\left(
X^{\prime },X\right) \right\rangle }{1-\left\langle \hat{S}\left( X^{\prime
},X\right) \right\rangle }}\right) ^{2}-\frac{1}{2}A\left( 1+\Delta \hat{f}%
\left( X\right) \right) F_{1}^{2}}\left( \hat{f}\left( X\right) -\bar{r}%
\right)
\end{equation*}%
with:%
\begin{eqnarray}
F_{1} &=&\left( 1-\hat{S}\right) \hat{f}\left( X\right) +\left\langle \hat{S}%
\left( X^{\prime },X\right) \right\rangle _{X^{\prime }}\left\langle \hat{f}%
\right\rangle \\
&&+\left( \left\langle \hat{S}_{E}^{B}\left( X,X^{\prime }\right)
\right\rangle _{X^{\prime }}+\left\langle \hat{S}_{L}^{B}\left( X,X^{\prime
}\right) \right\rangle _{X^{\prime }}+\left\langle \hat{S}\left( X^{\prime
},X\right) \right\rangle _{X^{\prime }}\frac{\left\langle \hat{S}%
_{E}^{B}\right\rangle +\left\langle \hat{S}_{L}^{B}\right\rangle }{%
1-\left\langle \hat{S}\right\rangle }\right) \frac{\left\langle \bar{K}%
\right\rangle \left\Vert \bar{\Psi}\right\Vert ^{2}}{\left\langle \hat{K}%
\right\rangle \left\Vert \hat{\Psi}\right\Vert ^{2}}\left\langle \bar{f}%
\right\rangle  \notag
\end{eqnarray}%
and $F_{1}$ can be obtained to the first approximation: 
\begin{eqnarray}
&&F_{1} \\
&\rightarrow &\left( 1-2z\right) \hat{f}\left( X\right) +2z\left\langle \hat{%
f}\right\rangle  \notag \\
&&+\left( 4x+\left\langle \hat{S}_{L}^{B}\left( X,X\right) \right\rangle
_{X}+2z\frac{4x+\left\langle \hat{S}_{L}^{B}\right\rangle }{1-2z}\right) 
\frac{\left\langle \bar{K}\right\rangle \left\Vert \bar{\Psi}\right\Vert ^{2}%
}{\left\langle \hat{K}\right\rangle \left\Vert \hat{\Psi}\right\Vert ^{2}}%
\left\langle \bar{f}\right\rangle  \notag \\
&\rightarrow &\left( 1-2z\right) \hat{f}\left( X\right) +2z\left\langle \hat{%
f}\right\rangle +\left( 4x+\left( 1-2x\right) \kappa z+2z\frac{4x+\left(
1-2x\right) \kappa z}{1-2z}\right) \frac{\left\langle \bar{K}\right\rangle
\left\Vert \bar{\Psi}\right\Vert ^{2}}{\left\langle \hat{K}\right\rangle
\left\Vert \hat{\Psi}\right\Vert ^{2}}\left\langle \bar{f}\right\rangle 
\notag \\
&=&\left( 1-2z\right) \hat{f}\left( X\right) +2z\left\langle \hat{f}%
\right\rangle +\frac{4x+\left( 1-2x\right) \kappa z}{1-2z}\frac{\left\langle 
\bar{K}\right\rangle \left\Vert \bar{\Psi}\right\Vert ^{2}}{\left\langle 
\hat{K}\right\rangle \left\Vert \hat{\Psi}\right\Vert ^{2}}\left\langle \bar{%
f}\right\rangle  \notag \\
&\rightarrow &\left( 1-2z\right) \left( \hat{f}\left( X\right) -r\right)
+2z\left( \left\langle \hat{f}\right\rangle -r\right) +C\left( \left\langle 
\bar{f}\right\rangle -\left( \kappa +1\right) r\right) +\left( C\left(
\kappa +1\right) +1\right) r  \notag
\end{eqnarray}%
We also have: 
\begin{eqnarray*}
&&\left\langle \hat{f}\right\rangle +\frac{\left\langle \hat{S}%
_{E}^{B}\right\rangle +\left\langle \hat{S}_{L}^{B}\right\rangle }{1-\bar{S}}%
\frac{\left\langle \bar{K}\right\rangle \left\Vert \bar{\Psi}\right\Vert ^{2}%
}{\left\langle \hat{K}\right\rangle \left\Vert \hat{\Psi}\right\Vert ^{2}}%
\left\langle \bar{f}\right\rangle \\
&\rightarrow &\left\langle \hat{f}\right\rangle +\frac{4x+\kappa z}{1-2x}%
\frac{\left\langle \bar{K}\right\rangle \left\Vert \bar{\Psi}\right\Vert ^{2}%
}{\left\langle \hat{K}\right\rangle \left\Vert \hat{\Psi}\right\Vert ^{2}}%
\left\langle \bar{f}\right\rangle \\
&=&\left\langle \hat{f}\right\rangle -r+C\left( \left\langle \bar{f}%
\right\rangle -\left( \kappa +1\right) r\right) +\left( C\left( \kappa
+1\right) +1\right) r
\end{eqnarray*}%
We will define:%
\begin{equation*}
C=\frac{4x+\kappa z}{1-2x}\frac{\left\langle \bar{K}\right\rangle \left\Vert 
\bar{\Psi}\right\Vert ^{2}}{\left\langle \hat{K}\right\rangle \left\Vert 
\hat{\Psi}\right\Vert ^{2}}
\end{equation*}%
and: 
\begin{eqnarray*}
L &=&C\left( \kappa +1\right) +1=\frac{4x+\kappa z}{1-2x}\frac{\left\langle 
\bar{K}\right\rangle \left\Vert \bar{\Psi}\right\Vert ^{2}}{\left\langle 
\hat{K}\right\rangle \left\Vert \hat{\Psi}\right\Vert ^{2}}\left( \kappa
+1\right) +1 \\
&=&1+\frac{2\left( 1-\bar{S}\right) ^{2}\left\Vert \bar{\Psi}_{0}\right\Vert
^{4}}{\left( 1-\hat{S}\right) ^{2}\left\Vert \hat{\Psi}_{0}\right\Vert
^{4}\left( 1+\kappa \right) ^{2}\left( 1+2\frac{\kappa \left( 1-2z\right) 2z%
}{\left( 1+\kappa \right) }\frac{\left\Vert \bar{\Psi}_{0}\right\Vert ^{2}}{%
\left\Vert \hat{\Psi}_{0}\right\Vert ^{2}}+\sqrt{1+4\frac{\kappa \left(
1-2z\right) 2z}{\left( 1+\kappa \right) }\frac{\left\Vert \bar{\Psi}%
_{0}\right\Vert ^{2}}{\left\Vert \hat{\Psi}_{0}\right\Vert ^{2}}}\right) }%
\frac{4x+\kappa z}{1-2x}\left( \kappa +1\right) \\
&\simeq &1+\frac{2\left( 1-2x\right) ^{2}\left\Vert \bar{\Psi}%
_{0}\right\Vert ^{4}}{\left( 1-2z\right) ^{2}\left\Vert \hat{\Psi}%
_{0}\right\Vert ^{4}\left( 1+\kappa \right) \left( 1+2\frac{\kappa \left(
1-2z\right) 2z}{\left( 1+\kappa \right) }\frac{\left\Vert \bar{\Psi}%
_{0}\right\Vert ^{2}}{\left\Vert \hat{\Psi}_{0}\right\Vert ^{2}}+\sqrt{1+4%
\frac{\kappa \left( 1-2z\right) 2z}{\left( 1+\kappa \right) }\frac{%
\left\Vert \bar{\Psi}_{0}\right\Vert ^{2}}{\left\Vert \hat{\Psi}%
_{0}\right\Vert ^{2}}}\right) }\frac{4x+\kappa z}{1-2x}
\end{eqnarray*}%
we can approximate:%
\begin{equation}
L\simeq 1+\frac{2z\left( 1-2x\right) \left\Vert \bar{\Psi}_{0}\right\Vert
^{4}}{\left( 1-2z\right) ^{2}\left\Vert \hat{\Psi}_{0}\right\Vert ^{4}\left(
2+8z\frac{\left\Vert \bar{\Psi}_{0}\right\Vert ^{2}}{\left\Vert \hat{\Psi}%
_{0}\right\Vert ^{2}}\right) }  \label{PRM}
\end{equation}%
As a consequence, the investors' return equation is:%
\begin{eqnarray*}
&&0=\left\{ \left( \left\langle \hat{f}\right\rangle -r+C\left( \left\langle 
\bar{f}\right\rangle -\left( \kappa +1\right) r\right) +\left( L\right)
r\right) ^{2}\left( \frac{\frac{\left\langle \hat{S}\left( X^{\prime
},X\right) \right\rangle _{X}}{1-\left\langle \hat{S}\left( X^{\prime
},X\right) \right\rangle _{X}\frac{\left\langle \hat{K}\right\rangle
\left\Vert \hat{\Psi}\right\Vert ^{2}}{\hat{K}_{X}\left\vert \hat{\Psi}%
\left( X\right) \right\vert ^{2}}}}{\frac{\left\langle \hat{S}\left(
X^{\prime },X\right) \right\rangle }{1-\left\langle \hat{S}\left( X^{\prime
},X\right) \right\rangle }}\right) ^{2}\right. \\
&&\left. -z\left( 1+\Delta \hat{f}\left( X\right) \right) B^{2}\right\}
^{-1}\left\{ \left( \left\langle \hat{f}\right\rangle -r+C\left(
\left\langle \bar{f}\right\rangle -\left( \kappa +1\right) r\right) +\left(
L\right) r\right) ^{2}\left( \frac{\frac{\left\langle \hat{S}\left(
X^{\prime },X\right) \right\rangle _{X}}{1-\left\langle \hat{S}\left(
X^{\prime },X\right) \right\rangle _{X}\frac{\left\langle \hat{K}%
\right\rangle \left\Vert \hat{\Psi}\right\Vert ^{2}}{\hat{K}_{X}\left\vert 
\hat{\Psi}\left( X\right) \right\vert ^{2}}}}{\frac{\left\langle \hat{S}%
\left( X^{\prime },X\right) \right\rangle }{1-\left\langle \hat{S}\left(
X^{\prime },X\right) \right\rangle }}\right) ^{2}\right. \\
&&\left. -2z\left( 1+\frac{\Delta \hat{f}\left( X\right) +\Delta \hat{r}%
\left( X\right) }{2}\right) B^{2}\right\} \left( \hat{f}\left( X\right) -%
\bar{r}\right) \\
&&-z\left( 1-\left( 1-2z\right) \Delta \left( \frac{f\left( X\right)
+r\left( X\right) }{2}\right) +\frac{\left\langle \hat{f}\left( X^{\prime
}\right) \right\rangle -\left\langle \hat{r}\left( X^{\prime }\right)
\right\rangle _{\hat{w}_{2}}}{2}\right) \frac{1-2z}{1-z}\left( \left\langle 
\hat{f}\left( X^{\prime }\right) \right\rangle -\left\langle \bar{r}%
\right\rangle \right) \\
&&-\frac{1-2z}{2}\left( f\left( X^{\prime }\right) -r\right)
\end{eqnarray*}

\bigskip with:%
\begin{equation*}
B=\left( \left( 1-2z\right) \left( \hat{f}\left( X\right) -r\right)
+2z\left( \left\langle \hat{f}\right\rangle -r\right) +C\left( \left\langle 
\bar{f}\right\rangle -\left( \kappa +1\right) r\right) +\left( L\right)
r\right)
\end{equation*}%
Replacing for $\frac{\Delta \hat{f}\left( X\right) +\Delta \hat{r}\left(
X\right) }{2}$ and $\Delta \hat{f}\left( X\right) $, we find:

\begin{eqnarray*}
&&0=\left\{ \left( \frac{\left\langle \hat{f}\right\rangle -r+C\left(
\left\langle \bar{f}\right\rangle -\left( \kappa +1\right) r\right) }{\left(
L\right) r}+1\right) ^{2}\left( \frac{\frac{\left\langle \hat{S}\left(
X^{\prime },X\right) \right\rangle _{X}}{1-\left\langle \hat{S}\left(
X^{\prime },X\right) \right\rangle _{X}\frac{\left\langle \hat{K}%
\right\rangle \left\Vert \hat{\Psi}\right\Vert ^{2}}{\hat{K}_{X}\left\vert 
\hat{\Psi}\left( X\right) \right\vert ^{2}}}}{\frac{\left\langle \hat{S}%
\left( X^{\prime },X\right) \right\rangle }{1-\left\langle \hat{S}\left(
X^{\prime },X\right) \right\rangle }}\right) ^{2}\right. \\
&&\left. -z\left( 1+\left( \hat{f}\left( X^{\prime }\right) -\bar{r}\right)
-z\left( \left\langle \hat{f}\left( X^{\prime }\right) \right\rangle
-r\right) -\frac{1-2z}{2}\left( \left\langle f\left( X\right) \right\rangle
-r\right) \right) \left( \frac{B}{Lr}\right) ^{2}\right\} ^{-1} \\
&&\times \left\{ \left( \frac{\left\langle \hat{f}\right\rangle -r+C\left(
\left\langle \bar{f}\right\rangle -\left( \kappa +1\right) r\right) }{\left(
L\right) r}+1\right) ^{2}\left( \frac{\frac{\left\langle \hat{S}\left(
X^{\prime },X\right) \right\rangle _{X}}{1-\left\langle \hat{S}\left(
X^{\prime },X\right) \right\rangle _{X}\frac{\left\langle \hat{K}%
\right\rangle \left\Vert \hat{\Psi}\right\Vert ^{2}}{\hat{K}_{X}\left\vert 
\hat{\Psi}\left( X\right) \right\vert ^{2}}}}{\frac{\left\langle \hat{S}%
\left( X^{\prime },X\right) \right\rangle }{1-\left\langle \hat{S}\left(
X^{\prime },X\right) \right\rangle }}\right) ^{2}\right. \\
&&\left. -2z\left( 1+\frac{1}{2}\left( \hat{f}\left( X^{\prime }\right) -%
\bar{r}\right) -z\left( \left\langle \hat{f}\left( X^{\prime }\right)
\right\rangle -\left\langle \bar{r}\right\rangle \right) -\frac{1-2z}{2}%
\left( \left\langle f\left( X\right) \right\rangle -r\right) \right) \left( 
\frac{B}{Lr}\right) ^{2-1}\right\} \left( \hat{f}\left( X\right) -\bar{r}%
\right) \\
&&-z\left( 1-\left( 1-2z\right) \Delta \left( \frac{f\left( X\right)
+r\left( X\right) }{2}\right) +\frac{\left\langle \hat{f}\left( X^{\prime
}\right) \right\rangle -\left\langle \hat{r}\left( X^{\prime }\right)
\right\rangle _{\hat{w}_{2}}}{2}\right) \\
&&\times \frac{1-2z}{1-z}\left( \left\langle \hat{f}\left( X^{\prime
}\right) \right\rangle -\left\langle \bar{r}\right\rangle \right) -\frac{1-2z%
}{2}\left( f\left( X^{\prime }\right) -r\right)
\end{eqnarray*}%
We replace:%
\begin{eqnarray*}
x &=&\hat{f}\left( X\right) -r \\
t &=&\left\langle \hat{f}\left( X^{\prime }\right) \right\rangle
-\left\langle \bar{r}\right\rangle \\
v &=&\left\langle \bar{f}\right\rangle -r \\
v^{\prime } &=&\left\langle f\right\rangle -r
\end{eqnarray*}%
\begin{eqnarray*}
\hat{D} &=&\left( \frac{\frac{\left\langle \hat{S}\left( X^{\prime
},X\right) \right\rangle _{X}}{1-\left\langle \hat{S}\left( X^{\prime
},X\right) \right\rangle _{X}\frac{\left\langle \hat{K}\right\rangle
\left\Vert \hat{\Psi}\right\Vert ^{2}}{\hat{K}_{X}\left\vert \hat{\Psi}%
\left( X\right) \right\vert ^{2}}}}{\frac{\left\langle \hat{S}\left(
X^{\prime },X\right) \right\rangle }{1-\left\langle \hat{S}\left( X^{\prime
},X\right) \right\rangle }}\right) ^{2} \\
&\rightarrow &\left( \frac{2z\left( 1+\frac{1}{2}x-zt-\frac{1-2z}{2}%
v^{\prime }\right) \frac{1-2z\left( 1+\frac{1-2z}{2}t-\frac{1-2z}{2}%
v^{\prime }\right) }{2z\left( 1+\frac{1-2z}{2}t-\frac{1-2z}{2}v^{\prime
}\right) }}{1-2z\left( 1+\frac{1}{2}x-zt-\frac{1-2z}{2}v^{\prime }\right)
\left( \left( \frac{\left( \left( 1-2z\right) \left( \hat{f}\left( X\right)
-r\right) +2z\left( \left\langle \hat{f}\right\rangle -r\right) +C\left(
\left\langle \bar{f}\right\rangle -\left( \kappa +1\right) r\right) +\left(
L\right) r\right) }{\left\langle \hat{f}\right\rangle -r+C\left(
\left\langle \bar{f}\right\rangle -\left( \kappa +1\right) r\right) +\left(
L\right) r}\right) \frac{\frac{2z\left( 1+\frac{1-2z}{2}t-\frac{1-2z}{2}%
v^{\prime }\right) }{1-2z\left( 1+\frac{1-2z}{2}t-\frac{1-2z}{2}v^{\prime
}\right) }}{\frac{2z\left( 1+\frac{1}{2}x-zt-\frac{1-2z}{2}v^{\prime
}\right) }{1-2z\left( 1+\frac{1}{2}x-zt-\frac{1-2z}{2}v^{\prime }\right) }}%
\right) ^{2}}\right) ^{2} \\
&=&\left( \frac{2z\left( 1+\frac{1}{2}x-zt-\frac{1-2z}{2}v^{\prime }\right) 
\frac{1-2z\left( 1+\frac{1-2z}{2}t-\frac{1-2z}{2}v^{\prime }\right) }{%
2z\left( 1+\frac{1-2z}{2}t-\frac{1-2z}{2}v^{\prime }\right) }}{1-2z\left( 1+%
\frac{1}{2}x-zt-\frac{1-2z}{2}v^{\prime }\right) \left( \left( \frac{\left(
1-2z\right) x+2zt+Cv+L}{t+Cv+L}\right) \frac{\frac{2z\left( 1+\frac{1-2z}{2}%
t-\frac{1-2z}{2}v^{\prime }\right) }{1-2z\left( 1+\frac{1-2z}{2}t-\frac{1-2z%
}{2}v^{\prime }\right) }}{\frac{2z\left( 1+\frac{1}{2}x-zt-\frac{1-2z}{2}%
v^{\prime }\right) }{1-2z\left( 1+\frac{1}{2}x-zt-\frac{1-2z}{2}v^{\prime
}\right) }}\right) ^{2}}\right) ^{2}
\end{eqnarray*}%
and the equation writes:%
\begin{eqnarray*}
&&0=\frac{\left( \frac{t+Cv}{\left( L\right) }+1\right) ^{2}\hat{D}-2z\left(
1+\frac{1}{2}x-zt-\frac{1-2z}{2}v^{\prime }\right) \left( \frac{\left(
1-2z\right) x+2zt+Cv}{\left( L\right) }+1\right) ^{2}}{\left( \frac{t+Cv}{%
\left( L\right) }+1\right) ^{2}\hat{D}-z\left( 1+x-zt-\frac{1-2z}{2}%
v^{\prime }\right) \left( \frac{\left( 1-2z\right) x+2zt+Cv}{\left( L\right) 
}+1\right) ^{2}} \\
&&\times \left( \hat{f}\left( X\right) -\bar{r}\right) -z\left( 1-\left(
1-2z\right) \Delta \left( \frac{f\left( X\right) +r\left( X\right) }{2}%
\right) +\frac{\left\langle \hat{f}\left( X^{\prime }\right) \right\rangle
-\left\langle \hat{r}\left( X^{\prime }\right) \right\rangle _{\hat{w}_{2}}}{%
2}\right) \\
&&\times \frac{1-2z}{1-z}\left( \left\langle \hat{f}\left( X^{\prime
}\right) \right\rangle -\left\langle \bar{r}\right\rangle \right) -\frac{1-2z%
}{2}\left( f\left( X^{\prime }\right) -r\right)
\end{eqnarray*}%
In first approximation, we can consider $\nu \simeq 0$. A first order
expansion yields:%
\begin{equation*}
0=a\left( z,L\right) x^{2}+\left( \frac{2z-1}{z-1}+z\left( b\left(
z,L\right) t+\frac{1}{2}\frac{1-2z}{\left( z-1\right) ^{2}}v^{\prime
}\right) \right) x-z\frac{1-2z}{1-z}t-\frac{1-2z}{2}\left( f\left( X^{\prime
}\right) -r\right)
\end{equation*}%
with:%
\begin{equation*}
a\left( z,L\right) =-z\frac{%
-L-24z-26Lz^{2}+12Lz^{3}-8Lz^{4}+9Lz+112z^{2}-224z^{3}+160z^{4}+2}{L\left(
z-1\right) ^{2}\left( 1-2z\right) ^{3}}
\end{equation*}%
\begin{equation*}
b\left( z,L\right) =\frac{2L+28z+4Lz^{2}-4Lz^{3}-13Lz-88z^{2}+80z^{3}-2}{%
L\left( 2z-1\right) ^{2}\left( z-1\right) ^{2}}>0
\end{equation*}%
\begin{equation*}
0=2L+28z+4Lz^{2}-4Lz^{3}-13Lz-88z^{2}+80z^{3}-2
\end{equation*}%
\begin{equation*}
2\left( 1+4z\right) +28z+4\left( 1+4z\right) z^{2}-4\left( 1+4z\right)
z^{3}-13\left( 1+4z\right) z-88z^{2}+80z^{3}-2
\end{equation*}%
The coefficient $a\left( z\right) $ is positive for:%
\begin{equation*}
L>\frac{\left( -24z+112z^{2}-224z^{3}+160z^{4}+2\right) }{%
-9z+26z^{2}-12z^{3}+8z^{4}+1}
\end{equation*}%
expressed in terms of average fields, this condition writes:%
\begin{equation*}
1+\frac{z\left\Vert \bar{\Psi}_{0}\right\Vert ^{4}\left( 1-2x\right) }{%
\left( 1-2z\right) ^{2}\left\Vert \hat{\Psi}_{0}\right\Vert ^{4}\left( 1+4z%
\frac{\left\Vert \bar{\Psi}_{0}\right\Vert ^{2}}{\left\Vert \hat{\Psi}%
_{0}\right\Vert ^{2}}\right) }>A
\end{equation*}%
where:%
\begin{equation*}
A=\frac{2-24z+112z^{2}-224z^{3}+160z^{4}}{1-9z+26z^{2}-12z^{3}+8z^{4}}
\end{equation*}

\bigskip Thus, the condition on the ratio of average densities for $a\left(
z\right) >0$ is:%
\begin{equation*}
\frac{\left\Vert \bar{\Psi}_{0}\right\Vert ^{2}}{\left\Vert \hat{\Psi}%
_{0}\right\Vert ^{2}}>\frac{2\left( 2z-1\right) ^{2}\left( A-1\right) }{%
\left( 1-2x\right) }+\frac{\sqrt{\left( 2\left( 2z-1\right) ^{2}\left(
A-1\right) \right) ^{2}-\left( 1-2x\right) \left( \left( A-1\right) \left(
4-4z-\frac{1}{z}\right) \right) }}{\left( 1-2x\right) }
\end{equation*}

In this case, $\allowbreak $the positive solution is:%
\begin{eqnarray*}
x &=&\sqrt{\frac{\left( \frac{2z-1}{z-1}+z\left( b\left( z,L\right) t+\frac{1%
}{2}\frac{1-2z}{\left( z-1\right) ^{2}}v^{\prime }\right) \right) ^{2}}{%
4a^{2}\left( z,L\right) }+\frac{\left( z\frac{1-2z}{1-z}t+\frac{1-2z}{2}%
\left( f\left( X^{\prime }\right) -r\right) \right) }{a\left( z,L\right) }}-%
\frac{\frac{2z-1}{z-1}+z\left( b\left( z,L\right) t+\frac{1}{2}\frac{1-2z}{%
\left( z-1\right) ^{2}}v^{\prime }\right) }{2a\left( z,L\right) } \\
&\simeq &\frac{2z\frac{1-2z}{1-z}t+\frac{\left( 1-2z\right) }{2}\left(
f\left( X^{\prime }\right) -r\right) }{\frac{2z-1}{z-1}-z\left( b\left(
z,L\right) t-\frac{1}{2}\frac{1-2z}{\left( z-1\right) ^{2}}v^{\prime
}\right) }
\end{eqnarray*}%
Coming back to the initial variables this becomes:%
\begin{eqnarray}
\hat{f}\left( X\right) -\left\langle \bar{r}\right\rangle &\simeq &\frac{z%
\frac{1-2z}{1-z}\left( \left\langle \hat{f}\left( X^{\prime }\right)
\right\rangle -\left\langle \bar{r}\right\rangle \right) +\frac{\left(
1-2z\right) }{2}\left( f\left( X^{\prime }\right) -\left\langle \bar{r}%
\right\rangle \right) }{\frac{1-2z}{1-z}-z\left( b\left( z,L\right) \frac{%
\left\langle \hat{f}\left( X^{\prime }\right) \right\rangle -\left\langle 
\bar{r}\right\rangle }{\left\langle \bar{r}\right\rangle }-\frac{1}{2}\frac{%
1-2z}{\left( z-1\right) ^{2}}\frac{f\left( X^{\prime }\right) -\left\langle 
\bar{r}\right\rangle }{\left\langle \bar{r}\right\rangle }\right) }
\label{Sln} \\
&\simeq &z\left( \left\langle \hat{f}\left( X^{\prime }\right) \right\rangle
-\left\langle \bar{r}\right\rangle \right) +\frac{\left( 1-z\right) }{2}%
\left( f\left( X^{\prime }\right) -\left\langle \bar{r}\right\rangle \right)
\notag
\end{eqnarray}%
with:%
\begin{equation*}
f\left( X^{\prime }\right) =f_{1}^{dr}\left( X\right) =\frac{f_{1}\left(
X\right) }{K_{X}^{r}}-\frac{C}{K_{X}}
\end{equation*}%
\begin{equation*}
K_{X}=\left( 1-\left( S\left( X,X\right) \frac{\hat{K}_{X}\left\vert \hat{%
\Psi}\left( X\right) \right\vert ^{2}}{K_{X}\left\vert \Psi \left( X\right)
\right\vert ^{2}}+S^{B}\left( X,X\right) \frac{\bar{K}_{X}\left\vert \bar{%
\Psi}\left( X\right) \right\vert ^{2}}{K_{X}\left\vert \Psi \left( X\right)
\right\vert ^{2}}\right) \right) \left( \left( \frac{2\epsilon }{3\sigma _{%
\hat{K}}^{2}}\right) ^{\frac{r}{2}}\frac{f_{1}\left( X\right) }{C_{0}+\frac{%
S_{L}\left( X\right) }{1-S_{E}\left( X\right) }\bar{r}}\right) ^{\frac{1}{r}}
\end{equation*}%
Solution (\ref{Sln}) is close to the average solution.

When:%
\begin{equation*}
\frac{\left\Vert \bar{\Psi}_{0}\right\Vert ^{2}}{\left\Vert \hat{\Psi}%
_{0}\right\Vert ^{2}}<\frac{2\left( 2z-1\right) ^{2}\left( A-1\right) }{%
\left( 1-2x\right) }+\frac{\sqrt{\left( 2\left( 2z-1\right) ^{2}\left(
A-1\right) \right) ^{2}-\left( 1-2x\right) \left( \left( A-1\right) \left(
4-4z-\frac{1}{z}\right) \right) }}{\left( 1-2x\right) }
\end{equation*}%
coefficient $a\left( z,L\right) <0$ and there are two positive solutions:%
\begin{eqnarray*}
&&\hat{f}\left( X^{\prime }\right) -\left\langle \bar{r}\right\rangle \\
&=&\frac{\frac{2z-1}{z-1}+z\left( b\left( z,L\right) \left( \left\langle 
\hat{f}\left( X^{\prime }\right) \right\rangle -\left\langle \bar{r}%
\right\rangle \right) +\frac{1}{2}\frac{1-2z}{\left( z-1\right) ^{2}}\left(
f\left( X^{\prime }\right) -\left\langle \bar{r}\right\rangle \right)
\right) }{2\left\vert a\left( z,L\right) \right\vert } \\
&&+\left\{ \frac{\left( \frac{2z-1}{z-1}+z\left( b\left( z,L\right) \left(
\left\langle \hat{f}\left( X^{\prime }\right) \right\rangle -\left\langle 
\bar{r}\right\rangle \right) +\frac{1}{2}\frac{1-2z}{\left( z-1\right) ^{2}}%
\left( f\left( X^{\prime }\right) -\left\langle \bar{r}\right\rangle \right)
\right) \right) ^{2}}{4a^{2}\left( z,L\right) }\right. \\
&&\left. -\frac{\left( z\frac{1-2z}{1-z}\left( \left\langle \hat{f}\left(
X^{\prime }\right) \right\rangle -\left\langle \bar{r}\right\rangle \right) +%
\frac{1-2z}{2}\left( f\left( X^{\prime }\right) -r\right) \right) }{%
\left\vert a\left( z,L\right) \right\vert }\right\} ^{\frac{1}{2}}
\end{eqnarray*}%
for the high-return solution, and:%
\begin{eqnarray*}
&&\hat{f}\left( X^{\prime }\right) -\left\langle \bar{r}\right\rangle \\
&=&\frac{\frac{2z-1}{z-1}+z\left( b\left( z,L\right) \left( \left\langle 
\hat{f}\left( X^{\prime }\right) \right\rangle -\left\langle \bar{r}%
\right\rangle \right) +\frac{1}{2}\frac{1-2z}{\left( z-1\right) ^{2}}\left(
f\left( X^{\prime }\right) -\left\langle \bar{r}\right\rangle \right)
\right) }{2\left\vert a\left( z,L\right) \right\vert } \\
&&-\left\{ \frac{\left( \frac{2z-1}{z-1}+z\left( b\left( z,L\right) \left(
\left\langle \hat{f}\left( X^{\prime }\right) \right\rangle -\left\langle 
\bar{r}\right\rangle \right) +\frac{1}{2}\frac{1-2z}{\left( z-1\right) ^{2}}%
\left( f\left( X^{\prime }\right) -\left\langle \bar{r}\right\rangle \right)
\right) \right) ^{2}}{4a^{2}\left( z,L\right) }\right. \\
&&\left. -\frac{\left( z\frac{1-2z}{1-z}\left( \left\langle \hat{f}\left(
X^{\prime }\right) \right\rangle -\left\langle \bar{r}\right\rangle \right) +%
\frac{1-2z}{2}\left( f\left( X^{\prime }\right) -r\right) \right) }{%
\left\vert a\left( z,L\right) \right\vert }\right\} ^{\frac{1}{2}}
\end{eqnarray*}%
for the low-return solution.

\subsubsection*{A10.5.2 Banks}

We look for solutions such that $\left\langle \bar{f}\left( X^{\prime
}\right) \right\rangle \simeq \left\langle \hat{f}\left( X^{\prime }\right)
\right\rangle \simeq \left\langle f_{1}\left( X\right) \right\rangle $ and
these average returns are close to $\bar{r}$.

Using \ (\ref{Gbr}):%
\begin{equation}
\bar{G}=S_{E}^{B}\left( X,X\right) \left( f_{1}\left( X\right) -\bar{r}%
\right)
\end{equation}%
and (\ref{Brs}) or (\ref{Brv}) write in first approximation:%
\begin{eqnarray}
0 &=&\frac{\left\langle \bar{f}\right\rangle ^{2}D-\bar{A}\left( 1+\frac{%
\bar{f}\left( X\right) +\Delta \bar{r}\left( X\right) -\tilde{f}}{2}\right)
\left( \bar{r}+\left( \bar{f}\left( X\right) -\bar{r}\right) \left(
1-2x\right) +2x\left( \left\langle \bar{f}\left( X\right) \right\rangle -%
\bar{r}\right) \right) ^{2}}{\left\langle \bar{f}\right\rangle ^{2}D-\frac{1%
}{2}\bar{A}\left( 1+\bar{f}\left( X\right) -\tilde{f}\right) \left( \bar{r}%
+\left( \bar{f}\left( X\right) -\bar{r}\right) \left( 1-2x\right) +2x\left(
\left\langle \bar{f}\left( X\right) \right\rangle -\bar{r}\right) \right)
^{2}} \\
&&-\left\langle \bar{S}_{E}\left( X^{\prime },X\right) \right\rangle
_{X^{\prime }}\frac{1-\left\langle \bar{S}\left( X^{\prime }\right)
\right\rangle }{1-\left\langle \bar{S}_{E}\left( X^{\prime }\right)
\right\rangle }\left( \left\langle \bar{f}\left( X^{\prime }\right)
\right\rangle -\bar{r}\right)  \notag \\
&&-\left\langle \hat{S}_{E}^{B}\left( X^{\prime },X\right) \right\rangle
_{X^{\prime }}\frac{1-\left\langle \hat{S}\left( X^{\prime }\right)
\right\rangle +\left\langle \hat{S}_{E}^{B}\left( X^{\prime }\right)
\right\rangle +\left\langle \hat{S}_{L}^{B}\left( X^{\prime }\right)
\right\rangle }{1-\left\langle \hat{S}_{E}\left( X^{\prime }\right)
\right\rangle +\left\langle \hat{S}_{E}^{B}\left( X^{\prime }\right)
\right\rangle }\left( \left\langle \hat{f}\left( X^{\prime }\right)
\right\rangle -\bar{r}\right) -S_{E}^{B}\left( X,X\right) \left( f_{1}\left(
X\right) _{dr}-\bar{r}\right)  \notag
\end{eqnarray}%
that is:%
\begin{eqnarray}
0 &=&\frac{\left\langle \bar{f}\right\rangle ^{2}D-\bar{A}\left( 1+\frac{%
\Delta \bar{f}\left( X\right) +\Delta \bar{r}\left( X\right) }{2}\right)
\left( \bar{r}+\left( \bar{f}\left( X\right) -\bar{r}\right) \left(
1-2x\right) +2x\left( \left\langle \bar{f}\left( X\right) \right\rangle -%
\bar{r}\right) \right) ^{2}}{\left\langle \bar{f}\right\rangle ^{2}D-\frac{1%
}{2}\bar{A}\left( 1+\Delta \bar{f}\left( X\right) \right) \left( \bar{r}%
+\left( \bar{f}\left( X\right) -\bar{r}\right) \left( 1-2x\right) +2x\left(
\left\langle \bar{f}\left( X\right) \right\rangle -\bar{r}\right) \right)
^{2}} \\
&&-\left\langle \bar{S}_{E}\left( X^{\prime },X\right) \right\rangle
_{X^{\prime }}\frac{1-\left\langle \bar{S}\left( X^{\prime }\right)
\right\rangle }{1-\left\langle \bar{S}_{E}\left( X^{\prime }\right)
\right\rangle }\left( \left\langle \bar{f}\left( X^{\prime }\right)
\right\rangle -\bar{r}\right)  \notag \\
&&-\left\langle \hat{S}_{E}^{B}\left( X^{\prime },X\right) \right\rangle
_{X^{\prime }}\frac{1-\left\langle \hat{S}\left( X^{\prime }\right)
\right\rangle +\left\langle \hat{S}_{E}^{B}\left( X^{\prime }\right)
\right\rangle +\left\langle \hat{S}_{L}^{B}\left( X^{\prime }\right)
\right\rangle }{1-\left\langle \hat{S}_{E}\left( X^{\prime }\right)
\right\rangle +\left\langle \hat{S}_{E}^{B}\left( X^{\prime }\right)
\right\rangle }\left( \left\langle \hat{f}\left( X^{\prime }\right)
\right\rangle -\bar{r}\right) -S_{E}^{B}\left( X,X\right) \left( f_{1}\left(
X\right) _{dr}-\bar{r}\right)  \notag
\end{eqnarray}%
where:%
\begin{equation*}
D=\frac{\frac{\left\langle \bar{S}\left( X^{\prime },X\right) \right\rangle
_{X^{\prime }}}{1-\left\langle \bar{S}\left( X^{\prime },X\right)
\right\rangle _{X^{\prime }}\left( \frac{\left( \left( 1-\left\langle \bar{S}%
\right\rangle \right) \bar{f}\left( X\right) +\left\langle \bar{S}\left(
X^{\prime },X\right) \right\rangle _{X^{\prime }}\left\langle \bar{f}%
\right\rangle \right) \frac{\left\langle \bar{S}\left( X^{\prime },X\right)
\right\rangle }{1-\left\langle \bar{S}\left( X^{\prime },X\right)
\right\rangle }}{\left\langle \bar{f}\right\rangle \frac{\left\langle \bar{S}%
\left( X^{\prime },X\right) \right\rangle _{X^{\prime }}}{1-\left\langle 
\bar{S}\left( X^{\prime },X\right) \right\rangle _{X^{\prime }}}}\right) ^{2}%
}}{\frac{\left\langle \bar{S}\left( X^{\prime },X\right) \right\rangle }{%
1-\left\langle \bar{S}\left( X^{\prime },X\right) \right\rangle }}
\end{equation*}%
\begin{equation*}
\Delta \bar{f}\left( X^{\prime }\right) =\bar{f}\left( X^{\prime }\right)
-\left( \left\langle \bar{w}\left( X\right) \right\rangle \frac{\left\langle 
\bar{f}\left( X^{\prime }\right) \right\rangle _{\bar{w}_{1}}+\left\langle 
\bar{r}\left( X^{\prime }\right) \right\rangle _{\bar{w}_{2}}}{2}%
+\left\langle \hat{w}_{1}^{B}\left( X\right) \right\rangle \left\langle \hat{%
f}\left( X^{\prime }\right) \right\rangle _{\hat{w}_{1}}+\left\langle
w_{1}^{B}\left( X\right) \right\rangle \left\langle f\left( X\right)
\right\rangle \right)
\end{equation*}%
where:%
\begin{eqnarray*}
\Delta \bar{f}\left( X^{\prime }\right) &=&\bar{f}\left( X^{\prime }\right)
-\left( x\left( \left\langle \bar{f}\left( X^{\prime }\right) \right\rangle
_{\bar{w}_{1}}+\left\langle \bar{r}\left( X^{\prime }\right) \right\rangle _{%
\bar{w}_{2}}\right) +4x\left\langle \hat{f}\left( X^{\prime }\right)
\right\rangle _{\hat{w}_{1}}+\left( 1-6x\right) \left\langle f\left(
X\right) \right\rangle \right) \\
&=&\bar{f}\left( X^{\prime }\right) -\left\langle \bar{r}\left( X^{\prime
}\right) \right\rangle -\left( x\left( \left\langle \bar{f}\left( X^{\prime
}\right) \right\rangle _{\bar{w}_{1}}-\left\langle \bar{r}\left( X^{\prime
}\right) \right\rangle \right) +4x\left( \left\langle \hat{f}\left(
X^{\prime }\right) \right\rangle -\left\langle \bar{r}\left( X^{\prime
}\right) \right\rangle \right) +\left( 1-6x\right) \left( \left\langle
f\left( X\right) \right\rangle -\left\langle \bar{r}\left( X^{\prime
}\right) \right\rangle \right) \right)
\end{eqnarray*}%
\begin{eqnarray*}
\Delta \left\langle \bar{r}\left( X^{\prime }\right) \right\rangle
&=&\left\langle \bar{r}\left( X^{\prime }\right) \right\rangle -\left(
x\left( \left\langle \bar{f}\left( X^{\prime }\right) \right\rangle _{\bar{w}%
_{1}}+\left\langle \bar{r}\left( X^{\prime }\right) \right\rangle _{\bar{w}%
_{2}}\right) +4x\left\langle \hat{f}\left( X^{\prime }\right) \right\rangle
_{\hat{w}_{1}}+\left( 1-6x\right) \left\langle f\left( X\right)
\right\rangle \right) \\
&=&-\left( x\left( \left\langle \bar{f}\left( X^{\prime }\right)
\right\rangle _{\bar{w}_{1}}-\left\langle \bar{r}\left( X^{\prime }\right)
\right\rangle \right) +4x\left( \left\langle \hat{f}\left( X^{\prime
}\right) \right\rangle -\left\langle \bar{r}\left( X^{\prime }\right)
\right\rangle \right) +\left( 1-6x\right) \left( \left\langle f\left(
X\right) \right\rangle -\left\langle \bar{r}\left( X^{\prime }\right)
\right\rangle \right) \right)
\end{eqnarray*}%
and:%
\begin{eqnarray*}
&&\frac{\Delta \bar{f}\left( X^{\prime }\right) +\Delta \left\langle \bar{r}%
\left( X^{\prime }\right) \right\rangle }{2} \\
&=&\frac{\bar{f}\left( X^{\prime }\right) -\left\langle \bar{r}\left(
X^{\prime }\right) \right\rangle }{2}-\left( x\left( \left\langle \bar{f}%
\left( X^{\prime }\right) \right\rangle _{\bar{w}_{1}}-\left\langle \bar{r}%
\left( X^{\prime }\right) \right\rangle \right) +4x\left( \left\langle \hat{f%
}\left( X^{\prime }\right) \right\rangle -\left\langle \bar{r}\left(
X^{\prime }\right) \right\rangle \right) +\left( 1-6x\right) \left(
\left\langle f\left( X\right) \right\rangle -\left\langle \bar{r}\left(
X^{\prime }\right) \right\rangle \right) \right)
\end{eqnarray*}

Replacing:%
\begin{eqnarray*}
\frac{1-\left\langle \bar{S}\left( X^{\prime }\right) \right\rangle }{%
1-\left\langle \bar{S}_{E}\left( X^{\prime }\right) \right\rangle }
&\rightarrow &\left\langle \overline{DF}\left( X^{\prime }\right)
\right\rangle \\
\frac{1-\left\langle \hat{S}\left( X^{\prime }\right) \right\rangle
+\left\langle \hat{S}_{E}^{B}\left( X^{\prime }\right) \right\rangle
+\left\langle \hat{S}_{L}^{B}\left( X^{\prime }\right) \right\rangle }{%
1-\left\langle \hat{S}_{E}\left( X^{\prime }\right) \right\rangle
+\left\langle \hat{S}_{E}^{B}\left( X^{\prime }\right) \right\rangle }
&\rightarrow &\left\langle \widehat{DF}\left( X^{\prime }\right)
\right\rangle
\end{eqnarray*}%
this leads to the expanded formula:

\begin{eqnarray}
0 &=&\frac{\left\langle \bar{f}\right\rangle ^{2}D-\bar{A}\left( 1+\frac{%
\bar{f}\left( X^{\prime }\right) -\left\langle \bar{r}\left( X^{\prime
}\right) \right\rangle }{2}-D\right) \left( \bar{r}+\left( \bar{f}\left(
X\right) -\bar{r}\right) \left( 1-2x\right) +2x\left( \left\langle \bar{f}%
\left( X\right) \right\rangle -\bar{r}\right) \right) ^{2}\frac{\left\Vert 
\bar{\Psi}_{0}\right\Vert ^{2}}{\left\Vert \bar{\Psi}_{0}\left( X\right)
\right\Vert ^{2}}}{\left\langle \bar{f}\right\rangle ^{2}D-\frac{1}{2}\bar{A}%
\left( 1+\bar{f}\left( X^{\prime }\right) -\left\langle \bar{r}\left(
X^{\prime }\right) \right\rangle -B\right) \left( \bar{r}+\left( \bar{f}%
\left( X\right) -\bar{r}\right) \left( 1-2x\right) +2x\left( \left\langle 
\bar{f}\left( X\right) \right\rangle -\bar{r}\right) \right) ^{2}\frac{%
\left\Vert \bar{\Psi}_{0}\right\Vert ^{2}}{\left\Vert \bar{\Psi}_{0}\left(
X\right) \right\Vert ^{2}}} \\
&&-\left\langle \bar{S}_{E}\left( X^{\prime },X\right) \right\rangle
\left\langle \overline{DF}\left( X^{\prime }\right) \right\rangle \left(
\left\langle \bar{f}\left( X^{\prime }\right) \right\rangle -\bar{r}\right) 
\notag \\
&&-\left\langle \hat{S}_{E}^{B}\left( X^{\prime },X\right) \right\rangle
_{X^{\prime }}\left\langle \widehat{DF}\left( X^{\prime }\right)
\right\rangle \left( \left\langle \hat{f}\left( X^{\prime }\right)
\right\rangle -\bar{r}\right) -S_{E}^{B}\left( X,X\right) \left( f_{1}\left(
X\right) _{dr}-\bar{r}\right)  \notag
\end{eqnarray}

\bigskip with:%
\begin{equation*}
B=x\left( \left\langle \bar{f}\left( X^{\prime }\right) \right\rangle _{\bar{%
w}_{1}}-\left\langle \bar{r}\left( X^{\prime }\right) \right\rangle \right)
+4x\left( \left\langle \hat{f}\left( X^{\prime }\right) \right\rangle
-\left\langle \bar{r}\left( X^{\prime }\right) \right\rangle \right) +\left(
1-6x\right) \left( \left\langle f\left( X\right) \right\rangle -\left\langle 
\bar{r}\left( X^{\prime }\right) \right\rangle \right)
\end{equation*}%
We then replace:%
\begin{eqnarray*}
\frac{\bar{f}\left( X^{\prime }\right) -\left\langle \bar{r}\left( X^{\prime
}\right) \right\rangle }{\left\langle \bar{r}\left( X^{\prime }\right)
\right\rangle } &\rightarrow &y+\kappa \\
\frac{\left\langle \bar{f}\left( X^{\prime }\right) \right\rangle
-\left\langle \bar{r}\left( X^{\prime }\right) \right\rangle }{\left\langle 
\bar{r}\left( X^{\prime }\right) \right\rangle } &\rightarrow &v+\kappa \\
\frac{\left\langle \hat{f}\left( X^{\prime }\right) \right\rangle
-\left\langle \bar{r}\left( X^{\prime }\right) \right\rangle }{\left\langle 
\bar{r}\left( X^{\prime }\right) \right\rangle } &\rightarrow &t \\
\left( \left\langle f\left( X\right) \right\rangle -\left\langle \bar{r}%
\left( X^{\prime }\right) \right\rangle \right) &\rightarrow &v^{\prime }
\end{eqnarray*}%
which leads to the equation:%
\begin{eqnarray}
0 &=&\frac{\left\langle \bar{f}\right\rangle ^{2}D-\bar{A}\left( 1+\frac{y}{2%
}-\left( xv+2xt+\frac{1-6x}{2}v^{\prime }\right) \right) \left( \bar{r}%
\left( 1+\kappa \right) +y\left( 1-2x\right) +2xv\right) ^{2}}{\left\langle 
\bar{f}\right\rangle ^{2}D-\frac{1}{2}\bar{A}\left( 1+y-\left( xv+2xt+\frac{%
1-6x}{2}v^{\prime }\right) \right) \left( \bar{r}\left( 1+\kappa \right)
+y\left( 1-2x\right) +2xv\right) ^{2}} \\
&&-\left\langle \bar{S}_{E}\left( X^{\prime },X\right) \right\rangle
_{X^{\prime }}\left\langle \overline{DF}\left( X^{\prime }\right)
\right\rangle \left( \left\langle \bar{f}\left( X^{\prime }\right)
\right\rangle -\bar{r}\right)  \notag \\
&&-\left\langle \hat{S}_{E}^{B}\left( X^{\prime },X\right) \right\rangle
_{X^{\prime }}\left\langle \widehat{DF}\left( X^{\prime }\right)
\right\rangle \left\langle \overline{DF}\left( X^{\prime }\right)
\right\rangle \left( \left\langle \hat{f}\left( X^{\prime }\right)
\right\rangle -\bar{r}\right) -S_{E}^{B}\left( X,X\right) \left( f_{1}\left(
X\right) _{dr}-\bar{r}\right)  \notag
\end{eqnarray}%
under the assumption:%
\begin{equation*}
\left( \left\langle f\left( X\right) \right\rangle -\left\langle \bar{r}%
\left( X^{\prime }\right) \right\rangle \right) <<1
\end{equation*}

\bigskip The equation rewrites:%
\begin{eqnarray}
0 &=&\frac{\left( \kappa +1+v\right) ^{2}D-2x\left( 1+\kappa \frac{1-6x}{2}+%
\frac{y}{2}-\left( xv+2xt+\frac{1-6x}{2}v^{\prime }\right) \right) \left(
\kappa +1+y\left( 1-2x\right) +2xv\right) ^{2}}{\left( \kappa +1+v\right)
^{2}D-x\left( 1+\kappa \frac{2-6x}{2}+y-\left( xv+2xt+\frac{1-6x}{2}%
v^{\prime }\right) \right) \left( \kappa +1+y\left( 1-2x\right) +2xv\right)
^{2}}  \label{qts} \\
&&-\left\langle \bar{S}_{E}\left( X^{\prime },X\right) \right\rangle
_{X^{\prime }}\left\langle \overline{DF}\left( X^{\prime }\right)
\right\rangle \left( \left\langle \bar{f}\left( X^{\prime }\right)
\right\rangle -\bar{r}\right)  \notag \\
&&-\left\langle \hat{S}_{E}^{B}\left( X^{\prime },X\right) \right\rangle
_{X^{\prime }}\left\langle \widehat{DF}\left( X^{\prime }\right)
\right\rangle \left( \left\langle \hat{f}\left( X^{\prime }\right)
\right\rangle -\bar{r}\right) -S_{E}^{B}\left( X,X\right) \left( f_{1}\left(
X\right) _{dr}-\bar{r}\right)  \notag
\end{eqnarray}%
where:%
\begin{eqnarray*}
D &=&\left( \frac{\frac{\left\langle \bar{S}\left( X^{\prime },X\right)
\right\rangle _{X^{\prime }}}{1-\left\langle \bar{S}\left( X^{\prime
},X\right) \right\rangle _{X^{\prime }}\left( \frac{\left( \left(
1-\left\langle \bar{S}\right\rangle \right) \bar{f}\left( X\right)
+\left\langle \bar{S}\left( X^{\prime },X\right) \right\rangle _{X^{\prime
}}\left\langle \bar{f}\right\rangle \right) \frac{\left\langle \bar{S}\left(
X^{\prime },X\right) \right\rangle }{1-\left\langle \bar{S}\left( X^{\prime
},X\right) \right\rangle }}{\left\langle \bar{f}\right\rangle \frac{%
\left\langle \bar{S}\left( X^{\prime },X\right) \right\rangle _{X^{\prime }}%
}{1-\left\langle \bar{S}\left( X^{\prime },X\right) \right\rangle
_{X^{\prime }}}}\right) ^{2}}}{\frac{\left\langle \bar{S}\left( X^{\prime
},X\right) \right\rangle }{1-\left\langle \bar{S}\left( X^{\prime },X\right)
\right\rangle }}\right) ^{2} \\
&\rightarrow &\left( \frac{\frac{2x\left( 1+\kappa \frac{1-6x}{2}+\frac{y}{2}%
-\left( xv+2xt+\frac{1-6x}{2}v^{\prime }\right) \right) }{1-2x\left(
1+\kappa \frac{1-6x}{2}+\frac{y}{2}-\left( xv+2xt+\frac{1-6x}{2}v^{\prime
}\right) \right) \left( \frac{\left( \left( 1-\left\langle \bar{S}%
\right\rangle \right) \bar{f}\left( X\right) +\left\langle \bar{S}\left(
X^{\prime },X\right) \right\rangle _{X^{\prime }}\left\langle \bar{f}%
\right\rangle \right) \frac{\left\langle \bar{S}\left( X^{\prime },X\right)
\right\rangle }{1-\left\langle \bar{S}\left( X^{\prime },X\right)
\right\rangle }}{\left\langle \bar{f}\right\rangle \frac{\left\langle \bar{S}%
\left( X^{\prime },X\right) \right\rangle _{X^{\prime }}}{1-\left\langle 
\bar{S}\left( X^{\prime },X\right) \right\rangle _{X^{\prime }}}}\right) ^{2}%
}}{\frac{2x\left( 1+\kappa \frac{1-6x}{2}+\frac{v}{2}-\left( xv+2xt+\frac{%
1-6x}{2}v^{\prime }\right) \right) }{1-2x\left( 1+\kappa \frac{1-6x}{2}+%
\frac{v}{2}-\left( xv+2xt+\frac{1-6x}{2}v^{\prime }\right) \right) }}\right)
^{2}
\end{eqnarray*}

and:%
\begin{eqnarray*}
&&\left( \frac{\left( \left( 1-\left\langle \bar{S}\right\rangle \right) 
\bar{f}\left( X\right) +\left\langle \bar{S}\left( X^{\prime },X\right)
\right\rangle _{X^{\prime }}\left\langle \bar{f}\right\rangle \right) \frac{%
\left\langle \bar{S}\left( X^{\prime },X\right) \right\rangle }{%
1-\left\langle \bar{S}\left( X^{\prime },X\right) \right\rangle }}{%
\left\langle \bar{f}\right\rangle \frac{\left\langle \bar{S}\left( X^{\prime
},X\right) \right\rangle _{X^{\prime }}}{1-\left\langle \bar{S}\left(
X^{\prime },X\right) \right\rangle _{X^{\prime }}}}\right) \\
&=&\frac{\left( 1-2x\left( 1+\kappa \frac{1-6x}{2}+\frac{v}{2}-\left( xv+2xt+%
\frac{1-6x}{2}v^{\prime }\right) \right) \right) \left( 1+\kappa +y\right)
+\left( 1+\kappa \frac{1-6x}{2}+\frac{y}{2}-\left( xv+2xt+\frac{1-6x}{2}%
v^{\prime }\right) \right) \left( 1+\kappa +v\right) }{\left( 1+\kappa
+v\right) \frac{2x\left( 1+\kappa \frac{1-6x}{2}+\frac{y}{2}-\left( xv+2xt+%
\frac{1-6x}{2}v^{\prime }\right) \right) }{1-2x\left( 1+\kappa \frac{1-6x}{2}%
+y-\left( xv+2xt+\frac{1-6x}{2}v^{\prime }\right) \right) }} \\
&&\times \frac{2x\left( 1+\kappa \frac{1-6x}{2}+\frac{v}{2}-\left( xv+2xt+%
\frac{1-6x}{2}v^{\prime }\right) \right) }{1-2x\left( 1+\kappa \frac{1-6x}{2}%
+\frac{v}{2}-\left( xv+2xt+\frac{1-6x}{2}v^{\prime }\right) \right) }
\end{eqnarray*}%
We approximate the first term in (\ref{qts}) by:$\bigskip $%
\begin{eqnarray*}
&&\frac{\left( \kappa +1+v\right) ^{2}D-2x\left( 1+\kappa \frac{1-6x}{2}+%
\frac{y}{2}-\left( xv+2xt+\frac{1-6x}{2}v^{\prime }\right) \right) \left(
\kappa +1+y\left( 1-2x\right) +2xv\right) ^{2}}{\left( \kappa +1+v\right)
^{2}D-x\left( 1+\kappa \frac{2-6x}{2}+y-\left( xv+2xt+\frac{1-6x}{2}%
v^{\prime }\right) \right) \left( \kappa +1+y\left( 1-2x\right) +2xv\right)
^{2}} \\
&\simeq &\frac{\left( D\left( \kappa +1\right) ^{2}-2x\left( \kappa -y\left(
2x-1\right) +1\right) ^{2}\left( \frac{1}{2}y-\kappa \left( 3x-\frac{1}{2}%
\right) +v^{\prime }\left( 3x-\frac{1}{2}\right) -2tx+1\right) \allowbreak
\right) }{D\left( \kappa +1\right) ^{2}-x\left( \kappa -y\left( 2x-1\right)
+1\right) ^{2}\left( y-\kappa \left( 3x-1\right) +v^{\prime }\left( 3x-\frac{%
1}{2}\right) -2tx+1\right) \allowbreak }
\end{eqnarray*}%
For $v\simeq 0$, \ and $x<<1$, equation (\ref{qts}) becomes at the lowest
order:$\allowbreak \allowbreak $%
\begin{eqnarray*}
&&0=\left( -2x\frac{y}{\left( \kappa +2\right) \left( \kappa +1\right) }%
+\allowbreak 1-x+\frac{1}{2}v^{\prime }x+2tx^{2}\right) y \\
&&-\left\langle \bar{S}_{E}\left( X^{\prime },X\right) \right\rangle
_{X^{\prime }}\left\langle \overline{DF}\left( X^{\prime }\right)
\right\rangle \left( \left\langle \bar{f}\left( X^{\prime }\right)
\right\rangle -\bar{r}\right) \\
&&-\left\langle \hat{S}_{E}^{B}\left( X^{\prime },X\right) \right\rangle
_{X^{\prime }}\left\langle \widehat{DF}\left( X^{\prime }\right)
\right\rangle \left( \left\langle \hat{f}\left( X^{\prime }\right)
\right\rangle -\bar{r}\right) -S_{E}^{B}\left( X,X\right) \left( f_{1}\left(
X\right) _{dr}-\bar{r}\right)
\end{eqnarray*}%
with solutions:%
\begin{equation*}
\bar{f}\left( X^{\prime }\right) -\left( \kappa +1\right) \left\langle \bar{r%
}\left( X^{\prime }\right) \right\rangle =\frac{\bar{R}_{exc}^{H,0}\left(
X\right) }{2}\pm \sqrt{\left( \left( \frac{\bar{R}_{exc}^{H,0}\left(
X\right) }{2}\right) ^{2}-\left( \kappa +2\right) \left( \kappa +1\right) 
\frac{1-x}{x}\bar{R}_{exc}^{L,0}\left( X\right) \right) ^{2}}
\end{equation*}%
where:%
\begin{equation*}
\bar{R}_{exc}^{H,0}\left( X\right) =\left( \kappa +2\right) \left( \kappa
+1\right) \left( \frac{1-x}{2x}\left\langle \bar{r}\left( X^{\prime }\right)
\right\rangle +\frac{\left( \left\langle f\left( X\right) \right\rangle
-\left\langle \bar{r}\left( X^{\prime }\right) \right\rangle \right) }{4}%
+\left( \left\langle \hat{f}\left( X^{\prime }\right) \right\rangle
-\left\langle \bar{r}\left( X^{\prime }\right) \right\rangle \right) x\right)
\end{equation*}%
and:%
\begin{eqnarray*}
&&\bar{R}_{exc}^{L,0}\left( X\right) =\left\langle \bar{S}_{E}\left(
X^{\prime },X\right) \right\rangle _{X^{\prime }}\left\langle \overline{DF}%
\left( X^{\prime }\right) \right\rangle \left( \left\langle \bar{f}\left(
X^{\prime }\right) \right\rangle -\bar{r}\right) \\
&&+\left\langle \hat{S}_{E}^{B}\left( X^{\prime },X\right) \right\rangle
_{X^{\prime }}\left\langle \widehat{DF}\left( X^{\prime }\right)
\right\rangle \left( \left\langle \hat{f}\left( X^{\prime }\right)
\right\rangle -\bar{r}\right) +S_{E}^{B}\left( X,X\right) \left( f_{1}\left(
X\right) _{dr}-\bar{r}\right)
\end{eqnarray*}%
Since $\frac{2x}{\left( \kappa +2\right) \left( \kappa +1\right) }<<1$, for
all level of uncertainty, the high-return solution corresponds to very high
returns, and the low-return solution is:%
\begin{eqnarray*}
&&\bar{f}\left( X^{\prime }\right) -\left( \kappa +1\right) \bar{r} \\
&\simeq &\left\langle \bar{S}_{E}\left( X^{\prime },X\right) \right\rangle
_{X^{\prime }}\left\langle \overline{DF}\left( X^{\prime }\right)
\right\rangle \left( \left\langle \bar{f}\left( X^{\prime }\right)
\right\rangle -\bar{r}\right) \\
&&+\left\langle \hat{S}_{E}^{B}\left( X^{\prime },X\right) \right\rangle
_{X^{\prime }}\left\langle \overline{DF}\left( X^{\prime }\right)
\right\rangle \left( \left\langle \hat{f}\left( X^{\prime }\right)
\right\rangle -\bar{r}\right) +S_{E}^{B}\left( X,X\right) \left( f_{1}\left(
X\right) _{dr}-\bar{r}\right) \\
&\simeq &x\left( \left\langle \bar{f}\left( X^{\prime }\right) \right\rangle
-\bar{r}\right) +4x\left( \left\langle \hat{f}\left( X^{\prime }\right)
\right\rangle -\bar{r}\right) +\left( 1-6x\right) \left( f_{1}\left(
X\right) _{dr}-\bar{r}\right)
\end{eqnarray*}

\section*{Appendix 11 Evaluation of stakes and return}

Once inward aggregate stakes of investment $\hat{S}_{\eta }\left( X\right) $%
, $\bar{S}_{\eta }\left( X\right) $... are found, we can retrieve the stakes 
$\hat{S}_{\eta }\left( X^{\prime },X\right) $, $\hat{S}_{\eta }\left(
X^{\prime },X\right) $... between two sectors.

\subsection{A11.1 Investors' stakes}

As in part one, the uncertainty coefficients are given by:%
\begin{equation*}
\frac{\hat{w}\left( X^{\prime },X\right) }{2}=\frac{\left( 1-\left( \gamma
\left\langle \hat{S}_{E}\left( X\right) \right\rangle \right) ^{2}\right) 
\hat{w}_{1}^{\left( 0\right) }\left( X^{\prime },X\right) }{1+\hat{w}%
_{1}^{\left( 0\right) }\left( X^{\prime },X\right) \left( 1-\left( \gamma
\left\langle \hat{S}_{E}\left( X\right) \right\rangle \right) ^{2}\right)
+\left( \gamma \left\langle \hat{S}_{E}\left( X_{1},X^{\prime }\right)
\right\rangle _{X_{1}}\right) ^{2}-\left( \gamma \left\langle \hat{S}%
_{E}\left( X\right) \right\rangle \right) ^{2}}
\end{equation*}%
and the shares are:%
\begin{eqnarray}
&&\hat{S}_{E}\left( X^{\prime },X\right) \\
&\rightarrow &\frac{\left( 1-\left( \gamma \left\langle \hat{S}_{E}\left(
X\right) \right\rangle \right) ^{2}\right) \hat{w}_{1}^{\left( 0\right)
}\left( X^{\prime },X\right) \left( 1+\Delta \hat{f}\left( X^{\prime
}\right) \right) }{1+\hat{w}_{1}^{\left( 0\right) }\left( X^{\prime
},X\right) \left( 1-\left( \gamma \left\langle \hat{S}_{E}\left( X\right)
\right\rangle \right) ^{2}\right) +\left( \gamma \left\langle \hat{S}%
_{E}\left( X_{1},X^{\prime }\right) \right\rangle _{X_{1}}\right)
^{2}-\left( \gamma \left\langle \hat{S}_{E}\left( X\right) \right\rangle
\right) ^{2}}  \notag
\end{eqnarray}

\begin{eqnarray*}
&&\hat{S}\left( X^{\prime },X\right) \\
&\rightarrow &\frac{2\left( 1-\left( \gamma \left\langle \hat{S}_{E}\left(
X\right) \right\rangle \right) ^{2}\right) \hat{w}_{1}^{\left( 0\right)
}\left( X^{\prime },X\right) \left( 1+\frac{\Delta \hat{f}\left( X^{\prime
}\right) +\Delta \hat{r}\left( X^{\prime }\right) }{2}\right) }{1+\hat{w}%
_{1}^{\left( 0\right) }\left( X^{\prime },X\right) \left( 1-\left( \gamma
\left\langle \hat{S}_{E}\left( X\right) \right\rangle \right) ^{2}\right)
+\left( \gamma \left\langle \hat{S}_{E}\left( X_{1},X^{\prime }\right)
\right\rangle _{X_{1}}\right) ^{2}-\left( \gamma \left\langle \hat{S}%
_{E}\left( X\right) \right\rangle \right) ^{2}}
\end{eqnarray*}%
To include decreasing returns, we consider:

\begin{equation*}
f\left( X\right) \rightarrow \frac{f_{1}\left( X\right) +\tau \Delta F_{\tau
}\left( \bar{R}\left( K,X\right) \right) }{K_{X}^{r}}-\frac{C}{K_{X}}-C_{0}
\end{equation*}

\subsection*{A11.2 Banks' stakes}

Using (\ref{hbn}), (\ref{hbt}):

\begin{eqnarray}
&&\left( \bar{w}\left( X^{\prime },X\right) \right) ^{-1} \\
&=&1+\frac{4}{\zeta ^{2}\bar{w}_{1}^{\left( 0\right) }\left( X^{\prime
},X\right) }\left\{ \frac{\bar{\zeta}^{2}\zeta ^{2}\left( 1+\frac{\left(
\gamma \left\langle \hat{S}_{E}\left( X_{1},\left( X^{\prime }\right)
^{\prime }\right) \right\rangle \right) ^{2}}{1-\left( \gamma \left\langle 
\hat{S}_{E}\left( X^{\prime },\left( X^{\prime }\right) ^{\prime }\right)
\right\rangle \right) ^{2}}\right) }{\left\langle \hat{w}_{1}^{\left(
0\right) B}\left( \left( X^{\prime }\right) ^{\prime },X^{\prime }\right)
\right\rangle _{\left( X^{\prime }\right) ^{\prime }}}+\xi ^{2}\right\} 
\notag \\
&&\times \left( \frac{1+\frac{\left( \bar{\gamma}\left\langle \bar{S}%
_{E}\left( X_{1},X^{\prime }\right) \right\rangle _{X_{1}}\right) ^{2}}{%
1-\left( \bar{\gamma}\left\langle \bar{S}_{E}\left( \left( X^{\prime
}\right) ^{\prime },X^{\prime }\right) \right\rangle \right) ^{2}}}{1+\frac{%
\left( \gamma \left\langle \hat{S}_{E}\left( X_{1},X^{\prime }\right)
\right\rangle _{X_{1}}\right) ^{2}}{1-\left( \gamma \left\langle \hat{S}%
_{E}\left( X^{\prime },\left( X^{\prime }\right) ^{\prime }\right)
\right\rangle \right) ^{2}}}\right) \left( w_{1}^{\left( 0\right) B}\left(
X^{\prime },X\right) +\frac{\zeta ^{2}}{\xi ^{2}}\left( 1+\frac{\left(
\gamma \left\langle \hat{S}_{E}\left( X_{1},X^{\prime }\right) \right\rangle
_{X_{1}}\right) ^{2}}{1-\left( \gamma \left\langle \hat{S}_{E}\left(
X^{\prime },\left( X^{\prime }\right) ^{\prime }\right) \right\rangle
\right) ^{2}}\right) \right)  \notag
\end{eqnarray}%
\begin{eqnarray}
&&\left( \hat{w}_{1}^{B}\left( X^{\prime },X\right) \right) ^{-1} \\
&=&1+\frac{\left\langle \hat{w}_{1}^{\left( 0\right) B}\left( \left(
X^{\prime }\right) ^{\prime },X^{\prime }\right) \right\rangle _{\left(
X^{\prime }\right) ^{\prime }}\frac{\zeta ^{2}\bar{w}_{1}^{\left( 0\right)
}\left( X^{\prime },X\right) }{w_{1}^{\left( 0\right) B}\left( X^{\prime
},X\right) }\left( \frac{1+\frac{\left( \gamma \left\langle \hat{S}%
_{E}\left( X_{1},X^{\prime }\right) \right\rangle _{X_{1}}\right) ^{2}}{%
1-\left( \gamma \left\langle \hat{S}_{E}\left( X^{\prime },\left( X^{\prime
}\right) ^{\prime }\right) \right\rangle \right) ^{2}}}{1+\frac{\left( \bar{%
\gamma}\left\langle \bar{S}_{E}\left( X_{1},X^{\prime }\right) \right\rangle
_{X_{1}}\right) ^{2}}{1-\left( \bar{\gamma}\left\langle \bar{S}_{E}\left(
\left( X^{\prime }\right) ^{\prime },X^{\prime }\right) \right\rangle
\right) ^{2}}}\right) }{4\left( \bar{\zeta}^{2}\zeta ^{2}\left( 1+\frac{%
\left( \gamma \left\langle \hat{S}_{E}\left( X_{1},\left( X^{\prime }\right)
^{\prime }\right) \right\rangle \right) ^{2}}{1-\left( \gamma \left\langle 
\hat{S}_{E}\left( X^{\prime },\left( X^{\prime }\right) ^{\prime }\right)
\right\rangle \right) ^{2}}\right) +\xi ^{2}\left\langle \hat{w}_{1}^{\left(
0\right) B}\left( \left( X^{\prime }\right) ^{\prime },X^{\prime }\right)
\right\rangle _{\left( X^{\prime }\right) ^{\prime }}\right) }  \notag \\
&&+\frac{\zeta ^{2}\left\langle \hat{w}_{1}^{\left( 0\right) B}\left( \left(
X^{\prime }\right) ^{\prime },X^{\prime }\right) \right\rangle _{\left(
X^{\prime }\right) ^{\prime }}}{\xi ^{2}w_{1}^{\left( 0\right) B}\left(
X^{\prime },X\right) }\left( 1+\frac{\left( \gamma \left\langle \hat{S}%
_{E}\left( X_{1},X^{\prime }\right) \right\rangle _{X_{1}}\right) ^{2}}{%
1-\left( \gamma \left\langle \hat{S}_{E}\left( X^{\prime },\left( X^{\prime
}\right) ^{\prime }\right) \right\rangle \right) ^{2}}\right)  \notag
\end{eqnarray}%
\begin{equation*}
\bar{w}^{B}\left( X,X\right) =1-\left\langle \bar{w}\left( X^{\prime
},X\right) \right\rangle _{X^{\prime }}-\left\langle \hat{w}_{1}^{B}\left(
X^{\prime },X\right) \right\rangle
\end{equation*}%
and $\bar{S}_{E}\left( X^{\prime },X\right) ,$ $\bar{S}_{L}\left( X^{\prime
},X\right) $, $\bar{S}\left( X^{\prime },X\right) $, $\hat{S}_{E}^{B}\left(
X^{\prime },X\right) $, $S_{E}^{B}\left( X,X\right) $ given by (\ref{SBN}), (%
\ref{SBT}), (\ref{SBF}), (\ref{scb}), (\ref{SFN}).

The coeffcients $\frac{\zeta ^{2}}{\xi ^{2}}$ $\bar{\zeta}^{2}$ are given by
(\ref{ZTR}), (\ref{ZTB}).%
\begin{equation}
\hat{w}_{2}^{B}\left( X^{\prime },X\right) =\frac{\left( 1-\left( \gamma
\left\langle \hat{S}_{E}\left( X_{1},X^{\prime }\right) \right\rangle
\right) ^{2}\right) \hat{w}_{2}^{\left( 0\right) }\left( X^{\prime
},X\right) \left( 1+\Delta \hat{r}\left( X^{\prime }\right) \right) }{1+\hat{%
w}_{2}^{\left( 0\right) }\left( X^{\prime },X\right) \left( 1-\left( \gamma
\left\langle \hat{S}_{E}\left( X_{1},X^{\prime }\right) \right\rangle
\right) ^{2}\right) +\left( \gamma \left\langle \hat{S}_{E}\left(
X_{1},X^{\prime }\right) \right\rangle _{X_{1}}\right) ^{2}-\left( \gamma
\left\langle \hat{S}_{E}\left( X_{1},X^{\prime }\right) \right\rangle
\right) ^{2}}
\end{equation}%
and:%
\begin{equation}
w_{2}^{B}\left( X\right) =1-\hat{w}_{2}^{B}\left( X^{\prime },X\right)
\end{equation}%
see (\ref{sdb}), (\ref{sdt}), and shares of loans are:%
\begin{equation*}
\frac{\hat{S}_{L}^{B}\left( X^{\prime },X\right) }{\kappa \left( 1-\bar{S}%
\left( X\right) \right) }=\hat{w}_{2}^{B}\left( X^{\prime },X\right) \left\{
1+\hat{w}_{2}^{B}\left( X^{\prime },X\right) \left( \hat{r}\left( X^{\prime
}\right) -\left\langle \hat{f}\left( X^{\prime }\right) \right\rangle _{\hat{%
w}_{1}}\right) +w_{2}^{B}\left( X\right) \left( \hat{r}\left( X^{\prime
}\right) -f\left( X\right) \right) \right\}
\end{equation*}%
\begin{equation*}
\frac{S_{L}^{B}\left( X,X\right) }{\kappa \left( 1-\bar{S}\left( X\right)
\right) }=w_{2}^{B}\left( X\right) \left[ 1+\hat{w}_{2}^{B}\left( X\right)
\left( r\left( X\right) -\left\langle \hat{r}\left( X^{\prime }\right)
\right\rangle _{\hat{w}_{2}}\right) \right]
\end{equation*}%
see (\ref{frl}), (\ref{sfB}).

\subsection*{A11.2 Firms' returns}

Returns for investors are obtained by replacing in the shares given above,
the average return for firms:%
\begin{equation*}
f\left( X\right) \rightarrow f_{1}^{dr}\left( X\right) =\frac{f_{1}\left(
X\right) +\tau \Delta F_{\tau }\left( \bar{R}\left( K,X\right) \right) }{%
K_{X}^{r}}-\frac{C}{K_{X}}-C_{0}
\end{equation*}%
\begin{equation}
\left\langle f\left( X\right) \right\rangle \rightarrow \left\langle
f_{1}^{dr}\left( X\right) \right\rangle =\frac{f_{1}\left( X\right) +\tau
\Delta F_{\tau }\left( \bar{R}\left( K,X\right) \right) }{\left\langle
K\right\rangle ^{r}}-\frac{C}{\left\langle K\right\rangle }-C_{0}
\label{Frn}
\end{equation}%
where constnt $C_{0}$ has been reintroduced.

The returns for firms are given, as in Appendix 8, by:%
\begin{equation}
f_{1}^{\left( e\right) }\left( X\right) =\left( 1+\underline{k}_{2}\left(
X\right) +\kappa \left[ \frac{\underline{k}_{2}^{B}}{1+\bar{k}}\right]
\left( X\right) \right) f_{1}^{\prime }\left( X\right) -\left( \underline{k}%
_{2}\left( X\right) +\kappa \left[ \frac{\underline{k}_{2}^{B}}{1+\bar{k}}%
\right] \left( X\right) \right) \bar{r}
\end{equation}%
Now using that:%
\begin{eqnarray*}
1+\underline{k}_{2}\left( X^{\prime }\right) +\kappa \left[ \frac{\underline{%
k}_{2}^{B}}{1+\bar{k}}\right] \left( X^{\prime }\right) &=&\frac{1-\left(
S_{E}\left( X^{\prime },X^{\prime }\right) \frac{\hat{K}_{X}\left\vert \hat{%
\Psi}\left( X^{\prime }\right) \right\vert ^{2}}{K_{X}\left\vert \Psi \left(
X^{\prime }\right) \right\vert ^{2}}+S_{E}^{B}\left( X^{\prime },X^{\prime
}\right) \frac{\bar{K}_{X^{\prime }}\left\vert \bar{\Psi}\left( X^{\prime
}\right) \right\vert ^{2}}{K_{X^{\prime }}\left\vert \Psi \left( X^{\prime
}\right) \right\vert ^{2}}\right) }{1-\left( \left\langle S\left( X^{\prime
},X^{\prime }\right) \right\rangle \frac{\hat{K}_{X}\left\vert \hat{\Psi}%
\left( X^{\prime }\right) \right\vert ^{2}}{K_{X}\left\vert \Psi \left(
X^{\prime }\right) \right\vert ^{2}}+\left\langle S^{B}\left( X^{\prime
},X^{\prime }\right) \right\rangle \frac{\bar{K}_{X^{\prime }}\left\vert 
\bar{\Psi}\left( X^{\prime }\right) \right\vert ^{2}}{K_{X^{\prime
}}\left\vert \Psi \left( X^{\prime }\right) \right\vert ^{2}}\right) } \\
&=&\frac{1-\left( S_{E}\left( X\right) +S_{E}^{B}\left( X\right) \right) }{%
1-\left( S\left( X\right) +S^{B}\left( X\right) \right) }
\end{eqnarray*}%
We find:%
\begin{eqnarray}
f_{1}^{\prime }\left( X\right) &=&\frac{1-\left( S_{E}\left( X\right)
+S_{E}^{B}\left( X\right) \right) }{1-\left( S\left( X\right) +S^{B}\left(
X\right) \right) }\left( \left( 1-\left( S\left( X\right) +S^{B}\left(
X\right) \right) \right) ^{r}\frac{f_{1}\left( X\right) +\tau \Delta F_{\tau
}\left( \bar{R}\left( K,X\right) \right) }{K_{X}^{r}}\right. \\
&&\left. -C_{0}-\frac{C}{K_{X}}\right) -\frac{S_{L}\left( X\right)
+S_{L}^{B}\left( X\right) }{1-\left( S\left( X\right) +S^{B}\left( X\right)
\right) }\bar{r}  \notag
\end{eqnarray}%
which corresponds to the net return of production plus variation in stocks,
from which the paiements of loans is substracted.

\subsection*{A11.3 States with default}

\subsubsection*{A11.3.1 initial default}

As explained in appendix 1, the equations for return have to be modified
when defaults occur. It may start from firms such that:%
\begin{equation*}
1+f_{1}^{\prime }\left( X\right) <0
\end{equation*}%
This case arises if the previous solutions%
\begin{equation}
\frac{1-S_{E}\left( X\right) }{1-S\left( X\right) }\left( \frac{f_{1}\left(
X\right) +\tau \Delta F_{\tau }\left( \bar{R}\left( K,X\right) \right) }{%
K_{X}^{r}}-\frac{C}{K_{X}}-C_{0}\right) -\frac{S\left( X\right) -S_{E}\left(
X\right) }{1-S\left( X\right) }\bar{r}<-1  \label{Cv}
\end{equation}%
where $S\left( X\right) $ stand for $S\left( X\right) +S^{B}\left( X\right) $
and $S_{E}\left( X\right) $ for $S_{E}\left( X\right) +S_{E}^{B}\left(
X\right) $.

If this condition (\ref{Cv}) is satisfied, the resolution has to be modified
to include\ defaults. In this case, the firms returns is:%
\begin{equation*}
f_{1}^{\prime }\left( X\right) \rightarrow \hat{f}\left( X\right) -\left( 1+%
\bar{r}\right) S_{L}\left( X\right)
\end{equation*}%
due to the loans to be paid back.

Since (\ref{Cv}) leads to:%
\begin{equation*}
\frac{f_{1}\left( X\right) +\tau \Delta F_{\tau }\left( \bar{R}\left(
K,X\right) \right) +\tau \Delta F_{\tau }\left( \bar{R}\left( K,X\right)
\right) }{K_{X}^{r}}-\frac{C}{K_{X}}-C_{0}<\frac{\left( S\left( X\right)
-S_{E}\left( X\right) \right) \bar{r}-\left( 1-S\left( X\right) \right) }{%
1-S_{E}\left( X\right) }
\end{equation*}%
the condition for default writes:%
\begin{equation*}
f\left( X\right) -\left( 1+\bar{r}\right) S_{L}\left( X\right) <\frac{\left(
S\left( X\right) -S_{E}\left( X\right) \right) \bar{r}-\left( 1-S\left(
X\right) \right) }{1-S_{E}\left( X\right) }-\left( 1+\bar{r}\right)
S_{L}\left( X\right)
\end{equation*}%
To detail this condition, we use that the various shares write:%
\begin{eqnarray}
&&S_{E}\left( X,X\right) \\
&=&\frac{w\left( X\right) }{2}\left( 1+\left( \hat{w}\left( X\right) \left(
f\left( X\right) -\frac{\left\langle \hat{f}\left( X^{\prime }\right)
\right\rangle _{\hat{w}_{1}}+\left\langle \hat{r}\left( X^{\prime }\right)
\right\rangle _{\hat{w}_{2}}}{2}\right) +\frac{w\left( X\right) }{2}\left(
f\left( X\right) -\bar{r}\left( X\right) \right) \right) \right)  \notag
\end{eqnarray}%
\begin{equation}
S\left( X,X\right) =w\left( X\right) \left( 1-\hat{w}\left( X\right) \frac{%
1+\left\langle \hat{f}\left( X^{\prime }\right) \right\rangle _{\hat{w}_{1}}%
}{2}\right)
\end{equation}%
\begin{eqnarray*}
&&S_{E}^{B}\left( X,X\right) \\
&=&w_{1}^{B}\left( X\right) \left\{ 1+\left\langle \bar{w}\left( X\right)
\right\rangle \left( f\left( X\right) -\frac{\left\langle \bar{f}\left(
X^{\prime }\right) \right\rangle _{\bar{w}_{1}}+\left\langle \bar{r}\left(
X^{\prime }\right) \right\rangle _{\bar{w}_{2}}}{2}\right) +\left\langle 
\hat{w}_{1}^{B}\left( X\right) \right\rangle \left( f\left( X\right)
-\left\langle \hat{f}\left( X^{\prime }\right) \right\rangle _{\hat{w}%
_{1}}\right) \right\}
\end{eqnarray*}%
\begin{equation*}
\frac{S_{L}^{B}\left( X,X\right) }{\kappa \left( 1-\bar{S}\left( X\right)
\right) }=w_{2}^{B}\left( X\right) \left[ 1+\hat{w}_{2}^{B}\left( X\right)
\left( r\left( X\right) -\left\langle \hat{r}\left( X^{\prime }\right)
\right\rangle _{\hat{w}_{2}}\right) \right]
\end{equation*}%
\begin{eqnarray*}
w_{1}^{B}\left( X\right) &=&w_{2}^{B}\left( X\right) =\frac{1}{2}w^{B}\left(
X\right) \\
\left\langle \hat{w}_{1}^{B}\left( X\right) \right\rangle &=&\left\langle 
\hat{w}_{2}^{B}\left( X\right) \right\rangle =\frac{1}{2}\left\langle \hat{w}%
^{B}\left( X\right) \right\rangle
\end{eqnarray*}%
\begin{equation*}
S^{B}\left( X,X\right) \simeq w^{B}\left( X\right) \left( 1-\frac{%
1+\left\langle \hat{f}\left( X^{\prime }\right) \right\rangle _{\hat{w}%
_{1}}+\left\langle \bar{r}\left( X^{\prime }\right) \right\rangle _{\bar{w}%
_{2}}}{2}\right)
\end{equation*}%
Then, given the return equations: 
\begin{eqnarray*}
\hat{G} &=&\frac{\left( 1-\left( \gamma \left\langle \hat{S}_{E}\left(
X\right) \right\rangle \right) ^{2}\right) }{2-\left( \gamma \left\langle 
\hat{S}_{E}\left( X\right) \right\rangle \right) ^{2}}\left( 1-\frac{\Delta
\left( \frac{f\left( X\right) +r\left( X\right) }{2}\right) }{2-\left(
\gamma \left\langle \hat{S}_{E}\left( X\right) \right\rangle \right) ^{2}}+%
\frac{\left\langle \hat{f}\left( X^{\prime }\right) \right\rangle
-\left\langle \hat{r}\left( X^{\prime }\right) \right\rangle _{\hat{w}_{2}}}{%
2}\right) \\
&&\times \frac{1-\left\langle \hat{S}\left( X^{\prime }\right) \right\rangle 
}{1-\left\langle \hat{S}_{E}\left( X^{\prime }\right) \right\rangle }\left(
\left\langle \hat{f}\left( X^{\prime }\right) \right\rangle -\left\langle 
\bar{r}\right\rangle \right) -S_{E}\left( X,X\right) \left( f\left( X\right)
-r\right)
\end{eqnarray*}%
and:

\begin{eqnarray}
\bar{G} &=&\left\langle \bar{S}_{E}\left( X^{\prime },X\right) \right\rangle
_{X^{\prime }}\frac{1-\left\langle \bar{S}\left( X^{\prime }\right)
\right\rangle }{1-\left\langle \bar{S}_{E}\left( X^{\prime }\right)
\right\rangle }\left( \left\langle \bar{f}\left( X^{\prime }\right)
\right\rangle -\bar{r}\right) \\
&&+\left\langle \hat{S}_{E}^{B}\left( X^{\prime },X\right) \right\rangle
_{X^{\prime }}\frac{1-\left\langle \hat{S}\left( X^{\prime }\right)
\right\rangle +\left\langle \hat{S}_{E}^{B}\left( X^{\prime }\right)
\right\rangle +\left\langle \hat{S}_{L}^{B}\left( X^{\prime }\right)
\right\rangle }{1-\left\langle \hat{S}_{E}\left( X^{\prime }\right)
\right\rangle +\left\langle \hat{S}_{E}^{B}\left( X^{\prime }\right)
\right\rangle }\left( \left\langle \hat{f}\left( X^{\prime }\right)
\right\rangle -\bar{r}\right) +S_{E}^{B}\left( X,X\right) \left( f_{1}\left(
X\right) -\bar{r}\right)  \notag
\end{eqnarray}%
We use that (see Appendix 8):%
\begin{equation*}
\hat{G}\simeq \left( \hat{f}\left( X\right) -\bar{r}\right)
\end{equation*}%
\begin{equation*}
\bar{G}\simeq \left( \bar{f}\left( X\right) -\bar{r}\right)
\end{equation*}%
This leads to the threshold for investors' default:%
\begin{eqnarray*}
&&f\left( X\right) -r \\
&<&-\frac{4\left( 2-\left( \gamma \left\langle \hat{S}_{E}\left( X\right)
\right\rangle \right) ^{2}\right) \left( 1+\frac{\left( 1-\left( \gamma
\left\langle \hat{S}_{E}\left( X\right) \right\rangle \right) ^{2}\right) }{%
2\left( 2-\left( \gamma \left\langle \hat{S}_{E}\left( X\right)
\right\rangle \right) ^{2}\right) ^{2}}\frac{1-\left\langle \hat{S}\left(
X^{\prime }\right) \right\rangle }{1-\left\langle \hat{S}_{E}\left(
X^{\prime }\right) \right\rangle }\left( \left\langle \hat{f}\left(
X^{\prime }\right) \right\rangle -\left\langle \bar{r}\right\rangle \right)
\right) }{\frac{1-\left( 1-\left( \gamma \left\langle \hat{S}_{E}\left(
X\right) \right\rangle \right) ^{2}\right) \left\langle \hat{f}\left(
X^{\prime }\right) \right\rangle _{\hat{w}_{1}}}{2-\left( \gamma
\left\langle \hat{S}_{E}\left( X\right) \right\rangle \right) ^{2}}-\bar{r}%
\left( X\right) }\frac{1-\left( \gamma \left\langle \hat{S}_{E}\left(
X\right) \right\rangle \right) ^{2}}{2-\left( \gamma \left\langle \hat{S}%
_{E}\left( X\right) \right\rangle \right) ^{2}} \\
&&\times \left( \left( 1+\bar{r}\right) +\frac{1+\frac{\left\langle \hat{f}%
\left( X^{\prime }\right) \right\rangle -\left\langle \hat{r}\left(
X^{\prime }\right) \right\rangle _{\hat{w}_{2}}}{2}-\frac{\Delta r\left(
X\right) }{2-\left( \gamma \left\langle \hat{S}_{E}\left( X\right)
\right\rangle \right) ^{2}}}{1+\frac{\left( 1-\left( \gamma \left\langle 
\hat{S}_{E}\left( X\right) \right\rangle \right) ^{2}\right) }{2\left(
2-\left( \gamma \left\langle \hat{S}_{E}\left( X\right) \right\rangle
\right) ^{2}\right) ^{2}}\frac{1-\left\langle \hat{S}\left( X^{\prime
}\right) \right\rangle }{1-\left\langle \hat{S}_{E}\left( X^{\prime }\right)
\right\rangle }\left( \left\langle \hat{f}\left( X^{\prime }\right)
\right\rangle -\left\langle \bar{r}\right\rangle \right) }\frac{%
1-\left\langle \hat{S}\left( X^{\prime }\right) \right\rangle }{%
1-\left\langle \hat{S}_{E}\left( X^{\prime }\right) \right\rangle }\left(
\left\langle \hat{f}\left( X^{\prime }\right) \right\rangle -\left\langle 
\bar{r}\right\rangle \right) \right)
\end{eqnarray*}%
that is:%
\begin{eqnarray*}
&&f\left( X\right) -r \\
&<&-\frac{4\left( 1+\frac{\hat{w}\left( X\right) w\left( X\right) }{2}\frac{%
1-\left\langle \hat{S}\left( X^{\prime }\right) \right\rangle }{%
1-\left\langle \hat{S}_{E}\left( X^{\prime }\right) \right\rangle }\left(
\left\langle \hat{f}\left( X^{\prime }\right) \right\rangle -\left\langle 
\bar{r}\right\rangle \right) \right) }{w\left( X\right) -\hat{w}\left(
X\right) \left\langle \hat{f}\left( X^{\prime }\right) \right\rangle _{\hat{w%
}_{1}}-\bar{r}\left( X\right) }\frac{\hat{w}\left( X\right) }{w\left(
X\right) } \\
&&\times \left( \left( 1+\bar{r}\right) +\frac{1+\frac{\left\langle \hat{f}%
\left( X^{\prime }\right) \right\rangle -\left\langle \hat{r}\left(
X^{\prime }\right) \right\rangle _{\hat{w}_{2}}}{2}-w\left( X\right) \Delta
r\left( X\right) }{1+\frac{\hat{w}\left( X\right) w\left( X\right) }{2}\frac{%
1-\left\langle \hat{S}\left( X^{\prime }\right) \right\rangle }{%
1-\left\langle \hat{S}_{E}\left( X^{\prime }\right) \right\rangle }\left(
\left\langle \hat{f}\left( X^{\prime }\right) \right\rangle -\left\langle 
\bar{r}\right\rangle \right) }\frac{1-\left\langle \hat{S}\left( X^{\prime
}\right) \right\rangle }{1-\left\langle \hat{S}_{E}\left( X^{\prime }\right)
\right\rangle }\left( \left\langle \hat{f}\left( X^{\prime }\right)
\right\rangle -\left\langle \bar{r}\right\rangle \right) \right)
\end{eqnarray*}%
The threshold for banks' default is obtained similarly by considering $\bar{f%
}\left( X\right) <-1$. This leads to:%
\begin{eqnarray}
-1 &>&\left\langle \bar{S}_{E}\left( X^{\prime },X\right) \right\rangle
_{X^{\prime }}\frac{1-\left\langle \bar{S}\left( X^{\prime }\right)
\right\rangle }{1-\left\langle \bar{S}_{E}\left( X^{\prime }\right)
\right\rangle }\left( \left\langle \bar{f}\left( X^{\prime }\right)
\right\rangle -\bar{r}\right) \\
&&+\left\langle \hat{S}_{E}^{B}\left( X^{\prime },X\right) \right\rangle
_{X^{\prime }}\frac{1-\left\langle \hat{S}\left( X^{\prime }\right)
\right\rangle +\left\langle \hat{S}_{E}^{B}\left( X^{\prime }\right)
\right\rangle +\left\langle \hat{S}_{L}^{B}\left( X^{\prime }\right)
\right\rangle }{1-\left\langle \hat{S}_{E}\left( X^{\prime }\right)
\right\rangle +\left\langle \hat{S}_{E}^{B}\left( X^{\prime }\right)
\right\rangle }\left( \left\langle \hat{f}\left( X^{\prime }\right)
\right\rangle -\bar{r}\right)  \notag \\
&&+S_{E}^{B}\left( X,X\right) \left( f_{1}\left( X\right) -\bar{r}\right) +%
\bar{r}  \notag
\end{eqnarray}%
and this corresponds to:%
\begin{eqnarray*}
f\left( X\right) -\bar{r} &<&-\frac{1}{S_{E}^{B}\left( X,X\right) }\left( 1+%
\bar{r}+\left\langle \bar{S}_{E}\left( X^{\prime },X\right) \right\rangle
_{X^{\prime }}\frac{1-\left\langle \bar{S}\left( X^{\prime }\right)
\right\rangle }{1-\left\langle \bar{S}_{E}\left( X^{\prime }\right)
\right\rangle }\left( \left\langle \bar{f}\left( X^{\prime }\right)
\right\rangle -\bar{r}\right) \right. \\
&&+\left. \left\langle \hat{S}_{E}^{B}\left( X^{\prime },X\right)
\right\rangle _{X^{\prime }}\frac{1-\left\langle \hat{S}\left( X^{\prime
}\right) \right\rangle +\left\langle \hat{S}_{E}^{B}\left( X^{\prime
}\right) \right\rangle +\left\langle \hat{S}_{L}^{B}\left( X^{\prime
}\right) \right\rangle }{1-\left\langle \hat{S}_{E}\left( X^{\prime }\right)
\right\rangle +\left\langle \hat{S}_{E}^{B}\left( X^{\prime }\right)
\right\rangle }\left( \left\langle \hat{f}\left( X^{\prime }\right)
\right\rangle -\bar{r}\right) \right)
\end{eqnarray*}%
where:%
\begin{eqnarray*}
&&S_{E}^{B}\left( X,X\right) \\
&=&w_{1}^{B}\left( X\right) \left\{ 1-\left\langle \bar{w}\left( X\right)
\right\rangle \left( 1+\frac{\left\langle \bar{f}\left( X^{\prime }\right)
\right\rangle _{\bar{w}_{1}}+\left\langle \bar{r}\left( X^{\prime }\right)
\right\rangle _{\bar{w}_{2}}}{2}\right) -\left\langle \hat{w}_{1}^{B}\left(
X\right) \right\rangle \left( 1+\left\langle \hat{f}\left( X^{\prime
}\right) \right\rangle _{\hat{w}_{1}}\right) \right\}
\end{eqnarray*}%
for $f\left( X\right) =-1$.

\subsubsection*{A11.3.2 Default Propagation}

\paragraph*{\protect\bigskip A11.3.2.1 Investors' modified returns equation}

We consider the zeroth ordr resolution. First order corrections can be found
straightforwardly as in Appendix 8 of Gosselin and Lotz (2025). Given (\ref%
{dfn}) and considering full default only: 
\begin{equation*}
\max \left( -1,\left( 1+f\left( X^{\prime }\right) \right) \frac{1-\left( 
\hat{S}\left( X^{\prime }\right) +\hat{S}_{E}^{B}\left( X^{\prime }\right) +%
\hat{S}_{L}^{B}\left( X^{\prime }\right) \right) }{\hat{S}_{L}\left(
X^{\prime }\right) }\right) =-1
\end{equation*}%
along with $\hat{S}=S$:

\begin{eqnarray*}
0 &=&\int \left( \delta \left( X-X^{\prime }\right) -\hat{S}_{E}\left(
X^{\prime },X\right) \right) \frac{1-\hat{S}\left( X^{\prime }\right) }{1-%
\hat{S}_{E}\left( X^{\prime }\right) }\left( f\left( X^{\prime }\right) -%
\bar{r}\right) dX^{\prime } \\
&&+\int \hat{S}_{L}\left( X^{\prime },X\right) dX^{\prime }+\int S_{L}\left(
X^{\prime },X\right) dX^{\prime } \\
&&-\int S_{E}\left( X^{\prime },X\right) \frac{1-\left( S\left( X^{\prime
}\right) +\left( S_{E}^{B}\left( X^{\prime }\right) +S_{L}^{B}\left(
X^{\prime }\right) \right) \right) }{1-S_{E}\left( X^{\prime }\right)
-S_{E}^{B}\left( X^{\prime }\right) }f_{1}\left( X^{\prime }\right)
dX^{\prime }
\end{eqnarray*}%
As before, we consider investors that invest in neighbor firms, which leads
to:

\begin{eqnarray}
&&\int \left( \Delta \left( X,X^{\prime }\right) -\hat{S}_{E}\left(
X^{\prime },X\right) \right) \frac{1-\hat{S}\left( X^{\prime }\right) }{1-%
\hat{S}_{E}\left( X^{\prime }\right) }f\left( X^{\prime }\right) dX^{\prime }
\\
&=&S_{E}\left( X,X\right) f_{1}^{\prime }\left( X\right) -\int_{\hat{S}_{-}}%
\hat{S}_{L}\left( X^{\prime },X\right) dX^{\prime }  \notag
\end{eqnarray}%
so that the resolution is obtained by shifting $f_{1}^{\prime }\left(
X\right) $ of:%
\begin{equation*}
-\frac{1}{S_{E}\left( X,X\right) }\int_{\hat{S}_{-}}\hat{S}_{L}\left(
X^{\prime },X\right) dX^{\prime }
\end{equation*}%
and by reevaluating the values of $f\left( X^{\prime }\right) $ for the
remaining investors that were above the threshld and that may now switch
below the threshold.

\paragraph*{A11.3.2.2 Banks' modified returns equation}

Considering banks leads to consider the following returns equations:%
\begin{eqnarray}
0 &=&\left( 1-\bar{S}_{E}\left( X^{\prime },X\right) \right) \left( \bar{f}%
\left( X^{\prime }\right) -\bar{r}\right) \frac{1-\bar{S}\left( X^{\prime
}\right) }{1-\bar{S}_{E}\left( X^{\prime }\right) }  \label{RbApp} \\
&&-\hat{S}_{E}^{B}\left( X^{\prime },X\right) \left( \hat{f}\left( X^{\prime
}\right) -\bar{r}\right) \frac{1-\left( \hat{S}\left( X^{\prime }\right) +%
\hat{S}_{E}^{B}\left( X^{\prime }\right) +\hat{S}_{L}^{B}\left( X^{\prime
}\right) \right) }{1-\left( \hat{S}_{E}\left( X^{\prime }\right) +\hat{S}%
_{E}^{B}\left( X^{\prime }\right) \right) }  \notag \\
&&+\left( 1+\bar{f}\left( X^{\prime }\right) \right) H\left( -\left( 1+\bar{f%
}\left( X^{\prime }\right) \right) \right) \bar{S}_{L}\left( X^{\prime
},X\right) \frac{\left( 1-\bar{S}\left( X^{\prime }\right) \right) }{\bar{S}%
_{L}\left( X^{\prime }\right) }  \notag \\
&&+\left( 1+\hat{f}\left( X^{\prime }\right) \right) H\left( -\left( 1+\hat{f%
}\left( X^{\prime }\right) \right) \right) \hat{S}_{L}^{B}\left( X^{\prime
},X\right) \frac{1-\left( S\left( X^{\prime }\right) +\left( S_{E}^{B}\left(
X^{\prime }\right) +S_{L}^{B}\left( X^{\prime }\right) \right) \right) }{%
S_{L}\left( X^{\prime }\right) +S_{L}^{B}\left( X^{\prime }\right) }  \notag
\\
&&+\left( 1+f_{1}^{\prime }\left( X^{\prime }\right) \right) H\left( -\left(
1+f_{1}^{\prime }\left( X^{\prime }\right) \right) \right) S_{L}^{B}\left(
X^{\prime },X\right) \frac{1-\left( \hat{S}\left( X^{\prime },X\right)
+\left( \hat{S}_{E}^{B}\left( X^{\prime }\right) +\hat{S}_{L}^{B}\left(
X^{\prime }\right) \right) \right) }{\hat{S}_{L}\left( X^{\prime }\right) +%
\hat{S}_{L}^{B}\left( X^{\prime }\right) }  \notag \\
&&-S_{E}^{B}\left( X^{\prime },X\right) \left\{ \frac{1-\left( S\left(
X^{\prime }\right) +\left( S_{E}^{B}\left( X^{\prime }\right)
+S_{L}^{B}\left( X^{\prime }\right) \right) \right) }{1-S_{E}\left(
X^{\prime }\right) -S_{E}^{B}\left( X^{\prime }\right) }\left( \left(
f_{1}^{\prime }\left( X^{\prime }\right) -\bar{r}\right) +\Delta F_{\tau
}\left( \bar{R}\left( K,X\right) \right) \right) \right\}  \notag
\end{eqnarray}%
\begin{eqnarray}
0 &=&\left( 1-\bar{S}_{E}\left( X^{\prime },X\right) \right) \left( \bar{f}%
\left( X^{\prime }\right) -\bar{r}\right) \frac{1-\bar{S}\left( X^{\prime
}\right) }{1-\bar{S}_{E}\left( X^{\prime }\right) } \\
&&-\hat{S}_{E}^{B}\left( X^{\prime },X\right) \left( \hat{f}\left( X^{\prime
}\right) -\bar{r}\right) \frac{1-\left( \hat{S}\left( X^{\prime }\right) +%
\hat{S}_{E}^{B}\left( X^{\prime }\right) +\hat{S}_{L}^{B}\left( X^{\prime
}\right) \right) }{1-\left( \hat{S}_{E}\left( X^{\prime }\right) +\hat{S}%
_{E}^{B}\left( X^{\prime }\right) \right) }  \notag \\
&&+\bar{S}_{L}\left( X^{\prime },X\right) +\hat{S}_{L}^{B}\left( X^{\prime
},X\right) +S_{L}^{B}\left( X^{\prime },X\right)  \notag \\
&&-S_{E}^{B}\left( X^{\prime },X\right) \left\{ \frac{1-\left( S\left(
X^{\prime }\right) +\left( S_{E}^{B}\left( X^{\prime }\right)
+S_{L}^{B}\left( X^{\prime }\right) \right) \right) }{1-S_{E}\left(
X^{\prime }\right) -S_{E}^{B}\left( X^{\prime }\right) }\left( \left(
f_{1}^{\prime }\left( X^{\prime }\right) -\bar{r}\right) +\Delta F_{\tau
}\left( \bar{R}\left( K,X\right) \right) \right) \right\}  \notag
\end{eqnarray}

\paragraph*{A11.3.2.3 Shifts in returns}

For investors, the returns are shifted by the followng amount:%
\begin{eqnarray*}
f\left( X\right) &\rightarrow &f\left( X\right) -\left( \frac{1}{S_{E}\left(
X,X\right) }\left( \int_{\hat{S}_{-}}\left( \hat{S}_{L}\left( X^{\prime
},X\right) \right) dX^{\prime }+S_{L}\left( X,X\right) \right) +df\left(
X\right) \right) \\
&\rightarrow &f\left( X\right) -\left( \frac{1}{S_{E}^{B}\left( X,X\right) }%
\left( \left\langle \hat{S}_{L}\left( X^{\prime },X\right) \right\rangle \mu
+S_{L}\left( X,X\right) \right) +df\left( X\right) \right)
\end{eqnarray*}%
where $\mu $ is the proportion of deflt investrs.in the return equation.

In first approximation, we will write:%
\begin{eqnarray*}
\hat{f}\left( X\right) &\rightarrow &\hat{f}\left( X\right) -\frac{%
1-\left\langle \hat{S}_{E}\left( X^{\prime }\right) \right\rangle }{%
1-\left\langle \hat{S}\left( X^{\prime }\right) \right\rangle }\frac{%
S_{E}\left( X,X\right) }{1-\left\langle \hat{S}_{E}\left( X\right)
\right\rangle }\frac{1}{S_{E}\left( X,X\right) }\left( \int_{\hat{S}%
_{-}}\left( \hat{S}_{L}\left( X^{\prime },X\right) \right) dX^{\prime
}+S_{L}\left( X,X\right) +df\left( X\right) \right) \\
&\rightarrow &\hat{f}\left( X\right) -\frac{1}{1-\left\langle \hat{S}\left(
X^{\prime }\right) \right\rangle }\left( \left( \left\langle \hat{S}%
_{L}\left( X^{\prime },X\right) \right\rangle +S_{L}\left( X,X\right)
\right) \mu +S_{E}\left( X,X\right) df\left( X\right) \right)
\end{eqnarray*}%
Similarly, for banks, returns are modified by the formla:%
\begin{eqnarray*}
\bar{f}\left( X^{\prime }\right) &\rightarrow &\bar{f}\left( X^{\prime
}\right) -\frac{1}{1-\bar{S}\left( X^{\prime }\right) }\left( \int_{\bar{S}%
_{-}}\bar{S}_{L}\left( X^{\prime },X\right) dX^{\prime }+\int_{\hat{S}_{-}}%
\hat{S}_{L}^{B}\left( X^{\prime },X\right) dX^{\prime }+S_{L}^{B}\left(
X,X\right) +S_{E}^{B}\left( X,X\right) df\left( X\right) \right) \\
&&-\frac{S_{E}^{B}\left( X,X\right) }{1-\bar{S}\left( X^{\prime }\right) }%
\frac{\hat{S}_{E}^{B}\left( X^{\prime },X\right) }{1-\left\langle \hat{S}%
\left( X^{\prime }\right) \right\rangle }\frac{1-\left( \hat{S}\left(
X^{\prime }\right) +\hat{S}_{E}^{B}\left( X^{\prime }\right) +\hat{S}%
_{L}^{B}\left( X^{\prime }\right) \right) }{1-\left( \hat{S}_{E}\left(
X^{\prime }\right) +\hat{S}_{E}^{B}\left( X^{\prime }\right) \right) } \\
&&\times \left( \int_{\hat{S}_{-}}\left( \hat{S}_{L}\left( X^{\prime
},X\right) \right) dX^{\prime }+S_{L}\left( X,X\right) +S_{E}\left(
X,X\right) df\left( X\right) \right)
\end{eqnarray*}%
\begin{eqnarray*}
&\rightarrow &\bar{f}\left( X^{\prime }\right) -\frac{1}{1-\bar{S}\left(
X^{\prime }\right) }\left( \left( \left\langle \bar{S}_{L}\left( X^{\prime
},X\right) \right\rangle _{X^{\prime }}+\left\langle \hat{S}_{L}^{B}\left(
X^{\prime },X\right) \right\rangle _{X^{\prime }}+S_{L}^{B}\left( X,X\right)
\right) \mu +S_{E}^{B}\left( X,X\right) df\left( X\right) \right) \\
&&-\frac{S_{E}^{B}\left( X,X\right) }{1-\bar{S}\left( X^{\prime }\right) }%
\frac{\hat{S}_{E}^{B}\left( X^{\prime },X\right) }{1-\left\langle \hat{S}%
\left( X^{\prime }\right) \right\rangle }\frac{1-\left( \hat{S}\left(
X^{\prime }\right) +\hat{S}_{E}^{B}\left( X^{\prime }\right) +\hat{S}%
_{L}^{B}\left( X^{\prime }\right) \right) }{1-\left( \hat{S}_{E}\left(
X^{\prime }\right) +\hat{S}_{E}^{B}\left( X^{\prime }\right) \right) } \\
&&\times \left( \left( \left\langle \left( \hat{S}_{L}\left( X^{\prime
},X\right) \right) \right\rangle _{X^{\prime }}+S_{L}\left( X,X\right)
\right) \mu +S_{E}\left( X,X\right) df\left( X\right) \right)
\end{eqnarray*}%
This can be regrouped as:%
\begin{eqnarray*}
&&\bar{f}\left( X^{\prime }\right) -\left( \frac{S_{E}^{B}\left( X,X\right) 
}{1-\bar{S}\left( X^{\prime }\right) }+S_{E}\left( X,X\right) \frac{%
S_{E}^{B}\left( X,X\right) }{1-\bar{S}\left( X^{\prime }\right) }\frac{\hat{S%
}_{E}^{B}\left( X^{\prime },X\right) }{1-\left\langle \hat{S}\left(
X^{\prime }\right) \right\rangle }\frac{1-\left( \hat{S}\left( X^{\prime
}\right) +\hat{S}_{E}^{B}\left( X^{\prime }\right) +\hat{S}_{L}^{B}\left(
X^{\prime }\right) \right) }{1-\left( \hat{S}_{E}\left( X^{\prime }\right) +%
\hat{S}_{E}^{B}\left( X^{\prime }\right) \right) }\right) df\left( X\right)
\\
&&-\left( \frac{1}{1-\bar{S}\left( X^{\prime }\right) }+\frac{%
S_{E}^{B}\left( X,X\right) }{1-\bar{S}\left( X^{\prime }\right) }\frac{\hat{S%
}_{E}^{B}\left( X^{\prime },X\right) }{1-\left\langle \hat{S}\left(
X^{\prime }\right) \right\rangle }\frac{1-\left( \hat{S}\left( X^{\prime
}\right) +\hat{S}_{E}^{B}\left( X^{\prime }\right) +\hat{S}_{L}^{B}\left(
X^{\prime }\right) \right) }{1-\left( \hat{S}_{E}\left( X^{\prime }\right) +%
\hat{S}_{E}^{B}\left( X^{\prime }\right) \right) }\right) \left(
\left\langle \left( \hat{S}_{L}\left( X^{\prime },X\right) \right)
\right\rangle _{X^{\prime }}+S_{L}\left( X,X\right) \right) \mu \\
&&-\frac{1}{1-\bar{S}\left( X^{\prime }\right) }\left( \left\langle \bar{S}%
_{L}\left( X^{\prime },X\right) \right\rangle _{X^{\prime }}+\left\langle 
\hat{S}_{L}^{B}\left( X^{\prime },X\right) \right\rangle _{X^{\prime
}}+S_{L}^{B}\left( X,X\right) \right) \mu
\end{eqnarray*}%
where $\mu $ is the proportion of default among investors, assumed to be the
same for other types of agents, firms and banks. In the sequel:%
\begin{equation*}
\frac{1-\left( S\left( X^{\prime }\right) +\left( S_{E}^{B}\left( X^{\prime
}\right) +S_{L}^{B}\left( X^{\prime }\right) \right) \right) }{1-S_{E}\left(
X^{\prime }\right) -S_{E}^{B}\left( X^{\prime }\right) }\left( \left(
f_{1}^{\prime }\left( X^{\prime }\right) -\bar{r}\right) +\Delta F_{\tau
}\left( \bar{R}\left( K,X\right) \right) \right)
\end{equation*}%
computes the firms returns. It will be given by (\ref{Frn}):%
\begin{equation*}
\left( \left( 1-\left( S\left( X\right) +S^{B}\left( X\right) \right)
\right) \right) ^{r}\left\langle f_{1}\left( X^{\prime }\right)
\right\rangle -C_{0}-\left( \left( 1-\left( S\left( X\right) +S^{B}\left(
X\right) \right) \right) \right) C
\end{equation*}

\paragraph*{A11.3.2.4 Capital ratios}

The capital ratio are given by (\ref{cpn}) and (\ref{cpt}):%
\begin{eqnarray}
&&\frac{\hat{K}_{X}\left\vert \hat{\Psi}\left( X\right) \right\vert ^{2}}{%
K_{X}\left\vert \Psi \left( X\right) \right\vert ^{2}}  \label{dkn} \\
&\simeq &\frac{18\sigma _{\hat{K}}^{2}}{\hat{\mu}}\frac{\left(
1-\left\langle \hat{S}\right\rangle \right) ^{2}\left( 1-\frac{S\left(
X,X\right) _{cr}\hat{K}_{X}\left\vert \hat{\Psi}\left( X\right) \right\vert
^{2}+S^{B}\left( X,X\right) _{cr}\bar{K}_{X}\left\vert \bar{\Psi}\left(
X\right) \right\vert ^{2}}{\left( \left( \frac{2\epsilon }{3\sigma _{\hat{K}%
}^{2}}\right) ^{\frac{r}{2}}\frac{f_{1}\left( X\right) }{C_{0}+\frac{%
S_{L}\left( X\right) }{1-S_{E}\left( X\right) }\bar{r}}\right) ^{\frac{2}{r}}%
}\right) ^{--2}\left( \left( \frac{2\epsilon }{3\sigma _{\hat{K}}^{2}}%
\right) ^{\frac{r}{2}}\frac{f_{1}\left( X\right) }{C_{0}+\frac{S_{L}\left(
X\right) }{1-S_{E}\left( X\right) }\bar{r}}\right) ^{-\frac{2}{r}}}{\left(
\left( 1-\left\langle \hat{S}\right\rangle \right) \left( \hat{f}\left(
X\right) +\frac{\left\langle \hat{S}_{E}^{B}\left( X,X^{\prime }\right)
\right\rangle _{X^{\prime }}+\left\langle \hat{S}_{L}^{B}\left( X,X^{\prime
}\right) \right\rangle _{X^{\prime }}\left\langle \bar{S}\right\rangle }{%
1-\left\langle \bar{S}\right\rangle }\left\langle \bar{f}\right\rangle
\right) +\left\langle \hat{S}\left( X^{\prime },X\right) \right\rangle
_{X^{\prime }}\left( \left\langle \hat{f}\right\rangle +\frac{\left\langle 
\hat{S}_{E}^{B}\right\rangle +\left\langle \hat{S}_{L}^{B}\right\rangle
\left\langle \bar{S}\right\rangle }{1-\left\langle \bar{S}\right\rangle }%
\left\langle \bar{f}\right\rangle \right) \right) ^{2}}  \notag
\end{eqnarray}%
\begin{eqnarray}
\frac{\bar{K}_{X}\left\vert \bar{\Psi}\left( X\right) \right\vert ^{2}}{%
K_{X}\left\vert \Psi \left( X\right) \right\vert ^{2}} &\simeq &72\frac{%
\sigma _{\hat{K}}^{2}V\left( 1-\left\langle \bar{S}\right\rangle \right)
^{2}\left( \left\langle \hat{f}\right\rangle +\frac{\left\langle \hat{S}%
_{E}^{B}\right\rangle +\left\langle \hat{S}_{L}^{B}\right\rangle }{%
1-\left\langle \bar{S}\right\rangle }\left\langle \bar{f}\right\rangle
\right) ^{4}}{\left( \bar{f}\left( X\right) \left( 1-\left\langle \bar{S}%
\right\rangle \right) +\left\langle \bar{S}\left( X^{\prime },X\right)
\right\rangle _{X^{\prime }}\left\langle \bar{f}\right\rangle \right) \left(
1-\left\langle \hat{S}\right\rangle \right) ^{4}\hat{\mu}}\left( \left\Vert 
\bar{\Psi}_{0}\right\Vert \right) ^{4}  \label{dkt} \\
&&\times \left( 1-\frac{S\left( X,X\right) _{cr}\hat{K}_{X}\left\vert \hat{%
\Psi}\left( X\right) \right\vert ^{2}+S^{B}\left( X,X\right) _{cr}\bar{K}%
_{X}\left\vert \bar{\Psi}\left( X\right) \right\vert ^{2}}{\left( \left( 
\frac{2\epsilon }{3\sigma _{\hat{K}}^{2}}\right) ^{\frac{r}{2}}\frac{%
f_{1}\left( X\right) }{C_{0}+\frac{S_{L}\left( X\right) }{1-S_{E}\left(
X\right) }\bar{r}}\right) ^{\frac{2}{r}}}\right) ^{-2}\left( \left( \frac{%
2\epsilon }{3\sigma _{\hat{K}}^{2}}\right) ^{\frac{r}{2}}\frac{f_{1}\left(
X\right) }{C_{0}+\frac{S_{L}\left( X\right) }{1-S_{E}\left( X\right) }\bar{r}%
}\right) ^{\frac{2}{r}}  \notag
\end{eqnarray}%
As in Appendix 8 of Gosselin and Lotz (2025), the dominant part of the
derivatives comes from:

\begin{eqnarray*}
\delta \frac{\hat{K}_{X}\left\vert \hat{\Psi}\left( X\right) \right\vert ^{2}%
}{K_{X}\left\vert \Psi \left( X\right) \right\vert ^{2}} &=&2\left( \frac{%
\delta S\left( X\right) +\delta S^{B}\left( X\right) }{1-\left( S\left(
X\right) +S^{B}\left( X\right) \right) }\right) \frac{\hat{K}_{X}\left\vert 
\hat{\Psi}\left( X\right) \right\vert ^{2}}{K_{X}\left\vert \Psi \left(
X\right) \right\vert ^{2}} \\
\delta \frac{\bar{K}_{X}\left\vert \bar{\Psi}\left( X\right) \right\vert ^{2}%
}{K_{X}\left\vert \Psi \left( X\right) \right\vert ^{2}} &=&2\left( \frac{%
\delta S\left( X\right) +\delta S^{B}\left( X\right) }{1-\left( S\left(
X\right) +S^{B}\left( X\right) \right) }\right) \frac{\bar{K}_{X}\left\vert 
\bar{\Psi}\left( X\right) \right\vert ^{2}}{K_{X}\left\vert \Psi \left(
X\right) \right\vert ^{2}}
\end{eqnarray*}

\paragraph*{A11.3.2.5 Modifications of stakes}

We can also compute the shares variation due to default:%
\begin{eqnarray*}
&&S_{E}^{B}\left( X,X\right) \\
&=&w_{1}^{B}\left( X\right) \left\{ 1+\left\langle \bar{w}\left( X\right)
\right\rangle \left( f\left( X\right) -\frac{\left\langle \bar{f}\left(
X^{\prime }\right) \right\rangle _{\bar{w}_{1}}+\left\langle \bar{r}\left(
X^{\prime }\right) \right\rangle _{\bar{w}_{2}}}{2}\right) +\left\langle 
\hat{w}_{1}^{B}\left( X\right) \right\rangle \left( f\left( X\right)
-\left\langle \hat{f}\left( X^{\prime }\right) \right\rangle _{\hat{w}%
_{1}}\right) \right\}
\end{eqnarray*}%
\begin{equation*}
\frac{S_{L}^{B}\left( X,X\right) }{\kappa \left( 1-\bar{S}\left( X\right)
\right) }=w_{2}^{B}\left( X\right) \left[ 1+\hat{w}_{2}^{B}\left( X\right)
\left( r\left( X\right) -\left\langle \hat{r}\left( X^{\prime }\right)
\right\rangle _{\hat{w}_{2}}\right) \right]
\end{equation*}%
As a consequence:%
\begin{eqnarray}
&&\delta S^{B}\left( X,X\right)  \label{dst} \\
&\rightarrow &-w_{1}^{B}\left( X\right) \left\{ \left( \left\langle \bar{w}%
\left( X\right) \right\rangle +\left\langle \hat{w}_{1}^{B}\left( X\right)
\right\rangle \right) \left( df\left( X\right) -\frac{1}{2}\frac{%
\left\langle \bar{w}\left( X\right) \right\rangle \delta \left\langle \bar{f}%
\left( X^{\prime }\right) \right\rangle _{\bar{w}_{1}}}{\left\langle \bar{w}%
\left( X\right) \right\rangle +\left\langle \hat{w}_{1}^{B}\left( X\right)
\right\rangle }-\frac{\left\langle \hat{w}_{1}^{B}\left( X\right)
\right\rangle \delta \left\langle \hat{f}\left( X^{\prime }\right)
\right\rangle _{\hat{w}_{1}}}{\left\langle \bar{w}\left( X\right)
\right\rangle +\left\langle \hat{w}_{1}^{B}\left( X\right) \right\rangle }%
\right) \right\}  \notag \\
&\rightarrow &-w_{1}^{B}\left( X\right) \left\{ \left\langle \bar{w}\left(
X\right) \right\rangle _{f\left( X\right) }df\left( X\right) -\left\langle 
\bar{w}\left( X\right) \right\rangle _{\bar{S}}\left( \left\langle \bar{S}%
_{L}\left( X^{\prime },X\right) \right\rangle _{X^{\prime }}+\left\langle 
\hat{S}_{L}^{B}\left( X^{\prime },X\right) \right\rangle _{X^{\prime
}}+S_{L}^{B}\left( X,X\right) \right) \mu \right.  \notag \\
&&\left. -\left\langle \hat{w}_{1}^{B}\left( X\right) \right\rangle \frac{%
\left( \left\langle \left( \hat{S}_{L}\left( X^{\prime },X\right) \right)
\right\rangle _{X^{\prime }}+S_{L}\left( X,X\right) \right) \mu }{%
1-\left\langle \hat{S}\left( X^{\prime }\right) \right\rangle }\right\} 
\notag
\end{eqnarray}%
where we define:%
\begin{eqnarray*}
\left\langle \bar{w}\left( X\right) \right\rangle _{f\left( X\right) }
&=&\left( \left\langle \bar{w}\left( X\right) \right\rangle +\left\langle 
\hat{w}_{1}^{B}\left( X\right) \right\rangle \right) -\left\langle \hat{w}%
_{1}^{B}\left( X\right) \right\rangle \frac{S_{E}\left( X,X\right) }{%
1-\left\langle \hat{S}\left( X^{\prime }\right) \right\rangle } \\
&&-\frac{\left\langle \bar{w}\left( X\right) \right\rangle }{2}\left( \frac{%
S_{E}^{B}\left( X,X\right) }{1-\bar{S}\left( X^{\prime }\right) }%
+S_{E}\left( X,X\right) \frac{S_{E}^{B}\left( X,X\right) }{1-\bar{S}\left(
X^{\prime }\right) }\frac{\hat{S}_{E}^{B}\left( X^{\prime },X\right) }{%
1-\left\langle \hat{S}\left( X^{\prime }\right) \right\rangle }\frac{%
1-\left( \hat{S}\left( X^{\prime }\right) +\hat{S}_{E}^{B}\left( X^{\prime
}\right) +\hat{S}_{L}^{B}\left( X^{\prime }\right) \right) }{1-\left( \hat{S}%
_{E}\left( X^{\prime }\right) +\hat{S}_{E}^{B}\left( X^{\prime }\right)
\right) }\right)
\end{eqnarray*}%
\begin{equation*}
\left\langle \bar{w}\left( X\right) \right\rangle _{\bar{S}}=\frac{%
\left\langle \bar{w}\left( X\right) \right\rangle }{2}\left( \frac{1}{1-\bar{%
S}\left( X^{\prime }\right) }+\frac{S_{E}^{B}\left( X,X\right) }{1-\bar{S}%
\left( X^{\prime }\right) }\frac{\hat{S}_{E}^{B}\left( X^{\prime },X\right) 
}{1-\left\langle \hat{S}\left( X^{\prime }\right) \right\rangle }\frac{%
1-\left( \hat{S}\left( X^{\prime }\right) +\hat{S}_{E}^{B}\left( X^{\prime
}\right) +\hat{S}_{L}^{B}\left( X^{\prime }\right) \right) }{1-\left( \hat{S}%
_{E}\left( X^{\prime }\right) +\hat{S}_{E}^{B}\left( X^{\prime }\right)
\right) }\right)
\end{equation*}%
Gathering (\ref{dkn}), (\ref{dkt}), (\ref{dst}), we obtain: 
\begin{eqnarray*}
\delta S^{B}\left( X\right) &=&-w_{1}^{B}\left( X\right) \left\{
\left\langle \bar{w}\left( X\right) \right\rangle _{f\left( X\right)
}df\left( X\right) -\left\langle \bar{w}\left( X\right) \right\rangle _{\bar{%
S}}\left( \left\langle \bar{S}_{L}\left( X^{\prime },X\right) \right\rangle
_{X^{\prime }}+\left\langle \hat{S}_{L}^{B}\left( X^{\prime },X\right)
\right\rangle _{X^{\prime }}+S_{L}^{B}\left( X,X\right) \right) \mu \right.
\\
&&\left. -\left\langle \hat{w}_{1}^{B}\left( X\right) \right\rangle \frac{%
\left( \left\langle \left( \hat{S}_{L}\left( X^{\prime },X\right) \right)
\right\rangle _{X^{\prime }}+S_{L}\left( X,X\right) \right) \mu }{%
1-\left\langle \hat{S}\left( X^{\prime }\right) \right\rangle }\right\} 
\frac{\bar{K}_{X}\left\vert \bar{\Psi}\left( X\right) \right\vert ^{2}}{%
K_{X}\left\vert \Psi \left( X\right) \right\vert ^{2}} \\
&&+2S^{B}\left( X,X\right) \left( \frac{\delta S\left( X\right) +\delta
S^{B}\left( X\right) }{1-\left( S\left( X\right) +S^{B}\left( X\right)
\right) }\right) \frac{\bar{K}_{X}\left\vert \bar{\Psi}\left( X\right)
\right\vert ^{2}}{K_{X}\left\vert \Psi \left( X\right) \right\vert ^{2}}
\end{eqnarray*}%
As in Appendix 8 of Gosselin and Lotz (2025):

\begin{eqnarray*}
\delta S\left( X\right) &=&-\frac{w\left( X\right) \hat{w}\left( X\right) }{2%
}\frac{\hat{K}_{X}\left\vert \hat{\Psi}\left( X\right) \right\vert ^{2}}{%
K_{X}\left\vert \Psi \left( X\right) \right\vert ^{2}}\left( df\left(
X\right) \left( 1-\frac{\left\langle S_{E}\left( X,X\right) \right\rangle }{%
1-\left\langle \hat{S}\left( X\right) \right\rangle }\right) -\frac{\left(
\left\langle \left( \hat{S}_{L}\left( X^{\prime },X\right) \right)
\right\rangle _{X^{\prime }}+S_{L}\left( X,X\right) \right) \mu }{%
1-\left\langle \hat{S}\left( X\right) \right\rangle }\right) \\
&&+2S\left( X,X\right) \left( \frac{\delta S\left( X\right) +\delta
S^{B}\left( X\right) }{1-\left( S\left( X\right) +S^{B}\left( X\right)
\right) }\right) \frac{\hat{K}_{X}\left\vert \hat{\Psi}\left( X\right)
\right\vert ^{2}}{K_{X}\left\vert \Psi \left( X\right) \right\vert ^{2}}
\end{eqnarray*}%
and we can compute the total variation of invested shares in firms:

\begin{eqnarray*}
\delta S^{B}\left( X\right) +\delta S\left( X\right) &=&-\frac{\left( \frac{%
w\left( X\right) \hat{w}\left( X\right) }{2}\left( 1-\frac{\left\langle
S_{E}\left( X,X\right) \right\rangle }{1-\left\langle \hat{S}\left( X\right)
\right\rangle }\right) \frac{\hat{K}_{X}\left\vert \hat{\Psi}\left( X\right)
\right\vert ^{2}}{K_{X}\left\vert \Psi \left( X\right) \right\vert ^{2}}%
+w_{1}^{B}\left( X\right) \left\langle \bar{w}\left( X\right) \right\rangle
_{f\left( X\right) }\frac{\bar{K}_{X}\left\vert \bar{\Psi}\left( X\right)
\right\vert ^{2}}{K_{X}\left\vert \Psi \left( X\right) \right\vert ^{2}}%
\right) df\left( X\right) }{1-2\frac{S\left( X\right) +S^{B}\left( X\right) 
}{1-\left( S\left( X\right) +S^{B}\left( X\right) \right) }} \\
&&+\frac{\left( \frac{w\left( X\right) \hat{w}\left( X\right) }{2}\frac{\hat{%
K}_{X}\left\vert \hat{\Psi}\left( X\right) \right\vert ^{2}}{K_{X}\left\vert
\Psi \left( X\right) \right\vert ^{2}}+w_{1}^{B}\left( X\right) \left\langle 
\hat{w}_{1}^{B}\left( X\right) \right\rangle \frac{\bar{K}_{X}\left\vert 
\bar{\Psi}\left( X\right) \right\vert ^{2}}{K_{X}\left\vert \Psi \left(
X\right) \right\vert ^{2}}\right) \frac{\left( \left\langle \left( \hat{S}%
_{L}\left( X^{\prime },X\right) \right) \right\rangle _{X^{\prime
}}+S_{L}\left( X,X\right) \right) \mu }{1-\left\langle \hat{S}\left(
X\right) \right\rangle }}{1-2\frac{S\left( X\right) +S^{B}\left( X\right) }{%
1-\left( S\left( X\right) +S^{B}\left( X\right) \right) }} \\
&&+\frac{w_{1}^{B}\left( X\right) \left\langle \bar{w}\left( X\right)
\right\rangle _{\bar{S}}\frac{\bar{K}_{X}\left\vert \bar{\Psi}\left(
X\right) \right\vert ^{2}}{K_{X}\left\vert \Psi \left( X\right) \right\vert
^{2}}\left( \left\langle \bar{S}_{L}\left( X^{\prime },X\right)
\right\rangle _{X^{\prime }}+\left\langle \hat{S}_{L}^{B}\left( X^{\prime
},X\right) \right\rangle _{X^{\prime }}+S_{L}^{B}\left( X,X\right) \right)
\mu }{1-2\frac{S\left( X\right) +S^{B}\left( X\right) }{1-\left( S\left(
X\right) +S^{B}\left( X\right) \right) }}
\end{eqnarray*}

\paragraph*{A11.3.2.6 Solving for modifications of returns}

Given that:%
\begin{equation*}
df\left( X\right) =-\left( C-\frac{r\left( f_{1}\left( X\right) +\tau \Delta
F_{\tau }\left( \bar{R}\left( K,X\right) \right) \right) }{\left( 1-S\left(
X\right) \right) ^{1-r}}\right) \left( \delta S^{B}\left( X\right) +\delta
S\left( X\right) \right)
\end{equation*}%
the variation $df\left( X\right) $ satisfies:%
\begin{eqnarray}
df\left( X\right) &=&\left( C-\frac{r\left( f_{1}\left( X\right) +\tau
\Delta F_{\tau }\left( \bar{R}\left( K,X\right) \right) \right) }{\left(
1-S\left( X\right) \right) ^{1-r}}\right)  \label{Dfb} \\
&&\left\{ \frac{\left( \frac{w\left( X\right) \hat{w}\left( X\right) }{2}%
\left( 1-\frac{\left\langle S_{E}\left( X,X\right) \right\rangle }{%
1-\left\langle \hat{S}\left( X\right) \right\rangle }\right) \frac{\hat{K}%
_{X}\left\vert \hat{\Psi}\left( X\right) \right\vert ^{2}}{K_{X}\left\vert
\Psi \left( X\right) \right\vert ^{2}}+w_{1}^{B}\left( X\right) \left\langle 
\bar{w}\left( X\right) \right\rangle _{f\left( X\right) }\frac{\bar{K}%
_{X}\left\vert \bar{\Psi}\left( X\right) \right\vert ^{2}}{K_{X}\left\vert
\Psi \left( X\right) \right\vert ^{2}}\right) df\left( X\right) }{1-2\frac{%
S\left( X\right) +S^{B}\left( X\right) }{1-\left( S\left( X\right)
+S^{B}\left( X\right) \right) }}\right.  \notag \\
&&-\frac{\left( \frac{w\left( X\right) \hat{w}\left( X\right) }{2}\frac{\hat{%
K}_{X}\left\vert \hat{\Psi}\left( X\right) \right\vert ^{2}}{K_{X}\left\vert
\Psi \left( X\right) \right\vert ^{2}}+w_{1}^{B}\left( X\right) \left\langle 
\hat{w}_{1}^{B}\left( X\right) \right\rangle \frac{\bar{K}_{X}\left\vert 
\bar{\Psi}\left( X\right) \right\vert ^{2}}{K_{X}\left\vert \Psi \left(
X\right) \right\vert ^{2}}\right) \frac{\left( \left\langle \left( \hat{S}%
_{L}\left( X^{\prime },X\right) \right) \right\rangle _{X^{\prime
}}+S_{L}\left( X,X\right) \right) \mu }{1-\left\langle \hat{S}\left(
X\right) \right\rangle }}{1-2\frac{S\left( X\right) +S^{B}\left( X\right) }{%
1-\left( S\left( X\right) +S^{B}\left( X\right) \right) }}  \notag \\
&&\left. -\frac{w_{1}^{B}\left( X\right) \left\langle \bar{w}\left( X\right)
\right\rangle _{\bar{S}}\frac{\bar{K}_{X}\left\vert \bar{\Psi}\left(
X\right) \right\vert ^{2}}{K_{X}\left\vert \Psi \left( X\right) \right\vert
^{2}}\left( \left\langle \bar{S}_{L}\left( X^{\prime },X\right)
\right\rangle _{X^{\prime }}+\left\langle \hat{S}_{L}^{B}\left( X^{\prime
},X\right) \right\rangle _{X^{\prime }}+S_{L}^{B}\left( X,X\right) \right)
\mu }{1-2\frac{S\left( X\right) +S^{B}\left( X\right) }{1-\left( S\left(
X\right) +S^{B}\left( X\right) \right) }}\right\}  \notag
\end{eqnarray}%
and the solution has the following form:%
\begin{equation}
df\left( X\right) =-\frac{A}{B}\mu dC  \label{Dfc}
\end{equation}%
with:%
\begin{equation*}
dC=\left( C-\frac{r\left( f_{1}\left( X\right) +\tau \Delta F_{\tau }\left( 
\bar{R}\left( K,X\right) \right) \right) }{\left( 1-S\left( X\right) \right)
^{1-r}}\right)
\end{equation*}%
and where:%
\begin{eqnarray*}
A &=&\left( \frac{w\left( X\right) \hat{w}\left( X\right) }{2}\frac{\hat{K}%
_{X}\left\vert \hat{\Psi}\left( X\right) \right\vert ^{2}}{K_{X}\left\vert
\Psi \left( X\right) \right\vert ^{2}}+w_{1}^{B}\left( X\right) \left\langle 
\hat{w}_{1}^{B}\left( X\right) \right\rangle \frac{\bar{K}_{X}\left\vert 
\bar{\Psi}\left( X\right) \right\vert ^{2}}{K_{X}\left\vert \Psi \left(
X\right) \right\vert ^{2}}\right) \frac{\left( \left\langle \left( \hat{S}%
_{L}\left( X^{\prime },X\right) \right) \right\rangle _{X^{\prime
}}+S_{L}\left( X,X\right) \right) }{1-\left\langle \hat{S}\left( X\right)
\right\rangle } \\
&&+w_{1}^{B}\left( X\right) \left\langle \bar{w}\left( X\right)
\right\rangle _{\bar{S}}\frac{\bar{K}_{X}\left\vert \bar{\Psi}\left(
X\right) \right\vert ^{2}}{K_{X}\left\vert \Psi \left( X\right) \right\vert
^{2}}\left( \left\langle \bar{S}_{L}\left( X^{\prime },X\right)
\right\rangle _{X^{\prime }}+\left\langle \hat{S}_{L}^{B}\left( X^{\prime
},X\right) \right\rangle _{X^{\prime }}+S_{L}^{B}\left( X,X\right) \right) \\
&=&A_{1}\left( \left\langle \left( \hat{S}_{L}\left( X^{\prime },X\right)
\right) \right\rangle _{X^{\prime }}+S_{L}\left( X,X\right) \right)
+A_{2}\left( \left\langle \bar{S}_{L}\left( X^{\prime },X\right)
\right\rangle _{X^{\prime }}+\left\langle \hat{S}_{L}^{B}\left( X^{\prime
},X\right) \right\rangle _{X^{\prime }}+S_{L}^{B}\left( X,X\right) \right)
\end{eqnarray*}%
\begin{equation*}
B=\left( 1-2\frac{S\left( X\right) +S^{B}\left( X\right) }{1-\left( S\left(
X\right) +S^{B}\left( X\right) \right) }\right) \left( 1-dC\frac{\left( 
\frac{w\left( X\right) \hat{w}\left( X\right) }{2}\left( 1-\frac{%
\left\langle S_{E}\left( X,X\right) \right\rangle }{1-\left\langle \hat{S}%
\left( X\right) \right\rangle }\right) \frac{\hat{K}_{X}\left\vert \hat{\Psi}%
\left( X\right) \right\vert ^{2}}{K_{X}\left\vert \Psi \left( X\right)
\right\vert ^{2}}+w_{1}^{B}\left( X\right) \left\langle \bar{w}\left(
X\right) \right\rangle _{f\left( X\right) }\frac{\bar{K}_{X}\left\vert \bar{%
\Psi}\left( X\right) \right\vert ^{2}}{K_{X}\left\vert \Psi \left( X\right)
\right\vert ^{2}}\right) }{1-2\frac{S\left( X\right) +S^{B}\left( X\right) }{%
1-\left( S\left( X\right) +S^{B}\left( X\right) \right) }}\right)
\end{equation*}%
We compute fraction $\mu $ by considering: 
\begin{equation*}
\bar{f}\left( X\right) <-1+d\bar{f}
\end{equation*}%
This is an approximation, since we should consider different fraction for
banks and firms, but it allows nethertheless to obtain an order of magnitude
for the fraction of defaults. We find:%
\begin{equation*}
\mu =\frac{1}{\frac{\left\langle d\bar{f}\right\rangle }{\mu }-2\left\langle 
\bar{f}\right\rangle }
\end{equation*}%
with solution:%
\begin{equation*}
\mu =\frac{1}{2}-\frac{1-\left\langle d\bar{f}\right\rangle +\left\langle 
\bar{f}\right\rangle }{2\left\langle \bar{f}\right\rangle }=\frac{%
\left\langle d\bar{f}\right\rangle -1}{2\left\langle \bar{f}\right\rangle }
\end{equation*}%
We then find the variation for banks returns:

\begin{eqnarray*}
\frac{\left\langle d\bar{f}\right\rangle }{\mu } &=&\left( \frac{%
S_{E}^{B}\left( X,X\right) }{1-\bar{S}\left( X^{\prime }\right) }%
+S_{E}\left( X,X\right) \frac{S_{E}^{B}\left( X,X\right) }{1-\bar{S}\left(
X^{\prime }\right) }\frac{\hat{S}_{E}^{B}\left( X^{\prime },X\right) }{%
1-\left\langle \hat{S}\left( X^{\prime }\right) \right\rangle }\frac{%
1-\left( \hat{S}\left( X^{\prime }\right) +\hat{S}_{E}^{B}\left( X^{\prime
}\right) +\hat{S}_{L}^{B}\left( X^{\prime }\right) \right) }{1-\left( \hat{S}%
_{E}\left( X^{\prime }\right) +\hat{S}_{E}^{B}\left( X^{\prime }\right)
\right) }\right) \frac{df\left( X\right) }{\mu } \\
&&+\left( \frac{1}{1-\bar{S}\left( X^{\prime }\right) }+\frac{%
S_{E}^{B}\left( X,X\right) }{1-\bar{S}\left( X^{\prime }\right) }\frac{\hat{S%
}_{E}^{B}\left( X^{\prime },X\right) }{1-\left\langle \hat{S}\left(
X^{\prime }\right) \right\rangle }\frac{1-\left( \hat{S}\left( X^{\prime
}\right) +\hat{S}_{E}^{B}\left( X^{\prime }\right) +\hat{S}_{L}^{B}\left(
X^{\prime }\right) \right) }{1-\left( \hat{S}_{E}\left( X^{\prime }\right) +%
\hat{S}_{E}^{B}\left( X^{\prime }\right) \right) }\right)  \\
&&\times \left( \left\langle \left( \hat{S}_{L}\left( X^{\prime },X\right)
\right) \right\rangle _{X^{\prime }}+S_{L}\left( X,X\right) \right)  \\
&&+\frac{1}{1-\bar{S}\left( X^{\prime }\right) }\left( \left\langle \bar{S}%
_{L}\left( X^{\prime },X\right) \right\rangle _{X^{\prime }}+\left\langle 
\hat{S}_{L}^{B}\left( X^{\prime },X\right) \right\rangle _{X^{\prime
}}+S_{L}^{B}\left( X,X\right) \right) 
\end{eqnarray*}%
and the overall loss of every remaining banks is:%
\begin{equation*}
\left\langle d\bar{f}\right\rangle =\frac{1}{\frac{\left\langle d\bar{f}%
\right\rangle }{\mu }-2\left\langle \bar{f}\right\rangle }\frac{\left\langle
d\bar{f}\right\rangle }{\mu }
\end{equation*}

\paragraph*{A11.3.2.7 Banks default condition}

Consequently, there is default if $\mu >0$ that is:%
\begin{equation*}
\frac{\left\langle d\bar{f}\right\rangle }{\mu }-2\left\langle \bar{f}%
\right\rangle >0
\end{equation*}%
and this translates by the condition:%
\begin{eqnarray}
&&\left( \frac{1}{1-\bar{S}\left( X^{\prime }\right) }+\frac{S_{E}^{B}\left(
X,X\right) }{1-\bar{S}\left( X^{\prime }\right) }\frac{\hat{S}_{E}^{B}\left(
X^{\prime },X\right) }{1-\left\langle \hat{S}\left( X^{\prime }\right)
\right\rangle }\frac{1-\left( \hat{S}\left( X^{\prime }\right) +\hat{S}%
_{E}^{B}\left( X^{\prime }\right) +\hat{S}_{L}^{B}\left( X^{\prime }\right)
\right) }{1-\left( \hat{S}_{E}\left( X^{\prime }\right) +\hat{S}%
_{E}^{B}\left( X^{\prime }\right) \right) }\right)  \label{Cdt} \\
&&\times \left( \left\langle \left( \hat{S}_{L}\left( X^{\prime },X\right)
\right) \right\rangle _{X^{\prime }}+S_{L}\left( X,X\right) \right)  \notag
\\
&&+\frac{1}{1-\bar{S}\left( X^{\prime }\right) }\left( \left\langle \bar{S}%
_{L}\left( X^{\prime },X\right) \right\rangle _{X^{\prime }}+\left\langle 
\hat{S}_{L}^{B}\left( X^{\prime },X\right) \right\rangle _{X^{\prime
}}+S_{L}^{B}\left( X,X\right) \right)  \notag \\
&>&2\left\langle \bar{f}\right\rangle -\left( \frac{S_{E}^{B}\left(
X,X\right) }{1-\bar{S}\left( X^{\prime }\right) }+S_{E}\left( X,X\right) 
\frac{S_{E}^{B}\left( X,X\right) }{1-\bar{S}\left( X^{\prime }\right) }\frac{%
\hat{S}_{E}^{B}\left( X^{\prime },X\right) }{1-\left\langle \hat{S}\left(
X^{\prime }\right) \right\rangle }\frac{1-\left( \hat{S}\left( X^{\prime
}\right) +\hat{S}_{E}^{B}\left( X^{\prime }\right) +\hat{S}_{L}^{B}\left(
X^{\prime }\right) \right) }{1-\left( \hat{S}_{E}\left( X^{\prime }\right) +%
\hat{S}_{E}^{B}\left( X^{\prime }\right) \right) }\right) df\left( X\right) 
\notag
\end{eqnarray}%
Using (\ref{Dfc}), we write the firms loss: 
\begin{equation*}
\frac{df\left( X\right) }{\mu }=-\frac{A_{1}\left( \left\langle \left( \hat{S%
}_{L}\left( X^{\prime },X\right) \right) \right\rangle _{X^{\prime
}}+S_{L}\left( X,X\right) \right) +A_{2}\left( \left\langle \bar{S}%
_{L}\left( X^{\prime },X\right) \right\rangle _{X^{\prime }}+\left\langle 
\hat{S}_{L}^{B}\left( X^{\prime },X\right) \right\rangle _{X^{\prime
}}+S_{L}^{B}\left( X,X\right) \right) }{B}dC
\end{equation*}

\paragraph*{A11.3.2.8 dependency of default condition on parameters}

To study the default condition we use that (\ref{Cdt}) writes: 
\begin{eqnarray*}
&&\mathbf{H}\left( \left\langle \left( \hat{S}_{L}\left( X^{\prime
},X\right) \right) \right\rangle _{X^{\prime }}+S_{L}\left( X,X\right)
\right) \\
&&+\mathbf{G}\left( \frac{1-S_{E}^{B}\left( X,X\right) \frac{A_{2}}{B}dC}{1-%
\bar{S}\left( X^{\prime }\right) }\right) \left( \left\langle \bar{S}%
_{L}\left( X^{\prime },X\right) \right\rangle _{X^{\prime }}+\left\langle 
\hat{S}_{L}^{B}\left( X^{\prime },X\right) \right\rangle _{X^{\prime
}}+S_{L}^{B}\left( X,X\right) \right) >2\left\langle \bar{f}\right\rangle
\end{eqnarray*}%
where:%
\begin{equation*}
\mathbf{H=}\frac{1-S_{E}^{B}\left( X,X\right) \frac{A_{1}}{B}dC}{1-\bar{S}%
\left( X^{\prime }\right) }+H\left( 1-S_{E}\left( X,X\right) \frac{A_{1}}{B}%
dC\right)
\end{equation*}%
\begin{equation*}
\mathbf{G=}\left( \frac{1-S_{E}^{B}\left( X,X\right) \frac{A_{2}}{B}dC}{1-%
\bar{S}\left( X^{\prime }\right) }\right)
\end{equation*}%
The term:%
\begin{equation*}
H=\frac{S_{E}^{B}\left( X,X\right) }{1-\bar{S}\left( X^{\prime }\right) }%
\frac{\hat{S}_{E}^{B}\left( X^{\prime },X\right) }{1-\left\langle \hat{S}%
\left( X^{\prime }\right) \right\rangle }\frac{1-\left( \hat{S}\left(
X^{\prime }\right) +\hat{S}_{E}^{B}\left( X^{\prime }\right) +\hat{S}%
_{L}^{B}\left( X^{\prime }\right) \right) }{1-\left( \hat{S}_{E}\left(
X^{\prime }\right) +\hat{S}_{E}^{B}\left( X^{\prime }\right) \right) }
\end{equation*}%
increases with participations $\hat{S}_{E}\left( X^{\prime }\right) +\hat{S}%
_{E}^{B}\left( X^{\prime }\right) $ and $\bar{S}\left( X^{\prime }\right) $
and $\left\langle \hat{S}\left( X^{\prime }\right) \right\rangle $, while it
decreases with loans $\hat{S}_{L}\left( X^{\prime }\right) +$ $\hat{S}%
_{L}^{B}\left( X^{\prime }\right) $.

The coefficients $A_{1}$ and $A_{2}$: 
\begin{equation*}
A_{1}=\frac{\frac{w\left( X\right) \hat{w}\left( X\right) }{2}\frac{\hat{K}%
_{X}\left\vert \hat{\Psi}\left( X\right) \right\vert ^{2}}{K_{X}\left\vert
\Psi \left( X\right) \right\vert ^{2}}+w_{1}^{B}\left( X\right) \left\langle 
\hat{w}_{1}^{B}\left( X\right) \right\rangle \frac{\bar{K}_{X}\left\vert 
\bar{\Psi}\left( X\right) \right\vert ^{2}}{K_{X}\left\vert \Psi \left(
X\right) \right\vert ^{2}}}{1-\left\langle \hat{S}\left( X\right)
\right\rangle }\text{, }A_{2}=w_{1}^{B}\left( X\right) \left\langle \bar{w}%
\left( X\right) \right\rangle _{\bar{S}}\frac{\bar{K}_{X}\left\vert \bar{\Psi%
}\left( X\right) \right\vert ^{2}}{K_{X}\left\vert \Psi \left( X\right)
\right\vert ^{2}}
\end{equation*}%
increase with $\left\langle \hat{S}\left( X\right) \right\rangle $, with
capital ratios $\frac{\hat{K}_{X}\left\vert \hat{\Psi}\left( X\right)
\right\vert ^{2}}{K_{X}\left\vert \Psi \left( X\right) \right\vert ^{2}}$ $%
\frac{\bar{K}_{X}\left\vert \bar{\Psi}\left( X\right) \right\vert ^{2}}{%
K_{X}\left\vert \Psi \left( X\right) \right\vert ^{2}}$ and, decrease with
uncertainty since $w_{1}^{B}\left( X\right) \left\langle \hat{w}%
_{1}^{B}\left( X\right) \right\rangle $ and $w_{1}^{B}\left( X\right)
\left\langle \bar{w}\left( X\right) \right\rangle _{\bar{S}}$ decrease with
uncertainty $\gamma $.

\begin{eqnarray*}
B &=&\left( 1-2\frac{S\left( X\right) +S^{B}\left( X\right) }{1-\left(
S\left( X\right) +S^{B}\left( X\right) \right) }\right) \\
&&\times \left( 1-dC\frac{\left( \frac{w\left( X\right) \hat{w}\left(
X\right) }{2}\left( 1-\frac{\left\langle S_{E}\left( X,X\right)
\right\rangle }{1-\left\langle \hat{S}\left( X\right) \right\rangle }\right) 
\frac{\hat{K}_{X}\left\vert \hat{\Psi}\left( X\right) \right\vert ^{2}}{%
K_{X}\left\vert \Psi \left( X\right) \right\vert ^{2}}+w_{1}^{B}\left(
X\right) \left\langle \bar{w}\left( X\right) \right\rangle _{f\left(
X\right) }\frac{\bar{K}_{X}\left\vert \bar{\Psi}\left( X\right) \right\vert
^{2}}{K_{X}\left\vert \Psi \left( X\right) \right\vert ^{2}}\right) }{1-2%
\frac{S\left( X\right) +S^{B}\left( X\right) }{1-\left( S\left( X\right)
+S^{B}\left( X\right) \right) }}\right)
\end{eqnarray*}%
$B$ increases with uncertainty $\gamma $. Then using that 
\begin{equation*}
\frac{\left\langle S_{E}\left( X,X\right) \right\rangle }{1-\left\langle 
\hat{S}\left( X\right) \right\rangle }=\frac{1-\left\langle \hat{S}\left(
X\right) \right\rangle -\left\langle S_{L}\left( X,X\right) \right\rangle }{%
1-\left\langle \hat{S}\left( X\right) \right\rangle }=1-\frac{\left\langle
S_{L}\left( X,X\right) \right\rangle }{1-\left\langle \hat{S}\left( X\right)
\right\rangle }
\end{equation*}%
decreases with $\left\langle \hat{S}\left( X\right) \right\rangle $, we
obtain that $B$ decreases with $\left\langle \hat{S}\left( X\right)
\right\rangle $. $B$ also decreases with capital ratios $\frac{\hat{K}%
_{X}\left\vert \hat{\Psi}\left( X\right) \right\vert ^{2}}{K_{X}\left\vert
\Psi \left( X\right) \right\vert ^{2}}$ $\frac{\bar{K}_{X}\left\vert \bar{%
\Psi}\left( X\right) \right\vert ^{2}}{K_{X}\left\vert \Psi \left( X\right)
\right\vert ^{2}}$.

As a consequence:%
\begin{equation*}
\frac{1-S_{E}^{B}\left( X,X\right) \frac{A_{1}}{B}dC}{1-\bar{S}\left(
X^{\prime }\right) }
\end{equation*}%
\begin{equation*}
\mathbf{G}=\frac{1-S_{E}^{B}\left( X,X\right) \frac{A_{2}}{B}dC}{1-\bar{S}%
\left( X^{\prime }\right) }
\end{equation*}%
increase as function of $\bar{S}\left( X^{\prime }\right) $, $\left\langle 
\hat{S}\left( X\right) \right\rangle $, with capital ratios $\frac{\hat{K}%
_{X}\left\vert \hat{\Psi}\left( X\right) \right\vert ^{2}}{K_{X}\left\vert
\Psi \left( X\right) \right\vert ^{2}}$ $\frac{\bar{K}_{X}\left\vert \bar{%
\Psi}\left( X\right) \right\vert ^{2}}{K_{X}\left\vert \Psi \left( X\right)
\right\vert ^{2}}$ \ and decrease as a function of $S_{E}^{B}\left(
X,X\right) $ and uncertaint $\gamma $.

Moreover:%
\begin{equation*}
\mathbf{H=}\frac{1-S_{E}^{B}\left( X,X\right) \frac{A_{1}}{B}dC}{1-\bar{S}%
\left( X^{\prime }\right) }+H\left( 1-S_{E}\left( X,X\right) \frac{A_{1}}{B}%
dC\right)
\end{equation*}%
increases as function of $\bar{S}\left( X^{\prime }\right) $, $\left\langle 
\hat{S}\left( X\right) \right\rangle $, $\hat{S}_{E}\left( X^{\prime
}\right) $, $\hat{S}_{E}^{B}\left( X^{\prime }\right) $, with capital ratios 
$\frac{\hat{K}_{X}\left\vert \hat{\Psi}\left( X\right) \right\vert ^{2}}{%
K_{X}\left\vert \Psi \left( X\right) \right\vert ^{2}}$, $\frac{\bar{K}%
_{X}\left\vert \bar{\Psi}\left( X\right) \right\vert ^{2}}{K_{X}\left\vert
\Psi \left( X\right) \right\vert ^{2}}$ while it decreases with loans $\hat{S%
}_{L}\left( X^{\prime }\right) +$ $\hat{S}_{L}^{B}\left( X^{\prime }\right) $%
\ and decrease as a function of $S_{E}^{B}\left( X,X\right) $, $S_{E}\left(
X,X\right) $ and uncertainty $\gamma $.

\section*{Appendix 12 Capital circulation}

Since we consider fluctuations and return dynamics, we can normalize private
capital and replace as in (\ref{FDNr}):%
\begin{eqnarray*}
&&f_{1}\left( X\right) -C \\
&\rightarrow &\left( 1-\left( \left\langle S\left( X^{\prime },X^{\prime
}\right) \right\rangle \frac{\hat{K}_{X^{\prime }}\left\Vert \hat{\Psi}%
\right\Vert ^{2}}{K_{X^{\prime }}\left\vert \Psi \left( X^{\prime }\right)
\right\vert ^{2}}+\left\langle S^{B}\left( X^{\prime },X^{\prime }\right)
\right\rangle \frac{\bar{K}_{X^{\prime }}\left\vert \bar{\Psi}\left(
X^{\prime }\right) \right\vert ^{2}}{K_{X^{\prime }}\left\vert \Psi \left(
X^{\prime }\right) \right\vert ^{2}}\right) \right) ^{r}f_{1}\left( X\right)
\\
&&-\left( 1-\left( \left\langle S\left( X^{\prime },X^{\prime }\right)
\right\rangle \frac{\hat{K}_{X^{\prime }}\left\Vert \hat{\Psi}\right\Vert
^{2}}{K_{X^{\prime }}\left\vert \Psi \left( X^{\prime }\right) \right\vert
^{2}}+\left\langle S^{B}\left( X^{\prime },X^{\prime }\right) \right\rangle 
\frac{\hat{K}_{X^{\prime }}\left\Vert \hat{\Psi}\right\Vert ^{2}}{%
K_{X^{\prime }}\left\vert \Psi \left( X^{\prime }\right) \right\vert ^{2}}%
\right) \right) C \\
&\rightarrow &\left( 1-\left( S\left( X,,\theta -1\right) +S_{E}^{B}\left(
X,,\theta -1\right) +S_{L}^{B}\left( X,\theta -1\right) \right) \right)
^{r}f_{1}\left( X\right) \\
&&-\left( 1-\left( S\left( X,,\theta -1\right) +S_{E}^{B}\left( X,,\theta
-1\right) +S_{L}^{B}\left( X,\theta -1\right) \right) \right) C
\end{eqnarray*}%
We start with the return equations for investrs and banks with decreasing
returns written as:%
\begin{eqnarray*}
0 &=&\frac{1-\hat{S}\left( X\right) }{1-\hat{S}_{E}\left( X\right) }\left( 
\hat{f}\left( X^{\prime }\right) -\bar{r}\right) -\left\langle \hat{S}%
_{E}\left( X^{\prime },X\right) \right\rangle _{X^{\prime }}\frac{%
1-\left\langle \hat{S}\left( X^{\prime }\right) \right\rangle }{%
1-\left\langle \hat{S}_{E}\left( X^{\prime }\right) \right\rangle }\left(
\left\langle \hat{f}\left( X^{\prime }\right) \right\rangle -\left\langle 
\bar{r}\right\rangle \right) \\
&&-S_{E}\left( X,X\right) \left( \left( 1-S^{T}\left( X,,\theta -1\right)
\right) ^{r}f_{1}\left( X\right) -C_{0}-\left( 1-S^{T}\left( X,,\theta
-1\right) \right) C-\bar{r}\right. \\
&&\left. +\left( 1-S^{T}\left( X,,\theta -1\right) \right) ^{r}\tau \left(
\left\langle f_{1}\left( X\right) \right\rangle -\left\langle f_{1}\left(
X^{\prime }\right) \right\rangle \right) \right)
\end{eqnarray*}%
\begin{eqnarray}
0 &=&\frac{1-\bar{S}\left( X\right) }{1-\bar{S}_{E}\left( X\right) }\left( 
\bar{f}\left( X\right) -\bar{r}\right) -\left\langle \bar{S}_{E}\left(
X^{\prime },X\right) \right\rangle _{X^{\prime }}\frac{1-\left\langle \bar{S}%
\left( X^{\prime }\right) \right\rangle }{1-\left\langle \bar{S}_{E}\left(
X^{\prime }\right) \right\rangle }\left( \left\langle \bar{f}\left(
X^{\prime }\right) \right\rangle -\bar{r}\right) \\
&&-\left\langle \hat{S}_{E}^{B}\left( X^{\prime },X\right) \right\rangle
_{X^{\prime }}\frac{1-\left\langle \hat{S}\left( X^{\prime }\right)
\right\rangle +\left\langle \hat{S}_{E}^{B}\left( X^{\prime }\right)
\right\rangle +\left\langle \hat{S}_{L}^{B}\left( X^{\prime }\right)
\right\rangle }{1-\left\langle \hat{S}_{E}\left( X^{\prime }\right)
\right\rangle +\left\langle \hat{S}_{E}^{B}\left( X^{\prime }\right)
\right\rangle }\left( \left\langle \hat{f}\left( X^{\prime }\right)
\right\rangle -\bar{r}\right)  \notag \\
&&-S_{E}^{B}\left( X,X\right) \left( \left( 1-S^{T}\left( X,,\theta
-1\right) \right) ^{r}\left( f_{1}\left( X\right) +\tau \left( \left\langle
f_{1}\left( X\right) \right\rangle -\left\langle f_{1}\left( X^{\prime
}\right) \right\rangle \right) \right) -C_{0}-\left( 1-S^{T}\left( X,,\theta
-1\right) \right) C-\bar{r}\right)  \notag
\end{eqnarray}%
where we defind $S^{T}\left( X,,\theta -1\right) $, the total share of
investd captl in firms:%
\begin{equation*}
S^{T}\left( X,,\theta -1\right) =S\left( X,,\theta -1\right)
+S_{E}^{B}\left( X,,\theta -1\right) +S_{L}^{B}\left( X,\theta -1\right)
\end{equation*}%
We derive the variations of each terms in this equations.

\subsection*{A12.1 Variations of total stakes}

We first estimate:%
\begin{equation}
\delta S^{T}\left( X,,\theta -1\right) =\delta S\left( X,,\theta -1\right)
+\delta S_{E}^{B}\left( X,,\theta -1\right) +\delta S_{L}^{B}\left( X,\theta
-1\right)  \label{DS}
\end{equation}%
and compute separately each term. We have:%
\begin{eqnarray*}
&&S_{E}^{B}\left( X,X\right) \\
&=&w_{1}^{B}\left( X\right) \left\{ 1+\left\langle \bar{w}\left( X\right)
\right\rangle \left( f\left( X\right) -\frac{\left\langle \bar{f}\left(
X^{\prime }\right) \right\rangle _{\bar{w}_{1}}+\left\langle \bar{r}\left(
X^{\prime }\right) \right\rangle _{\bar{w}_{2}}}{2}\right) +\left\langle 
\hat{w}_{1}^{B}\left( X\right) \right\rangle \left( f\left( X\right)
-\left\langle \hat{f}\left( X^{\prime }\right) \right\rangle _{\hat{w}%
_{1}}\right) \right\}
\end{eqnarray*}%
which implies:%
\begin{eqnarray*}
&&\delta S_{E}^{B}\left( X,X\right) \\
&=&w_{1}^{B}\left( X\right) \left( \left\langle \bar{w}\left( X\right)
\right\rangle +\left\langle \hat{w}_{1}^{B}\left( X\right) \right\rangle
\right) \delta f\left( X\right)
\end{eqnarray*}%
using:%
\begin{eqnarray*}
\frac{\bar{K}_{X}\left\vert \bar{\Psi}\left( X\right) \right\vert ^{2}}{%
K_{X}\left\vert \Psi \left( X\right) \right\vert ^{2}} &\simeq &\frac{72%
\frac{\sigma _{\hat{K}}^{2}V\left\langle \hat{g}\right\rangle ^{4}}{\bar{g}%
^{2}\left( X\right) \hat{\mu}}\left( \left\Vert \bar{\Psi}_{0}\right\Vert
\right) ^{4}}{\epsilon \sqrt{\sigma _{\hat{K}}^{2}\frac{\left\vert \Psi
_{0}\left( X\right) \right\vert ^{2}}{\epsilon }}X^{3}}\sigma _{\hat{K}%
}^{2}\left( Z\left( X\right) \left( f_{1}\left( X\right) -r\left( X\right)
\right) +r\left( X\right) \right) ^{2} \\
&\simeq &\frac{72\frac{\sigma _{\hat{K}}^{2}V\left( 1-\left\langle \bar{S}%
\right\rangle \right) ^{2}\left( \left\langle \hat{f}\right\rangle +\frac{%
\left\langle \hat{S}_{E}^{B}\right\rangle +\left\langle \hat{S}%
_{L}^{B}\right\rangle }{1-\left\langle \bar{S}\right\rangle }\left\langle 
\bar{f}\right\rangle \right) ^{4}}{\left( \bar{f}\left( X\right) \left(
1-\left\langle \bar{S}\right\rangle \right) +\left\langle \bar{S}\left(
X^{\prime },X\right) \right\rangle _{X^{\prime }}\left\langle \bar{f}%
\right\rangle \right) \left( 1-\left\langle \hat{S}\right\rangle \right) ^{4}%
\hat{\mu}}\left( \left\Vert \bar{\Psi}_{0}\right\Vert \right) ^{4}}{\epsilon 
\sqrt{\sigma _{\hat{K}}^{2}\frac{\left\vert \Psi _{0}\left( X\right)
\right\vert ^{2}}{\epsilon }}X^{3}}\sigma _{\hat{K}}^{2}\left( Z\left(
X\right) \left( f_{1}\left( X\right) -r\left( X\right) \right) +r\left(
X\right) \right) ^{2}
\end{eqnarray*}%
to compute $\delta f\left( X\right) $, leads ultimately to:

\begin{eqnarray*}
&&\delta S_{E}^{B}\left( X,\theta -1\right) \\
&=&w_{1}^{B}\left( X\right) \left( \left\langle \bar{w}\left( X\right)
\right\rangle +\left\langle \hat{w}_{1}^{B}\left( X\right) \right\rangle
\right) \delta f\left( X,\theta -1\right) \\
&&+S_{E}^{B}\left( X,X\right) \left( \frac{\partial _{\bar{f}\left( X\right)
}\bar{K}_{X}\left\vert \bar{\Psi}\left( X\right) \right\vert ^{2}}{\bar{K}%
_{X}\left\vert \bar{\Psi}\left( X\right) \right\vert ^{2}}\delta \bar{f}%
\left( X,\theta -1\right) -\frac{\partial _{f\left( X\right)
}K_{X}\left\vert \Psi \left( X\right) \right\vert ^{2}}{K_{X}\left\vert \Psi
\left( X\right) \right\vert ^{2}}\delta f\left( X,\theta -1\right) \right) \\
&=&\left( w_{1}^{B}\left( X\right) \left( \left\langle \bar{w}\left(
X\right) \right\rangle +\left\langle \hat{w}_{1}^{B}\left( X\right)
\right\rangle \right) -S_{E}^{B}\left( X,\theta -1\right) \frac{\partial
_{f\left( X\right) }K_{X}\left\vert \Psi \left( X\right) \right\vert ^{2}}{%
K_{X}\left\vert \Psi \left( X\right) \right\vert ^{2}}\right) \frac{\partial
f\left( X\right) }{\partial S^{T}\left( X,\theta -2\right) }\delta
S^{T}\left( X,\theta -2\right) \\
&&+S_{E}^{B}\left( X,\theta -1\right) \frac{\partial _{\bar{f}\left(
X\right) }\bar{K}_{X}\left\vert \bar{\Psi}\left( X\right) \right\vert ^{2}}{%
\bar{K}_{X}\left\vert \bar{\Psi}\left( X\right) \right\vert ^{2}}\delta \bar{%
f}\left( X,\theta -1\right)
\end{eqnarray*}%
The last term in (\ref{DS}) is estimated by writing:%
\begin{equation*}
S_{L}^{B}\left( X\right) =S_{L}^{B}\left( X,X\right) \frac{\bar{K}%
_{X}\left\vert \bar{\Psi}\left( X\right) \right\vert ^{2}}{K_{X}\left\vert
\Psi \left( X\right) \right\vert ^{2}}
\end{equation*}%
and:%
\begin{equation}
\frac{S_{L}^{B}\left( X,X\right) }{\kappa \left( 1-\left\langle \bar{S}%
\left( X\right) \right\rangle \right) }=\frac{1}{2-\left( \gamma
\left\langle \hat{S}_{E}\left( X^{\prime },X\right) \right\rangle \right)
^{2}}\left[ 1+\frac{1-\left( \gamma \left\langle \hat{S}_{E}\left( X^{\prime
},X\right) \right\rangle \right) ^{2}}{2-\left( \gamma \left\langle \hat{S}%
_{E}\left( X^{\prime },X\right) \right\rangle \right) ^{2}}\left( r\left(
X\right) -\left\langle \hat{r}\left( X^{\prime }\right) \right\rangle _{\hat{%
w}_{2}}\right) \right]
\end{equation}%
which yields:%
\begin{eqnarray*}
\delta S_{L}^{B}\left( X,\theta -1\right) &=&S_{L}^{B}\left( X,\theta
-1\right) \left( \frac{\partial _{\bar{f}\left( X\right) }\bar{K}%
_{X}\left\vert \bar{\Psi}\left( X\right) \right\vert ^{2}}{\bar{K}%
_{X}\left\vert \bar{\Psi}\left( X\right) \right\vert ^{2}}\delta \bar{f}%
\left( X,\theta -1\right) -\frac{\partial _{f\left( X\right)
}K_{X}\left\vert \Psi \left( X\right) \right\vert ^{2}}{K_{X}\left\vert \Psi
\left( X\right) \right\vert ^{2}}\delta f\left( X,\theta -1\right) \right) \\
&=&S_{L}^{B}\left( X,\theta -1\right) \frac{\partial _{\bar{f}\left(
X\right) }\bar{K}_{X}\left\vert \bar{\Psi}\left( X\right) \right\vert ^{2}}{%
\bar{K}_{X}\left\vert \bar{\Psi}\left( X\right) \right\vert ^{2}}\delta \bar{%
f}\left( X,\theta -1\right) \\
&&-S_{L}^{B}\left( X,\theta -1\right) \frac{\partial _{f\left( X\right)
}K_{X}\left\vert \Psi \left( X\right) \right\vert ^{2}}{K_{X}\left\vert \Psi
\left( X\right) \right\vert ^{2}}\frac{\partial f\left( X\right) }{\partial
S^{T}\left( X,\theta -2\right) }\delta S^{T}\left( X,\theta -2\right)
\end{eqnarray*}%
The first term in (\ref{DS}) is computed directly:%
\begin{eqnarray*}
\delta S\left( X,\theta -1\right) &=&\frac{\frac{\hat{w}\left( X\right) }{2}%
\frac{\partial f\left( X\right) }{\partial S\left( X,\theta -2\right) }%
\delta S\left( X,\theta -2\right) }{1+\left( \hat{w}\left( X\right) \left( 
\frac{f\left( X\right) +\bar{r}\left( X\right) }{2}-\frac{\left\langle \hat{f%
}\left( X^{\prime }\right) \right\rangle _{\hat{w}_{1}}+\left\langle \hat{r}%
\left( X^{\prime }\right) \right\rangle _{\hat{w}_{2}}}{2}\right) \right) }%
S\left( X\right) +\frac{\delta \frac{\hat{K}_{X}\left\vert \hat{\Psi}\left(
X\right) \right\vert ^{2}}{K_{X}\left\vert \Psi \left( X\right) \right\vert
^{2}}}{\frac{\hat{K}_{X}\left\vert \hat{\Psi}\left( X\right) \right\vert ^{2}%
}{K_{X}\left\vert \Psi \left( X\right) \right\vert ^{2}}}S\left( X\right) \\
&=&\frac{\frac{\hat{w}\left( X\right) }{2}\frac{\partial f\left( X\right) }{%
\partial S\left( X,\theta -2\right) }\delta S\left( X,\theta -2\right) }{%
1+\left( \hat{w}\left( X\right) \left( \frac{f\left( X\right) +\bar{r}\left(
X\right) }{2}-\frac{\left\langle \hat{f}\left( X^{\prime }\right)
\right\rangle _{\hat{w}_{1}}+\left\langle \hat{r}\left( X^{\prime }\right)
\right\rangle _{\hat{w}_{2}}}{2}\right) \right) }S\left( X\right) \\
&&+\frac{\partial _{\hat{f}\left( X\right) }\left( \frac{\hat{K}%
_{X}\left\vert \hat{\Psi}\left( X\right) \right\vert ^{2}}{K_{X}\left\vert
\Psi \left( X\right) \right\vert ^{2}}\right) S\left( X\right) }{\frac{\hat{K%
}_{X}\left\vert \hat{\Psi}\left( X\right) \right\vert ^{2}}{K_{X}\left\vert
\Psi \left( X\right) \right\vert ^{2}}}\delta \hat{f}\left( X,\theta
-1\right) \\
&&+\frac{\partial _{f\left( X\right) }\left( \frac{\hat{K}_{X}\left\vert 
\hat{\Psi}\left( X\right) \right\vert ^{2}}{K_{X}\left\vert \Psi \left(
X\right) \right\vert ^{2}}\right) S\left( X\right) }{\frac{\hat{K}%
_{X}\left\vert \hat{\Psi}\left( X\right) \right\vert ^{2}}{K_{X}\left\vert
\Psi \left( X\right) \right\vert ^{2}}}\frac{\partial f\left( X\right) }{%
\partial S^{T}\left( X,\theta -2\right) }\delta S^{T}\left( X,\theta
-2\right)
\end{eqnarray*}%
\begin{eqnarray*}
&&\delta S\left( X,\theta -1\right) \\
&=&\left( \frac{\frac{\hat{w}\left( X\right) }{2}S\left( X\right) }{1+\left( 
\hat{w}\left( X\right) \left( \frac{f\left( X\right) +\bar{r}\left( X\right) 
}{2}-\frac{\left\langle \hat{f}\left( X^{\prime }\right) \right\rangle _{%
\hat{w}_{1}}+\left\langle \hat{r}\left( X^{\prime }\right) \right\rangle _{%
\hat{w}_{2}}}{2}\right) \right) }+\frac{\partial _{f\left( X\right) }\left( 
\frac{\hat{K}_{X}\left\vert \hat{\Psi}\left( X\right) \right\vert ^{2}}{%
K_{X}\left\vert \Psi \left( X\right) \right\vert ^{2}}\right) S\left(
X\right) }{\frac{\hat{K}_{X}\left\vert \hat{\Psi}\left( X\right) \right\vert
^{2}}{K_{X}\left\vert \Psi \left( X\right) \right\vert ^{2}}}\right) \frac{%
\partial f\left( X\right) }{\partial S\left( X,\theta -2\right) }\delta
S\left( X,\theta -2\right) \\
&&+\frac{\partial _{\hat{f}\left( X\right) }\left( \frac{\hat{K}%
_{X}\left\vert \hat{\Psi}\left( X\right) \right\vert ^{2}}{K_{X}\left\vert
\Psi \left( X\right) \right\vert ^{2}}\right) S\left( X\right) }{\frac{\hat{K%
}_{X}\left\vert \hat{\Psi}\left( X\right) \right\vert ^{2}}{K_{X}\left\vert
\Psi \left( X\right) \right\vert ^{2}}}\delta \hat{f}\left( X,\theta
-1\right)
\end{eqnarray*}%
and we find ultimately:%
\begin{eqnarray}
&&\delta S^{T}\left( X,,\theta -1\right)  \label{DSd} \\
&=&\left( S_{E}^{B}\left( X,\theta -1\right) +S_{L}^{B}\left( X,\theta
-1\right) \right) \frac{\partial _{\bar{f}\left( X\right) }K_{X}\left\vert
\Psi \left( X\right) \right\vert ^{2}}{K_{X}\left\vert \Psi \left( X\right)
\right\vert ^{2}}\delta \bar{f}\left( X,\theta -1\right) +\frac{\partial _{%
\hat{f}\left( X\right) }\left( \frac{\hat{K}_{X}\left\vert \hat{\Psi}\left(
X\right) \right\vert ^{2}}{K_{X}\left\vert \Psi \left( X\right) \right\vert
^{2}}\right) S\left( X\right) }{\frac{\hat{K}_{X}\left\vert \hat{\Psi}\left(
X\right) \right\vert ^{2}}{K_{X}\left\vert \Psi \left( X\right) \right\vert
^{2}}}\delta \hat{f}\left( X,\theta -1\right)  \notag \\
&&+\left\{ w_{1}^{B}\left( X\right) \left( \left\langle \bar{w}\left(
X\right) \right\rangle +\left\langle \hat{w}_{1}^{B}\left( X\right)
\right\rangle \right) -\left( S_{E}^{B}\left( X,\theta -1\right)
+S_{L}^{B}\left( X,\theta -1\right) \right) \frac{\partial _{f\left(
X\right) }K_{X}\left\vert \Psi \left( X\right) \right\vert ^{2}}{%
K_{X}\left\vert \Psi \left( X\right) \right\vert ^{2}}\right.  \notag \\
&&\left. +\left( \frac{\frac{\hat{w}\left( X\right) }{2}S\left( X\right) }{%
1+\left( \hat{w}\left( X\right) \left( \frac{f\left( X\right) +\bar{r}\left(
X\right) }{2}-\frac{\left\langle \hat{f}\left( X^{\prime }\right)
\right\rangle _{\hat{w}_{1}}+\left\langle \hat{r}\left( X^{\prime }\right)
\right\rangle _{\hat{w}_{2}}}{2}\right) \right) }+\frac{\partial _{f\left(
X\right) }\left( \frac{\hat{K}_{X}\left\vert \hat{\Psi}\left( X\right)
\right\vert ^{2}}{K_{X}\left\vert \Psi \left( X\right) \right\vert ^{2}}%
\right) S\left( X\right) }{\frac{\hat{K}_{X}\left\vert \hat{\Psi}\left(
X\right) \right\vert ^{2}}{K_{X}\left\vert \Psi \left( X\right) \right\vert
^{2}}}\right) \right\} \frac{\delta S^{T}\left( X,\theta -2\right) \partial
f\left( X\right) }{\partial S^{T}\left( X,\theta -2\right) }  \notag
\end{eqnarray}

\subsection*{A12.2 Variations of return equations}

\subsubsection*{A12.2.1 Computation of $\protect\delta f\left( X,\protect%
\theta -1\right) $}

We first compute $\delta f\left( X,\theta -1\right) $%
\begin{equation*}
\frac{\partial f\left( X\right) }{\partial S^{T}\left( X,\theta -2\right) }%
=-\left( r\left( 1-S^{T}\left( X,\theta -2\right) \right) ^{r-1}\left(
f_{1}\left( X\right) +\Delta F_{\tau }\left( \bar{R}\left( K,X\right)
\right) \right) -C\right)
\end{equation*}%
so that:%
\begin{equation*}
\delta f\left( X,\theta -1\right) =-\left( r\left( 1-S^{T}\left( X,\theta
-2\right) \right) ^{r-1}\left( f_{1}\left( X\right) +\Delta F_{\tau }\left( 
\bar{R}\left( K,X\right) \right) \right) -C\right) \delta S^{T}\left(
X,\theta -2\right)
\end{equation*}%
Then using:

\begin{equation*}
\frac{1-\left( \hat{S}\left( X,\theta \right) +\delta \hat{S}\left( X,\theta
\right) \right) }{1-\left( \hat{S}_{E}\left( X^{\prime },\theta \right)
+\delta \hat{S}_{E}\left( X,\theta \right) \right) }=\bigskip -\frac{\delta 
\hat{S}\left( X,\theta \right) }{1-\left( \hat{S}_{E}\left( X^{\prime
},\theta \right) \right) }+\frac{\left( 1-\left( \hat{S}\left( X,\theta
\right) \right) \right) }{\left( 1-\left( \hat{S}_{E}\left( X^{\prime
},\theta \right) \right) \right) ^{2}}\delta \hat{S}_{E}\left( X,\theta
\right)
\end{equation*}%
and:

\begin{eqnarray}
&&\delta \hat{S}_{E}\left( X^{\prime }\right) \\
&\rightarrow &\frac{1}{2}\frac{\left( 1-\left( \gamma \left\langle \hat{S}%
_{E}\left( X\right) \right\rangle \right) ^{2}\right) \left( 1+\Delta \hat{f}%
\left( X^{\prime }\right) \right) }{2-\left( \gamma \left\langle \hat{S}%
_{E}\left( X\right) \right\rangle \right) ^{2}-\gamma \left\langle \hat{S}%
_{E}\left( X\right) \right\rangle \gamma \left\langle \hat{w}\left(
X^{\prime },X\right) \right\rangle \left\langle w\left( X\right)
\right\rangle \Delta \left( \frac{f\left( X^{\prime }\right) +r\left(
X^{\prime }\right) }{2}\right) }\frac{\left\langle \hat{K}\right\rangle
\left\Vert \hat{\Psi}\right\Vert ^{2}}{\hat{K}_{X^{\prime }}\left\vert \hat{%
\Psi}\left( X^{\prime }\right) \right\vert ^{2}}  \notag
\end{eqnarray}%
where:%
\begin{equation}
\Delta \hat{f}\left( X^{\prime }\right) =\hat{f}\left( X^{\prime }\right)
-\left( \left\langle \hat{w}\left( X\right) \right\rangle \frac{\left\langle 
\hat{f}\left( X^{\prime }\right) \right\rangle _{\hat{w}_{1}}+\left\langle 
\hat{r}\left( X^{\prime }\right) \right\rangle _{\hat{w}_{2}}}{2}%
+\left\langle w\left( X\right) \right\rangle \frac{\left\langle f\left(
X\right) \right\rangle +\left\langle r\left( X\right) \right\rangle }{2}%
\right)
\end{equation}%
In first approximation: 
\begin{equation*}
\delta \hat{S}_{E}\left( X,\theta -1\right) \simeq \frac{\delta \Delta \hat{f%
}\left( X,\theta -1\right) }{1+\Delta \hat{f}\left( X,\theta -1\right) }\hat{%
S}_{E}\left( X\right) -\frac{\frac{\partial \hat{K}_{X^{\prime }}\left\vert 
\hat{\Psi}\left( X^{\prime }\right) \right\vert ^{2}}{\partial \hat{f}\left(
X,\theta -1\right) }}{\hat{K}_{X^{\prime }}\left\vert \hat{\Psi}\left(
X^{\prime }\right) \right\vert ^{2}}\hat{S}_{E}\left( X\right) \delta \hat{f}%
\left( X,\theta -1\right)
\end{equation*}%
with:%
\begin{eqnarray}
&&\frac{\hat{K}_{X}\left\vert \hat{\Psi}\left( X\right) \right\vert ^{2}}{%
K_{X}\left\vert \Psi \left( X\right) \right\vert ^{2}} \\
&\simeq &\frac{\frac{18\sigma _{\hat{K}}^{2}}{\hat{\mu}\hat{g}^{2}\left(
X\right) }V\left\Vert \hat{\Psi}_{0}\left( X\right) \right\Vert ^{4}}{%
\epsilon \sqrt{\sigma _{\hat{K}}^{2}\frac{\left\vert \Psi _{0}\left(
X\right) \right\vert ^{2}}{\epsilon }}X^{3}}\sigma _{\hat{K}}^{2}\left(
Z\left( X\right) \left( f_{1}\left( X\right) -r\left( X\right) \right)
+r\left( X\right) \right) ^{2}  \notag \\
&\simeq &\frac{\frac{18\sigma _{\hat{K}}^{2}}{\hat{\mu}\epsilon \sqrt{\sigma
_{\hat{K}}^{2}\frac{\left\vert \Psi _{0}\left( X\right) \right\vert ^{2}}{%
\epsilon }}}\left( 1-\left\langle \hat{S}\right\rangle \right)
^{2}V\left\Vert \hat{\Psi}_{0}\right\Vert ^{4}\sigma _{\hat{K}}^{2}\left(
Z\left( X\right) \left( f_{1}\left( X\right) -r\left( X\right) \right)
+r\left( X\right) \right) ^{2}}{\left( \left( 1-\left\langle \hat{S}%
\right\rangle \right) \hat{f}\left( X\right) +\frac{\left\langle \hat{S}%
_{E}^{B}\left( X,X^{\prime }\right) \right\rangle _{X^{\prime
}}+\left\langle \hat{S}_{L}^{B}\left( X,X^{\prime }\right) \right\rangle
_{X^{\prime }}\left\langle \bar{S}\right\rangle }{1-\left\langle \bar{S}%
\right\rangle }\left\langle \bar{f}\right\rangle +\left\langle \hat{S}\left(
X^{\prime },X\right) \right\rangle _{X^{\prime }}\left( \left\langle \hat{f}%
\right\rangle +\frac{\left\langle \hat{S}_{E}^{B}\right\rangle +\left\langle 
\hat{S}_{L}^{B}\right\rangle \left\langle \bar{S}\right\rangle }{%
1-\left\langle \bar{S}\right\rangle }\left\langle \bar{f}\right\rangle
\right) \right) ^{2}X^{3}}  \notag
\end{eqnarray}%
Then, we obtain the variation $\delta \hat{f}\left( X,\theta \right) $ by
gathering the different contributions: 
\begin{eqnarray*}
&&\frac{1-\left( \hat{S}\left( X\right) \right) }{1-\left( \hat{S}_{E}\left(
X\right) \right) }\delta \hat{f}\left( X,\theta \right) \\
&=&\left\{ \left( \frac{\hat{S}\left( X\right) }{2\left( 1-\left( \hat{S}%
_{E}\left( X^{\prime }\right) \right) \right) }\left( \frac{1}{1+\frac{%
\Delta \hat{f}\left( X^{\prime }\right) +\Delta \hat{r}\left( X^{\prime
}\right) }{2}}-\frac{\frac{\partial \hat{K}_{X^{\prime }}\left\vert \hat{\Psi%
}\left( X^{\prime }\right) \right\vert ^{2}}{\partial \hat{f}\left( X,\theta
-1\right) }}{\hat{K}_{X^{\prime }}\left\vert \hat{\Psi}\left( X^{\prime
}\right) \right\vert ^{2}}\right) \right. \right. \\
&&\left. -\frac{\left( 1-\left( \hat{S}\left( X\right) \right) \right) \hat{S%
}_{E}\left( X\right) }{\left( 1-\left( \hat{S}_{E}\left( X^{\prime }\right)
\right) \right) ^{2}}\left( \frac{1}{1+\Delta \hat{f}\left( X\right) }-\frac{%
\frac{\partial \hat{K}_{X^{\prime }}\left\vert \hat{\Psi}\left( X^{\prime
}\right) \right\vert ^{2}}{\partial \hat{f}\left( X,\theta -1\right) }}{\hat{%
K}_{X^{\prime }}\left\vert \hat{\Psi}\left( X^{\prime }\right) \right\vert
^{2}}\right) \right) \left( \hat{f}\left( X\right) -\bar{r}\right) \\
&&\left. +S_{E}\left( X,X,\theta -1\right) \frac{\partial f\left( X\right) }{%
\partial S^{T}\left( X,\theta -1\right) }\frac{\partial _{\hat{f}\left(
X\right) }\left( \frac{\hat{K}_{X}\left\vert \hat{\Psi}\left( X\right)
\right\vert ^{2}}{K_{X}\left\vert \Psi \left( X\right) \right\vert ^{2}}%
\right) }{\frac{\hat{K}_{X}\left\vert \hat{\Psi}\left( X\right) \right\vert
^{2}}{K_{X}\left\vert \Psi \left( X\right) \right\vert ^{2}}}\right\} \delta 
\hat{f}\left( X,\theta -1\right) \\
&&-S_{E}\left( X,X,\theta -1\right) \left( S_{E}^{B}\left( X,\theta
-1\right) +S_{L}^{B}\left( X,\theta -1\right) \right) \frac{\partial f\left(
X\right) }{\partial S^{T}\left( X,\theta -1\right) }\frac{\partial _{\bar{f}%
\left( X\right) }K_{X}\left\vert \Psi \left( X\right) \right\vert ^{2}}{%
K_{X}\left\vert \Psi \left( X\right) \right\vert ^{2}}\delta \bar{f}\left(
X,\theta -1\right) \\
&&+\left\{ -\frac{\left( 1-\left\langle \hat{S}\left( X^{\prime }\right)
\right\rangle \right) \left( \left\langle \hat{f}\left( X^{\prime }\right)
\right\rangle -\bar{r}\right) }{1-\left\langle \hat{S}_{E}\left( X^{\prime
}\right) \right\rangle }\frac{\left\langle \hat{w}\left( X^{\prime
},X\right) \right\rangle \left\langle w\left( X\right) \right\rangle }{4}%
\right. \\
&&+\frac{\frac{w\left( X\right) }{4}\left( 1+\hat{w}\left( X\right) \right)
S_{E}\left( X\right) }{1+\left( \hat{w}\left( X\right) \left( f\left(
X\right) -\frac{\left\langle \hat{f}\left( X^{\prime }\right) \right\rangle
_{\hat{w}_{1}}+\left\langle \hat{r}\left( X^{\prime }\right) \right\rangle _{%
\hat{w}_{2}}}{2}\right) +\frac{w\left( X\right) }{2}\left( f\left( X\right) -%
\bar{r}\left( X\right) \right) \right) }\left( f\left( X\right) -\bar{r}%
\right) \\
&&+S_{E}\left( X,X,\theta -1\right) \frac{\partial f\left( X\right) }{%
\partial S^{T}\left( X,\theta -1\right) } \\
&&\times \left( w_{1}^{B}\left( X\right) \left( \left\langle \bar{w}\left(
X\right) \right\rangle +\left\langle \hat{w}_{1}^{B}\left( X\right)
\right\rangle \right) -\left( S_{E}^{B}\left( X,\theta -1\right)
+S_{L}^{B}\left( X,\theta -1\right) \right) \frac{\partial _{f\left(
X\right) }K_{X}\left\vert \Psi \left( X\right) \right\vert ^{2}}{%
K_{X}\left\vert \Psi \left( X\right) \right\vert ^{2}}\right. \\
&&\left. \left. +\frac{\frac{\hat{w}\left( X\right) }{2}S\left( X\right) }{%
1+\left( \hat{w}\left( X\right) \left( \frac{f\left( X\right) +\bar{r}\left(
X\right) }{2}-\frac{\left\langle \hat{f}\left( X^{\prime }\right)
\right\rangle _{\hat{w}_{1}}+\left\langle \hat{r}\left( X^{\prime }\right)
\right\rangle _{\hat{w}_{2}}}{2}\right) \right) }+\frac{\partial _{f\left(
X\right) }\left( \frac{\hat{K}_{X}\left\vert \hat{\Psi}\left( X\right)
\right\vert ^{2}}{K_{X}\left\vert \Psi \left( X\right) \right\vert ^{2}}%
\right) S\left( X\right) }{\frac{\hat{K}_{X}\left\vert \hat{\Psi}\left(
X\right) \right\vert ^{2}}{K_{X}\left\vert \Psi \left( X\right) \right\vert
^{2}}}\right) \right\} \\
&&\times \frac{\partial f\left( X\right) }{\partial S^{T}\left( X,\theta
-2\right) }\delta S^{T}\left( X,\theta -2\right)
\end{eqnarray*}

\subsubsection*{A12.2.2 Computation of $\protect\delta \bar{f}$}

As for investors, the variation of the return equation is obtained by
estimatin the various average shares connecting the agents of the group and
their variations. We found previously for banks shares in investors:%
\begin{eqnarray*}
&&\left\langle \hat{S}_{E}^{B}\left( X^{\prime },X\right) \right\rangle
_{X^{\prime }} \\
&\simeq &\left\langle \hat{w}_{1}^{B}\left( X^{\prime },X\right)
\right\rangle \left[ 1+\left\langle \bar{w}\left( X\right) \right\rangle
\left( \left\langle \hat{f}\left( X^{\prime }\right) \right\rangle -\frac{%
\left\langle \bar{f}\left( X^{\prime }\right) \right\rangle _{\bar{w}%
_{1}}+\left\langle \bar{r}\left( X^{\prime }\right) \right\rangle _{\bar{w}%
_{2}}}{2}\right) +\left\langle w_{1}^{B}\left( X\right) \right\rangle \left(
\left\langle \hat{f}\left( X^{\prime }\right) \right\rangle -f\left(
X\right) \right) \right]
\end{eqnarray*}%
so that:%
\begin{eqnarray*}
&&\delta \left\langle \hat{S}_{E}^{B}\left( X^{\prime },X,\theta -1\right)
\right\rangle _{X^{\prime }} \\
&\simeq &-\left\langle \hat{w}_{1}^{B}\left( X^{\prime },X,\theta -1\right)
\right\rangle \left\langle w_{1}^{B}\left( X\right) \right\rangle \delta
f\left( X,\theta -1\right) \\
&=&-\left\langle \hat{w}_{1}^{B}\left( X^{\prime },X\right) \right\rangle
\left\langle w_{1}^{B}\left( X\right) \right\rangle \frac{\partial f\left(
X\right) }{\partial S^{T}\left( X,\theta -2\right) }\delta S^{T}\left(
X,\theta -2\right)
\end{eqnarray*}%
Banks shares in other banks are given by:%
\begin{eqnarray*}
&&\left\langle \bar{S}_{E}\left( X^{\prime },X\right) \right\rangle
_{X^{\prime }} \\
&=&\frac{\left\langle \bar{w}\left( X^{\prime },X\right) \right\rangle }{2}
\\
&&\times \left( 1+\bar{w}\left( X\right) \left( \frac{\left\langle \bar{f}%
\left( X^{\prime }\right) \right\rangle _{\bar{w}_{1}}-\left\langle \bar{r}%
\left( X^{\prime }\right) \right\rangle _{\bar{w}_{2}}}{2}\right) +\hat{w}%
_{1}^{B}\left( X\right) \left( \left\langle \bar{f}\left( X^{\prime }\right)
\right\rangle -\left\langle \hat{f}\left( X^{\prime }\right) \right\rangle _{%
\hat{w}_{1}}\right) +w_{1}^{B}\left( X\right) \left( \left\langle \bar{f}%
\left( X^{\prime }\right) \right\rangle -f\left( X\right) \right) \right) \\
&\simeq &\frac{\left\langle \bar{w}\left( X^{\prime },X\right) \right\rangle 
}{2} \\
&&\times \left( 1+\left\langle \bar{w}\left( X\right) \right\rangle \left( 
\frac{\left\langle \bar{f}\left( X^{\prime }\right) \right\rangle _{\bar{w}%
_{1}}-\left\langle \bar{r}\left( X^{\prime }\right) \right\rangle _{\bar{w}%
_{2}}}{2}\right) \right. \\
&&\left. +\left\langle \hat{w}_{1}^{B}\left( X\right) \right\rangle \left(
\left\langle \bar{f}\left( X^{\prime }\right) \right\rangle -\left\langle 
\hat{f}\left( X^{\prime }\right) \right\rangle _{\hat{w}_{1}}\right)
+\left\langle w_{1}^{B}\left( X\right) \right\rangle \left( \left\langle 
\bar{f}\left( X^{\prime }\right) \right\rangle -f\left( X\right) \right)
\right)
\end{eqnarray*}%
\begin{eqnarray*}
\left\langle \bar{S}_{E}\left( X^{\prime },X\right) \right\rangle _{X}
&=&\left\langle \frac{\bar{w}\left( X^{\prime },X\right) }{2}\right\rangle
_{X}\left( 1+\left\{ \left\langle \bar{w}\left( X\right) \right\rangle
\left( \bar{f}\left( X^{\prime }\right) -\frac{\left\langle \bar{f}\left(
X^{\prime }\right) \right\rangle _{\bar{w}_{1}}+\left\langle \bar{r}\left(
X^{\prime }\right) \right\rangle _{\bar{w}_{2}}}{2}\right) \right. \right. \\
&&\left. \left. +\left\langle \hat{w}_{1}^{B}\left( X\right) \right\rangle
\left( \bar{f}\left( X^{\prime }\right) -\left\langle \hat{f}\left(
X^{\prime }\right) \right\rangle _{\hat{w}_{1}}\right) +\left\langle
w_{1}^{B}\left( X\right) \right\rangle \left( \bar{f}\left( X^{\prime
}\right) -\left\langle f\left( X\right) \right\rangle \right) \right\}
\right)
\end{eqnarray*}%
\begin{eqnarray*}
&&\left\langle \bar{S}_{L}\left( X^{\prime },X\right) \right\rangle _{X} \\
&=&\left\langle \frac{\bar{w}\left( X^{\prime },X\right) }{2}\right\rangle
\left( 1+\left\{ \bar{w}\left( X\right) \left( \bar{r}\left( X^{\prime
}\right) -\frac{\left\langle \bar{f}\left( X^{\prime }\right) \right\rangle
_{\bar{w}_{1}}+\left\langle \bar{r}\left( X^{\prime }\right) \right\rangle _{%
\bar{w}_{2}}}{2}\right) \right. \right. \\
&&\left. \left. +\left\langle \hat{w}_{1}^{B}\left( X\right) \right\rangle
\left( \bar{r}\left( X^{\prime }\right) -\left\langle \hat{f}\left(
X^{\prime }\right) \right\rangle _{\hat{w}_{1}^{B}}\right) +\left\langle
w_{1}^{B}\left( X\right) \right\rangle \left( \bar{r}\left( X^{\prime
}\right) -f\left( X\right) \right) \right\} \right)
\end{eqnarray*}%
\begin{eqnarray*}
\left\langle \bar{S}\left( X^{\prime },X\right) \right\rangle _{X}
&=&\left\langle \bar{S}_{E}\left( X^{\prime },X\right) \right\rangle
_{X}+\left\langle \bar{S}_{L}\left( X^{\prime },X\right) \right\rangle _{X}
\\
&=&\left\langle \bar{w}\left( X^{\prime },X\right) \right\rangle \left[
1+\left\{ \bar{w}\left( X\right) \left( \frac{\bar{f}\left( X^{\prime
}\right) +\bar{r}\left( X^{\prime }\right) }{2}-\frac{\left\langle \bar{f}%
\left( X^{\prime }\right) \right\rangle _{\bar{w}_{1}}+\left\langle \bar{r}%
\left( X^{\prime }\right) \right\rangle _{\bar{w}_{2}}}{2}\right) \right.
\right. \\
&&\left. \left. +\hat{w}_{1}^{B}\left( X\right) \left( \frac{\bar{f}\left(
X^{\prime }\right) +\bar{r}\left( X^{\prime }\right) }{2}-\left\langle \hat{f%
}\left( X^{\prime }\right) \right\rangle _{\hat{w}_{1}}\right)
+w_{1}^{B}\left( X\right) \left( \frac{\bar{f}\left( X^{\prime }\right) +%
\bar{r}\left( X^{\prime }\right) }{2}-\left\langle f\left( X\right)
\right\rangle \right) \right\} \right]
\end{eqnarray*}%
so that the variations becomes:%
\begin{eqnarray*}
\delta \left\langle \bar{S}_{E}\left( X^{\prime },X\right) ,\theta
-1\right\rangle _{X^{\prime }} &=&-\frac{\left\langle \bar{w}\left(
X^{\prime },X\right) \right\rangle }{2}w_{1}^{B}\left( X\right) \delta
f\left( X,\theta -1\right) \\
&=&-\frac{\left\langle \bar{w}\left( X^{\prime },X\right) \right\rangle }{2}%
w_{1}^{B}\left( X\right) \frac{\partial f\left( X\right) }{\partial
S^{T}\left( X,\theta -2\right) }\delta S^{T}\left( X,\theta -2\right)
\end{eqnarray*}%
and:%
\begin{equation*}
\delta \left\langle \bar{S}_{E}\left( X^{\prime },X\right) \right\rangle
_{X}=\left\langle \frac{\bar{w}\left( X^{\prime },X\right) }{2}\right\rangle
_{X}\left( \left\langle \bar{w}\left( X\right) \right\rangle +\left\langle 
\hat{w}_{1}^{B}\left( X\right) \right\rangle +\left\langle w_{1}^{B}\left(
X\right) \right\rangle \right) \delta \bar{f}\left( X^{\prime }\right)
\end{equation*}

\begin{eqnarray*}
\delta \left\langle \bar{S}\left( X^{\prime },X,\theta -1\right)
\right\rangle _{X} &=&\left\langle \bar{w}\left( X^{\prime },X\right)
\right\rangle \left\{ \frac{1}{2}\left( \bar{w}\left( X\right) +\hat{w}%
_{1}^{B}\left( X\right) +w_{1}^{B}\left( X\right) \right) -w_{1}^{B}\left(
X\right) \right\} \delta f\left( X,\theta -1\right) \\
&=&\left\langle \bar{w}\left( X^{\prime },X\right) \right\rangle \left\{ 
\frac{1}{2}\left( \bar{w}\left( X\right) +\hat{w}_{1}^{B}\left( X\right)
+w_{1}^{B}\left( X\right) \right) \right\} \delta \bar{f}\left( X^{\prime
},\theta -1\right)
\end{eqnarray*}%
To compute average perturbations, we also consider the half averaged
quantities that measure the average capital flow that comes in a given
sector:%
\begin{eqnarray*}
\bar{S}_{E}\left( X^{\prime }\right) &=&\left\langle \bar{S}_{E}\left(
X^{\prime },X\right) \right\rangle _{X}\frac{\left\langle \bar{K}%
\right\rangle \left\Vert \bar{\Psi}\right\Vert ^{2}}{\left\langle \bar{K}%
_{X^{\prime }}\right\rangle \left\vert \bar{\Psi}\left( X^{\prime }\right)
\right\vert ^{2}} \\
\bar{S}_{L}\left( X\right) &=&\left\langle \bar{S}_{L}\left( X^{\prime
},X\right) \right\rangle _{X}\frac{\left\langle \bar{K}\right\rangle
\left\Vert \bar{\Psi}\right\Vert ^{2}}{\left\langle \bar{K}_{X^{\prime
}}\right\rangle \left\vert \bar{\Psi}\left( X^{\prime }\right) \right\vert
^{2}} \\
\bar{S}\left( X^{\prime }\right) &=&\left\langle \bar{S}\left( X^{\prime
},X\right) \right\rangle _{X}\frac{\left\langle \bar{K}\right\rangle
\left\Vert \bar{\Psi}\right\Vert ^{2}}{\left\langle \bar{K}_{X^{\prime
}}\right\rangle \left\vert \bar{\Psi}\left( X^{\prime }\right) \right\vert
^{2}}
\end{eqnarray*}

\begin{eqnarray*}
\left\langle \bar{S}\left( X^{\prime },X\right) \right\rangle _{X}
&=&\left\langle \bar{S}_{E}\left( X^{\prime },X\right) \right\rangle
_{X}+\left\langle \bar{S}_{L}\left( X^{\prime },X\right) \right\rangle _{X}
\\
&=&\left\langle \bar{w}\left( X^{\prime },X\right) \right\rangle \left[
1+\left\{ \bar{w}\left( X\right) \left( \frac{\bar{f}\left( X^{\prime
}\right) +\bar{r}\left( X^{\prime }\right) }{2}-\frac{\left\langle \bar{f}%
\left( X^{\prime }\right) \right\rangle _{\bar{w}_{1}}+\left\langle \bar{r}%
\left( X^{\prime }\right) \right\rangle _{\bar{w}_{2}}}{2}\right) \right.
\right. \\
&&\left. \left. +\hat{w}_{1}^{B}\left( X\right) \left( \frac{\bar{f}\left(
X^{\prime }\right) +\bar{r}\left( X^{\prime }\right) }{2}-\left\langle \hat{f%
}\left( X^{\prime }\right) \right\rangle _{\hat{w}_{1}}\right)
+w_{1}^{B}\left( X\right) \left( \frac{\bar{f}\left( X^{\prime }\right) +%
\bar{r}\left( X^{\prime }\right) }{2}-\left\langle f\left( X\right)
\right\rangle \right) \right\} \right]
\end{eqnarray*}%
where:%
\begin{equation*}
\frac{\left\langle \bar{K}\right\rangle \left\Vert \bar{\Psi}\right\Vert ^{2}%
}{\bar{K}_{X}\left\vert \bar{\Psi}\left( X\right) \right\vert ^{2}}=\frac{%
\bar{g}^{2}\left( X\right) \left( \left\Vert \bar{\Psi}_{0}\right\Vert
\right) ^{4}}{\left\langle \bar{g}\right\rangle ^{2}\left( \left\vert \bar{%
\Psi}_{0}\left( X\right) \right\vert \right) ^{4}}=\frac{\left( \bar{f}%
\left( X\right) \left( 1-\left\langle \bar{S}\right\rangle \right)
+\left\langle \bar{S}\left( X^{\prime },X\right) \right\rangle _{X^{\prime
}}\left\langle \bar{f}\right\rangle \right) ^{2}}{\left\langle \bar{f}%
\right\rangle ^{2}}
\end{equation*}%
leading to the variations:%
\begin{equation*}
\delta \left\langle \bar{S}\left( X^{\prime },X\right) \right\rangle
_{X}=H\left( X^{\prime }\right) \left( \delta \frac{\bar{f}\left( X^{\prime
}\right) +\bar{r}\left( X^{\prime }\right) }{2}\right) \left\langle \bar{S}%
\left( X^{\prime },X\right) \right\rangle _{X}
\end{equation*}%
where:%
\begin{equation*}
H\left( X^{\prime }\right) =\frac{\left( \left\langle \bar{w}\left( X\right)
\right\rangle +\left\langle \hat{w}_{1}^{B}\left( X\right) \right\rangle
+\left\langle w_{1}^{B}\left( X\right) \right\rangle \right) }{1+\bar{w}%
\left( X\right) \Delta _{1}\left( \frac{\bar{f}\left( X^{\prime }\right) +%
\bar{r}\left( X^{\prime }\right) }{2}\right) +\hat{w}_{1}^{B}\left( X\right)
\Delta _{2}\left( \frac{\bar{f}\left( X^{\prime }\right) +\bar{r}\left(
X^{\prime }\right) }{2}\right) +w_{1}^{B}\left( X\right) \Delta _{3}\left( 
\frac{\bar{f}\left( X^{\prime }\right) +\bar{r}\left( X^{\prime }\right) }{2}%
\right) }
\end{equation*}%
\begin{eqnarray*}
\Delta _{1}\left( \frac{\bar{f}\left( X^{\prime }\right) +\bar{r}\left(
X^{\prime }\right) }{2}\right) &=&\left( \frac{\bar{f}\left( X^{\prime
}\right) +\bar{r}\left( X^{\prime }\right) }{2}-\frac{\left\langle \bar{f}%
\left( X^{\prime }\right) \right\rangle _{\bar{w}_{1}}+\left\langle \bar{r}%
\left( X^{\prime }\right) \right\rangle _{\bar{w}_{2}}}{2}\right) \\
\Delta _{2}\left( \frac{\bar{f}\left( X^{\prime }\right) +\bar{r}\left(
X^{\prime }\right) }{2}\right) &=&\left( \frac{\bar{f}\left( X^{\prime
}\right) +\bar{r}\left( X^{\prime }\right) }{2}-\left\langle \hat{f}\left(
X^{\prime }\right) \right\rangle _{\hat{w}_{1}}\right) \\
\Delta _{3}\left( \frac{\bar{f}\left( X^{\prime }\right) +\bar{r}\left(
X^{\prime }\right) }{2}\right) &=&\left( \frac{\bar{f}\left( X^{\prime
}\right) +\bar{r}\left( X^{\prime }\right) }{2}-\left\langle f\left(
X\right) \right\rangle \right)
\end{eqnarray*}%
\begin{equation*}
\delta \bar{S}\left( X^{\prime }\right) =\frac{1}{2}H\left( X^{\prime
}\right) \bar{S}\left( X^{\prime }\right) \delta \bar{f}\left( X^{\prime
}\right) -\frac{\partial _{\bar{f}\left( X\right) }\left\langle \bar{K}%
_{X^{\prime }}\right\rangle \left\vert \bar{\Psi}\left( X^{\prime }\right)
\right\vert ^{2}}{\left\langle \bar{K}_{X^{\prime }}\right\rangle \left\vert 
\bar{\Psi}\left( X^{\prime }\right) \right\vert ^{2}}\bar{S}\left( X^{\prime
}\right) \delta \bar{f}\left( X^{\prime }\right)
\end{equation*}%
\begin{eqnarray*}
&&\delta \bar{S}_{E}\left( X^{\prime }\right) =H\left( X\right) \bar{S}%
_{E}\left( X^{\prime }\right) \delta \bar{f}\left( X^{\prime }\right) \\
&&-\frac{\partial _{\bar{f}\left( X\right) }\left\langle \bar{K}_{X^{\prime
}}\right\rangle \left\vert \bar{\Psi}\left( X^{\prime }\right) \right\vert
^{2}}{\left\langle \bar{K}_{X^{\prime }}\right\rangle \left\vert \bar{\Psi}%
\left( X^{\prime }\right) \right\vert ^{2}}\bar{S}_{E}\left( X^{\prime
}\right) \delta \bar{f}\left( X^{\prime }\right)
\end{eqnarray*}%
Ultimately, expanding the following expression to the lowest order:%
\begin{equation*}
\frac{1-\left( \bar{S}\left( X,\theta \right) +\delta \bar{S}\left( X,\theta
\right) \right) }{1-\left( \bar{S}_{E}\left( X^{\prime },\theta \right)
+\delta \bar{S}_{E}\left( X,\theta \right) \right) }=-\frac{\delta \bar{S}%
\left( X,\theta \right) }{1-\left( \bar{S}_{E}\left( X^{\prime },\theta
\right) \right) }+\frac{\left( 1-\left( \bar{S}\left( X,\theta \right)
\right) \right) }{\left( 1-\left( \bar{S}_{E}\left( X^{\prime },\theta
\right) \right) \right) ^{2}}\delta \bar{S}_{E}\left( X,\theta \right)
\end{equation*}%
leads to the average variation equation of return equation:%
\begin{eqnarray}
0 &=&\frac{1-\bar{S}\left( X\right) }{1-\bar{S}_{E}\left( X\right) }\delta 
\bar{f}\left( X\right) +\left( -\frac{\delta \bar{S}\left( X,\theta \right) 
}{1-\left( \bar{S}_{E}\left( X^{\prime },\theta \right) \right) }+\frac{%
\left( 1-\left( \bar{S}\left( X,\theta \right) \right) \right) }{\left(
1-\left( \bar{S}_{E}\left( X^{\prime },\theta \right) \right) \right) ^{2}}%
\delta \bar{S}_{E}\left( X,\theta \right) \right) \left( \bar{f}\left(
X\right) -\bar{r}\right) \\
&&-\delta \left\langle \bar{S}_{E}\left( X^{\prime },X\right) \right\rangle
_{X^{\prime }}\frac{1-\left\langle \bar{S}\left( X^{\prime }\right)
\right\rangle }{1-\left\langle \bar{S}_{E}\left( X^{\prime }\right)
\right\rangle }\left( \left\langle \bar{f}\left( X^{\prime }\right)
\right\rangle -\bar{r}\right)  \notag \\
&&-\delta \left\langle \hat{S}_{E}^{B}\left( X^{\prime },X\right)
\right\rangle _{X^{\prime }}\frac{1-\left\langle \hat{S}\left( X^{\prime
}\right) \right\rangle +\left\langle \hat{S}_{E}^{B}\left( X^{\prime
}\right) \right\rangle +\left\langle \hat{S}_{L}^{B}\left( X^{\prime
}\right) \right\rangle }{1-\left\langle \hat{S}_{E}\left( X^{\prime }\right)
\right\rangle +\left\langle \hat{S}_{E}^{B}\left( X^{\prime }\right)
\right\rangle }\left( \left\langle \hat{f}\left( X^{\prime }\right)
\right\rangle -\bar{r}\right)  \notag \\
&&-\delta S_{E}^{B}\left( X,X\right) \left( f\left( X\right) -\bar{r}\right)
-S_{E}^{B}\left( X,X\right) \frac{\partial f\left( X\right) }{\partial
S^{T}\left( X,\theta -1\right) }\delta S^{T}\left( X,,\theta -1\right) 
\notag
\end{eqnarray}%
that is expanded in the followin form:%
\begin{eqnarray}
0 &=&\frac{1-\bar{S}\left( X\right) }{1-\bar{S}_{E}\left( X\right) }\delta 
\bar{f}\left( X,\theta -1\right) +\left( -\frac{\delta \bar{S}\left(
X,\theta -1\right) }{1-\left( \bar{S}_{E}\left( X^{\prime },\theta -1\right)
\right) }+\frac{\left( 1-\left( \bar{S}\left( X,\theta -1\right) \right)
\right) }{\left( 1-\left( \bar{S}_{E}\left( X^{\prime },\theta -1\right)
\right) \right) ^{2}}\delta \bar{S}_{E}\left( X,\theta -1\right) \right)
\left( \bar{f}\left( X\right) -\bar{r}\right) \\
&&+\frac{\left\langle \bar{w}\left( X^{\prime },X\right) \right\rangle }{2}%
w_{1}^{B}\left( X\right) \frac{1-\left\langle \bar{S}\left( X^{\prime
}\right) \right\rangle }{1-\left\langle \bar{S}_{E}\left( X^{\prime }\right)
\right\rangle }\left( \left\langle \bar{f}\left( X^{\prime }\right)
\right\rangle -\bar{r}\right) \frac{\partial f\left( X\right) }{\partial
S^{T}\left( X,\theta -2\right) }\delta S^{T}\left( X,\theta -2\right)  \notag
\\
&&+\left\langle \hat{w}_{1}^{B}\left( X^{\prime },X\right) \right\rangle
\left\langle w_{1}^{B}\left( X\right) \right\rangle \frac{1-\left\langle 
\hat{S}\left( X^{\prime }\right) \right\rangle +\left\langle \hat{S}%
_{E}^{B}\left( X^{\prime }\right) \right\rangle +\left\langle \hat{S}%
_{L}^{B}\left( X^{\prime }\right) \right\rangle }{1-\left\langle \hat{S}%
_{E}\left( X^{\prime }\right) \right\rangle +\left\langle \hat{S}%
_{E}^{B}\left( X^{\prime }\right) \right\rangle }\left( \left\langle \hat{f}%
\left( X^{\prime }\right) \right\rangle -\bar{r}\right) \frac{\partial
f\left( X\right) }{\partial S^{T}\left( X,\theta -2\right) }\delta
S^{T}\left( X,\theta -2\right)  \notag \\
&&-w_{1}^{B}\left( X\right) \left( \left\langle \bar{w}\left( X\right)
\right\rangle +\left\langle \hat{w}_{1}^{B}\left( X\right) \right\rangle
\right) \left( f\left( X\right) -\bar{r}\right) \frac{\partial f\left(
X\right) }{\partial S^{T}\left( X,\theta -2\right) }\delta S^{T}\left(
X,\theta -2\right)  \notag \\
&&-S_{E}^{B}\left( X,X\right) \frac{\partial f\left( X\right) }{\partial
S^{T}\left( X,\theta -1\right) }\delta S^{T}\left( X,\theta -1\right)  \notag
\end{eqnarray}%
Using (\ref{DSd}) to write $\delta S^{T}\left( X,\theta -1\right) $ yields
then:%
\begin{eqnarray*}
&&\frac{1-\bar{S}\left( X\right) }{1-\bar{S}_{E}\left( X\right) }\delta \bar{%
f}\left( X,\theta -1\right) \\
&=&\left\{ \left( \frac{1}{2}H\left( X\right) \bar{S}\left( X\right) -\frac{%
\bar{S}\left( X^{\prime },\theta -1\right) \partial _{\bar{f}\left( X\right)
}\left\langle \bar{K}_{X}\right\rangle \left\vert \bar{\Psi}\left( X\right)
\right\vert ^{2}}{\left\langle \bar{K}_{X^{\prime }}\right\rangle \left\vert 
\bar{\Psi}\left( X^{\prime }\right) \right\vert ^{2}}\right) \frac{\left( 
\bar{f}\left( X\right) -\bar{r}\right) }{1-\left( \bar{S}_{E}\left(
X^{\prime },\theta -1\right) \right) }\right. \\
&&-\left( H\left( X\right) \bar{S}_{E}\left( X\right) -\frac{\bar{S}%
_{E}\left( X^{\prime },\theta -1\right) \partial _{\bar{f}\left( X\right)
}\left\langle \bar{K}_{X}\right\rangle \left\vert \bar{\Psi}\left( X\right)
\right\vert ^{2}}{\left\langle \bar{K}_{X^{\prime }}\right\rangle \left\vert 
\bar{\Psi}\left( X^{\prime }\right) \right\vert ^{2}}\right) \frac{\left(
1-\left( \bar{S}\left( X,\theta -1\right) \right) \right) \left( \bar{f}%
\left( X\right) -\bar{r}\right) }{\left( 1-\left( \bar{S}_{E}\left(
X^{\prime },\theta -1\right) \right) \right) ^{2}} \\
&&\left. +\left( S_{E}^{B}\left( X,\theta -1\right) +S_{L}^{B}\left(
X,\theta -1\right) \right) \frac{\partial _{\bar{f}\left( X\right) }\bar{K}%
_{X}\left\vert \bar{\Psi}\left( X\right) \right\vert ^{2}}{\bar{K}%
_{X}\left\vert \bar{\Psi}\left( X\right) \right\vert ^{2}}S_{E}^{B}\left(
X,X\right) \frac{\partial f\left( X\right) }{\partial S^{T}\left( X,\theta
-1\right) }\right\} \delta \bar{f}\left( X,\theta -1\right) \\
&&+S_{E}^{B}\left( X,X\right) \frac{\partial f\left( X\right) }{\partial
S^{T}\left( X,\theta -1\right) }\frac{\partial _{\hat{f}\left( X\right)
}\left( \frac{\hat{K}_{X}\left\vert \hat{\Psi}\left( X\right) \right\vert
^{2}}{K_{X}\left\vert \Psi \left( X\right) \right\vert ^{2}}\right) S\left(
X\right) }{\frac{\hat{K}_{X}\left\vert \hat{\Psi}\left( X\right) \right\vert
^{2}}{K_{X}\left\vert \Psi \left( X\right) \right\vert ^{2}}}\delta \hat{f}%
\left( X,\theta -1\right) \\
&&+\left\{ w_{1}^{B}\left( X\right) \left( \left\langle \bar{w}\left(
X\right) \right\rangle +\left\langle \hat{w}_{1}^{B}\left( X\right)
\right\rangle \right) \left( f\left( X\right) -\bar{r}\right) \frac{\partial
f\left( X\right) }{\partial S^{T}\left( X,\theta -2\right) }-\frac{%
\left\langle \bar{w}\left( X^{\prime },X\right) \right\rangle }{2}%
w_{1}^{B}\left( X\right) \frac{1-\left\langle \bar{S}\left( X^{\prime
}\right) \right\rangle }{1-\left\langle \bar{S}_{E}\left( X^{\prime }\right)
\right\rangle }\left( \left\langle \bar{f}\left( X^{\prime }\right)
\right\rangle -\bar{r}\right) \right. \\
&&-\left\langle \hat{w}_{1}^{B}\left( X^{\prime },X\right) \right\rangle
\left\langle w_{1}^{B}\left( X\right) \right\rangle \frac{1-\left\langle 
\hat{S}\left( X^{\prime }\right) \right\rangle +\left\langle \hat{S}%
_{E}^{B}\left( X^{\prime }\right) \right\rangle +\left\langle \hat{S}%
_{L}^{B}\left( X^{\prime }\right) \right\rangle }{1-\left\langle \hat{S}%
_{E}\left( X^{\prime }\right) \right\rangle +\left\langle \hat{S}%
_{E}^{B}\left( X^{\prime }\right) \right\rangle }\left( \left\langle \hat{f}%
\left( X^{\prime }\right) \right\rangle -\bar{r}\right) \\
&&+S_{E}^{B}\left( X,X\right) \frac{\partial f\left( X\right) }{\partial
S^{T}\left( X,\theta -1\right) } \\
&&\times \left\{ w_{1}^{B}\left( X\right) \left( \left\langle \bar{w}\left(
X\right) \right\rangle +\left\langle \hat{w}_{1}^{B}\left( X\right)
\right\rangle \right) -\left( S_{E}^{B}\left( X,\theta -1\right)
+S_{L}^{B}\left( X,\theta -1\right) \right) \frac{\partial _{\bar{f}\left(
X\right) }\bar{K}_{X}\left\vert \bar{\Psi}\left( X\right) \right\vert ^{2}}{%
\bar{K}_{X}\left\vert \bar{\Psi}\left( X\right) \right\vert ^{2}}\right. \\
&&\left. \left. +\left( \frac{\frac{\hat{w}\left( X\right) }{2}S\left(
X\right) }{1+\left( \hat{w}\left( X\right) \left( \frac{f\left( X\right) +%
\bar{r}\left( X\right) }{2}-\frac{\left\langle \hat{f}\left( X^{\prime
}\right) \right\rangle _{\hat{w}_{1}}+\left\langle \hat{r}\left( X^{\prime
}\right) \right\rangle _{\hat{w}_{2}}}{2}\right) \right) }+\frac{\partial
_{f\left( X\right) }\left( \frac{\hat{K}_{X}\left\vert \hat{\Psi}\left(
X\right) \right\vert ^{2}}{K_{X}\left\vert \Psi \left( X\right) \right\vert
^{2}}\right) S\left( X\right) }{\frac{\hat{K}_{X}\left\vert \hat{\Psi}\left(
X\right) \right\vert ^{2}}{K_{X}\left\vert \Psi \left( X\right) \right\vert
^{2}}}\right) \right\} \right\} \\
&&\frac{\partial f\left( X\right) }{\partial S^{T}\left( X,\theta -2\right) }%
\delta S^{T}\left( X,\theta -2\right)
\end{eqnarray*}

\subsection*{A12.3 Matricial form of the system of variations}

The three variations can be compactly written through a system of
coefficients:%
\begin{equation*}
\frac{1-\bar{S}\left( X\right) }{1-\bar{S}_{E}\left( X\right) }\delta \bar{f}%
\left( X,\theta \right) =a\delta \bar{f}\left( X,\theta -1\right) +b\delta 
\hat{f}\left( X,\theta -1\right) +c\delta S^{T}\left( X,\theta -2\right)
\end{equation*}%
\begin{eqnarray*}
a &=&\left\{ \left( \frac{1}{2}H\left( X\right) \bar{S}\left( X\right) -%
\frac{\bar{S}\left( X^{\prime },\theta -1\right) \partial _{\bar{f}\left(
X\right) }\left\langle \bar{K}_{X}\right\rangle \left\vert \bar{\Psi}\left(
X\right) \right\vert ^{2}}{\left\langle \bar{K}_{X^{\prime }}\right\rangle
\left\vert \bar{\Psi}\left( X^{\prime }\right) \right\vert ^{2}}\right)
\right. \frac{\left( \bar{f}\left( X\right) -\bar{r}\right) }{1-\left( \bar{S%
}_{E}\left( X^{\prime },\theta -1\right) \right) } \\
&&-\left( H\left( X\right) \bar{S}_{E}\left( X\right) -\frac{\bar{S}%
_{E}\left( X^{\prime },\theta -1\right) \partial _{\bar{f}\left( X\right)
}\left\langle \bar{K}_{X}\right\rangle \left\vert \bar{\Psi}\left( X\right)
\right\vert ^{2}}{\left\langle \bar{K}_{X^{\prime }}\right\rangle \left\vert 
\bar{\Psi}\left( X^{\prime }\right) \right\vert ^{2}}\right) \frac{\left(
1-\left( \bar{S}\left( X,\theta -1\right) \right) \right) \left( \bar{f}%
\left( X\right) -\bar{r}\right) }{\left( 1-\left( \bar{S}_{E}\left(
X^{\prime },\theta -1\right) \right) \right) ^{2}} \\
&&\left. +\left( S_{E}^{B}\left( X,\theta -1\right) +S_{L}^{B}\left(
X,\theta -1\right) \right) \frac{\partial _{\bar{f}\left( X\right) }\bar{K}%
_{X}\left\vert \bar{\Psi}\left( X\right) \right\vert ^{2}}{\bar{K}%
_{X}\left\vert \bar{\Psi}\left( X\right) \right\vert ^{2}}S_{E}^{B}\left(
X,X\right) \frac{\partial f\left( X\right) }{\partial S^{T}\left( X,\theta
-1\right) }\right\}
\end{eqnarray*}%
\begin{equation*}
b=S_{E}^{B}\left( X,X\right) \frac{\partial f\left( X\right) }{\partial
S^{T}\left( X,\theta -1\right) }\frac{\partial _{\hat{f}\left( X\right)
}\left( \frac{\hat{K}_{X}\left\vert \hat{\Psi}\left( X\right) \right\vert
^{2}}{K_{X}\left\vert \Psi \left( X\right) \right\vert ^{2}}\right) S\left(
X\right) }{\frac{\hat{K}_{X}\left\vert \hat{\Psi}\left( X\right) \right\vert
^{2}}{K_{X}\left\vert \Psi \left( X\right) \right\vert ^{2}}}
\end{equation*}%
\begin{eqnarray*}
c &=&\left\{ w_{1}^{B}\left( X\right) \left( \left\langle \bar{w}\left(
X\right) \right\rangle +\left\langle \hat{w}_{1}^{B}\left( X\right)
\right\rangle \right) \left( f\left( X\right) -\bar{r}\right) \frac{\partial
f\left( X\right) }{\partial S^{T}\left( X,\theta -2\right) }-\frac{%
\left\langle \bar{w}\left( X^{\prime },X\right) \right\rangle }{2}%
w_{1}^{B}\left( X\right) \frac{1-\left\langle \bar{S}\left( X^{\prime
}\right) \right\rangle }{1-\left\langle \bar{S}_{E}\left( X^{\prime }\right)
\right\rangle }\left( \left\langle \bar{f}\left( X^{\prime }\right)
\right\rangle -\bar{r}\right) \right. \\
&&-\left\langle \hat{w}_{1}^{B}\left( X^{\prime },X\right) \right\rangle
\left\langle w_{1}^{B}\left( X\right) \right\rangle \frac{1-\left\langle 
\hat{S}\left( X^{\prime }\right) \right\rangle +\left\langle \hat{S}%
_{E}^{B}\left( X^{\prime }\right) \right\rangle +\left\langle \hat{S}%
_{L}^{B}\left( X^{\prime }\right) \right\rangle }{1-\left\langle \hat{S}%
_{E}\left( X^{\prime }\right) \right\rangle +\left\langle \hat{S}%
_{E}^{B}\left( X^{\prime }\right) \right\rangle }\left( \left\langle \hat{f}%
\left( X^{\prime }\right) \right\rangle -\bar{r}\right) \\
&&+S_{E}^{B}\left( X,X\right) \frac{\partial f\left( X\right) }{\partial
S^{T}\left( X,\theta -1\right) }\left\{ w_{1}^{B}\left( X\right) \left(
\left\langle \bar{w}\left( X\right) \right\rangle +\left\langle \hat{w}%
_{1}^{B}\left( X\right) \right\rangle \right) \right. \\
&&-\left( S_{E}^{B}\left( X,\theta -1\right) +S_{L}^{B}\left( X,\theta
-1\right) \right) \frac{\partial _{f\left( X\right) }K_{X}\left\vert \Psi
\left( X\right) \right\vert ^{2}}{K_{X}\left\vert \Psi \left( X\right)
\right\vert ^{2}} \\
&&\left. \left. +\left( \frac{\frac{\hat{w}\left( X\right) }{2}S\left(
X\right) }{1+\left( \hat{w}\left( X\right) \left( \frac{f\left( X\right) +%
\bar{r}\left( X\right) }{2}-\frac{\left\langle \hat{f}\left( X^{\prime
}\right) \right\rangle _{\hat{w}_{1}}+\left\langle \hat{r}\left( X^{\prime
}\right) \right\rangle _{\hat{w}_{2}}}{2}\right) \right) }+\frac{\partial
_{f\left( X\right) }\left( \frac{\hat{K}_{X}\left\vert \hat{\Psi}\left(
X\right) \right\vert ^{2}}{K_{X}\left\vert \Psi \left( X\right) \right\vert
^{2}}\right) S\left( X\right) }{\frac{\hat{K}_{X}\left\vert \hat{\Psi}\left(
X\right) \right\vert ^{2}}{K_{X}\left\vert \Psi \left( X\right) \right\vert
^{2}}}\right) \right\} \right\} \frac{\partial f\left( X\right) }{\partial
S^{T}\left( X,\theta -2\right) }
\end{eqnarray*}%
\begin{equation*}
\frac{1-\left( \hat{S}\left( X\right) \right) }{1-\left( \hat{S}_{E}\left(
X\right) \right) }\delta \hat{f}\left( X,\theta \right) =d\delta \bar{f}%
\left( X,\theta -1\right) +e\delta \hat{f}\left( X,\theta -1\right) +f\delta
S^{T}\left( X,\theta -2\right)
\end{equation*}

\begin{equation*}
d=S_{E}\left( X,X,\theta -1\right) \left( S_{E}^{B}\left( X,\theta -1\right)
+S_{L}^{B}\left( X,\theta -1\right) \right) \frac{\partial f\left( X\right) 
}{\partial S^{T}\left( X,\theta -1\right) }\frac{\partial _{\bar{f}\left(
X\right) }\bar{K}_{X}\left\vert \bar{\Psi}\left( X\right) \right\vert ^{2}}{%
\bar{K}_{X}\left\vert \bar{\Psi}\left( X\right) \right\vert ^{2}}
\end{equation*}%
\begin{eqnarray*}
e &=&\left\{ \left( \frac{\hat{S}\left( X\right) }{2\left( 1-\left( \hat{S}%
_{E}\left( X^{\prime }\right) \right) \right) }\left( \frac{1}{1+\frac{%
\Delta \hat{f}\left( X^{\prime }\right) +\Delta \hat{r}\left( X^{\prime
}\right) }{2}}-\frac{\frac{\partial \hat{K}_{X^{\prime }}\left\vert \hat{\Psi%
}\left( X^{\prime }\right) \right\vert ^{2}}{\partial \hat{f}\left( X,\theta
-1\right) }}{\hat{K}_{X^{\prime }}\left\vert \hat{\Psi}\left( X^{\prime
}\right) \right\vert ^{2}}\right) \right. \right. \\
&&\left. -\frac{\left( 1-\left( \hat{S}\left( X\right) \right) \right) \hat{S%
}_{E}\left( X\right) }{\left( 1-\left( \hat{S}_{E}\left( X^{\prime }\right)
\right) \right) ^{2}}\left( \frac{1}{1+\Delta \hat{f}\left( X\right) }-\frac{%
\frac{\partial \hat{K}_{X^{\prime }}\left\vert \hat{\Psi}\left( X^{\prime
}\right) \right\vert ^{2}}{\partial \hat{f}\left( X,\theta -1\right) }}{\hat{%
K}_{X^{\prime }}\left\vert \hat{\Psi}\left( X^{\prime }\right) \right\vert
^{2}}\right) \left( \hat{f}\left( X\right) -\bar{r}\right) \right) \\
&&\left. +S_{E}\left( X,X,\theta -1\right) \frac{\partial f\left( X\right) }{%
\partial S^{T}\left( X,\theta -1\right) }\frac{\partial _{\hat{f}\left(
X\right) }\left( \frac{\hat{K}_{X}\left\vert \hat{\Psi}\left( X\right)
\right\vert ^{2}}{K_{X}\left\vert \Psi \left( X\right) \right\vert ^{2}}%
\right) }{\frac{\hat{K}_{X}\left\vert \hat{\Psi}\left( X\right) \right\vert
^{2}}{K_{X}\left\vert \Psi \left( X\right) \right\vert ^{2}}}\right\}
\end{eqnarray*}%
\bigskip 
\begin{eqnarray*}
&&f=\left\{ -\frac{\left( 1-\left\langle \hat{S}\left( X^{\prime }\right)
\right\rangle \right) \left( \left\langle \hat{f}\left( X^{\prime }\right)
\right\rangle -\bar{r}\right) }{1-\left\langle \hat{S}_{E}\left( X^{\prime
}\right) \right\rangle }\frac{\left\langle \hat{w}\left( X^{\prime
},X\right) \right\rangle \left\langle w\left( X\right) \right\rangle }{4}%
\right. \\
&&+\frac{\frac{w\left( X\right) }{4}\left( 1+\hat{w}\left( X\right) \right)
S_{E}\left( X\right) }{1+\left( \hat{w}\left( X\right) \left( f\left(
X\right) -\frac{\left\langle \hat{f}\left( X^{\prime }\right) \right\rangle
_{\hat{w}_{1}}+\left\langle \hat{r}\left( X^{\prime }\right) \right\rangle _{%
\hat{w}_{2}}}{2}\right) +\frac{w\left( X\right) }{2}\left( f\left( X\right) -%
\bar{r}\left( X\right) \right) \right) }\left( f\left( X\right) -\bar{r}%
\right) \\
&&+S_{E}\left( X,X,\theta -1\right) \frac{\partial f\left( X\right) }{%
\partial S^{T}\left( X,\theta -1\right) } \\
&&\times \left( w_{1}^{B}\left( X\right) \left( \left\langle \bar{w}\left(
X\right) \right\rangle +\left\langle \hat{w}_{1}^{B}\left( X\right)
\right\rangle \right) -\left( S_{E}^{B}\left( X,\theta -1\right)
+S_{L}^{B}\left( X,\theta -1\right) \right) \frac{\partial _{f\left(
X\right) }K_{X}\left\vert \Psi \left( X\right) \right\vert ^{2}}{%
K_{X}\left\vert \Psi \left( X\right) \right\vert ^{2}}\right. \\
&&\left. \left. +\frac{\frac{\hat{w}\left( X\right) }{2}S\left( X\right) }{%
1+\left( \hat{w}\left( X\right) \left( \frac{f\left( X\right) +\bar{r}\left(
X\right) }{2}-\frac{\left\langle \hat{f}\left( X^{\prime }\right)
\right\rangle _{\hat{w}_{1}}+\left\langle \hat{r}\left( X^{\prime }\right)
\right\rangle _{\hat{w}_{2}}}{2}\right) \right) }+\frac{\partial _{f\left(
X\right) }\left( \frac{\hat{K}_{X}\left\vert \hat{\Psi}\left( X\right)
\right\vert ^{2}}{K_{X}\left\vert \Psi \left( X\right) \right\vert ^{2}}%
\right) S\left( X\right) }{\frac{\hat{K}_{X}\left\vert \hat{\Psi}\left(
X\right) \right\vert ^{2}}{K_{X}\left\vert \Psi \left( X\right) \right\vert
^{2}}}\right) \right\} \\
&&\times \frac{\partial f\left( X\right) }{\partial S^{T}\left( X,\theta
-2\right) }
\end{eqnarray*}

\begin{equation*}
\delta S^{T}\left( X,,\theta -1\right) =g\delta \bar{f}\left( X,\theta
-1\right) +h\delta \hat{f}\left( X,\theta -1\right) +i\delta S^{T}\left(
X,\theta -2\right)
\end{equation*}%
\begin{eqnarray*}
g &=&\left( S_{E}^{B}\left( X,\theta -1\right) +S_{L}^{B}\left( X,\theta
-1\right) \right) \frac{\partial _{\bar{f}\left( X\right) }\bar{K}%
_{X}\left\vert \bar{\Psi}\left( X\right) \right\vert ^{2}}{\bar{K}%
_{X}\left\vert \bar{\Psi}\left( X\right) \right\vert ^{2}} \\
h &=&\frac{\partial _{\hat{f}\left( X\right) }\left( \frac{\hat{K}%
_{X}\left\vert \hat{\Psi}\left( X\right) \right\vert ^{2}}{K_{X}\left\vert
\Psi \left( X\right) \right\vert ^{2}}\right) S\left( X\right) }{\frac{\hat{K%
}_{X}\left\vert \hat{\Psi}\left( X\right) \right\vert ^{2}}{K_{X}\left\vert
\Psi \left( X\right) \right\vert ^{2}}} \\
i &=&\left\{ w_{1}^{B}\left( X\right) \left( \left\langle \bar{w}\left(
X\right) \right\rangle +\left\langle \hat{w}_{1}^{B}\left( X\right)
\right\rangle \right) -\left( S_{E}^{B}\left( X,\theta -1\right)
+S_{L}^{B}\left( X,\theta -1\right) \right) \frac{\partial _{f\left(
X\right) }K_{X}\left\vert \Psi \left( X\right) \right\vert ^{2}}{%
K_{X}\left\vert \Psi \left( X\right) \right\vert ^{2}}\right. \\
&&\left. +\left( \frac{\frac{\hat{w}\left( X\right) }{2}S\left( X\right) }{%
1+\left( \hat{w}\left( X\right) \left( \frac{f\left( X\right) +\bar{r}\left(
X\right) }{2}-\frac{\left\langle \hat{f}\left( X^{\prime }\right)
\right\rangle _{\hat{w}_{1}}+\left\langle \hat{r}\left( X^{\prime }\right)
\right\rangle _{\hat{w}_{2}}}{2}\right) \right) }+\frac{\partial _{f\left(
X\right) }\left( \frac{\hat{K}_{X}\left\vert \hat{\Psi}\left( X\right)
\right\vert ^{2}}{K_{X}\left\vert \Psi \left( X\right) \right\vert ^{2}}%
\right) S\left( X\right) }{\frac{\hat{K}_{X}\left\vert \hat{\Psi}\left(
X\right) \right\vert ^{2}}{K_{X}\left\vert \Psi \left( X\right) \right\vert
^{2}}}\right) \right\} \frac{\partial f\left( X\right) }{\partial
S^{T}\left( X,\theta -2\right) }
\end{eqnarray*}

The effect of fluctuations in investors stakes as been studied in Gosselin
and Lotz 25 a and we focus on banks only. To evaluate the effect of
introducing banks, we consider:%
\begin{eqnarray*}
S_{L}^{B}\left( X,\theta -1\right) &>&>S_{E}^{B}\left( X,\theta -1\right) \\
S_{L}^{B}\left( X,\theta -1\right) &>&>S\left( X\right) \\
&&\hat{S}\left( X\right)
\end{eqnarray*}%
\begin{eqnarray*}
a &>&>b \\
d &>&>e \\
h &>&>
\end{eqnarray*}%
so that in first approximation we consider the three dimensional system:%
\begin{equation*}
\left( 
\begin{array}{c}
\delta \bar{f}\left( X,\theta \right) \\ 
\delta \hat{f}\left( X,\theta \right) \\ 
\delta S^{T}\left( X,,\theta -1\right)%
\end{array}%
\right) =\left[ 
\begin{array}{ccc}
a & 0 & c \\ 
d & -1 & f \\ 
g & 0 & i%
\end{array}%
\right] \left( 
\begin{array}{c}
\delta \bar{f}\left( X,\theta -1\right) \\ 
\delta \hat{f}\left( X,\theta -1\right) \\ 
\delta S^{T}\left( X,,\theta -2\right)%
\end{array}%
\right)
\end{equation*}

\subsection*{A12.4 Eigenvalues and stability}

The eigenvalues of this system are:%
\begin{equation*}
\left( -1,\frac{1}{2}a+\frac{1}{2}i-\frac{1}{2}\sqrt{\left( a-i\right)
^{2}+4cg},\frac{1}{2}a+\frac{1}{2}i+\frac{1}{2}\sqrt{\left( a-i\right)
^{2}+4cg}\right)
\end{equation*}%
and eigenvectors:%
\begin{eqnarray*}
&&\left( 
\begin{array}{c}
0 \\ 
1 \\ 
0%
\end{array}%
\right) ,\left( 
\begin{array}{c}
-\frac{\left( 2-\sqrt{\left( a-D\right) ^{2}+4cg}+D+a\right) c}{2\left(
f+af-cd\right) } \\ 
\frac{f\left( \sqrt{a^{2}-2aD+D^{2}+4cg}-D+a\right) -2cd}{2\left(
f+af-cd\right) } \\ 
\frac{1}{2f+2af-2cd}\left( a-D+aD-D^{2}+D\sqrt{a^{2}-2aD+D^{2}+4cg}+\sqrt{%
a^{2}-2aD+D^{2}+4cg}-2cg\right)%
\end{array}%
\right) , \\
&&\left( 
\begin{array}{c}
-\frac{1}{2f+2af-2cd}\left( 2c+c\sqrt{a^{2}-2aD+D^{2}+4cg}+cD+ac\right) \\ 
-\frac{1}{2f+2af-2cd}\left( f\sqrt{a^{2}-2aD+D^{2}+4cg}+fD-af+2cd\right) \\ 
-\frac{1}{2f+2af-2cd}\left( D-a-aD+D^{2}+D\sqrt{a^{2}-2aD+D^{2}+4cg}+\sqrt{%
a^{2}-2aD+D^{2}+4cg}+2cg\right)%
\end{array}%
\right)
\end{eqnarray*}%
where in first approximation:%
\begin{eqnarray*}
a &=&\left\{ \left( \frac{\frac{1}{2}\left( \bar{w}\left( X\right) +\hat{w}%
_{1}^{B}\left( X\right) +w_{1}^{B}\left( X\right) \right) \bar{S}\left(
X\right) }{\left( 1+\bar{w}\left( X\right) \Delta _{1}\left( \frac{\bar{f}%
\left( X^{\prime }\right) +\bar{r}\left( X^{\prime }\right) }{2}\right) +%
\hat{w}_{1}^{B}\left( X\right) \Delta _{2}\left( \frac{\bar{f}\left(
X^{\prime }\right) +\bar{r}\left( X^{\prime }\right) }{2}\right)
+w_{1}^{B}\left( X\right) \Delta _{3}\left( \frac{\bar{f}\left( X^{\prime
}\right) +\bar{r}\left( X^{\prime }\right) }{2}\right) \right) }\right.
\right. \\
&&\left. -\frac{\bar{S}\left( X^{\prime },\theta -1\right) \partial _{\bar{f}%
\left( X\right) }\left\langle \bar{K}_{X}\right\rangle \left\vert \bar{\Psi}%
\left( X\right) \right\vert ^{2}}{\left\langle \bar{K}_{X^{\prime
}}\right\rangle \left\vert \bar{\Psi}\left( X^{\prime }\right) \right\vert
^{2}}\right) \frac{\left( \bar{f}\left( X\right) -\bar{r}\right) }{1-\left( 
\bar{S}_{E}\left( X^{\prime },\theta -1\right) \right) } \\
&&-\left( \frac{\left( \bar{w}\left( X\right) +\hat{w}_{1}^{B}\left(
X\right) +w_{1}^{B}\left( X\right) \right) \bar{S}_{E}\left( X\right) }{%
\left( 1+\bar{w}\left( X\right) \Delta _{1}\left( \frac{\bar{f}\left(
X^{\prime }\right) +\bar{r}\left( X^{\prime }\right) }{2}\right) +\hat{w}%
_{1}^{B}\left( X\right) \Delta _{2}\left( \frac{\bar{f}\left( X^{\prime
}\right) +\bar{r}\left( X^{\prime }\right) }{2}\right) +w_{1}^{B}\left(
X\right) \Delta _{3}\left( \frac{\bar{f}\left( X^{\prime }\right) +\bar{r}%
\left( X^{\prime }\right) }{2}\right) \right) }\right. \\
&&\left. -\frac{\bar{S}_{E}\left( X^{\prime },\theta -1\right) \partial _{%
\bar{f}\left( X\right) }\left\langle \bar{K}_{X}\right\rangle \left\vert 
\bar{\Psi}\left( X\right) \right\vert ^{2}}{\left\langle \bar{K}_{X^{\prime
}}\right\rangle \left\vert \bar{\Psi}\left( X^{\prime }\right) \right\vert
^{2}}\right) \frac{\left( 1-\left( \bar{S}\left( X,\theta -1\right) \right)
\right) \left( \bar{f}\left( X\right) -\bar{r}\right) }{\left( 1-\left( \bar{%
S}_{E}\left( X^{\prime },\theta -1\right) \right) \right) ^{2}} \\
&&\left. +\left( S_{E}^{B}\left( X,\theta -1\right) +S_{L}^{B}\left(
X,\theta -1\right) \right) \frac{\partial _{\bar{f}\left( X\right) }\bar{K}%
_{X}\left\vert \bar{\Psi}\left( X\right) \right\vert ^{2}}{\bar{K}%
_{X}\left\vert \bar{\Psi}\left( X\right) \right\vert ^{2}}S_{E}^{B}\left(
X,X\right) \frac{\partial f\left( X\right) }{\partial S^{T}\left( X,\theta
-1\right) }\right\} -1
\end{eqnarray*}

\begin{eqnarray}
i &=&\left\{ w_{1}^{B}\left( X\right) \left( \left\langle \bar{w}\left(
X\right) \right\rangle +\left\langle \hat{w}_{1}^{B}\left( X\right)
\right\rangle \right) -\left( S_{E}^{B}\left( X,\theta -1\right)
+S_{L}^{B}\left( X,\theta -1\right) \right) \frac{\partial _{f\left(
X\right) }K_{X}\left\vert \Psi \left( X\right) \right\vert ^{2}}{%
K_{X}\left\vert \Psi \left( X\right) \right\vert ^{2}}\right.  \label{FI} \\
&&\left. +\left( \frac{\frac{\hat{w}\left( X\right) }{2}S\left( X\right) }{%
1+\left( \hat{w}\left( X\right) \left( \frac{f\left( X\right) +\bar{r}\left(
X\right) }{2}-\frac{\left\langle \hat{f}\left( X^{\prime }\right)
\right\rangle _{\hat{w}_{1}}+\left\langle \hat{r}\left( X^{\prime }\right)
\right\rangle _{\hat{w}_{2}}}{2}\right) \right) }+\frac{\partial _{f\left(
X\right) }\left( \frac{\hat{K}_{X}\left\vert \hat{\Psi}\left( X\right)
\right\vert ^{2}}{K_{X}\left\vert \Psi \left( X\right) \right\vert ^{2}}%
\right) S\left( X\right) }{\frac{\hat{K}_{X}\left\vert \hat{\Psi}\left(
X\right) \right\vert ^{2}}{K_{X}\left\vert \Psi \left( X\right) \right\vert
^{2}}}\right) \right\} \frac{\partial f\left( X\right) }{\partial
S^{T}\left( X,\theta -2\right) }  \notag
\end{eqnarray}%
\begin{eqnarray*}
&&c\rightarrow \left\{ w_{1}^{B}\left( X\right) \left( \left\langle \bar{w}%
\left( X\right) \right\rangle +\left\langle \hat{w}_{1}^{B}\left( X\right)
\right\rangle \right) \left( f\left( X\right) -\bar{r}\right) \frac{\partial
f\left( X\right) }{\partial S^{T}\left( X,\theta -2\right) }\right. \\
&&-\frac{\left\langle \bar{w}\left( X^{\prime },X\right) \right\rangle }{2}%
w_{1}^{B}\left( X\right) \frac{1-\left\langle \bar{S}\left( X^{\prime
}\right) \right\rangle }{1-\left\langle \bar{S}_{E}\left( X^{\prime }\right)
\right\rangle }\left( \left\langle \bar{f}\left( X^{\prime }\right)
\right\rangle -\bar{r}\right) \\
&&-\left\langle \hat{w}_{1}^{B}\left( X^{\prime },X\right) \right\rangle
\left\langle w_{1}^{B}\left( X\right) \right\rangle \frac{1-\left\langle 
\hat{S}\left( X^{\prime }\right) \right\rangle +\left\langle \hat{S}%
_{E}^{B}\left( X^{\prime }\right) \right\rangle +\left\langle \hat{S}%
_{L}^{B}\left( X^{\prime }\right) \right\rangle }{1-\left\langle \hat{S}%
_{E}\left( X^{\prime }\right) \right\rangle +\left\langle \hat{S}%
_{E}^{B}\left( X^{\prime }\right) \right\rangle }\left( \left\langle \hat{f}%
\left( X^{\prime }\right) \right\rangle -\bar{r}\right) \\
&&\left. -S_{E}^{B}\left( X,X\right) \frac{\partial f\left( X\right) }{%
\partial S^{T}\left( X,\theta -1\right) }\left( S_{E}^{B}\left( X,\theta
-1\right) +S_{L}^{B}\left( X,\theta -1\right) \right) \frac{\partial
_{f\left( X\right) }K_{X}\left\vert \Psi \left( X\right) \right\vert ^{2}}{%
K_{X}\left\vert \Psi \left( X\right) \right\vert ^{2}}\right\} \frac{%
\partial f\left( X\right) }{\partial S^{T}\left( X,\theta -2\right) }
\end{eqnarray*}%
\begin{equation*}
g=\left( S_{E}^{B}\left( X,\theta -1\right) +S_{L}^{B}\left( X,\theta
-1\right) \right) \frac{\partial _{\bar{f}\left( X\right) }\bar{K}%
_{X}\left\vert \bar{\Psi}\left( X\right) \right\vert ^{2}}{\bar{K}%
_{X}\left\vert \bar{\Psi}\left( X\right) \right\vert ^{2}}<0
\end{equation*}%
\begin{eqnarray*}
\frac{\partial f\left( X\right) }{\partial S^{T}\left( X,\theta -2\right) }
&=&-r\left( 1-S^{T}\left( X,\theta -2\right) \right) ^{-1}f\left( X\right) \\
&\simeq &-r\left( f\left( X\right) -\left\langle \bar{f}\left( X^{\prime
}\right) \right\rangle \right) f\left( X\right)
\end{eqnarray*}%
For banks in general $\bar{f}\left( X\right) -\bar{r}<<\hat{f}\left(
X\right) -\bar{r}$ and:%
\begin{equation*}
a<\alpha
\end{equation*}%
Compared to $\beta $, the coefficient $i$ includes: 
\begin{equation*}
i\rightarrow -\left( S_{E}^{B}\left( X,\theta -1\right) +S_{L}^{B}\left(
X,\theta -1\right) \right) \frac{\partial _{f\left( X\right)
}K_{X}\left\vert \Psi \left( X\right) \right\vert ^{2}}{K_{X}\left\vert \Psi
\left( X\right) \right\vert ^{2}}\frac{\partial f\left( X\right) }{\partial
S^{T}\left( X,\theta -2\right) }<0
\end{equation*}%
and $c$ includes additional term:%
\begin{eqnarray*}
&&\left( w_{1}^{B}\left( X\right) \left( \left\langle \bar{w}\left( X\right)
\right\rangle +\left\langle \hat{w}_{1}^{B}\left( X\right) \right\rangle
\right) \left( f\left( X\right) -\bar{r}\right) \frac{\partial f\left(
X\right) }{\partial S^{T}\left( X,\theta -2\right) }-\frac{\left\langle \bar{%
w}\left( X^{\prime },X\right) \right\rangle }{2}w_{1}^{B}\left( X\right) 
\frac{1-\left\langle \bar{S}\left( X^{\prime }\right) \right\rangle }{%
1-\left\langle \bar{S}_{E}\left( X^{\prime }\right) \right\rangle }\left(
\left\langle \bar{f}\left( X^{\prime }\right) \right\rangle -\bar{r}\right)
\right) \\
&&\times \frac{\partial f\left( X\right) }{\partial S^{T}\left( X,\theta
-2\right) }
\end{eqnarray*}%
which is positive when $\left\langle \bar{f}\left( X^{\prime }\right)
\right\rangle -\bar{r}>0$

Both terms lead to more stable eigenvalues than in the case of sole
investors. We consider several cases as befor.

\subsubsection*{A12.4.1 Case 1 $\bar{f}\left( X\right) -\bar{r}<<1$}

In general, for $\left( \bar{f}\left( X\right) -\bar{r}\right) <<1$ 
\begin{eqnarray*}
a+i &\simeq &-\left( S_{E}^{B}\left( X,\theta -1\right) +S_{L}^{B}\left(
X,\theta -1\right) \right) \frac{1}{f\left( X\right) }S_{E}^{B}\left(
X,X\right) \frac{\partial f\left( X\right) }{\partial S^{T}\left( X,\theta
-1\right) } \\
&&+\left\{ w_{1}^{B}\left( X\right) \left( \left\langle \bar{w}\left(
X\right) \right\rangle +\left\langle \hat{w}_{1}^{B}\left( X\right)
\right\rangle \right) +\left( S_{E}^{B}\left( X,\theta -1\right)
+S_{L}^{B}\left( X,\theta -1\right) \right) \frac{1}{f\left( X\right) }%
\right. \\
&&\left. +\frac{\frac{\hat{w}\left( X\right) }{2}S\left( X\right) }{1+\left( 
\hat{w}\left( X\right) \left( \frac{f\left( X\right) +\bar{r}\left( X\right) 
}{2}-\frac{\left\langle \hat{f}\left( X^{\prime }\right) \right\rangle _{%
\hat{w}_{1}}+\left\langle \hat{r}\left( X^{\prime }\right) \right\rangle _{%
\hat{w}_{2}}}{2}\right) \right) }\right\} \frac{\partial f\left( X\right) }{%
\partial S^{T}\left( X,\theta -2\right) } \\
&<&0
\end{eqnarray*}%
\begin{eqnarray*}
-ai &\simeq &\left( S_{E}^{B}\left( X,\theta -1\right) +S_{L}^{B}\left(
X,\theta -1\right) \right) \frac{1}{f\left( X\right) }S_{E}^{B}\left(
X,X\right) \frac{\partial f\left( X\right) }{\partial S^{T}\left( X,\theta
-1\right) } \\
&&\times \left\{ w_{1}^{B}\left( X\right) \left( \left\langle \bar{w}\left(
X\right) \right\rangle +\left\langle \hat{w}_{1}^{B}\left( X\right)
\right\rangle \right) +\left( S_{E}^{B}\left( X,\theta -1\right)
+S_{L}^{B}\left( X,\theta -1\right) \right) \frac{1}{f\left( X\right) }%
\right. \\
&&\left. +\frac{\frac{\hat{w}\left( X\right) }{2}S\left( X\right) }{1+\left( 
\hat{w}\left( X\right) \left( \frac{f\left( X\right) +\bar{r}\left( X\right) 
}{2}-\frac{\left\langle \hat{f}\left( X^{\prime }\right) \right\rangle _{%
\hat{w}_{1}}+\left\langle \hat{r}\left( X^{\prime }\right) \right\rangle _{%
\hat{w}_{2}}}{2}\right) \right) }\right\} \frac{\partial f\left( X\right) }{%
\partial S^{T}\left( X,\theta -2\right) } \\
&\rightarrow &\left( S_{E}^{B}\left( X,\theta -1\right) +S_{L}^{B}\left(
X,\theta -1\right) \right) S_{E}^{B}\left( X,X\right) \left( r\left(
1-S^{T}\left( X,\theta -2\right) \right) ^{-1}\right) ^{2}
\end{eqnarray*}%
\begin{eqnarray*}
-cg &\simeq &\left( S_{E}^{B}\left( X,\theta -1\right) +S_{L}^{B}\left(
X,\theta -1\right) \right) \frac{1}{\bar{f}\left( X\right) } \\
&&\times \left\{ w_{1}^{B}\left( X\right) \left( \left\langle \bar{w}\left(
X\right) \right\rangle +\left\langle \hat{w}_{1}^{B}\left( X\right)
\right\rangle \right) \left( f\left( X\right) -\bar{r}\right) \frac{\partial
f\left( X\right) }{\partial S^{T}\left( X,\theta -2\right) }\right. \\
&&-\left\langle \hat{w}_{1}^{B}\left( X^{\prime },X\right) \right\rangle
\left\langle w_{1}^{B}\left( X\right) \right\rangle \frac{1-\left\langle 
\hat{S}\left( X^{\prime }\right) \right\rangle +\left\langle \hat{S}%
_{E}^{B}\left( X^{\prime }\right) \right\rangle +\left\langle \hat{S}%
_{L}^{B}\left( X^{\prime }\right) \right\rangle }{1-\left\langle \hat{S}%
_{E}\left( X^{\prime }\right) \right\rangle +\left\langle \hat{S}%
_{E}^{B}\left( X^{\prime }\right) \right\rangle }\left( \left\langle \hat{f}%
\left( X^{\prime }\right) \right\rangle -\bar{r}\right) \\
&&\left. +S_{E}^{B}\left( X,X\right) \frac{\partial f\left( X\right) }{%
\partial S^{T}\left( X,\theta -1\right) }\left( S_{E}^{B}\left( X,\theta
-1\right) +S_{L}^{B}\left( X,\theta -1\right) \right) \frac{1}{\bar{f}\left(
X\right) }\right\} \frac{\partial f\left( X\right) }{\partial S^{T}\left(
X,\theta -2\right) } \\
&\rightarrow &\left( \left( S_{E}^{B}\left( X,\theta -1\right)
+S_{L}^{B}\left( X,\theta -1\right) \right) \frac{1}{\bar{f}\left( X\right) }%
\right) ^{2}\frac{\partial f\left( X\right) }{\partial S^{T}\left( X,\theta
-1\right) }\frac{\partial f\left( X\right) }{\partial S^{T}\left( X,\theta
-2\right) } \\
&&\left( \left( S_{E}^{B}\left( X,\theta -1\right) +S_{L}^{B}\left( X,\theta
-1\right) \right) \frac{f\left( X\right) }{\bar{f}\left( X\right) }\right)
^{2}\left( r\left( 1-S^{T}\left( X,\theta -2\right) \right) ^{-1}\right) ^{2}
\end{eqnarray*}%
nd $-cg>-ai$ ldng t stbl.

\subsubsection*{A12.4.2 Case 2 $\bar{f}\left( X\right) -\bar{r}>>1$, $%
f\left( X\right) $ finite}

The coefficients are estimated: 
\begin{eqnarray}
a &\simeq &\left( \frac{\frac{1}{2}\left( \bar{w}\left( X\right) +\hat{w}%
_{1}^{B}\left( X\right) +w_{1}^{B}\left( X\right) \right) \bar{S}\left(
X,\theta -1\right) }{\left( 1+\bar{w}\left( X\right) \Delta _{1}\left( \frac{%
\bar{f}\left( X^{\prime }\right) +\bar{r}\left( X^{\prime }\right) }{2}%
\right) +\hat{w}_{1}^{B}\left( X\right) \Delta _{2}\left( \frac{\bar{f}%
\left( X^{\prime }\right) +\bar{r}\left( X^{\prime }\right) }{2}\right)
+w_{1}^{B}\left( X\right) \Delta _{3}\left( \frac{\bar{f}\left( X^{\prime
}\right) +\bar{r}\left( X^{\prime }\right) }{2}\right) \right) }\right.
\label{Fa} \\
&&\left. -\frac{\bar{S}\left( X^{\prime },\theta -1\right) \partial _{\bar{f}%
\left( X\right) }\left\langle \bar{K}_{X}\right\rangle \left\vert \bar{\Psi}%
\left( X\right) \right\vert ^{2}}{\left\langle \bar{K}_{X^{\prime
}}\right\rangle \left\vert \bar{\Psi}\left( X^{\prime }\right) \right\vert
^{2}}\right) \frac{\left( \bar{f}\left( X\right) -\bar{r}\right) }{1-\left( 
\bar{S}_{E}\left( X^{\prime },\theta -1\right) \right) }  \notag \\
&&-\left( \frac{\left( \bar{w}\left( X\right) +\hat{w}_{1}^{B}\left(
X\right) +w_{1}^{B}\left( X\right) \right) \bar{S}_{E}\left( X^{\prime
},\theta -1\right) }{\left( 1+\bar{w}\left( X\right) \Delta _{1}\left( \frac{%
\bar{f}\left( X^{\prime }\right) +\bar{r}\left( X^{\prime }\right) }{2}%
\right) +\hat{w}_{1}^{B}\left( X\right) \Delta _{2}\left( \frac{\bar{f}%
\left( X^{\prime }\right) +\bar{r}\left( X^{\prime }\right) }{2}\right)
+w_{1}^{B}\left( X\right) \Delta _{3}\left( \frac{\bar{f}\left( X^{\prime
}\right) +\bar{r}\left( X^{\prime }\right) }{2}\right) \right) }\right. 
\notag \\
&&\left. -\frac{\bar{S}_{E}\left( X^{\prime },\theta -1\right) \partial _{%
\bar{f}\left( X\right) }\left\langle \bar{K}_{X}\right\rangle \left\vert 
\bar{\Psi}\left( X\right) \right\vert ^{2}}{\left\langle \bar{K}_{X^{\prime
}}\right\rangle \left\vert \bar{\Psi}\left( X^{\prime }\right) \right\vert
^{2}}\right) \frac{\left( 1-\left( \bar{S}\left( X,\theta -1\right) \right)
\right) \left( \bar{f}\left( X\right) -\bar{r}\right) }{\left( 1-\left( \bar{%
S}_{E}\left( X^{\prime },\theta -1\right) \right) \right) ^{2}}  \notag
\end{eqnarray}%
Since:%
\begin{equation*}
\frac{\bar{S}\left( X,\theta -1\right) }{1-\left( \bar{S}_{E}\left(
X^{\prime },\theta -1\right) \right) }-\frac{\bar{S}_{E}\left( X^{\prime
},\theta -1\right) \left( 1-\left( \bar{S}\left( X,\theta -1\right) \right)
\right) }{\left( 1-\left( \bar{S}_{E}\left( X^{\prime },\theta -1\right)
\right) \right) ^{2}}=\frac{1-\left( \bar{S}\left( X,\theta -1\right)
\right) }{1-\left( \bar{S}_{E}\left( X^{\prime },\theta -1\right) \right) }%
\frac{\bar{S}_{L}\left( X,\theta -1\right) }{\left( 1-\left( \bar{S}%
_{E}\left( X^{\prime },\theta -1\right) \right) \right) ^{2}}
\end{equation*}%
we find as in appendix 10 that:%
\begin{equation*}
a\simeq \left( 1+\Delta \bar{r}\right) \left( 1+\Delta \bar{f}\left(
X\right) \right) \left( \bar{f}\left( X\right) -\bar{r}\right) >0
\end{equation*}%
We use that:%
\begin{equation*}
\frac{\partial _{f\left( X\right) }K_{X}\left\vert \Psi \left( X\right)
\right\vert ^{2}}{K_{X}\left\vert \Psi \left( X\right) \right\vert ^{2}}%
\simeq \frac{1}{f\left( X\right) }
\end{equation*}%
so that:%
\begin{eqnarray}
i &\simeq &\left\{ w_{1}^{B}\left( X\right) \left( \left\langle \bar{w}%
\left( X\right) \right\rangle +\left\langle \hat{w}_{1}^{B}\left( X\right)
\right\rangle \right) +\left( S_{E}^{B}\left( X,\theta -1\right)
+S_{L}^{B}\left( X,\theta -1\right) \right) \frac{1}{f\left( X\right) }%
\right.  \label{Fi} \\
&&\left. +\left( \frac{\frac{\hat{w}\left( X\right) }{2}S\left( X\right) }{%
1+\left( \hat{w}\left( X\right) \left( \frac{f\left( X\right) +\bar{r}\left(
X\right) }{2}-\frac{\left\langle \hat{f}\left( X^{\prime }\right)
\right\rangle _{\hat{w}_{1}}+\left\langle \hat{r}\left( X^{\prime }\right)
\right\rangle _{\hat{w}_{2}}}{2}\right) \right) }\right) \right\} \frac{%
\partial f\left( X\right) }{\partial S^{T}\left( X,\theta -2\right) }  \notag
\\
&\simeq &\left\{ w_{1}^{B}\left( X\right) \left( \left\langle \bar{w}\left(
X\right) \right\rangle +\left\langle \hat{w}_{1}^{B}\left( X\right)
\right\rangle \right) +\left( S_{E}^{B}\left( X,\theta -1\right)
+S_{L}^{B}\left( X,\theta -1\right) \right) \frac{1}{f\left( X\right) }%
\right.  \notag \\
&&\left. +\frac{S\left( X\right) }{f\left( X\right) -\bar{f}\left( X\right) }%
\right\} r\left( 1-S^{T}\left( X,\theta -2\right) \right) ^{-1}f\left(
X\right)  \notag \\
&<&0  \notag
\end{eqnarray}%
and:%
\begin{eqnarray*}
&&a+i \\
&\rightarrow &\left( 1+\Delta \bar{r}\right) \left( 1+\Delta \bar{f}\left(
X\right) \right) \left( \bar{f}\left( X\right) -\bar{r}\right) \\
&&-\left\{ w_{1}^{B}\left( X\right) \left( \left\langle \bar{w}\left(
X\right) \right\rangle +\left\langle \hat{w}_{1}^{B}\left( X\right)
\right\rangle \right) +\left( S_{E}^{B}\left( X,\theta -1\right)
+S_{L}^{B}\left( X,\theta -1\right) \right) \frac{1}{f\left( X\right) }%
\right. \\
&&\left. +\frac{S\left( X\right) }{f\left( X\right) -\bar{f}\left( X\right) }%
\right\} r\left( 1-S^{T}\left( X,\theta -2\right) \right) ^{-1}f\left(
X\right) \\
&\simeq &\left( 1+\Delta \bar{r}\right) \left( 1+\Delta \bar{f}\left(
X\right) \right) \left( \bar{f}\left( X\right) -\bar{r}\right) \\
&&-\left\{ w_{1}^{B}\left( X\right) \left( \left\langle \bar{w}\left(
X\right) \right\rangle +\left\langle \hat{w}_{1}^{B}\left( X\right)
\right\rangle \right) +\left( S_{E}^{B}\left( X,\theta -1\right)
+S_{L}^{B}\left( X,\theta -1\right) \right) \frac{1}{f\left( X\right) }%
\right\} \\
&&\times r\left( 1-S^{T}\left( X,\theta -2\right) \right) ^{-1}f\left(
X\right) \\
&>&0
\end{eqnarray*}

which corresponds to an unstable case.

\subsubsection*{A12.4.2 Case 2 bis $\bar{f}\left( X\right) -\bar{r}>>1$, $%
f\left( X\right) >>1$}

\begin{equation*}
a\simeq \left( 1+\Delta \bar{r}\right) \left( 1+\Delta \bar{f}\left(
X\right) \right) \left( \bar{f}\left( X\right) -\bar{r}\right) >0
\end{equation*}%
\begin{eqnarray}
i &=&-\left\{ w_{1}^{B}\left( X\right) \left( \left\langle \bar{w}\left(
X\right) \right\rangle +\left\langle \hat{w}_{1}^{B}\left( X\right)
\right\rangle \right) \right. \\
&&\left. +\frac{S\left( X\right) }{f\left( X\right) -\left\langle \hat{f}%
\left( X^{\prime }\right) \right\rangle }\right\} r\left( 1-S^{T}\left(
X,\theta -2\right) \right) ^{-1}f\left( X\right)  \notag \\
&\simeq &-w_{1}^{B}\left( X\right) \left( \left\langle \bar{w}\left(
X\right) \right\rangle +\left\langle \hat{w}_{1}^{B}\left( X\right)
\right\rangle \right) r\left( 1-S^{T}\left( X,\theta -2\right) \right)
^{-1}f\left( X\right)  \notag
\end{eqnarray}%
\begin{equation*}
a+i<0
\end{equation*}

\begin{eqnarray*}
-ai &\simeq &\left( 1+\Delta \bar{r}\right) \left( 1+\Delta \bar{f}\left(
X\right) \right) \left( \bar{f}\left( X\right) -\bar{r}\right) \\
&&\times w_{1}^{B}\left( X\right) \left( \left\langle \bar{w}\left( X\right)
\right\rangle +\left\langle \hat{w}_{1}^{B}\left( X\right) \right\rangle
\right) r\left( 1-S^{T}\left( X,\theta -2\right) \right) ^{-1}f\left(
X\right)
\end{eqnarray*}%
\begin{equation*}
c\simeq w_{1}^{B}\left( X\right) \left( \left\langle \bar{w}\left( X\right)
\right\rangle +\left\langle \hat{w}_{1}^{B}\left( X\right) \right\rangle
\right) \left( f\left( X\right) -\bar{r}\right) \left( r\left( 1-S^{T}\left(
X,\theta -2\right) \right) ^{-1}\right) ^{2}
\end{equation*}%
\begin{equation*}
-g\simeq \left( S_{E}^{B}\left( X,\theta -1\right) +S_{L}^{B}\left( X,\theta
-1\right) \right) \frac{1}{\bar{f}\left( X\right) }
\end{equation*}%
\begin{eqnarray*}
-cg &\simeq &w_{1}^{B}\left( X\right) \left( \left\langle \bar{w}\left(
X\right) \right\rangle +\left\langle \hat{w}_{1}^{B}\left( X\right)
\right\rangle \right) \left( f\left( X\right) -\bar{r}\right) \\
&&\times \left( r\left( 1-S^{T}\left( X,\theta -2\right) \right)
^{-1}\right) ^{2}\left( S_{E}^{B}\left( X,\theta -1\right) +S_{L}^{B}\left(
X,\theta -1\right) \right) \frac{1}{\bar{f}\left( X\right) } \\
&\simeq &w_{1}^{B}\left( X\right) \left( \left\langle \bar{w}\left( X\right)
\right\rangle +\left\langle \hat{w}_{1}^{B}\left( X\right) \right\rangle
\right) \left( S_{E}^{B}\left( X,\theta -1\right) +S_{L}^{B}\left( X,\theta
-1\right) \right) \\
&&\times \left( r\left( f\left( X\right) -\left\langle \bar{f}\left(
X^{\prime }\right) \right\rangle \right) f\left( X\right) \right) ^{2}\frac{%
\left( f\left( X\right) -\bar{r}\right) }{\bar{f}\left( X\right) }
\end{eqnarray*}%
and:%
\begin{equation*}
-ai\simeq \left( 1+\Delta \bar{r}\right) \left( \bar{f}\left( X\right) -\bar{%
r}\right)
\end{equation*}%
\begin{equation*}
\left( S_{E}^{B}\left( X,\theta -1\right) +S_{L}^{B}\left( X,\theta
-1\right) \right) rf\left( X\right) \frac{\left( f\left( X\right) -\bar{r}%
\right) }{\bar{f}\left( X\right) }
\end{equation*}%
Thus, when $rf\left( X\right) \frac{\left( f\left( X\right) -\bar{r}\right) 
}{\bar{f}\left( X\right) }$ is large enough and:%
\begin{equation*}
-ai<-ch
\end{equation*}%
this case corresponds to stabilty. Otherwise, it is unstable.

\bigskip

\subsubsection*{A12.4.3 Formula for perturbations}

The initial variations $\delta \bar{f}$, $\delta \hat{f}$, $\delta S$, write
in terms of eigenvectrs:

\begin{equation*}
\left( 
\begin{array}{c}
\delta \bar{f} \\ 
\delta \hat{f} \\ 
\delta S%
\end{array}%
\right) =aV_{1}+bV_{2}+cV_{3}
\end{equation*}%
with:%
\begin{equation*}
V_{1}=\left( 
\begin{array}{c}
0 \\ 
1 \\ 
0%
\end{array}%
\right)
\end{equation*}%
\begin{eqnarray*}
V_{2} &=&\left( 
\begin{array}{c}
-\frac{\left( 2-\sqrt{\left( a-D\right) ^{2}+4cg}+D+a\right) c}{2\left(
f+af-cd\right) } \\ 
\frac{f\left( \sqrt{a^{2}-2aD+D^{2}+4cg}-D+a\right) -2cd}{2\left(
f+af-cd\right) } \\ 
\frac{1}{2f+2af-2cd}\left( a-D+aD-D^{2}+D\sqrt{a^{2}-2aD+D^{2}+4cg}+\sqrt{%
a^{2}-2aD+D^{2}+4cg}-2cg\right)%
\end{array}%
\right) , \\
V_{3} &=&\left( 
\begin{array}{c}
-\frac{1}{2f+2af-2cd}\left( 2c+c\sqrt{a^{2}-2aD+D^{2}+4cg}+cD+ac\right) \\ 
-\frac{1}{2f+2af-2cd}\left( f\sqrt{a^{2}-2aD+D^{2}+4cg}+fD-af+2cd\right) \\ 
-\frac{1}{2f+2af-2cd}\left( D-a-aD+D^{2}+D\sqrt{a^{2}-2aD+D^{2}+4cg}+\sqrt{%
a^{2}-2aD+D^{2}+4cg}+2cg\right)%
\end{array}%
\right)
\end{eqnarray*}%
The coefficients $a$, $b$ and $c$ depend on the initial variations and on
the coordinates of the $V_{i}$. We find the following formula:%
\begin{equation*}
\left( 
\begin{array}{c}
a \\ 
b \\ 
c%
\end{array}%
\right) =\left( V_{1},V_{2},V_{3}\right) ^{-1}\left( 
\begin{array}{c}
\delta \bar{f} \\ 
\delta \hat{f} \\ 
\delta S%
\end{array}%
\right)
\end{equation*}%
where $\left( V_{1},V_{2},V_{3}\right) $ is the matrix with columns given by
the $V_{i}$.

\bigskip

and the formula for perturbation in the continuous approximation writes: 
\begin{equation*}
\left( 
\begin{array}{c}
\delta \bar{f}\left( X,\theta \right) \\ 
\delta \hat{f}\left( X,\theta \right) \\ 
\delta S\left( X,\theta -1\right)%
\end{array}%
\right) \rightarrow a\exp \left( -\theta \right) V_{1}+b\exp \left( \lambda
_{-}\theta \right) V_{2}+c\exp \left( \lambda _{+}\theta \right) V_{3}
\end{equation*}%
To the first approximation, this leads to:%
\begin{equation*}
\left( 
\begin{array}{c}
\delta \bar{f}\left( X,\theta \right) \\ 
\delta \hat{f}\left( X,\theta \right) \\ 
\delta S\left( X,\theta -1\right)%
\end{array}%
\right) \rightarrow c\exp \left( \lambda _{+}\theta \right) V_{3}
\end{equation*}%
where:%
\begin{equation*}
c=\left( 0,0,1\right) \left( \left( V_{1},V_{2},V_{3}\right) ^{-1}\left( 
\begin{array}{c}
\delta \bar{f} \\ 
\delta \hat{f} \\ 
\delta S%
\end{array}%
\right) \right)
\end{equation*}

\subsection*{A12.5 Deviations from averages perturbations equations}

Introducing deviations from averages interaction amounts as in the first
part, to introduce\ additional terms that are in first approximation:%
\begin{eqnarray*}
\frac{1-\bar{S}\left( X\right) }{1-\bar{S}_{E}\left( X\right) }\delta \bar{f}%
\left( X,\theta -1\right) &=&a\delta \bar{f}\left( X,\theta -1\right)
+b\delta \hat{f}\left( X,\theta -1\right) +c\delta S^{T}\left( X,\theta
-2\right) \\
&&+\int V\left( X,X^{\prime }\right) \delta \bar{f}\left( X^{\prime },\theta
-1\right) +\int W\left( X,X^{\prime }\right) \delta \hat{f}\left( X^{\prime
},\theta -1\right)
\end{eqnarray*}%
\begin{eqnarray*}
\frac{1-\left( \hat{S}\left( X\right) \right) }{1-\left( \hat{S}_{E}\left(
X\right) \right) }\delta \hat{f}\left( X,\theta \right) &=&d\delta \bar{f}%
\left( X,\theta -1\right) +e\delta \hat{f}\left( X,\theta -1\right) +f\delta
S^{T}\left( X,\theta -2\right) \\
&&+\int T\left( X,X^{\prime }\right) \delta \bar{f}\left( X^{\prime },\theta
-1\right)
\end{eqnarray*}%
\begin{equation*}
\delta S^{T}\left( X,,\theta -1\right) =g\delta \bar{f}\left( X,\theta
-1\right) +h\delta \hat{f}\left( X,\theta -1\right) +i\delta S^{T}\left(
X,\theta -2\right)
\end{equation*}%
\begin{eqnarray*}
V\left( X,X^{\prime }\right) &=&\bar{S}_{E}\left( X^{\prime },X\right) \frac{%
1-\left\langle \bar{S}\left( X^{\prime }\right) \right\rangle }{%
1-\left\langle \bar{S}_{E}\left( X^{\prime }\right) \right\rangle } \\
W\left( X,X^{\prime }\right) &=&\hat{S}_{E}^{B}\left( X^{\prime },X\right) 
\frac{1-\left\langle \hat{S}\left( X^{\prime }\right) \right\rangle
+\left\langle \hat{S}_{E}^{B}\left( X^{\prime }\right) \right\rangle
+\left\langle \hat{S}_{L}^{B}\left( X^{\prime }\right) \right\rangle }{%
1-\left\langle \hat{S}_{E}\left( X^{\prime }\right) \right\rangle
+\left\langle \hat{S}_{E}^{B}\left( X^{\prime }\right) \right\rangle }
\end{eqnarray*}

\begin{equation*}
T\left( X,X^{\prime }\right) =\hat{S}_{E}\left( X^{\prime },X\right) \frac{%
1-\left\langle \hat{S}\left( X^{\prime }\right) \right\rangle }{%
1-\left\langle \hat{S}_{E}\left( X^{\prime }\right) \right\rangle }
\end{equation*}%
This modifies the matrix of the system by an additional contribution:%
\begin{equation*}
\left( 
\begin{array}{ccc}
a & 0 & c \\ 
d & -1 & f \\ 
g & 0 & i%
\end{array}%
\right) +\left( 
\begin{array}{ccc}
V\left( X,X^{\prime }\right) & W\left( X,X^{\prime }\right) & 0 \\ 
0 & T\left( X,X^{\prime }\right) & 0 \\ 
0 & 0 & 0%
\end{array}%
\right)
\end{equation*}%
Using that for block matrices:%
\begin{equation*}
\left[ 
\begin{array}{cc}
P^{-1} & 0 \\ 
0 & N^{-1}%
\end{array}%
\right] \left[ 
\begin{array}{cc}
A & R \\ 
R^{\prime } & A^{\prime }%
\end{array}%
\right] \left[ 
\begin{array}{cc}
P & 0 \\ 
0 & N%
\end{array}%
\right] =\left[ 
\begin{array}{cc}
A & P^{-1}RN \\ 
N^{-1}R^{\prime }P & A^{\prime }%
\end{array}%
\right]
\end{equation*}%
so that after block diagonalization interactions are given by:%
\begin{equation*}
P^{-1}\left[ 
\begin{array}{ccc}
V\left( X^{\prime },X\right) & W\left( X,X^{\prime }\right) &  \\ 
0 & T\left( X^{\prime },X\right) &  \\ 
0 & 0 & 
\end{array}%
\right] N
\end{equation*}%
and: 
\begin{equation*}
N^{-1}\left[ 
\begin{array}{ccc}
V\left( X^{\prime },X\right) & W\left( X,X^{\prime }\right) &  \\ 
0 & T\left( X^{\prime },X\right) &  \\ 
0 & 0 & 
\end{array}%
\right] P
\end{equation*}%
We assume that, for the average system:%
\begin{equation*}
\left[ 
\begin{array}{ccc}
a & 0 & c \\ 
d & -1 & f \\ 
g & 0 & D%
\end{array}%
\right]
\end{equation*}%
the diagonalization matrices can be approximated by their avergs, s that $%
P=N $, nd the transformation of connection terms becomes:

\begin{eqnarray*}
&&\left( 
\begin{array}{ccc}
\alpha & 0 & \eta \\ 
\beta & 1 & \theta \\ 
\gamma & 0 & \rho%
\end{array}%
\right) ^{-1}\left( 
\begin{array}{ccc}
V & W & 0 \\ 
0 & T & 0 \\ 
0 & 0 & 0%
\end{array}%
\right) \left( 
\begin{array}{ccc}
\alpha & 0 & \eta \\ 
\beta & 1 & \theta \\ 
\gamma & 0 & \rho%
\end{array}%
\right) \\
&=&\left( 
\begin{array}{ccc}
V\alpha \frac{\rho }{\alpha \rho -\gamma \eta }+W\beta \frac{\rho }{\alpha
\rho -\gamma \eta } & W\frac{\rho }{\alpha \rho -\gamma \eta } & W\theta 
\frac{\rho }{\alpha \rho -\gamma \eta }+V\eta \frac{\rho }{\alpha \rho
-\gamma \eta } \\ 
\beta \left( T+W\frac{\theta \gamma -\beta \rho }{\alpha \rho -\gamma \eta }%
\right) +V\alpha \frac{\theta \gamma -\beta \rho }{\alpha \rho -\gamma \eta }
& T+W\frac{\theta \gamma -\beta \rho }{\alpha \rho -\gamma \eta } & \theta
\left( T+W\frac{\theta \gamma -\beta \rho }{\alpha \rho -\gamma \eta }%
\right) +V\eta \frac{\theta \gamma -\beta \rho }{\alpha \rho -\gamma \eta }
\\ 
-V\alpha \frac{\gamma }{\alpha \rho -\gamma \eta }-W\beta \frac{\gamma }{%
\alpha \rho -\gamma \eta } & -W\frac{\gamma }{\alpha \rho -\gamma \eta } & 
-W\theta \frac{\gamma }{\alpha \rho -\gamma \eta }-V\gamma \frac{\eta }{%
\alpha \rho -\gamma \eta }%
\end{array}%
\right)
\end{eqnarray*}%
where:%
\begin{equation*}
\left( 
\begin{array}{c}
\alpha \\ 
\beta \\ 
\gamma%
\end{array}%
\right) =\left( 
\begin{array}{c}
-\frac{\left( 2-\sqrt{\left( a-D\right) ^{2}+4cg}+D+a\right) c}{2\left(
f+af-cd\right) } \\ 
\frac{f\left( \sqrt{a^{2}-2aD+D^{2}+4cg}-D+a\right) -2cd}{2\left(
f+af-cd\right) } \\ 
\frac{1}{2f+2af-2cd}\left( a-D+aD-D^{2}+D\sqrt{a^{2}-2aD+D^{2}+4cg}+\sqrt{%
a^{2}-2aD+D^{2}+4cg}-2cg\right)%
\end{array}%
\right)
\end{equation*}%
\begin{equation*}
\left( 
\begin{array}{c}
c \\ 
f \\ 
D%
\end{array}%
\right) =\left( 
\begin{array}{c}
-\frac{\left( 2+a+D-\sqrt{\left( a-D\right) ^{2}+4cg}\right) c}{2\left(
f+af-cd\right) } \\ 
\frac{f\left( a-D+\sqrt{\left( a-D\right) ^{2}+4cg}\right) -2cd}{2\left(
f+af-cd\right) } \\ 
\frac{\left( a-D+\sqrt{\left( a-D\right) ^{2}+4cg}\right) \left( 1+D\right)
-2cg}{2f+2af-2cd}%
\end{array}%
\right)
\end{equation*}

\begin{eqnarray*}
&&\frac{1}{2}\left( a+D+\sqrt{\left( a-D\right) ^{2}+4cg}\right) \\
&\rightarrow &\frac{1}{2}\left( a+D+\sqrt{\left( a-D\right) ^{2}+4cg}\right)
+\left( W\theta \frac{\gamma }{\alpha \rho -\gamma \eta }+V\gamma \frac{\eta 
}{\alpha \rho -\gamma \eta }\right)
\end{eqnarray*}

\subsection*{A12.6 Modifications of eigenvalues}

As in the first part, for interactions without loops, eigenvalues are not
modified and fluctuations do not modify stable states. To include loops, we
frst consider the case of reciprocal interactions.

In first approxmtn we can replace eigenvectrs by their averages and
modifications for eigenvalues:%
\begin{eqnarray*}
\lambda _{+} &\rightarrow &\lambda _{+}+\left( 
\begin{array}{ccc}
-V\alpha \frac{\gamma }{\alpha \rho -\gamma \eta }-W\beta \frac{\gamma }{%
\alpha \rho -\gamma \eta } & -W\frac{\gamma }{\alpha \rho -\gamma \eta } & 
-W\theta \frac{\gamma }{\alpha \rho -\gamma \eta }-V\gamma \frac{\eta }{%
\alpha \rho -\gamma \eta }%
\end{array}%
\right) \\
&&\times \left( 
\begin{array}{ccc}
\frac{1}{\lambda _{+}-\lambda _{-}^{\prime }} & 0 & 0 \\ 
0 & \frac{1}{\lambda _{+}+1} & 0 \\ 
0 & 0 & \frac{1}{\lambda _{+}-\lambda _{+-}^{\prime }}%
\end{array}%
\right) \left( 
\begin{array}{c}
W^{\prime }\theta \frac{\rho }{\alpha \rho -\gamma \eta }+V^{\prime }\eta 
\frac{\rho }{\alpha \rho -\gamma \eta } \\ 
\theta \left( T^{\prime }+V^{\prime }\frac{\theta \gamma -\beta \rho }{%
\alpha \rho -\gamma \eta }\right) +V^{\prime }\eta \frac{\theta \gamma
-\beta \rho }{\alpha \rho -\gamma \eta } \\ 
-W^{\prime }\theta \frac{\gamma }{\alpha \rho -\gamma \eta }-V^{\prime
}\gamma \frac{\eta }{\alpha \rho -\gamma \eta }%
\end{array}%
\right)
\end{eqnarray*}%
leading to the variation in eignvalues:%
\begin{equation*}
\frac{1}{\lambda _{+}-\lambda _{+-}^{\prime }}\gamma ^{2}\left( W^{\prime
}\theta +V^{\prime }\eta \right) \frac{W\theta +V\eta }{\left( \alpha \rho
-\gamma \eta \right) ^{2}}-\frac{1}{\lambda _{+}-\lambda _{-}^{\prime }}\rho
\left( W^{\prime }\theta +V^{\prime }\eta \right) \gamma \frac{V\alpha
+W\beta }{\left( \alpha \rho -\gamma \eta \right) ^{2}}-\frac{W}{\lambda
_{+}+1}\gamma \frac{\theta \left( T^{\prime }+W^{\prime }\frac{\theta \gamma
-\beta \rho }{\alpha \rho -\gamma \eta }\right) +V^{\prime }\eta \frac{%
\theta \gamma -\beta \rho }{\alpha \rho -\gamma \eta }}{\alpha \rho -\gamma
\eta }
\end{equation*}%
If the interaction is symetric:%
\begin{eqnarray*}
\lambda _{+} &\rightarrow &\lambda _{+}+\left( 
\begin{array}{ccc}
-V\alpha \frac{\gamma }{\alpha \rho -\gamma \eta }-W\beta \frac{\gamma }{%
\alpha \rho -\gamma \eta } & -W\frac{\gamma }{\alpha \rho -\gamma \eta } & 
-W\theta \frac{\gamma }{\alpha \rho -\gamma \eta }-V\gamma \frac{\eta }{%
\alpha \rho -\gamma \eta }%
\end{array}%
\right) \\
&&\times \left( 
\begin{array}{ccc}
\frac{1}{\lambda _{+}-\lambda _{-}^{\prime }} & 0 & 0 \\ 
0 & \frac{1}{\lambda _{+}+1} & 0 \\ 
0 & 0 & \frac{1}{\lambda _{+}-\lambda _{+-}^{\prime }}%
\end{array}%
\right) \left( 
\begin{array}{c}
W\theta \frac{\rho }{\alpha \rho -\gamma \eta }+V\eta \frac{\rho }{\alpha
\rho -\gamma \eta } \\ 
\theta \left( T+W\frac{\theta \gamma -\beta \rho }{\alpha \rho -\gamma \eta }%
\right) +V\eta \frac{\theta \gamma -\beta \rho }{\alpha \rho -\gamma \eta }
\\ 
-W\theta \frac{\gamma }{\alpha \rho -\gamma \eta }-V\gamma \frac{\eta }{%
\alpha \rho -\gamma \eta }%
\end{array}%
\right) \\
&=&\lambda _{+}+\frac{1}{\lambda _{+}-\lambda _{+-}^{\prime }}\gamma ^{2}%
\frac{\left( W\theta +V\eta \right) ^{2}}{\left( \alpha \rho -\gamma \eta
\right) ^{2}}-\frac{1}{\lambda _{+}-\lambda _{-}^{\prime }}\rho \left(
W\theta +V\eta \right) \gamma \frac{V\alpha +W\beta }{\left( \alpha \rho
-\gamma \eta \right) ^{2}}-\frac{W}{\lambda _{+}+1}\gamma \frac{\theta
\left( T+W\frac{\theta \gamma -\beta \rho }{\alpha \rho -\gamma \eta }%
\right) +V\eta \frac{\theta \gamma -\beta \rho }{\alpha \rho -\gamma \eta }}{%
\alpha \rho -\gamma \eta }
\end{eqnarray*}%
Similarly:

\begin{eqnarray*}
\lambda _{-} &\rightarrow &\lambda _{-}+\left( 
\begin{array}{ccc}
\beta \left( T+W\frac{\theta \gamma -\beta \rho }{\alpha \rho -\gamma \eta }%
\right) +V\alpha \frac{\theta \gamma -\beta \rho }{\alpha \rho -\gamma \eta }
& T+W\frac{\theta \gamma -\beta \rho }{\alpha \rho -\gamma \eta } & \theta
\left( T+W\frac{\theta \gamma -\beta \rho }{\alpha \rho -\gamma \eta }%
\right) +V\eta \frac{\theta \gamma -\beta \rho }{\alpha \rho -\gamma \eta }%
\end{array}%
\right) \\
&&\times \left( 
\begin{array}{ccc}
\frac{1}{\lambda _{-}-\lambda _{-}^{\prime }} & 0 & 0 \\ 
0 & \frac{1}{\lambda _{-}+1} & 0 \\ 
0 & 0 & \frac{1}{\lambda _{-}-\lambda _{+-}^{\prime }}%
\end{array}%
\right) \left( 
\begin{array}{c}
W\theta \frac{\rho }{\alpha \rho -\gamma \eta }+V\eta \frac{\rho }{\alpha
\rho -\gamma \eta } \\ 
\theta \left( T+W\frac{\theta \gamma -\beta \rho }{\alpha \rho -\gamma \eta }%
\right) +V\eta \frac{\theta \gamma -\beta \rho }{\alpha \rho -\gamma \eta }
\\ 
-W\theta \frac{\gamma }{\alpha \rho -\gamma \eta }-V\gamma \frac{\eta }{%
\alpha \rho -\gamma \eta }%
\end{array}%
\right)
\end{eqnarray*}

\begin{eqnarray*}
&&\lambda _{+}+\left( 
\begin{array}{ccc}
-V\alpha \frac{\gamma }{\alpha \rho -\gamma \eta }-W\beta \frac{\gamma }{%
\alpha \rho -\gamma \eta } & -W\frac{\gamma }{\alpha \rho -\gamma \eta } & 
-W\theta \frac{\gamma }{\alpha \rho -\gamma \eta }-V\gamma \frac{\eta }{%
\alpha \rho -\gamma \eta }%
\end{array}%
\right) \\
&&\times \left( 
\begin{array}{ccc}
\frac{1}{\lambda _{+}-\lambda _{-}^{\prime }} & 0 & 0 \\ 
0 & \frac{1}{\lambda _{+}+1} & 0 \\ 
0 & 0 & \frac{1}{\lambda _{+}-\lambda _{+-}^{\prime }}%
\end{array}%
\right) \left( 
\begin{array}{c}
W^{\prime }\theta \frac{\rho }{\alpha \rho -\gamma \eta }+V^{\prime }\eta 
\frac{\rho }{\alpha \rho -\gamma \eta } \\ 
\theta \left( T^{\prime }+W^{\prime }\frac{\theta \gamma -\beta \rho }{%
\alpha \rho -\gamma \eta }\right) +V^{\prime }\eta \frac{\theta \gamma
-\beta \rho }{\alpha \rho -\gamma \eta } \\ 
-W^{\prime }\theta \frac{\gamma }{\alpha \rho -\gamma \eta }-V^{\prime
}\gamma \frac{\eta }{\alpha \rho -\gamma \eta }%
\end{array}%
\right)
\end{eqnarray*}%
More general, considr cycles of interactions. the loop yields modfication:

\begin{equation*}
\Delta \lambda _{i\alpha }=\sum \prod \frac{W_{j_{k+1}\alpha
_{k+1},j_{k}\alpha _{k}}}{\lambda _{i\alpha }-\lambda _{j_{k+1}\alpha _{k+1}}%
}
\end{equation*}%
where $\alpha _{k}$ takes their values. where $W_{j_{k+1}\alpha
_{k+1},j_{k}\alpha _{k}}$ coefficients of:%
\begin{eqnarray*}
&&\left( 
\begin{array}{ccc}
V\alpha \frac{\rho }{\alpha \rho -\gamma \eta }+W\beta \frac{\rho }{\alpha
\rho -\gamma \eta } & W\frac{\rho }{\alpha \rho -\gamma \eta } & W\theta 
\frac{\rho }{\alpha \rho -\gamma \eta }+V\eta \frac{\rho }{\alpha \rho
-\gamma \eta } \\ 
\beta \left( T+W\frac{\theta \gamma -\beta \rho }{\alpha \rho -\gamma \eta }%
\right) +V\alpha \frac{\theta \gamma -\beta \rho }{\alpha \rho -\gamma \eta }
& T+W\frac{\theta \gamma -\beta \rho }{\alpha \rho -\gamma \eta } & \theta
\left( T+W\frac{\theta \gamma -\beta \rho }{\alpha \rho -\gamma \eta }%
\right) +V\eta \frac{\theta \gamma -\beta \rho }{\alpha \rho -\gamma \eta }
\\ 
-V\alpha \frac{\gamma }{\alpha \rho -\gamma \eta }-W\beta \frac{\gamma }{%
\alpha \rho -\gamma \eta } & -W\frac{\gamma }{\alpha \rho -\gamma \eta } & 
-W\theta \frac{\gamma }{\alpha \rho -\gamma \eta }-V\gamma \frac{\eta }{%
\alpha \rho -\gamma \eta }%
\end{array}%
\right) \\
&=&\left( 
\begin{array}{ccc}
C_{1} & C_{2} & C_{3}%
\end{array}%
\right) =\left( 
\begin{array}{c}
L_{1} \\ 
L_{2} \\ 
L_{3}%
\end{array}%
\right)
\end{eqnarray*}%
and the modification is:%
\begin{eqnarray*}
&&L_{0\alpha }\left( 
\begin{array}{ccc}
\frac{1}{\lambda _{\alpha }-\lambda _{n-}^{\prime }} & 0 & 0 \\ 
0 & \frac{1}{\lambda _{\alpha }+1} & 0 \\ 
0 & 0 & \frac{1}{\lambda _{\alpha }-\lambda _{k+-}^{\prime }}%
\end{array}%
\right) \\
&&\times \prod\limits_{1}^{n-1}\left[ 
\begin{array}{ccc}
V\alpha \frac{\rho }{\alpha \rho -\gamma \eta }+W\beta \frac{\rho }{\alpha
\rho -\gamma \eta } & W\frac{\rho }{\alpha \rho -\gamma \eta } & W\theta 
\frac{\rho }{\alpha \rho -\gamma \eta }+V\eta \frac{\rho }{\alpha \rho
-\gamma \eta } \\ 
\beta \left( T+W\frac{\theta \gamma -\beta \rho }{\alpha \rho -\gamma \eta }%
\right) +V\alpha \frac{\theta \gamma -\beta \rho }{\alpha \rho -\gamma \eta }
& T+W\frac{\theta \gamma -\beta \rho }{\alpha \rho -\gamma \eta } & \theta
\left( T+W\frac{\theta \gamma -\beta \rho }{\alpha \rho -\gamma \eta }%
\right) +V\eta \frac{\theta \gamma -\beta \rho }{\alpha \rho -\gamma \eta }
\\ 
-V\alpha \frac{\gamma }{\alpha \rho -\gamma \eta }-W\beta \frac{\gamma }{%
\alpha \rho -\gamma \eta } & -W\frac{\gamma }{\alpha \rho -\gamma \eta } & 
-W\theta \frac{\gamma }{\alpha \rho -\gamma \eta }-V\gamma \frac{\eta }{%
\alpha \rho -\gamma \eta }%
\end{array}%
\right] _{k} \\
&&\times \left( 
\begin{array}{ccc}
\frac{1}{\lambda _{\alpha }-\lambda _{k-}^{\prime }} & 0 & 0 \\ 
0 & \frac{1}{\lambda _{\alpha }+1} & 0 \\ 
0 & 0 & \frac{1}{\lambda _{\alpha }-\lambda _{k+-}^{\prime }}%
\end{array}%
\right) C_{n\alpha }
\end{eqnarray*}%
that is:%
\begin{eqnarray*}
&&L_{0\alpha }\left( 
\begin{array}{ccc}
\frac{1}{\lambda _{-}+1} & 0 & 0 \\ 
0 & \frac{1}{\lambda _{-}-\lambda _{n-}^{\prime }} & 0 \\ 
0 & 0 & \frac{1}{\lambda _{-}-\lambda _{k+-}^{\prime }}%
\end{array}%
\right) \\
&&\times \prod\limits_{1}^{n-1}\left( 
\begin{array}{ccc}
\frac{V_{k}\alpha \frac{\rho }{\alpha \rho -\gamma \eta }+W_{k}\beta \frac{%
\rho }{\alpha \rho -\gamma \eta }}{\lambda _{\alpha }+1} & \frac{W_{k}\frac{%
\rho }{\alpha \rho -\gamma \eta }}{\lambda _{\alpha }-\lambda _{k-}^{\prime }%
} & \frac{W_{k}\theta \frac{\rho }{\alpha \rho -\gamma \eta }+V_{k}\eta 
\frac{\rho }{\alpha \rho -\gamma \eta }}{\lambda _{\alpha }-\lambda
_{k+-}^{\prime }} \\ 
\frac{\beta \left( T_{k}+W_{k}\frac{\theta \gamma -\beta \rho }{\alpha \rho
-\gamma \eta }\right) +V_{k}\alpha \frac{\theta \gamma -\beta \rho }{\alpha
\rho -\gamma \eta }}{\lambda _{\alpha }+1} & \frac{T_{k}+W_{k}\frac{\theta
\gamma -\beta \rho }{\alpha \rho -\gamma \eta }}{\lambda _{\alpha }-\lambda
_{k-}^{\prime }} & \frac{\theta \left( T_{k}+W_{k}\frac{\theta \gamma -\beta
\rho }{\alpha \rho -\gamma \eta }\right) +V_{k}\eta \frac{\theta \gamma
-\beta \rho }{\alpha \rho -\gamma \eta }}{\lambda _{\alpha }-\lambda
_{k+-}^{\prime }} \\ 
-\frac{V_{k}\alpha \frac{\gamma }{\alpha \rho -\gamma \eta }+W_{k}\beta 
\frac{\gamma }{\alpha \rho -\gamma \eta }}{\lambda _{\alpha }+1} & -\frac{%
W_{k}\frac{\gamma }{\alpha \rho -\gamma \eta }}{\lambda _{\alpha }-\lambda
_{k-}^{\prime }} & -\frac{W_{k}\theta \frac{\gamma }{\alpha \rho -\gamma
\eta }+V_{k}\gamma \frac{\eta }{\alpha \rho -\gamma \eta }}{\lambda _{\alpha
}-\lambda _{k+-}^{\prime }}%
\end{array}%
\right) C_{n\alpha }
\end{eqnarray*}%
where:%
\begin{equation*}
L_{0+}\rightarrow \left( 
\begin{array}{ccc}
-V\alpha \frac{\gamma }{\alpha \rho -\gamma \eta }-W\beta \frac{\gamma }{%
\alpha \rho -\gamma \eta } & -W\frac{\gamma }{\alpha \rho -\gamma \eta } & 
-W\theta \frac{\gamma }{\alpha \rho -\gamma \eta }-V\gamma \frac{\eta }{%
\alpha \rho -\gamma \eta }%
\end{array}%
\right)
\end{equation*}%
\begin{equation*}
C_{n+}\rightarrow \left( 
\begin{array}{c}
W\theta \frac{\rho }{\alpha \rho -\gamma \eta }+V\eta \frac{\rho }{\alpha
\rho -\gamma \eta } \\ 
\theta \left( T+W\frac{\theta \gamma -\beta \rho }{\alpha \rho -\gamma \eta }%
\right) +V\eta \frac{\theta \gamma -\beta \rho }{\alpha \rho -\gamma \eta }
\\ 
-W\theta \frac{\gamma }{\alpha \rho -\gamma \eta }-V\gamma \frac{\eta }{%
\alpha \rho -\gamma \eta }%
\end{array}%
\right)
\end{equation*}%
\begin{equation*}
L_{0-}\rightarrow \left( 
\begin{array}{ccc}
\beta \left( T+W\frac{\theta \gamma -\beta \rho }{\alpha \rho -\gamma \eta }%
\right) +V\alpha \frac{\theta \gamma -\beta \rho }{\alpha \rho -\gamma \eta }
& T+W\frac{\theta \gamma -\beta \rho }{\alpha \rho -\gamma \eta } & \theta
\left( T+W\frac{\theta \gamma -\beta \rho }{\alpha \rho -\gamma \eta }%
\right) +V\eta \frac{\theta \gamma -\beta \rho }{\alpha \rho -\gamma \eta }%
\end{array}%
\right)
\end{equation*}%
\begin{equation*}
C_{n+}\rightarrow \left( 
\begin{array}{c}
W\frac{\rho }{\alpha \rho -\gamma \eta } \\ 
T+W\frac{\theta \gamma -\beta \rho }{\alpha \rho -\gamma \eta } \\ 
-W\frac{\gamma }{\alpha \rho -\gamma \eta }%
\end{array}%
\right)
\end{equation*}

\subsection*{A12.7 Variations in terms of stakes}

The variations of the system nd transitns can directly be written in terms
of stakes between investors and banks.

\subsubsection*{A12.7.1 Investors}

As in the first prt:%
\begin{eqnarray*}
\delta \hat{S}_{E}\left( X^{\prime },X,\theta \right) &=&\frac{\hat{w}\left(
X^{\prime },X\right) }{2}\left( 1+\left( \delta \hat{f}\left( X^{\prime
}\right) -\left\langle w\left( X\right) \right\rangle \delta \frac{f\left(
X\right) +r\left( X\right) }{2}\right) \right) \\
&\rightarrow &\frac{\hat{w}\left( X^{\prime },X\right) }{2}\delta \hat{f}%
\left( X^{\prime },\theta \right)
\end{eqnarray*}%
\begin{eqnarray*}
\delta \hat{S}\left( X^{\prime },X,\theta \right) &=&\hat{w}\left( X^{\prime
},X\right) \left( 1+\delta \Delta \left( \frac{\hat{f}\left( X^{\prime
}\right) +\hat{r}\left( X^{\prime }\right) }{2}\right) \right) \\
&\rightarrow &\hat{w}\left( X^{\prime },X\right) \frac{\delta \hat{f}\left(
X^{\prime },\theta \right) +\hat{r}\left( X^{\prime },,\theta \right) }{2}
\end{eqnarray*}%
where:%
\begin{equation*}
\hat{w}\left( X^{\prime },X\right) \rightarrow \frac{\left( 1-\left( \gamma
\left\langle \hat{S}_{E}\left( X\right) \right\rangle \right) ^{2}\right) 
\hat{w}_{1}^{\left( 0\right) }\left( X^{\prime },X\right) }{1+\hat{w}%
_{1}^{\left( 0\right) }\left( X^{\prime },X\right) \left( 1-\left( \gamma
\left\langle \hat{S}_{E}\left( X\right) \right\rangle \right) ^{2}\right)
+\left( \gamma \left\langle \hat{S}_{E}\left( X_{1},X^{\prime }\right)
\right\rangle _{X_{1}}\right) ^{2}-\left( \gamma \left\langle \hat{S}%
_{E}\left( X\right) \right\rangle \right) ^{2}}
\end{equation*}%
The formula for $\hat{w}\left( X^{\prime },X\right) $ shows that this
parameter is mainly exogeneous.

\subsubsection*{A12.7.2 Banks}

For banks, we have $\bar{S}_{E}\left( X^{\prime },X\right) ,$ $\bar{S}%
_{L}\left( X^{\prime },X\right) $, $\bar{S}\left( X^{\prime },X\right) $, $%
\hat{S}_{E}^{B}\left( X^{\prime },X\right) $, $S_{E}^{B}\left( X,X\right) $
given by (\ref{SBN}), (\ref{SBT}), (\ref{SBF}), (\ref{scb}), (\ref{SFN}):%
\begin{eqnarray}
&&\bar{S}_{E}\left( X^{\prime },X\right) \\
&=&\frac{\bar{w}\left( X^{\prime },X\right) }{2}\left( 1+\left\{ \bar{w}%
\left( X\right) \left( \bar{f}\left( X^{\prime }\right) -\frac{\left\langle 
\bar{f}\left( X^{\prime }\right) \right\rangle _{\bar{w}_{1}}+\left\langle 
\bar{r}\left( X^{\prime }\right) \right\rangle _{\bar{w}_{2}}}{2}\right)
\right. \right.  \notag \\
&&\left. \left. +\hat{w}_{1}^{B}\left( X\right) \left( \bar{f}\left(
X^{\prime }\right) -\left\langle \hat{f}\left( X^{\prime }\right)
\right\rangle _{\hat{w}_{1}}\right) +w_{1}^{B}\left( X\right) \left( \bar{f}%
\left( X^{\prime }\right) -f\left( X\right) \right) \right\} \right)  \notag
\end{eqnarray}%
\begin{eqnarray}
&&\bar{S}_{L}\left( X^{\prime },X\right) \\
&=&\frac{\bar{w}\left( X^{\prime },X\right) }{2}\left( 1+\left\{ \bar{w}%
\left( X\right) \left( \bar{r}\left( X^{\prime }\right) -\frac{\left\langle 
\bar{f}\left( X^{\prime }\right) \right\rangle _{\bar{w}_{1}}+\left\langle 
\bar{r}\left( X^{\prime }\right) \right\rangle _{\bar{w}_{2}}}{2}\right)
\right. \right.  \notag \\
&&\left. \left. +\hat{w}_{1}^{B}\left( X\right) \left( \bar{r}\left(
X^{\prime }\right) -\left\langle \hat{f}\left( X^{\prime }\right)
\right\rangle _{\hat{w}_{1}^{B}}\right) +w_{1}^{B}\left( X\right) \left( 
\bar{r}\left( X^{\prime }\right) -f\left( X\right) \right) \right\} \right) 
\notag
\end{eqnarray}

\begin{eqnarray}
\bar{S}\left( X^{\prime },X\right) &=&\bar{S}_{E}\left( X^{\prime },X\right)
+\bar{S}_{L}\left( X^{\prime },X\right)  \label{SBFApp} \\
&=&\bar{w}\left( X^{\prime },X\right) \left[ 1+\left\{ \bar{w}\left(
X\right) \left( \frac{\bar{f}\left( X^{\prime }\right) +\bar{r}\left(
X^{\prime }\right) }{2}-\frac{\left\langle \bar{f}\left( X^{\prime }\right)
\right\rangle _{\bar{w}_{1}}+\left\langle \bar{r}\left( X^{\prime }\right)
\right\rangle _{\bar{w}_{2}}}{2}\right) \right. \right.  \notag \\
&&\left. \left. +\hat{w}_{1}^{B}\left( X\right) \left( \frac{\bar{f}\left(
X^{\prime }\right) +\bar{r}\left( X^{\prime }\right) }{2}-\left\langle \hat{f%
}\left( X^{\prime }\right) \right\rangle _{\hat{w}_{1}}\right)
+w_{1}^{B}\left( X\right) \left( \frac{\bar{f}\left( X^{\prime }\right) +%
\bar{r}\left( X^{\prime }\right) }{2}-f\left( X\right) \right) \right\} %
\right]  \notag
\end{eqnarray}

\begin{eqnarray}
&&\hat{S}_{E}^{B}\left( X^{\prime },X\right)  \label{SHNApp} \\
&=&\hat{w}_{1}^{B}\left( X^{\prime },X\right) \left[ 1+\bar{w}\left(
X\right) \left( \hat{f}\left( X^{\prime }\right) -\frac{\left\langle \bar{f}%
\left( X^{\prime }\right) \right\rangle _{\bar{w}_{1}}+\left\langle \bar{r}%
\left( X^{\prime }\right) \right\rangle _{\bar{w}_{2}}}{2}\right) \right. 
\notag \\
&&\left. +\hat{w}_{1}^{B}\left( X\right) \left( \hat{f}\left( X^{\prime
}\right) -\left\langle \hat{f}\left( X^{\prime }\right) \right\rangle _{\hat{%
w}_{1}}\right) +w_{1}^{B}\left( X\right) \left( \hat{f}\left( X^{\prime
}\right) -f\left( X\right) \right) \right]  \notag
\end{eqnarray}%
\begin{equation}
\frac{\hat{S}_{L}^{B}\left( X^{\prime },X\right) }{\kappa \left( 1-\bar{S}%
\left( X\right) \right) }=\hat{w}_{2}^{B}\left( X^{\prime },X\right) \left\{
1+\hat{w}_{2}^{B}\left( X\right) \left( \hat{r}\left( X^{\prime }\right)
-\left\langle \hat{f}\left( X^{\prime }\right) \right\rangle _{\hat{w}%
_{1}}\right) +w_{2}^{B}\left( X\right) \left( \hat{r}\left( X^{\prime
}\right) -f\left( X\right) \right) \right\}  \label{sdbApp}
\end{equation}

\begin{eqnarray}
&&S_{E}^{B}\left( X,X\right) \\
&=&w_{1}^{B}\left( X\right) \left\{ 1+\bar{w}\left( X\right) \left( f\left(
X\right) -\frac{\left\langle \bar{f}\left( X^{\prime }\right) \right\rangle
_{\bar{w}_{1}}+\left\langle \bar{r}\left( X^{\prime }\right) \right\rangle _{%
\bar{w}_{2}}}{2}\right) +\hat{w}_{1}^{B}\left( X\right) \left( f\left(
X\right) -\left\langle \hat{f}\left( X^{\prime }\right) \right\rangle _{\hat{%
w}_{1}}\right) \right\}  \notag
\end{eqnarray}%
\begin{equation}
\frac{S_{L}^{B}\left( X,X\right) }{\kappa \left( 1-\bar{S}\left( X\right)
\right) }=w_{2}^{B}\left( X\right) \left[ 1+\hat{w}_{2}^{B}\left( X\right)
\left( r\left( X\right) -\left\langle \hat{r}\left( X^{\prime }\right)
\right\rangle _{\hat{w}_{2}}\right) \right]
\end{equation}%
with:

\begin{eqnarray}
&&\left( \bar{w}\left( X^{\prime },X\right) \right) ^{-1}=1+\frac{4}{\zeta
^{2}\bar{w}_{1}^{\left( 0\right) }\left( X^{\prime },X\right) }\left\{ \frac{%
\bar{\zeta}^{2}\zeta ^{2}\left( 1+\frac{\left( \gamma \left\langle \hat{S}%
_{E}\left( X_{1},\left( X^{\prime }\right) ^{\prime }\right) \right\rangle
\right) ^{2}}{1-\left( \gamma \left\langle \hat{S}_{E}\left( X^{\prime
},\left( X^{\prime }\right) ^{\prime }\right) \right\rangle \right) ^{2}}%
\right) }{\left\langle \hat{w}_{1}^{\left( 0\right) B}\left( \left(
X^{\prime }\right) ^{\prime },X^{\prime }\right) \right\rangle _{\left(
X^{\prime }\right) ^{\prime }}}+\xi ^{2}\right\} \\
&&\times \left( \frac{1+\frac{\left( \bar{\gamma}\left\langle \bar{S}%
_{E}\left( X_{1},X^{\prime }\right) \right\rangle _{X_{1}}\right) ^{2}}{%
1-\left( \bar{\gamma}\left\langle \bar{S}_{E}\left( \left( X^{\prime
}\right) ^{\prime },X^{\prime }\right) \right\rangle \right) ^{2}}}{1+\frac{%
\left( \gamma \left\langle \hat{S}_{E}\left( X_{1},X^{\prime }\right)
\right\rangle _{X_{1}}\right) ^{2}}{1-\left( \gamma \left\langle \hat{S}%
_{E}\left( X^{\prime },\left( X^{\prime }\right) ^{\prime }\right)
\right\rangle \right) ^{2}}}\right) \left( w_{1}^{\left( 0\right) B}\left(
X^{\prime },X\right) +\frac{\zeta ^{2}}{\xi ^{2}}\left( 1+\frac{\left(
\gamma \left\langle \hat{S}_{E}\left( X_{1},X^{\prime }\right) \right\rangle
_{X_{1}}\right) ^{2}}{1-\left( \gamma \left\langle \hat{S}_{E}\left(
X^{\prime },\left( X^{\prime }\right) ^{\prime }\right) \right\rangle
\right) ^{2}}\right) \right)  \notag
\end{eqnarray}%
\begin{eqnarray}
\left( \hat{w}_{1}^{B}\left( X^{\prime },X\right) \right) ^{-1} &=&1+\frac{%
\left\langle \hat{w}_{1}^{\left( 0\right) B}\left( \left( X^{\prime }\right)
^{\prime },X^{\prime }\right) \right\rangle _{\left( X^{\prime }\right)
^{\prime }}\frac{\zeta ^{2}\bar{w}_{1}^{\left( 0\right) }\left( X^{\prime
},X\right) }{w_{1}^{\left( 0\right) B}\left( X^{\prime },X\right) }\left( 
\frac{1+\frac{\left( \gamma \left\langle \hat{S}_{E}\left( X_{1},X^{\prime
}\right) \right\rangle _{X_{1}}\right) ^{2}}{1-\left( \gamma \left\langle 
\hat{S}_{E}\left( X^{\prime },\left( X^{\prime }\right) ^{\prime }\right)
\right\rangle \right) ^{2}}}{1+\frac{\left( \bar{\gamma}\left\langle \bar{S}%
_{E}\left( X_{1},X^{\prime }\right) \right\rangle _{X_{1}}\right) ^{2}}{%
1-\left( \bar{\gamma}\left\langle \bar{S}_{E}\left( \left( X^{\prime
}\right) ^{\prime },X^{\prime }\right) \right\rangle \right) ^{2}}}\right) }{%
4\left( \bar{\zeta}^{2}\zeta ^{2}\left( 1+\frac{\left( \gamma \left\langle 
\hat{S}_{E}\left( X_{1},\left( X^{\prime }\right) ^{\prime }\right)
\right\rangle \right) ^{2}}{1-\left( \gamma \left\langle \hat{S}_{E}\left(
X^{\prime },\left( X^{\prime }\right) ^{\prime }\right) \right\rangle
\right) ^{2}}\right) +\xi ^{2}\left\langle \hat{w}_{1}^{\left( 0\right)
B}\left( \left( X^{\prime }\right) ^{\prime },X^{\prime }\right)
\right\rangle _{\left( X^{\prime }\right) ^{\prime }}\right) } \\
&&+\frac{\zeta ^{2}\left\langle \hat{w}_{1}^{\left( 0\right) B}\left( \left(
X^{\prime }\right) ^{\prime },X^{\prime }\right) \right\rangle _{\left(
X^{\prime }\right) ^{\prime }}}{\xi ^{2}w_{1}^{\left( 0\right) B}\left(
X^{\prime },X\right) }\left( 1+\frac{\left( \gamma \left\langle \hat{S}%
_{E}\left( X_{1},X^{\prime }\right) \right\rangle _{X_{1}}\right) ^{2}}{%
1-\left( \gamma \left\langle \hat{S}_{E}\left( X^{\prime },\left( X^{\prime
}\right) ^{\prime }\right) \right\rangle \right) ^{2}}\right)  \notag
\end{eqnarray}%
\begin{equation*}
\bar{w}^{B}\left( X,X\right) =1-\left\langle \bar{w}\left( X^{\prime
},X\right) \right\rangle _{X^{\prime }}-\left\langle \hat{w}_{1}^{B}\left(
X^{\prime },X\right) \right\rangle
\end{equation*}%
and $\frac{\zeta ^{2}}{\xi ^{2}}$ $\bar{\zeta}^{2}$ gvn by (\ref{ZTR}), (\ref%
{ZTB}).%
\begin{equation}
\hat{w}_{2}^{B}\left( X^{\prime },X\right) =\frac{\left( 1-\left( \gamma
\left\langle \hat{S}_{E}\left( X_{1},X^{\prime }\right) \right\rangle
\right) ^{2}\right) \hat{w}_{2}^{\left( 0\right) }\left( X^{\prime
},X\right) \left( 1+\Delta \hat{r}\left( X^{\prime }\right) \right) }{1+\hat{%
w}_{2}^{\left( 0\right) }\left( X^{\prime },X\right) \left( 1-\left( \gamma
\left\langle \hat{S}_{E}\left( X_{1},X^{\prime }\right) \right\rangle
\right) ^{2}\right) +\left( \gamma \left\langle \hat{S}_{E}\left(
X_{1},X^{\prime }\right) \right\rangle _{X_{1}}\right) ^{2}-\left( \gamma
\left\langle \hat{S}_{E}\left( X_{1},X^{\prime }\right) \right\rangle
\right) ^{2}}
\end{equation}%
and%
\begin{equation}
w_{2}^{B}\left( X\right) =1-\hat{w}_{2}^{B}\left( X^{\prime },X\right)
\end{equation}%
givn by(\ref{sdb}), (\ref{sdt})

Moreover:%
\begin{equation*}
\frac{\hat{S}_{L}^{B}\left( X^{\prime },X\right) }{\kappa \left( 1-\bar{S}%
\left( X\right) \right) }=\hat{w}_{2}^{B}\left( X^{\prime },X\right) \left\{
1+\hat{w}_{2}^{B}\left( X^{\prime },X\right) \left( \hat{r}\left( X^{\prime
}\right) -\left\langle \hat{f}\left( X^{\prime }\right) \right\rangle _{\hat{%
w}_{1}}\right) +w_{2}^{B}\left( X\right) \left( \hat{r}\left( X^{\prime
}\right) -f\left( X\right) \right) \right\}
\end{equation*}%
\begin{equation*}
\frac{S_{L}^{B}\left( X,X\right) }{\kappa \left( 1-\bar{S}\left( X\right)
\right) }=w_{2}^{B}\left( X\right) \left[ 1+\hat{w}_{2}^{B}\left( X\right)
\left( r\left( X\right) -\left\langle \hat{r}\left( X^{\prime }\right)
\right\rangle _{\hat{w}_{2}}\right) \right]
\end{equation*}%
As for investors the formula for coefficients $\bar{w}^{B}$, $\hat{w}%
_{1}^{B} $,... show that thes parameter are mainly exogeneous. Consequently:%
\begin{equation*}
\delta \bar{S}_{E}\left( X^{\prime },X\right) =\frac{\bar{w}\left( X^{\prime
},X\right) }{2}\delta \bar{f}\left( X^{\prime }\right) \text{, }\delta \bar{S%
}_{L}\left( X^{\prime },X\right) =\frac{\bar{w}\left( X^{\prime },X\right) }{%
2}\delta \bar{r}\left( X^{\prime }\right)
\end{equation*}

\begin{equation*}
\delta \bar{S}\left( X^{\prime },X\right) =\bar{w}\left( X^{\prime
},X\right) \frac{\delta \bar{f}\left( X^{\prime }\right) +\delta \bar{r}%
\left( X^{\prime }\right) }{2}
\end{equation*}

\begin{equation*}
\delta \hat{S}_{E}^{B}\left( X^{\prime },X\right) =\frac{\hat{w}%
_{1}^{B}\left( X^{\prime },X\right) }{2}\delta \hat{f}\left( X^{\prime
}\right) \text{, }\frac{\delta \hat{S}_{L}^{B}\left( X^{\prime },X\right) }{%
\kappa \left( 1-\bar{S}\left( X\right) \right) }=\frac{\hat{w}_{2}^{B}\left(
X^{\prime },X\right) }{2}\delta \hat{r}\left( X^{\prime }\right)
\end{equation*}

\begin{equation}
\delta S_{E}^{B}\left( X,X\right) =w_{1}^{B}\left( X\right) \delta f\left(
X\right) \text{, }\frac{\delta S_{L}^{B}\left( X,X\right) }{\kappa \left( 1-%
\bar{S}\left( X\right) \right) }=w_{2}^{B}\left( X\right) \hat{w}%
_{2}^{B}\left( X\right) \delta r\left( X\right)
\end{equation}%
The dynamic equations for $\delta \bar{f}\left( X,\theta \right) $, $\delta 
\hat{f}\left( X,\theta \right) $, $\delta S^{T}\left( X,,\theta -1\right) $
can be replaced by a dynamic equation: 
\begin{equation*}
\left( 
\begin{array}{c}
\delta \bar{S}\left( X^{\prime },X,\theta \right) \\ 
\delta \hat{S}\left( X^{\prime },X,\theta \right) \\ 
\delta S^{T}\left( X,,\theta -1\right)%
\end{array}%
\right) =\left[ 
\begin{array}{ccc}
a & 0 & \frac{\bar{w}\left( X^{\prime },X\right) }{2}c \\ 
\frac{\hat{w}\left( X^{\prime },X\right) }{\bar{w}\left( X^{\prime
},X\right) }d & -1 & \frac{\hat{w}\left( X^{\prime },X\right) }{2}f \\ 
\frac{g}{\frac{\bar{w}\left( X^{\prime },X\right) }{2}} & 0 & i%
\end{array}%
\right] \left( 
\begin{array}{c}
\delta \bar{S}\left( X^{\prime },X,\theta -1\right) \\ 
\delta \hat{S}\left( X^{\prime },X,\theta -1\right) \\ 
\delta S^{T}\left( X,,\theta -2\right)%
\end{array}%
\right)
\end{equation*}%
with the same eigenvalues as the system for $\delta \bar{f}\left( X,\theta
\right) $, $\delta \hat{f}\left( X,\theta \right) $, $\delta S^{T}\left(
X,,\theta -1\right) $.

\subsection*{A12.8 Transitions induced by group connections}

As seen above, the banks tend to stablize flctuatns, when they behave as
lenders. When they behave as investrs, unstablt may arise. This is mainly
the case when $\bar{f}\left( X\right) -\bar{r}>>1$, $f\left( X\right)
=O\left( 1\right) $. In this case connexions between different grps may
induce transition:%
\begin{equation*}
\bar{f}\left( X\right) -\bar{r}=O\left( 1\right) ,f\left( X\right) =O\left(
1\right)
\end{equation*}%
towards instability:%
\begin{equation*}
\bar{f}\left( X\right) -\bar{r}>>1,f\left( X\right) =O\left( 1\right)
\end{equation*}%
which correspnds to connect to groups with high $\left\langle \bar{f}\left(
X\right) \right\rangle $ or high $\left\langle \hat{f}\left( X\right)
\right\rangle $, the bank behavng as investrs in this sectors with high
profitabilty.

\end{document}